\documentclass[rmp,aps,reprint,amsmath,amssymb,superscriptaddress,showpacs,nofootinbib]{revtex4-1}

\usepackage{bm}
\usepackage{graphicx}
\usepackage{color}
\usepackage{bbm}
\usepackage[utf8]{inputenc}
\usepackage{dcolumn}
\usepackage[colorlinks=true,citecolor=blue,urlcolor=blue,linkcolor=blue,hyperfigures=true]{hyperref}
\usepackage{textcomp}
\usepackage[american]{babel}
\usepackage{amssymb}
\usepackage{amsmath}
\usepackage{pstricks}
\usepackage{dsfont}
\usepackage{upgreek}
\usepackage{lmodern}
\usepackage{bookmark}

\def \Im {\mathop {\rm Im}}

\DeclareMathAlphabet{\pazocal}{OMS}{zplm}{m}{n}
\newcommand\Let{\mathrel{\mathop:\!\!=}}

\newcommand{\kv}{\ensuremath{\mathbf{k}}}

\newcommand{\QV}{\ensuremath{\mathbf{Q}}}

\newcommand{\qv}{\ensuremath{\mathbf{q}}}

\newcommand{\vc}[1]{\ensuremath{\mathbf{#1}}}

\newcommand{\av}[1]{\ensuremath{\left\langle #1 \right\rangle}}

\newcommand\varpm{\mathbin{\vcenter{\hbox{\oalign{\hfil$\scriptstyle+$\hfil\cr\noalign{\kern-.3ex}$\scriptscriptstyle({-})$\cr}}}}}
\newcommand\varmp{\mathbin{\vcenter{\hbox{\oalign{$\scriptstyle({+})$\cr\noalign{\kern-.3ex}\hfil$\scriptscriptstyle-$\hfil\cr}}}}}

\begin{document}

\title{Diagrammatic routes to nonlocal correlations\texorpdfstring{\\}{} beyond  dynamical mean field theory}

\author{G. Rohringer}
\affiliation{\mbox{Russian Quantum Center, 143025 Skolkovo, Russia}}
\affiliation{\mbox{Institute for Solid State Physics, TU Wien, 1040 Vienna, Austria}}

\author{H. Hafermann}
\affiliation{\mbox{Mathematical and Algorithmic Sciences Lab, Paris Research Center, Huawei Technologies}\\ \mbox{France SASU, 92100 Boulogne-Billancourt, France}}

\author{A. Toschi}
\affiliation{\mbox{Institute for Solid State Physics, TU Wien, 1040 Vienna, Austria}}

\author{A. A. Katanin}
\affiliation{\mbox{M. N.~Mikheev Institute of Metal Physics, Russian Academy of Sciences, 620108 Ekaterinburg, Russia}}

\author{A. E. Antipov}
\affiliation{\mbox{Station Q, Microsoft Research, Santa Barbara, California 93106-6105}}
\affiliation{\mbox{Department of Physics University of Michigan, Randall Laboratory,  Ann Arbor, Michigan 48109-1040}}

\author{M. I. Katsnelson}
\affiliation{\mbox{Radboud University Nijmegen, Institute for Molecules and Materials, NL-6525 AJ Nijmegen, The Netherlands}}
\affiliation{\mbox{Ural Federal University, 620002 Ekaterinburg, Russia}}

\author{A. I. Lichtenstein}
\affiliation{\mbox{I. Institut f\"ur Theoretische Physik, Universit\"at Hamburg, Jungiusstra\ss e 9, D-20355 Hamburg, Germany}}
\affiliation{\mbox{Ural Federal University, 620002 Ekaterinburg, Russia}}

\author{A. N. Rubtsov}
\affiliation{\mbox{Russian Quantum Center, 143025 Skolkovo, Russia}}
\affiliation{\mbox{Department of Physics, M.V. Lomonosov Moscow State University, 119991 Moscow, Russia}}

\author{K. Held}
\affiliation{\mbox{Institute for Solid State Physics, TU Wien, 1040 Vienna, Austria}}
\date{\today}

\pacs{
71.10.-w,
71.10.Fd,
71.27.+a
}

\begin{abstract}
Strong  electronic correlations pose one of the biggest challenges to solid state theory. Recently developed methods that address this problem by
starting with the local, eminently important correlations of dynamical mean field theory (DMFT) are reviewed. In addition, nonlocal correlations on all length scales are generated through Feynman diagrams, with  a local  two-particle vertex instead of the bare Coulomb interaction as a building block. With these diagrammatic extensions of DMFT long-range charge-, magnetic-, and superconducting fluctuations as well as (quantum) criticality can be addressed in strongly correlated electron systems. An overview is provided of the successes and results achieved mainly for model Hamiltonians and an outline is given of future prospects for realistic material calculations.
\end{abstract}

\maketitle
\tableofcontents{}

\section{Introduction}

The understanding of strongly correlated systems counts among the most difficult problems of solid state physics, since standard perturbation theory in terms of the bare Coulomb interaction breaks down. Dynamical mean field theory (DMFT) represents a breakthrough in this respect as it includes a major part of electronic correlations: the local ones. It does so in a nonperturbative way. For a three-dimensional ($3d$) lattice at elevated temperature and in the absence of a close-by second-order phase transition, the local correlations (as described by DMFT) prevail. They bring forth, among others, quasiparticle renormalizations, Mott-Hubbard metal-insulator transitions, orbital, charge and magnetic ordering, see~\onlinecite{Georges1996} for a review. Building on its success, DMFT nowadays is routinely employed for realistic material and non-equilibrium calculations, for reviews see \onlinecite{Held2006}, \onlinecite{Kotliar2006},  \onlinecite{Held2007}, \onlinecite{Katsnelson2008} and~\onlinecite{Aoki2014}, respectively. It also fostered the development of  new impurity solvers~\cite{Bulla2008,Gull2011}.

Non-local correlations, on the other hand, are at the heart of some of the most fascinating physical phenomena such as high-temperature superconductivity \cite{Bednorz1986} and quantum criticality \cite{Loehneysen2007}. They are also responsible for the long-range correlations in the vicinity of phase transitions or Lifshitz transitions \cite{Lifshitz1960} and play a crucial role in the physics of graphene~\cite{Kutov2012} to name but a few. These nonlocal correlations are missing in DMFT, which is mean field in space but takes into account correlations in time. Often we can understand nonlocal physics in terms of perturbation theory or the ladder replication thereof. 

\begin{figure}[tb]
  \includegraphics[width=0.85\columnwidth]{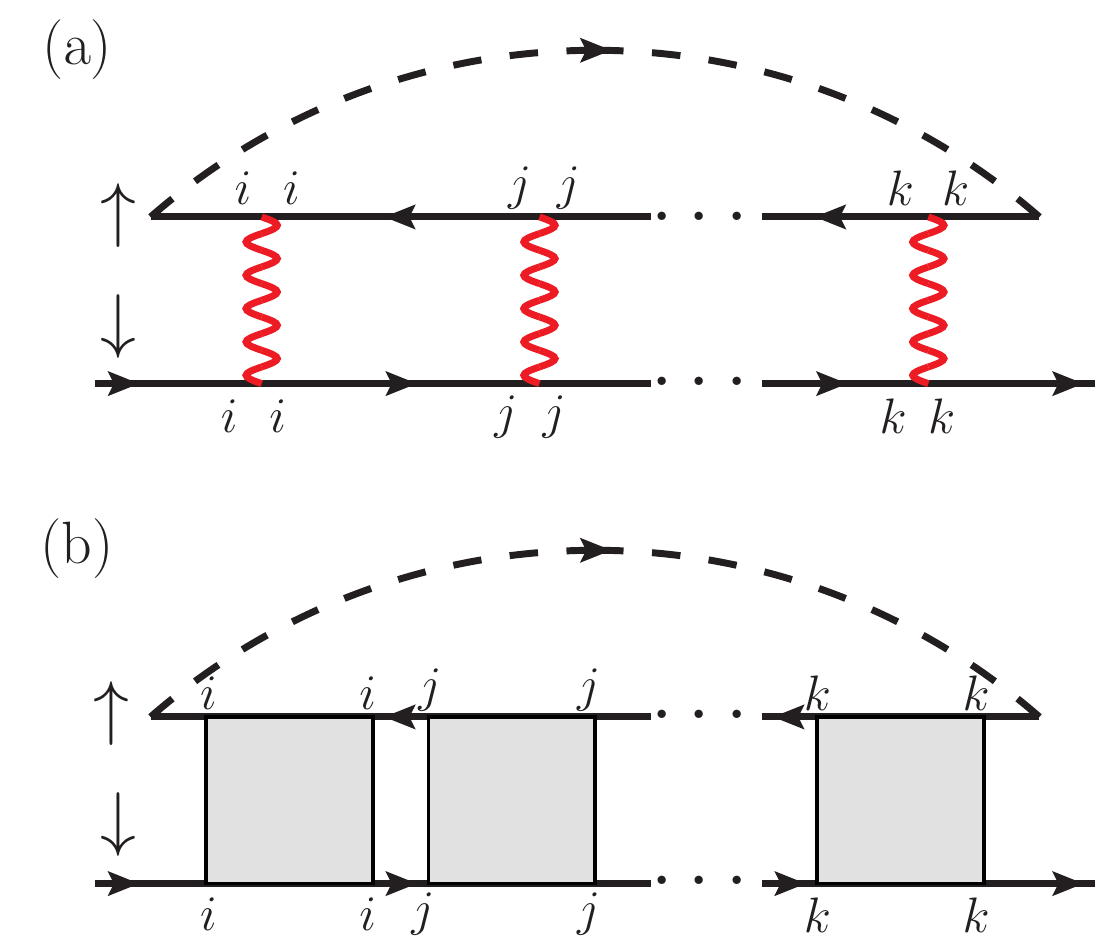}
    \caption{(Color online) (a) Ladder series of Feynman diagrams describing the repeated scattering of a minority (down) spin at the majority (up) spins. 	Wiggly line: (local) interaction at sites $i$, $j$, $k$ etc.; straight line: interacting Green's function. Closing the diagram by the dashed Green's function line yields the magnon  contribution to the self-energy. (b) In diagrammatic extensions of DMFT the same kind of diagrams are generated  with the non-perturbative local vertex (gray boxes) instead of the bare interaction as a building block. This local vertex  contains the bare interaction and all local vertex corrections.}
  \label{Fig:magnon}
\end{figure}

Let us take, as a specific and illustrative example, the elementary excitations of a ferromagnet: magnons. These can be described by the repeated scattering of a minority-spin electron at the prevalent majority-spin electrons; see Fig.~\ref{Fig:magnon} (a). This corresponds to ladder-type Feynman diagrams which allow us to calculate the magnetic susceptibility or to identify its spin wave poles as the collective (bosonic) excitations of the system: the magnons. As described by~\onlinecite{Hertz1973} one can diagrammatically ``close'' the Green's function in the majority-spin channel by adding the dashed Green's function line in  Fig.~\ref{Fig:magnon}, which yields the minority-spin self-energy. This self-energy  describes the scattering of electronic quasiparticles with the particle-hole excitations (magnons). \footnote{Such feedback of collective excitations on the fermionic degrees of freedom is crucially important for the nonquasiparticle states in the spin gap of half-metallic ferromagnets. These are an important limiting factor for  spintronics  applications; see~\onlinecite{Katsnelson2008} for a review.}

In DMFT such magnon contributions to the self-energy are contained only in their local version, where all sites in  Fig.~\ref{Fig:magnon} are the same,  $i=j=\ldots=k$. In $\mathbf k$ space, this translates into a  $\mathbf k$-independent contribution. Instead of a magnon dispersion relation $E(\mathbf k)$, in DMFT we have  a single magnon energy and a gap in the magnon spectrum. Consequently, the important physics of low-energy  long-range magnon fluctuations is {\sl not} captured correctly by  the DMFT self-energy.

The same kind of diagrams, if one also includes the SU(2)-related transversal spin fluctuations, describes the paramagnons in the paramagnetic phase~\cite{Moriya1985}. These are nothing but the spin fluctuations dominating in the vicinity of a magnetic phase transition. Their effect on the  spectrum and self-energy may be dramatic and may alter a metallic into a (pseudo) gapped phase. Such physics is missing in DMFT which does not feature any precursors of the incipient  magnetic ordering. The spin fluctuations may also serve as a pairing glue, an attractive interaction in the particle-particle or cooperon channel,  possibly leading to high-temperature superconductivity~\cite{Scalapino12}. Also at a quantum critical point the paramagnon contribution is important. Indeed, it is at the basis of the~\onlinecite{Hertz1976},~\onlinecite{Millis1993} and~\onlinecite{Moriya1985} theory of quantum criticality.

The aim of the diagrammatic extensions of DMFT is to describe the physics of long-ranged collective excitations, but beyond the weak-coupling ladder diagrams of  Fig.~\ref{Fig:magnon}(a) now also for strongly correlated systems. In fact, the key to such physics lies in Feynman diagrams such as those in  Fig.~\ref{Fig:magnon} (a), but with the bare interaction replaced by a strongly renormalized, local two-particle vertex, as illustrated in Fig.~\ref{Fig:magnon} (b). This way the important local  correlations can be fully included through the local  two-particle DMFT vertex from the beginning and, through this vertex, salso affect the short- and long-range correlations. As we will see, spin fluctuations and other nonlocal correlations such as the critical fluctuations in the vicinity of a (quantum) critical point can be described this way, even in strongly correlated systems.

\subsection{Brief History}
Let us start with a brief synopsis of the various methods and approaches  that aim at extending  DMFT to include nonlocal correlations. We recall that DMFT  becomes exact in the limit of high coordination number or alternatively for dimension $d\rightarrow \infty$~\cite{Metzner1989}. DMFT maps a lattice model onto the self-consistent solution of an Anderson impurity model (AIM) \cite{Georges1992a}, allowing for an essentially exact solution, e.g., by quantum Monte Carlo (QMC) simulations \cite{Jarrell1992}. 

From the very beginning, there have been attempts to include nonlocal correlations beyond the local ones of DMFT. The first such approach was
the $1/d$  approach of~\onlinecite{Schiller1995} which includes all diagrams  to next-to-leading order in $1/d$ and results
in a two-site impurity model. This way nonlocal correlations between neighboring sites are included. A systematic expansion of DMFT has also been proposed in the strong-coupling limit by~\onlinecite{Stanescu2004}, following the lines of~\onlinecite{Pairault1998}.

\begin{figure}[t]
  \includegraphics[scale=0.725,angle=0]{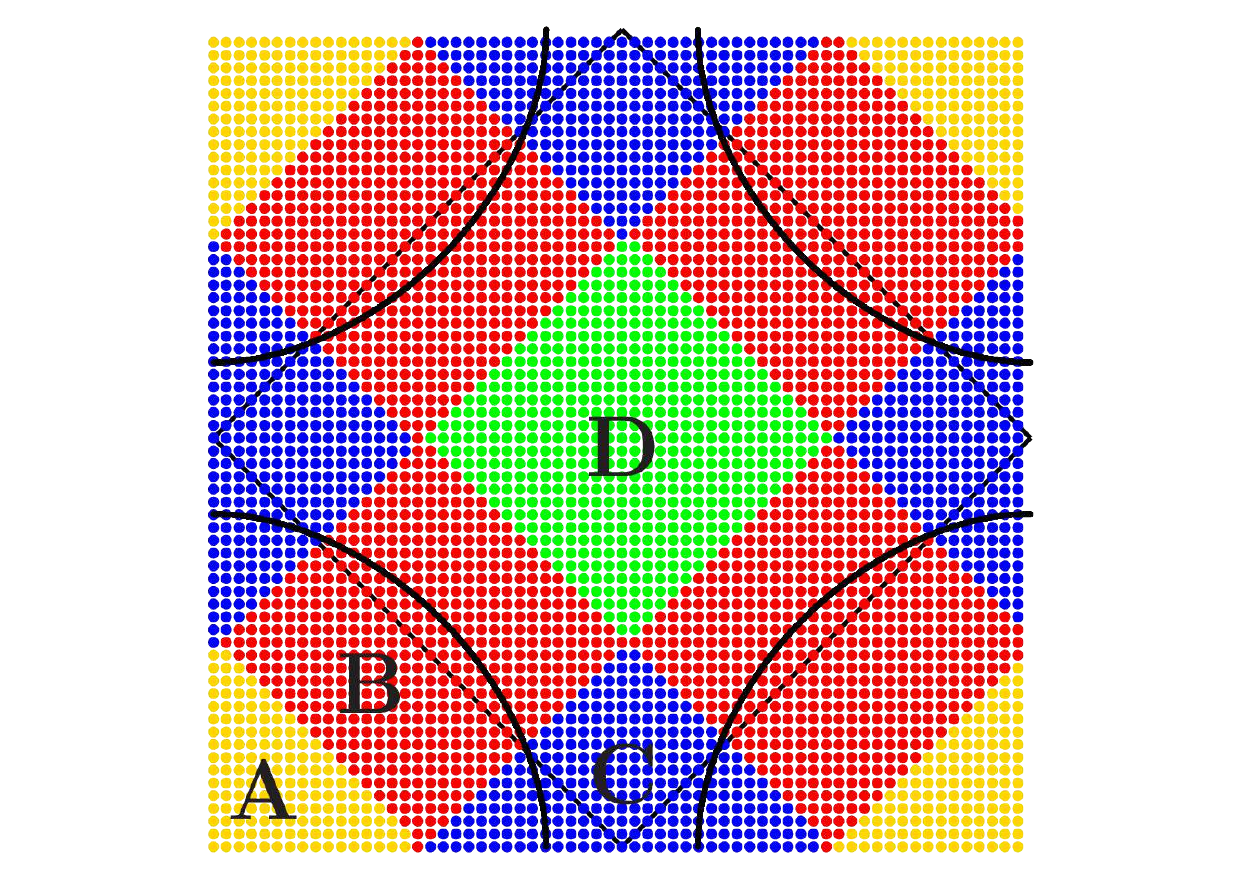} 
    \caption{(Color online) Typical momentum-space discretization for diagrammatic (individual dots) and cluster extensions of DMFT (colored/gray-scaled patches); Black lines show noninteracting Fermi surfaces for the square lattice with hopping parameters $t'/t=0$ and $t'/t=-0.3$. Diagrammatic extensions allow to resolve fine details along the Fermi surface.}
  \label{fig:dcapatches}
\end{figure}

Particularly important and widely employed are cluster extensions of DMFT: the dynamical cluster approximation (DCA) by~\onlinecite{Hettler1998} and the cellular DMFT (CDMFT) by~\onlinecite{Lichtenstein2000} and~\onlinecite{Kotliar2001}. These map a lattice model onto a cluster of sites embedded in a dynamical mean field. Thereby  nonlocal correlations within the cluster are accounted for, and those to the outside (described by a generalized DMFT bath) are neglected. Impressive successes of these approaches are  the description of pseudogap physics and unconventional superconductivity in the Hubbard model. Indeed, cluster extensions of DMFT became an integral part of  the theory of high-temperature superconductivity [for more recent results and larger clusters see~\onlinecite{Gull2013},~\onlinecite{Harland2016},~\onlinecite{Sordi2011} and~\onlinecite{Sakai12}]. A particular advantage of cluster extensions of DMFT is that they systematically allow for studying larger and larger clusters, providing a controlled way of approaching the exact result (infinite cluster limit) with the cluster size as a control parameter. In practice, numerical limitations due to the exponential growth of the cluster Hilbert space restrict the cluster extensions however to relatively small clusters of about $10\times 10$ sites. While correlations are included nonperturbatively, they remain  short-ranged even in two dimensions (2D) and for a single orbital. Cluster extensions have been reviewed by \onlinecite{Maier2005a}. In this review we focus instead on the complementary, diagrammatic extensions of DMFT. In these approaches, corrections to the DMFT self-energy are computed through Feynman diagrams, which  allows one to reach significantly larger lattice sizes, as illustrated in  Fig.~\ref{fig:dcapatches}.

Motivated by identifying particularly important contributions missing in DMFT, the first diagrammatic extensions supplemented the local DMFT self-energy by the nonlocal one of another approach. For example, in the $GW$+DMFT approach~\cite{Sun02,Biermann2003}, this is the nonlocal  screened exchange. \onlinecite{Sadovskii2005} added spin fluctuations contained in the spin-fermion model and~\onlinecite{Kitatani2015} those of the fluctuation exchange approximation (FLEX).

Dynamical vertex approaches on the other hand generate both, local and nonlocal, electronic correlations from a common, underlying entity: the local but frequency-dependent (i.e., dynamical) two-particle vertex. This development started with the  dynamical vertex approximation (D$\Gamma $A), see~\onlinecite{Toschi2007} and the closely related work by~\onlinecite{Kusunose2006}. D$\Gamma $A assumes  the locality of the $n$-particle irreducible vertex, recovering DMFT for $n=1$ and generating a nonlocal self-energy and susceptibility corrections for $n=2$. One can view this as a resummation of Feynman diagrams not in terms of orders in the interaction, but in terms of the locality of diagrams -- an approach which reestablishes the exact solution for $n\rightarrow \infty$. In an independent development,~\onlinecite{Rubtsov2008} introduced the dual fermion (DF) approach in which the lattice problem is  expressed in terms of a local reference system and a coupling to the nonlocal degrees of freedom. A perturbation theory around this solvable reference system is obtained by decoupling the impurity by means of dual fields through a Hubbard-Stratonovich transformation. The dual fermions interact through the $n\geq 2$-particle vertex functions of the local reference system. In practice the three-particle and all higher-order vertices are neglected in both D$\Gamma$A and DF, except for error estimates. \onlinecite{Slezak2009} devised a multiscale approach where short-range correlations are treated on a DCA cluster and long-range correlations diagrammatically. These groundbreaking works have laid the foundation for further generalizations and developments of the methods and various applications, of which we provide a brief overview in the following.

The one-particle irreducible (1PI) approach by~\onlinecite{Rohringer2013} is based on a functional in terms of the one-particle irreducible vertex; it inherits properties of both D$\Gamma$A and DF. The dynamical mean field theory to the functional renormalization group (DMF$^2$RG) approach by \onlinecite{Taranto2014} exploits the functional renormalization group (fRG) to generate the nonlocal diagrams beyond DMFT.  The  triply irreducible local expansion (TRILEX) of~\onlinecite{Ayral2015} is based on the three-point fermion-boson vertex. The nonlocal expansion scheme of~\onlinecite{Li2015} is a framework for expanding around a local reference problem which includes DF and the cumulant expansion as  special cases.

Extensions to nonequilibrium~\cite{Munoz2013} and real-space formulations~\cite{Valli2010,Takemori2016} are also possible. All of these approaches are closely related and rely on the same concept of tak\-ing the local vertex and generating nonlocal interactions from it as illustrated in Fig.~\ref{Fig:magnon} (b). They differ in the building blocks of the new perturbation expansion, in particular, the vertex (e.g., irreducible or full), the type of diagrams generated (e.g., ladder or parquet) and the details of the self-consistency schemes; cf.\  Table~\ref{Table:alg} on p.~\pageref{Table:alg} for an overview. They allow us to describe the same kind of physics contained in weak-coupling ladder diagrams [Fig.~\ref{Fig:magnon} (a)], but now  strong DMFT correlations are included through the vertex [Fig.~\ref{Fig:magnon} (b)].

In a complementary development,~\onlinecite{Si1996} and~\onlinecite{Chitra00} devised the extended DMFT (EDMFT), which describes the local correlations induced by nonlocal interactions, which can actually be mapped onto local bosonic degrees of freedom. The dual boson (DB) approach of~\onlinecite{Rubtsov12} also addresses nonlocal interactions, but it treats, in the spirit of the DF approach,  single- and two-particle excitations on the same footing. DB explicitly includes long-range bosonic modes and hence goes much beyond EDMFT. In D$\Gamma$A the nonlocal interaction can also be taken into account, in the form of a bare nonlocal vertex which allows for realistic \emph{ab initio} D$\Gamma$A material calculations~\cite{Toschi2011,Galler2016}. This naturally includes $GW$, DMFT, and nonlocal spin fluctuations. It is the aim of this review to provide in Sec.~\ref{Sec:Methods} a comprehensive overview of the different approaches as well as to draw a clear picture of the physics they can describe.

In the following we mention a few highlights and applications and refer the interested reader to Sec.~\ref{Sec:results} for a more detailed discussion. The physical results obtained using the diagrammatic extensions of DMFT are similar as for cluster extensions regarding short-range nonlocal correlations. However, the diagrammatic extensions also include long-range correlations, and hence allow us to address physical problems that were not accessible before. This is illustrated by Fig.~\ref{fig:dcapatches} which shows the typical momentum resolution in momentum space for cluster and diagrammatic extensions of DMFT. The improved momentum resolution allowed~\onlinecite{Rohringer2011} and~\onlinecite{Hirschmeier2015} to calculate the critical exponents of the antiferromagnetic (AF) phase transition in the three-dimensional (3D) Hubbard model in D$\Gamma$A and DF, respectively. Here the long-range correlations are of particular importance in the critical region close to a second-order phase transition. As one may expect from universality, these critical exponents are numerically compatible with those of the Heisenberg model. Similarly, the critical exponents of the Falicov-Kimball model as determined by~\onlinecite{Antipov2014} are of the Ising universality class. \onlinecite{Schaefer2016}  analyzed the quantum critical point in the Hubbard model which emerges when antiferromagnetism is suppressed by doping and find unusual critical exponents because of Kohn lines on the Fermi surface. The diagrammatic extensions of DMFT also show that spin fluctuations suppress the N\'eel temperature significantly in 3D~\cite{Katanin2009,Rohringer2011,Otsuki2014}. In 2D, the Mott-Hubbard transition can be significantly reshaped or even completely suppressed since the paramagnetic phase becomes always insulating at sufficiently low temperature in the unfrustrated case~\cite{Schaefer2015-2}. Pertinent steps have also been taken toward our understanding of high-temperature superconductivity: \onlinecite{Otsuki2014,Kitatani2015} studied superconducting instabilities and~\onlinecite{Gunnarsson2015} performed a diagnostics of the fluctuations responsible for the pseudogap. Further highlights are the renormalization of the plasmon dispersion by electronic correlations~\cite{vanLoon2014}, disorder-induced weak localization~\cite{Yang2014}, Lifshitz transitions in dipolar ultracold gases~\cite{vanLoon2016-2} and the flat band formation (Fermi condensation) near Van Hove filling~\cite{Yudin2014}.

\subsection{Outline}
\label{sec:outline}
This review is organized along the following lines: We first focus, in Sec.~\ref{sec:vertex}, on the two-particle vertex function as it is the building block of the diagrammatic approaches. In particular, Sec.~\ref{sec:formalism} sets the stage and introduces the notation used throughout the review. We define the various vertex functions, discuss their symmetries and introduce the Bethe-Salpeter and parquet equations. Section~\ref{sec:dmft}  briefly recapitulates the DMFT, which serves as the starting point for the diagrammatic extensions. In Sec.~\ref{sec:vertexphysics} we discuss the physical contents of the two-particle vertex and the origin of its asymptotic behavior for large frequencies. Finally, Sec.~\ref{sec:calver} summarizes the various methods for calculating the local two-particle vertex numerically from the AIM.

In Sec.~\ref{Sec:Methods} we review the various methods developed in recent years for calculating nonlocal correlations beyond DMFT. Most of these have a two-particle vertex as a starting point.  We start, in Sec.~\ref{sec:dga}, with the historically first vertex extension: the D$\Gamma$A approach. Its parquet and ladder variants are introduced in Secs.~\ref{sec:parquetDGA} and \ref{sec:dgaladder}, respectively. Extensions to nonlocal interactions and multiorbital models are discussed in Sec.~\ref{sec:abinitioDGA}, before turning to the closely related  functional integral formalism of the quadruply irreducible local expansion (QUADRILEX) in Sec.~\ref{sec:quadrilex}. In Sec.~\ref{sec:funcint} we present the DF approach, which performs a diagrammatic expansion around a local reference system in terms of dual fermions. We discuss in particular the DF diagrammatics, the choice of the local reference system, as well as scaling and convergence. The approach can be viewed as a particular diagrammatic resummation in the nonlocal expansion scheme discussed in Sec.~\ref{sec:nonlocalexp}. We also discuss  the related superperturbation theory in Sec.~\ref{sec:spert}. The 1PI approach can be considered as an intermediate approach in between D$\Gamma$A, which is based on the irreducible vertex, and DF, which is based on the reducible vertex. It inherits properties from both methods. The one-particle irreducible formalism is obtained through a Legendre transformation of the DF generating functional, as described in Sec.~\ref{subsec:1pi}. In Sec.~\ref{subsec:DMFT2fRG} we present a sophisticated alternative to generate nonlocal correlations and vertices with the DMFT vertex as a starting point: the fRG. As we discuss in Sec.~\ref{sec:cluster}, all these diagrammatic extensions can naturally be formulated using a cluster instead of a single DMFT site as a starting point. Section~\ref{subsec:DMFTplus} is devoted to diagrammatic extensions of DMFT that are based on a perturbation in the bare interaction instead of the two-particle vertex. These approaches supplement the DMFT self-energy with a nonlocal one. Diagrammatic extensions of EDMFT are finally discussed in Sec.~\ref{sec:edmftpp}: the EDMFT+$GW$ approach in Sec.~\ref{sec:edmftGW}, the dual boson approach in Sec.~\ref{sec:db}, and TRILEX in Sec.~\ref{sec:TRILEX}. A separate section, Sec.~\ref{sec:conservation}, is devoted to conservation laws  and crossing symmetry.

In Sec.~\ref{Sec:results} we review the main results achieved hitherto with diagrammatic extensions of DMFT. The application to the Hubbard model in three down to zero dimensions in Sec.~\ref{sec:hubbard} illustrates the physics these methods can describe and provides, at the same time, a unified picture for this fundamental model of electronic correlations. The application to the Kondo lattice model (KLM) in Sec.~\ref{sec:Results_HF} requires one to account for the interplay between local Kondo physics and long-range antiferromagnetic fluctuations and therefore is an ideal playground for diagrammatic extensions. Applications to models for annealed and quenched disorder, i.e., the Falicov-Kimball model in Sec.~\ref{sec:FKM} and the Anderson-Hubbard model in Sec.~\ref{sec:disorder}, illustrate the versatility of diagrammatic extensions. Finally, Sec.~\ref{sec:nonlocalv} discusses results for models and realistic material  calculations with nonlocal interactions and multiple orbitals.

In Sec.~\ref{Sec:opensource} we provide an overview of open source codes that are available for solving the  AIM, the computation of the two-particle vertex and for diagrammatic extensions of DMFT.
Finally, in Sec.~\ref{sec:conclusion} we close with a summary and outlook and with Table~\ref{Table:alg} providing a comparison of the various diagrammatic extensions.

\section{Diagrammatics at the two-particle level}
\label{sec:vertex}

\subsection{Formalism and symmetries}
\label{sec:formalism}

In the following we provide a concise overview of the two-particle formalism. For further details and derivations we refer the reader to  \onlinecite{Rohringer2012}.

The starting point for deriving the Feynman diagrammatic formalism at the one- and two-particle level is the general definition of the $n$-particle imaginary time Green's function:
\begin{align}
 \label{equ:defngreengeneral}
 {G}^{(n)}_{i_1\ldots i_{2n}}&(\tau_1,\ldots,\tau_{2n})=\\&(-1)^n\left\langle \mbox{T}_\tau \left[c^{\phantom{\dagger}}_{i_1}(\tau_1)c^{\dagger}_{i_2}(\tau_2)c^{\phantom{\dagger}}_{i_3}(\tau_3) \dots c^{\dagger}_{i_{2n}}(\tau_{2n})\right]\right\rangle,\nonumber
\end{align}
where even indices correspond to creation ($c^{\dagger}$) and odd indices to annihilation operators ($c$). Here $\langle\ldots\rangle=\mbox{Tr}(e^{-\beta\pazocal{H}}\ldots\rangle/Z$ denotes the thermal average with $Z=\mbox{Tr}(e^{-\beta\pazocal{H}})$ being the partition function for Hamiltonian $\pazocal{H}$, $\beta=1/T$ is the inverse temperature, and $T_\tau$ denotes the time ordering operator. The indices $i_j\widehat{=}(\mathbf{r}_j/\mathbf{k}_j,l_j,\sigma_j,\ldots)$ encode the set of all degrees of freedom of the system, e.g., space coordinate (lattice site)/momentum, orbital, spin, etc. In the following we will consider mostly  single-orbital systems.

\begin{figure}
  \centering
  \includegraphics[width=0.85\columnwidth]{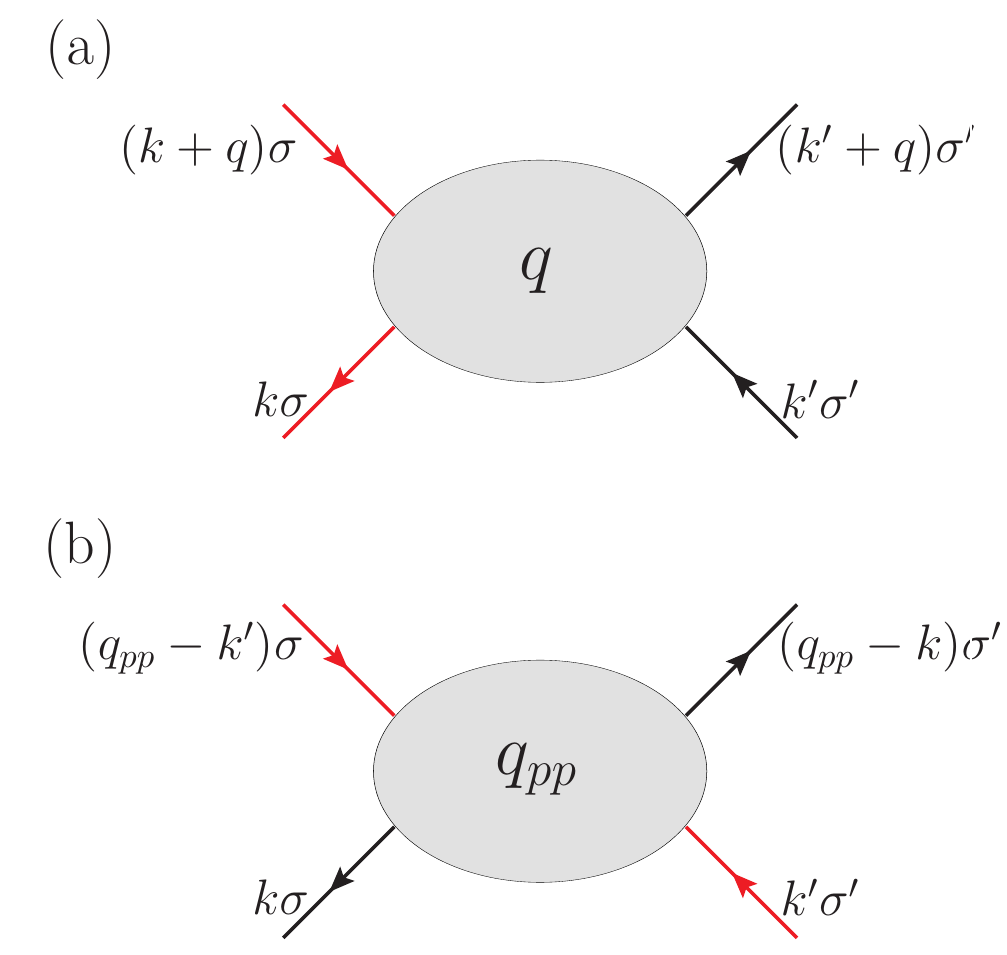}
    \caption{(Color online) Illustration of a two-particle Green's function from the (a) particle-hole and (b) particle-particle perspectives, described by the corresponding frequency notations. The red (gray) arrows denote the  particle and hole in (a) and the two  particles  in  (b), which are considered as the ``incoming'' ones. The total frequency/momentum transferred in the scattering process is then given by the generalized bosonic momentum $q=(\mathbf{q},\omega)$ and $q_{pp}=(\mathbf{q}_{pp},\omega_{pp})$, respectively. Representation (b)  can be obtained from (a) by a mere coordinate transformation in the space of the three frequencies/momenta, i.e., $q\rightarrow q_{pp}= q+k+k'$ ($k_{pp}=k$ and $k'_{pp}=k'$). }
  \label{fig:ph_pp_scattering}
\end{figure}

From the general case, the usual one-particle Green's function in momentum space is derived as
\begin{align}
 \label{equ:def1pgf}
 &G_{\mathbf{k}}(\tau)={G}^{(1)}_{\mathbf{k}\sigma,\mathbf{k}\sigma}(\tau,0) = - \left\langle \mbox{T}_\tau \left[c^{\phantom{\dagger}}_{\mathbf{k}\sigma}(\tau)c^\dagger_{\mathbf{k}\sigma}(0)\right]\right\rangle\nonumber \; ,\\
 &G_{\mathbf{k}\nu}=\int_0^{\beta}d\tau\;e^{i\nu\tau}G_{\mathbf{k}}(\tau),
\end{align}
where $\nu=(2n+1)\pi/\beta$ with $n\in\mathds{Z}$ is a fermionic Matsubara frequency [later $\omega=(2m)\pi/\beta$ denotes a bosonic Matsubara frequency]. Whenever convenient, we adopt the more compact four-vector notation $ G_k\equiv G_{\mathbf{k}\nu}$ with the generalized fermionic $k\widehat{=}(\nu,\mathbf{k})$ and bosonic  momentum $q\widehat{=}(\omega,\mathbf{q})$. For conciseness, we restrict ourselves here and in the following to the time- and lattice-translationally invariant, SU(2)-symmetric (paramagnetic) case. Consequently, the one-particle Green's function is diagonal in generalized momentum and spin space with $G_{\mathbf{k}\nu\uparrow\uparrow}(\tau)= G_{\mathbf{k}\nu\downarrow\downarrow}(\tau)= G_{\mathbf{k}\nu}(\tau)$. From $G_{\mathbf{k}\nu}$ and its noninteracting counterpart $G_{0,\mathbf{k}\nu}$, the {\sl one-particle irreducible} self-energy is calculated via the standard Dyson equation 
\begin{equation}
\Sigma_{\mathbf{k}\nu}\!=[G_{0,\mathbf{k}\nu}]^{-1}\!-\![G_{\mathbf{k}\nu}]^{-1} .
\label{Eq:Dyson}
\end{equation}
For the two-particle Green's function [$n\!=\!2$ in Eq.~\eqref{equ:defngreengeneral}] we can drop one momentum and time index due to time and lattice translational invariance and arrive at the compact form
\begin{align}
 \label{equ:def2pgf}
 &G^{(2)}_{\sigma\sigma',\mathbf{k}\mathbf{k'}\mathbf{q}}(\tau_1,\tau_2,\tau_3)\equiv G^{(2)}_{\mathbf{k}\sigma,\mathbf{k}+\mathbf{q}\sigma,\mathbf{k'}+\mathbf{q}\sigma',\mathbf{k'}\sigma'}(\tau_1,\tau_2,\tau_3,0)\nonumber \; ,\\
&G_{\sigma\sigma',\mathbf{k}\mathbf{k'}\mathbf{q}}^{(2),\nu\nu'\omega}=\int_0^{\beta}d\tau_1d\tau_2d\tau_3\;e^{i\nu\tau_1}e^{-i(\nu+\omega)\tau_2}e^{i(\nu'+\omega)\tau_3}\nonumber\\&\hspace{4cm}\times G^{(2)}_{\sigma\sigma',\mathbf{k}\mathbf{k'}\mathbf{q}}(\tau_1,\tau_2,\tau_3).
\end{align}
The way the frequencies are assigned to the Matsubara times and, hence, to the creation and annihilation operators in Eq.~\eqref{equ:defngreengeneral} is referred to as particle-hole notation. In this notation the two-particle Green's function can be viewed as the scattering amplitude of an incoming particle and  hole with  total energy $\omega$ and total momentum $\mathbf{q}$; see the red (gray) lines in Fig.~\ref{fig:ph_pp_scattering}(a). It is particularly convenient for describing systems where particle-hole (e.g., spin or charge) fluctuations dominate. Systems with strong particle-particle fluctuations, on the other hand, are more easily described exploiting the so-called particle-particle representation of the two-particle Green's function that is illustrated in Fig.~\ref{fig:ph_pp_scattering}(b). In this notation the two-particle Green's function can be interpreted as scattering amplitude between two particles with total energy and momentum $q_{pp}=q+k+k'$. Let us stress that the two-particle Green's function contains both ($ph$ and $pp$) scattering processes independent of its representation. The choice of the representation corresponds only to selecting the most convenient ``coordinate system'' for the description of the problem (see, e.g., \onlinecite{Gunnarsson2015} and \onlinecite{Bickers2004}).

The two-particle Green's function depends on four spin indices corresponding to $2^4=16$ spin components. Because of the conservation of the total spin, $10$ of them vanish and, from the remaining $6$, the two components $\sigma(-\sigma)\sigma(-\sigma)$ can be expressed via $\sigma\sigma(-\sigma)(-\sigma)$ by means of the crossing symmetry (see the last line in Table~\ref{tab:symmetries}; it originates from the fact that we have the same Feynman diagrams when exchanging the two incoming lines  in Fig.~\ref{fig:ph_pp_scattering}). For the remaining four components $\sigma\sigma\sigma'\sigma'$ we introduced the shorthand notation $\sigma\sigma'$ in Eq.~\eqref{equ:def2pgf}. There are additional relations between these due to  SU(2) symmetry (see the second line in Table~\ref{tab:symmetries}). However, as these relations involve shifts of frequency and momenta, it is more convenient to work with two ($\uparrow\uparrow$ and $\uparrow\downarrow$) components explicitly.

From the one- and two-particle Green's functions, the generalized  susceptibilities are readily obtained as
\begin{align} 
 \label{equ:defgensusc} 
 &\chi_{\sigma\sigma',\mathbf{k}\mathbf{k'}\mathbf{q}}^{\nu\nu'\omega}=G_{\sigma\sigma',\mathbf{k}\mathbf{k'}\mathbf{q}}^{(2),\nu\nu'\omega}-\beta G_{\mathbf{k}\nu}G_{\mathbf{k'}\nu'}\delta_{\omega 0}\delta_{\mathbf{q}\mathbf{0}}\nonumber\\
 &\chi_{c/s,\mathbf{k}\mathbf{k'}\mathbf{q}}^{\nu\nu'\omega}=\chi_{\uparrow\uparrow,\mathbf{k}\mathbf{k'}\mathbf{q}}^{\nu\nu'\omega}\pm\chi_{\uparrow\downarrow,\mathbf{k}\mathbf{k'}\mathbf{q}}^{\nu\nu'\omega}. 
\end{align}
In the second line we introduced the charge ($c$) and spin ($s$)  components of the generalized susceptibility.\footnote{These components have a definite spin $S$ and projection $S_z$ of the incoming particle-hole pair: The charge channel corresponds to $S=0$, $S_{z}=0$, and the spin channel to $S=1$, $S_{z}=0$. The components with $\uparrow\downarrow\downarrow\uparrow$ and $\downarrow\uparrow\uparrow\downarrow$ correspond to $S=1$, $S_{z}=\pm 1$, and must be equal to $S=1$, $S_{z}=0$ due to SU(2). It is hence convenient to work with the two components  ($c$/$s$) only. A similar decomposition into singlet and triplet channels applies for the particle-particle channel.} From these  the corresponding physical susceptibilities (response functions) are computed in the particle-hole sector by performing the summation over all the fermionic variables: 
\begin{align} 
 \label{equ:defphyssusc} 
 \chi_{r,\mathbf{q}}^{\omega}=\sum_{\underset{\mathbf{k}\mathbf{k'}}{\nu\nu'}}\chi_{r,\mathbf{k}\mathbf{k'}\mathbf{q}}^{\nu\nu'\omega} &  & \mbox{with} \; \; r = c,s,
\end{align}
where a proper normalization of the momentum and frequency sums is implicitly assumed [i.e., $\sum_{\mathbf{k}} 1= 1$ and $\sum_{\nu}\widehat{=}\frac{1}{\beta}\sum_\nu$]. An analogous definition holds for the physical particle-particle susceptibility where the corresponding summations have to be performed in particle-particle notation.

\begin{figure} 
  \centering 
  \includegraphics[width=0.50\textwidth]{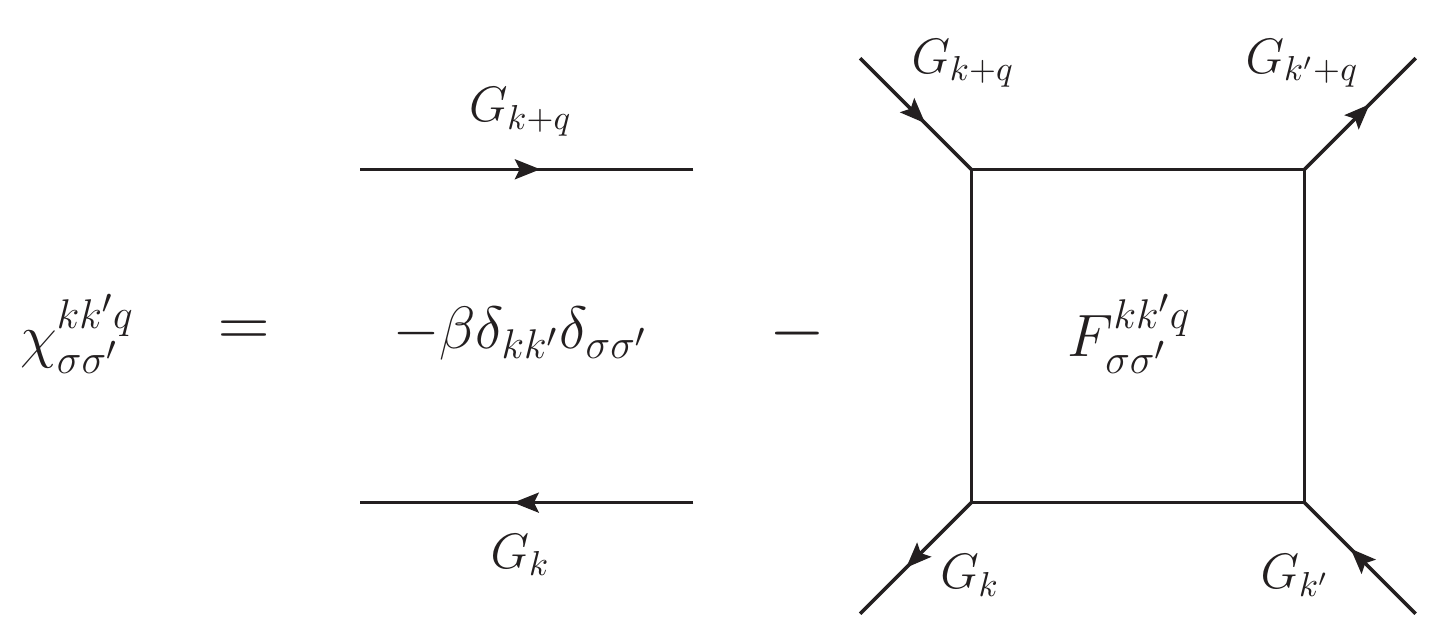} 
    \caption{Decomposition of the generalized susceptibility into a disconnected and a connected part. The first part describes the independent propagation of the particle-hole pair in the interacting system, while the second originates from all possible scattering processes between them. For readability of the diagram we have adopted the four-vector notation.}
  \label{fig:decomp1PIF} 
\end{figure} 

In order to classify the different two-particle processes diagrammatically, we can decompose the generalized susceptibility into two parts (see  Fig.~\ref{fig:decomp1PIF}): (i) a product of two one-particle Green's functions corresponding to an independent propagation of the particle and the hole and (ii) vertex corrections to the susceptibility. The latter  describes all the particle-hole scattering processes, which give rise to collective excitations. The corresponding equation, depicted diagrammatically in Fig.~\ref{fig:decomp1PIF}, reads
\begin{align} 
 \label{equ:defvertex} 
 \chi_{r,\mathbf{k}\mathbf{k'}\mathbf{q}}^{\nu\nu'\omega}&=\underset{\chi_{0,\mathbf{k}\mathbf{k'}\mathbf{q}}^{\nu\nu'\omega}}{\underbrace{-\beta G_{\mathbf{k}\nu}G_{(\mathbf{k}+\mathbf{q})(\nu+\omega)}\delta_{\nu\nu'}\delta_{\mathbf{k}\mathbf{k'}}}}\nonumber\\- G_{\mathbf{k}\nu}&G_{(\mathbf{k}+\mathbf{q})(\nu+\omega)}F_{r,\mathbf{k}\mathbf{k'}\mathbf{q}}^{\nu\nu'\omega}G_{\mathbf{k'}\nu'}G_{(\mathbf{k'}+\mathbf{q})(\nu'+\omega)},
\end{align}
with $r=c,s$ and the signs have been chosen in such a way that $F_{\uparrow\downarrow,\mathbf{k}\mathbf{k'}\mathbf{q}}^{\nu\nu'\omega}\rightarrow +U$ when the local interaction $U\rightarrow 0$. $F_{r,\mathbf{k}\mathbf{k'}\mathbf{q}}^{\nu\nu'\omega}$ is the two-particle vertex function, which  contains all Feynman diagrams connecting all four external Green's functions. In the Fermi-liquid regime, $F$ is proportional to the scattering amplitude between quasiparticles~\cite{Abrikosov1975}.
\begin{figure*}[t]
  \centering 
  \includegraphics[width=\textwidth]{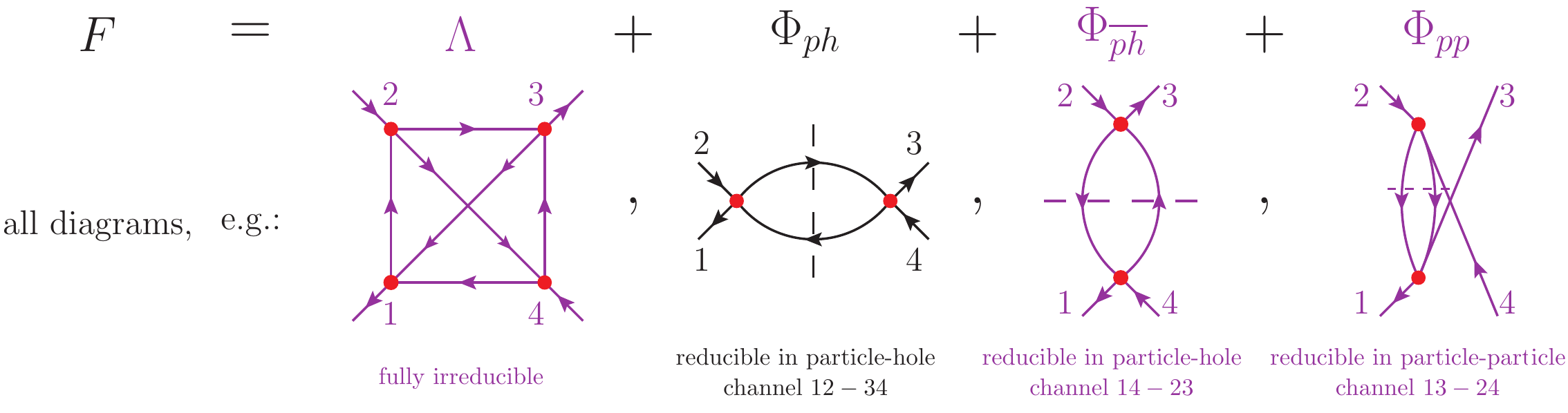} 
    \caption {(Color online) Parquet decomposition of the one-particle irreducible vertex $F$ into its two-particle fully irreducible contribution $\Lambda$ and the three  contributions $\Phi_{\ell}$ reducible in the particle-hole ($ph$), vertical particle-hole ($\overline{ph}$), and particle-particle channels ($pp$). The latter can be separated into two parts by cutting two Green's functions as indicated by the dashed lines. For instance, for the  $\ell=ph$ channel, the legs  12 and 34 are separated. The subsets of diagrams marked in violet (light gray) are part of the irreducible $ph$ vertex $\Gamma_{\ell=ph}$ which contains all diagrams that cannot be separated in channel $\ell=ph$. Note that all diagrams in this figure are meant as so-called {\sl skeleton} diagrams, i.e., all lines correspond to full interacting Green's functions, except for the external legs that mark only the incoming/outgoing generalized momenta. Red dots denote the bare (Hubbard) interaction $U$.}
  \label{fig:decomp2PILambda} 
\end{figure*} 

A refined classification is obtained by categorizing the Feynman diagrams of $F$ in terms of their {\sl two-particle reducibility}. All Feynman diagrams contributing to $F$ can be split into four topologically distinct classes. They are either fully two-particle irreducible or reducible in one of three channels: particle-hole ($ph$), vertical particle-hole ($\overline{ph}$), or particle-particle ($pp$). For example, a diagram is said to be reducible in the particle-hole channel when it can be split into two parts by cutting two lines corresponding to a particle-hole pair; see Fig.~\ref{fig:decomp2PILambda}. This decomposition is at the heart of the so-called {\sl parquet} equations which were first introduced by~\onlinecite{Diatlov1957} [cf.\ \onlinecite{DeDominicis1962,DeDominicis1964,Janis2001,Janis2017,Bickers1991,Bickers2004}]. Denoting by $\Phi_{\ell,r,\mathbf{k}\mathbf{k}'\mathbf{q},\uparrow\downarrow}^{\nu\nu'\omega}$ the set of diagrams which are two-particle reducible (2PR) in channel $\ell$ and by $\Lambda_{r,\mathbf{k}\mathbf{k}'\mathbf{q}}^{\nu\nu'\omega}$ the set of all fully irreducible diagrams, we have the unique decomposition (cf.\ Fig.~\ref{fig:decomp2PILambda})
\begin{align}
 \label{equ:parquedecompF}
 F_{r,\mathbf{k}\mathbf{k}'\mathbf{q}}^{\nu\nu'\omega}=\Lambda_{r,\mathbf{k}\mathbf{k}'\mathbf{q}}^{\nu\nu'\omega}+\Phi_{ph,r,\mathbf{k}\mathbf{k}'\mathbf{q}}^{\nu\nu'\omega}+\Phi_{\overline{ph},r,\mathbf{k}\mathbf{k}'\mathbf{q}}^{\nu\nu'\omega}+&\Phi_{pp,r,\mathbf{k}\mathbf{k}'\mathbf{q}}^{\nu\nu'\omega}.
\end{align}
We stress, that one has to clearly discriminate between the index $\ell$ which refers to a subset of diagrams for the full vertex with a certain topology (reducible or irreducible in a given channel) and the index $r$ which represents just the spin arguments of the vertex [specifically the linear combination as in  Eq.~(\ref{equ:defgensusc}) so that without vertex corrections and a Hubbard interaction $U$: $F_d=U$, $F_s=-U$]. In the literature both,  $\ell$ and $r$, are often referred to as ``channels'' although these are completely different concepts. In fact, the decomposition \eqref{equ:parquedecompF} holds independently of the considered spin combination $r$.

Alternatively, the contributions to $F$ can be divided into only two parts, i.e., those which are reducible and those which are irreducible in a given channel $\ell$:
\begin{align}
\label{equ:decompF}
F_{r,\mathbf{k}\mathbf{k}'\mathbf{q}}^{\nu\nu'\omega} = \Gamma_{\ell,r,\mathbf{k}\mathbf{k}'\mathbf{q}}^{\nu\nu\omega} + \Phi_{\ell,r,\mathbf{k}\mathbf{k}'\mathbf{q}}^{\nu\nu'\omega}.
\end{align}
This defines the vertices $\Gamma_{\ell,r,\mathbf{k}\mathbf{k'}\mathbf{q}}^{\nu\nu'\omega}$ which are two-particle irreducible in channel $\ell$ (see Fig. \ref{fig:decomp2PILambda} for $\ell=ph$). They are related to the full vertex $F$ through the {\sl Bethe-Salpeter} equations (BSEs)\footnote{The BSEs can be equivalently formulated for the generalized susceptibilities:
\begin{align*} 
 \chi_{r,\mathbf{k}\mathbf{k'}\mathbf{q}}^{\nu\nu'\omega}=\chi_{0,\mathbf{k}\mathbf{k'}\mathbf{q}}^{\nu\nu'\omega}-\sum_{\substack{\nu_1\nu_2\\\mathbf{k_1}\mathbf{k_2}}}\chi_{0,\mathbf{k}\mathbf{k_1}\mathbf{q}}^{\nu\nu_1\omega}\Gamma_{\ell,r,\mathbf{k_1}\mathbf{k_2}\mathbf{q}}^{\nu_1\nu_2\omega}\chi_{r,\mathbf{k_2}\mathbf{k'}\mathbf{q}}^{\nu_2\nu'\omega},
\end{align*}
where the bare bubble $\chi_{0,\mathbf{k}\mathbf{k'}\mathbf{q}}^{\nu\nu'\omega}$ has been defined in Eq.~(\ref{equ:defvertex}).}. For the $\ell=ph$ channel \cite{Bickers1991,Bickers2004}, the BSE explicitly reads
\begin{align} 
 \label{equ:decomp2PIGamma} 
  F_{r,\mathbf{k}\mathbf{k'}\mathbf{q}}^{\nu\nu'\omega}=\Gamma_{ph,r,\mathbf{k}\mathbf{k'}\mathbf{q}}^{\nu\nu'\omega}&+\sum_{\mathbf{k_1}\nu_1}\Gamma_{ph,r,\mathbf{k}\mathbf{k_1}\mathbf{q}}^{\nu\nu_1\omega}\\&\times G_{\mathbf{k_1}\nu_1}G_{(\mathbf{k_1}+\mathbf{q})(\nu_1+\omega)}F_{r,\mathbf{k_1}\mathbf{k'}\mathbf{q}}^{\nu_1\nu'\omega}.\nonumber
\end{align}
Note that due to SU(2) symmetry, the charge ($r\!=\!c$) and the spin ($r\!=\!s$) sectors do not couple. From a diagrammatic perspective the BSEs correspond to an infinite summation of ladder diagrams. Physically, they describe collective excitations in the different scattering channels  while the parquet equation~\eqref{equ:parquedecompF} provides for their mutual renormalization.

\begin{figure}[tb]
  \centering
  \includegraphics[width=\columnwidth]{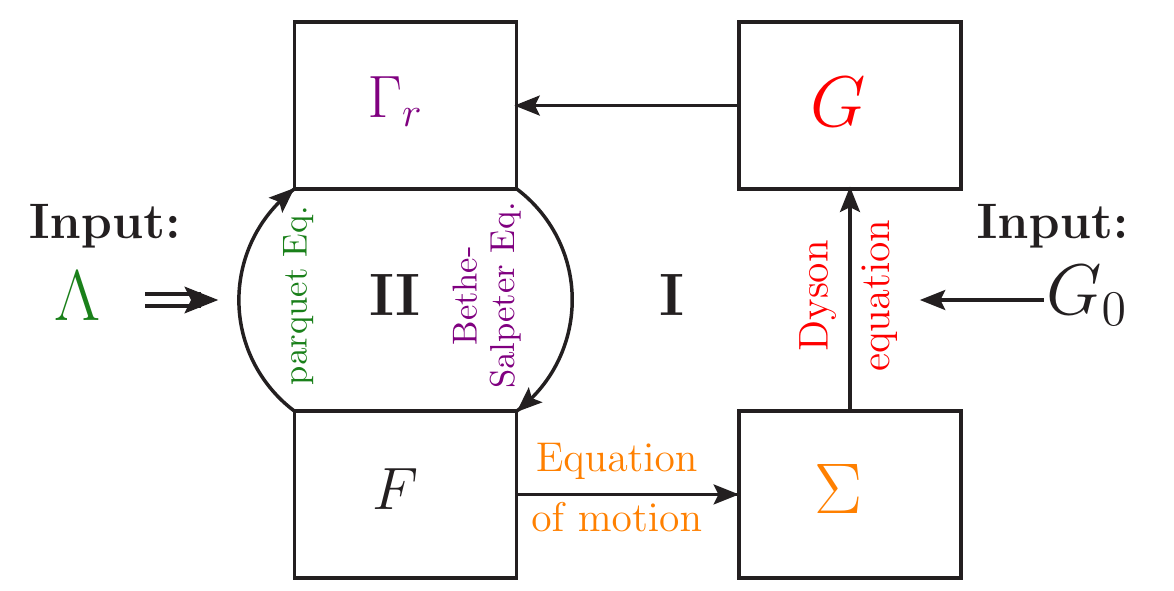}
    \caption{(Color online) Flow diagram for solving the parquet equations. Left: If the fully irreducible vertex $\Lambda$ is given, the parquet equation~(\ref{equ:parquedecompF}) and the three BSEs~(\ref{equ:decomp2PIGamma}) for $\ell=ph,\overline{ph},pp$ allow us to calculate the four unknowns $F$, $\Gamma_\ell$. Right: As in the BSE~(\ref{equ:decomp2PIGamma}) also the interacting Green's function $G$ enters, we need to extend the self-consistency loop by two additional unknowns ($G$ and $\Sigma$) and equations [the equation of motion~(\ref{equ:equofmotion})  and the Dyson equation~(\ref{Eq:Dyson})]. The latter has the noninteracting Green's function $G_0$ as input.}
  \label{fig:parquet}
\end{figure}

Equations.~\eqref{equ:parquedecompF}-\eqref{equ:decomp2PIGamma} form a closed set of four equations for $F$, $\Gamma_\ell$ ($\ell=pp,ph,\overline{ph}$) and $\Lambda$, which can be solved self-consistently, provided one of these five quantities and the one-particle Green's function are known (for the case in which $\Lambda$ is given, see the left part of Fig.~\ref{fig:parquet}). As we usually do not know the exact vertex, we have to consider approximations. For instance, the so-called parquet approximation assumes that the fully irreducible vertex is replaced by the constant bare interaction, i.e., $\Lambda^{kk'q}=U$ \cite{Bickers2004}; or in parquet D$\Gamma$A, $\Lambda^{kk'q}$ is approximated by its local counterpart ($\Lambda^{\nu\nu'\omega}$). 

The above set of four parquet equations corresponds to loop {\sc II} in  Fig.~\ref{fig:parquet} and needs to be supplemented by the self-consistent calculation of the one-particle Green's function and self-energy (loop {\sc I} in Fig.~\ref{fig:parquet}).  For obtaining these one-particle functions from the two-particle  vertex, we exploit the relation between Green's functions of different particle number in  the (Heisenberg) equation of motion. This leads to the Schwinger-Dyson equation, which connects the vertex $F$ with the self-energy $\Sigma$ and reads for a Hubbard-like model with a local interaction $U$ [cf.\  Hamiltonian (\ref{Eq:Hubbard}) below]:
\begin{align}
\label{equ:equofmotion}
 \Sigma_{\mathbf{k}\nu}=\frac{Un}{2}-{U}\sum_{\substack{\nu'\omega\\\mathbf{k'}\mathbf{q}}}F_{\uparrow\downarrow,\mathbf{kk'q}}^{\nu\nu'\omega}&G_{\mathbf{k'}\nu'}G_{(\mathbf{k'}+\mathbf{q})(\nu'+\omega)}\nonumber\\[-0.5cm]&\hspace{0.3cm}\times G_{(\mathbf{k}+\mathbf{q})(\nu+\omega)}. 
\end{align}
Here $n$ denotes the particle density of the system. For the generalization to multiple orbitals and nonlocal interactions, see, for example,~\onlinecite{Galler2016}. Equation~\eqref{equ:equofmotion} represents an exact relation between the two- and one-particle correlation functions.  Hence for a given $\Lambda$ we have altogether five equations and five unknowns which can be calculated self-consistently as indicated in Fig.~\ref{fig:parquet}.

In diagrammatic extensions of DMFT discussed in Sec.~\ref{Sec:Methods}, the Schwinger-Dyson  equation~(\ref{equ:equofmotion}) is also often used when obtaining $F$  via other (e.g., ladder) resummations of Feynman diagrams. The Schwinger-Dyson equation also provides the basis for the fluctuation diagnostics method. By performing partial summations over $\mathbf{k'}$ and $\nu'$ in Eq.~\eqref{equ:equofmotion}, the physical origin of the spectral features in the self-energy can be identified~\cite{Gunnarsson2015}.

The dependence of two-particle Green's- and vertex functions on several indices makes their numerical calculations, postprocessing, and storage evidently much more challenging than that of the single-particle  Green's functions. Hence exploiting all the symmetries of the system is vital to reduce the numerical and memory storage requirements. Various symmetry relations are summarized in Table~\ref{tab:symmetries} for Hubbard-type models.
\begin{table}[t!]
\centering
\begin{tabular}{|m{0.3\columnwidth}|m{0.7\columnwidth}|}
\hline
 \begin{center}{\bf Symmetry}\end{center} & \begin{center}{\bf Symmetry relation}\end{center}   \\
  \hline
  \begin{center}Complex conjugation\end{center} & \begin{center}$ 
  (F_{\sigma\sigma'}^{kk'q})^*=F_{\sigma\sigma'}^{(-k')(-k)(-q)}$\end{center} \\ 
  \hline
  \begin{center}SU(2) symmetry\end{center} & \begin{center}$F_{\sigma\sigma'}^{kk'q}=F_{(-\sigma)(-\sigma')}^{kk'q}=F_{\sigma'\sigma}^{kk'q}$ \\[0.5cm]  
  $F^{kk'q}_{\sigma\sigma}=F^{kk'q}_{\sigma(-\sigma)}-F^{k(k+q)
  (k'-k)}_{\sigma(-\sigma)}$ \end{center}  \\ \hline
  \begin{center} Time reversal symmetry\end{center} & \begin{center} $ F_{\sigma\sigma'}^{kk'q}=F_{\sigma'\sigma}^{k'kq}\overset{\text{SU(2)}}{=}F_{\sigma\sigma'}^{k'kq}$ \end{center}  \\ \hline
  \begin{center}Particle-hole symmetry\\$\left(\mu=\frac{U}{2}\;\text{only}\right)$\end{center} & \begin{center} $\left(F_{\sigma\sigma'}^{kk'q}\right)^*=F_{\sigma\sigma'}^{kk'q}$ \end{center}  \\ \hline
  \begin{center}SU(2)$_{\text{P}}$ symmetry \\$\left(\mu=\frac{U}{2}\;\text{only}\right)$\end{center} & \begin{center} $F^{kk'q}_{\sigma\sigma}=F^{k(-k'-q)(k'-k)}_{\sigma(-\sigma)}-F^{k(-k'-q)
  q}_{\sigma(-\sigma)}$\end{center} \\ \hline
  \begin{center} Crossing symmetry\end{center} &
\begin{eqnarray*}F_{\sigma\sigma'\sigma'\sigma}^{kk'q}&=&-F_{\sigma\sigma\sigma'\sigma'}^{k(k+q) (k'-k)}\\ &\equiv&-F_{\sigma\sigma'}^{k(k+q) (k'-k)} \end{eqnarray*}\\ \hline
\end{tabular} 
\caption{Summary of the symmetry relations for the vertex function $F$ for single-orbital Hubbard-type models 
[in four-vector notation $k\widehat{=}(\nu,\mathbf{k})$]. $U$ denotes the local interaction parameter and $\mu$ the chemical potential. The crossing symmetry can be understood by considering the invariance under exchanging  the two incoming lines 
in Fig.~\ref{fig:ph_pp_scattering} (a); for a more detailed discussion and an illustration of the crossing symmetry see \onlinecite{Rohringer2013a} and \onlinecite{Galler2016}, respectively.
\label{tab:symmetries} }
\end{table} 
While the symmetry properties reported there are valid for the vertices $F$ and $\Lambda$, they do not hold in general for the explicitly channel dependent quantities  $\Phi_{\ell}$ and $\Gamma_{\ell}$ since the symmetry relations will mix one channel with the others. For an exhaustive discussion of the specific symmetry properties of $\Phi_{\ell}$ and $\Gamma_{\ell}$, see~\onlinecite{Rohringer2012,Rohringer2013a}.

Starting with the next section, we also consider local vertex functions, Green's functions and self-energies of an AIM problem. These quantities are  frequency- but not  momentum-dependent.  In the following we distinguish such local vertices  from the lattice  vertices  by dropping the momentum index, i.e., we write  $F_{r}^{\nu\nu'\omega}$ for the full local vertex instead of $F_{r,\mathbf{k}\mathbf{k}'\mathbf{q}}^{\nu\nu'\omega}$ for the lattice quantity defined in Eq.~(\ref{equ:defvertex}), and the same holds for $\Gamma_{r}^{\nu\nu'\omega}$ and $\Lambda_{r}^{\nu\nu'\omega}$. For the one-particle Green's functions and self-energies we add a label ``loc'', i.e.,   $G^{\rm loc}_\nu$ ($\Sigma^{\rm loc}_\nu$) for the local problem [i.e., the associated AIM, see Sec.~\ref{sec:dmft}] instead of $G_{{\mathbf k}\nu}$ ($\Sigma_{{\mathbf k}\nu}$ for the lattice problem. As  $G^{\rm loc}_{{\mathbf k}\nu}$ we denote the lattice Green's  function that we obtained from the  Dyson equation~(\ref{Eq:Dyson}) with $\Sigma^{\rm loc}_\nu$ of the local problem as input.

\subsection{Synopsis of dynamical mean-field theory}
\label{sec:dmft}

For completeness, and for setting the stage for the diagrammatic extensions that follow, let us briefly outline the DMFT approach here. For more details of the method and its multiorbital extensions, see the reviews of~\onlinecite{Georges1996} and~\onlinecite{Held2007}, and for a first reading \onlinecite{Kotliar2004}. A series of lecture notes on the occasion of 25 years of DMFT can be found in \onlinecite{DMFT25};  further reviews with a focus on DFT+DMFT are \onlinecite{Held2006} and \onlinecite{Kotliar2006}.

For simplicity, we consider here the single-band Hubbard model with Hamiltonian
\begin{equation}
\label{Eq:Hubbard}
  \mathcal{H} = \sum_{ij,\sigma} t_{ij}^{\phantom{\dagger}} c_{i\sigma}^{\dagger}c_{j\sigma}^{\phantom{\dagger}} + U\sum_{i}n^{\phantom{\dagger}}_{i\uparrow}n^{\phantom{\dagger}}_{i\downarrow},
\end{equation}
where $t_{ij}$ denotes the hopping amplitude between lattice sites $i$ and $j$,  $U$ the local Coulomb repulsion,  and  $\sigma \in\{{\uparrow,} \downarrow\}$  the spin; $n^{\phantom{\dagger}}_{i\sigma}\equiv c_{i\sigma}^{\dagger}c^{\phantom{\dagger}}_{i\sigma}$.

DMFT is a self-consistent theory at the  one-particle level, which approximates the  one-particle vertex  $\Sigma$ to be local. This local self-energy  $\Sigma^{\rm loc}_\nu$ and a local Green's function  $G^{\rm loc}_\nu$  are determined  self-consistently.

In particular, the {\em first DMFT self-consistency equation} calculates the local Green's function from the local self-energy $\Sigma_\nu^{\text{loc}}$
\begin{equation}
  G^{\rm loc}_\nu = \sum_{\mathbf k} G^{\text{loc}}_{\mathbf{k}\nu},
\end{equation}
where $G^{\text{loc}}_{\mathbf{k}\nu}$ is the DMFT lattice Green's function obtained from  $\Sigma^{\rm loc}_\nu$ via the Dyson equation~(\ref{Eq:Dyson}) reformulated as
\begin{equation}
\label{equ:defDMFTgf}
  G^{\text{loc}}_{\mathbf{k}\nu} =  [i\nu + \mu -\varepsilon_{\mathbf k} - \Sigma^{\rm loc}_\nu]^{-1}.
\end{equation}
Here $\varepsilon_{\mathbf k}$ is the Fourier-transform of  $t_{ij}$. This step allows us to calculate $G^{\rm loc}_\nu$ from the local, i.e.,  ${\mathbf k}$-independent $\Sigma^{\rm loc}_\nu$. As we will see in Sec.~\ref{Sec:Methods}, the DMFT lattice Green's function $G_{\mathbf{k}\nu}^\text{loc}$ and the difference $G^{\text{loc}}_{\mathbf{k}\nu}-G^{\rm loc}_\nu$ appear prominently in the context of diagrammatic extensions of DMFT.

The {\em second DMFT self-consistency equation} is defined by summing {\em all} skeleton Feynman diagrams in terms of the local $U$  and $G^{\rm loc}_\nu$ to obtain the local DMFT self-energy   $\Sigma^{\rm loc}_\nu$ again [cf.\ Fig.~\ref{fig:dmftvsdga}(a) in Sec.~\ref{sec:dga} below]. These two steps are iterated until self-consistency.

In practice, this second step is achieved through the numerical solution of an AIM
\begin{eqnarray}
\mathcal{H}&=& \sum_{\ell\sigma} \epsilon^{\phantom{\dagger}}_{\ell} \; a_{{\ell}\sigma}^{\dagger}a_{{\ell}\sigma}^{\phantom{\dagger}}  +  \sum_{\ell\sigma}  V^{\phantom{\dagger}}_{\ell} \;  a_{{\ell}\sigma}^{\dagger} c_{\sigma}^{\phantom{\dagger}}  + {\rm h.c.}  \nonumber \\ && + U n^{\phantom{\dagger}}_{\uparrow}n^{\phantom{\dagger}}_{\downarrow},
\label{Eq:AIM}
\end{eqnarray}
which has the same interaction $U$ as the Hubbard model (\ref{Eq:Hubbard}) but only on one site. This site ($c^{(\dagger)}$) is coupled through the hybridization $V_{\ell}$ to a bath of conduction electrons  $a_{\ell\sigma}^{\dagger}$ at energies  $\epsilon_{\ell}$. If the interacting Green's function of the AIM is the same as $G^{\rm loc}_\nu$, it yields the same Feynman diagrammatic contribution to the self-energy as DMFT: all local terms. To achieve the latter (at self-consistency) one first calculates the local noninteracting Green's function ${\cal G}_\nu$ of the auxiliary AIM (at  the interacting site)
via the Dyson equation for the AIM
\begin{eqnarray}
\label{Eq:DMFTDelta}
({\cal G}_\nu)^{-1} &=& (G^{\rm loc}_\nu)^{-1}+\Sigma^{\rm loc}_\nu \label{Eq:AIMDyson}  \\&=&i\nu+\mu-\underbrace{\sum_{\ell}\frac{|V_\ell|^2}{i\nu+\mu- \epsilon_{\ell}}}_{\equiv \Delta_{\nu}},
\end{eqnarray}
which is directly related to a  corresponding hybridization function $\Delta_{\nu}$. Then one solves the AIM for its Green's function  $G^{\rm loc}_\nu$ and uses the AIM Dyson equation~(\ref{Eq:AIMDyson}) again to obtain the AIM self-energy 
\begin{equation}
\Sigma^{\rm loc}_\nu = ({\cal G}_\nu)^{-1}- (G^{\rm loc}_\nu)^{-1}.
\end{equation}
This  closes the DMFT iteration loop and, at self-consistency, $\Sigma^{\rm loc}_\nu$ is the DMFT self-energy.
The DMFT solution yields the local $\Sigma^{\rm loc}_\nu$, $G^{\rm loc}_\nu$ and DMFT lattice Green's function
$G^{\rm loc}_{\mathbf k \nu}$. At self-consistency we can also calculate  local two-particle Green's and vertex functions as discussed in Sec.~\ref{sec:calver} below.

\subsection{Physical contents of the local vertex}
\label{sec:vertexphysics}

The physical meaning of the {\sl one-}particle Green's function and its 1PI counterpart, the self-energy, is nowadays standard textbook knowledge; see, e.g., \onlinecite{Abrikosov1975,Mahan2000}. Less information is available about the physical content of the {\sl two-}particle Green's function and its 1PI and 2PI counterparts, the so-called vertex functions, whose definitions are given in Sec.~\ref{sec:formalism}. Yet, in recent years, the development of diagrammatic extensions of DMFT has triggered significant progress in this direction.
  
In this section we discuss the frequency dependence of the local DMFT vertex functions for models with a constant bare interaction $U$ and their microscopic interpretation. This requires the analysis of the frequency structure of all possible Feynman diagrams~\cite{Rohringer2012,Rohringer2013a,Wentzell2016}. The dependence of a certain diagram on the three external frequencies ($\nu,\nu',\omega$) is controlled by its topology, i.e., by the way the particles/holes enter. If two particle lines, or one particle and one hole line, are attached to the same bare vertex, the entire diagram will depend only on the sum/difference of their frequencies. This can be seen, for example, in the diagram of Fig.~\ref{fig:3vertextypes}(a) which depends only on $\omega$ but not on $\nu$ and $\nu'$. For a fixed $\omega$ it will, hence, remain constant for arbitrarily large values of $\nu$ and $\nu'$.

On the contrary, if an external particle/hole is connected by the bare interaction to three internal lines, the corresponding diagram will explicitly depend on its frequency. This is illustrated in Fig.~\ref{fig:3vertextypes}(c) where the lower right-most part of the diagram (circle) gives rise to the expression $\sum_{\nu_1\nu_2}G^{\text{loc}}_{\nu'+\nu_1}G^{\text{loc}}_{\nu_2+\nu_1}G^{\text{loc}}_{\nu_2}\ldots$ which, hence, explicitly depends on $\nu'$ and will decay for large values of $\nu'$ as $1/\nu'$.   

Excluding the trivial situation when all four external lines enter at the same interaction vertex (which gives just the constant contribution $U$), the considerations above suggest the following threefold classification of the diagrams of $F$:

\begin{figure*}[t!]
  \centering
  \includegraphics[width=1.0\textwidth]{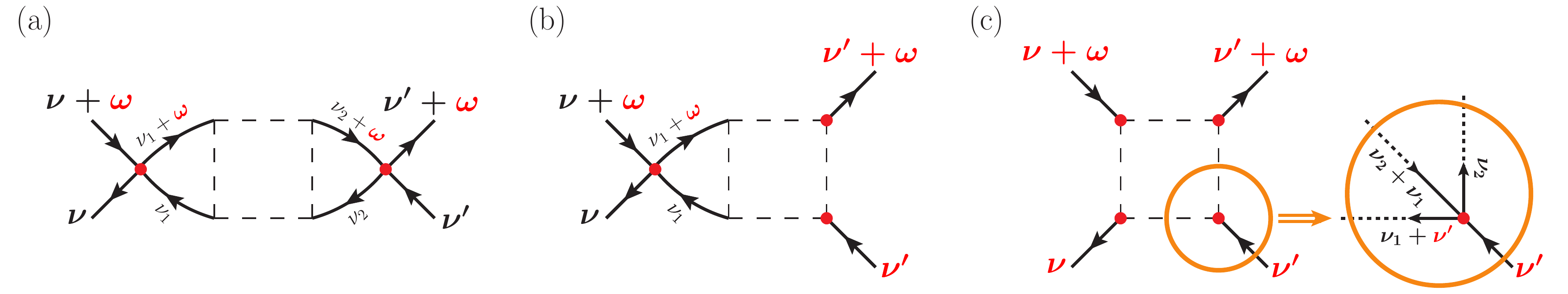}
    \caption{(Color online) Categorization of diagrams according to their frequency dependence. (a) Diagram where the left and right pairs of external lines are attached to the same two bare interaction vertices, (b) diagram where only the left external lines are connected to the same bare vertex, and (c) diagram where all external lines enter at different bare vertices. The external frequencies, on which the diagram depends explicitly, are marked in red (gray).}
  \label{fig:3vertextypes}
\end{figure*}

\begin{figure*}[tb]
\begin{minipage}{0.67\textwidth}
  \includegraphics[width=0.92\textwidth]{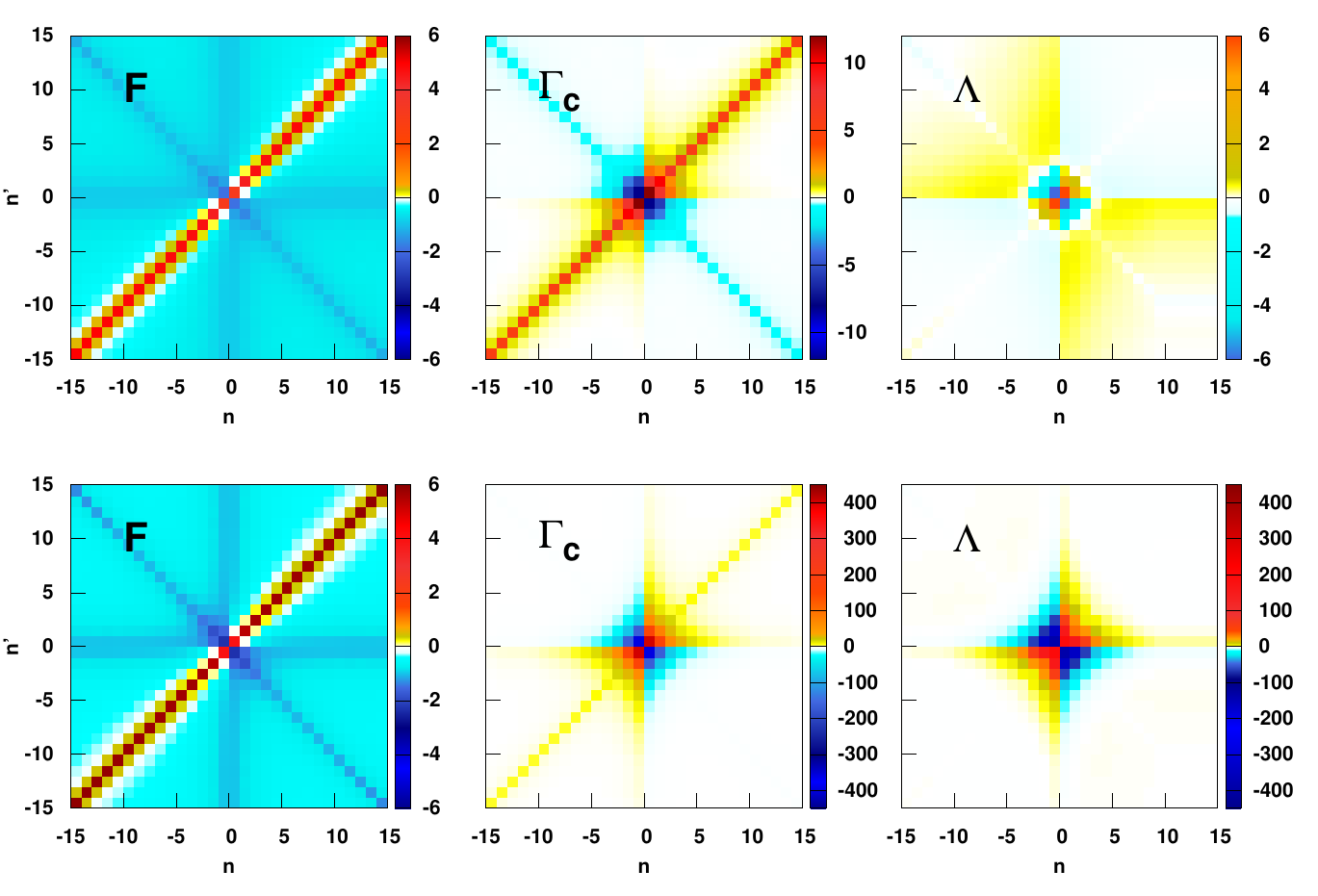}
\end{minipage} \hfill
\begin{minipage}{0.28\textwidth}
    \caption{(Color online) DMFT results for the full local vertex ($F_c^{\nu\nu'(\omega=0)}$-U, left), the 2PI vertex in the $ph$ charge channel ($\Gamma_c^{\nu\nu'(\omega=0)}-U$, middle), and the fully 2PI vertex ($\Lambda_c^{\nu\nu'(\omega=0)}$-U, right) at Matsubara frequencies $\nu^{(\prime)}=(2n^{(\prime)}+1)\pi/\beta$. The calculations have been performed for the  Hubbard model on a square lattice with nearest-neighbor hopping $t$ at $T=0.4t$, $U=4.8t$ (lower panel) and $U= 5.08 t$ (upper panel). The intensity (color bar) is given in units of $4t$. Adapted from~\onlinecite{Schaefer2016c}.}
  \label{fig:PlotDMFTvertex}
\end{minipage}
\end{figure*}

(i) The first group includes all diagrams where both the incoming and outgoing particle-hole (or particle-particle) pairs enter at the respective same bare interaction vertex; see Fig.~\ref{fig:3vertextypes}(a). Such diagrams depend only on the corresponding frequency differences between the particle and hole entering the diagram at the respective same bare vertex. In the case of  Fig.~\ref{fig:3vertextypes}(a) this frequency is $\omega$, which one can see from the fact that all internal frequency summations in Fig.~\ref{fig:3vertextypes}(a) depend only on $\omega$ but not on $\nu$ and $\nu'$.  Note that the diagram in Fig.~\ref{fig:3vertextypes}(a) is reducible in the $ph$ channel; it belongs to $\Phi_{ph}$ in  Fig.\ref{fig:decomp2PILambda}. The two other  possibilities of how two external legs can be pairwise attached to bare vertices are diagrams reducible in the $\overline{ph}$ and $pp$ channels. These depend on only {\sl one} (bosonic) frequency(combination) $\nu-\nu'$ and $\nu+\nu'+\omega$, respectively. In Fig.~\ref{fig:PlotDMFTvertex} (left), these diagrams are responsible for the constant background ($\omega=0$), the main ($\nu-\nu'=0$), and secondary ($\nu+\nu'+\omega=0$) diagonals of the DMFT vertex $F$. From a physical perspective, diagrams of type (i) correspond to physical susceptibilities. For example, the contribution to $F$ originating from the sum of all diagrams of type (a) in  Fig.~\ref{fig:3vertextypes} corresponds to a $ph$ (charge or spin) susceptibility \cite{Rohringer2012,Rohringer2013,Wentzell2016}.

(ii) The second class includes all diagrams where only {\sl one} pair of external lines is attached to the same bare vertex. Their contribution depends on {\sl two} (one bosonic and one fermionic) Matsubara frequencies. For example, Fig.~\ref{fig:3vertextypes}(b) depends on $\omega$ and $\nu'$ but not on $\nu$. Such diagrammatic contributions remain constant along one-dimensional {\sl lines} in the three-dimensional frequency space. For instance, for the fixed value $\omega=0$ and $\nu'=\pi/\beta$ in the density plot of $F$ [Fig.~\ref{fig:PlotDMFTvertex} (left)], such diagrams are responsible for the welldefined ``$+$''-shaped structure, i.e., the enhanced scattering rate along $\nu', \nu=\pi/\beta$. From a physical perspective, such ``eye''-shaped  diagrams are related\footnote{Precisely, the fermion-boson vertices are obtained by eliminating all internal susceptibilities, i.e., all internal subdiagrams of type (i), i.e., dividing these diagrams by $1-U\chi^{\omega}$\cite{Rohringer2016}.} to a fermion-boson coupling, as it  appears, e.g., in ladder D$\Gamma$A~\cite{Katanin2009,Rohringer2016}, DB~\cite{Rubtsov12,vanLoon2014a} and TRILEX \cite{Ayral2015,Ayral2016a}.

(iii) The third class consists of all diagrams where all four external lines enter at different bare vertices; see Fig.~\ref{fig:3vertextypes}(c). Their contribution to $F$ depends on {\sl all three} frequencies (as illustrated by the  circle for the $\nu'$ dependence) and hence decay in {\sl all} directions of frequency space. This is important for many-body algorithms based on local DMFT vertex functions, because diagrams of this type need to be considered for small frequencies only~\cite{Wentzell2016}. 

As diagrams of type (i) and (ii) in Fig.~\ref{fig:3vertextypes}(a) and (b) are two-particle reducible, all diagrams of the fully 2PI vertex $\Lambda$ belong to  class (iii). Hence,  $\Lambda^{\nu\nu'\omega}$ decays in all three directions of frequency space except for a constant background $U$; cf.\ Fig.~\ref{fig:PlotDMFTvertex} (right). In other words, the asymptotic behavior of $F$ originates from reducible diagrams only. Consequently, when considering 2PI diagrams in one channel ($\Gamma_r$), one ``loses'' all the asymptotic structures generated by the two-particle reducible  diagrams in this channel, keeping only the high-frequency features  from the reducible diagrams in the complementary channels (cf.\ Sec.~\ref{sec:formalism}). This is nicely illustrated by comparing the DMFT data for $F$ and $\Gamma_c$ in Fig.~\ref{fig:PlotDMFTvertex} (upper panels), where the disappearance of the $ph$  $\omega=0$ structure   corresponds to the vanishing background.

The simplification of the high-frequency asymptotics is a helpful factor in the numerical manipulation of the 2PI vertex functions. However, we should note that, at the same time, the low-frequency structure of the (2PI) vertices can become very complicated---in certain parameter regimes. Specifically, as reported in recent DMFT (and DCA) studies at the two-particle level \cite{Schafer2013, Janis2014,Gunnarsson2016,Schaefer2016c,Ribic2016}, $\Gamma_c$ and $\Lambda$ acquire strong low-frequency dependencies, and even become {\sl divergent} in certain cases.  This can be seen in the DMFT results of Fig.~\ref{fig:PlotDMFTvertex} (lower panel), computed for a  $U$  just before a divergence (note the  large values for $\Gamma_c$ and $\Lambda$  in the color scale). These divergences occur already for rather moderate $U$ in  DMFT  for the Hubbard model \cite{Schafer2013,Gunnarsson2016,Schaefer2016c}.

\begin{figure}[tb]
  \centering
  \includegraphics[width=\columnwidth]{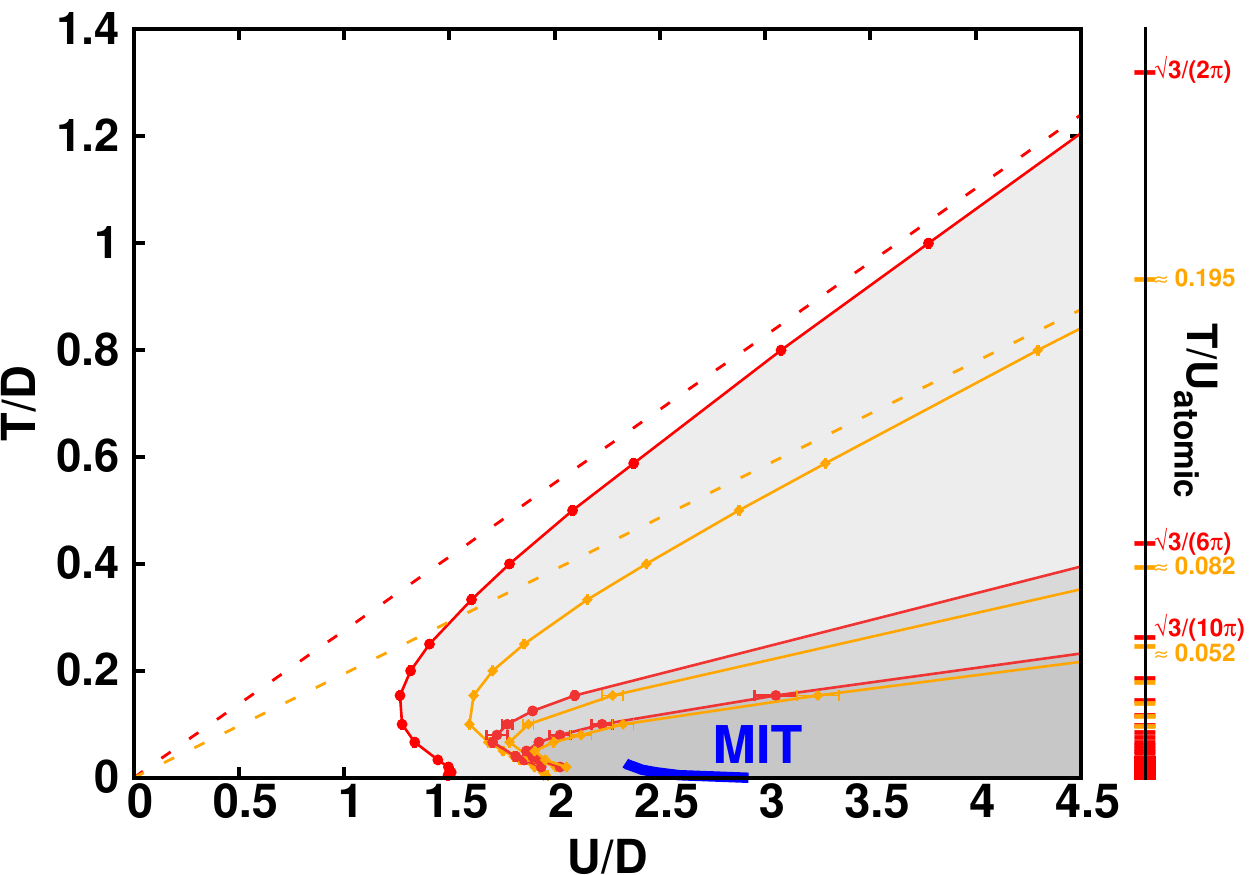}
    \caption{(Color online) DMFT phase-diagram of the half-filled Hubbard model (on a square lattice with half bandwidth $D=4t=1$): Here the first seven lines, where the 2PI vertex $\Gamma_{c}$  alone (red/dark gray) [or simultaneously with $\Gamma_{pp}$ (orange/light gray)] diverges for zero transfer frequency ($\omega=0$) have been reported. The dashed straight lines correspond to the divergence conditions of the atomic limit (scale on the right, \onlinecite{Rohringer2013a,Schaefer2016c}), which is approached by the DMFT data for $U, T >> D$. At lower $T$, the lines display a clear reentrance, roughly resembling the shape of the Mott-Hubbard metal-insulator transition (MIT) (blue line). Note that the first divergence line, marking the end of the perturbative regime, is located well inside the correlated metallic region. Adapted from~\cite{Schaefer2016c}.}
  \label{fig:vertexdivHubb}
\end{figure}

Figure~\ref{fig:vertexdivHubb} explicitly shows the multiple lines where $\Gamma_c$, $\Gamma_{\rm pp}$, and $\Lambda$ diverge in the DMFT phase diagram. Their presence has been demonstrated also in CDMFT or DCA calculations of the Hubbard model \cite{Gunnarsson2016,Vucicevic2018} as well as, more recently, for a pure AIM with a constant electronic bath \cite{Chalupa2017}.    
 
Analytical (or semianalytical) calculations for the  Falicov-Kimball model \cite{Schafer2013,Janis2014,Ribic2016,Schaefer2016c}, for the one-point model \cite{Stan2015} or in the atomic limit \cite{Schafer2013,Schaefer2016c} prove that the observation of such divergences is not a numerical artifact, but  rather a general manifestation of the breakdown of perturbation theory in correlated systems \cite{Gunnarsson2017}. In fact, it has been shown  \cite{Kozik2015, Stan2015, Schaefer2016c, Gunnarsson2017,Vucicevic2018} and later rigorously proved in the Supplemental Material of \onlinecite{Gunnarsson2017} that, for the same parameter values where the vertex divergences are observed, crossings between the physical and different unphysical branches of the self-energy functional $\Sigma[G]$ can occur. This reflects the  intrinsic multivaluedness of the Luttinger-Ward formalism in the many-body theory \cite{Keiter2000,Kozik2015}, unless physical constraints for $\Sigma$ are  explicitly considered \cite{Pruschke2001,Potthoff2006,Tarantino2017}.

While the frequency structure of the full vertex $F$ is now well understood (see Fig.~\ref{fig:3vertextypes}) \cite{Rohringer2012,Wentzell2016}, the theoretical implications of the multiple divergences of the 2PI vertex are not fully clarified yet, representing a subject of ongoing discussion and research~\cite{Kozik2015,Stan2015,Rossi2016,Gunnarsson2016,Schaefer2016c,Tarantino2017,Gunnarsson2017, Vucicevic2018,Chalupa2017}. Let us note, however, that from the algorithmic point of view, the divergences of the 2PI vertex $\Gamma$ can be easily circumvented for the Bethe-Salpeter ladder resummations of the diagrammatic extensions of DMFT (see Sec.~\ref{sec:dgaladder}).

\subsection{Calculation of the local vertex}
\label{sec:calver}

In recent years, we have witnessed a staggering increase in applications of two-particle Green's functions and a correspondingly rapid development of efficient algorithms to compute them. To large extent this has been driven by the development of diagrammatic extensions of DMFT, as well as increased computational resources. Here we provide a summary of the different numerical methods for calculating the local vertex and related response functions and provide references for further reading. 
 
The static ($\omega=0$) component of the vertex has already been calculated in the early days of DMFT for obtaining susceptibilities~\cite{Jarrell1992}. Pioneering attempts to compute the dynamical and momentum structure of a two-particle vertex were made by \onlinecite{Jarrell1997} and \onlinecite{Maier2006} using QMC methods. \onlinecite{Kusunose2006} computed generalized susceptibilities using second-order iterated perturbation theory (IPT). More recently the vertex has also been obtained within functional renormalization group calculations~\cite{Kinza2013}. The most frequently used methods to calculate the vertex however are exact diagonalization (ED) and QMC based implementations, including continuous-time (CTQMC) variants. These methods allow one to control the error. We discuss them in the following with a focus on the developments regarding the calculation of the two-particle vertex. A detailed review of CTQMC methods is beyond the scope of this article; see the comprehensive review of \onlinecite{Gull2011a}. Open source implementations are publicly available for most methods nowadays. We provide an overview in Sec.~\ref{Sec:opensource}.

Impurity solvers typically evaluate the two-particle Green's function from which vertex functions are extracted as described in Sec.~\ref{sec:formalism}. Their generalization to $n$-particle correlation functions is, in principle, straightforward. Their computation however quickly exhausts available computation time and memory resources. While the Green's functions can be measured as a function of imaginary time, they are usually evaluated on Matsubara frequencies. The number of required sampling points for three independent times (or frequencies) in the case of the two-particle Green's function scales as $\beta^{3}$, which quickly becomes a computational and memory burden. In addition, the treatment of the various discontinuities arising from the anticommutation relations of the fermionic operators is cumbersome.

\subsubsection{Exact Diagonalization}

In ED, the two-particle Green's function \eqref{equ:defngreengeneral} can be computed with a relatively simple algorithmic extension of the standard ED implementation used in the context of DMFT~\cite{Georges1996}. It is most conveniently evaluated by applying the Fourier transform \eqref{equ:def2pgf} (from imaginary time to Matsubara frequencies) to the Lehmann representation \cite{Abrikosov1975,Mahan2000}. The two-particle Green's function can then be written as a sum over products of the matrix elements of the four creation and annihilation operators. The sums in the terms are weighted with a function that depends on the eigenvalues and the three external Matsubara frequencies, see the appendixes of \onlinecite{Toschi2007} and \onlinecite{Hafermann2009a}.  Because of the exponential increase of the Hilbert space and the required fourfold nested sum over the eigenstates, ED is essentially limited to single-orbital calculations and a maximal number of $5$  sites ($4$ bath + $1$ impurity site). Calculations can be sped up through parallelization, by using a lookup table for the exponential factors $\exp(-\beta E_{i})$, by truncating the sums for terms in which the exponentials are negligible, and most importantly, by exploiting symmetries. The Lanczos algorithm \cite{Georges1996}, which simplifies the calculation of the single-particle Green's function, cannot directly be applied to the two-particle Green's function. The reason is that transitions between two nearby states at arbitrarily high energies contribute; this might be circumvented, however, using the correction vector method~\cite{TanakaPC2016}.

\subsubsection{Quantum Monte Carlo}

\paragraph{Hirsch-Fye and continuous-time auxiliary field algorithm}

While ED and CTQMC are more commonly used nowadays, the Hirsch-Fye QMC algorithm~ \cite{Hirsch1986} has been also employed to calculate $n$-particle Green's functions. Here observables are intrinsically affected by the Trotter decomposition error. Nevertheless, even the numerically  delicate (diverging) fully 2PI vertex $\Lambda$ can be extracted with sufficient accuracy in physically relevant parameter regimes of the single-band Hubbard model in DMFT and DCA calculations~\cite{Maier2006, Gunnarsson2016}. In addition, working on a discrete-time grid avoids dealing with equal-time discontinuities which arise in modern continuous-time algorithms.

At low temperatures, when controlling the Trotter error becomes impractical, a CTQMC algorithm  offers superior performance. For instance, the continuous-time auxiliary-field (CT-AUX) algorithm~\cite{Gull2008a} is based on an auxiliary-field decoupling of the interaction vertices similar to the Hirsch-Fye algorithm, but samples a varying number of fields at arbitrary times. CT-AUX is particularly efficient for large cluster problems. The Fourier transform of the Green's function measurement to Matsubara frequencies can be sped up significantly using a nonequidistant fast Fourier transform (NFFT) algorithm~\cite{Staar2012}  when the perturbation order is sufficiently large $(\gtrsim 20)$. This applies to other continuous-time methods as well.

\paragraph{Continuous-time expansion in the interaction}

In  the  continuous-time expansion in the interaction (CT-INT), the measurement of two-particle Green's functions amounts to performing a Monte Carlo average over ratios of determinants which differ by two rows and columns~\cite{Rubtsov2005,Gull2011} (instead of one for the single-particle function). Similar to  Hirsch-Fye QMC and CT-AUX, the measurement for a particular correlation function can symbolically be obtained by enumerating all Wick contractions of the operators appearing in the definition of the correlation function and replacing them by configuration-dependent quantities $G_{\tilde{\tau}_{1},\ldots\tilde{\tau}_{N}}(\tau,\tau')^{\sigma\sigma'}$. Here $N$ denotes the CT-INT perturbation order, and $\tilde{\tau}_{1},\ldots\tilde{\tau}_{N}$ denotes the continuous QMC times of the Monte Carlo configurations. The Wick contraction yields:
\begin{align}
\langle T_{\tau}c_{\sigma}(\tau_{1})c^{\dagger}_{\sigma}(\tau_{2}) &c_{\sigma'}(\tau_{3})  c^{\dagger}_{\sigma'}(0)\rangle \to
\\
& G_{\tilde{\tau}_{1},\ldots\tilde{\tau}_{N}}^{\sigma\sigma}(\tau_{1},\tau_{2})G_{\tilde{\tau}_{1},\ldots\tilde{\tau}_{N}}^{\sigma'\sigma'}(\tau_{3},0)\notag\\
&-\delta_{\sigma\sigma'} G_{\tilde{\tau}_{1},\ldots\tilde{\tau}_{N}}^{\sigma\sigma}(\tau_{1},0)G_{\tilde{\tau}_{1},\ldots\tilde{\tau}_{N}}^{\sigma\sigma}(\tau_{3},\tau_{2}). \nonumber
\end{align}
By Fourier transform a measurement directly in frequency is straightforwardly obtained. The latter can be factorized into $G_{\tilde{\tau}_{1},\ldots\tilde{\tau_{N}}}(\nu,\nu')^{\sigma\sigma'}$ factors to speed up the calculation. With $N$, $N_{\nu}$, and $N_{\omega}$ denoting the perturbation order, the number of fermionic and bosonic frequencies, respectively, the measurement scales as $\mathcal{O}(N^{2}N_{\nu}^{2})+\mathcal{O}(N_{\nu}^{2}N_{\omega})$.

\paragraph{Continuous-time expansion in the hybridization}

In the case of CTQMC with hybridization expansion  (CT-HYB) ~\cite{Werner2006,Werner2006a}, the partition function is expanded in terms of the AIM hybridization function. Here, the Monte Carlo weight is a product of the determinant of hybridization functions and a trace over the atomic states. In the original implementation, the self-energies and vertex functions exhibit relatively large fluctuations at intermediate to high frequencies \cite{Gull2007,Hafermann2012}. This problem can be cured by expressing the self-energy as a ratio of two correlation functions $G$ and $F$ (improved estimator), a relation which follows from the equation of motion and corresponds to $F=\Sigma G$. This trick was first introduced in the numerical renormalization group (NRG) context~\cite{Bulla1998,Bulla2008}. Improved estimators exist for the reducible vertex function as well~\cite{Hafermann2012}, including impurity models with spin-boson coupling~\cite{Otsuki2013}, retarded interactions~\cite{Hafermann2014} and multiorbital interactions beyond density-density terms~\cite{Gunacker2016}. Note also that in CT-HYB, the conventional approach to obtain the Green's function and vertex is ``removing'' hybridization lines. This procedure does not allow us to calculate all components of the multiorbital vertex function, a limitation that was overcome by \onlinecite{Gunacker15} using worm sampling. 

Let us also note a first calculation of the local three-particle vertex using  CT-INT \cite{Hafermann2009}, CT-AUX, and CT-HYB \cite{Ribic2017b}. Slices through this three-particle vertex show similar structures as the two-particle vertex shown above.

\subsubsection{Handling vertex asymptotics} 

A generic problem of the numerical treatment of vertex functions is the large memory size required to store the three-frequency-dependent vertex. This limits the size of the frequency box where the vertex can be treated exactly. However, similar to single-particle quantities, the vertex approaches an asymptotic behavior at high frequencies. This behavior can be characterized by diagrams similar to the ones discussed in Sec.~\ref{sec:vertexphysics} and exploited to simplify calculations.

Pioneering work in this direction was done by \onlinecite{Kunes2011} who, for calculating the DMFT susceptibility more accurately, expressed  the high-frequency asymptotic behavior of the $\omega=0$ vertex function irreducible in the particle-hole channel in terms of the local dynamical susceptibility. Extensions to more general cases can be found in \onlinecite{Tagliavini2018}. Starting from the diagrammatic considerations of \onlinecite{Rohringer2012}, \onlinecite{Li2016} and \onlinecite{Wentzell2016} derived more general relations for the asymptotics of the three-frequency  vertex based on the  parquet equations, and proposed a parametrization scheme of the full high-frequency behavior  of vertex functions based on a diagrammatic analysis. \onlinecite{Kaufmann2017} implemented the measurement of these asymptotics in CT-HYB. The corresponding ED expressions were reported by \onlinecite{Tagliavini2018}. The asymptotic behavior of the vertex depends on two frequencies and allows the calculation of the vertex in an  arbitrarily large frequency box with reduced  statistical noise, while taking a fraction of the numerical effort and storage required for the full three-frequency-dependent vertex. The latter however is still needed at low frequencies, where the vertex deviates from this asymptotics.

Alternatively, correlation functions can be represented in a Legendre polynomial basis~\cite{Boehnke2011} to obtain a compact representation. For vertex functions it is advantageous to use a mixed representation where the bosonic frequency dependence is kept whereas the fermionic ones are projected onto the Legendre polynomial basis. Provided a sufficiently large cutoff $N_{l}$ of polynomial coefficients, the Legendre representation allows the calculation of the vertex at arbitrarily high frequencies. The measurement scales as $\mathcal{O}(N^{2}N_{l}^{2}N_{\omega})$ \onlinecite{Shinaoka2017} introduced an intermediate representation between the imaginary-time and real-frequency domains. It is based on sparse modeling of data in a basis that is derived from the singular value decomposition of the kernel relating the data in these domains~\cite{Otsuki2017}. Interestingly, it includes the Legendre representation as the high-temperature limit, but requires even less coefficients $N_{l}$, in particular at low temperatures. As for the Legendre basis, the transformation is unitary, so that the entire calculation can in principle be performed in this basis and only final results need to be transformed back to Matsubara representation.

\section{Methods}
\label{Sec:Methods}

After discussing the diagrammatics and physics of the local vertex in the previous section, we are now ready to turn to the recently developed diagrammatic vertex extensions of DMFT (for an overview, see  Sec.~\ref{sec:outline} and  Table~\ref{Table:alg} in Sec.~\ref{sec:conclusion}). These have a common underlying principle, which follows two steps:

\begin{itemize}
\item A local approximation is performed at the 2P-level, which corresponds to identifying the building block of the specific approach. This  (highly nonperturbative) building block is one of the local vertices discussed in the previous section.

\item A diagrammatic approach is built around this local building block to include nonlocal correlations beyond DMFT into the self-energy and susceptibilities.
\end{itemize}

Out of this line fall diagrammatic extensions of DMFT which simply combine the local DMFT  with the nonlocal self-energy from another approach; such approaches are discussed in  Sec.~\ref{subsec:DMFTplus}.

\subsection{Dynamical vertex approximation (D\texorpdfstring{$\Gamma$}{G}A)}
\label{sec:dga}

\begin{figure}[t!]
 \centering
 \includegraphics[width=0.99\columnwidth]{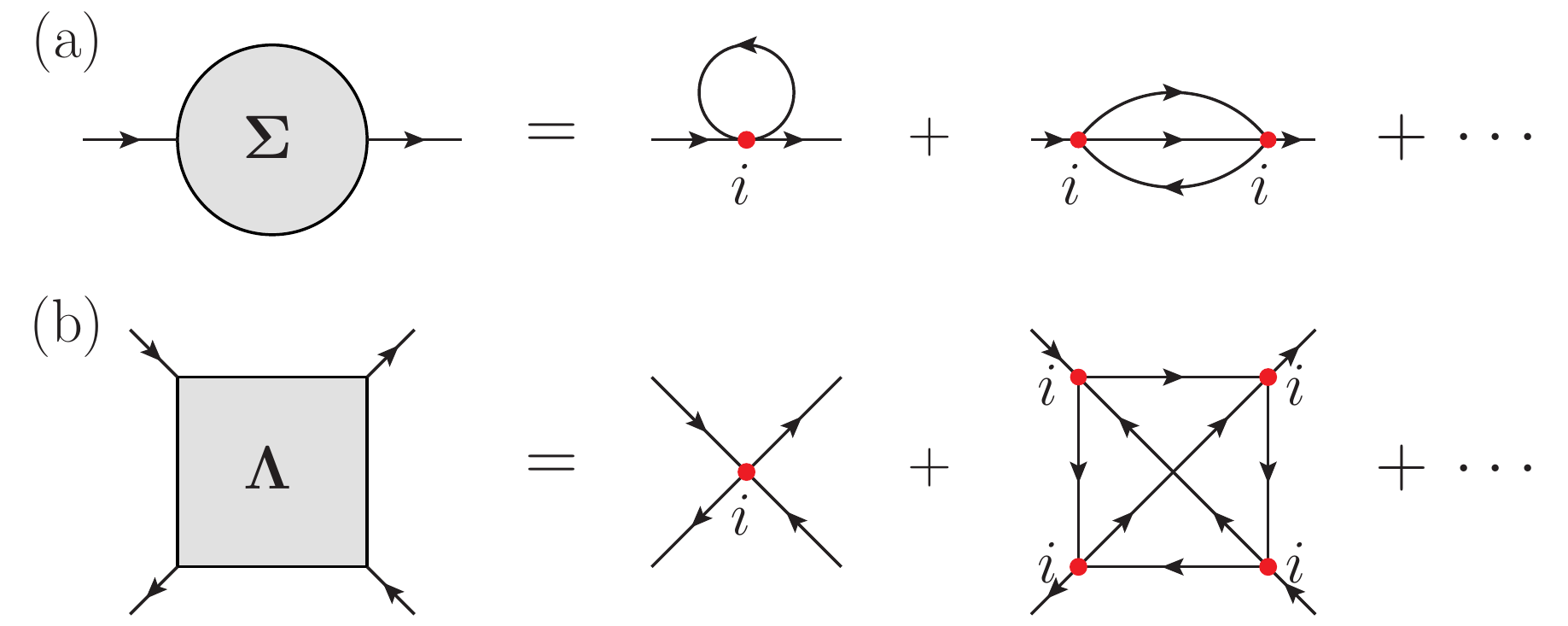}
  \caption{(Color online) (a) In DMFT the fully 1PI one-particle vertex, i.e., the self-energy $\Sigma$, is purely local. (b) In D$\Gamma$A instead the fully 2PI two-particle vertex $\Lambda$ is approximated to be local (lines: interacting Green's function; dots: bare Hubbard interaction $U$; $i$ represents a site index of the lattice).}
 \label{fig:dmftvsdga}
\end{figure}

The best way to understand the basic concepts of the D$\Gamma$A is to start by considering the diagrammatics of DMFT: DMFT assumes the {\sl locality} of all (skeleton)  diagrams for the self-energy, see Fig.~\ref{fig:dmftvsdga} (a). The self-energy, however, is nothing but the one-particle (irreducible) vertex. Hence, a systematic generalization of DMFT is directly  obtained by requiring the locality at the $n$-particle level: \onlinecite{Toschi2007} assume  the fully $n$-particle irreducible ($n$PI) $n$-particle vertex to be {\sl local}. Differently from DMFT,   the self-energy or other vertices with less than $n$ particles  do  acquire nonlocal contributions,  as does the full  $n$-particle vertex entering the susceptibility or generally the $k$PI  $n$-particle vertex for $k<n$. 

Taking the limit $n \rightarrow \infty$  (and a proper generalization of 2PI defined in Sec.~\ref{sec:formalism} to  $n$PI) corresponds to considering {\sl all} diagrams for the Hubbard model or any other model with a local interaction. Therefore the exact solution is recovered for  $n \rightarrow \infty$.
In practice one is however restricted to D$\Gamma$A on the $n=2$ particle level which is  illustrated  in Fig.~\ref{fig:dmftvsdga} (b). 
This assumes the fully 2PI {\em two}-particle vertex  $\Lambda$ to be {\em local} (for the definitions see Sec.~\ref{sec:formalism}).

In principle, one can systematically improve the approach by going to the  $n=3$ particle vertex etc. Actually doing so in practice is difficult, but at least  an error estimate for the standard two-particle calculations seems feasible  \cite{Ribic2017b}.

\paragraph*{Why a local  \texorpdfstring{$\Lambda$}{Lambda} ?}
Assuming that a {\em local}   $\Lambda$ is a good approximation can be understood first from a Feynman diagrammatic perspective: The fully irreducible diagrams are topologically very compact, and hence the most local ones. Each   $\Lambda$ diagram generates many diagrams for the full vertex $F$.  For example, the bare (local) interaction $U$ (as part of  $\Lambda$) generates, via the parquet equations, all Feynman diagrams of up to third order in $U$ as well as many higher-order diagrams; even a local  $\Lambda$ generates a nonlocal $F$ including the typical ladder diagrams for spin fluctuations of Fig.~\ref{Fig:magnon}. The locality of $\Lambda$  is also further supported by numerical data in $d=2$: Even in the parameter regions with strong nonlocal correlations in the self-energy $\Sigma$ and the full vertex $F$,  $\Lambda$ still remains local or ${\mathbf k}$ independent to a very good approximation as shown in DCA calculations by \onlinecite{Maier2006}. 

From a physical perspective, this numerical evidence of a purely local $\Lambda$ can be attributed to the absence of any ladder diagrams, which are typically associated with collective [spin density wave (SDW), charge density wave (CDW), etc.] modes of the system. As these modes are responsible for strong nonlocal correlations, the momentum dependence of  $\Lambda$ can be (and often is) particularly weak, consistent with the D$\Gamma$A assumption, even in situations  where $F$ is strongly momentum dependent.

Our understanding of nonlocal physics is also often based on ladder diagrams in terms of the bare $U$, e.g.\ the magnon self-energy~\cite{Hertz1973}, see Fig.~\ref{Fig:magnon}. Hitherto, however, such approaches were restricted to weak coupling \cite{Vilk1997}. Taking a local  $\Lambda$ instead of $U$ in D$\Gamma$A includes all the strong local DMFT correlations (responsible for quasiparticle renormalizations, Mott transitions, etc.), but at the same time allows us to study nonlocal correlations and collective excitations on all length scales.

\subsubsection{Parquet  D\texorpdfstring{$\Gamma$}{G}A}
\label{sec:parquetDGA}

The locality assumption for $\Lambda$ is the first step in the construction of the D$\Gamma$A. The second step is to define the diagrams to be constructed from this local building block. For the D$\Gamma$A, this second step is naturally the application of the parquet equations~\cite{Toschi2007,Held2014} (see  Fig.~\ref{fig:parquet}), which allow the calculation of the full vertex $F$, self-energy $\Sigma$  etc. from $\Lambda$.

\begin{figure*}[t!]
  \includegraphics[width=1.0\textwidth]{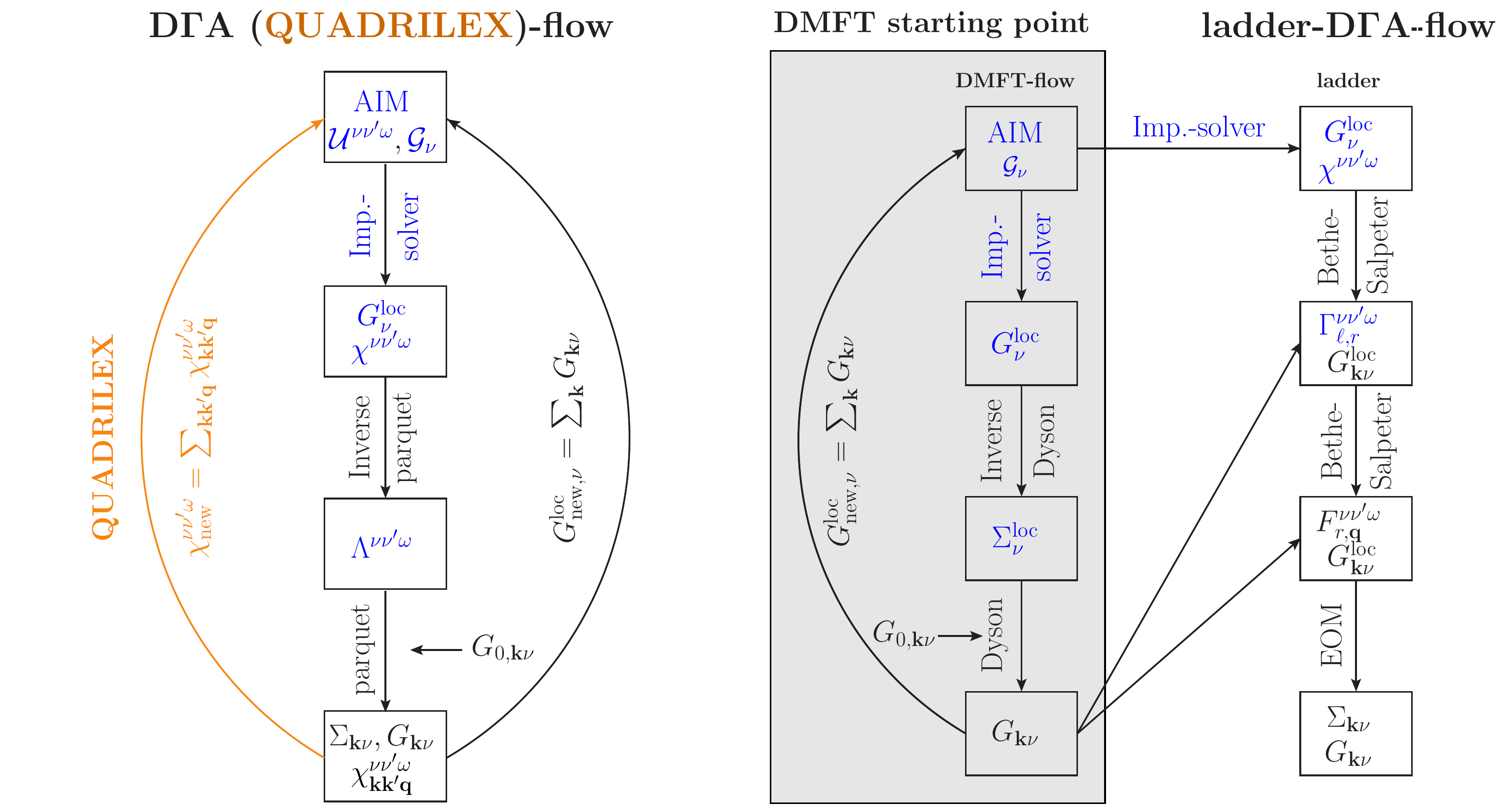}
    \caption{(Color online) Flow diagram for parquet D$\Gamma$A with or without the QUADRILEX self-consistency (in orange/light gray) for the interaction  (left),  DMFT (middle), and   ladder D$\Gamma$A (right).Quantities obtained from (or defining) an auxiliary AIM are indicated in blue/dark gray.}
 \label{fig:dgaladderflow}
\end{figure*}

The algorithmic implementation of D$\Gamma$A is realized through the following steps [see Fig.~\ref{fig:dgaladderflow} (left)], which we illustrate here by a comparison with the more common DMFT algorithm in Fig.~\ref{fig:dgaladderflow}  (middle): (i) First, we solve an AIM. In contrast to DMFT, where only the one-particle Green's function of the AIM is calculated, in D$\Gamma$A also the two-particle Green's function of the AIM needs to be computed. (ii) Second, we extract from the AIM the irreducible building block of our theory. In D$\Gamma$A, this requires one to invert the parquet equations of the AIM  to compute the fully 2PI local vertex, i.e., the three-frequency-dependent $\Lambda^{\nu\nu'\omega}$.   In DMFT this corresponds to calculating the AIM's  local self-energy. (iii) Third, as mentioned above, in D$\Gamma$A we use the $\Lambda^{\nu\nu'\omega}$ of the AIM as input for the parquet equation~(\ref{equ:parquedecompF}) of the finite-$d$ lattice system under consideration. The self-consistent solution of this equation together with the BSEs~(\ref{equ:decomp2PIGamma}) and the Schwinger-Dyson equation~(\ref{equ:equofmotion}) yields  the {\sl momentum}-dependent self-energy and correlation functions of the D$\Gamma$A. This step corresponds in DMFT to calculating the  Green's function, through the lattice Dyson equation with the local DMFT self-energy as an input. (iv) Finally, if the local D$\Gamma$A Green's function differs from the input local Green's function, the initial AIM is accordingly updated (yielding a new $\Lambda$). Steps (i)-(iv) are repeated until self-consistency, analogous to the DMFT self-consistency on the one-particle level.

The richer physical content of D$\Gamma$A is paid for by a higher numerical effort compared to DMFT. This applies, in particular, to steps (i) the calculation of the local vertex and (iii) solving the parquet plus BSE equations. For (i), one needs to perform an accurate numerical calculation of the two-particle Green's function of the AIM with its full dependence on three frequencies (cf.\ Sec.~\ref{sec:calver}), whereas in the DMFT loop only the one-particle Green's function is required. For (iii), we recall that a numerical solution of the parquet equations for lattice systems is highly demanding. Cutting-edge parquet algorithms have been developed~\cite{Yang2009,Tam2013,Li2016} and first  D$\Gamma$A  parquet calculations have been presented for a benzene ring~\cite{Valli2015} and the 2D Hubbard model~\cite{Li2016}.
 
\subsubsection{Ladder D\texorpdfstring{$\Gamma$}{G}A}
\label{sec:dgaladder}

A full parquet solution of the D$\Gamma$A equations is still not feasible (i) in the proximity of (quantum) phase transitions which requires a fine $\mathbf k$ grid or (ii) for {\em ab initio}  D$\Gamma$A~\cite{Toschi2011,Galler2016} calculations which require more orbitals. Hence, simplifications of the D$\Gamma$A scheme are often needed. Here we discuss the most commonly used approximation, the ``ladder approximation'', and discuss its physical justification. This route was followed by \onlinecite{Kusunose2006}, \onlinecite{Toschi2007}, and \onlinecite{Slezak2009}.

Starting from the parquet D$\Gamma$A [see Eq.~(\ref{equ:parquedecompF})], we recall that the momentum dependence of its two-particle reducible terms $\Phi_{r,\ell,\mathbf{k}\mathbf{k}'\mathbf{q}}^{\nu\nu'\omega}$ is crucial for describing second-order phase transitions, e.g., (anti)ferromagnetism ($r\!=\!s$) and charge density waves ($r\!=\!c$) in the particle-hole sector ($\ell\!=\!ph$ or $\overline{ph}$), or singlet ($r=\uparrow\downarrow$) superconductivity ($\ell\!=\!pp$). The proximity of such an instability is indicated by a strong enhancement of the momentum-dependent  $\Phi_{\ell,r,\mathbf{k}\mathbf{k}'\mathbf{q}}^{\nu\nu'\omega}$ in the corresponding scattering channel(s) defined by $\ell$ and $r$.  Hence, in all situations where the leading instability of the system is known {\sl a priori}, one can consider a momentum-dependent $\Phi_{\ell,r,\mathbf{k}\mathbf{k}'\mathbf{q}}^{\nu\nu'\omega}$ {\sl only} in the relevant channel, keeping in the other channels the  local $\Phi_{\ell,r}$'s of DMFT only. This approximation reduces the parquet equations to the BSE and hence represents a ``ladder approximation'' of D$\Gamma$A.

In the following, we explicitly recapitulate the ladder-D$\Gamma$A equations~\cite{Toschi2007,Rohringer2013a} for one of the most relevant situations: If magnetic fluctuations dominate, we can restrict ourselves to the two $ph$-channels $\ell=ph,\overline{ph}$.\footnote{In principle, we could restrict ourselves also to the spin sector $r\!=\!s$ neglecting nonlocal charge fluctuations $r\!=\!c$. However, such a simplification would break the crossing and SU(2) symmetry and, hence, we consider $r\!=\!c$ on the same footing as $r\!=\!s$.} This is also the  implementation that has been employed in most D$\Gamma$A calculations hitherto (see Sec.~\ref{Sec:results}); \onlinecite{Kusunose2006} considered the  $\ell=ph$ channel only.

As discussed above, by applying the ladder approximation to the two $ph$ channels, i.e., $ph$ and $\overline{ph}$, we assume both $\Lambda$ and $\Phi_{pp}$ in Eq.~\eqref{equ:parquedecompF} to be purely local. Hence, the expression for the full vertex entering in the equation of motion [Eq.~(\ref{equ:equofmotion})] for the D$\Gamma$A self-energy, reads 
\begin{equation}
\label{equ:dgaladderapprox}
 F_{\uparrow\downarrow,\mathbf{k}\mathbf{k}'\mathbf{q}}^{\nu\nu'\omega}=\Lambda_{\uparrow\downarrow}^{\nu\nu'\omega}+\Phi_{ph,\uparrow\downarrow,\mathbf{k}\mathbf{k}'\mathbf{q}}^{\nu\nu'\omega}+\Phi_{\overline{ph},\uparrow\downarrow,\mathbf{k}\mathbf{k}'\mathbf{q}}^{\nu\nu'\omega}+\Phi_{pp,\uparrow\downarrow}^{\nu\nu'\omega}.
\end{equation}
While the  momentum dependence of $\Phi_{pp}$ has been neglected, it  still needs to  be calculated for $\Phi_{ph}$ and  $\Phi_{\overline{ph}}$. This is possible through the BSE~(\ref{equ:decomp2PIGamma}). Within a full parquet approach the corresponding irreducible vertices $\Gamma_{ph,r}$ and $\Gamma_{\overline{ph},r}$ would be momentum dependent through mutual screening between the channels. In our ladder D$\Gamma$A approximation we do not consider such renormalization effects and, hence, assume $\Gamma_{ph,r}$ and $\Gamma_{\overline{ph},r}$ to be purely local:
\begin{align}
 \label{equ:dgabse}
  \! \!F_{\text{lad},r,\mathbf{q}}^{\nu\nu'\omega}&=\Gamma_{r}^{\nu\nu'\omega}+\Phi_{r,\mathbf{q}}^{\nu\nu'\omega}\nonumber\\&=  \Gamma_{ph,r}^{\nu\nu'\omega}+\!\sum_{\tilde{\mathbf{k}}\tilde{\nu}}\Gamma_{ph,r}^{\nu\tilde{\nu}\omega}G^{\text{loc}}_{\tilde{\mathbf{k}}\tilde{\nu}}G^{\text{loc}}_{(\tilde{\mathbf{k}}+\!\mathbf{q})(\tilde{\nu}\!+\!\omega)}F_{\text{lad},r,\mathbf{q}}^{\tilde{\nu}\nu'\omega},
\end{align}
where $r=c,s$ and  $G^{\text{loc}}_{\mathbf{k}\nu}=[G^{-1}_{0,\mathbf{k}\nu}-\Sigma^{\text{loc}}_\nu]^{-1}$ is the  DMFT Green's function of Eq.~(\ref{equ:defDMFTgf}). An analogous equation holds for the $\overline{ph}$ channel which is exactly equivalent to Eq.~(\ref{equ:dgabse}) for $r=s$ due to SU(2) and the crossing symmetry; see Table \ref{tab:symmetries}.

Equation~(\ref{equ:dgabse}) is a great algorithmic simplification with respect to  a full parquet treatment because $F_{\text{lad},r,\mathbf{q}}^{\nu\nu'\omega}$ depends on the transferred momentum $\mathbf{q}$ only, rather than on all three momenta in the parquet equation.
Combining Eqs.~\eqref{equ:dgaladderapprox} and \eqref{equ:dgabse}, the final expression for the full vertex reads
\begin{align}
\label{equ:dgaladderfinal}
 F_{\uparrow\downarrow,\mathbf{k}\mathbf{k}'\mathbf{q}}^{\nu\nu'\omega}=&\frac{1}{2}\left(F_{\text{lad},c,\mathbf{q}}^{\nu\nu'\omega}\right.\left.\!-\!F_{\text{lad},s,\mathbf{q}}^{\nu\nu'\omega}\right)-F_{\text{lad},s,\mathbf{k}'-\mathbf{k}}^{\nu(\nu+\omega)(\nu'-\nu)}+\nonumber\\&-\frac{1}{2}\left(F_{c}^{\nu\nu'\omega}\!-\!F_{s}^{\nu\nu'\omega}\right).
\end{align}
Here the purely local terms in the last line provide a double counting correction of local contributions~\cite{Toschi2007,Rohringer2013a}. Inserting Eq.~\eqref{equ:dgaladderfinal} for $F_{\uparrow\downarrow,\mathbf{k}\mathbf{k}'\mathbf{q}}^{\nu\nu'\omega}$ into the Schwinger-Dyson Eq.~(\ref{equ:equofmotion}), we finally obtain the following explicit expression for the ladder D$\Gamma$A self-energy:  
\begin{align}
\label{equ:sigmadgaladder}
 \Sigma_{\mathbf{k}\nu}=&\frac{Un}{2}-{U}\sum_{\nu'\omega}\sum_{\mathbf{k}'\mathbf{q}}F_{\uparrow\downarrow,\mathbf{k}\mathbf{k}'\mathbf{q}}^{\nu\nu'\omega}\times\nonumber\\&\times G_{\mathbf{k}'\nu'}^{\text{loc}}G^{\text{loc}}_{(\mathbf{k}'+\mathbf{q})(\nu'+\omega)}G^{\text{loc}}_{(\mathbf{k}+\mathbf{q})(\nu+\omega)}.
\end{align}
The practical implementation of Eqs.~\eqref{equ:dgaladderapprox}-\eqref{equ:sigmadgaladder} is illustrated by the flow diagram Fig.~\ref{fig:dgaladderflow} (right); cf.\ \onlinecite{Held2008}.  The calculation steps are the following: (i) A complete DMFT self-consistency cycle is performed as outlined in Fig.~\ref{fig:dgaladderflow} (middle). (ii) After  DMFT convergence the local one- and two-particle Green's functions of the AIM are computed. (iii) The irreducible local vertices in the channel(s) under consideration, $\Gamma_{\ell,r}^{\nu\nu'\omega}$, are computed via the inversion of BSEs for the AIM. (iv) The  local irreducible vertex functions $\Gamma_{\ell,r}^{\nu\nu'\omega}$ and the (momentum-dependent) DMFT Green's function $G^{\text{loc}}_{\mathbf{k}\nu}$ serve as an input for the lattice BSEs which, in turn, yield the ladder vertex $F_{\text{lad},r,\mathbf{q}}^{\nu\nu'\omega}$ and the full $F$ via Eq.~\eqref{equ:dgaladderfinal}. (v) The self-energy is derived from the DMFT Green's function $G^{\text{loc}}_{\mathbf{k}\nu}$ and the vertex $F_{\text{lad},r,\mathbf{q}}^{\nu\nu'\omega}$ by means of the equation of motion~\eqref{equ:sigmadgaladder}. 

\paragraph*{Moriyaesque \texorpdfstring{$\lambda$}{lambda} correction:}
The ladder D$\Gamma$A algorithm described above enormously reduces the numerical effort with respect to the full parquet implementation. However,  ladder D$\Gamma$A leads to violations of several sum rules for the susceptibilities which are obtained from the BSE~(\ref{equ:dgaladderapprox}), since mutual screening effects between different scattering channels are not taken into account. One example of such a sum rule, which is not fulfilled within ladder D$\Gamma$A, is related to the total density of the system and reads:
\begin{align}
\label{equ:sumchi}
  &\sum_{\omega\mathbf{q}}\chi_{\uparrow\uparrow,\mathbf{q}}^{\omega}=\sum_{\omega\mathbf{q}}\frac{1}{2}\left(\chi_{c,\mathbf{q}}^{\omega}+\chi_{s,\mathbf{q}}^{\omega}\right)=\nonumber\\&=\langle n_{\uparrow}n_{\uparrow}\rangle-\langle n_{\uparrow}\rangle\langle n_{\uparrow}\rangle=\frac{n}{2}\left(1-\frac{n}{2}\right).
\end{align}
This relation automatically guarantees the correct ($1/\nu$) high-frequency asymptotics of the self-energy in any scheme based on the equation of motion (\ref{equ:equofmotion}). A corresponding violation of Eq.~(\ref{equ:sumchi}) hence leads to an incorrect asymptotic behavior which can be indeed observed in ladder D$\Gamma$A \cite{Toschi2007,Katanin2009}.

To overcome this problem the ladder D$\Gamma$A susceptibilities obtained from Eq.~(\ref{equ:dgaladderapprox}) are supplemented by means of a Moriyasque $\lambda$ correction. Considering the Ornstein-Zernike form of charge and spin modes at momentum $\mathbf{Q_r}$
\begin{equation}
\label{equ:spinmode}
 \chi_{r,\mathbf{q}}^{\omega=0}=\frac{A}{\left(\mathbf{q}-\mathbf{Q_r}\right)^2+\xi_r^{-2}}, 
\end{equation}
and following the Moriya theory of itinerant magnetism~\cite{Moriya1985} it is natural to apply a correction to $\chi_{r,\mathbf{q}}^{\omega}$ by modifying the correlation length $\xi_r$ (i.e., the mass) of the propagator. This is consistent with the well-known fact that a mean-field theory such as DMFT {\sl overestimates} the correlation length of the system. It  accounts for nonlocal  contributions to the particle-hole-irreducible  vertices. 

Since within the ladder D$\Gamma$A scheme the propagator $\chi_{r,\mathbf{q}}^{\omega}$ (without any correction and self-consistency) corresponds exactly to the DMFT one, it is reasonable to reduce this overrated DMFT correlation length of the mode $r$, fixing it to a value such that condition (\ref{equ:sumchi}) is fulfilled. In practice this is done by applying the transformation 
\begin{equation}
 \label{equ:lambdachi}
 \left(\chi_{r,\mathbf{q}}^{\omega}\right)^{-1}\rightarrow\left(\chi_{r,\mathbf{q}}^{\omega}\right)^{-1}+\lambda_r=(\chi_{r,\mathbf{q}}^{\lambda,\omega})^{-1}.
\end{equation}
Rewriting the ladder D$\Gamma$A equation of motion in such a way that it explicitly contains the physical susceptibility and inserting  the $\lambda$-corrected susceptibilities $\chi_{r,\mathbf{q}}^{\lambda_r,\omega}$ into it leads to the $\lambda$-corrected self-energy; see \onlinecite{Katanin2009} and \onlinecite{Rohringer2016}. The relation of this procedure to the dual boson approach is discussed in Sec.~\ref{sec:db}.

Let us point out  that the  divergencies of vertex functions $\Gamma_{r}$ mentioned in  Sec.~\ref{sec:vertexphysics}  do not affect  the ladder  D$\Gamma$A algorithm. In fact, $\Gamma_{r} = F_r/(1 + G_{loc}G_{loc}F_r)$ and the Bethe-Salpter equation~(\ref{equ:dgabse})  can be reformulated in terms of the full vertex $F_r$  [see, e.g., \onlinecite{Rohringer2013a} and for  multiorbital and {\em ab initio}   D$\Gamma$A calculations, see \onlinecite{Galler2016}]:
\begin{equation}
\label{equ:ladderdganodiv} 
 F_{\text{lad},r,\mathbf{q}}^{\nu\nu'\omega} = F_r^{\nu\nu'\omega}+\!\sum_{\tilde{\mathbf{k}}\tilde{\nu}}F_{r}^{\nu\tilde{\nu}\omega}
 \tilde{G}_{0,\tilde{\mathbf{k}}\tilde{\nu}}\tilde{G}_{0,(\tilde{\mathbf{k}}+\!\mathbf{q})(\tilde{\nu}\!+\!\omega)}F_{\text{lad},r,\mathbf{q}}^{\tilde{\nu}\nu'\omega} \; ,
\end{equation}
where $\tilde{G}_{0,\mathbf{k}\nu}= G^{\text{loc}}_{\mathbf{k}\nu}-  G^{\text{loc}}_{\nu}$; cf.\ Eq.~(\ref{equ:dualgaussfinal}). This circumvents the occurrence of any 2PI vertex divergence in the ladder-D$\Gamma$A scheme (and in the calculation of DMFT susceptibilities that exploit identical Bethe-Salpeter expressions). At present,  it remains unclear whether the divergences of the 2PI vertex functions  can be circumvented similarly in parquet-based algorithms, such as the parquet D$\Gamma$A and QUADRILEX. 

\subsubsection{{\em Ab initio} D\texorpdfstring{$\Gamma$}{G}A for materials calculations}
\label{sec:abinitioDGA}

Up to this point, we have considered a single orbital and a local interaction $U$ in the D$\Gamma$A approach. An extension to  nonlocal interactions and multiple orbitals has  been developed and implemented by~\onlinecite{Galler2016} and \onlinecite{Galler2017b}, cf. \onlinecite{Galler2017a}, building upon earlier ideas put forward by~\onlinecite{Toschi2011}. Because of this {\em ab initio} material calculations are also possible and have been performed for SrVO$_3$, this variant is coined AbinitioD$\Gamma$A. As a full parquet  D$\Gamma$A for multiple orbitals is beyond of what is feasible with present-day computational resources, the key quantity in AbinitioD$\Gamma$A is the  irreducible vertex $\Gamma$ in the particle-hole (and transversal particle-hole) channel just as in ladder D$\Gamma$A. 

\begin{figure}
  \includegraphics[width=0.96\columnwidth,clip]{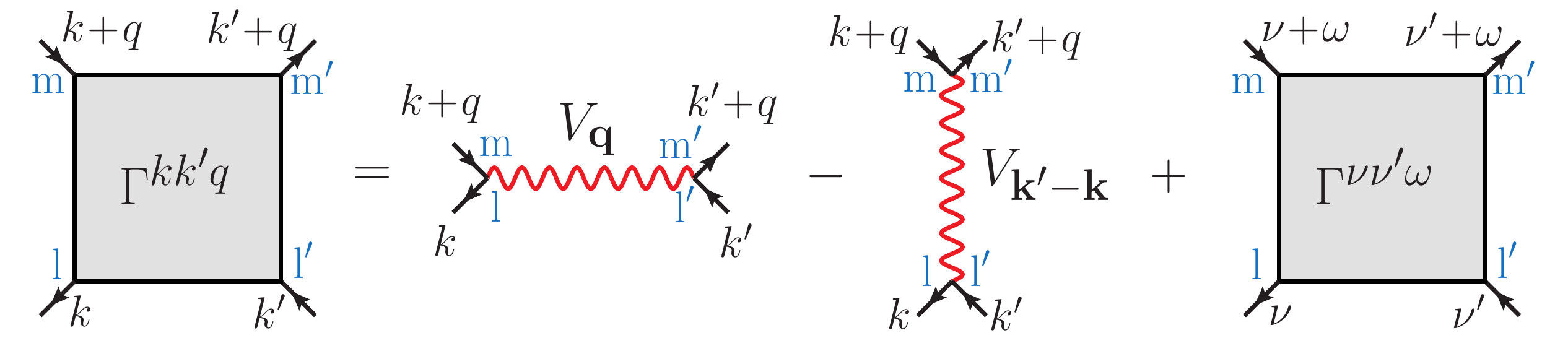}
    \caption{\label{Fig:AbinitioDGA} (Color online) In AbinitioD$\Gamma$A the irreducible vertex $\Gamma$ in the  particle-hole  and transversal particle-hole channel is approximated by the bare {\em nonlocal} Coulomb interaction $V^{\mathbf q}$ and the {\em local} vertex $\Gamma_{\rm loc}$ which depends on orbitals ($l$, $m$ ...)  and  frequencies ($\nu$, $\nu'$, $\omega$) but not momenta ($k$, $k'$, $q$) and  includes the local Coulomb interaction $U$.}
\end{figure}

The key assumption of AbinitioD$\Gamma$A is to approximate  $\Gamma$ by the corresponding local vertex plus the nonlocal Coulomb interaction ${{V}}^{\mathbf q}$, see Fig.~\ref{Fig:AbinitioDGA}:
\begin{eqnarray}
\label{eqn:gammastart}
  \Gamma^{{k}{k}'q}_{ph, \sigma\sigma',  lmm'l'} & \equiv & \Gamma^{\nu\nu'\omega}_{ ph, \sigma\sigma', lmm' l'} \nonumber \\ && + {{V}}^{{\mathbf q}}_{\sigma\sigma',lmm' l'}- \delta_{\sigma\sigma'} V^{{\mathbf k}'-{\mathbf k}}_{\sigma\sigma, mm' ll'} \; . 
\end{eqnarray}
Here, $l, m, m',$ and $l'$ denote the orbital indices, and  $\Gamma^{\nu\nu'\omega}_{ ph,\sigma\sigma',lmm' l'}$ includes the local bare interaction $U$ plus all purely local vertex corrections. In calculations with strongly and weakly correlated, say $d$ and $p$ orbitals, one can also approximate the local vertex of the $p$ orbital by $U$. This allows calculations for more orbitals since the calculation of the local vertex remains a large numerical effort. Alternatively one can take the screening of an outer window of orbitals into account, which translates into an additional frequency dependence for $U$ and  ${{V}}^{{\mathbf q}}_{\sigma\sigma',lmm' l'}$.

As in ladder D$\Gamma$A [Eqs.~(\ref{equ:dgabse}) and (\ref{equ:dgaladderapprox})], the full vertex is constructed from the vertex (\ref{eqn:gammastart}) using the BSE in the particle-hole and transversal particle-hole channel,with a reformulation in terms of $F$ instead of $\Gamma$ to avoid numerical obstacles associated with the divergences in $\Gamma$ discussed in Sec.~\ref{sec:vertexphysics}.

The multiorbital extension \cite{Galler2016} of the Schwinger-Dyson equation~(\ref{equ:sigmadgaladder})  is employed to obtain $\Sigma$ from the calculated $F$, following the  ladder D$\Gamma$A scheme. This  Schwinger-Dyson equation has various $U$ and $V^ {\mathbf q}$ terms and is not restricted any longer to the $\uparrow\downarrow$-spin combination of $F$. 

What is still neglected in AbinitioD$\Gamma$A  are nonlocal irreducible vertex corrections. But since from $\Gamma^{{k}{k}'q}$ in Eq.~(\ref{eqn:gammastart}) the particle-hole ladder and the transversal particle-hole ladder diagrams are constructed, we still retrieve many correlations originating from ${{V}}^{\mathbf q}$: (i) Inserting the  $V^{\mathbf q}$ term of Fig.~\ref{Fig:AbinitioDGA} and Eq.~(\ref{eqn:gammastart})  into the particle-hole  BSE ladder yields the random phase approximation (RPA) screening, and  from this screened $F$ the  Schwinger-Dyson equation of motion assembles the $GW$ self-energy. (ii) From the local $\Gamma^{\nu\nu'\omega}$ and local Green's function in the subsequent equations, the DMFT self-energy is recovered. Hence,  AbinitioD$\Gamma$A is a unifying framework which naturally generates all  $GW$ diagrams and all DMFT diagrams. (iii) Beyond  $GW$+DMFT, further nonlocal correlations  are included: the nonlocal spin fluctuations of ladder D$\Gamma$A, the transversal particle-hole ladder in terms of  ${{V}}^{\mathbf q}$, and mixed terms.

\subsubsection{QUADRILEX: a functional integral perspective}
\label{sec:quadrilex}

The quadruply irreducible local expansion scheme (QUADRILEX) by \onlinecite{Ayral2016} is closely related to the D$\Gamma$A. It is essentially an extension of D$\Gamma$A in two respects: (i) it provides the framework of a functional integral formalism and (ii) from this functional an additional self-consistency for the two-particle vertex. This self-consistency enters in addition to the one for the one-particle Green's function considered in D$\Gamma$A before, see Fig.~\ref{fig:dgaladderflow}. As D$\Gamma$A,  QUADRILEX is based on the fully 2PI vertex and builds Feynman diagrams around it.\footnote{Note that in contrast to the notation of this review \onlinecite{Ayral2016} denoted this as the  four-particle irreducible level since there are four (incoming and outgoing) legs associated with the two-particle (2PI in our notation) vertex, hence the name QUADRILEX.}

For constructing a functional integral,  \onlinecite{Ayral2016} employed an idea by \onlinecite{DeDominicis1964b}: The standard functional of \onlinecite{BaymKadanoff}  depends on the interacting one-particle Green's function $G$ and the bare  interaction $U$.  \onlinecite{DeDominicis1964b} proposed a Legendre transformation to a functional that depends on  $G$ {\em and} the two-particle Green's function $G^{(2)}$ instead.  As a nontrivial term it contains the set of all 2PI energy diagrams $\mathcal{K}_{4}[G,G^{(2)}]$ (\onlinecite{Ayral2016}, coined  $\mathcal{K}_{2}$ by \onlinecite{DeDominicis1964b}). In the  QUADRILEX formalism, \onlinecite{Ayral2016} approximated this functional by its local counterpart
\begin{equation}
\label{Eq:K2}
  \mathcal{K}_{4}[G_{\mathbf{k}\nu},G_{\sigma\sigma',\mathbf{k}\mathbf{k'}\mathbf{q}}^{(2),\nu\nu'\omega}] \approx \mathcal{K}_{4}[G^{\rm loc}_\nu,G_{\sigma\sigma'}^{(2),\nu\nu'\omega}].
\end{equation}
This naturally extends the DMFT, which corresponds to a Baym-Kadanoff functional that depends on $G^{\rm loc}_\nu$ only, see \onlinecite{Janis92}. The fully  2PI vertex $\Lambda$ can be obtained from the functional derivative of $\mathcal{K}_{4}$  with respect to  $G^{(2)}$: $\Lambda=U-2{\delta}\mathcal{K}_{4}/({\delta}G^{2})$. The approximation $\mathcal{K}_{4}=0$ yields $\Lambda=U$ and generates the parquet approximation. Using the local approximation (\ref{Eq:K2}) instead yields by construction the same $\Lambda$ as in D$\Gamma$A, namely all local fully 2PI diagrams. But the functional formalism also leads to an improved outer self-consistency, which now includes both the one-particle and two-particle level on an equal footing.

This can be understood as follows: As in DMFT, the approximated functional (\ref{Eq:K2}) can be determined by an auxiliary impurity model. In order to match the local part of the lattice $G^{(2)}$ to the corresponding impurity quantity with its full frequency dependence, one has to introduce an adjustable, three-frequency-dependent interaction vertex  $\mathcal{U}^{\nu\nu'\omega}$ into the impurity model in addition to the familiar Weiss field ${\cal G}$. $\mathcal{U}^{\nu\nu'\omega}$ allows for including the feedback of collective modes into the impurity model. The  D$\Gamma$A can be understood as a special case of QUADRILEX with the additional approximation $\mathcal{U}^{\nu\nu'\omega}\approx U$. For the AIM in the {\sl outer} self-consistency cycle Fig.~\ref{fig:dgaladderflow} envisages this additional step in the self-consistency loop. For more details see \onlinecite{Ayral2016}.

The QUADRILEX scheme has not yet been implemented. Notwithstanding possible convergence issues of the self-consistency cycle or a possible sign problem incurred by the retarded interaction, the implementation might be feasible with today's technology—--at least for
a single orbital.

\subsection{Dual fermion (DF) theory}
\label{sec:funcint}

It is common wisdom that models of strongly correlated electrons, such as the Hubbard model, are difficult to treat at large scale and in the thermodynamic limit. The challenge however is not the interaction per se, but the fact that the hopping term and the interaction term are diagonal in different bases, namely in momentum- and lattice-space. In the physically most interesting regimes, both terms are generally of similar order, so that an expansion around the weak- or strong-coupling limit is not applicable. A powerful idea to  approach the problem is to {\sl separate} it into nontrivial subproblems that can be treated efficiently and (numerically) exactly and a coupling between them. The splitting is ideally done in a way that this coupling represents a small parameter of the theory, which can be treated perturbatively. The simplest example of such a theory is DMFT (see Sec.~\ref{sec:dmft}). Here the lattice problem is decomposed into a collection of local Anderson impurity models (AIMs).  In DMFT the coupling between impurities is neglected. As we will see in the following, DMFT can therefore be viewed as the lowest-order perturbative treatment of the coupling between the local impurity and nonlocal degrees of freedom of the system. A perturbative treatment of the coupling reintroduces nonlocal correlations.

From a formal perspective, an action-based formalism provides the most natural basis to achieve  a decoupling into local and nonlocal degrees of freedom. In the following we will discuss this for the Hubbard model, although the concept is more general and can be applied also to other models. See, for example,~\onlinecite{Rubtsov2006} for an application to $\phi^{4}$ theory. The action of the Hubbard model \eqref{Eq:Hubbard} is given by
\begin{align}
\label{equ:actionhubbard}
 \pazocal{S}[c^+,c]=&\sum_{\mathbf{k}\nu\sigma}\left[-i\nu+\varepsilon_{\mathbf{k}}-\mu\right]c^+_{\mathbf{k}\nu\sigma}c^{\phantom +}_{\mathbf{k}\nu\sigma}\nonumber \\ +U&\sum_i\int_0^{\beta}d\tau\;c^+_{i\uparrow}(\tau)c^{\phantom +}_{i\uparrow}(\tau)c^+_{i\downarrow}(\tau)c^{\phantom +}_{i\downarrow}(\tau),
\end{align}
where $c^{(+)}_{i\sigma}(\tau)$ and $c^{(+)}_{\kv\nu\sigma}$ are the fermionic Grassmann fields corresponding to the annihilation (creation) operators $\hat{c}^{(\dagger)}_{i\sigma}$ and $\hat{c}^{(\dagger)}_{\mathbf{k}\sigma}$, respectively. In the spirit of the previous discussion we introduce a local reference action which is diagonal in lattice space:
\begin{align}
\label{equ:actionref}
 \pazocal{S}_{\text{loc}}[c^+,c]=&\sum_{\nu\sigma}\left[-i\nu+\Delta_{\nu}-\mu\right]c^+_{\nu\sigma}c^{\phantom +}_{\nu\sigma}\nonumber\\&+U\int_0^{\beta}d\tau\;c^+_{\uparrow}(\tau)c_{\uparrow}(\tau)c^+_{\downarrow}(\tau)c_{\downarrow}(\tau).
\end{align}
The Gaussian part $\Delta_{\nu}$ of the reference system is frequency and in general site dependent, but we limit our presentation to the homogeneous case. The rationale behind introducing an AIM as the reference system is that we can compute its one- and two-particle Green's functions numerically exactly as described in Sec.~\ref{sec:calver}.

Keeping in mind the idea of a separation into solvable parts and a supposedly weak coupling between them, we express the lattice action~Eq.~\eqref{equ:actionhubbard} in terms of the local reference system Eq.~\eqref{equ:actionref} by formally adding and subtracting an arbitrary hybridization function $\Delta_{\nu}$:
\begin{equation}
\label{equ:actionrew}
  \pazocal{S}[c^+,c]=\sum_{i}\pazocal{S}_{\text{loc}}[c_{i}^+,c_{i}]+\sum_{\mathbf{k}\nu\sigma}\left[\varepsilon_{\mathbf{k}}-\Delta_{\nu}\right]c^+_{\mathbf{k}\nu\sigma}c^{\phantom +}_{\mathbf{k}\nu\sigma}.
\end{equation}
The generating functional $W[\eta^+,\eta,\widetilde{\eta}^+,\widetilde{\eta}]=\ln Z[\eta^+,\eta,\widetilde{\eta}^+,\widetilde{\eta}]$ for the action~\eqref{equ:actionrew} reads:
\begin{align}
\label{equ:partit}
  Z&[\eta^+,\eta,\widetilde{\eta}^+,\widetilde{\eta}]=\int D[c^+,c]\exp\Bigl[ -\pazocal{S}_{\text{loc}}[c^+,c]\\&-\sum_{\mathbf{k}\nu\sigma}\left[\varepsilon_{\mathbf{k}}-\Delta_{\nu}\right]C^+_{\mathbf{k}\nu\sigma}C^{\phantom +}_{\mathbf{k}\nu\sigma}+ \sum_{\mathbf{k}\nu\sigma}c^+_{\mathbf{k}\nu\sigma}\eta^{\phantom +}_{\mathbf{k}\nu\sigma}+\eta^+_{\mathbf{k}\nu\sigma}c^{\phantom +}_{\mathbf{k}\nu\sigma}\Bigr],\notag
\end{align}
where $C^{(+)}_{\mathbf{k}\nu\sigma}=c^{(+)}_{\mathbf{k}\nu\sigma}+b^{-1}_{\nu\sigma}\widetilde{\eta}^{(+)}_{\mathbf{k}\nu\sigma}$, and $\widetilde{\eta}^{(+)}_{\mathbf{k}\nu\sigma}$ represent so-called dual source fields which have been introduced to better reveal the connection between dual and physical fermion correlation functions. The coupling $b_{\nu\sigma}$  denotes a --in principle arbitrary-- function of spin and Matsubara frequencies. The derivatives of the functional \eqref{equ:partit} with respect to the source fields $\eta^{+},\eta$ yield the connected physical correlation functions of the system. For example, the single-particle Green's function is obtained as $G=\left.\partial^{2}W/\partial\eta^{+}\partial\eta\right|_{\eta^{(+)}=\widetilde{\eta}^{(+)}=0}$.

The central step of the DF derivation is to {\sl decouple} the reference system of impurities which are connected through the term  $\varepsilon_{\mathbf{k}}-\Delta_{\nu}$ by introducing new fields $\widetilde{c}^{(+)}$ representing the so-called {\sl dual fermions}. This is achieved through the Hubbard-Stratonovich transformation \cite{Rubtsov2008}:
\begin{widetext}
\begin{align}
\label{equ:hs}
  e^{-\left[\varepsilon_{\mathbf{k}}-\Delta_{\nu}\right]C^+_{\mathbf{k}\nu\sigma}C^{\phantom +}_{\mathbf{k}\nu\sigma}}=
  \frac{-1}{\prod_{\mathbf{k}\nu\sigma} b^{2}_{\nu\sigma}\left[\varepsilon_{\mathbf{k}}-\Delta_{\nu}\right]^{-1}}
  \int  \!D[\widetilde{c}^+,\widetilde{c}] \, e^{b^2_{\nu\sigma}\left[\varepsilon_{\mathbf{k}}-\Delta_{\nu}\right]^{-1}\widetilde{c}^+_{\mathbf{k}\nu\sigma}\widetilde{c}^{\phantom +}_{\mathbf{k}\nu\sigma} + b^{\phantom +}_{\nu\sigma}\left[c^+_{\mathbf{k}\nu\sigma}\widetilde{c}^{\phantom +}_{\mathbf{k}\nu\sigma}+\widetilde{c}^+_{\mathbf{k}\nu\sigma}c^{\phantom +}_{\mathbf{k}\nu\sigma}\right]+ {\widetilde{\eta}^+_{\mathbf{k}\nu\sigma}\widetilde{c}^{\phantom +}_{\mathbf{k}\nu\sigma}+\widetilde{c}^+_{\mathbf{k}\nu\sigma}\widetilde{\eta}^{\phantom +}_{\mathbf{k}\nu\sigma}}}.
\end{align}
\end{widetext}
The label ``dual'' emphasizes that no approximation is made in this step, analogous to the transformation of a vector to the dual vector space. Note that we use here the Hubbard-Stratonovich transformation in a rather unconventional way, namely to decouple the {\sl Gaussian} rather than the interacting part of the action. 

When applying the transformation~\eqref{equ:hs} to~\eqref{equ:partit}, the combination of the terms $e^{-\mathcal{S}_{\text{loc}}[c^+,c]}$ and $e^{b^{\phantom +}_{\nu\sigma}\left[c^+_{\mathbf{k}\nu\sigma}\widetilde{c}^{\phantom +}_{\mathbf{k}\nu\sigma}+\widetilde{c}^+_{\mathbf{k}\nu\sigma}c^{\phantom +}_{\mathbf{k}\nu\sigma}\right]}$, integrated over the physical fields $c^+$ and $c$, yields the functional $W_{\text{loc}}[b\widetilde{c}^+,b\widetilde{c}] = \ln Z_{\text{loc}}[b\widetilde{c}^+,b\widetilde{c}]$ which is  diagonal in real space:
\begin{align}
 \label{equ:defgenref}
 W_{\text{loc}}[b\widetilde{c}^+,b\widetilde{c}]=\ln\int &D[c^+,c]\prod_{i}e^{-\pazocal{S}_{\text{loc}}[c_{i}^+,c_{i}]}\nonumber\\ \times&e^{\sum_{\nu\sigma}b_{\nu\sigma}\left[c^+_{i\nu\sigma}\widetilde{c}_{i\nu\sigma}+\widetilde{c}_{i\nu\sigma}^+c_{i\nu\sigma}\right]}.
\end{align}
We aim to obtain a theory which contains dual (fermionic) variables only. We therefore expand $W_{\text{loc}}$ in terms of the local coupling between dual and physical fermions and formally integrate out the latter. Because of the exponential containing $\pazocal{S}_{\text{loc}}$, this integral corresponds to an average over the reference system. At expansion order $2n$ in the fields $c^+$ and $c$, one therefore obtains the connected part of the $n$-particle impurity Green's function. We use the freedom to choose $b_{\nu\sigma}$ to obtain a particularly convenient form of the result. Setting $b_{\nu\sigma}=(G^\text{loc}_\nu)^{-1}$, where $G^\text{loc}_\nu$ is the single-particle impurity Green's function, removes the external Green's function legs of the $n$-particle (connected) correlation functions. The local generating functional can then be expressed in the form
\begin{align}
\label{equ:genrefexpand}
  &W_{\text{loc}}[b\widetilde{c}^+,b\widetilde{c}]=\notag\\&\sum_i\left[\ln Z_{\text{loc}}-\sum_{\nu\sigma}(G^\text{loc}_\nu)^{-1}\widetilde{c}_{i\nu\sigma}^+\widetilde{c}_{i\nu\sigma}+V_{\text{eff}}[\widetilde{c}^+_i,\widetilde{c}_i]
  \right],
\end{align}
where we have defined the effective interaction between dual fermions:
 \begin{align}
\label{equ:vdef}  
  & V_{\text{eff}}[\widetilde{c}^+_i,\widetilde{c}_i]=\notag\\&
  \frac{1}{4}\sum_{\substack{\nu\nu'\omega \\ \sigma\sigma'}}\left(2-\delta_{\sigma\sigma'}\right)F_{\sigma\sigma'}^{\nu\nu'\omega}\;\widetilde{c}^+_{i\nu\sigma}\widetilde{c}_{i(\nu+\omega)\sigma}\widetilde{c}^+_{i(\nu'+\omega)\sigma'}\widetilde{c}_{i\nu'\sigma'}+\ldots
\end{align}
The interaction contains the local $n$-particle {\sl vertex} functions of the reference system. In particular, $F_{\sigma\sigma'}^{\nu\nu'\omega}$ is the two-particle vertex function. We have omitted three-particle and higher-order terms as these are often neglected in practical calculations. 

We define the dual action $\widetilde{\pazocal{S}}$ through
\begin{align}
\label{equ:defdualaction}
 \widetilde{\pazocal{S}}[\widetilde{c}^+,\widetilde{c}]=
 -\sum_{\mathbf{k}\nu\sigma}\widetilde{G}_{0,\mathbf{k}\nu}^{-1}\widetilde{c}^+_{\mathbf{k}\nu\sigma}
 \widetilde{c}_{\mathbf{k}\nu\sigma}+\sum_iV_{\text{eff}}[\widetilde{c}^+_i,\widetilde{c}_i],
\end{align}
and introduce the bare dual Green's function
\begin{eqnarray}
\label{equ:dualgaussfinal}
  \widetilde{G}_{0,\mathbf{k}\nu}&=& \left[(G^\text{loc}_\nu)^{-1} + (\Delta_{\nu}-\varepsilon_{\mathbf{k}})\right]^{-1} - G^\text{loc}_\nu.
\end{eqnarray}
With these definitions, the final form of the generating functional in terms of dual fields is given by
\begin{widetext}
\begin{equation}
\label{equ:partitdual}
  \!\! W[\eta^+,\eta,\widetilde{\eta}^+,\widetilde{\eta}]=\ln\!\!\int\! \!D[\widetilde{c}^+,\widetilde{c}]\,e^{- \widetilde{\pazocal{S}}[\widetilde{c}^+,\widetilde{c}]-\sum_{\mathbf{k}\nu\sigma}[\varepsilon_{\mathbf{k}}-\Delta_{\nu}]^{-1}(G^\text{loc}_\nu)^{-1}[\widetilde{c}^+_{\mathbf{k}\nu\sigma}\eta^{\phantom +}_{\mathbf{k}\nu\sigma}+\eta^+_{\mathbf{k}\nu\sigma}\widetilde{c}^{\phantom +}_{\mathbf{k}\nu\sigma}]+[\varepsilon_{\mathbf{k}}-\Delta_{\nu}]^{-1}\eta^+_{\mathbf{k}\nu\sigma}\eta^{\phantom +}_{\mathbf{k}\nu\sigma} +\widetilde{\eta}^+_{\mathbf{k}\nu\sigma}\widetilde{c}^{\phantom +}_{\mathbf{k}\nu\sigma}+\widetilde{c}^+_{\mathbf{k}\nu\sigma}\widetilde{\eta}^{\phantom +}_{\mathbf{k}\nu\sigma}}.
\end{equation}
\end{widetext}
From Eq.~\eqref{equ:partitdual}, the relation between the correlation functions for real and dual fields is easily derived: the single-particle propagator $ \widetilde{G}_{\mathbf{k}\nu}=-\langle\widetilde{c}^{\phantom +}_{\mathbf{k}\nu\sigma}\widetilde{c}^+_{\mathbf{k}\nu\sigma}\rangle$ of the dual fields is obtained from a functional derivative with respect to $\widetilde{\eta}^+$ and $\widetilde{\eta}$ with the sources set to zero.

A closer look at the term in square brackets in Eq.~(\ref{equ:dualgaussfinal}) reveals that it equals  $G^\text{loc}_{{\mathbf k}\nu}$, the lattice Green's function with the self-energy taken from the local reference problem, as also defined in the context of DMFT [Eq.~\eqref{equ:defDMFTgf}]. The bare dual Green's function $\widetilde{G}_{0,\mathbf{k}\nu}=G^\text{loc}_{{\mathbf k}\nu}-G^\text{loc}_{\nu}$ can be interpreted as its nonlocal part. A diagrammatic expansion in terms of the dual Green's function accounts for nonlocal contributions, while the local ones are taken into account on the level of the impurity model. It is intuitively clear that double counting of local contributions is avoided with this construction. The derivative of $W[\eta^+,\eta,\widetilde{\eta}^+,\widetilde{\eta}]$ with respect to the sources $\eta^{+}$ and $\eta$ yields, on the other hand, the physical Green's function $G_{\mathbf{k}\nu}$. Applying this functional derivative to Eq.~(\ref{equ:partitdual}) straightforwardly leads to
\begin{eqnarray}
G_{\mathbf{k}\nu} &=& [\varepsilon_{\mathbf{k}}-\Delta_{\nu}]^{-1}(G^\text{loc}_\nu)^{-1}\widetilde{G}_{\mathbf{k}\nu}(G^\text{loc}_\nu)^{-1}[\varepsilon_{\mathbf{k}}-\Delta_{\nu}]^{-1}\nonumber\\&& -[\varepsilon_{\mathbf{k}}-\Delta_{\nu}]^{-1},
\label{equ:relationrealdual}
\end{eqnarray}
which can be rewritten as a relation between the dual and the physical self-energy in the following form:
\begin{equation}
 \label{equ:relationrealdualsigma}
 \Sigma_{\mathbf{k}\nu}=\Sigma_{\nu}^{\text{loc}}+\frac{\widetilde{\Sigma}_{\mathbf{k}\nu}}{1+G^{\text{loc}}_{\nu}\widetilde{\Sigma}_{\mathbf{k}\nu}}.
\end{equation}
Analogous relations between higher-order correlation functions are obtained similarly from higher-order derivatives, see \onlinecite{Brener08} and \onlinecite{Rubtsov2009}. Using these exact relations, any result obtained in dual space can be transformed back to the physical fermion space.

Because of the complicated form of the dual interaction $V_{\text{eff}}$, the benefits of the transformation to dual variables are not immediately obvious. The idea is that the bare dual propagator $\widetilde{G}$ and the bare dual interaction $V_{\text{eff}}$ implicitly contain the local physics through the underlying AIM and  represent a much better starting point for any kind of perturbative expansion than the original action.

Note that since the hybridization is arbitrary, DF provides an expansion around a generic AIM. In the particular case where $\Delta_{\nu}$ equals its DMFT value, the DF approach represents a diagrammatic expansion around DMFT.
In this case it is easy to see that inserting the corresponding $\widetilde{G}_{0,\mathbf{k}\nu}$ into~Eq.~(\ref{equ:relationrealdual}) indeed yields the DMFT Green's function. DMFT therefore corresponds to a system of noninteracting dual fermions and appears as the lowest order in the approach. It is believed that the DF series delivers a good practical convergence even in cases where standard Feynman diagrammatic techniques fail.

\begin{figure}[t!]
  \centering
  \includegraphics[width=0.35\textheight]{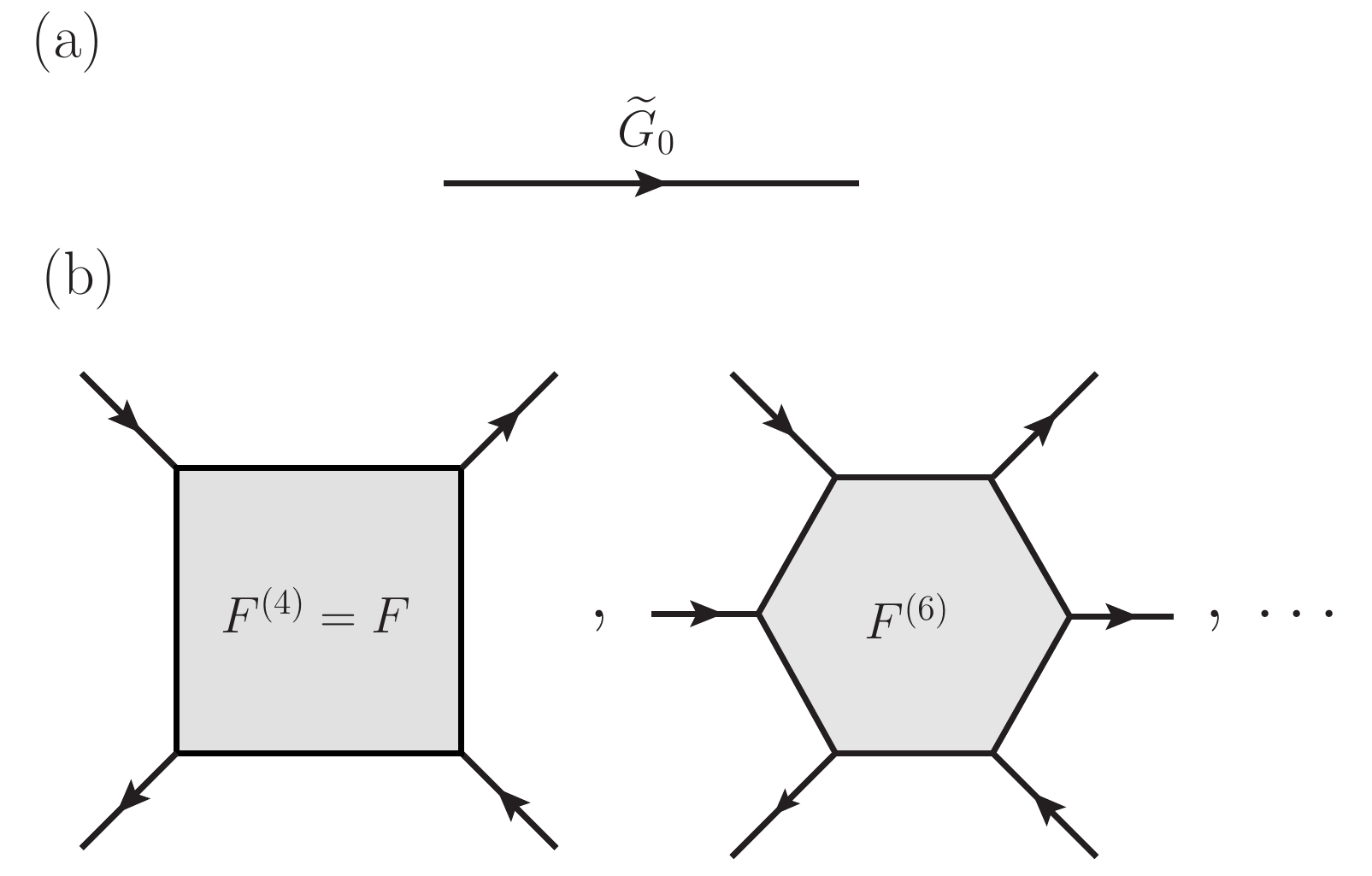}
    \caption{Diagrammatic elements of the DF approach: (a) The bare DF propagator $\widetilde{G}_0$ [Eq.~(\ref{equ:dualgaussfinal})] given by the purely {\sl nonlocal} Green's function and (b) the DF interaction represented by the local connected $n$-particle vertex functions $F^{(2n)}$.}
  \label{fig:dfdiagramelements}
\end{figure}

In practice, two approximations are performed for the action in Eq.~(\ref{equ:defdualaction}): (i) The dual interaction is terminated at some finite-order, typically only the local two-particle vertex function $F^{(4)}=F$ is taken into account and (ii) an approximation to the dual self-energy is constructed using Feynman diagrams. 

To which extent (i) the truncation at the two-particle level is justified is an open question.  \onlinecite{Hafermann2009} and \onlinecite{Ribic2017b} calculated the three-particle vertex $F^{(6)}$, allowing for an error estimate of a DF calculation truncated at the two-particle vertex level. Their results for the 2D Hubbard model indicate that  DF corrections from $F^{(6)}$ are small for some parameters but can be sizable for others. Numerical results and scaling considerations (see Sec.~\ref{sec:dfscaling}) suggest that contributions from higher-order vertices can often be neglected. Certainly a more systematic study of the whole parameter space is mandatory as is a similar analysis for the other diagrammatic extensions of DMFT.

\subsubsection{Selection of diagrams}
\label{sec:diagramchoice}

\begin{figure}[t!]
  \centering
  \includegraphics[width=0.5\textwidth]{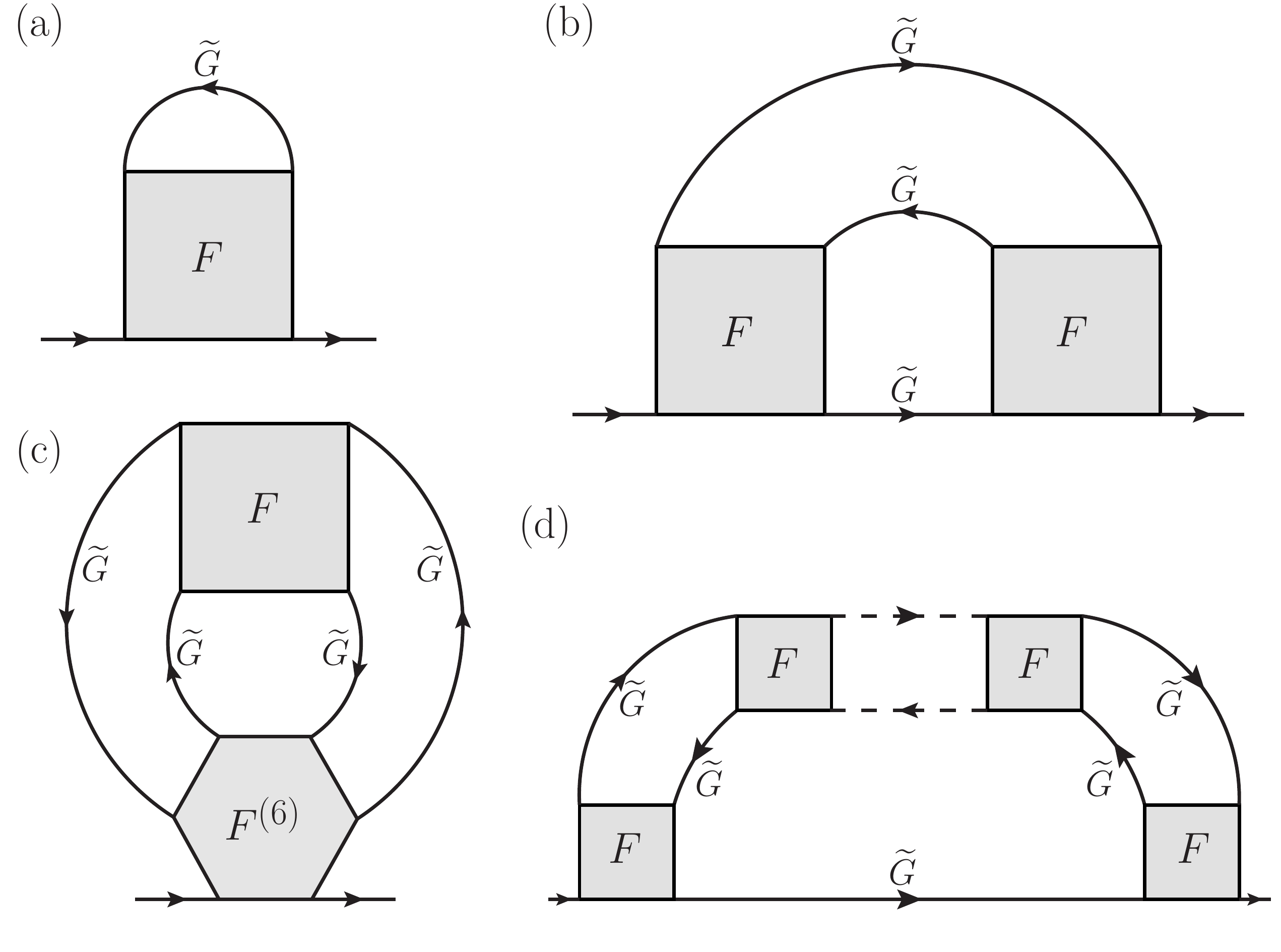}
    \caption{(a) DF Hartree-Fock and (b) second-order diagram constructed from the local two-particle vertex. (c) An example of a diagram containing the three-particle vertex. (d) Generic DF ladder diagram constructed from two-particle vertices.}
  \label{fig:dfsecondorder}
\end{figure}

The diagrammatic elements of the expansion are the dual propagators $\widetilde{G}_{0}$ and the $n$-particle vertices of the local reference model; see Fig.~\ref{fig:dfdiagramelements}. In practice, the dual self-energy is constructed from a subset of finite or infinite-order dual diagrams. Generic examples are shown in Fig.~\ref{fig:dfsecondorder}. Since the $n$-particle vertex functions in the dual interaction are fully antisymmetric by virtue of the fermionic anticommutation relations, the diagrammatic rules of the perturbation theory~\cite{Negele1998,Hafermannphd} are similar to those of \onlinecite{Hugenholtz1957}. Because the coupling $b_{\nu}$ introduced in Eq.~\eqref{equ:hs} between physical and dual fermions is local and spin-diagonal, the choice of diagrams is very similar to regular perturbation theory and can be guided by the physics. The nonlocal expansion scheme discussed later shows that the DF approach is in fact an efficient scheme to resum certain classes of diagrams in lattice fermion space. Exact relations between dual and physical fermions further guarantee that the poles corresponding to two-particle excitations (and higher-order processes) are the same for dual and physical fermions~\cite{Brener08}.

The leading-order diagram of the expansion is the local Hartree-Fock-type diagram shown in Fig.~\ref{fig:dfsecondorder} (a). This diagram gives a local contribution which can be eliminated by a specific choice of the hybridization function (see Sec.~\ref{sec:choosedeltanu}). The next-leading diagram shown in Fig.~\ref{fig:dfsecondorder} (b) defines the second-order DF approximation DF$^{(2)}$. Figure~\ref{fig:dfsecondorder}(c) is the leading diagram containing the three-particle vertex. DF$^{(2)}$ gives rise to a pseudogaplike behavior in the weak-to-intermediate coupling regime of the 2D Hubbard model in the symmetry broken phase~\cite{Rubtsov2008}. An important approximation is the ladder DF approximation~\cite{Hafermann2009}, which sums generic ladder diagrams shown in Fig.~\ref{fig:dfsecondorder}(d) to all orders. The ladder approximation describes antiferromagnetic fluctuations and the pseudogap in the paramagnetic phase and yields accurate results in practice over a wide parameter range~\cite{Gukelberger2016}. The particle-hole ladder DF self-energy is given by
\begin{align}
  \widetilde{\Sigma}_{\mathbf{k}\nu} =& - \sum_{k'q r} A_r F_{r}^{\nu\nu'\omega}\widetilde{G}_{\mathbf{k}'\nu'}\widetilde{G}_{(\mathbf{k}'+\qv)(\nu'+\omega)}\widetilde{G}_{(\mathbf{k}+\qv)(\nu+\omega)}\notag\\
  &\times[F_{\text{lad},r,\qv}^{\nu\nu'\omega}-\frac{1}{2}F_{r}^{\nu\nu'\omega}] \; .
\label{eq:sigmadf}
\end{align}
Here we have introduced the notations $A_\text{c}=1/2$, $A_\text{s}=3/2$ for $r=c,s$, where the factor of $3$ accounts for the spin degeneracy. The expression is obtained by incorporating ladder diagrams from the $ph$ and $\overline{ph}$ channels (cf. Sec.~\ref{sec:formalism}) into the lattice vertex in the dual Schwinger-Dyson equation. Since the interaction here is fully antisymmetric, both channels give identical contributions. $F_{\text{lad}}$ is the vertex in the ladder approximation as defined in Eq.~\eqref{equ:dgabse}. The latter can equivalently be obtained by solving a Bethe-Salpeter equation written in terms of $F$ and $\widetilde{G}$ [c.f.~(\ref{equ:ladderdganodiv})]~\cite{Brener08}.

Particle-particle fluctuations can be added straightforwardly. When it is known a priori in which channel the dominant instability occurs (magnetic, charge, superconducting) it is sufficient to construct ladder diagrams in this channel, whereas for competing instabilities one has to resort to parquet diagrams in dual space. To avoid a possible bias through the restriction to a certain subset of diagrams, ~\onlinecite{Iskakov2016} and~\onlinecite{Gukelberger2016} developed a method to sample DF diagrams with the two-particle vertex using diagrammatic Monte Carlo, which shows good agreement with diagrammatic determinant Monte Carlo benchmarks. 

\subsubsection{Choice of hybridization function}
\label{sec:choosedeltanu}

The hybridization function $\Delta_{\nu}$ can be chosen arbitrarily, and DMFT may or may not be the optimal starting point in the presence of strong nonlocal correlations and in low dimensions. Since the underlying AIM can be solved numerically exactly, it is  desirable to include a major part of the correlations into this reference system. The hybridization of the AIM is updated iteratively until it fulfills either the condition
\begin{equation}
\label{equ:DMFTcondition2}
  \sum_{\mathbf{k}} G_{\mathbf{k}\nu}=G^\text{loc}_\nu,
\end{equation}
with $G_{\mathbf{k}\nu}$ from Eq.~\eqref{equ:relationrealdual}, or the more commonly employed condition
\begin{equation}
\label{equ:DMFTconditionDual2}
  \sum_{\mathbf{k}}\widetilde{G}_{\mathbf{k}\nu}=0.
\end{equation}
The latter condition implies that nonlocal dual self-energy contributions that yield contributions to the local dual Green's function will effectively be absorbed into the impurity problem. Furthermore, the Hartree diagram, Fig.\ref{fig:dfsecondorder}(a), and all diagrams which contain a local loop vanish.

The effect of  updating the impurity  is particularly important in low dimensions, where DMFT and the exact solution can differ qualitatively. It is, for example, essential to capture the Mott phase in one and two dimensions in the parameter region where DMFT yields a metallic solution. The insulating solution shown in Fig.~\ref{fig:gloc1d} has been obtained from a metallic DMFT starting point using condition~\eqref{equ:DMFTconditionDual2}, while~Eq.(\ref{equ:DMFTcondition2}) yields a strongly renormalized metal in this regime. Note that the final hybridization in the former case is qualitatively different from the DMFT one, because the impurity model itself becomes insulating.
When to apply which self-consistency condition has so far not been investigated systematically.

\begin{figure}[t!]
  \centering
  \includegraphics[width=0.5\columnwidth]{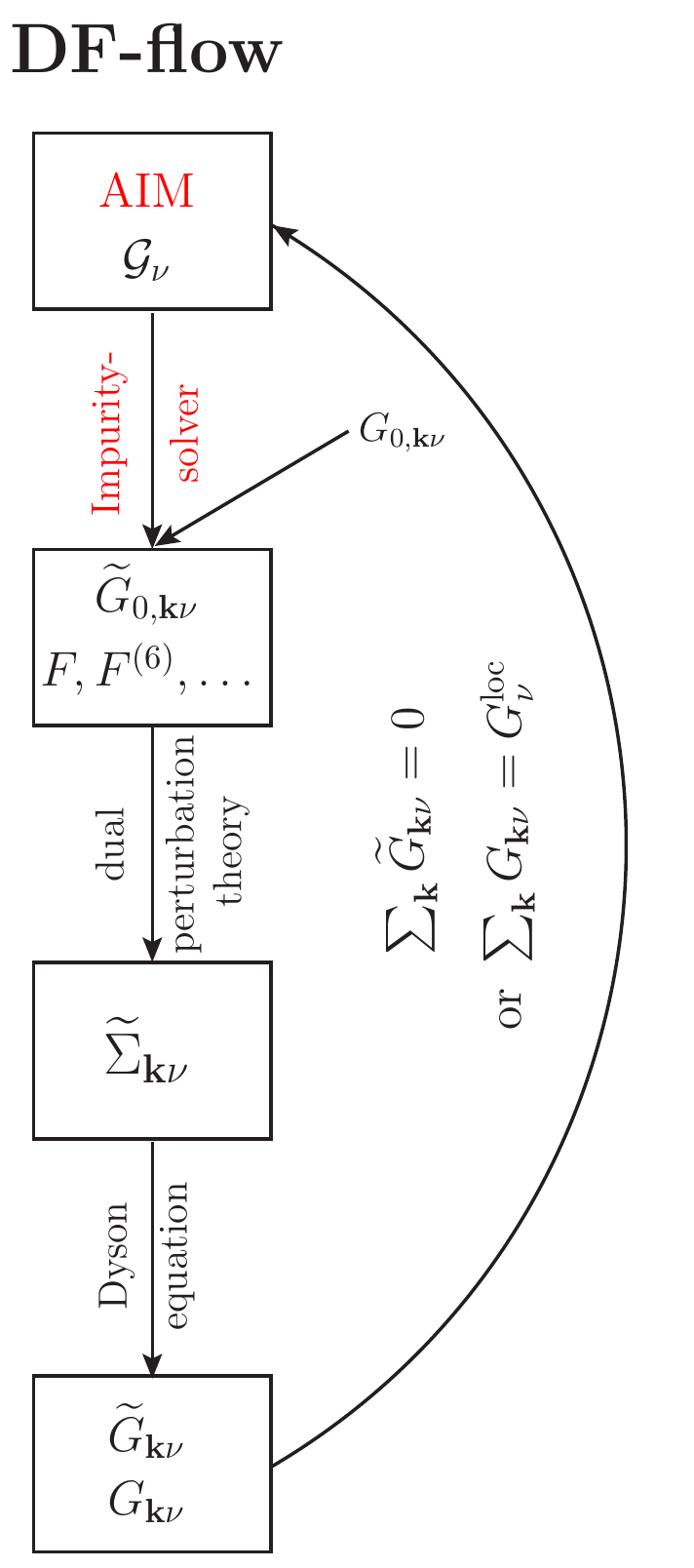}
    \caption{Flow diagram for DF calculations.}
  \label{fig:df_flow}
\end{figure}

In practice, the condition (\ref{equ:DMFTcondition2}) or (\ref{equ:DMFTconditionDual2}) is implemented in a self-consistent scheme as shown in Fig.~\ref{fig:df_flow}: (i) First, the local self-energy $\Sigma^{\text{loc}}$ and the local two-particle and possibly higher-order vertex functions ($F,F^{(6)},\ldots$) are extracted from an initial solution of the AIM. From the former, the bare dual propagator $\widetilde{G}_{0,\mathbf{k}\nu}$ is constructed according to Eq.~(\ref{equ:dualgaussfinal}). Here the bare dispersion $\varepsilon_{{\mathbf k}}$ or the  noninteracting lattice Green's function $G_{0,\mathbf{k}\nu}$ enters  the flow diagram. (ii) The dual self-energy $\widetilde{\Sigma}_{\mathbf{k}\nu}$ is calculated by means of dual perturbation theory, which can include a self-consistent renormalization of the dual Green's function in the selected diagrams. (iii) From the dual self-energy the dual Green's function $\widetilde{G}_{\mathbf{k}\nu}$ is obtained via the Dyson equation and finally the lattice Green's function $G_{\mathbf{k}\nu}$ via the transformation \eqref{equ:relationrealdual}. (iv) The hybridization is updated using either $\widetilde{G}_{\mathbf{k}\nu}$ or $G_{\mathbf{k}\nu}$ to fulfill the condition \eqref{equ:DMFTcondition2} or \eqref{equ:DMFTconditionDual2}. This closes the self-consistency cycle. 

\subsubsection{Scaling considerations and convergence}
\label{sec:dfscaling}

Since DF can be viewed as an expansion around DMFT, it is further instructive to consider the scaling from the perspective of an expansion in $1/d$. In large dimensions, the hopping scales as $t\sim\mathcal{O}(1/\sqrt{d})$ and hence the same holds for the purely nonlocal $\tilde{G}\sim\mathcal{O}(1/\sqrt{d})$. The leading second-order, ladder diagrams [Fig.~\ref{fig:dfsecondorder} (d)] and parquet scale as $\mathcal{O}(1/\sqrt{d}^{3})$ in real space. The diagram Fig.~\ref{fig:dfsecondorder} (c) and other diagrams with higher-order vertices are of  $\mathcal{O}(1/d^{2})$.

The DF approach further converges quickly around both noninteracting and strong-coupling limits. It essentially inherits this property from DMFT. In the weak-coupling limit $U\to0$ the vertex functions are small. In the atomic limit $\epsilon_{\mathbf{k}}\to 0$, the dual Green's function for $\Delta\equiv 0$ becomes the small parameter: $\tilde{G}_{\mathbf{k}\nu}\approx \epsilon_{\mathbf{k}}(G^{\text{loc}}_\nu)^{2}$. For finite $\Delta\neq 0$ the leading eigenvalue of the ladder in dual space is significantly reduced compared to the one of the physical fermions also at intermediate coupling, indicating faster convergence~\cite{Hafermann2009}.

\subsubsection{Generalizations}
We have discussed the derivation of the DF approach for translationally invariant lattices. The method can also be straightforwardly generalized to other scenarios. It can be derived in real space~\cite{Takemori2016a} to address spatially inhomogeneous and finite systems, to disordered systems  (Sec.~\ref{sec:disorder}) and the  symmetry broken phase~\cite{Rubtsov2009}. Susceptibilities can be calculated to detect symmetry broken phases~\cite{Li2008,Brener08}. The formalism can be generalized straightforwardly also to the multiorbital case and to clusters, where the DF expansion is performed around a small cluster as the reference system (see Sec.~\ref{sec:clusterdf}).

\subsubsection{DF as a cluster solver}
DF can be employed as an approximate solver for large clusters occurring in the context of cluster extensions of DMFT~\cite{Maier2005}. We note that this is conceptually different from expanding around a cluster reference system, which we discuss in Sec.~\ref{sec:cluster}. Here DF implementations often work with a discretized grid and Fourier transforms and hence solve large but finite lattices with periodic boundary conditions. From this perspective, the DF cluster must be embedded into a mean field for including the third length scale beyond the extension of the cluster~\cite{Yang2013}. Performing the DCA coarse graining on the DF lattice significantly enhances the convergence with system size and facilitates reaching the thermodynamic limit.

\paragraph{Superperturbation theory}
\label{sec:spert}

While DF can be viewed as a diagrammatic extension of DMFT, the DF idea of performing a diagrammatic  expansion around a reference problem that can be easily solved is more general. We refer to this kind of perturbation theory around a nontrivial, for example interacting, starting point as a superperturbation.

We have the freedom to expand around a local reference system \eqref{equ:actionref} with a discrete hybridization $\Delta_{\nu}\to\Delta^{(N)}_{\nu} = \sum_{k=1}^{N}\lvert V_{k}\rvert^{2}/(i\nu - \epsilon_{k})$ and a small number $N$ of bath sites, which can be solved with ED; see Sec.~\ref{sec:calver}. For $N=0$ we obtain the strong-coupling expansion of the Hubbard model~\cite{Pairault1998,Pairault2000}. For very small $N=1,2$ the DF perturbation also accounts for the local physics not captured in the simple reference system. The choice of the hybridization parameters is not unique~\cite{Jung2010}. The solution however takes only seconds on a modern PC and allows for analytical continuation using Pad\'e approximants (see Sec.~\ref{sec:hubbard}). 

The approach can also be applied as a solver for the local impurity problem~\eqref{equ:actionref} itself, taking an impurity with hybridization $\Delta^{(N)}_{\nu}$ as the reference~\cite{Hafermann2009b}. One can analytically show that a first-order approximation to the dual self-energy reproduces the strongly hybridized weak-coupling and the weakly hybridized strong-coupling limits. For $\Delta^{(N)}_{\nu}\equiv 0$  and in the limit of small hybridization $\Delta_{\nu}$, it reproduces the noninteracting limit and the result of a first-order expansion of the Green's function in the hybridization: $G_{12} \approx g_{12} + g_{12}\, \beta\, {\rm Tr}[g\,\Delta] + \chi_{1234}\Delta_{43}$~\cite{Dai2005}. The one- and two-particle Green's functions $g_{12}$ and $\chi_{1234}$ of the reference system can be expressed in terms of the ED eigenvalues and matrix elements and can be analytically continued to the real axis~\cite{Jung2011}. The naive expansion exhibits a causality problem, which can however be cured~\cite{Jung2010} by introducing a renormalization parameter~\cite{Krivenko2010}.

Superperturbation theory can further be formulated on the Keldysh contour~\cite{Jung2012}, allowing 
for the nonequilibrium solution of the AIM. \onlinecite{Jung2012} studied the time evolution after switching on the hybridization of the AIM. \onlinecite{Munoz2013} addressed the nonlinear conductance through a quantum dot. They developed a first-order DF expansion around the symmetric, interacting AIM as the reference system in terms of the level energy $E_{d}$ of the dot up to order $\mathcal{O}(V^{2})$ in the coupling $V$ to the leads. The reference system was solved within renormalized perturbation theory around the strong-coupling fixed point. Contrary to perturbation theory in $U$, which preserves current conservation only in the particle-hole symmetric AIM, the DF scheme is current-conserving beyond $\mathcal{O}(V^{2})$. \onlinecite{Merker2013} found good agreement with numerical renormalization group calculations for the linear conductance.

\paragraph{Non-local expansion scheme}
\label{sec:nonlocalexp}

The nonlocal expansion scheme~\cite{Li2015}  is a general framework to construct approximations for strongly correlated systems that includes fluctuations at all length scales. Closely related to the DF approach, the action $\mathcal{S}$ [Eq.~(\ref{equ:actionhubbard})] of a model with local interaction  is separated into an arbitrary local reference system $\mathcal{S}^{\text{loc}}$ (not necessarily an impurity model) and a term containing a nonlocal hybridization $V_{ij,\nu}\Let [\mathcal{G}^{-1}_{\nu}]_{ij}$ (which equals the nonlocal part of the bare propagator):
\begin{align}
  \mathcal{S}[c^{+},c]
  &= \sum_{i}\mathcal{S}^{\text{loc}}[c^{+}_{i},c_{i}]+\sum_{i\neq j}\sum_{\nu\sigma}c_{i\nu\sigma}^{+}V_{ij\nu}c_{j\nu\sigma}
  \label{eq:nlexpschemeaction} \; .
\end{align}
Instead of introducing dual variables, the nonlocal expansion scheme is generated by expanding the lattice Green's function directly in the nonlocal hybridization:
\begin{align}
  G_{kl,\nu}&=-\frac{1}{\mathcal{Z}}\prod_{i=1}^{N}\int D[c_{i}^{+},c^{\phantom +}_{i}]e^{-\mathcal{S}^{\text{loc}}[c_{i}^{+},c_{i}]} \sum_{n=0}^{\infty}\frac{(-1)^{n}}{n!}\nonumber \\ &\times\Big[\sum_{i\neq j}\sum_{\nu'\sigma'}c_{i\nu'\sigma'}^{+}V_{ij,\nu'}c^{\phantom +}_{j\nu'\sigma'}\Big]^n c_{k\nu'\sigma'}^{\phantom +}c_{l\nu'\sigma'}^{+}.
\end{align}
The cumulant expansion~\cite{Sarker1988,Metzner1991} is obtained for $V_{ij}=t_{ij}$. Similar to DF, the path integral over the fermionic fields weighted with the exponential of the local action generates the local correlation functions of the local interacting system. The DF approach can be understood as a particular diagrammatic resummation scheme in the nonlocal expansion scheme: If we take the nonlocal hybridization  $V_{ij}$ as the  Fourier transform of $-(\Delta_{\nu}-\epsilon_{\mathbf{k}})$, the actions \eqref{equ:actionrew} and \eqref{eq:nlexpschemeaction} take the same form. The bare dual propagator then corresponds to the renormalized hybridization $\tilde{V}_{ij}=([V^{-1}-G^{\text{loc}}\mathbbm{1}]^{-1})_{ij}$ [note that~\onlinecite{Li2015} used couplings $b_{\nu\sigma}=1$]. Using the resummation $G_{\mathbf{k}\nu}=\Lambda_{\mathbf{k}\nu}/(\mathbbm{1}-V_{\mathbf{k}\nu}\Lambda_{\mathbf{k}\nu})$ where $\Lambda_{\mathbf{k}\nu}$ contains 1PI diagrams in terms of $V$, one can show that a DF approximation with given dual self-energy $\tilde{\Sigma}_{\mathbf{k}\nu}$ is equivalent to the nonlocal expansion scheme with $\Lambda_{\mathbf{k}\nu}=G^\text{loc}_\nu+\tilde{\Sigma}_{\mathbf{k}\nu}$. The scheme therefore provides further justification for the choice of diagrams based on physical considerations. 

\subsection{One-particle irreducible (1PI) approach}
\label{subsec:1pi} 

\begin{figure}[t]
  \centering
  \includegraphics[width=0.7\columnwidth]{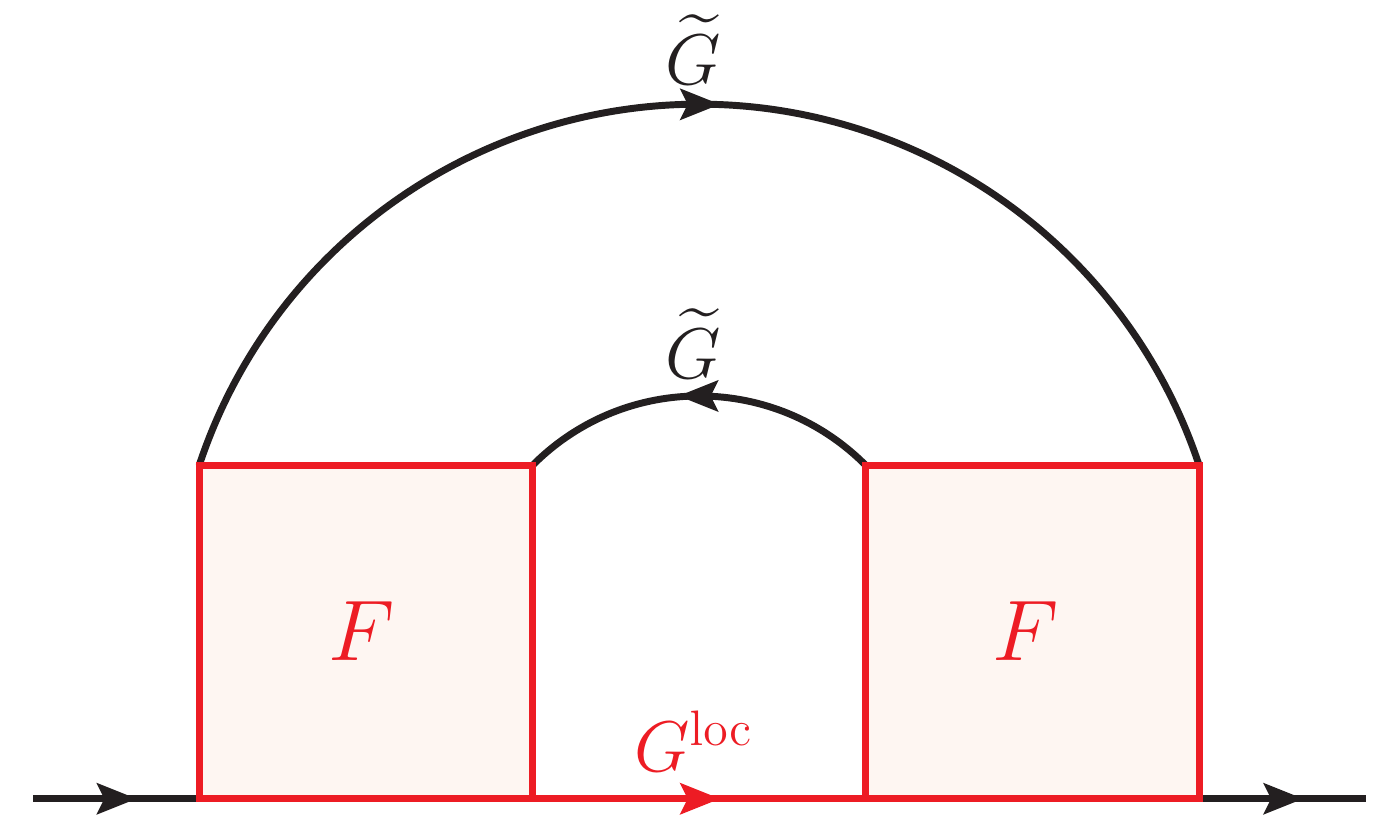}
    \caption{(Color online) First-order DF diagram that includes the local three-particle vertex (red/light gray part of the diagram). We show the particular example of a three-particle vertex that is {\em one-particle reducible} [cutting the $G^{\rm loc}$ line separates the red (light gray) part into two pieces]. Such a vertex is included in DF but not in 1PI which is  {\em one-particle irreducible}.}	\label{fig:dual_fermion_secondorder_red}
  \centering
  \includegraphics[width=\columnwidth]{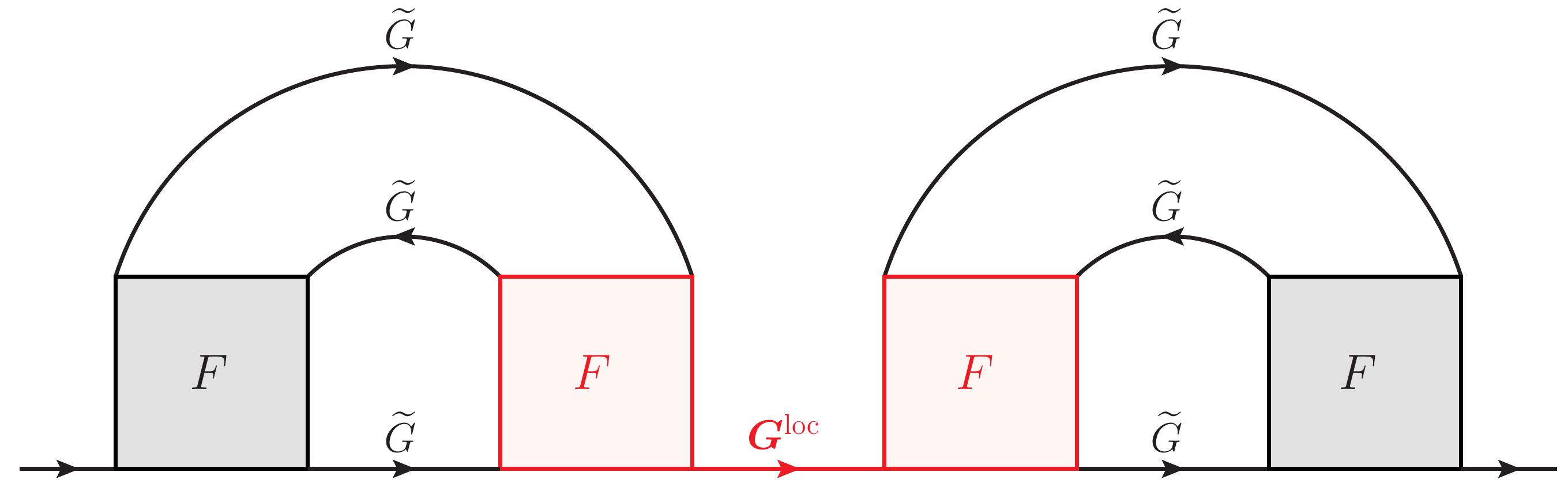}
    \caption{(Color online) Third-order diagram for the dual self-energy including the local three-particle vertex (red/light gray). In terms of real electrons this contribution appears to be spurious as it corresponds to a one-particle reducible contribution to the self-energy.}
  \label{fig:dual_fermion_reducible}
\end{figure}

In the DF theory the interaction $V_{\rm eff}$ [Eq.~(\ref{equ:vdef})] between the dual electrons is given by the local $n$-particle vertices $F$ of the reference system. Except for the two-particle vertex, these vertices in general contain {\sl one-particle reducible} contributions. There are two questions associated with  such terms in the DF approach: (i) Consider the second-order diagram of DF in Fig.~\ref{fig:dfsecondorder} (b), which contains only dual, i.e., nonlocal [see Eq.~(\ref{equ:dualgaussfinal})] propagators $\widetilde{G}$. A corresponding diagram where one of the three lines is the local Green's function $G^{\rm loc}$ is shown in Fig.~\ref{fig:dual_fermion_secondorder_red}. This diagram is included in DF at the level of the local {\sl three-particle} vertex, which is typically neglected. It is not obvious why such diagrams should not contribute to the dual self-energy with the same order of magnitude as second-order and ladder diagrams. (ii) The inclusion of the local three-particle vertex within DF leads to seemingly spurious {\sl one-particle reducible} contributions to the self-energy as depicted in Fig.~\ref{fig:dual_fermion_reducible}. Such contributions are canceled by the transformation Eq.~(\ref{equ:relationrealdualsigma}) from the dual to the physical fermions~\cite{Katanin2013}. However, when three-particle local vertices are neglected, Eq.~(\ref{equ:relationrealdual}) {\sl introduces} rather than removes such spurious contributions in the lattice Green's functions of the physical electrons. 
 
The 1PI approach aims at eliminating all terms which stem from the one-particle reducible vertices in the perturbative expansion of the self-energy. We can switch  from the one-particle reducible of DF to a 1PI formalism by a  Legendre transformation of the generating functional $W_{\text{loc}}$ [Eqs.~(\ref{equ:defgenref},\ref{equ:genrefexpand})] from the Grassmann field for the dual fermions $\widetilde{c}^{(+)}$ to new 1PI fields 
 $\phi^{(+)}$:
\begin{align}
\label{equ:cap4legendre}
  \Gamma_{\text{loc}}&[\phi^+,\phi] =  W_{\text{loc}}[\widetilde{c}^+,\widetilde{c}]
  +\sum_{i\nu\sigma}\left[\widetilde{c}_{i\nu\sigma}^+\phi_{i\nu\sigma}+\phi^+_{i\nu\sigma}\widetilde{c}_{i\nu\sigma}\right],\nonumber\\& 
  \phi^+_{i\nu\sigma}=\frac{\delta W_{\text{loc}}[\widetilde{c}^+,\widetilde{c}]
  }{\delta\widetilde{c}_{i\nu\sigma}},  \,
  \phi_{i\nu\sigma}=-\frac{\delta W_{\text{loc}}[\widetilde{c}^+,\widetilde{c}]
  }{\delta \widetilde{c}^{+}_{i\nu\sigma}}.
\end{align}
The new functional $\Gamma_{\text{loc}}$ generates the local 1PI vertex functions of the local problem, see \onlinecite{Negele1998}. From the very beginning this precludes the appearance of 1PI contributions in the three-particle and higher-order vertex functions. Replacing $W_{\text{loc}}$ in the generating functional of the DF approach, Eq.~\eqref{equ:defgenref}, by Eq.~\eqref{equ:cap4legendre} and expanding $\Gamma_{\text{loc}}$ up to fourth order in $\phi^+_i$ and $\phi_i$
one can show~\cite{Rohringer2013,Rohringer2013a} that one obtains the action $\mathcal{S}^{\text{1PI}}=\mathcal{S}^{\text{1PI}}_0+\mathcal{S}^{\text{1PI}}_{\text{I}}+\mathcal{S}^{\text{1PI}}_{\text{s}}$ of the 1PI approach with 
\begin{widetext}
\begin{subequations}
\label{equ:cap4action1PI} 
\begin{align}
 \label{equ:cap4action1PIbare}
 &\mathcal{S}^{\text{1PI}}_0[\phi^+,\phi,\psi^+,\psi]=-\sum_{k\sigma}\left\{G^{-1}_k\left(\phi_{k\sigma}^+\phi_{k\sigma}+\psi_{k\sigma}^+\phi_{k\sigma}+\phi_{k\sigma}^+\psi_{k\sigma}\right)+\left[G^{-1}_k-(G^{\text{loc}}_{\nu})^{-1}\right]\psi_{k\sigma}^+\psi_{k\sigma}\right\},\\
 &\mathcal{S}^{\text{1PI}}_{\text{I}}[\phi^+,\phi,\psi^+,\psi]=\frac{1}{4}\sum_{kk'q}\sum_{\sigma\sigma'}\left(2-\delta_{\sigma\sigma'}\right)F_{\sigma\sigma'}^{\nu\nu'\omega}\left[\phi^+_{k\sigma}\phi_{(k+q)\sigma}\phi^+_{(k'+q)\sigma'}\phi_{k'\sigma'}+\right.\nonumber\\ &\hspace{1.0cm}\left.+2\psi^+_{k\sigma}\phi_{(k+q)\sigma}\phi^+_{(k'+q)\sigma'}\phi_{k'\sigma'}+2\phi^+_{k\sigma}\phi_{(k+q)\sigma}\phi^+_{(k'+q)\sigma'}\psi_{k'\sigma'}\right]+\sum_i
 {\mathrm{Tr}}\ln M[\phi^+_i,\phi_i],\label{equ:cap4action1PIint}\\
 &\mathcal{S}^{\text{1PI}}_{\text{s}}[\phi^+,\phi,\psi^+,\psi]=-\sum_{k\sigma}\left[(\phi_{k\sigma}^+ + \psi^{+}_{k\sigma})\eta_{k\sigma}+\eta^{+}_{k\sigma}(\phi_{k\sigma}+\psi_{k\sigma})\right],
 \end{align}
 \end{subequations}
 \end{widetext}
where the four-vector notation has been adopted. $M[\phi^+_i,\phi_i]$ is the Jacobian of the transformation from the variables $\widetilde{c}^+$,$\widetilde{c}$ to $\phi^+$,$\phi$. Additional fields $\psi^+$ and $\psi$ have been introduced  in Eqs.~(\ref{equ:cap4action1PI}) that decouple three-particle interaction terms for $\phi^+$ and $\phi$ by means of a Hubbard-Stratonovich transformation. The latter arise from the application of the Legendre transform~\eqref{equ:cap4legendre} to the Gaussian term in the dual fields in Eq.~\eqref{equ:hs} [for details see~\onlinecite{Rohringer2013} and~\onlinecite{Rohringer2013a}]. They describe the one-particle reducible contributions of the three-particle and higher-order vertices to the self-energy.

\begin{figure}
  \includegraphics[width=0.45\textwidth]{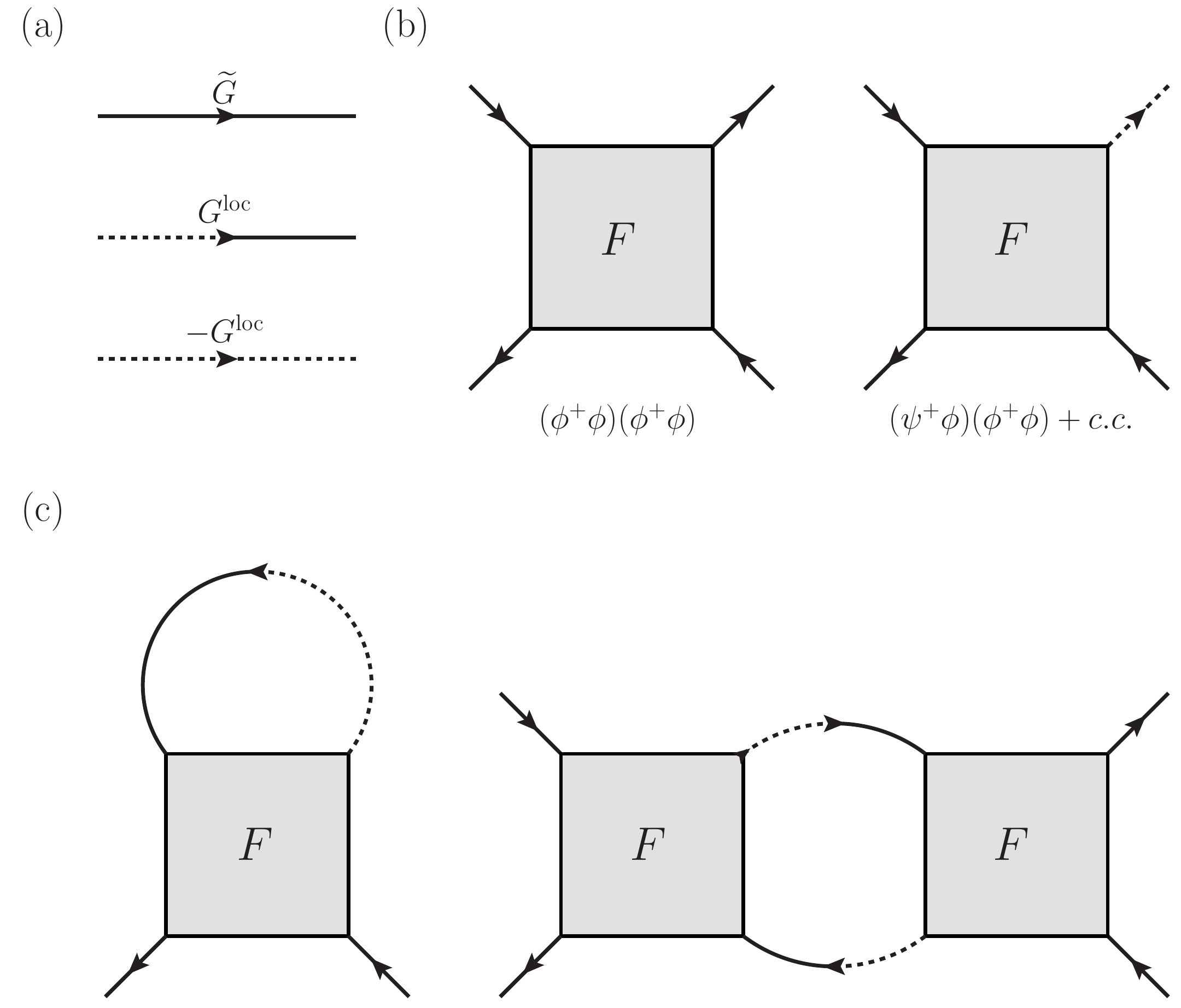}
    \caption{Diagrammatic elements of the 1PI approach: (a) The bare propagators Eq.~(\ref{equ:cap4nambubare}), (b) the interaction terms which are given by the local vertex function $F$, and (c) terms that stem from the Jacobian $M[\phi^+_i,\phi_i]$, providing for the cancellation of double-counted local contributions; for more details see \cite{Rohringer2013a}. Solid and dashed lines correspond to $\phi$ and $\psi$ fields, respectively.}
  \label{fig:diagramelements1pI}
\end{figure}

We can understand the $\phi^{(+)}$ and $\psi^{(+)}$ as different parts of a bare 1PI propagator:
\begin{equation}
 \label{equ:cap4nambubare}
\mathbf{G}_{0,k}=\frac{1}{\beta} 
\left(\!\!\begin{array}{cc}
\langle \phi^+_{k \sigma}\phi^{\phantom{+}}_{k \sigma} \rangle\! &\!\langle  \phi^+_{k \sigma}\psi^{\phantom{+}}_{k \sigma}\rangle \\
\!\langle \psi^+_{k \sigma}\phi^{\phantom{+}}_{k \sigma} \rangle\! &\!\langle  \psi^+_{k \sigma}\psi^{\phantom{+}}_{k \sigma}\rangle \\
\end{array}
\!\!\right)
=\left(\!\!
\begin{array}{cc}
G^{\text{loc}}_{\mathbf{k}}\!-G^{\text{loc}}_\nu \! &\!\! G^{\text{loc}}_\nu \\
G^{\text{loc}}_\nu\! &\!\! -G^{\text{loc}}_\nu
\end{array}
\!\!\!\right)\!,
\end{equation}
where $G^{\text{loc}}_{\mathbf{k}}$ is the lattice Green's function which including the local self-energy of the reference AIM model; see Eq.~\eqref{equ:defDMFTgf}. The propagator for the $\phi^{(+)}$ fields, $\widetilde{G}_{0,\mathbf{k}\nu}=G^{\text{loc}}_{\mathbf{k}\nu}\!-G^{\text{loc}}_\nu$ is purely nonlocal and equals the dual Green's function. The propagator for the $\psi^{(+)}$ fields, on the other hand, is given by the local Green's function $G^{\text{loc}}_\nu$ of the reference system. This diagrammatic element is {\sl absent} in DF. The diagrammatic elements of Eqs.~\eqref{equ:cap4action1PI} and~\eqref{equ:cap4nambubare}, which define the 1PI perturbation theory, are illustrated graphically in Fig. \ref{fig:diagramelements1pI}.

\begin{figure}
  \centering
  \includegraphics[width=0.5\textwidth]{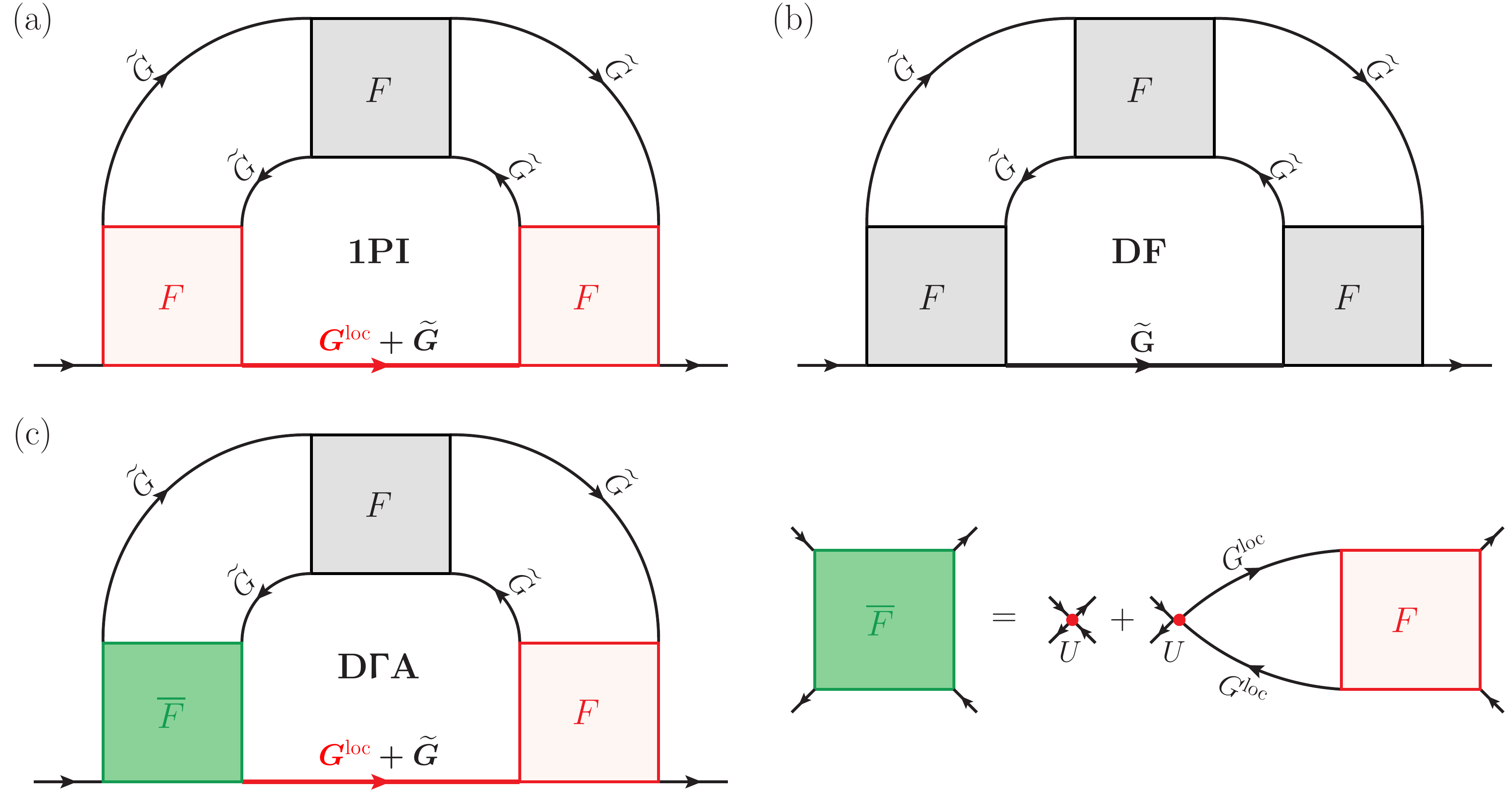}
    \caption{(Color online)  Third order diagrams in terms of the local vertex F for (a) 1PI, (b) DF, and (c) D$\Gamma$A. The red (light gray) part marks contributions that are included on the two-particle vertex level in 1PI and D$\Gamma$A while in DF they require the inclusion of the one-particle reducible three-particle vertex, cf. Fig.~\ref{fig:dual_fermion_reducible}. In D$\Gamma$A the leftmost bare vertex $U$ is only partially screened by $\overline{F}$ (instead of the full local vertex $F$) which contains all two-particle reducible diagrams independent of the incoming fermionic frequency $\nu$ (cf. Sec.~\ref{sec:vertexphysics}, Fig.~\ref{fig:3vertextypes}).}
  \label{fig:compmethods}
\end{figure}

\paragraph*{Ladder Approximation in the 1PI approach:}
Similarly to D$\Gamma$A and DF, we restrict our considerations to ladder diagrams for the 1PI action, assuming that they describe the most important physical processes. If we consider all possible bubble diagrams constructed from the diagrammatic elements depicted in Figs.~\ref{fig:diagramelements1pI} (a) and \ref{fig:diagramelements1pI} (b), but omit contributions that are canceled by counterterms of Fig.~\ref{fig:diagramelements1pI} (c) and terms that vanish because of the self-consistency condition~\eqref{equ:DMFTconditionDual2}, the 1PI self-energy is eventually obtained as \cite{Rohringer2013,Rohringer2013a}
\begin{align}
  \Sigma_{\text{1PI},k}=& \Sigma^{\text{loc}}_\nu 	-\sum_{k'qr}A_r\Gamma_{r}^{\nu\nu'\omega}\left[G_{k'}^{\text{loc}}G^{\text{loc}}_{k'+q}F_{\text{lad},r,\mathbf{q}}^{\nu'\nu\omega}\right. \notag \\
  &\phantom{\Sigma^{\text{loc}}_\nu -\frac{1}{\beta^2}\sum_{k'qr}}
  \left.-G^\text{loc}_{\nu'}G^\text{loc}_{\nu'+\omega}F_{r}^{\nu'\nu\omega}\right] G^{\text{loc}}_{k+q} \notag \\&\!+\!\frac{1}{2}\!\sum_{k'qr}\!A_rF_{r}^{\nu\nu'\omega}\widetilde{G}_{k'}\widetilde{G}_{k'+q}F_{r}^{\nu'\nu\omega}\widetilde{G}_{k+q}. \label{equ:cap4sigma1PI1}
\end{align}
with $A_\text{c}=1/2$, $A_\text{s}=3/2$ for $r=c,s$ as in Eq.~(\ref{eq:sigmadf}), and $F_{\text{lad},r,\mathbf{q}}^{\nu\nu'\omega}$ is the same BSE ladder vertex as in D$\Gamma$A [Eq.~\eqref{equ:dgabse}]. The last line of Eq.~(\ref{equ:cap4sigma1PI1}) subtracts the term counted twice in the  $\lambda=ph$ and $\lambda=\overline{ph}$ ladder. Figure~\ref{fig:compmethods}  provides an overview of the diagrammatic differences  between  the ladder versions of DF, 1PI, and D$\Gamma$A, by explicitly comparing the third-order diagrams in $F$.

The high-frequency behavior of the 1PI ladder self-energy $\Sigma_{\text{1PI},k}$ exhibits the very same violation of the exact $1/i\nu$ asymptotics as the corresponding ladder D$\Gamma$A self-energy. The problem requires the inclusion of the Moriya $\lambda$ correction, Eq.~\eqref{equ:lambdachi}, as in D$\Gamma$A. For further details, see \onlinecite{Rohringer2013,Rohringer2013a}.

\subsection{DMFT plus functional renormalization group}
\label{subsec:DMFT2fRG}

The functional renormalization group (fRG) approach [for a review see \onlinecite{Metzner2012}] provides an alternative way for generating nonlocal correlations beyond DMFT --- through the fRG flow between the local problem and the corresponding lattice problem. To generate such a flow, the DMF$^2$RG approach of~\onlinecite{Taranto2014} considers the decomposition of the lattice action in the form~\eqref{equ:actionrew}, where the coupling to the local reference problem is controlled by the flow parameter $\mathit{\Lambda}$
\begin{equation}  
\label{equ:actionconnect}
  \pazocal{S}_\mathit{\Lambda} [c^+,c]=\sum_{i}\pazocal{S}_{\text{loc}}[c^+_{i},c_{i}]+\mathit{\Lambda}\sum_{\nu%
  \mathbf{k}\sigma}\left( \varepsilon_{\mathbf{k}}-\Delta_{\nu}\right)c^+_{\nu%
  \mathbf{k}\sigma}c_{\nu\mathbf{k}\sigma}.
\end{equation}
$\pazocal{S}_\mathit{\Lambda}$ interpolates between the local DMFT action Eq.~\eqref{equ:actionref} for $\mathit{\Lambda}=0$ and the full action of the model at hand Eq.~\eqref{equ:actionhubbard} for $\mathit{\Lambda}=1$. The  action (\ref{equ:actionconnect}) can be used to construct the flow equations in the 1PI fRG approach for the  $\mathit{\Lambda} $-dependent self-energy $\Sigma _{\mathit{\Lambda},k}$ and the two-particle vertex $F_{\mathit{\Lambda} }^{kk^{\prime }q}$. In schematic form, these flow equations read
\begin{subequations}
\label{eqn:DMF2RG_floweq}
\begin{eqnarray}
  \frac{\mathrm{d}\Sigma _{\mathbf{\mathit{\Lambda} }}}{\mathrm{d}\mathit{\Lambda} } &=&F_\mathit{\Lambda}\circ S_\mathit{\Lambda} \; ,
  \label{eqn:DMF2RG_floweq0}\\
  \frac{\mathrm{d}F_{\mathbf{\mathit{\Lambda} }}}{\mathrm{d}\mathit{\Lambda} } &=&F_\mathit{\Lambda} \circ (S_\mathit{\Lambda} \circ
  G_\mathit{\Lambda})\circ F_\mathit{\Lambda}\; ,
\end{eqnarray}%
\end{subequations}
where $\circ $ denotes a summation over intermediate momenta and frequencies according to the standard diagrammatic rules and 
\begin{equation}
\label{eq:single}
  S_{\mathit{\Lambda},k}\!=\! G^2_{\mathit{\Lambda},k}\!
  \left( \varepsilon_{\mathbf k}-\Delta_\nu \right)
\end{equation}
is the so-called single-scale propagator with
\begin{equation}
  G_{\mathit{\Lambda}, k}=\left[i\nu-\mathit{\Lambda}\varepsilon_{\mathbf{k}}-(1-\mathit{\Lambda}) \Delta_\nu-\Sigma_{\mathit{\Lambda},k}\right]^{-1}
\end{equation}
the flowing Green's function. As in conventional fRG calculations, in Eqs.~(\ref{eqn:DMF2RG_floweq})a and (\ref{eqn:DMF2RG_floweq})b we have truncated the infinite hierarchy of fRG equations by neglecting the three-particle vertices. The initial conditions for the differential equations (\ref{eqn:DMF2RG_floweq}) are determined through the local reference problem $\Sigma_{\mathit{\Lambda}=0,k}=\Sigma^{\rm loc}_\nu$ and $F_{\mathit{\Lambda}=0,r}^{kk'q}=F^{\nu \nu' \omega}_r$. By construction, the DMF$^2$RG method is free from  double counting: the local properties continuously evolve toward the corresponding lattice counterparts. Note that contrary to the dual fermion approach the method is formulated in terms of physical, and not dual fermions, although a similar method can also be applied in dual space~\cite{Wentzell2015,Katanin2015}.

Differently from  the parquet D$\Gamma$A, the DMF$^2$RG approach performs the summation of parquet-type diagrams via the solution of   differential  equations.  At the same time it is based on the local 1PI vertex $F$ instead of the more cumbersome (and potentially diverging) 2PI vertex ${\Lambda}$ (not to be confused here with the flow parameter $\mathit{\Lambda}$).  The vertex $F$ collects nonlocal components through the fRG flow. In standard fRG, $F$  grows fast with increasing interaction strength so that the truncation of the infinite hierarchy of fRG equations at the two-particle level is less justified. The hope is that with DMFT presenting a good correlated starting point, the actual fRG flow needs to cover less ground in DMF$^2$RG than in standard fRG.

First DMF$^2$RG calculations for the 2D Hubbard model at half filling were performed by \onlinecite{Taranto2014}. From a numerical point of view, the  bottleneck of the fRG flow in Eqs.~(\ref{eqn:DMF2RG_floweq}) is a memory restriction, because the vertex functions that depend on three momenta and frequencies in different channels are intertwined.  This is similar as for the parquet equations~(\ref{equ:parquedecompF} and \ref{equ:decomp2PIGamma}) and can be mitigated by using vertex asymptotics~\cite{Wentzell2016}.

At the same time, the latest theoretical advance \cite{Kugler2018,Kugler2018a} has shown how it is possible, with a reasonable numerical effort, to compute all $n$-loop contributions in the truncated fRG flow, which corresponds to {\sl fully} resumming all diagrams of the parquet approximation. Hence, exploiting this "multiloop" treatment within the DMF$^2$RG scheme might provide an elegant way to circumvent the 2PI-vertex divergences, even for parquet-based algortithms.

Recently, \onlinecite{Katanin2016} proposed to employ the 2PI fRG approach to consider nonlocal corrections beyond EDMFT (see Sec.~\ref{sec:edmftpp}). This generalization is especially useful for the treatment of nonlocal two-particle interactions in strongly correlated systems. Using this approach one can consider the evolution from the EDMFT local problem Eq.~(\ref{eq:imp_action0}) to the lattice problem Eq.~(\ref{eq:action}). The resulting equations are similar, but not identical to those in the DB$\gamma$ approach of~\onlinecite{Stepanov2016}: the 2PI approach includes the effect of the one-particle reducible six-point and higher-order vertices in the dual boson approach, cf. the discussion in Sec.~\ref{subsec:1pi}.

\subsection{Extending vertex approaches to clusters}
\label{sec:clusterdf}
\label{sec:cluster}  

In the previous sections, we reviewed extensions of DMFT that generate nonlocal correlations from a {\em local} vertex. All of these approaches can be generalized quite naturally by taking the vertex calculated on a small but finite cluster as the starting point. This has the advantage that certain short-range correlations that are difficult to capture diagrammatically, such as the formation of a compound singlet by neighboring spins, can be treated numerically in a more rigorous way. At the same time, correlations on length scales that exceed numerically feasible cluster sizes are treated diagrammatically. Using DCA or cluster DMFT for a small cluster and D$\Gamma$A, DF, etc.\, for a large  cluster, also offers the advantage  that it is possible to study the convergence with respect to the size of the small and large cluster systematically.

Such cluster-based calculations have been pioneered by~\onlinecite{Slezak2009} who proposed a multiscale many-body approach. Correlations on short length scales are incorporated by calculating the irreducible vertex $\Gamma_{ph,r}^{kk'q}$ on a small DCA cluster corresponding to a coarse ${\mathbf k}$ grid within QMC. Correlations on larger length scales are accounted for by solving the BSE equation for the approximate full vertex on a larger cluster, in analogy to the ladder D$\Gamma$A of Sec.~\ref{sec:dgaladder}. The self-energy is obtained through the Schwinger-Dyson equation. Correlations exceeding the larger cluster are accounted for on a mean field level. As calculating the vertex on a cluster is a formidable numerical task,~\onlinecite{Slezak2009} approximated $\Gamma_{ph,r}^{kk'q}\approx \Gamma_{ph,r}^{kk'0}$, neglecting the dependence on the bosonic momentum and frequency. They also considered a simplifying ansatz for the self-energy and more approximative solutions for the larger cluster. Further details of the approach, its validation, and application to the 1D Hubbard model can be found in~\onlinecite{Slezak2009}.

In the cluster DF (CDF) approach by~\onlinecite{Hafermann2008} the DF expansion is performed around the CDMFT solution as the reference system. For the 1D Hubbard model, the CDF is  considerably closer than DF to the benchmark of the density matrix renormalization group (DMRG)\cite{Hafermann2008}. A disadvantage is that CDMFT breaks translational invariance of the lattice. \onlinecite{Hafermannphd} showed that the dual corrections however tend to partially restore the translational invariance. Another consequence of broken translational invariance is that the two-particle vertex is a rank-4 tensor in the spatial indices. An alternative is to perform a diagrammatic expansion around a DCA cluster by embedding the latter in a DF lattice. The first version of such a DFDCA approach was introduced by  \onlinecite{Yang2011a}, where the vertex function depends on only three cluster momenta and requires less memory. The DFDCA results show that the second-order correction beyond a small $2\times 2$  cluster significantly reduces the N\'eel temperature. The results converge inversely with the linear cluster size $L_c$ in accordance with DCA convergence to the exact limit $L_c \to \infty$. \onlinecite{Iskakov2018} analyzed the approximations of the DFDCA method and identified DCA interaction coarse graining as a primary source of error.

While the accuracy of these methods is controlled through the cluster size, the recent cluster generalization of TRILEX~\cite{Ayral2017} employs a different control parameter. It is based on the Fierz ambiguity and that, with increasing cluster size, observables become independent of the ratio of spin to charge fluctuations.

\subsection{DMFT + nonlocal self-energy}
\label{subsec:DMFTplus}

In this section, we review  \emph{DMFT + nonlocal self-energy}  methods which supplement the local DMFT self-energy by nonlocal contributions obtained independently using another method, typically  within perturbation theory in the bare interaction. This implies an explicit separation of the local and nonlocal parts of the self-energy. These methods were proposed earlier and are generally simpler than the DMFT extensions described in the previous subsections, which incorporate nonlocal correlations, more systematically, on the basis of a local two-particle vertex.

The idea to augment DMFT with a nonlocal self-energy was first introduced for the nonlocal electron interaction in the context of the $GW$+DMFT and EDMFT+$GW$ approaches~\cite{Sun02,Biermann2003}. These supplement the local DMFT self-energy by the screened exchange diagram of $GW$, which offers an appealing route to realistic material calculations.  We discuss EDMFT+$GW$ together with other extensions of EDMFT  in Sec.~\ref{sec:edmftpp}, and refer the interested reader to \onlinecite{Tomczak2017} for a tutorial review of $GW$+DMFT.

In this context, let us briefly mention the self-energy embedding theory (SEET) by \onlinecite{Kamenka2015}. Similar to $GW$+DMFT, SEET defines one (or several) correlated set(s) of orbitals within which the interaction is treated more accurately, e.g., in ED, while the interaction with the rest of the orbitals (and between these sets of orbitals)  is treated in weak-coupling perturbation theory such as in $GW$. The main difference is that the correlated subspace in SEET is adjusted in terms of energy or the one-particle density matrix, whereas it is defined in terms of locality in  $GW$+DMFT; see \onlinecite{Zgid2017} for the interrelation.

Let us now return to the DMFT + nonlocal self-energy approaches. For  the Hubbard model, the summation of an infinite series of diagrams for the nonlocal self-energy was considered in DMFT+$\Sigma_{k}$~\cite{Sadovskii2005,Kuchinskii2005,Kuchinskii2006} and DMFT+FLEX~\cite{Gukelberger2015,Kitatani2015}. As with all  DMFT + nonlocal self-energy methods, they are based on the separation
\begin{equation}
\label{Eq:SigmaNonloc}
  \Sigma _{\mathbf{k}\nu }=\Sigma ^{\mathrm{loc}}_\nu+\Sigma^{\rm nloc}_{%
  \mathbf{k}\nu },
\end{equation}%
where $\Sigma^{\text{loc}}$ includes contributions exclusively built from local propagators and $\Sigma^{\rm nloc}_{\mathbf{k}\nu }$ represents the contribution of nonlocal correlations; it may or may not have its own local part.

The DMFT+$\Sigma_{k}$ method [see \onlinecite{Kuchinskii2012} for a review] considers the interaction of electrons via bosonic excitations, which originate from the same fermionic system. DMFT+$\Sigma_k$  performs an approximate summation of an infinite number of Feynman diagrams (beyond the ladder approximation), which is based on the combinatorial rules for some specific form of the bosonic propagator, relating diagrams of the same order with the corresponding noncrossing diagrams. In particular, assuming that in two dimensions the
static bosonic propagator in the magnetic channel  has the form of a product of one-dimensional propagators,%
\begin{equation}
  S(\mathbf{Q}+\mathbf{q})=\Delta ^{2}\frac{\xi ^{-1}}{q_{x}^{2}+\xi ^{-2}}%
  \frac{\xi ^{-1}}{q_{y}^{2}+\xi ^{-2}},
\end{equation}%
where $\mathbf{Q}$ is the order parameter wave vector, $\Delta $ characterizes the strength of the electron-boson interaction, $\xi$ corresponds to the correlation length and the relations $v_{\mathbf{k}}^{x}v_{\mathbf{k}+\mathbf{Q}}^{x}>0$ and $v_{\mathbf{k}}^{y}v_{\mathbf{k}+\mathbf{Q}}^{y}>0$ for the Fermi velocity components $\mathbf{v}_{\mathbf{k}}=(v_{\mathbf{k}}^{x},v_{\mathbf{k}}^{y})$ are fulfilled, one can show~\cite{Schmalian1999,Sadovskii1999} that in the static approximation for the bosonic propagator all diagrams of the same order give equal contributions. At $\xi \rightarrow \infty $ the problem can then be mapped to one with a single Gaussian field, while at finite $\xi $ recursion relations for the contribution of the diagrams of different orders can be obtained. In particular, the contribution of the orders $\geq n$ of the perturbation theory to the nonlocal self-energy are related through~\cite{Sadovskii1999,Sadovskii2005} 
\begin{equation}
\label{Eq:rec}
  \Sigma_{\mathbf{k}\nu}^{{\rm nloc}(n)}=\frac{s_{n}}{\nu-\varepsilon _{\mathbf{k}}^{(n)}+i n v_{\mathbf{k}}^{(n)}\xi ^{-1}-\Sigma
  ^{\mathrm{loc}}_\nu-{\Sigma}_{\mathbf{k}\nu}^{{\rm nloc}(n+1)}},
\end{equation}%
where $s_{n}$ are the appropriate combinatorial factors, $\varepsilon _{\mathbf{k}}^{(n)}=\varepsilon _{\mathbf{k}},$ $v_{\mathbf{k}}^{(n)}=|v_{\mathbf{k}}^{x}|+|v_{\mathbf{k}}^{y}|$ for odd $n,$ and $\varepsilon _{\mathbf{k}}^{(n)}=\varepsilon _{\mathbf{k+Q}},$ $v_{\mathbf{k}}^{(n)}=|v_{\mathbf{k+Q}}^{x}|+|v_{\mathbf{k+Q}}^{y}|$ for even $n.$ The physical nonlocal part of the self-energy is $\Sigma^{\rm nloc}_{\mathbf{k}\nu}=\Sigma_{\mathbf{k}\nu }^{{\rm nloc}(1)}$, i.e., the final self-energy of the recursion relation~(\ref{Eq:rec}). Diagrams of sufficiently high order $n\gg 1$ can be neglected, which provides the initial condition $\Sigma_{\mathbf{k}\nu}^{{\rm nloc}(n)}=0$ for Eq.~(\ref{Eq:rec}). The above mentioned inequalities $v_{\mathbf{k}}^{x,y}v_{\mathbf{k}+\mathbf{Q}}^{x,y}>0$ may not be fulfilled for realistic dispersions; the corresponding recursion relations are then  only approximate.

Equation~(\ref{Eq:rec}) represents a rough approximation for the spin propagator in two dimensions. Indeed, its generalization leads to a pseudogap as a precursor of antiferromagnetism even in higher dimensions. For the more physical Ornstein-Zernike form of the bosonic propagator near a (quantum) phase transition, 
\begin{equation}
  S(\mathbf{Q}+\mathbf{q})=\frac{\Delta ^{2}}{q^{2}+\xi ^{-2}},
\end{equation}
the recursion relations~\eqref{Eq:rec} become approximate for any electronic dispersion and in principle should be replaced by the corresponding integral equations~\cite{Katanin2005}. The DMFT+$\Sigma _{k}$ method using Eq.~\eqref{Eq:rec}  was applied to describe pseudogaps induced by antiferromagnetic correlations in 2D  \cite{Sadovskii2005,Kuchinskii2005,Kuchinskii2006} and spectra of high-$T_c$ superconductors  \cite{Nekrasov2008,Nekrasov2011}. Beyond the self-energy, \onlinecite{Kuchinskii2007} also calculated two-particle properties such as the  optical conductivity in the DMFT+$\Sigma_k$ framework, and  nonlocal (Debye) phonons  were taken into account in \onlinecite{Kuchinskii2009}.

An alternative method to augment DMFT with a nonlocal self-energy is to sample diagrams of the perturbation series contributing to $\Sigma^{\rm nloc}_{\kv}$ using a bold diagrammatic Monte Carlo (BDMC) algorithm~\cite{Pollet2011}. The sampling procedure removes the potential bias incurred by choosing a certain class of diagrams in the perturbation theory. In DMFT+BDMC the diagrams are constructed in terms of dressed or ``bold-line'' propagators $G_{\kv \nu}$ and bare interaction vertices. Since the propagators contain local self-energy insertions, only skeleton diagrams are sampled up to a given order. Double counting is avoided by requiring that at least two vertices with different site indices and hence at least one nonlocal propagator are accounted for in $\Sigma^{\rm nloc}_{\kv}$. An improvement of $10^{4}$ in efficiency compared to conventional BDMC has been demonstrated for the Anderson localization problem in the nonperturbative regime. For the Hubbard model, however, it has been shown that the bold-line diagrammatic series may converge to a wrong answer~\cite{Kozik2015}.

In approaches where the resulting ${\Sigma}^{\rm nloc}_{\mathbf{k}}$ has a nonzero local part as in DMFT+$\Sigma_{k}$ and DMFT+BDMC, an external self-consistency is assumed: the local Green's function calculated with the self-energy~\eqref{Eq:SigmaNonloc} is used to solve $\Sigma^{\text{loc}}=\Sigma[G^{\text{loc}}]$ --typically through the self-consistent solution of an Anderson impurity model-- and the process is iterated until convergence is reached.

In the DMFT+FLEX method~\cite{Gukelberger2015,Kitatani2015} the summation of the standard ladder and bubble diagrams of the FLEX type is performed to obtain ${\Sigma}^{\rm nloc}_{\mathbf{k}\nu}$; the local part of these diagrams is subtracted to avoid double counting. This corresponds to the lowest-order approximation for the fermionic self-energy (with respect to fermion-boson coupling) in the DMFT+$\Sigma_k$ approach, but with the bosonic propagator determined microscopically as a sum of RPA diagrams. The considered set of diagrams is similar to the ladder D$\Gamma$A approach of Sec.~\ref{sec:dga}, except that DMFT+FLEX uses the bare vertex in these diagrams instead of a local two-particle irreducible vertex.
Let us also note the work by~\onlinecite{Hague04}, which supplements the DCA on a small cluster by the FLEX solution on a large cluster.

\subsection{EDMFT++ theories}
\label{sec:edmftpp}

Many important effects regarding the physics of correlated systems are based on nonlocal interactions in solids and require a consistent description of collective excitations (plasmons, magnons, orbitons etc.), which can strongly affect the original electronic degrees of freedom. The aim of extended DMFT (EDMFT) \cite{Sengupta1995,Si1996,Kajueter96,Smith00,Chitra01,Sun02} is to include such nonlocal interactions  and collective (bosonic) degrees of freedom into the  DMFT framework. The central quantities of interest in EDMFT are the electronic self-energy and the polarization operator (bosonic self-energy). The latter includes nonlocal interaction effects and leads to a bosonic impurity problem with frequency-dependent interaction. Akin to DMFT, the fermionic and bosonic self-energies remain purely local. This  is often insufficient for describing fluctuations that are inherently nonlocal in character; for example, EDMFT  fails to correctly describe plasmons~\cite{Hafermann2014a}. Hence, there have been various attempts to include nonlocal corrections beyond  EDMFT. We sum them up under the term EDMFT++ theories. The cluster extension of EDMFT is an example~\cite{Pixley2015}. In the following, we focus on  diagrammatic EDMFT++  methods, which include the EDMFT+GW approach, the dual boson (DB) approach and the triply-irreducible local expansion (TRILEX). Before that, we recapitulate the EDMFT approach.

An extended Hubbard model with a nonlocal interaction can, e.g., be obtained from first-principles constrained RPA~\cite{Aryasetiawan2004}, where a frequency-dependent nonlocal interaction for the correlated subspace arises from screening by broad bands of conduction electrons.
A corresponding effective action takes the form
\begin{align}
\mathcal{S}=-\sum_{\mathbf{k}\nu\sigma} G_{0,\mathbf{k}\nu}^{-1}c^{+}_{\mathbf{k}\nu\sigma}c^{\phantom{+}}_{\mathbf{k}\nu\sigma}+\frac{1}{2}\sum_{{\bf q}\omega}U^{\phantom{+}}_{\bf q \omega }\rho^{*}_{\qv\omega}\rho^{\phantom{*}}_{\qv\omega}.
\label{eq:action}
\end{align}
Here $G^0_{\mathbf{k}\nu}=[i\nu+\mu-\varepsilon^{\phantom{+}}_{\bf k}]^{-1}$ is the bare lattice Green's function; the interaction $U_{\bf q\omega}=U_{\omega}+V_{\bf q\omega}$ consists of the on-site term $U$ and nonlocal long-range Coulomb interaction $V$.

For simplicity we consider only charge fluctuations given by the complex bosonic variable
$\rho^{\phantom{+}}_{\qv \omega} = \sum_{\mathbf{k} \nu\sigma}[c^{+}_{\mathbf{k} \nu\sigma}c^{\phantom{*}}_{(\mathbf{k}+\qv)(\nu+\omega)\sigma}-\av{c^{+}_{\mathbf{k}\nu\sigma}c^{\phantom{*}}_{\mathbf{k}\nu\sigma}}\delta_{\omega 0}\delta_{\qv \mathbf{0}}]$ in the following.

\subsubsection{EDMFT approach}

In EDMFT the kinetic terms are scaled as $1/\sqrt{D}$ as in DMFT, but the nonlocal interaction also scales as $1/\sqrt{D}$~\cite{Smith00} instead of  $1/{D}$ in DMFT~\cite{Muller-Hartmann1988}. This way, nonlocal quantum fluctuations arising from the intersite interaction survive and are treated on the same footing as the local ones. In particular, the self-energy  remains local since any nonlocal skeleton diagram that contains a nonlocal interaction also contains a nonlocal Green's function and  is  subleading.  As a consequence, the self-energy, as well as local higher-order correlation functions of the lattice problem, can be obtained from an effective impurity problem of the form
\begin{align}
\label{eq:imp_action0}
  \mathcal{S}_{\text{loc}}=&-\sum_{\nu\sigma} c^{+}_{\nu\sigma}[i\nu+\mu-\Delta^{\phantom{*}}_{\nu}]c^{\phantom{*}}_{\nu\sigma} 
  +\, \frac{1}{2}\,\sum_{\omega}\,{\cal U}^{\phantom{*}}_{\omega}\, \rho^*_{\omega} \rho^{\phantom{*}}_{\omega}.
\end{align}
While the fermionic hybridization $\Delta_{\nu}$ acts as a fermionic bath describing the dynamics of the local quantum fluctuations, the intersite interactions induce fluctuations that give rise to a  bosonic local bath that eventually modifies the bare  interaction $U_{\omega}\rightarrow {\cal U}_{\omega}$. It can be viewed as a dynamical mean field on the two-particle level. Note that the nonlocal interactions induce a frequency dependence even when the frequency dependence of the local and nonlocal interaction in Eq.~(\ref{eq:action}) is neglected.

The impurity model can be solved using suitably generalized standard impurity solvers (see Sec.~\ref{sec:calver}) to treat the retarded interactions. This allows one to obtain the local impurity Green's function $G^{\text{loc}}_{\nu}$, susceptibility $\chi_{\omega}$ and renormalized interaction ${\cal W}_{\omega}$ defined as follows:
\begin{align}
G^{\text{loc}}_\nu = -\av{c^{\phantom{*}}_{\nu}c^{+}_{\nu}}_\text{loc},
\label{eq:gdef} &&\\
\chi_{\omega} = -\av{\rho^{\phantom{*}}_{\omega}\rho^{*}_{\omega}}_\text{loc}, &&
{\cal W}_{\omega} = {\cal U}_{\omega} + {\cal U}_{\omega}\chi_{\omega}{\cal U}_{\omega}.
\label{eq:gchiwdef}
\end{align}
The average is taken with respect to the local action \eqref{eq:imp_action0}, and the functions \eqref{eq:gdef} and \eqref{eq:gchiwdef} are determined self-consistently in EDMFT. Here $G^{\text{loc}}_\nu$ is related to a local self-energy $\Sigma^{\rm loc}_{\nu}$ and $\chi_\omega$ to a local polarization operator $\Pi^{\rm loc}_{\omega}\equiv (\chi^{-1}_{\omega}+{\cal U}^{\phantom{*}}_{\omega})^{-1}$, respectively. With these {\em local} (fermionic and bosonic) EDMFT self-energies, the lattice Green's function $G^{\rm loc}_{\kv\nu}$ [as in Eq.~\ref{equ:defDMFTgf})] and screened interaction $W^{\rm loc}_{\qv\omega}$ are calculated according to 
\begin{align}
\label{eq:WEDMFT}
  \!\!(G^{\rm loc}_{\kv\nu})^{-1} &= \hspace{0.1cm}(G_{0,\kv\nu})^{-1}-\Sigma^{\rm loc}_{\nu} \, ,\\
  \!\!(W^{\rm loc}_{\qv\omega})^{-1} &= (W_{0,\qv\omega})^{-1}\!-\Pi^{\rm loc}_{\omega} \, .
\end{align}
Here $W_{0}$ denotes the bare interaction, which is equal to $U_{\qv\omega}$ or $V_{\qv\omega}$ in the case of $UV$ or $V$decoupling, respectively~\cite{Ayral2012,Ayral2013}. Finally, the local impurity problem is specified through the self-consistency conditions
\begin{align}
\label{eq:edmftselfc}
  G^{\text{loc}}_\nu = \sum_{\kv} G^{\rm loc}_{\kv\nu} \, , &&
  {\cal W}_{\omega} =\sum_{\qv} W^{\rm loc}_{\qv\omega}  \, .
\end{align}
EDMFT can be employed to describe the second-order transition to a charge-ordered insulator driven by the competition between a local and a nearest-neighbor intersite interaction $V$~\cite{Sun02} and signaled by a divergence of the susceptibility $\chi_{\mathbf{q}\omega }=1/(1/\Pi_\omega^{\text{loc}}+\Lambda_\omega-V_{{\bf q}\omega})$ at $\mathbf{q}=(\pi,\pi)$ and $\omega=0$. Here the difference to the corresponding DMFT phase transition \cite{Wahle1998} is the additional local bosonic bath  and a modified $\Pi_\omega^{\text{loc}}$.

EDMFT can be shown to be conserving in infinite dimensions to order $1/D$. The momentum dependence in the susceptibility stems from the nonlocal interaction only. To describe extended collective modes and to obtain a conserving approximation in finite dimensions that fulfills Ward identities however requires a momentum-dependent bosonic self-energy~\cite{Hafermann2014a}. Diagrammatic extensions of EDMFT supplement both the fermionic and bosonic self-energies with a nonlocal part that is obtained within perturbation theory, similarly as in diagrammatic extensions of DMFT. The separation into local and nonlocal contributions can be written in the form
\begin{eqnarray}
\label{Eq:SigmaPiNonloc}
  \Sigma _{\mathbf{k}\nu}=\Sigma ^{\mathrm{loc}}_\nu +{\Sigma }^{\rm nloc}_{\mathbf{k}\nu} , \;\;\;
  \Pi _{\mathbf{q}\omega}=\Pi ^{\mathrm{loc}}_\omega +{\Pi }^{\rm nloc}_{\mathbf{q}\omega}.
\end{eqnarray}
The goal of EDMFT++ theories is to approximate these nonlocal functions with EDMFT as a starting point. 

\subsubsection{EDMFT\texorpdfstring{$+GW$}{+GW} approach}
\label{sec:edmftGW}
 
\begin{figure}
  \centering
  \includegraphics[width=0.45\textwidth]{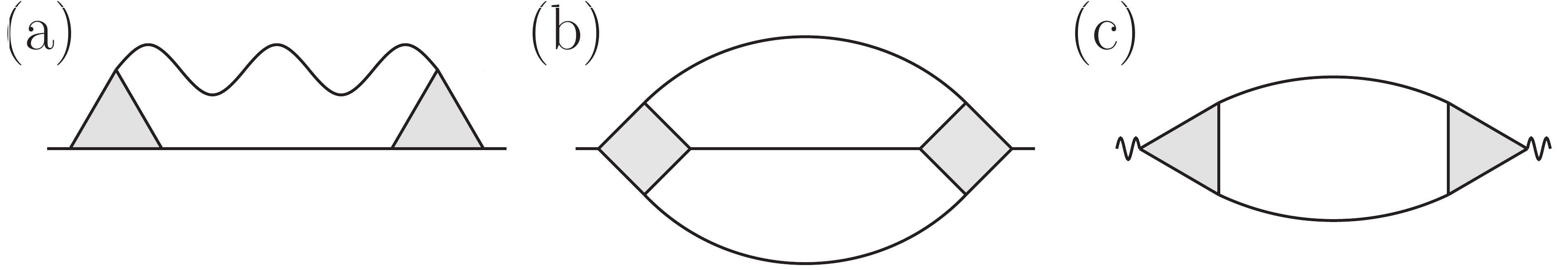}
    \caption{The lowest-order nonlocal self-energy diagrams, treated by the EDMFT+GW method describing (a) the interaction of the electron with the bosonic mode (wiggly line), (b) the inter-electron interaction, and (c) correction to the polarization operator. A shaded triangle denotes a fermion-boson vertex as defined in Eq.~(\ref{eq:3pvertex}), while the shaded diamond (square) corresponds to an electron-electron vertex [see Eq.~(\ref{equ:defvertex})]. All vertices are taken in the local approximation. Adapted from Ref.~\onlinecite{Sun02}.}
  \label{fig:DMFT_Plus_Fig1}
\end{figure}

In EDMFT$+GW$~\cite{Sun02,Biermann2003,Tomczak2012,Ayral2012,Ayral2013,Hansmann2013,Tomczak2014,Huang14,Boehnke2016}, $\widetilde{\Sigma}_{\kv\nu}$ and $\widetilde{\Pi}_{\qv\omega}$ are given in terms of second-order diagrams (see Fig.~\ref{fig:DMFT_Plus_Fig1}). In practice the bare interaction is taken instead of the local vertex. The self-energy and polarization operator diagrams from the $GW$ approximation~\cite{Hedin1965,Aryasetiawan1998,Hedin1999} are added to the dynamical mean-field solution treating nonlocal correlations. Double counting of the local impurity contributions is efficiently avoided by using only the nonlocal part of these diagrams. Since the local propagators are equal to those given through the local action (\ref{eq:imp_action0}) by virtue of the self-consistency conditions~\eqref{eq:edmftselfc}, it is possible to express the $GW$ corrections solely in terms of nonlocal propagators. The nonlocal parts ${\Sigma }^{\rm nloc}_{\kv\nu}$ and ${\Pi }^{\rm nloc}_{\qv\omega}$ of the self-energies are correspondingly replaced by
\begin{align}
\label{eq:Sigma_Pi_GW}
  {\Sigma}^{\rm GW}_{\kv\nu} &=-\sum\limits_{\qv\omega}\widetilde{G}_{(\kv-\qv)(\nu-\omega)}\widetilde{W}_{\qv\omega},\notag\\
  {\Pi}^{\rm GW}_{\qv\omega} &= \,2\,\sum\limits_{\kv\nu}\,\widetilde{G}_{(\kv+\qv)(\nu+\omega)}\widetilde{G}_{\kv\nu}.
\end{align}
Here the factor of $2$ in the second line of Eq.~\eqref{eq:Sigma_Pi_GW} accounts for the spin degeneracy, and  the  nonlocal propagators are explicitly given by
\begin{align}
\label{eq:WEMDFTGW}
  \widetilde{G}_{\kv\nu} = G_{\kv\nu} -G^{\text{loc}}_\nu\,, &&
  \widetilde{W}_{\qv\omega} = W_{\qv\omega} - {\cal W}_{\omega}\,.
\end{align}

In this construction, the local interaction $U$ has already been accounted for in the impurity problem. The bare nonlocal interaction $W_{0,\qv\omega}$ enters Eq.~\eqref{eq:WEMDFTGW} through $W_{\qv\omega}^{-1}=W_{0,\qv\omega}^{-1}-\Pi_{\qv\omega}$. For instance, it can  be taken in the form of $V$ decoupling ($W_{0,\qv\omega} = V_{\qv\omega}$), which leads to a simple separation of local and nonlocal contributions to the self-energy $\widetilde{\Sigma}_{\kv\nu}$. Unfortunately, this form of renormalized interaction leads to an overestimation of nonlocal correlation effects~\cite{Ayral2012,Ayral2013}. On the other hand, the $UV$-decoupling ($W_{0,\qv\omega} = U_{\omega} + V_{\qv\omega}$) is more consistent with standard perturbation theory for the full Coulomb interaction, but leads to formal problems with separation of local and nonlocal parts of the diagrams~\cite{Stepanov2016a}. A simplified treatment of the screening using Thomas-Fermi theory has been proposed by~\onlinecite{paris_sex}. The form of the renormalized interaction and the way to avoid the double counting in general remain  subject to discussion~\cite{Gukelberger2015}.

\subsubsection{Dual boson (DB) approach}
\label{sec:db}

The DB scheme by \onlinecite{Rubtsov12} aims to treat the action (\ref{eq:action}) in a similar  spirit as the DF approach. Analogously to the DF fields, dual bosonic fields and a corresponding bosonic bath are introduced. These fields decouple  nonlocal interaction terms in models with long-range interactions; the bosonic bath provides an effective treatment of collective excitations. 

In the following presentation of the DB method we exclude, for simplicity, exchange interactions and local spin degrees of freedom. They can be included with some care by introducing vector spin bosons~\cite{Ayral2015}. Moreover, we consider only the single-band case, but  a generalization of the formalism to several orbitals or bands is straightforward.

First, completely analogous to the representation Eq.~(\ref{equ:actionrew}) of the lattice action in DF, we rewrite Eq.~(\ref{eq:action}) in terms of a local reference action $\mathcal{S}_{\text{loc}}$ including a retarded interaction [Eq.~\eqref{eq:imp_action0}] and nonlocal correction terms:
\begin{align}
\label{eq:actionlatt}
  \mathcal{S}[c^+,c] =\sum_{i} \mathcal{S}_{\rm loc}[c^+_i,c_i] &+ \sum_{\mathbf{k}\nu\sigma}[\varepsilon_{\mathbf{k}}-\Delta_\nu]c^+_{\mathbf{k}\nu\sigma}c_{\mathbf{k}\nu\sigma}\nonumber\\+&\frac{1}{2}\sum_{\mathbf{q}\omega}[U_{\mathbf{q}\omega}-\mathcal{U}_\omega]\rho^*_{\mathbf{q}\omega}\rho_{\mathbf{q}\omega}.
\end{align}
The local bare interaction is given by the sum of the local part of the possibly frequency-dependent bare interaction of the lattice model ($U_\omega$) and the bosonic bath ($\Lambda_\omega$), i.e., $\mathcal{U}_\omega=U_\omega+\Lambda_\omega$.\footnote{Considering that the full interaction of the lattice system is given by $U_{\mathbf{q}\omega}=U_\omega+V_{\mathbf{q}\omega}$ one has ${\cal U}_{\omega}-U_{\bf q {\omega}}=\Lambda_{\omega}-V_{\bf q {\omega}}$, which shows that the method is independent of the selected decoupling scheme ($U$ or $UV$)~\cite{vanLoon2014a}.}

In the next step, we decouple the local and nonlocal degrees of freedom in $\mathcal{S}$ [Eq.~(\ref{eq:actionlatt})] by means of a  fermionic (for $\sum_{\mathbf{k}\nu\sigma}[\varepsilon_{\mathbf{k}}-\Delta_\nu]c^+_{\mathbf{k}\nu\sigma}c_{\mathbf{k}\nu\sigma}$) and a bosonic (for $\frac{1}{2}\sum_{\mathbf{q}\omega}[U_{\mathbf{q}\omega}-\mathcal{U}_\omega]\rho^*_{\mathbf{q}\omega}\rho_{\mathbf{q}\omega}$) Hubbard-Stratonovich transformation. The fermionic one is exactly the same as for the DF theory in Eq.~(\ref{equ:hs}) which introduces the dual fermionic variables $\widetilde{c}^+$ and $\widetilde{c}$. The decoupling of the bosonic degrees of freedom is done via the transform
\begin{align}
\label{equ:bhs}
  &e^{-\,\frac12 [U_{\mathbf{q}\omega}-\mathcal{U}_\omega]\rho^{*}_{{\bf q}\omega}\rho^{\phantom{*}}_{{\bf q}\omega}} =-\alpha_\omega^2 ([U_{\mathbf{q}\omega}-\mathcal{U}_\omega]/2)^{-1}\\ &\;\;\times\int D[ \widetilde{\rho}^*,\widetilde{\rho}]\,e^{\alpha_\omega^2([U_{\mathbf{q}\omega}-\mathcal{U}_\omega]/2)^{-1}\widetilde{\rho}^{*}_{{\bf q}\omega} \widetilde{\rho}^{\phantom{*}}_{{\bf q}\omega} + \alpha_\omega[\rho^{*}_{\omega}\widetilde{\rho}^{\phantom{*}}_{\omega} + \widetilde{\rho}^{*}_{\omega}\rho^{\phantom{*}}_{\omega}]},\notag
\end{align}
where the integration measure $D[ \widetilde{\rho}^*,\widetilde{\rho}]\equiv D\widetilde{\rho}^*D\widetilde{\rho}/\pi$ includes the normalization factor $1/\pi$, and $\alpha_\omega$ is a (for the moment arbitrary) function of $\omega$ [cf. $b_{\nu\sigma}$ in DF in Eq.~(\ref{equ:hs})]. The sign in front of $[\mathcal{U}_\omega-U_{\mathbf{q}\omega}]$ within the integral has to be properly chosen in order to guarantee the convergence of the integral~\cite{Rubtsov12}. Alternatively one can employ a decoupling introducing a real field~\cite{vanLoon2014a}.

The DB action can be written in the form
\begin{align}
\label{Eq:DBaction},
  \widetilde{\mathcal{S}}[\widetilde{c}^+,\widetilde{c},\widetilde{\rho}^*,\widetilde{\rho}] =&
  -\sum_{\mathbf{k}\nu} \widetilde{G}_{0,\mathbf{k}\nu}^{-1} \widetilde{c}^+_{\mathbf{k}\nu\sigma} \widetilde{c}^{\phantom{+}}_{\mathbf{k}\nu\sigma} - \frac{1}{2}\sum_{\mathbf{q}\omega} \widetilde{W}_{0,\mathbf{q}\omega}^{-1} \widetilde{\rho}^{*}_{\mathbf{q}\omega} \widetilde{\rho}^{\phantom{*}}_{\mathbf{q}\omega}\notag\\&+ \sum_{i}V_{\text{eff}}[\widetilde{c}^+_{i},\widetilde{c}_{i},\widetilde{\rho}^*_{i},\widetilde{\rho}_{i}]
\end{align}
where, similar to Eq.~(\ref{eq:WEMDFTGW}), the bare dual fermion and boson propagators are given by
\begin{align}
\label{eq:tildeG0}
  \widetilde{G}_{0,\kv\nu} &= [(G^{\text{loc}}_{\nu})^{-1}+\Delta_{\nu}-\varepsilon_{\bf k}]^{-1} - G^{\text{loc}}_{\nu} = G^{\rm loc}_{\kv\nu} - G^{\text{loc}}_{\nu} ,
\\ \label{eq:tildeW0}
  \widetilde{W}_{0,\qv\omega} &= \alpha_{\omega}^{-1}\left[[U_{\bf q\omega}\!-{\cal U}_{\omega}]^{-1} \!-\! \chi_{\omega}\right]^{-1}\!\!\alpha_{\omega}^{-1} = W^{\rm loc}_{\qv\omega} -{\cal W}^{\text{loc}}_{\omega}.
\end{align}
For convenience, we choose $\alpha_{\omega}={\cal W}_{\omega}/{\cal U}_{\omega}=(1+{\cal U}_{\omega}\chi_{\omega})$  as the local renormalization factor.

The explicit form of the dual interaction $V_{\text{eff}}$ is obtained analogously to DF by expanding the $c^{+}, c$-dependent part of the partition function into a series and integrating out these degrees of freedom ($\rho^*$, $\rho$ are built from  $c^+$, $c$). In addition to purely fermionic vertex functions, the result contains also fermion-boson vertices. The corresponding lowest order terms in $V_{\text{eff}}$ are given by 
\begin{align}
  V_{\text{eff}}[\widetilde{c}^+,\widetilde{c},\widetilde{\rho}^*]
  =&\sum_{\nu\nu'\omega}\gamma^{\nu\omega}\,\widetilde{c}^+_{\nu+\omega}\widetilde{c}^{\phantom{*}}_{\nu} \widetilde{\rho}^{*}_{\omega} + \gamma^{\nu\omega\,*}\,\widetilde{c}^+_{\nu}\widetilde{c}^{\phantom{*}}_{\nu+\omega} \widetilde{\rho}_{\omega} \notag \\
  & +\frac14\,\sum_{\nu\nu'\omega}F^{\nu\nu'\omega}\,\widetilde{c}^{+}_{\nu}\widetilde{c}^{\phantom{*}}_{\nu+\omega}\widetilde{c}^{+}_{\nu'+\omega}\widetilde{c}^{\phantom{*}}_{\nu}.
\end{align}
The spin dependence, which is the same as in Eq.~(\ref{equ:vdef}), is suppressed for clarity. The three-point fermion-boson vertex $\gamma^{\nu\omega}$ can be expressed as in TRILEX (Sec.\ref{sec:TRILEX}) in terms of the original variables of the impurity reference system as~\cite{Ayral2016,Rohringer2016}
\begin{align}
\label{eq:3pvertex}
  \gamma^{\nu\omega}&=G^{-1}_{\nu}G^{-1}_{\nu+\omega}\alpha^{-1}_{\omega}\av{c^{\phantom{*}}_{\nu}c^{+}_{\nu+\omega}\rho^{\phantom{+}}_{\omega}},
\end{align}
and the four-point vertex function $F^{\nu\nu'\omega}$ is the same as in the DF theory [Eq.~\eqref{equ:vdef}].
Note that the effective electron-boson interaction never vanishes even if the local electron-electron interaction goes to zero~\cite{vanLoon2014a}. The effective fermions always interact with the effective bosons through a three-leg vertex which is of order unity. From this viewpoint, DMFT appears to be a more robust approximation in finite dimensions than EDMFT, which requires at least additional GW-like diagrams. Even the static nonlocal Fock term cannot be neglected~\cite{Ayral2017a}.

\begin{figure}[t]
  \includegraphics[width=.95\columnwidth]{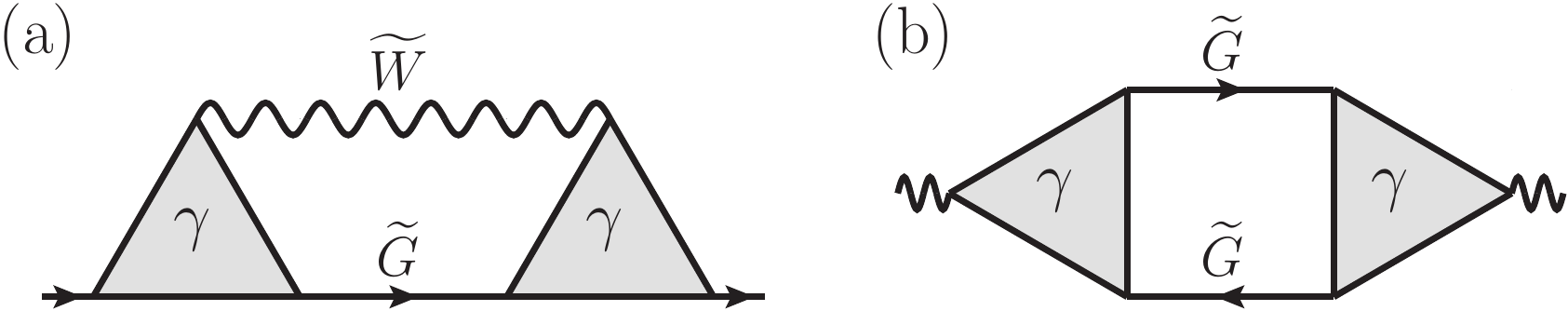}
    \caption{Second-order diagrams contributing to the nonlocal (a) fermionic and  (b) bosonic DB self-energies, i.e., to $\widetilde{\Sigma}_{\kv\nu\sigma}$ and $\widetilde{\Pi}_{\qv\omega}$, respectively.}
  \label{fig:dbdiag}
\end{figure}

Free dual boson propagators correspond to the EDMFT approximation. Corrections to EDMFT are obtained by constructing the dual self-energy $\widetilde{\Sigma}_{\kv\nu}$ and polarization operator $\widetilde{\Pi}_{\qv\omega}$ as well as renormalized dual propagators, i.e., the dual Green's function $\widetilde{G}_{\kv\nu} =~ -\av{\widetilde{c}^{\phantom{+}}_{\kv\nu}\widetilde{c}^{+}_{\kv\nu}}$ and the screened dual interaction $\widetilde{W}_{\qv\omega} =~ -\av{ \widetilde{\rho}^{\phantom{*}}_{\qv\omega} \widetilde{\rho}^{*}_{\qv\omega}}$ from these building blocks diagrammatically~\cite{Rubtsov12, vanLoon2014a, Stepanov2016}; see Fig.~\ref{fig:dbdiag}.

The physical Green's function $G_{\kv\nu}$ and the renormalized interaction $W_{\qv\omega}$ of the original model can be expressed in terms of the dual quantities as follows:
\begin{align}
\label{eq:DBsig}
  G_{\kv\nu}^{-1} &= \,(G^{\rm loc}_{\kv\nu})^{-1}\, - {\widetilde\Sigma_{\kv\nu}}({1+G^{\text{loc}}_{\nu}\widetilde\Sigma_{\kv\nu}})^{-1},\\
\label{eq:dual_X-Pi}  
  W_{\qv\omega}^{-1} &= (W^{\rm loc}_{\qv\omega})^{-1} - {\widetilde\Pi_{\qv\omega}}({1+{\cal W}^{\text{loc}}_{\omega}\widetilde\Pi_{\qv\omega}})^{-1}.
\end{align}

The fermionic and bosonic bath can be taken from a converged EDMFT calculation  (which is numerically less costly) or 
can be determined  self-consistently within DB theory. The latter is possible via the EDMFT self-consistency conditions~\eqref{eq:edmftselfc} but also other conditions are discussed \cite{Stepanov2016}. The  dual polarization operator $\widetilde{\Pi }_{\mathbf{q}\omega }$ can, e.g., be chosen as the sum of ladder diagrams~\cite{vanLoon2014,Hafermann2014a,vanLoon2014a,Stepanov2016}:
\begin{equation}
\label{eq:dualpolladder}
  \widetilde{\Pi }_{\mathbf{q}\omega }=\sum_{\nu\nu'}\gamma^{(\nu +\omega) -\omega }\widetilde{\chi }_{\mathbf{q}}^{\nu \omega }\left[I+F^{\nu \nu ^{\prime}\omega }\widetilde{\chi }_{\mathbf{q}}^{\nu^{\prime}\omega }\right] _{\nu \nu ^{\prime }}^{-1}\gamma^{\nu ^{\prime}\omega },
\end{equation}%
where $\widetilde{\chi }_{\mathbf{q}}^{\nu \omega } =\sum\limits_{\mathbf{k}}\widetilde{G}_{\mathbf{k}}(\nu )\widetilde{G}_{\mathbf{k+q}}(\nu +\omega )$.

Using ladder diagrams within the DB approach~\cite{vanLoon2014,Hafermann2014a,vanLoon2014a}, one obtains the following  physical (lattice) susceptibility for the extended Hubbard model [see \onlinecite{Stepanov2016} who used $X$ instead of $\chi$]:
\begin{equation}
\label{eq:Suscept_phys}
  \chi_{\mathbf{q}\omega }=\frac{1}{1/\Pi _{\mathbf{q}\omega }^{(1)}+\Lambda_\omega-V_{{\bf q}\omega}},
\end{equation}
where $\Pi _{\mathbf{q}\omega }^{(1)} =(\chi _{\omega }+\chi_{\omega }\widetilde{\Pi }_{\mathbf{q}\omega }\chi_{\omega })$. This physical susceptibility fulfills the important property of charge conservation $\lim_{\vc{q}\to \vc{0}}\chi_{\mathbf{q}\omega\neq 0}=0$ see~\onlinecite{Hafermann2014a} and \onlinecite{Stepanov2016}.	

Equation~\eqref{eq:Suscept_phys} also allows us to make a close connection to the Moriya $\lambda$ correction in ladder D$\Gamma$A (Sec.~\ref{sec:dgaladder}). When the interaction is local and nonlocal effects from the self-energy on the internal fermionic lines are neglected, \onlinecite{Hafermann2014a} showed that $\Pi _{\mathbf{q}\omega }^{(1)}$ is equivalent to the DMFT  susceptibility. Hence with $V_{{\bf q}\omega}=0$, the ladder DB susceptibility \eqref{eq:Suscept_phys} equals that of ladder D$\Gamma$A if the  D$\Gamma$A Moriya $\lambda$ correction is taken as frequency dependent and equal to the DB $\Lambda_{\omega}$.

Also note that the $\Lambda_\omega$ in the susceptibility \eqref{eq:Suscept_phys} allows us to restore a property that is broken in DMFT (with $\Lambda_\omega=0$), namely, that the double occupation calculated from the susceptibility of the impurity problem equals that calculated from the susceptibility of the lattice problem. The DB double occupancy is closer to DCA benchmarks than either DMFT result~\cite{vanLoon2016}. The momentum dependence of the polarization operator introduced in DB is important for a thermodynamically consistent treatment of the charge response~\cite{vanLoon2015}. More applications of the dual boson approach are discussed in Sec.~\ref{sec:nonlocalv}.

\subsubsection{TRILEX approach}
\label{sec:TRILEX}

The physical motivation for the triply irreducible local expansion (TRILEX) scheme by \onlinecite{Ayral2015,Ayral2016} is to include both Mott and spin-fluctuation physics (long-range bosonic modes), which are thought to be essential ingredients to describe high-temperature superconductivity~\cite{Dagotto94}. It is based on a similar functional construction as QUADILEX (Sec.~\ref{sec:quadrilex}), but now based on the functional $\mathcal{K}_{3}$, which contains all three- and two-particle  irreducible diagrams~\cite{DeDominicis1964}, corresponding to three Green's function legs. The TRILEX approximation restricts these diagrams to the local ones, i.e., approximates  $\mathcal{K}_{3}$  by its local counterpart \cite{Ayral2016}
\begin{align}
\label{eq:k3trilex}
  \mathcal{K}_{3}[G_{{\mathbf k}\nu},W_{{\mathbf q}\omega},\chi_{{\mathbf k}{\mathbf q}}^{\nu\omega}] \approx \mathcal{K}_{3}[G_\nu,W_{\omega},\chi^{\nu\omega}].
\end{align}
In addition to $G$ and $W$, $\mathcal{K}_{3}$ is also a functional of the three-point electron-boson correlation function $\chi$, which should not be confused with the local susceptibility.  As in EDMFT, the functional~\eqref{eq:k3trilex} can be obtained from a self-consistently determined quantum impurity model which now includes a dynamical electron-boson coupling $\lambda_{\nu\omega}$  related to  $\chi$.

In the general framework, the fermionic and bosonic self-energies are given by the \onlinecite{Hedin1965} equations
\begin{align}
  \Sigma_{\kv\nu} &= -\sum\limits_{\qv\omega}G^{\phantom{\kv}}_{(\kv-\qv)(\nu-\omega)}W^{\phantom{\kv}}_{\qv\omega}\gamma_{\kv\qv}^{\nu\omega} =
  \includegraphics[width=0.16\linewidth]{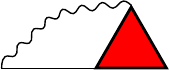}~, \notag\\
\label{eq:HedinPi}
  \Pi_{\qv\omega} &= \,2\,\sum\limits_{\kv\nu}G^{\phantom{\kv}}_{(\kv+\qv)(\nu+\omega)}G^{\phantom{\kv}}_{\kv\nu}\,\gamma_{\kv\qv}^{\nu\omega}=
  \includegraphics[width=0.16\linewidth]{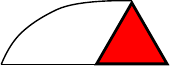}~, 
\end{align}
where $\gamma_{\kv\qv}^{\nu\omega}$ is the exact three-point lattice vertex. \onlinecite{Ayral2015} approximated  this quantity by its local counterpart
\begin{equation}
  \gamma_{\kv\qv}^{\nu\omega}\approx\gamma^{\nu\omega} \; .
\end{equation} 
This vertex is  computationally easier to handle than most diagrammatic extensions discussed in this review, because $\gamma^{\nu\omega}$ depends on only two instead of three independent frequencies. TRILEX bears some similarity to EDMFT+$GW$, but  the nonlocal $GW$ diagrams are now additionally dressed by $\gamma^{\nu\omega}$.

Application of the formalism to a purely fermionic model such as the Hubbard model requires the introduction of bosonic fields. To this end, the Coulomb interaction is (arbitrarily) decomposed into one or more channels (charge and spin in $x$, $y$, and $z$ directions). By construction, the method interpolates, for the charge channel, between the $GW$ approximation at weak coupling and the atomic limit at strong coupling. It yields  Fermi arcs,  spin fluctuations, and a reduction of the mean-field critical temperature (to what extent depends on the chosen decoupling),  but yields a slightly larger critical  $U$ for the Mott transition than DMFT \cite{Ayral2016}.

One may speculate that TRILEX underestimates spatial correlations because a local approximation to the three-leg vertex is more restrictive.
Indeed,  D$\Gamma$A has been formulated also in the form of a three-leg vertex similar to Eq.~\eqref{eq:HedinPi}; see \onlinecite{Katanin2009}. But in D$\Gamma$A  $\gamma_{\kv\qv}^{\nu\omega}$ is obtained from the BSE ladder [Eq.~\eqref{equ:dgabse}] constructed from the local two-particle (four leg) vertex  $\Gamma_{ph,r}^{\nu\nu'\omega}$ and is $\qv$ dependent (nonlocal). The same holds for DB.

There are also apparent similarities between the TRILEX and DB approaches. Both introduce bosonic modes. But while in  TRILEX the local three-leg vertex of the impurity model enters the calculation and the two-particle polarization is  included through Eq.~\eqref{eq:HedinPi}, in DB the local four-leg susceptibility directly enters the calculation. One can reshuffle the DB diagrams into a form with a three-leg vertex as in  Eq.~\eqref{eq:HedinPi}. This DB three-leg vertex then includes the nonlocal $\widetilde{W}$ and  $\widetilde{G}$ of DB and  is again nonlocal.

\subsection{Conservation laws and two-particle self-consistency}
\label{sec:conservation}

Within an ideally exact treatment of a correlated many-body problem, rigorous equations relate the one- and two-particle Green's functions in specific ways. Some of these relations are considered to be of particular importance, because they reflect fundamental aspects of the underlying physics, as the conservation laws and the Pauli principle. For instance, in an exact theory, the one- and two-particle Green's  functions must fulfill all the continuity equations for the conserved quantities of the theory  (particle number, energy, etc.), as well as all the crossing symmetries related to the Pauli principle. This implies, in turn, the fulfillment of famous sum-rules, such as the $f$-sum rule for the optical conductivity (an example for the first case) and the relation between the local/instantaneous charge and spin susceptibilities and the electronic densities or the double occupancies [for an example of the second case, see, e.g., Eq.~(\ref{equ:sumchi})]. 

For approximate theories, such as the diagrammatic extensions of DMFT, at least some of these exact relations are violated. In fact, when considering a given many-body approximation, it is always important to understand which  conservation laws and sum rules are preserved and which are not. Usually, and consistent with the above considerations, we distinguish between two different classes of approaches: (i) the ones which satisfy (all) conservation laws [and the related sum rules; see \onlinecite{BaymKadanoff}] and (ii) the ones which fulfill sum rules for one- {\sl and} two-particle Green's functions based on the Pauli principle \cite{Vilk1997}. 

The former class of approximations, defined as ``conserving'', can be derived from the Luttinger-Ward functional expressed in terms of the one-particle Green's function $\Phi[G]$ by taking the first and second functional derivatives with respect to  $G$ to define the 1PI self-energy and 2PI vertex functions, respectively. This procedure, referred to as $\Phi$ derivability (of a given  approach), guarantees that all conserved quantities at the microscopic level (e.g., at each scattering event in the diagrammatic theory) are translated into corresponding continuity equations and sum rules at the macroscopic level.

The latter class of sum rules, instead, is typically   satisfied in parquet-based \cite{Bickers2004} approaches or explicit two-particle self-consistent schemes \cite{Vilk1994}. It has been conjectured \cite{Janis1999,Bickers2004} that  no theoretical approach for correlated electrons should be able to  fulfill both requirements, except for the exact solution.  For instance, it has been stated \cite{Smith1992,Janis1999} that any parquet-based approximation, which in general preserves the Pauli principle, is never conserving, unless the exact fully irreducible vertex $\Lambda$ is used as an input (which generates the exact solution). Moreover, it has been demonstrated that the set of diagrams fulfilling Ward identities differs in finite orders from the set needed to preserve analytic properties (causality) of the self-energy \cite{Janis2004}. \onlinecite{Janis1998} and \onlinecite{Janis2016} suggested ways to overcome such difficulties.

Whether conservation laws or the two-particle self-consistent relations compatible with the Pauli principle are more important for an approximate theoretical treatment cannot be answered in general as it depends on the specific application for which the approach is adopted: For the calculation of transport properties or the description of plasmons the exact fulfillment of (charge) conservation laws is expected to be crucial, while for the analysis of (second-order) phase transitions and critical phenomena the two-particle self-consistent relations, as those of the parquet equations, might be more important. 

As prototypical example, let us consider the RPA calculation of the susceptibilities. If this calculation is performed together with a  Hartree  self-energy, the approximation is conserving. Hence, charge conservation (as well as the gauge invariance) of the results is guaranteed. At the same time, RPA violates per construction the crossing symmetries related to the Pauli principle and, in fact, it provides a rather poor (i.e.\ mean field) description of the transition temperatures and critical properties.

Let us turn now to the specific case of the diagrammatic extensions of DMFT. To this aim, it is convenient to start by recapitulating the situation for DMFT. DMFT can be derived from a Luttinger-Ward functional expressed in terms of a purely local one-particle Green's function $\Phi[G_{\text{loc}}]$ \cite{Janis92}. Therefore, DMFT is a conserving approach, and all quantities conserved at the microscopical level are translated, thanks to its $\Phi$ derivability \cite{BaymKadanoff}, into the corresponding continuity equations, e.g., for the particle number and the energy \footnote{Note that this is not the case for the momentum. In fact, because of the purely local nature of the DMFT diagrammatics no momentum conservation holds at the ``microscopic'' level (i.e., for each scattering process in the diagrammatics). Hence, the $\Phi$ derivability of DMFT does not guarantee the fulfillment of a continuity equation for the momentum \cite{Hettler2000}.} and the related sum rules (such as, the $f$-sum rule), although care has to be taken when deriving two- and more-particle Green's functions from $\Phi[G]$ in infinite dimensions \cite{Janis1999a}. On the other hand, whenever DMFT is used as an approximation for treating finite dimensional systems it breaks several two-particle self-consistent relations, including the ones preserving the Pauli principle: This leads, for example, to a violation of the $\chi$-sum rule Eq.~(\ref{equ:sumchi}) and to intrinsic inconsistencies in the calculation of the potential energy \cite{vanLoon2016,Rohringer2016} depending on whether it is computed directly from the two-particle Green's function or from one-particle quantities only via the Galitskii-Migdal formula (see, e.g., Fig.~2 in~\onlinecite{vanLoon2016}). An attempt to make DMFT two-particle self-consistent by adding a dynamic interaction in the impurity model leads to inconsistencies at the one-particle level and violation of conservation laws \cite{Krien2017}.

The D$\Gamma$A as well as the  DF (in their parquet implementation) is based on the solution of the parquet equations and, hence,  two-particle self-consistent. Per construction they preserve  all the crossing relations. Consistent with the conjecture of \onlinecite{Smith1992} both are however per se not conserving. Cutting Green's function lines of the  D$\Gamma$A or  DF self-energy (in the spirit of a  $\Phi[G]$ derivable theory) also implies cutting internal Green's function lines of the local, fully irreducible vertex; but such cuts include diagrams that are not taken into account in the susceptibilities. Similarly, the QUADRILEX  functional extension of  D$\Gamma$A cannot be written in terms of the (local) one-particle Green's function only, but it explicitly includes the (local) two-particle Green's function and is not $\Phi$-derivable.  It remains to be seen whether  some modifications as  e.g.\ along the lines of \onlinecite{Janis2017} can actually lead to a conservering approximation.

For the DB approach and similarly for the  $\lambda$ correction of the ladder  D$\Gamma$A additional equations arise which can be used to fulfill conservation laws. While EDMFT is conserving in infinite dimensions~\cite{Smith00}, it violates local conservation and breaks Ward identities in finite dimensions. This is a consequence of the fact that the polarization operator does not depend on momentum. In contrast, within the DB approach it is possible to include diagrammatic corrections that exactly restore the $q^{2}$ behavior of the polarization for small momenta as required by gauge invariance [while maintaining a local self-energy \onlinecite{Hafermann2014a}]; cf.~Sec. \ref{sec:nonlocalv}. On the other hand,  $\lambda$ corrections of the ladder  D$\Gamma$A  \cite{Katanin2009} have been employed to enforce the $\chi$ sum rule Eq.~(\ref{equ:sumchi}) and to guarantee the consistency of the potential energy at the one- and two-particle level as well as the fulfillment of the $f$-sum rule (related to charge conservation \onlinecite{Rohringer2016}).

\section{Applications and Results}
\label{Sec:results}

\subsection{Hubbard Model}
\label{sec:hubbard}
\label{sec:Results_HF}

The Hubbard model is arguably  the most fundamental model for strongly correlated electrons. Let us recall its  Hamiltonian, Eq. (\ref{Eq:Hubbard}):
\begin{equation}
\mathcal{H} = \sum_{ij,\sigma} t_{ij} c_{i\sigma}^{\dagger}c_{j\sigma}^{\phantom{\dagger}} + U\sum_{i}n_{i\uparrow}n_{i\downarrow},
\label{eq:Hubbard}
\end{equation}
with hopping amplitude  $t_{ij}$ and  local Coulomb repulsion $U$. The model provides the basic physical description of the Mott-Hubbard metal-insulator transitions (MIT) in bulk 3D systems~\cite{Gebhard1997,Imada1998} as well as of ferromagnetism~\cite{Mielke1993,Vollhardt98a} and antiferromagnetism~\cite{Jarrell1992}. In 2D, the Hubbard model~\eqref{eq:Hubbard} is believed to capture the low-energy physics of the superconducting cuprates~\cite{Dagotto94,Scalapino12}. Therefore the cases of 3D and 2D are  of particular interest. Despite its formal simplicity, an exact solution is known only in 1D through the Bethe ansatz~\cite{Lieb1968} and through DMFT in the limit of infinite dimensions~\cite{Metzner1989,Georges1996}.

In this section, we summarized the unified picture of the Hubbard model physics in finite dimensions as it emerges from the applications of diagrammatic extensions of DMFT; we compared the results to those of other methods, wherever available. Given that the starting point of diagrammatic extensions is a DMFT solution, we first considered the 3D system, which can be regarded as ``closer'' to the physics of $d=\infty$. We then subsequently lowered the dimension so that deviations from DMFT become progressively larger.  

\subsubsection{Three dimensions}
\label{sec:HM3d}

\begin{figure}[t]
  \includegraphics[width=\columnwidth,angle=0]{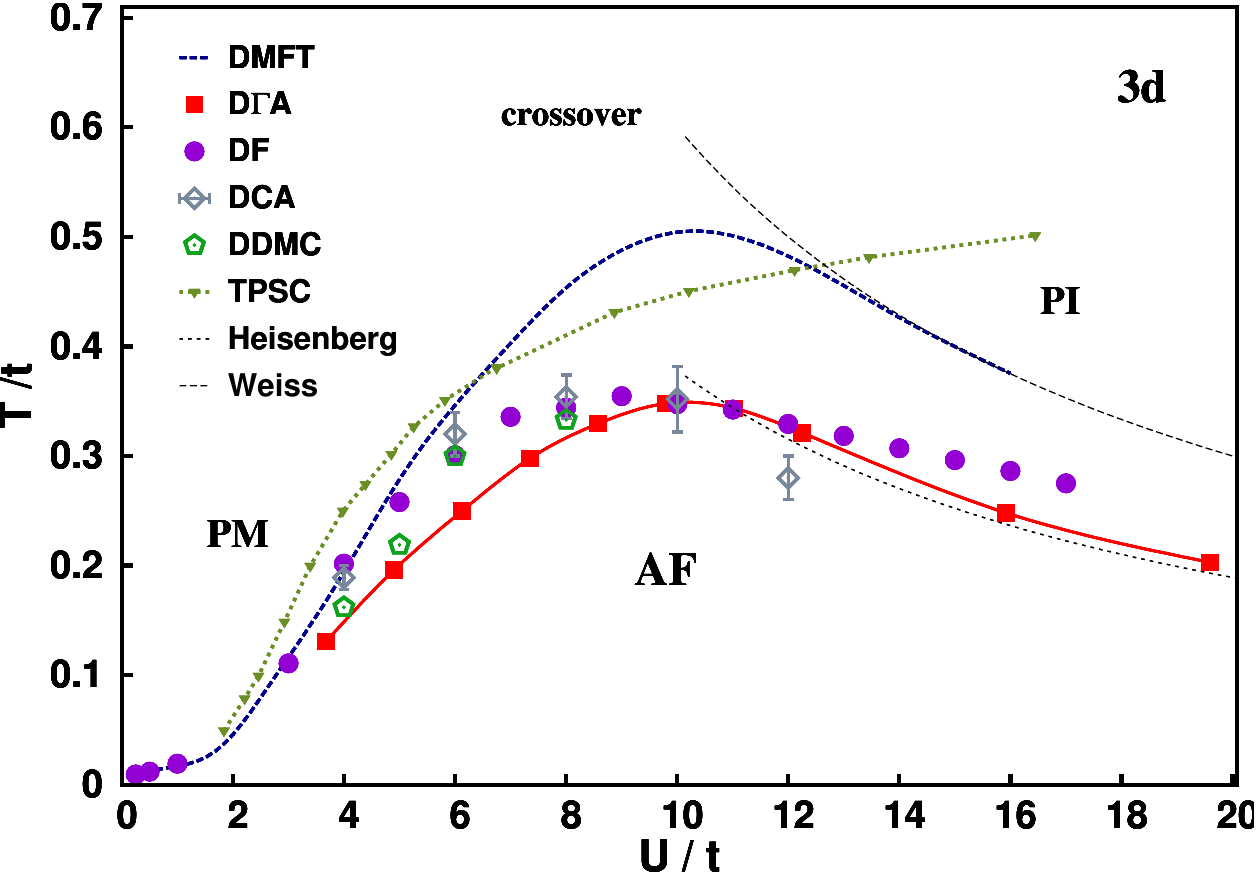} 
    \caption{(Color online) Phase-diagram of the half-filled Hubbard model in 3D with nearest-neighbor hopping ($t_{ij}=-t$), as obtained by various methods; see the text.}
  \label{fig:pd3d}
\end{figure}

The magnetic phase diagram of the unfrustrated 3D Hubbard model at half filling, obtained by several approaches, is shown in Fig.~\ref{fig:pd3d}. At high $T$, the model shows a crossover from a paramagnetic metal (PM) at weak coupling to a paramagnetic insulator (PI) at strong coupling. As the temperature is lowered, the model undergoes a second-order transition to an AF state; see \onlinecite{Georges1996,Kent2005} for the DMFT phase diagram. Non-local corrections beyond DMFT have been analyzed using D$\Gamma$A~\cite{Rohringer2011} and DF~\cite{Hirschmeier2015} in the ladder approximation. The two approaches yield a rather coherent picture of the underlying physics, for both the magnetic transition and its critical properties.

In particular, D$\Gamma$A and DF correctly predict a sizable reduction of the  N\'eel temperature $T_{N}$  w.r.t.~DMFT in Fig.~\ref{fig:pd3d}, which is a direct consequence of nonlocal spin fluctuations. The ratio of the reduction is, as expected, smaller at weak coupling~\cite{Schauerte2002}, and larger (more than $30\%$)  at intermediate to strong coupling. Notably, $T_{N}$ from D$\Gamma$A correctly approaches the  behavior of the Heisenberg model~\cite{Sandvik98} in the strong-coupling limit. This is a significant improvement over  DMFT which approaches the Weiss mean field theory of the Heisenberg model~\cite{Takahashi77} in this limit. It also improves over the two-particle self-consistent  theory (TPSC) \cite{Vilk1997} which reaches a plateau for $T_N$ at large $U$. Moreover, in the most interesting regime of intermediate coupling $8t\lesssim U \lesssim12t$, both methods agree remarkably well with the DCA results by \onlinecite{Kent2005} and diagrammatic determinant Monte Carlo (DDMC)  by \onlinecite{Kozik2013}, in spite of the intrinsic differences between these methods. Minor quantitative deviations between DF and D$\Gamma$A are observed only in the weak- and strong-coupling limits. At weak coupling, D$\Gamma$A has a smaller $T_N$ than DF, DCA~\cite{Kent2005}, TPSC~\cite{Dare2000} and perturbation theory~\cite{Schauerte2002}. On the other hand, it yields a result that is closer to  latest DDMC by \onlinecite{Kozik2013} which prevents a final judgment regarding the accuracy of the different approximations in this parameter regime. In the opposite limit, DF overestimates $T_{N}$~\cite{Hirschmeier2015}. We suspect that this result is not intrinsic to the method, but either due to the hybridization not being computed self-consistently or because the vertex function was computed at a single bosonic frequency. The latter approximation is sufficient to investigate critical behavior, but it can affect nonuniversal quantities such as $T_{N}$.  

\begin{figure}[t]
  \hspace{-.23em}\includegraphics[width=0.4\columnwidth,angle=0]{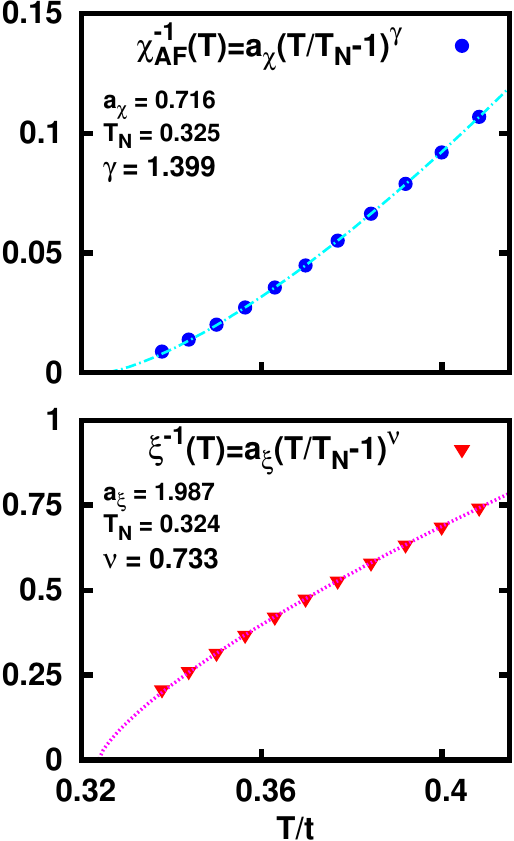}
  \hspace{.28cm} \includegraphics[width=0.54\columnwidth,angle=0]{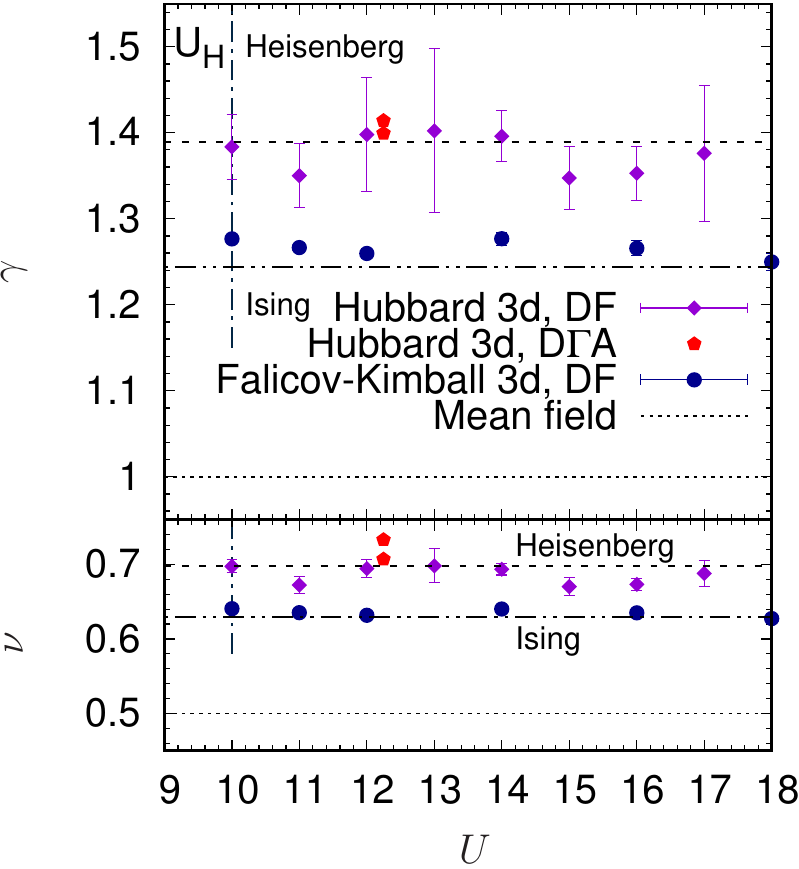}
    \caption{(Color online) Left panels:  D$\Gamma$A inverse AF susceptibility $\chi_{\text{AF}}^{-1}$ (top) and inverse correlation length $\xi^{-1}$ (bottom) for  the 3D Hubbard model at half filling and $U\approx 12.2t$. The corresponding critical exponents $\gamma$ and $\nu$ are obtained by the fit shown. Adapted from~\onlinecite{Rohringer2011}. Right panel: Critical exponents  $\gamma$ and $\nu$  vs   $U$ from  DF compared to the D$\Gamma$A result  from the left panels (and a second fit to estimate the error). DF results for the Falicov-Kimball model and the mean field critical exponents are shown for comparison. Adapted from~\onlinecite{Hirschmeier2015}.}
  \label{fig:exponents} 
\end{figure}

Diagrammatic extensions of DMFT have also been used to study the critical behavior of the 3D Hubbard model~\cite{Rohringer2011,Hirschmeier2015}. Results for the critical exponents $\gamma$ and $\nu$ governing the $T$ dependence of the diverging susceptibility ($\chi_{\text{AF}}$) and correlation length ($\xi$), respectively, are shown in Fig.~\ref{fig:exponents}. They differ significantly from the mean field exponents obtained within the DMFT description. Whether this is due to the dynamic (frequency-dependent) nature of the vertex functions or due to the type of self-consistency ($\lambda$ corrections in D$\Gamma$A or the inner/outer self-consistency in DF) is still a subject of investigation. In fact, a deviation from mean field exponents can also be observed in TPSC \cite{Dare2000}. The latter is based on the bare $U$ and  originates from enforcing two-particle self-consistency conditions. 

Within the present numerical precision, the DF and D$\Gamma$A critical exponents appear to be compatible with the universality class of the 3D Heisenberg model~\cite{Holm1993} and with the scaling relation $\gamma/\nu=2-\eta$ (the  exponent $\eta$ is small and could not yet be precisely extracted). This would be the expected result, as the Hubbard model maps onto the Heisenberg model in the strong-coupling regime, and the dimension and symmetry of the AF order parameter suggest the same universality class. At the same time, the current numerical uncertainty is already about 10$\%$ for a single-exponent fitting function [e.g., $\chi_{\text AF}^{-1}(T)= a (T- T_N)^\gamma$]. Allowing subleading order terms \cite{Semon2012} in the fitting function further increases the uncertainty: In this framework, the fitted exponents might also become compatible with the ratio of integers as, e.g., in the case of TPSC \cite{Dare2000}.

\begin{figure}[b]
  \includegraphics[width=0.8\columnwidth,angle=0]{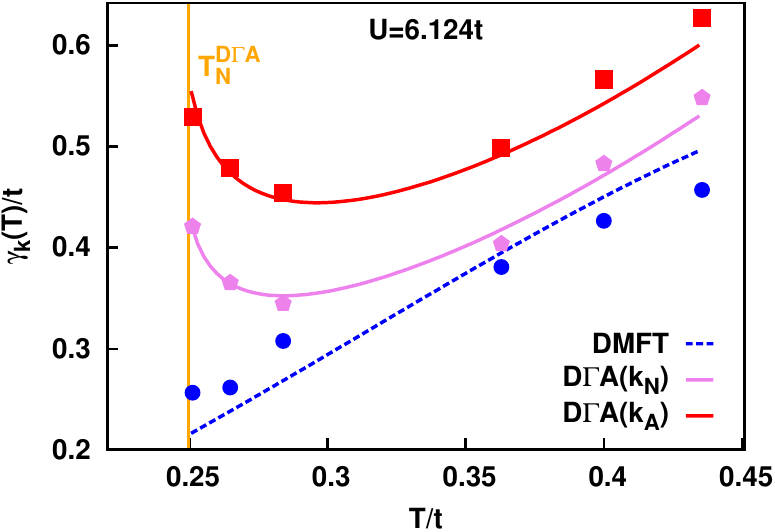} 
    \caption{(Color online) Scattering rates $\gamma_{\mathbf{k}}$ at the nodal $\mathbf{k}_N = (\pi/2,\pi/2,\pi/2)$ and antinodal $\mathbf{k}_A = (0, \pi/2, \pi)$ points in the 3D Hubbard model as obtained by D$\Gamma$A. Adapted from \onlinecite{Rohringer2016}.}
  \label{fig:scatt3d}
\end{figure}

The effects of antiferromagnetic fluctuations also show up in other quantities such as spectral functions~\cite{Katanin2009,Rohringer2011,Fuchs2011a,Rohringer2016} as well as in thermodynamic~\cite{Rohringer2016} and  transport properties~\cite{Gull2011}. This can be seen, e.g., in the $T$ dependence of the electronic scattering rate, defined as $\gamma_{\kv}=-\Im\Sigma(\kv,\omega=0)$. Figure~\ref{fig:scatt3d} shows results for $\gamma_{\kv}$ in DMFT and D$\Gamma$A~\cite{Rohringer2016} for an intermediate $U$ value. In DMFT the scattering rate decreases monotonously with decreasing $T$, as expected for a Fermi liquid. Instead the D$\Gamma$A results show a nonmonotonous behavior of $\gamma_{\mathbf{k}}$ with a minimum of the scattering rate at intermediate $T$. That is, as the phase transition is approached, $\gamma_{\mathbf{k}}$ increases due to the enhanced scattering at nonlocal spin fluctuations. An analogous behavior is predicted at weak coupling by TPSC~\cite{Vilk1997} with the significant difference that $\gamma_{\mathbf{k}}$ diverges (logarithmically) at $T_N$ while in D$\Gamma$A such a singularity is cut off by the local quasi-particle scattering rate of DMFT~\cite{Rohringer2016}. This demonstrates that the inclusion of DMFT physics qualitatively modifies the results obtained by perturbative approaches even at weak coupling.

\begin{figure}[t]
  \includegraphics[width=\columnwidth,angle=0]{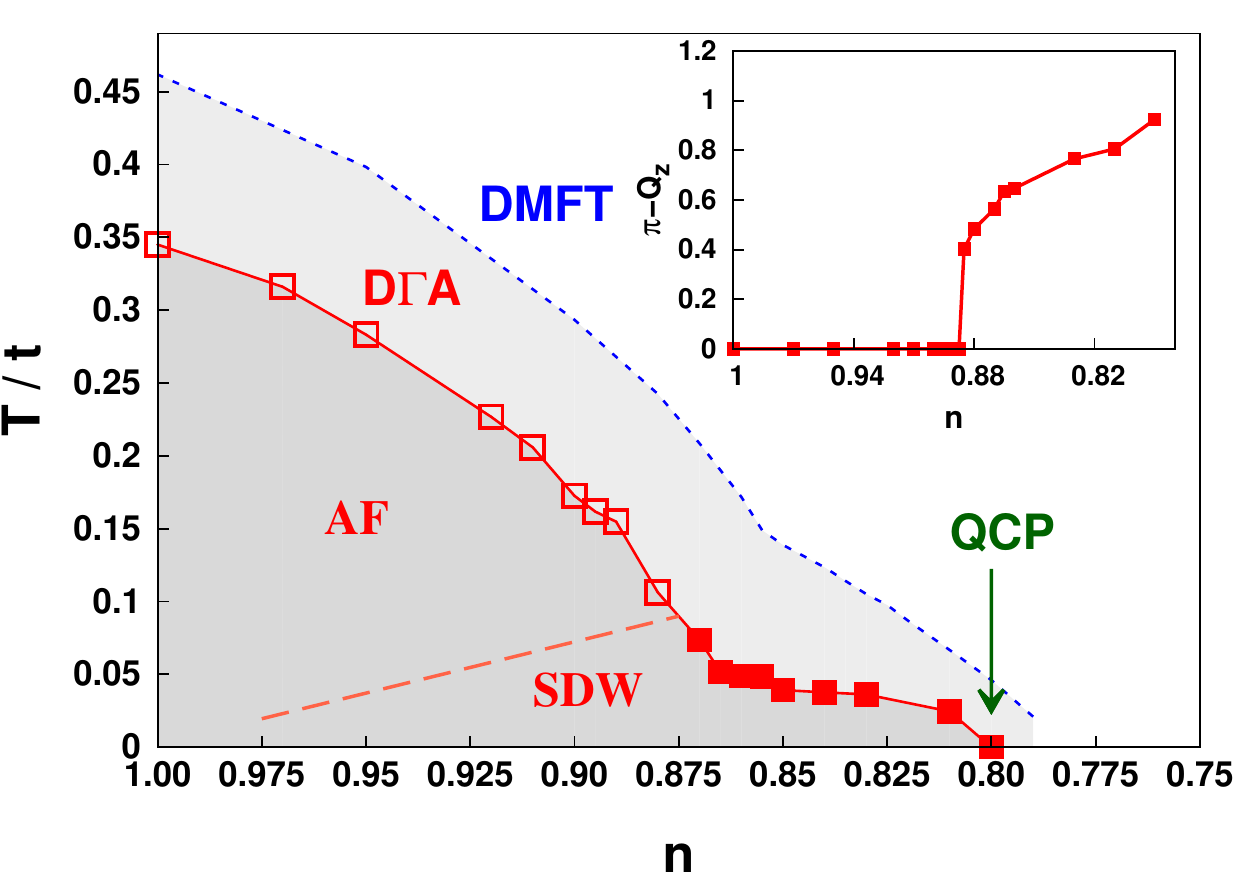} 
    \caption{(Color online) Magnetic phase diagram of the 3D Hubbard model as a function of filling $n$ and $T$ at fixed $U= 9.78 t$. D$\Gamma$A shows a transition from commensurate (open squares, AF)  to incommensurate ordering (filled square, SDW) as well as the emergence of a quantum critical point at $n\sim0.8$. Inset: Degree of incommensurability vs $n$. Adapted from~\onlinecite{Schaefer2016}.}
  \label{fig:pd3d-out}
\end{figure}

Motivated by the ability to describe the nontrivial physics of the finite-$T$ magnetic transition in the particle-hole symmetric case,  first calculations away from half filling have been recently performed in 3D within D$\Gamma$A by \onlinecite{Schaefer2016}. Beyond the extension of the magnetic phase diagram in 3D, the main interest here lies in the occurrence of a quantum critical point (QCP); see Fig.~\ref{fig:pd3d-out}. By progressively decreasing the electronic density $n$ at a fixed $U \simeq 10 t$, one finds that (i) $T_N$ is progressively reduced, (ii) the AF pattern becomes incommensurate with an ordering vector ${\bf Q}= (\pi,\pi,\pi-Q_z)$, and (iii) a QCP eventually emerges at about $20\%$ of hole-doping. The D$\Gamma$A and analytical calculations in \onlinecite{Schaefer2016} further show how the corresponding quantum critical properties are driven by Fermi-surface features (specifically by lines of Kohn points in the present case). These properties yield  quite different exponents and scaling relations ($\nu=1,\gamma=0.5$ as in an analytical RPA calculation) compared to those predicted by conventional Hertz-Millis-Moriya theory (where $\nu=0.75,\gamma=1.5$ \onlinecite{Loehneysen2007}).

A quantum phase transition between a paramagnetic semimetal and an antiferromagnetic insulator has also been analyzed most recently for the honeycomb lattice by \onlinecite{Hirschmeier2018}. Their crossover line between the non-Fermi-liquid regime and the renormalized classical regime agrees with a variety of other numerical methods.

\subsubsection{Two dimensions: Square lattice at half filling}
\label{sec:2dhubbsquare}

The two-dimensional case poses a stringent test on diagrammatic extensions of DMFT for two reasons: As a consequence of the theorem by \onlinecite{Mermin1966}, continuous symmetries cannot be spontaneously broken in 2D at a finite temperature in systems with short-range interaction. This means that (i) diagrammatic extensions have to account for fluctuations at all length scales, because they are essential for the proof of the theorem, and (ii) the physics of the DMFT starting point is qualitatively incorrect because it predicts a finite temperature mean-field transition to the AF-ordered state.

Cluster calculations struggle to account for long-range fluctuations and are hampered by the sign problem away from half filling. Diagrammatic extensions therefore provide a valuable complementary viewpoint, and fulfill the Mermin-Wagner theorem \cite{Katanin2009}. These methods have mainly been applied to the square and frustrated triangular lattices. Because these two lattices give rise to rather different physics, we review them separately.

\paragraph*{Metal-insulator transition vs crossover}

The occurrence of a MIT in the phase diagram of the 2D Hubbard model, together with its physical interpretation, has been intensely debated in the literature since the 1970's \cite{Castellani1979,Vekic1993,Vilk1996,Anderson1997,Vilk1997,Mancini2000,Moukouri2000,Avella2001,Hubbcollbook,Boies1995}. There is a general consensus about the AF ordering of the ground state, which smoothly evolves from a nesting-driven (Slater) to a superexchange-driven (Heisenberg) AF insulator with increasing $U$. In the Heisenberg limit, we have effective spin degrees of freedom coupled by the exchange interaction $J=4t^{2}/U$ governing the low-energy physics. In the view of \onlinecite{Anderson1997}, the 2D Hubbard physics would be similar to 1D and, thus, intrinsically non-perturbative, with a gap present for all $U>0$. In this heuristic picture, which is in contrast with more rigorous studies of the 1D $\rightarrow$ 2D crossover \cite{Castellani1992,Boies1995}, localized moments would form at sufficiently low temperatures, open a spectral gap, and finally order at $T=0$.

\begin{figure}[t]
  \includegraphics[width=\columnwidth,angle=0]{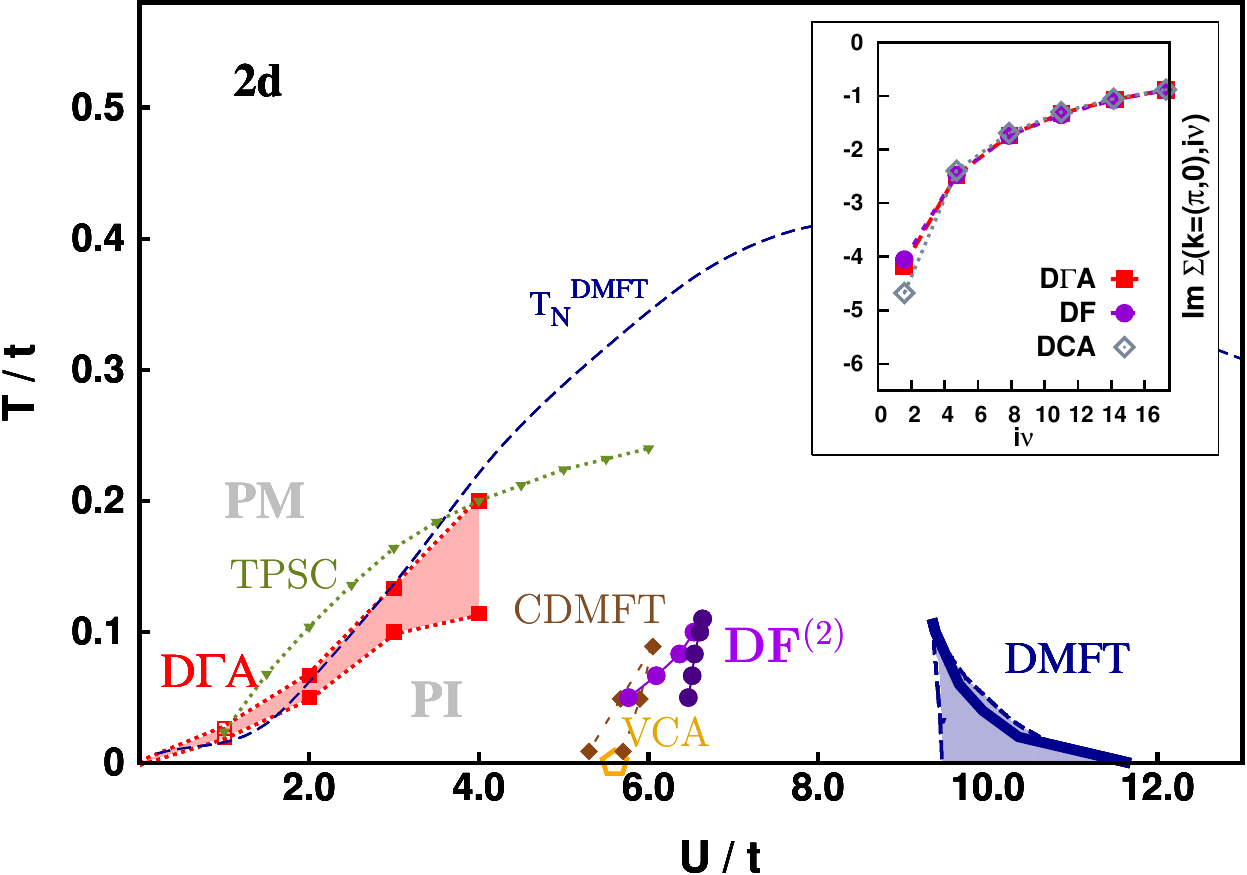}
    \caption{(Color online) Phase diagram of the square lattice 2D Hubbard model at half filling as obtained by various methods, see the text. The two lines of DMFT, CDMFT, and DF$^{(2)}$ represent the border of the corresponding coexistence regions (first-order transition), whereas there is a smooth crossover in TPSC and D$\Gamma$A (with a pseudogap region shaded in red/light gray). The DMFT N\'eel temperature is also given for reference (blue/dashed line). Inset (from \onlinecite{LeBlanc2015}): Comparison of the self-energy  at the antinodal point for $U/t=8$, and $T/t=0.5$.}
  \label{fig:pd2d}
\end{figure}

We discuss the phase diagram shown in Fig.~\ref{fig:pd2d} in view of this background, starting from the purely local description of DMFT, and adding the effect of nonlocal correlations over progressively larger length scales. In DMFT, by enforcing the PM solution, one finds the well-known first-order Mott MIT [with associated coexistence region, see \onlinecite{Georges1996}], ending with  a second-order critical endpoint at $U_c \simeq 10 t$ (taken from \onlinecite{Blumer2002} and \onlinecite{Park2008}).  The low-$T$ Mott PI is characterized by independent spin-$1/2$ magnetic moments with a high residual entropy of $\ln 2$ per site. As a result the transition line in Fig.~\ref{fig:pd2d} has a negative slope.

When short-range (AF) correlations between the moments are included, as in CDMFT~\cite{Park2008},\footnote{CDMFT \cite{Fratino2017} and DCA calculations \cite{Moukouri2001,WernerPC13,Merino2014} not shown in Fig.~\ref{fig:pd2d} also indicate a reduction of $U_c$ with increasing cluster size.} variational cluster approximation (VCA)~\cite{Schaefer2015-2}, or second-order DF (DF$^{(2)}$)~\onlinecite{Hafermannphd}), several changes are observed in Fig.~\ref{fig:pd2d}: (i)  $U_c$ is considerably reduced, (ii) the width of the coexistence region shrinks, and (iii) the entropy of the PI phase is strongly reduced, so that the slope of the transition line  is reversed. The position, as well as  the physical nature of the MIT, changes further by including AF correlations of progressively larger spatial extension. When even long-range AF fluctuations  are included, as in ladder D$\Gamma$A, extrapolated lattice QMC \cite{Schaefer2015-2}, ladder DF \cite{vanLoon2017}, TPSC~\cite{Vilk1996,Vilk1997}, and the nonlinear sigma model approach by \onlinecite{Borejsza2003,Borejsza2004}, the MIT is eventually transformed into a {\sl crossover} located at a very small $U$ value, compatible\footnote{The numerical determination of the crossover to a PI behavior is particularly challenging in the $U/t \rightarrow 0$ limit due to the  increasing length scales; for a specific discussion, see \onlinecite{vanLoon2017}. In this respect the D$\Gamma$A estimate at the lowest $U=t$ (empty symbol) should be regarded, most likely, as an upper bound limit for the crossover position.} with $U_c \rightarrow 0$ for $T\rightarrow 0$  (red colored/light gray region in Fig.~\ref{fig:pd2d}). As illustrated by the results presented next, the physical origin of the  low-$T$ insulating behavior in the 2D system is completely different from the one behind the Mott insulating phase described by DMFT.

\begin{figure}[t]
  \includegraphics[width=0.99\columnwidth,angle=0]{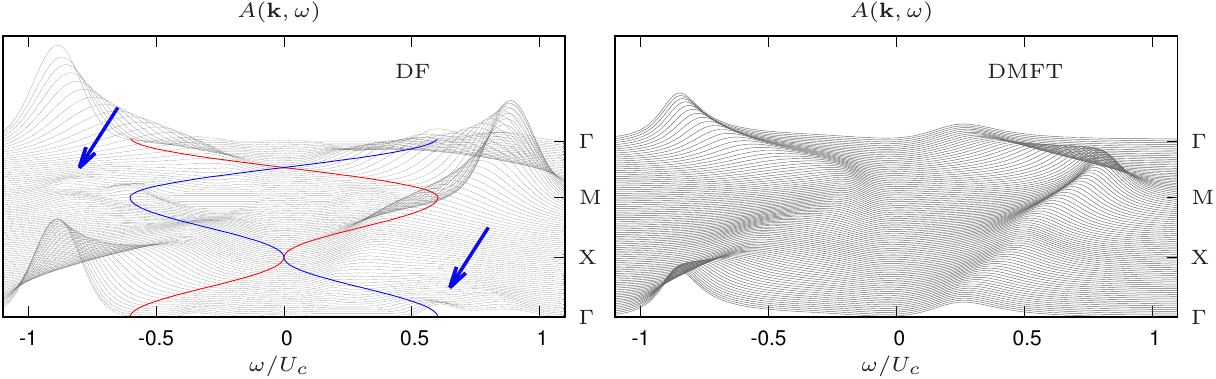} 
    \caption{(Color online) Spectral function $A(\kv,\omega)$ in DF$^{(2)}$ (left) and DMFT (right) at $T=0.22$ for $U=7$ and  $U=10$, respectively. The frequency has been rescaled by the respective critical $U_c$, which is finite $U_{c}=6.64t$ for DF$^{(2)}$ and $U_{c}=9.35t$ for DMFT. Colored lines show the bare dispersion $\varepsilon_{\kv}$ (red/light gray) and $\varepsilon_{\kv+\QV}=-\varepsilon_{\kv}$ (blue/dark gray) with $\QV=(\pi,\pi)$, which corresponds to a folding of the band at the effective magnetic zone boundary. The structures marked by arrows arise from dynamical short-range antiferromagnetic correlations captured in DF$^{(2)}$. From \onlinecite{Brener08}.}
\label{fig:df2dspec}
\end{figure}

\paragraph*{Physical observables and interpretation}

While at high $T$ and large $U$ the results for the one-particle self-energy (and spectral function) of D$\Gamma$A and DF are very similar to those of cluster methods (DCA) (see the inset of Fig.~\ref{fig:pd2d})~\cite{LeBlanc2015}, a more detailed discussion is necessary at low temperatures. In particular, we start by noting that in second-order  DF$^{(2)}$ the diagrammatic contributions decay rapidly in real space. Figure~\ref{fig:df2dspec} shows the second-order DF$^{(2)}$ results for the spectral function which demonstrates how short-range correlations affect this quantity compared to DMFT: At relatively large $U$ values, just above the Mott transition point, broad spectral structures appear in the vicinity of the $\Gamma$ point. \onlinecite{Brener08} attributed these ``shadow bands''  [akin to the (red and blue) dispersion in  Fig.~\ref{fig:df2dspec}] to short-ranged dynamical AF correlations.

\begin{figure}[t]
\begin{tabular}{cc}
  \includegraphics[width=0.51\columnwidth,angle=0]{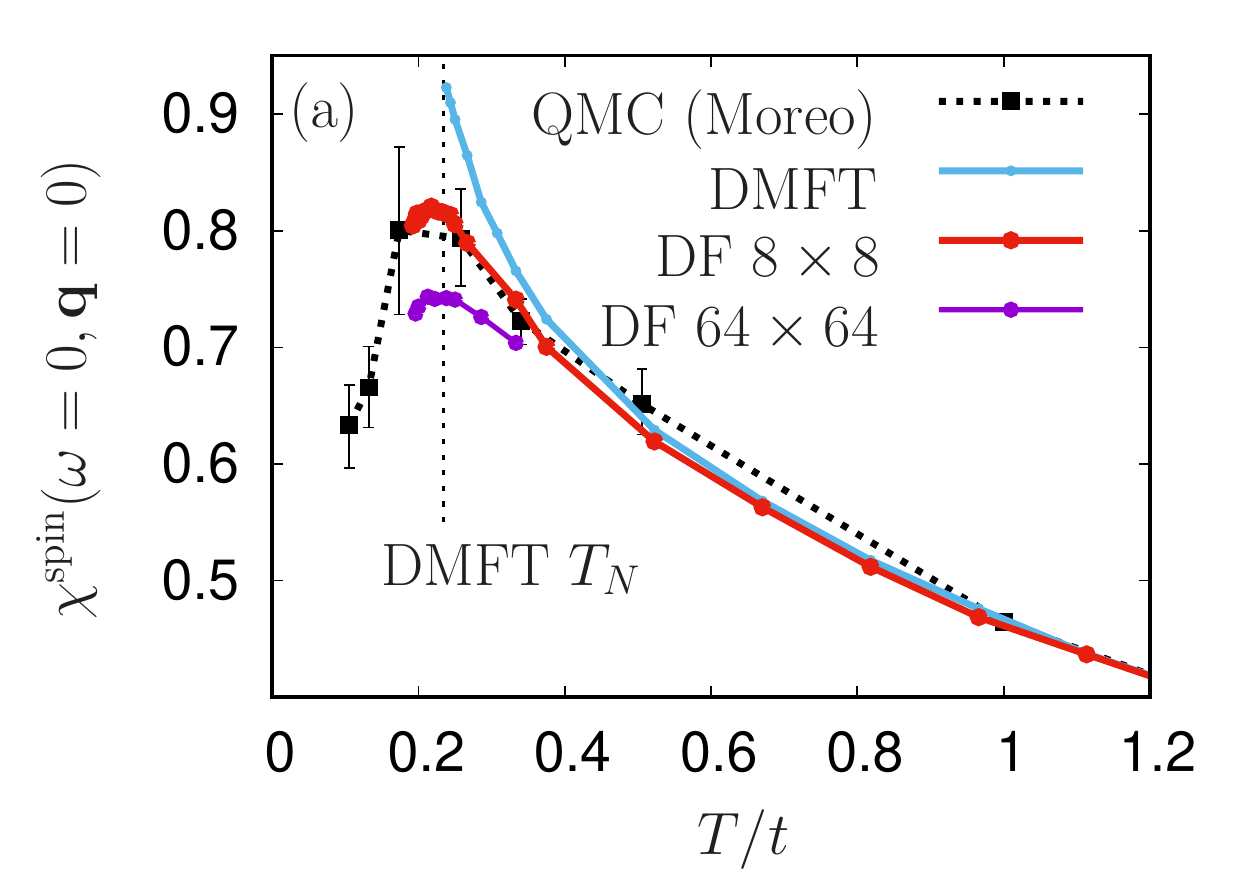} & \hspace{-0.75em}\includegraphics[width=0.51\columnwidth,angle=0]{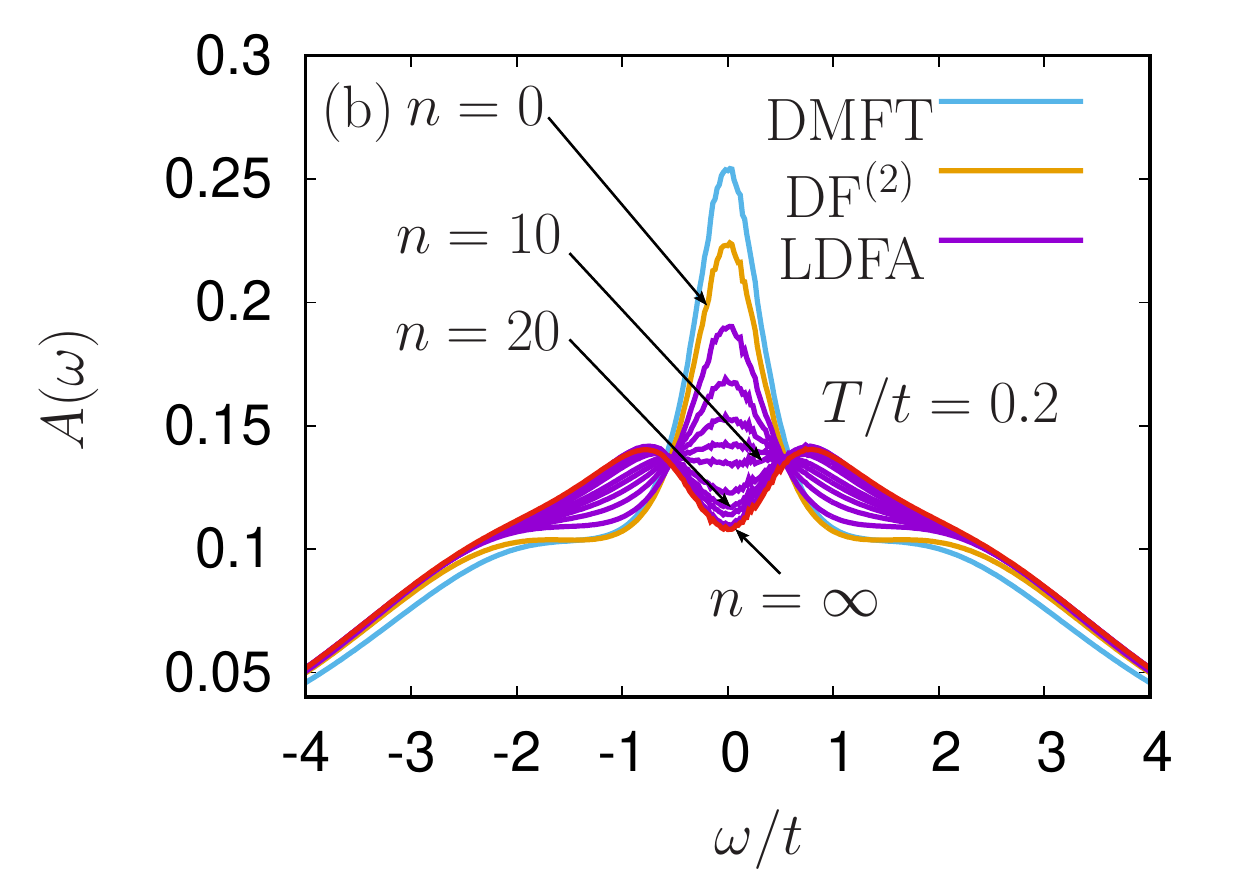}
\end{tabular}
  \includegraphics[width=\columnwidth,angle=0]{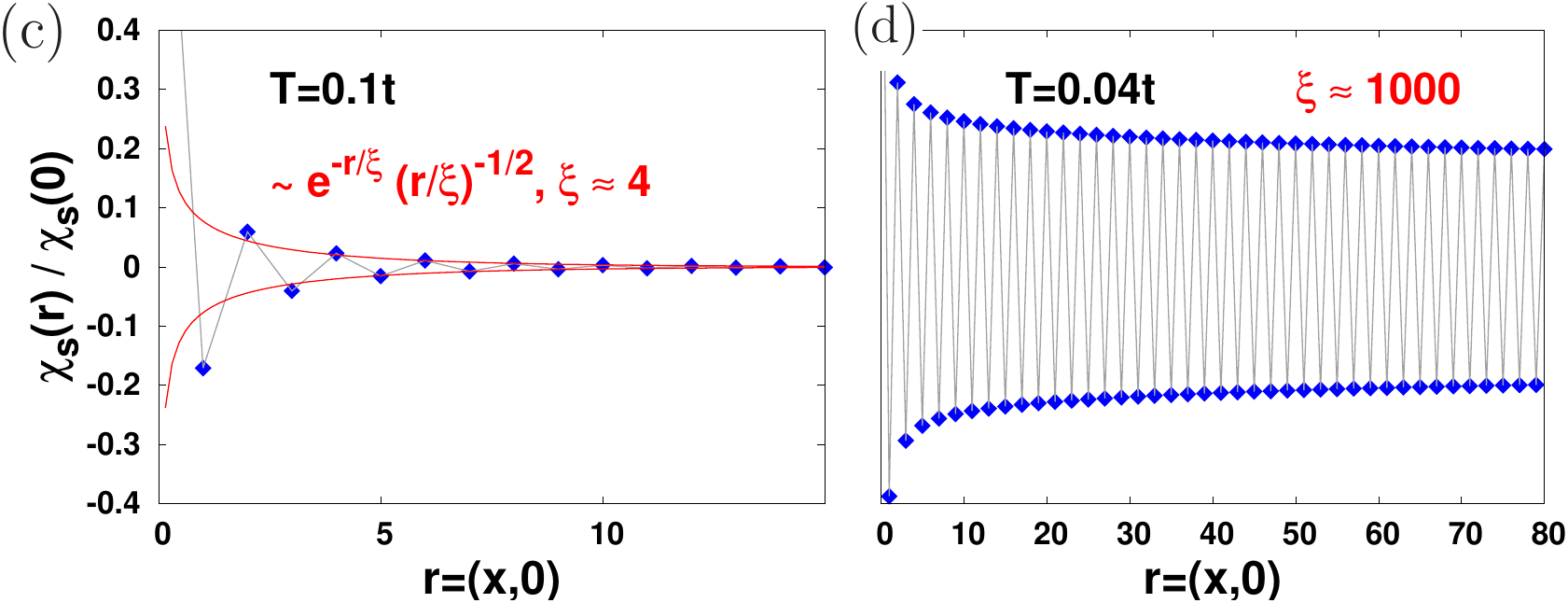} 
  \includegraphics[width=1.0\columnwidth,angle=0]{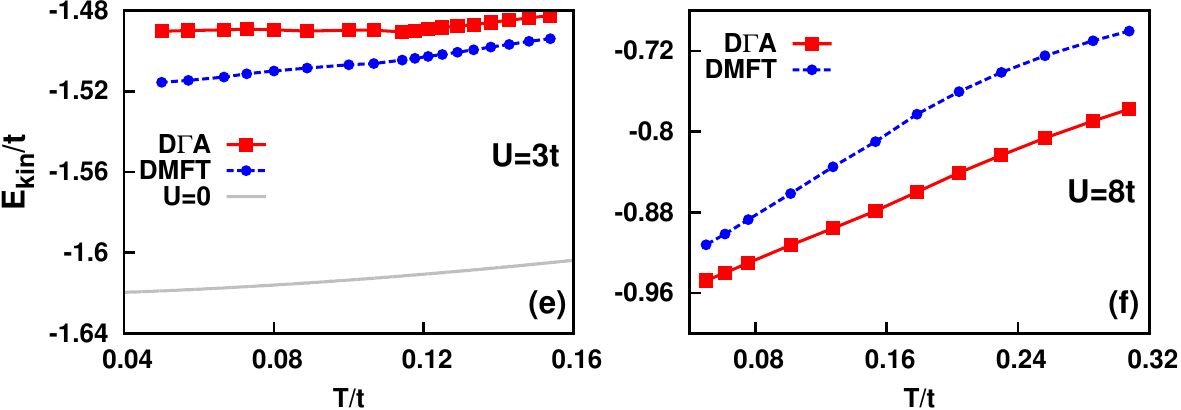} 
    \caption{(Color online) (a) $T$ dependence of the  ferromagnetic susceptibility for $U=4t$ in DMFT, lattice QMC \cite{Moreo1993} for an $8\times 8$ lattice, and ladder DF for an $8\times 8$ and  $64\times 64$ lattices. (b) Ladder DF local spectrum $A(\omega)$ at $T/t=0.2$ including ladder diagrams up to order $n+2$ in the vertex $F$ ($n=0$ corresponds to DF$^{(2)}$)~\cite{Hafermannphd}. (c), (d) Real-space spin susceptibility $\chi_{s}$ computed in ladder D$\Gamma$A for $U=2t$ in the PM ($T=0.1t$) (c) and in the PI  phase ($T=0.04t$) (d). (e), (f) Kinetic energy of the $2d$ Hubbard model at (e) $U=3t$  and (f) $U/t=8$ \cite{Rohringer2016}.}
  \label{fig:observ2d}
\end{figure}

In the weak-coupling regime, complementary ladder DF and D$\Gamma$A results allow us to draw a clear-cut picture. First we observe that the $T$ dependence of the uniform magnetic susceptibility computed in ladder DF in Fig.~\ref{fig:observ2d}(a) displays a downturn in the vicinity of $T_N^{\rm DMFT}$~\cite{vanLoon2017,LeBlanc2016_2}. This temperature approximately marks the onset of the PI phase, below which AF fluctuations become particularly strong and a pseudogap develops~\cite{Rost2012}, as can also be seen in Fig.~\ref{fig:observ2d}(b). The downturn is absent in DMFT and DCA~\cite{LeBlanc2016_2} but well matches lattice QMC results~\cite{Moreo1993}, which shows that extended AF fluctuations govern the physics in this regime and reduce the FM susceptibility. As is evident from Fig.~\ref{fig:observ2d}(b), only high-order diagrams ($n\gg2$) in the ladder expansion can describe such long-range fluctuations. By contrast, the second-order calculation does not include long-range correlations: there is no pseudogap in  DF$^{(2)}$, there are only very weak finite-size effects in the susceptibility in DF$^{(2)}$~\cite{Brener08,Li2008}, and  $U_c$ of the MIT is finite. The  importance of long-range AF fluctuations also in the intermediate coupling regime is confirmed by extracting   the magnon dispersion from ladder DF, which matches available inelastic neutron scattering data for La$_2$CuO$_4$~\cite{LeBlanc2016_2}. 
 
The D$\Gamma$A results for the spin susceptibility in real space confirm this picture [see Figs.~\ref{fig:observ2d}(c) and \ref{fig:observ2d}(d)]: In the weak-coupling regime they show the typical AF oscillation pattern. At high temperatures  in the PM phase (c), it decays over a short length scale $\xi$ of a few lattice spacings ($\xi \simeq 4$). The onset of PI behavior is instead associated with a large increase of $\xi$ to about $\xi \simeq 1000$ (d). We thus conclude that the weaker the coupling, the larger the length scale of the AF fluctuations that is needed to open the spectral gap. Consistent with the TPSC approach by \onlinecite{Vilk1996,Vilk1997}, the PI behavior for small $U$ emerges in the low-temperature regime, where $\xi$~\cite{Schaefer2015-2} and the antiferromagnetic susceptibility~\cite{Otsuki2014} grow exponentially with $1/T$ as required by the Mermin-Wagner theorem. Analytical approximations of the ladder-D$\Gamma$A equations in the limit of $\xi \rightarrow \infty$ by \onlinecite{Rohringer2016} demonstrate, however, that the suppression of the spectral weight at the Fermi energy is slower than the exponential behavior predicted by the TPSC, being consistent, instead, with an electronic scattering rate $\sim 1/T^2$. Further weakening of the spectral weight suppression at low-$T$ might eventually arise in a full self-consistent D$\Gamma$A or DF calculations, as suggested by the most recent DF results \cite{vanLoon2017} and by the comparison with the nonlinear sigma model \cite{Borejsza2004}. 

Figure~\ref{fig:observ2d}(f) shows that at strong coupling $U=8t$ antiferromagnetic fluctuations in D$\Gamma$A lead to a  kinetic energy gain in comparison with the PM DMFT solution. This is in agreement with the Heisenberg picture where it corresponds to a gain of superexchange energy. By contrast, at weak coupling  $U=3t$ as in Fig.~\ref{fig:observ2d}~(e), the D$\Gamma$A kinetic energy is smaller (in absolute value) than the DMFT one. Here we have, however, a gain of potential energy due to antiferromagnetic fluctuations, as in a Slater antiferromagnet~\cite{Rohringer2016,vanLoon2017}. This demonstrates that in the paramagnetic phase the AF fluctuations evolve gradually from Slater to Heisenberg paramagnons with increasing $U$, reflecting the properties of the underlying ground state \cite{Borejsza2004}. The crossover between the two regimes appears approximately located in the parameter region, where the MIT in the C-DMFT (plaquette) calculation is found \cite{Fratino2017}. 
 
\begin{figure}[t]
\begin{center}
  \includegraphics[width=\columnwidth,angle=0]{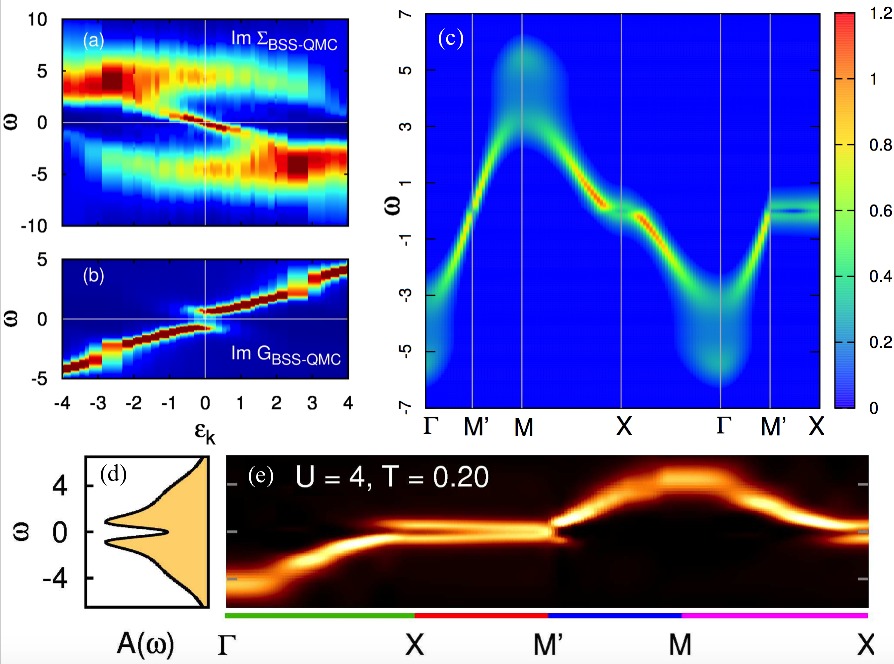} 
\end{center}
    \caption{(Color online) Imaginary part of (a) self-energy  and (b) Green's function vs energy-momentum dispersion $\epsilon_k$  as obtained within lattice QMC. From~\onlinecite{Pudleiner2016}. Momentum-resolved spectral function within (c) variational DF (from~\onlinecite{Jung2010}) and (e) lattice QMC (from \onlinecite{Rost2012}) along the high-symmetry lines of the Brillouin zone. The
latter shows a pseudogap at the Fermi surface in (d) the momentum-
integrated spectrum. The DF spectral function also exhibits a water-
fall. All panels are for the 2D Hubbard model  at $U=4t$, $T\approx 0.2t$, and half filling.}
\label{fig:pseudogap2d}
\end{figure}

Pseudogaps and the formation of Fermi arcs, i.e., the destruction of quasiparticles near the antinodal region, has been reproduced by various flavors of diagrammatic extensions, including DF~\cite{Rubtsov2009}, D$\Gamma$A~\cite{Katanin2009}, DMF$^{2}$RG~\cite{Taranto2014} and TRILEX~\cite{Ayral2015}. 

Figure~\ref{fig:pseudogap2d} shows the self-energy  as well as the momentum-resolved spectral functions in  the pseudogap regime. Figure~\ref{fig:pseudogap2d}(c)-\ref{fig:pseudogap2d}(e) reveal that the  Fermi surface is fully gapped in both variational DF~\cite{Jung2010} and lattice QMC extrapolated to the thermodynamic limit; for related pseudogaps in  DMFT+$\Sigma_{k}$ see \onlinecite{Nekrasov2008,Nekrasov2011}, for TPSC \onlinecite{Moukouri2000}, and for  D$\Gamma$A see \onlinecite{Katanin2009,Rohringer2016}. The opening of the pseudogap is reflected in a transition from a $z$ shape to an inverse $z$-shape structure in the self-energy; see Fig.~\ref{fig:pseudogap2d}(a) and \onlinecite{Pudleiner2016}. It is also connected to zeros in the Green's function \cite{Sakai2009}. The lattice QMC calculations of Fig.~\ref{fig:pseudogap2d}(a) and D$\Gamma$A also show that, except for the pseudogap itself, the $({ k_x},{ k_y)}$ dependence of the self-energy can be expressed by a single parameter: the energy-momentum dispersion $\varepsilon_{({ k_x},{ k_y})}$.

Taken together, D$\Gamma$A and DF  are compatible in the regimes of their applicability, with lattice QMC, TPSC and cluster extensions of DMFT and yield the following physics: With decreasing $T$, a pseudogap opens due to  AF fluctuations first in the antinodal and then in the nodal direction  \cite{Schaefer2015-3} marking the MIT of Fig.~\ref{fig:pd2d}. For the square lattice with perfect nesting this happens at arbitrarily small interaction $U$.

\subsubsection{Two dimensions: Square lattice off half filling}
\label{sec:2dhubbouthalf}

\begin{figure}[t]
\begin{center}
  \includegraphics[width=0.75\columnwidth,angle=0]{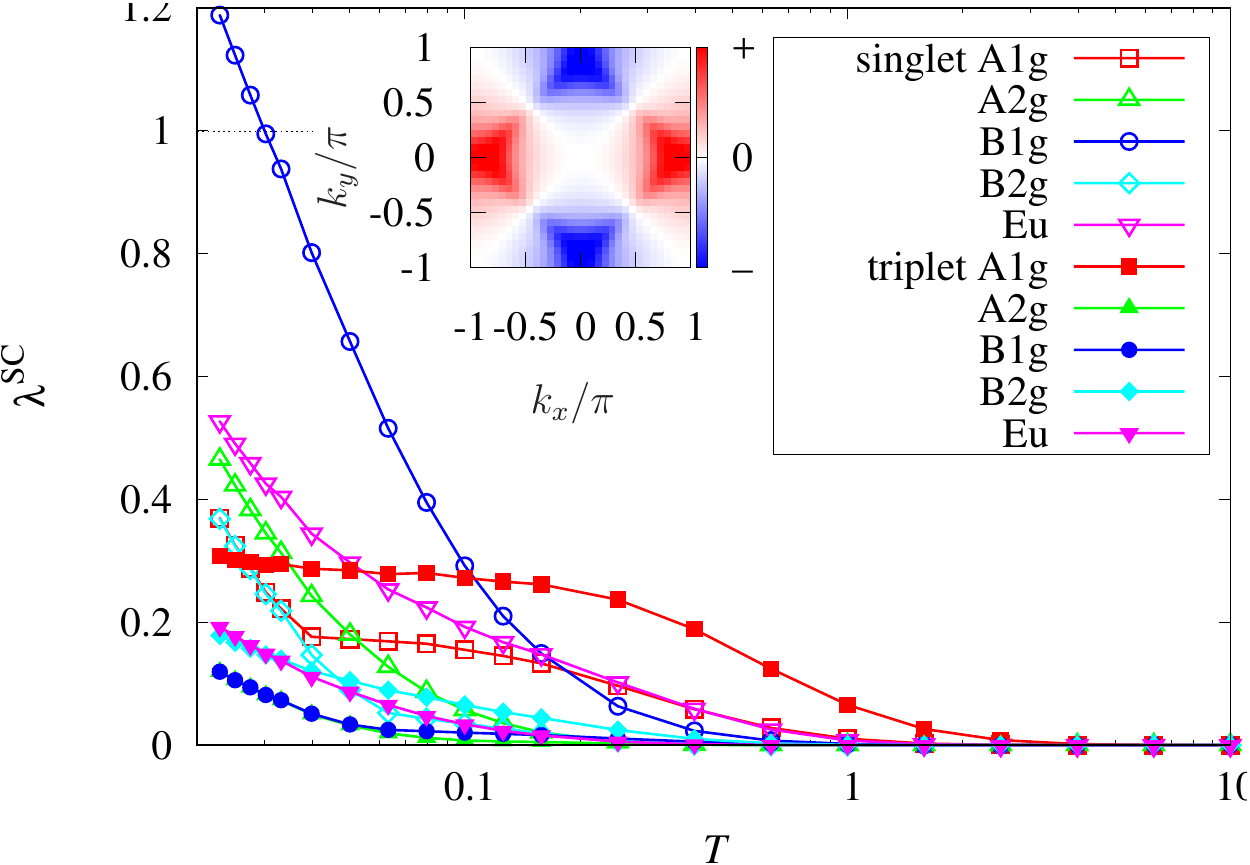}
  \includegraphics[width=0.65\columnwidth,angle=0]{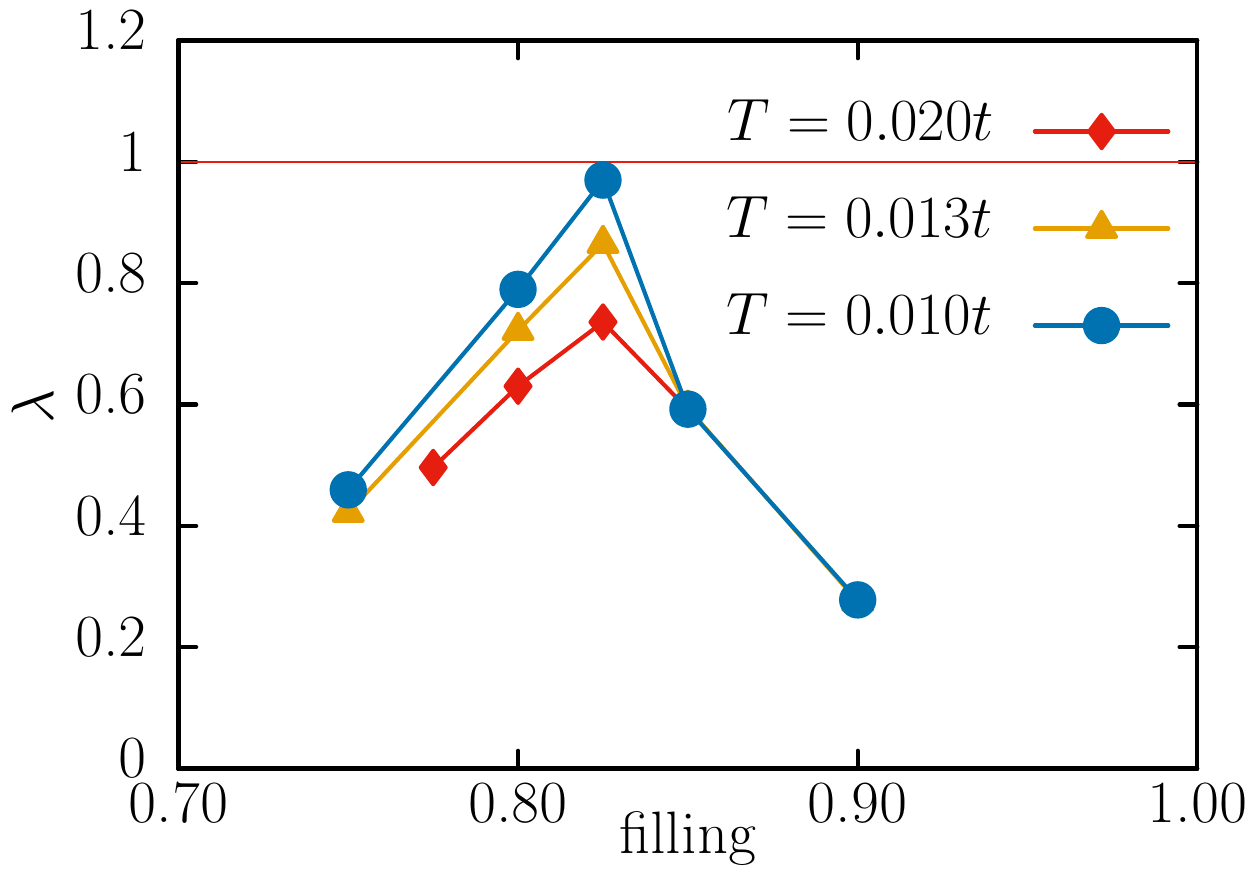}
  \includegraphics[width=0.257\columnwidth,angle=0,trim=5.3cm 10.85cm 8.5cm 5cm,clip]{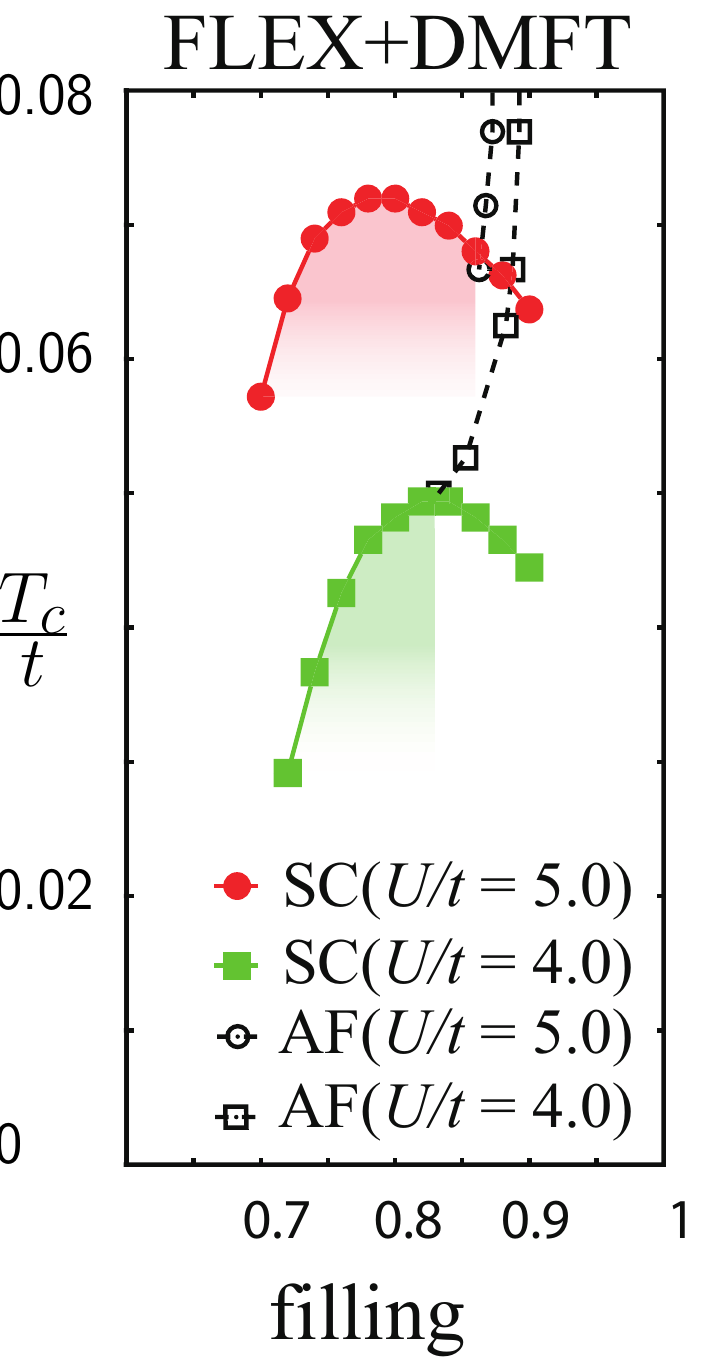} 
\end{center}
    \caption{(Color online) Top: Temperature dependence of the leading eigenvalue in the particle-particle channel for the hole-doped 2D Hubbard model in DF at $U/t=8$ and 14\% hole doping, separated into contributions for given symmetry. The eigenvalue associated with singlet B$_{1g}$ symmetry crosses 1, implying a transition to the superconducting state. The inset shows the momentum dependence of the eigenfunction $\phi_{\vc{k}}(\omega_{0})$ corresponding to this eigenvalue at $T=0.1$, with apparent $d$-wave symmetry. From~\onlinecite{Otsuki2014}. Bottom left:  Filling dependence of the leading D$\Gamma$A  $d$-wave eigenvector $\lambda$ in the particle-particle channel for the 2D Hubbard model  at $U=6t$ indicating superconducting order below $T=0.01 t$ ($\lambda=1$). From \onlinecite{Kitatani2017}. Bottom right: Superconducting and antiferromagnetic critical temperature vs electron filling in FLEX+DMFT. From \onlinecite{Kitatani2015}.}
  \label{fig:lambdasc}
\end{figure}

At finite doping, studies of superconductivity in the Hubbard model are of primary interest. In this respect, the arguably biggest success of  cluster DMFT and DCA calculations has been the observation of superconductivity~\cite{Lichtenstein2000,Maier2005a,Capone2006,Haule2007b,Sordi2012,Gull2013} which helped establishing the presence of superconductivity in the Hubbard model. Superconductivity requires a framework which captures both strong local dynamical correlations and spatial fluctuations as offered by diagrammatic extensions of DMFT. In DF, the effective pairing interaction has been constructed similarly as in FLEX, namely, by inserting the ladder diagrams of the horizontal and vertical  particle-hole channels (see Sec.~\ref{sec:funcint}) into the irreducible particle-particle vertex $\Gamma^{\nu\nu'\omega}_{pp,\alpha,\vc{k}\vc{k}'\vc{q}}$~\cite{Hafermann2009a,Otsuki2014}. The former incorporates the charge, as well as the longitudinal and transverse spin fluctuations that are expected to be predominant at moderate hole doping~\cite{Gunnarsson2015,Gunnarsson2016}. The transition temperature is found by computing the leading eigenvalue of a linearized Eliashberg-like equation,
\begin{equation}
\label{ref_to_eqn}
  \sum_{\vc{k}\omega}\Gamma^{\nu\nu'\omega=0}_{pp,r,\vc{k}\vc{k}'\vc{q}=\vc{0}}\widetilde{G}_{(-\vc{k}')(-\omega)}\widetilde{G}_{\vc{k}'\omega'}\phi_{\vc{k}'\omega'} = \phi_{\vc{k}\omega},
\end{equation}
where $r=s,t$ stands for the singlet and triplet channels. Results classified according to the different symmetries are shown in Fig.~\ref{fig:lambdasc} (top). The order parameter of the leading instability has $d$-wave symmetry as illustrated through the corresponding eigenfunction $\phi_{\vc{k}}$ which is essentially the gap function. Superconductivity is observed for hole doping $\delta\leq0.18$ for $U/t=8$. The transition temperature $T_{c}\lesssim 0.05 t$ at 10\% hole doping is compatible with the value $0.0518 t$ reported by \onlinecite{Staar2013a} for $U/t=7$. In the DF calculation, superconductivity might however not be realized for doping levels $\lesssim 15\%$ due to the presence of phase separation observed in the same study at higher $T$. DF shows no superconducting dome structure with a downturn toward half filling, even though the formation of the pseudogap at the Fermi level feeds back into the calculation of the effective pairing interaction through the dual Green’s functions.

In contrast, such a superconducting dome is found in other diagrammatic extensions of DMFT such as D$\Gamma$A \cite{Kitatani2017} [Fig.~\ref{fig:lambdasc} (bottom left)], FLEX+DMFT  \cite{Kitatani2015} [Fig.~\ref{fig:lambdasc} (bottom right)],  EDMFT+$GW$ and TRILEX~\cite{Vucicevic2017}, in agreement with DCA \cite{Maier2005} and TPSC \cite{Kyung2003}. Here  the superconducting dome emerges from the competition of two effects (see e.g., \cite{Kitatani2017}): (i)  antiferromagnetic fluctuations and hence the irreducible vertex in the $pp$ channel get stronger toward half filling, and (ii) the pseudogap around half filling suppresses the Green's function lines which connect these vertices in the particle-particle BSE ladder. In Fig.~\ref{fig:lambdasc} (bottom left and right), the   leading eigenvalue in the particle-particle channel was computed via Eq.~\eqref{ref_to_eqn}. While FLEX+DMFT and TRILEX overestimates the nonlocal self-energy and the critical temperature, D$\Gamma$A yields a reasonable $T_{c}\approx  40\,$K for $t\approx 0.35\,$eV.  As shown by \onlinecite{Kitatani2017}, the frequency structure of the local vertex is indeed very important for the actual value of  $T_{c}$, leading to a considerably lower  $T_{c}$ and  agreement with experiment.

Hidden order in the form of a staggered flux state (of $d$-density wave) is among  several candidates for the origin of the pseudogap~\cite{Chakravarty2001}. DF calculations show that the density wave with $d$-wave symmetry dominates over density waves with other symmetries at lower $T$. However, the susceptibility shows no divergence at the accessible temperatures. The $d$-density wave state also appears to be shadowed by the superconducting state, since its extrapolated $T_c$  is lower~\cite{Otsuki2014}.

\subsubsection{Two dimensions: Triangular lattice}
\label{sec:2dhubbtriang}

Frustrated  strongly correlated electron systems such as the Hubbard model on a triangular lattice are characterized by macroscopically degenerate ground states that lead to strong quantum fluctuations and a multitude of instabilities. Such systems hence exhibit very rich phase diagrams comprised of Mott insulating, superconducting, or resonating valence bond (RVB) states, commensurate or incommensurate SDW, or noncollinear magnetic order. Important experimental realizations are (i) the  stacked triangular CoO$_{2}$ layers in quasi-two-dimensional sodium cobaltate Na$_{x}$CoO$_{2}$, (ii) the  organic salt $\kappa$-(BEDT-TTF)$_{2}$Cu$_{2}$(CN)$_{3}$  where two BEDT-TTF molecules  form spin $S=1/2$ dimers which in turn constitute a triangular lattice, (iii) adatoms on a Si (111) surface, and (iv) bilayers of transition metal oxide heterostructures grown in the (111) direction. Diagrammatic extensions of DMFT are particularly suited in this context, because spatial correlations are highly relevant in the presence of frustration. Moreover, diagrammatic approaches are not affected by the sign problem which, in the presence of frustration, strongly hampers QMC simulations or cluster extensions of DMFT that employ QMC as a solver. 

Regarding the {\em metal-insulator transition}, \onlinecite{Lee2008} showed that, compared to DMFT, the critical $U$ of the Mott transition is reduced down to $U_{c}\sim 7t$ in DF$^{(2)}$, which agrees with DCA results~\cite{Lee2008}. This is analogous to the effect described in Sec.~\ref{sec:2dhubbsquare} for the square lattice. The difference is that here there is no perfect nesting for the triangular lattice so that  $U_{c}$ stays finite and even large because of the high frustration.

\begin{figure}[t]
\begin{center}
  \includegraphics[width=0.85\columnwidth,angle=0]{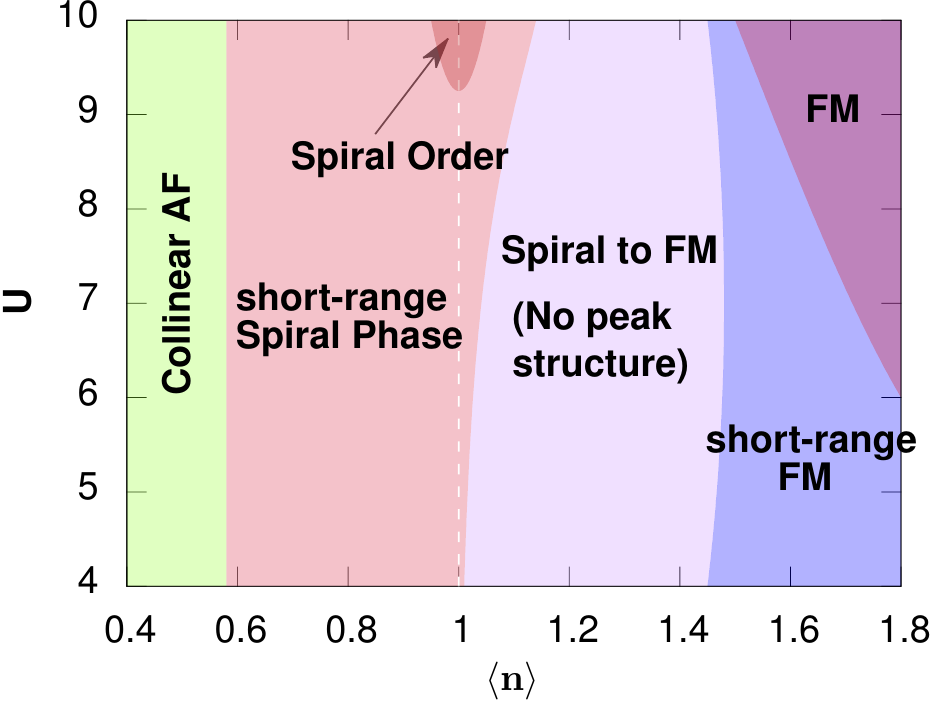} 
\end{center}
    \caption{(Color online) Magnetic phase diagram of the doped triangular-lattice Hubbard model at fixed temperature $T=0.1t$, as obtained within DF. Long-range spiral and FM order is found at sufficiently large values of $U$ and collinear AF at low electron fillings $\langle n\rangle$. From  \onlinecite{Li2014}.}
  \label{fig:phasedtriang}
\end{figure}

\paragraph*{Magnetism}
According to results from different many-body methods, the triangular lattice Hubbard model favors the 120$^{\circ}$ N\'eel state at large $U/t\sim 10$~\cite{Yoshioka2009,Ohashi2008}, and this is also the case in the  DF$^{(2)}$ calculation by~\onlinecite{Lee2008} as well as in the  DF$^{(2)}$ expansion around a DCA-like three-site cluster by~\onlinecite{Antipov2011}. The rich magnetic phase diagram obtained  by~\onlinecite{Li2014}  is shown in Fig.~\ref{fig:phasedtriang}. While the observation of finite temperature transitions is likely  an artifact of the extrapolation method for the inverse susceptibility, one can interpret the finite transition temperature as an upper bound for a quasi-2D system of layers coupled in the third dimension.  Close to half filling and at large $U$, spiral order is found which includes the  120$^{\circ}$ N\'eel state at half filling. Interestingly, the frustration pushes the critical $U$ for ordering above the aforementioned $U_c$ for the MIT so that a nonmagnetic insulating phase  is realized at half filling, possibly a spin-liquid state~\cite{Morita2002,Sahebsara2008,Yang2010}.

The spiral ordering is quickly destroyed upon doping, but short-range order marked by a peak in the static susceptibility at the corresponding wave vector survives. For electron doping ($\langle n\rangle>1$) the spiral correlations make way for FM short-range- and at large $U$ also long-range ordering. Quite asymmetrically, on the hole-doped side ($\langle n\rangle<1$) the susceptibility peaks at  wave vector $\vc{Q}=(\pi,\pi)$ corresponding to a collinear AF. \onlinecite{Laubach2015} tuned the lattice from triangular to  square by changing a ``vertical'' bond hopping $t'$ and found in DF an evolution from the 120$^{\circ}$ AF order of the half-filled triangular lattice to the  $\vc{Q}=(\pi,\pi)$ collinear AF order on the square lattice, in agreement with VCA.

\onlinecite{Li2011} reported DF$^{(2)}$ results for the Sn/Si(111) $\sqrt{3}\times\sqrt{3}R$30$^{\circ}$ surface system which can be mapped onto a triangular lattice Hubbard model with a band structure calculated within LDA.  For this system, additional frustration due to next-nearest-neighbor hopping however suppresses the 120$^{\circ}$ AF in favor of the collinear AF.  \onlinecite{Hansmann2013} used $GW$+DMFT instead and 
emphasized the importance of long-range Coulomb interactions and charge ordering.

\paragraph*{Energy and entropy}

\onlinecite{Antipov2011} showed that spatial correlations significantly lower the energy of the spin-liquid state at half filling, while leaving the energy of the N\'eel state essentially unaffected. As for the entropy, \onlinecite{Li2014} reported that  it increases with  $U$  at fixed $T$ and  ascribed this counterintuitive trend  (which is in contrast to that for the unfrustrated square lattice) to a significant increase in the spin entropy due to localization.  This opens a new possibility for adiabatic cooling in cold-atom experiments by tuning $U$. Note that the highest entropy occurs at a filling $\av{n}\approx 1.35$ which coincides with the optimal filling for superconductivity in sodium cobaltate and signals the competition between the localized spin and the charge degrees of freedom. The high entropy can be related to  the Seebeck coefficient through Kelvin's formula and might be an important contribution to the large thermopower of Na$_{x}$CoO$_{2}\cdot1.3$H$_{2}$O at $\av{n}\sim 1.5$ \cite{Terasaki1997}. Another factor favoring a large thermopower is the enhancement of the electron-hole asymmetry due to local electronic correlations which was found by \onlinecite{Wissgott2010,Wissgott2011} using density functional theory (DFT+DMFT).

\onlinecite{Wilhelm2015}  also used the DFT band structure of the relevant $a_{1g}$orbital as a starting point and found a spin-polaron peak in DF near van Hove filling -- possibly explaining a weak absorption feature observed in optics experiments of nearly ferromagnetic Na$_{0.7}$CoO$_{2}$ by \onlinecite{Wang2004}. The spin-polaron excitation at $\Gamma$ has been traced back to the binding of quasiparticles with an FM paramagnon~\cite{Boehnke2012} originating from the spin channel of ladder DF. The interplay of many-particle scattering incorporated through the DF self-energy and nesting also leads to a band flattening near van Hove singularities as reported by~\onlinecite{Yudin2014}:  In analogy to a Bose-Einstein condensate, this highly degenerate fermionic state is referred to as ``Fermi condensation'' and possibly signals an instability at lower $T$.

\subsubsection{One and zero dimensions}
\label{sec:0dHM}

One- and zero-dimensional systems are arguably the most challenging for diagrammatic approaches that start from DMFT, which as we recall is exact for $d=\infty$. In 1D the Fermi surface degenerates to two points. As a result of nesting, the ladder diagram series typically diverges in several channels, so that ladder approximations cannot be applied. The 1D Hubbard model can nevertheless be studied  using finite order diagrams as, e.g., in  DF$^{(2)}$~\cite{Hafermann2008} or using the parquet equations \cite{Valli2015}. A proper description of the insulating state, which is the known ground state for all $U>0$, requires a fully self-consistent calculation, in which the hybridization function changes from metallic character in DMFT to insulating in DF.

\begin{figure}[tb]
  \includegraphics[width=.8\columnwidth]{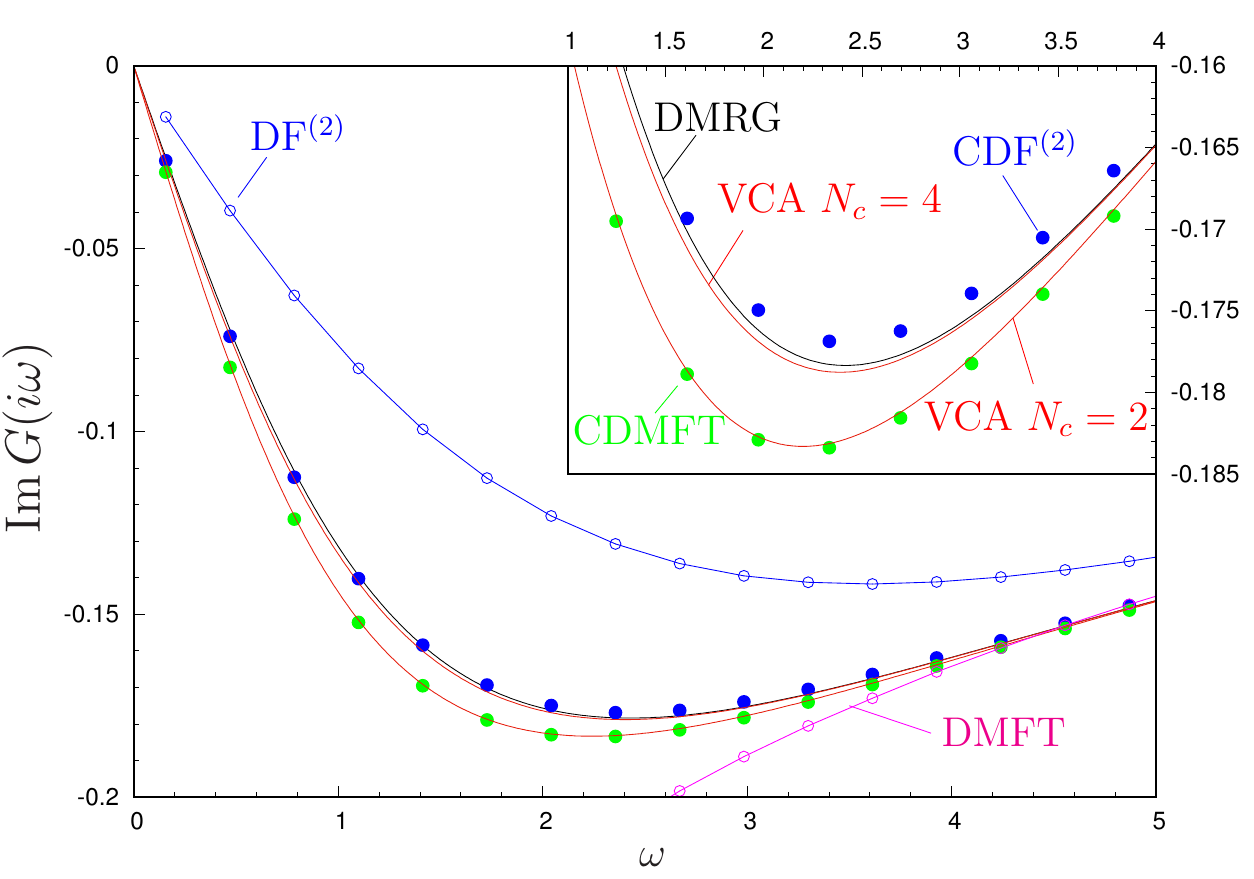}
    \caption{(Color online) Local Green's function of the 1D Hubbard model obtained within DF$^{(2)}$ and two-site cluster CDF$^{(2)}$, in comparison to numerically exact DMRG, CDMFT and zero temperature variational cluster approach (VCA) calculations as well as DMFT (inset: zoom in). Adapted from~\onlinecite{Hafermann2008}.}
  \label{fig:gloc1d}
\end{figure}

Figure~\ref{fig:gloc1d} shows such  DF$^{(2)}$ calculations  based on a single-site DMFT that capture the insulating state. The second-order diagrams however do not include the nonperturbative singlet correlations needed for an accurate description of the 1D physics. Hence, we see in Fig.~\ref{fig:gloc1d} significant deviations from the numerically exact density matrix renormalization group (DMRG).

An alternative route followed by  \onlinecite{Slezak2009} and \onlinecite{Hafermann2008} is to use a cluster instead of a DMFT solution as a starting point as discussed in Sec.~\ref{sec:cluster}. Already the expansion around a two-site  CDMFT solution captures crucial aspects of the 1D dimer physics and yields quantitative agreement with the DMRG benchmarks in Fig.~\ref{fig:gloc1d}. Whether spin-charge separation as in a Luttinger liquid is captured by such diagrammatic extensions is an open question, and 1D calculations off half filling are imperative.

\begin{figure}[t!]
\begin{center}
  \includegraphics[width=.8\columnwidth,angle=0]{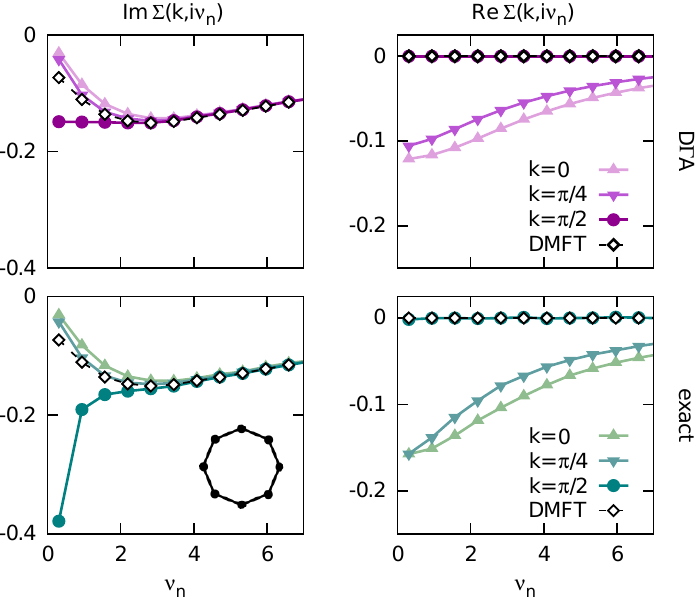} 
\end{center}
    \caption{(Color online) \label{fig:nanoring} (Left) Imaginary and (right) real part of the self-energy vs Matsubara frequency $\nu_n$ for an $8$-site Hubbard nanoring (see bottom left inset) with nearest-neighbor hopping $t$ and local interaction $U/t=2$ at $T/t=0.1$ at half filling. (Upper panels) Parquet D$\Gamma$A, (lower panels) the exact solution, and (both panels) DMFT are compared. From \onlinecite{Valli2015}.}
  \label{fig:phased_db}
\end{figure}

An equally challenging issue is the treatment of nonlocal correlations in 0D (molecularlike) systems. Progress toward the understanding of spatial correlations in finite, discrete systems was recently achieved by means of a comparison~\cite{Valli2012,Valli2015} between nanoscopic (or real-space) DMFT, nanoscopic D$\Gamma$A~\cite{Valli2010} and the exact solution of small correlated molecules (up to $10$ sites). In Fig.~\ref{fig:nanoring} we show the results of the first parquet D$\Gamma$A calculation, performed for a ring  molecule of 8-correlated equivalent sites (see the inset of Fig.~\ref{fig:nanoring}; it can also be considered as a 1D Hubbard model with periodic boundary conditions). The comparison with DMFT shows that a substantial, although not complete part of  the strong nonlocal correlations characterizing the exact solution of this 0D system is actually captured by the parquet D$\Gamma$A calculation. The improvement with respect to the parquet approximation solution and, thus, the importance of including the full frequency dependence of the 2PI vertex function depend strongly on the parameter regime considered. It remains to be investigated whether the realization of an external self-consistency cycle in D$\Gamma$A (see Sec.~\ref{sec:quadrilex}) can close the remaining gap to the exact solution. The systematic analysis of \onlinecite{Valli2012,Valli2015} further identified situations where a DMFT calculation is reliable. This can be used as a guide for the study of more complex systems, such as quantum points contacts with $\sim$ 100 atoms~\cite{Florens2007,Valli2010,Jacob2010} or transition metal oxide nanoclusters~\cite{Das2011,Valli2015a}.

\subsection{Heavy fermions and Kondo lattice model (KLM)}
\label{Results_HF}
Heavy fermion systems  are intermetallic compounds in which strongly correlated and localized electrons in partially filled $f$ shells of a rare earth or actinide element coexist with weakly correlated electrons of much broader bands provided by the other elements or orbitals. At elevated temperatures the $f$-electron magnetic moments are weakly coupled to a Fermi sea formed by $s$, $p$, or $d$ electrons. At low temperatures, however, localized moments and conduction electrons can form new entities below the Kondo temperature $T_K$,  and magnetic order or unconventional superconductivity may be realized along with quantum critical points. The local Kondo physics as well as magnetic ordering is already contained in DMFT, but without spatial correlations that become important at the critical point and that are the realm of diagrammatic extensions of DMFT. 

Let us consider the spin-1/2 KLM,
\begin{equation}
\label{Eq:KLM}
  \mathcal{H} = \sum_{ij,\sigma} t_{ij} c_{i\sigma}^{\dagger}c_{j\sigma} + J\sum_{i} \vc{S}_{i}\cdot \vc{s}_{i} \; ,
\end{equation}
(also known as the $s$-$d$  model) which is a minimal model for heavy fermion physics. Here $\vc{S}_{i}$ are the local and $\vc{s}_{i}=(1/2)\sum_{\sigma\sigma'}c_{i\sigma}^{\dagger}\boldsymbol{\sigma}_{\sigma\sigma'}c_{i\sigma'}$ the conduction electron spins, respectively.

The DMFT solution of this model reproduces the qualitatively different behavior at high and low temperatures. At low temperature, significantly smaller than $T_K$, a well-defined hybridization gap opens. Reducing $J$, \onlinecite{Otsuki2009} found a quantum phase transition from this Kondo insulator to an AF-ordered state at  $J_c \simeq 2.18$. If the conduction band is doped away from half filling,  the formation of heavy quasiparticles leads to a large Fermi-surface, which shows that the local moments in fact contribute to the Fermi surface volume~\cite{Otsuki2009}.

The  KLM has been studied using the DF method at first   by \onlinecite{Sweep2013}.  The corresponding implementation is essentially the same as for the Hubbard model, the only difference being the interaction term of the underlying impurity model.  \onlinecite{Sweep2013} employed a weak-coupling CTQMC impurity solver using two bands for the localized and conduction electrons. Their rough estimate of the critical exchange interaction $J_{c}$ in DF (at relatively high temperatures) yielded already a significant ($\sim 50\%$) reduction of  $J_{c}$  with respect to DMFT, induced by  nonlocal fluctuations.

In a more recent ladder DF study, \onlinecite{Otsuki2015} employed an interaction-expansion-type CTQMC algorithm specifically devised for the Coqblin-Schrieffer model and  addressed the competition between $d$- and $p$-wave superconductivity in the 2D  KLM. On the square lattice and for a half-filled conduction electron band, the perfect nesting of the Fermi surface favors an AF ordering of the localized magnetic moments through the  Ruderman-Kittel-Kasuya-Yoshida (RKKY) interaction. Hence similar to the 2D Hubbard model in Sec.~\ref{sec:2dhubbsquare}, it is natural to ask which type of superconductivity emerges in heavy fermion materials near the AF quantum critical point.

\begin{figure}[b]
\begin{center}
  \includegraphics[width=.8\columnwidth,angle=0]{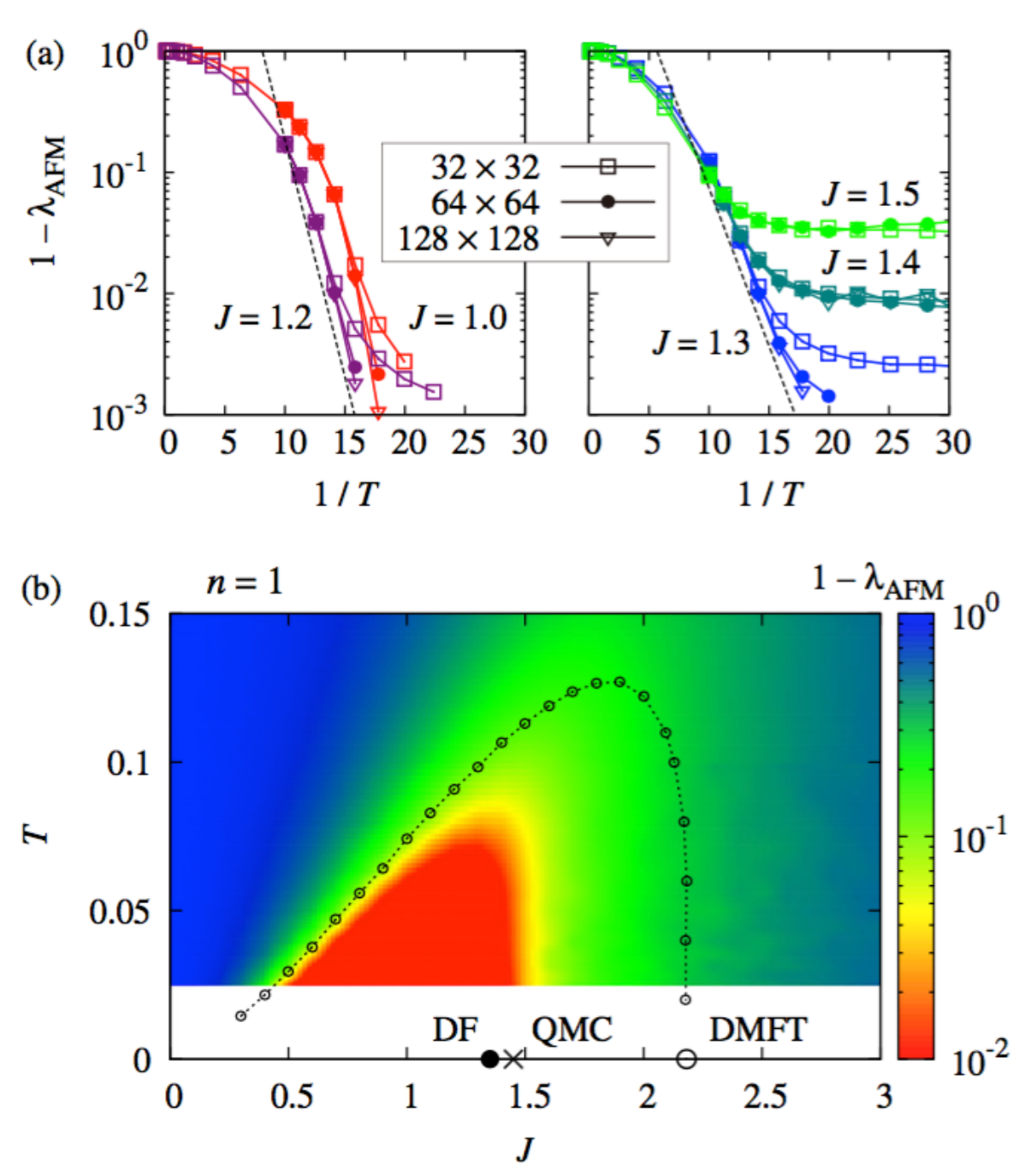} 
\end{center}
    \caption{(Color online)   (a) Critical scaling and lattice size dependence of the leading eigenvalue $\lambda_{\text{AFM}}$ for the KLM in DF theory, indicating the presence of an AF ground state (left panel) while for $J>J_{c}$  (right panel) the Kondo effect suppresses AF fluctuations. (b) Critical region defined by $1-\lambda_{\text{AFM}}\lesssim 10^{-2}$ exhibiting a dome shape similar to the DMFT phase boundary, albeit significantly shrunk. From~\onlinecite{Otsuki2015}.}
\label{fig:phased_klm}
\end{figure}

Figure~\ref{fig:phased_klm} (a) shows the  leading AF eigenvalue which has a critical behavior $1-\lambda_{\text{AFM}}\propto e^{-\Delta/T}$ and a strong size dependence for $J\leq 1.2$. This indicates an AF ground state. For sufficiently large $J$, AF is suppressed due to the Kondo effect and the leading eigenvalue approaches a constant at low $T$; see the right panel of Fig.~\ref{fig:phased_klm} (a). The estimated position of the quantum critical point $J_{c}^{\text{DF}}=1.35\pm 0.05$ is close to the lattice QMC value of \onlinecite{Assaad1999}; see Fig.~\ref{fig:phased_klm} (b). This is another example of the quantitative accuracy of results obtained by means of diagrammatic methods.

\onlinecite{Otsuki2015} determined the superconducting transition temperature by an eigenvalue analysis as described in Sec.~\ref{sec:2dhubbsquare}. The eigenvalues corresponding to eigenfunctions of B$_{1g}$ ($d$-wave) and $E_{u}$ ($p$-wave symmetry) are nearly degenerate in a wide doping range. While $d$-wave superconductivity is realized for $J\lesssim 0.9$, it is replaced by $p$-wave superconductivity as the leading instability for $J\gtrsim 0.9$. In both cases, AF fluctuations are likely to be the origin of the pairing. Remarkably, the crossover from $d$-wave to $p$-wave pairing correlates with the crossover from a small to the large Fermi surface. For weak coupling, $d$-wave pairing is favorable, because the regions of high intensity of the eigenfunction coincide with the van Hove points. As the hybridization band is formed and a large Fermi surface develops, low-energy excitations appear around $\vc{k}=(\pi/2,\pi/2)$ and $p$-wave symmetry emerges as a consequence of the scattering between them. The $p$-wave superconductivity is hence a consequence of the formation of heavy quasiparticles, which distinguishes the KLM from the Hubbard model.

\subsection{Falicov-Kimball (FK) model}
\label{sec:Results_FK}
\label{sec:FKM}

In the FK model  mobile electrons interact with localized ones via a Hubbard-type interaction $U$. This way it describes (annealed) disorder and represents one of the simplest systems where correlation effects can be observed. The Hamiltonian for the spinless FK model reads
\begin{equation}
\label{MethodFKM:eq:hamilt}
  \mathcal{H} = \sum_{k} \varepsilon^{\phantom{\dagger}}_k  c^\dagger_{k} c^{\phantom{\dagger}}_{k}  + \sum_{i} \varepsilon^{\phantom{\dagger}}_f f^\dagger_i f^{\phantom{\dagger}}_i + \sum_{i}U c^\dagger_i c^{\phantom{\dagger}}_i f^\dagger_i f^{\phantom{\dagger}}_i,
\end{equation}
where $U$ is the interaction strength, $\varepsilon_k$ is the dispersion relation for the mobile ($c$) electrons, and $\varepsilon_f$ is the local potential of the immobile ($f$) electrons; $i$ labels the lattice sites.

The simple nature of the FK model compared to the Hubbard model is seen from the Hamiltonian (\ref{MethodFKM:eq:hamilt}): every local $f$-electron occupation operator $w_i = f^\dagger_i f_i$ commutes with the Hamiltonian, providing an extensive number of conserved degrees of freedom. This has a number of important consequences. First, a set of mathematically rigorous results, including the existence of a phase transition to the $f$-$c$ checkerboard-ordered phase of the Ising universality class in the particle-hole symmetric model has been established \cite{Kennedy1986,Brandt1986}. Second, the model is amenable to a sign-problem free Monte Carlo sampling \cite{Maska2005, Maska2006, Zonda2009, Antipov2016}, providing exact predictions for finite-size systems. Finally, the ``impurity problem'' solved in DMFT is exactly solvable in both equilibrium \cite{Brandt1989} and nonequilibrium \cite{Eckstein2008,Eckstein2009}. 

For the FK model, the free-energy functional in $d=\infty$  and the self-consistent  equations, which  later became known as the DMFT equations, have also been derived for the first time in a seminal paper by  \onlinecite{Janis1991}. DMFT has been used to study, among others, thermodynamics and  spectral functions \cite{Brandt1989,Brandt1990,Brandt1991},  phase separation  upon doping \cite{Freericks2000c},  dynamical properties including a discontinuity at zero frequency \cite{Freericks2000a}, and the absence of thermalization \cite{Eckstein2008,Eckstein2009}. The majority of these results is summarized in the review by \onlinecite{Freericks2003}.

The combination of these factors makes the FK model an ideal test bed for computational approaches to strongly correlated systems, including cluster extensions \cite{Hettler1998,Hettler2000,Maier2005} and, of particular importance here, diagrammatic extensions of DMFT. The latter profit from the fact that the interacting single-particle and multiparticle Green's functions of the DMFT impurity problem can be obtained analytically. The local single-particle propagator reads 
\begin{equation}
\label{Eq:DMFTFKM}
  G^{\mathrm{loc}}_\nu = w \mathcal{G}_\nu + (1 - w) \left[\mathcal{G}^{-1}_\nu - U\right]^{-1},
\end{equation}
where $\mathcal{G}_\nu = \left[i\nu + \mu - \Delta_\nu\right]^{-1}$  is the noninteracting Green's function of the DMFT impurity problem, which can be calculated self-consistently together with the lattice Dyson equation and the $f$-electron occupation $w$. The DMFT susceptibility and the local irreducible vertex in the particle-hole channel can be calculated as well \cite{Freericks2003}. This corresponds to  the following full vertex 
\begin{equation}
  F^{\nu\nu'\omega} =  \beta\left( \delta_{\omega 0}- \delta_{\nu\nu'} \right) a_{\nu'}a_{\nu+\omega}
\end{equation}
where $a_\nu= (\Sigma_\nu^{\rm loc}-U) \Sigma_\nu^{\rm loc}/(\sqrt{w(1-w)}U)$; see  \onlinecite{Ribic2016}. Using these simplifications a set of further results for the analytical properties regarding the correlation functions of the FK model has been obtained: \onlinecite{Antipov2014} calculated the  antisymmetrized three-particle vertex $F^{(6),\nu \nu' \nu''} = w (1-w) (2w - 1) U^3 a_\nu a_{\nu'} a_{\nu''}$ which is zero at  particle-hole symmetry. \onlinecite{Ribic2016b} extended this to the full $n$-particle vertex for an arbitrary number of particles $n$ and  estimated the error of truncating the DF theory at the $n\!=\!2$ particle level. Diagrams emerging from the three-particle vertex in DF theory may yield contributions of the same magnitude. \onlinecite{Shvaika2000} and \onlinecite{Ribic2016} also calculated the exact two-particle reducible, irreducible in $ph$ and $pp$ channels, and the fully irreducible vertex of the FK model. \onlinecite{Schaefer2016c} identified the applicability of perturbation theory by studying divergences of the DMFT vertex functions; below a single energy scale $\nu^*(U)$, the low-energy spectral properties of the model have a nonperturbative nature. Related results have also been reported by \onlinecite{Janis2014} and \onlinecite{Ribic2016}.
 
\begin{figure}[t]
  \includegraphics[width=\columnwidth]{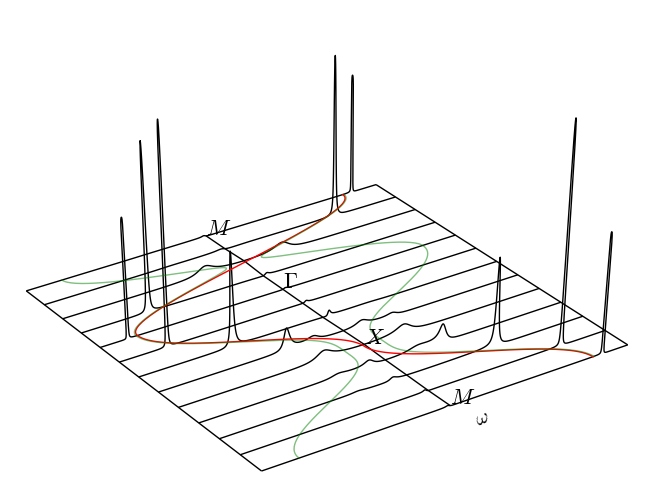}
    \caption{(Color online) Spectral functions of the FK model along a path through the Brillouin zone and as a function of the real frequency $\omega$ at $U=1t$, $T = 0.07t$, and half filling, which can be compared to the corresponding result for the Hubbard model, Fig.~\ref{fig:df2dspec}. The red (dark gray) line in the plane is the bare dispersion and the green (light gray) line that of a CDW checkerboard phase with a doubling of the unit cell. From \onlinecite{Ribic2016}.}
  \label{FKM:Spectrum}
\end{figure}

The application of two-particle methods has demonstrated important physical aspects of the model.  
Using 1PI and DF \onlinecite{Ribic2016} showed how nonlocal correlations emerge as precursors to the charge density wave (CDW) transition. These precursors lead to a more insulating  solution and a four-peak structure in the ${\mathbf k}$-resolved spectral function in parts of the Brillouin zone as seen in Fig.~\ref{FKM:Spectrum}. These peaks have been interpreted as a mixture of the DMFT metal-insulator  transition  caused  by  local  correlations and nonlocal checkerboard CDW correlations. \onlinecite{Yang2014} have analyzed the interaction-driven crossover into the Mott phase and  related it to the CDW correlations.

\begin{figure}[t]
  \includegraphics[width=\columnwidth]{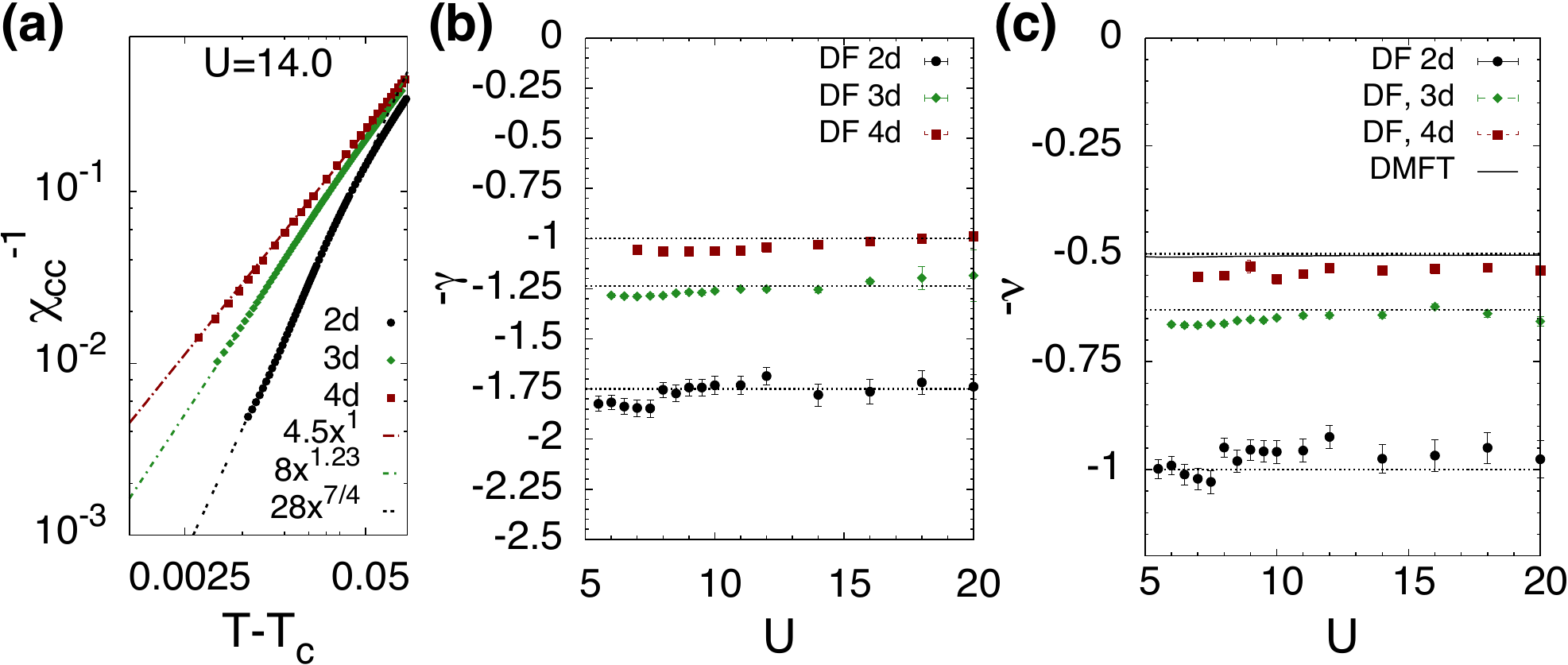}
    \caption{(Color online) (a) Inverse static $c$-electron charge susceptibility in $d=2,3$, and $4$ as a function of $T-T_c$ at $U=14t$. The slope of the double logarithmic plot yields the critical exponent $\gamma$  shown in (b) for different values of  $U > 5t$. Lines in (b) indicate the prediction for the Ising universality class. Error bars represent the regression errors. (c) Corresponding exponent $\nu$ for the correlation length  vs predictions for the Ising model (dashed lines) and DMFT results (solid line) as a function of $U$. From \onlinecite{Antipov2014}.}
  \label{Method_FKM:Antipov2014_fig2_3}
\end{figure}

\onlinecite{Antipov2014} studied the critical properties of the charge-ordering transition using the DF method. The inverse charge susceptibility of the $c$ electrons of the  FK model in Fig.~\ref{Method_FKM:Antipov2014_fig2_3} shows different power laws (i.e., different critical exponents) at the phase transition for different dimensions $d$.  These critical exponents are also different from those for the Hubbard model; see Fig.~\ref{fig:exponents}. But as for the Hubbard model in three dimensions,  the critical exponents $\gamma$ ([susceptibility, Fig.~\ref{Method_FKM:Antipov2014_fig2_3}(b)], $\nu$ [correlation length, Fig.~\ref{Method_FKM:Antipov2014_fig2_3}(c)], and anomalous dimension exponent $\eta = 2/\nu - \gamma$ extracted for $U > 5$ agree with the expected exact values, i.e., in this case the values for the Ising universality class. These results show that the diagrammatic extensions of DMFT can provide microscopic details of strongly correlated systems and at the same time correctly capture their critical properties. 

The interest in the FK model has reappeared at different times. From the initial proposals of describing metal-insulator transitions in $f$-electron systems~\cite{Falicov1969, Ramirez1970} and testing various methods for  strongly correlated electron systems, the recent interest is fueled by the progress in manufacturing artificial cold-atom systems and various manifestations of Anderson localization. In particular, multiband systems with large mass imbalance recently became available~\cite{Jotzu2015,Greif2015}. Monte Carlo simulations by \onlinecite{Liu2015} indicate an Ising-type AF ordering in case of a mass imbalance between the two spin species; \onlinecite{Philipp2016} concluded that there is a Kondo effect for an arbitrary small hopping of the more localized electrons. In other words, the mass imbalanced Hubbard model resembles the FK model regarding the symmetry of the ordering parameter, but the Hubbard model regarding the MIT. At the same time, the existence of Anderson localization in the absence of explicit disorder was recently shown by \onlinecite{Antipov2016}.

\subsection{Models of Disorder}
\label{sec:disorder}

Out of the many possible nonlocal effects, disorder plays a special role in condensed matter physics as it is ubiquitous in electronic materials. A spatially random potential reduces the extent of electronic wave functions to a localization length $\xi_l$, changing the motion of charge carriers and increasing the resistivity of the system. When $\xi_l$ becomes smaller than the linear system size $L$ the disorder renders an otherwise metallic system insulating \cite{Anderson1958,Thouless1974}, a phenomenon known as Anderson transition. 

On the technical side, disorder or at least a local disorder potential is closely related to the FK model of the previous section. The difference is that here the disorder distribution is externally given (quenched disorder) whereas in the FK model the localized electrons are thermodynamically distributed (annealed disorder). For disorder problems, DMFT  corresponds to the coherent potential approximation (CPA) of \onlinecite{Soven1969}  and \onlinecite{Taylor1967} as was  shown by~\onlinecite{Vlaming1992} whereas the relevance of the inverse of the coordination number was already pointed out by \onlinecite{Schwartz1972}. The CPA has the same averaging as in Eq.\ (\ref{Eq:DMFTFKM}) but for a fixed (quenched) $w$; see \onlinecite{Janis1992}. If the electrons also interact, a combination of CPA and DMFT is possible. Among others, it yields information about the local densities of states (DOSs) \cite{Byczuk2009} and the geometrically averaged density of states in the typical medium theory \cite{Dobrosavljevic2003}. The decay and the probability distribution of the latter quantity indicate the Anderson transition. Improved estimates of local and geometric density of states and critical disorder strengths can be obtained through cluster extensions of these theories \cite{Jarrell2001a,Terletska2014,Ekuma2014a}.

\begin{figure}[t]
\begin{center}
  \includegraphics[width=\columnwidth]{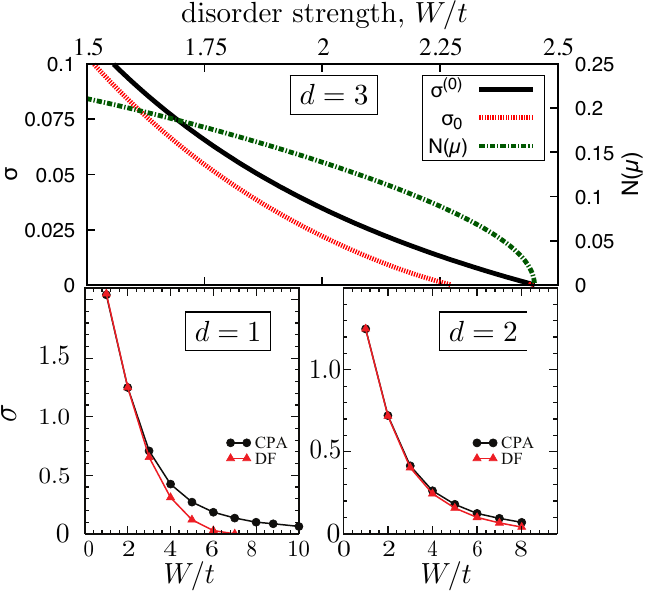}
\end{center}
    \caption{(Color online) Conductivity $\sigma$ as a function of the disorder strength $W$. Top: Results for 3D  at $T=0$ with leading-order vertex correction ($\sigma_0$) compared to the bubble contribution ($\sigma^{(0)}$) and the DOS at the Fermi level  ($\rho_F$). Adapted from \onlinecite{Pokorny2013}. Bottom: DF vs CPA results for $d=1$ (left) and $d=2$ (right) at $T=0.02t$. From \onlinecite{Terletska2013}.}
  \label{Results_Disorder:Terletska2013_fig4}
\end{figure}

The simplest model to incorporate the quenched disorder effects is the Anderson model, which reads
\begin{equation}
\label{eq:anderson_ins}
  H = \sum_{\mathrm{k}} \varepsilon_k  n_k + \sum_i {v_i n_i},
\end{equation}
where random potentials $v$ are distributed with a distribution $p(v)$. 

Similar to the case of clean systems, the self-consistent description at the single-particle level lacks some of the important physics present in the problem. For example, the vanishing conductivity in Anderson insulators needs a description at the two-particle level that is absent in CPA \cite{Jarrell2001a} and requires additional diagrammatic calculations \cite{Kroha1990,Janis2005}. Diagrammatic extensions of the CPA for model (\ref{eq:anderson_ins}) were pioneered by \onlinecite{Janis2001}, who developed a parquet approach to calculate the two-particle vertex and the conductivity. Using the local irreducible vertex and the local Green's function from the CPA, self-consistent equations for the full vertex, the nonlocal Green's function and the self-energy are derived  and evaluated. The vertex corrections to the conductivity from this method have been calculated by \onlinecite{Janis2010} and \onlinecite{Pokorny2013}. The results for 3D and binary disorder are shown in Fig.\ \ref{Results_Disorder:Terletska2013_fig4} (top). Vertex corrections always reduce the conductivity, and the leading-order correction renders the conductivity negative at a large disorder strength $W$. 

A complementary DF extension of CPA for the Anderson model (\ref{eq:anderson_ins}) was put forward by \onlinecite{Terletska2013}. Using the ``replica trick'' relation $\ln Z = \lim_{m\to 0} m^{-1}(Z^m - 1)$, the effect of disorder is replaced by a local elastic effective interaction between electrons in different replicas:  
\begin{equation}
  W^{\mathrm{dis}} = \sum_{l=2}^\infty \frac{\kappa_l (v)}{l!} \left[\sum_m \int d\tau n_i^m(\tau)\right]^l.
\end{equation} 
Here $\kappa_l (v)$ is the $l$-th order cumulant of the disorder distribution $p(v)$ and $m$ is the replica index reformulated in the language of multiple scattering theory. Since the disorder-induced interaction is local, the complexity of the problem at hand is similar to the Hubbard and Falicov-Kimball models, described in Secs.~\ref{sec:hubbard} and  \ref{sec:Results_FK}, respectively. The remaining ladder DF steps follow Sec.~\ref{sec:funcint}. \onlinecite{Terletska2013} used a box disorder distribution and showed that  weak localization effects suppress the conductivity in 1D and 2D; see the bottom panels of Fig.~\ref{Results_Disorder:Terletska2013_fig4}. Note that the DF extension of CPA can be done also without a replica trick as shown by \onlinecite{Osipov2013}.

The advantage of the diagrammatic extensions of CPA is that they can be straightforwardly extended to interacting systems with disorder. This was done by \onlinecite{Yang2014}, \onlinecite{Janis2014} and \onlinecite{Haase2016}, who considered the Anderson-Falicov-Kimball and Anderson-Hubbard models. Even a clean FK model exhibits localization effects due to the intrinsic annealed disorder, i.e., the scattering at the immobile $f$ electrons \cite{Antipov2016}. \onlinecite{Yang2014} showed that the added quenched disorder in the Anderson-FK model also localizes the system. \onlinecite{Janis2014} obtained the full phase diagram of the FK model in infinite dimensions, and showed that the critical disorder-driven metal-insulator transition shares its universal critical behavior with the interaction-driven Mott transition. \onlinecite{Haase2016} showed in DF that the disorder in the Hubbard model at small disorder strength tends to increase the impact of antiferromagnetism by raising the N\'eel temperature, increases the  $U$ value of the Mott transition, and at large disorder strengths brings the conductivity of the 3D system to zero in agreement with the picture of an Anderson transition. The phase diagram of this  Anderson-Hubbard model with metallic, Anderson insulating and Mott insulating phases was also determined before by \onlinecite{Byczuk2005} using DMFT and a geometrically averaged (”typical”) DOS as well as by  \onlinecite{Kuchinskii2010} employing  DMFT+$\Sigma_k$.

\subsection{Non-local interactions and multiorbitals}
\label{sec:nonlocalv}

Despite the success of the Hubbard model to capture important physical aspects of correlated electrons and materials, it misses some interesting physics such as plasmons and inhomogeneous charge density waves.
These effects are related to nonlocal interactions, which are included in the extended Hubbard model  
\begin{equation}
\label{eq:ehm}
  \mathcal{H} = \sum_{ij,\sigma} t_{ij} c_{i\sigma}^{\dagger}c^{\phantom{\dagger}}_{j\sigma} + U\sum_{i}n_{i\uparrow}n_{i\downarrow} + \frac{1}{2}\sum_{ij\sigma\sigma'} V_{ij} n_{i\sigma}n_{j\sigma'}.
\end{equation}
The nonlocal interaction  $V_{ij}$ can be sizable, with a magnitude reaching up to 60\% of the on-site Coulomb interaction. As a result graphene for example appears metallic, even though it would be on the verge of the insulating state if only the local Coulomb interaction was considered~\cite{Wehling11}. For certain surface systems, the nonlocal interaction may also exhibit a slow $1/r$ decay with distance $r$, rendering long-range contributions important~\cite{Hansmann2013,Hansmann2016}.

Addressing dynamical screening and long-range physics in correlated fermionic systems is challenging since QMC simulations typically suffer from the sign problem. In DMFT, on the other hand, the nonlocal interaction is restricted to its static Hartree contribution. Cluster extensions of DMFT naturally face the difficulty to treat interaction terms that extend beyond the finite cluster. Hence, intercluster interactions  are either truncated [e.g., in \onlinecite{Jiang2017}], coarse grained [e.g, in \onlinecite{Arita2004} and \onlinecite{Terletska2017}], or treated through a mean field decoupling [e.g., in \onlinecite{Bolech03} and \onlinecite{Reymbaut2016}] similar to the variational cluster perturbation theory~\cite{Aichhorn07}. In essence, the range of the interaction is limited by the size of the cluster.

This restriction is lifted in extended DMFT (EDMFT)~\cite{Si1996,Kajueter96,Smith00,Chitra00,Chitra01,Sengupta1995} and $GW$+DMFT~\cite{Sun02,Biermann2003,Sun2004}, where the nonlocal interaction can have arbitrary momentum dependence and range. These methods capture the Mott transition and at the same time the effects of screening and the charge-order transition driven by the intersite interaction. The $GW$+DMFT self-energy describes band renormalization effects and Hubbard satellites; EDMFT, on the other hand, captures the local correlations induced by the nonlocal interaction. Both methods, EDMFT and $GW$+DMFT, require the calculation of a local impurity problem with a frequency-dependent interaction, which is quite straightforward in CTQMC for a density-density type of interaction \cite{Werner07,Werner10}. 

\begin{figure}[tb]
\begin{center}
  \includegraphics[width=.54\columnwidth,angle=0]{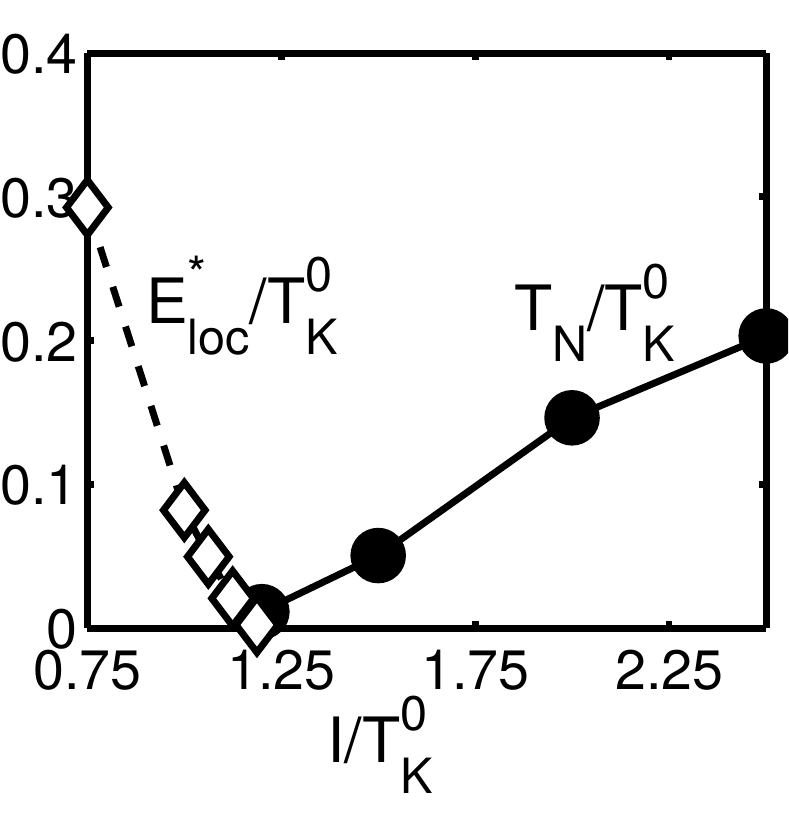}  \hfill
  \includegraphics[width=.33\columnwidth,angle=0]{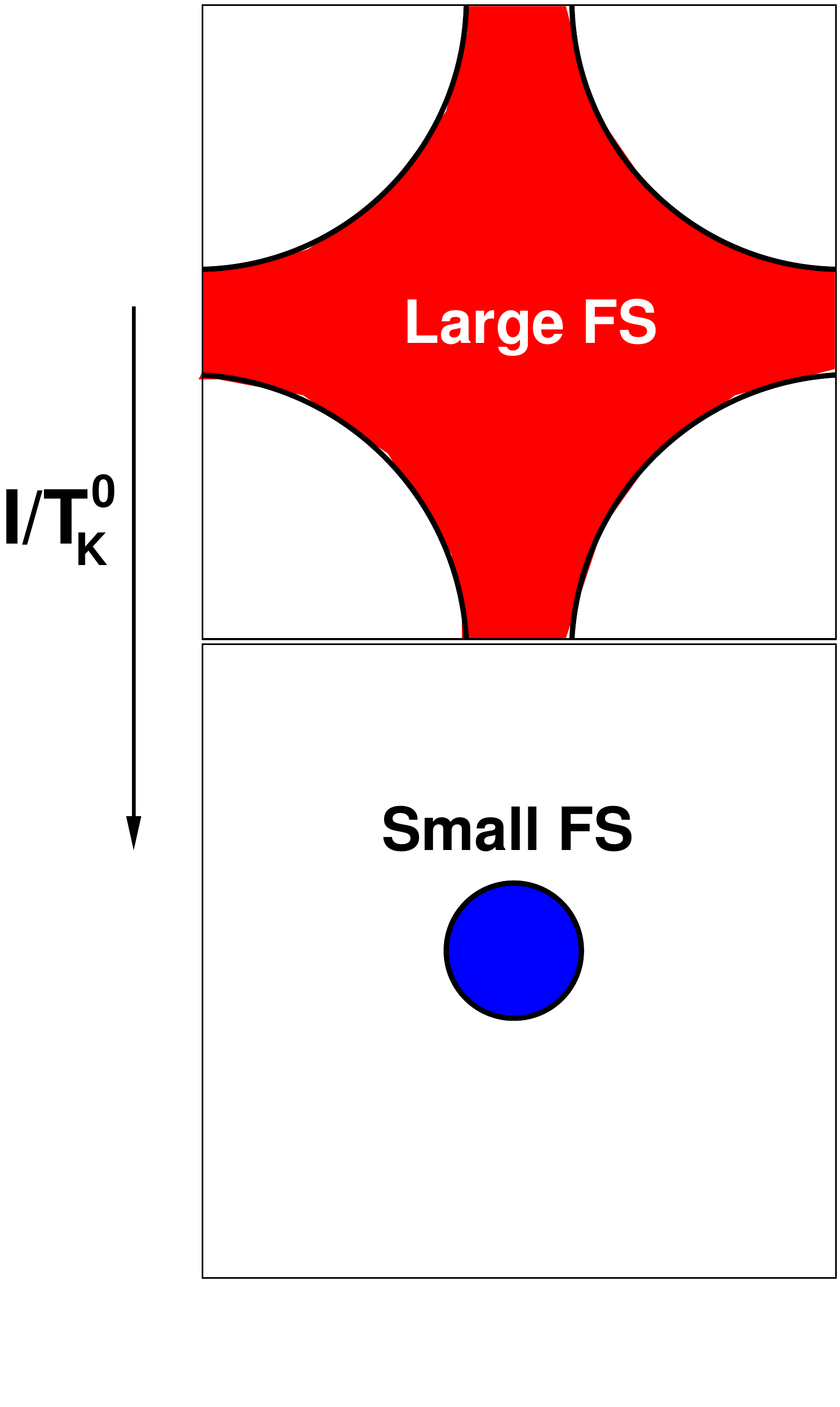}  
    \caption{(Color online) (Left) N\'eel temperature $T_N$ and Kondo breakdown scale $E^*_{\rm loc}$ as a function of the nonlocal interaction  $I$ relative to the Kondo scale $T_K^0$ for the Kondo lattice model with nonlocal spin interaction. The Kondo effect vanishes simultaneously with the magnetic phase transition. (Right) Fermi-surface volume collapse at the quantum critical point from a large Fermi surface where the $f$ electrons contribute to the Fermi surface because of the Kondo effect (for $E^*_{\rm loc}>0$) to a small Fermi surface  of the conduction electrons only.  From \onlinecite{Zhu2003}.}
  \label{fig:locQC}
\end{center}
\end{figure}

A highlight of EDMFT has been establishing the picture of local quantum criticality. Figure~\ref{fig:locQC} shows the results by \onlinecite{Zhu2003} for the Kondo lattice model (\ref{Eq:KLM}) with an additional spin-dependent nonlocal interaction $I$ of Ising type. At zero temperature there is a quantum critical point separating  the Kondo phase with a large Fermi surface at small  $I$ and the magnetic phase for large $I$. \onlinecite{Grempel2003} determined the corresponding critical exponent $\alpha\approx 0.7$, which is outside the standard Hertz-Millis-Moriya theory \cite{Loehneysen2007}.

Regarding $GW$+DMFT, \onlinecite{Ayral2012} showed that a fully self-consistent treatment of the $GW$+DMFT cures some of the deficiencies of self-consistent $GW$, such as the failure to describe plasmon satellites. \onlinecite{Werner2012} incorporated the frequency-dependent interaction obtained from constrained RPA into LDA+DMFT. This scheme may be viewed as a simplified version of $GW$+DMFT where the self-energy is local and the two-particle quantities are evaluated at a non-self-consistent level. Application to the normal phase of the iron pnictide superconductor BaFe$_{2}$As$_{2}$ showed that the  dynamical screening of the interaction significantly affects the low-energy electronic structure. A second effect beyond standard DFT+DMFT are exchange contributions to the self-energy that stem from the nonlocality of the interaction. These nonlocal self-energies were shown to be significant in BaFe$_2$As$_2$, other iron pnictides, and chalcogenides \cite{jmt_pnict}, as well as in transition metal oxides \cite{Miyake13,Tomczak2012}.

\begin{figure}[tb]
\begin{center}
  \includegraphics[width=1.0\columnwidth,angle=0]{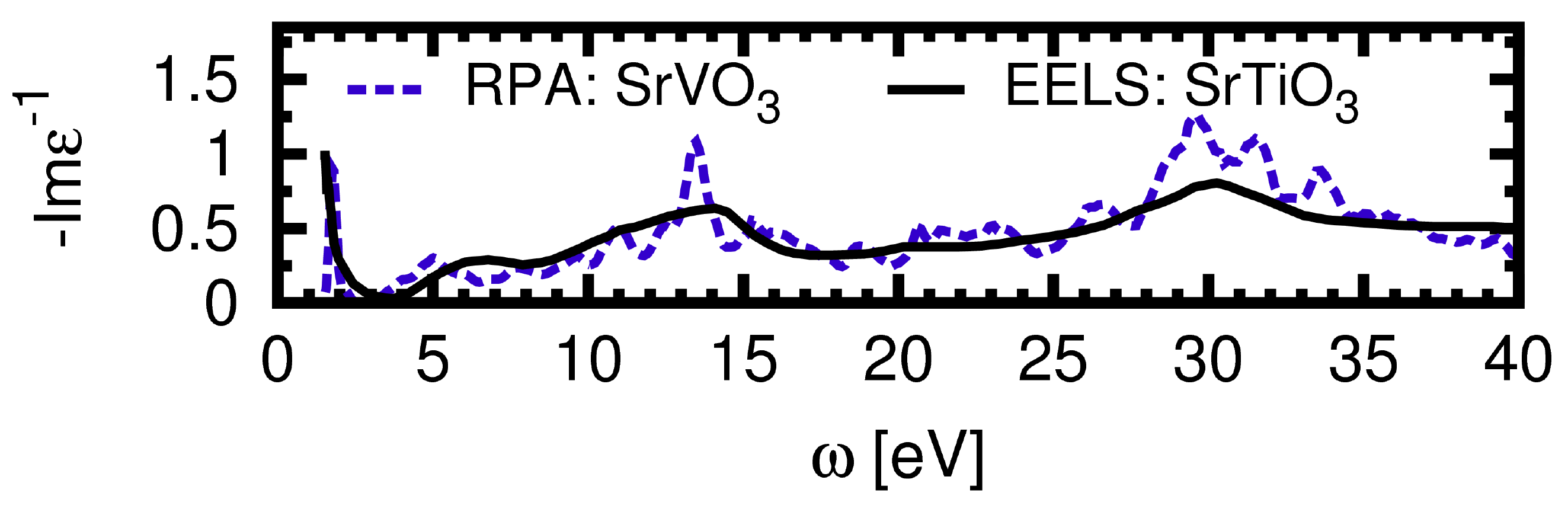}  
    \caption{(Color online) Inverse dielectric function of SrVO$_{3}$ in RPA as a function of frequency, showing  plasmon peaks at  $5$  and $14\,$eV. From~\onlinecite{Tomczak2014}.}
  \label{fig:SVOplasmons}
\end{center}
\end{figure}

These realistic calculations as well as $GW$+EDMFT model calculations for the extended Hubbard Hamiltonian~\eqref{eq:ehm} by \onlinecite{Ayral2013} show that the dynamical screening leads to plasmonic features in the spectral function. \onlinecite{Huang14} further found that including nonlocal interactions up to the third-nearest neighbors destabilizes the charge-ordered state, which may be viewed as a kind of frustration effect. \onlinecite{Tomczak2012,Tomczak2014} and \onlinecite{Boehnke2016} studied the spectral properties of the prototypical correlated metal SrVO$_{3}$ using $GW$+DMFT and self-consistent  $GW$+EDMFT, respectively.  They found that the effective local interaction is considerably reduced due to dynamical screening effects in RPA and conclude that the spectral function exhibits a plasmon satellite in the region of the previously reported upper Hubbard band. In the dielectric function, Fig.~\ref{fig:SVOplasmons}, this plasmon peak is around $5\,$eV. This and the larger plasmon peak around $14\,$eV agree with the electron-energy-loss spectrum (EELS) of~\onlinecite{Kohiki2000}.

Despite the success of EDMFT and $GW$+EDMFT, they do not provide a complete description of plasmons. This is due to an inconsistent treatment of the single- and two-particle properties, which breaks local charge conservation and gauge invariance. In particular, from the continuity equation $\omega^2 \av{nn^{*}}_{{\qv\omega}}=q^2 \av{jj^{*}}_{{\qv\omega}}$ it follows that the polarization behaves as $q^{2}/\omega^{2}$ for small $q$ and finite (Matsubara) frequencies $\omega_{m}$. This is the case for the Lindhardt function in RPA, non-self-consistent $GW$ (so-called $G_{0}W_{0}$), quasiparticle self-consistent $GW$ (QS$GW$) \cite{Faleev2004}, and  QS$GW$+DMFT \cite{jmt_sces14}, but not in  EDMFT and $GW$+DMFT. In EDMFT, the polarization is momentum independent and the plasmon dispersion diverges for $q\to 0$ in the presence of long-range Coulomb interaction. Vertex corrections from a local but frequency-dependent irreducible vertex are necessary to fulfill the Ward identity. In the dual boson (DB) approach~(Sec.~\ref{sec:db}), such vertex corrections  can be  constructed diagrammatically via nonlocal polarization corrections. The resulting polarization $\Pi_{\omega}$ vanishes for small $q$  as $q^2$ \cite{Rubtsov12,Hafermann2014,Stepanov2016}. Hence, the solution of the plasmon pole defined by $1+V(\qv)\Pi_{\omega}(\qv)=0$ yields the correct  dispersion relation $\omega(\qv) = \omega_p + a q ^2$ at small $q$. Here $a$ is a constant and $\omega_p$ is the plasma frequency.

\begin{figure}[bt]
\begin{center}
  \includegraphics[width=.8\columnwidth,angle=0]{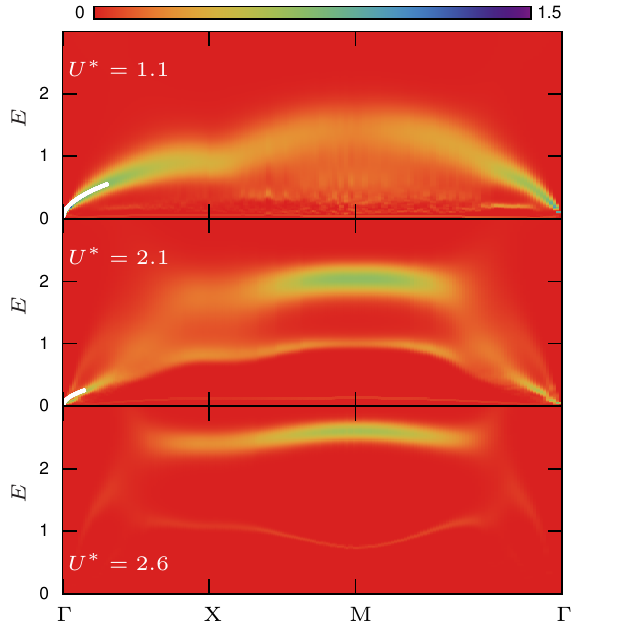} 
\end{center}
    \caption{(Color online) Inverse dielectric function of the extended 2D Hubbard model with long-range Coulomb interaction as a function of momentum and energy for three different values of the effective on-site interaction $U^{*}$. The two-particle excitations show a renormalized dispersion and spectral weight transfer. From~\onlinecite{vanLoon2014}.}
  \label{fig:2dplasmons}
\end{figure}

Using DB, \onlinecite{vanLoon2014} showed that the two-particle excitations exhibit both a renormalization of the dispersion and a spectral weight transfer. This is similar to the analogous interaction effects known for single-particle excitations. Figure~\ref{fig:2dplasmons} shows the inverse dielectric function $-\rm{Im}\epsilon_{\vc{q}}^{-1}(E)$ of 2D surface plasmons in the presence of long-range interaction $V(q)$. For weak interaction one observes a broad particle-hole continuum and the expected $\omega_{p}(q) \sim \sqrt{q}$ dispersion of the 2D plasmon  at small $q$.  As the interaction is increased (Fig.~\ref{fig:2dplasmons} middle), the plasmon dispersion is renormalized, and spectral weight is transferred to a  second branch which now becomes visible at larger energies. Above $U^{*}\sim 2.4$ the system is a Mott insulator, and only the weakly dispersing second band at energy $U^{*}$ associated with doublon-holon excitations survives.

The extended Hubbard model with nearest-neighbor interaction $V$ also shows a transition to a charge density wave (CDW) ordering, which was studied in   lattice Monte Carlo calculations by \onlinecite{Zhang89}, in DMFT by \onlinecite{Wahle1998}, in EDMFT by \onlinecite{Sun02}, and in the  two-particle self-consistent approach by \onlinecite{Davoudi07}. In DB, the momentum dependence of the polarization corrections is also included. This shifts the DB CDW transition to smaller $V$ values compared to EDMFT and agrees with RPA in the weak-coupling limit~\cite{vanLoon2014a}.

\begin{figure}[t]
\begin{center}
  \includegraphics[width=.9\columnwidth,angle=0]{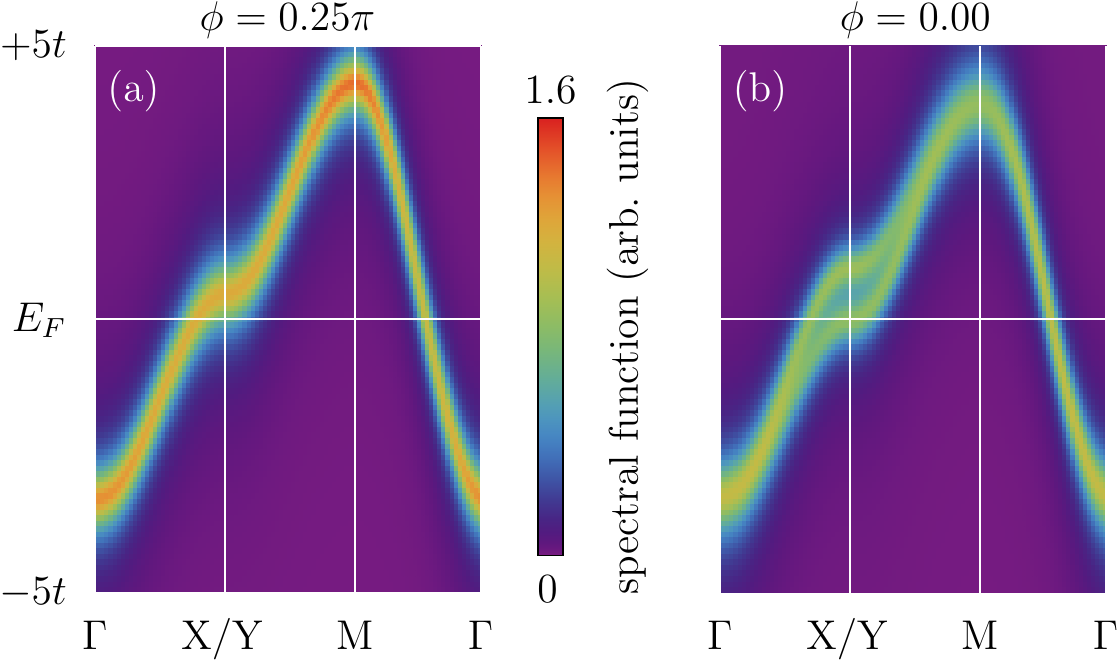} 
\end{center}
    \caption{(Color online) Spectral function of the dipolar fermion Hubbard model superimposed over two different paths in the Brillouin zone via the $X$ and $Y$ points, respectively. In the symmetric case, both paths are equivalent (left). The dipolar interaction drives the Lifshitz transition by breaking the symmetry between the $X$ and $Y$ points (right). From~\onlinecite{vanLoon2016-2}.}
  \label{fig:specfunclifshitz}
\end{figure}

Long-range nonlocal interactions can also play a crucial role for ultracold quantum gases in optical lattices~\cite{Lewenstein2012,Bloch2008,Bloch2012}. For example, these highly tunable systems allow one to realize the dipolar Fermi Hubbard model \cite{Lewenstein2012,Baranov2008,Lahaye2009,Zoller2012} which corresponds to the Hamiltonian \eqref{eq:ehm} with an anisotropic long-range dipolar interaction $V_{jk} \sim \left[1-3(\hat{r}_{jk}\cdot \hat{d})^2 \right]/(r_{jk})^3$. Here $\hat{r}_{jk}$ is the direction and $r_{jk}$ the magnitude of the lattice vector from site $j$ to $k$, and all dipoles are assumed to point in the same direction $\hat{d}$. Using the DB approach,~\onlinecite{vanLoon2015-2} found that for sufficiently large dipole strengths and dipoles oriented perpendicular to the lattice plane $(\phi=0)$, a transition to checkerboard order occurs, while a striped phase emerges when they point to nearest neighbors $(\phi=\pi/2)$. For dipoles pointing along the diagonal $(\phi=\pi/4)$, \onlinecite{vanLoon2016-2} found a novel ultra-long-range-ordered phase which can alter the topological properties and lead to a Lifshitz transition (Fig.~\ref{fig:specfunclifshitz}).

A major advantage of diagrammatic extensions of DMFT is that multiorbital and realistic material calculations are  much more feasible than in cluster extensions of DMFT which are restricted to a very few lattice sites \cite{Biermann2005,Lee2012}. Against this background it is maybe surprising that there is hitherto only a single multiorbital calculation by \onlinecite{Galler2016}.\footnote{Compare our discussion of~\onlinecite{Wilhelm2015} above for a one-band calculation with DFT-derived parameters and \onlinecite{Hirschmeier2018} for a one-band DF calculation with two lattice sites in the unit cell.} One reason for this is that the calculation of the local multiorbital vertex requires considerable effort, requiring worm sampling in CT-HYB \cite{Gunacker15} for actually  calculating all contributions of the vertex.  Within the D$\Gamma$A framework, nonlocal interactions $V_{\mathbf q}$ can be taken into account as part of the irreducible vertex  so that these are naturally included in  {\em ab initio} materials calculations; see Sec.~\ref{sec:abinitioDGA}. This way all DMFT and $GW$ Feynman diagrams are included as well as nonlocal correlations beyond both.

\begin{figure}[tb]
  \includegraphics[width=.95\columnwidth,clip]{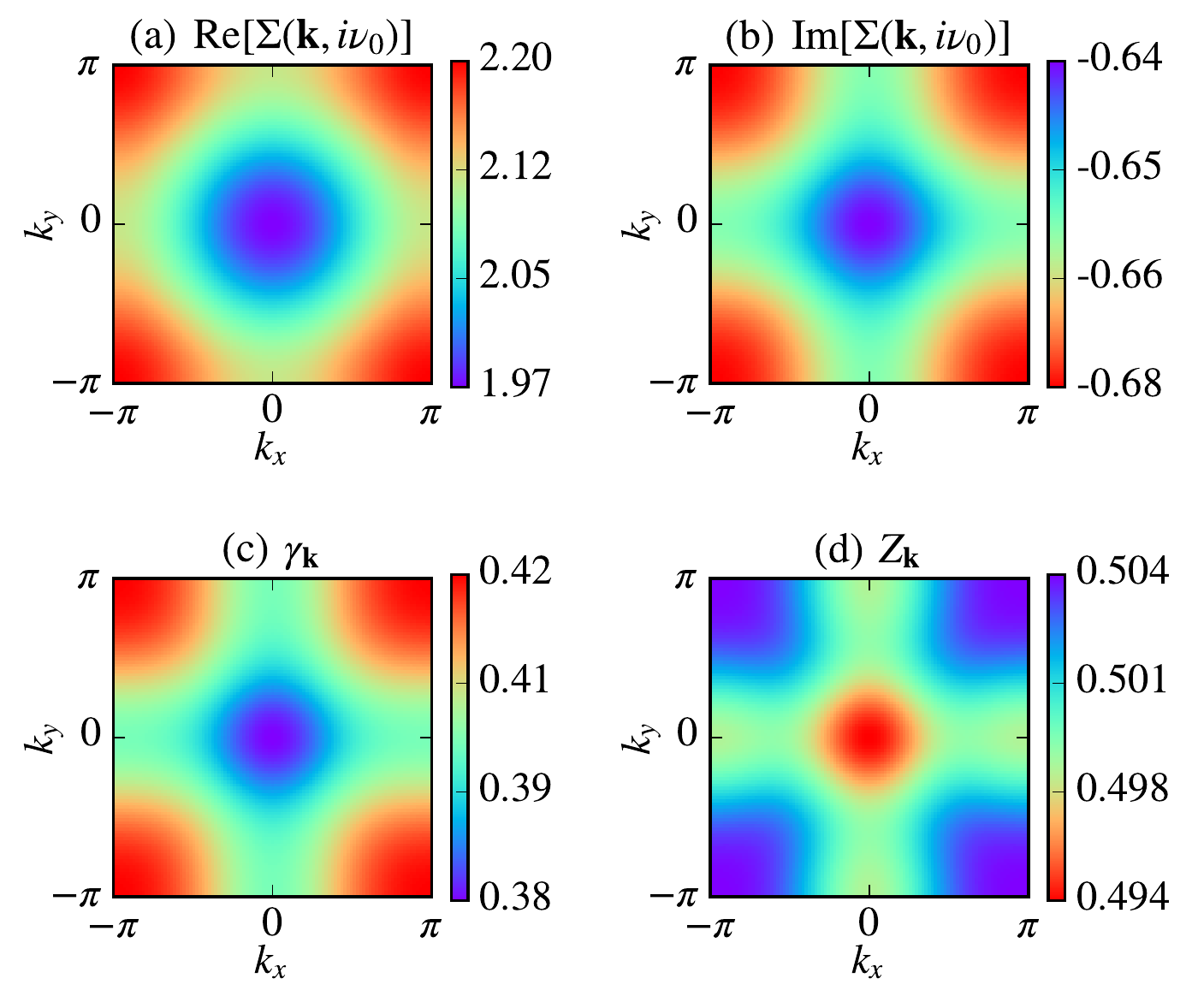} 
    \caption{(Color online) AbinitioD$\Gamma$A for SrVO$_3$ showing the momentum dependence of  (a)  the real and (b) imaginary parts of the self-energy at the lowest Matsubara frequency ($i\nu_0$) for the $x-y$ orbital in the $k_z=0$ plane; (c) corresponding scattering rate $\gamma_{\bf k}$ and (d) quasi particle weight $Z_{\bf k}$. From \onlinecite{Galler2016}.}
  \label{FigDGASrVO3}  
\end{figure}

In their AbinitioD$\Gamma$A calculation of  SrVO$_3$,  \onlinecite{Galler2016}  took the local vertex of all three vanadium $t_{2g}$ orbitals  into account and calculate from it through ladder D$\Gamma$A diagrams the self-energy which becomes momentum dependent; for computational details cf.~\onlinecite{Galler2017b}. Bulk SrVO$_3$ is a strongly correlated metal and, at least at elevated temperatures, far away from any (e.g., magnetic) ordering. Hence, one would expect for such a 3D  material rather weak effects of nonlocal correlations, with the exception of $GW$-like screening effects. But even for $V_{\mathbf q}=0$ (i.e., without such $GW$ contributions), \onlinecite{Galler2016} found a momentum differentiation larger than $0.2$eV in the real part  of the self-energy [see Fig.~\ref{FigDGASrVO3}(a)], while the ${\mathbf k}$ dependence of its imaginary part and of the quasiparticle weight and scattering rate is much weaker [cf.\ related findings for the 3D Hubbard model \cite{Schaefer2015}]. In essence, the  momentum (and orbital) differentiation of the real part of the self-energy  pushes the occupied and unoccupied states farther away from each other. This shows that even far away from a phase transition and even beyond $GW$-type diagrams ($V_{\mathbf q}$=0), nonlocal correlations play a role in actual materials. Much larger effects are to be expected  in the vicinity of second-order phase transitions and for 2D or layered materials. The  AbinitioD$\Gamma$A approach presents a promising route to model such materials. We hence expect realistic multi orbital calculations that include nonlocal charge and spin fluctuations beyond $GW$+DMFT to thrive in the future.

\section{Open source implementations}
\label{Sec:opensource}

The increasing complexity of numerical methods requires expert knowledge and leads to increasing implementation efforts. For these reasons, and to ensure reproducibility of results, it is vital to make codes publicly accessible. We encourage such efforts and believe they should be rewarded. In recent years, a number of open source libraries and codes have appeared or will appear in the near future. We can separate these into program packages that (i) solve a local impurity problem and allow calculating the two-particle vertex and that (ii) calculate nonlocal correlations beyond DMFT diagrammatically.

For step (i), solving the AIM, let us mention six program packages: {\tt ALPS}~\cite{ALPS2,ALPSCORE}, {\tt iQIST}~\cite{Huang2015},  {\tt pomerol} \cite{Antipov2015a}, {\tt TRIQS}~\cite{TRIQS}, {\tt EDMFTF}~\cite{Haule2007}, and  {\tt w2dynamics}~\cite{w2dynamics,w2dynamics2018}.  The {\tt ALPS} and {\tt TRIQS} libraries aim to provide a reusable set of components to facilitate the implementation of algorithms for strongly correlated systems. Strong-coupling CTQMC impurity solvers based on {\tt ALPS} are available~\cite{Gull2011b,Hafermann2013,Shinaoka2016}. The {\tt iQIST} package~\cite{Huang2015} provides a collection of impurity solvers and preprocessing and postprocessing tools, allowing also for the computation of two-particle functions. The  {\tt pomerol} code~\cite{Antipov2015a} provides an optimized implementation of the ED method to compute vertex functions. The {\tt TRIQS} package includes weak- and strong-coupling CTQMC~\cite{Seth2016} and has been extended to two-particle quantities with the development of TRILEX \cite{Ayral2015,Ayral2016a}  and QUADRILEX \cite{Ayral2016}. The  {\tt EDMFTF} code~\cite{Haule2007} aims at DFT+DMFT materials calculations with a special emphasis on realizing the Luttinger-Ward functional of \onlinecite{Haule2015}. The {\tt w2dynamics} package is also based on the hybridization expansion CTQMC and uses improved estimators and worm sampling for calculating all components of the two-particle multiorbital vertex   \cite{Gunacker15,Gunacker2016} as well as vertex asymptotics~\cite{Kaufmann2017}.

As for (ii), diagrammatic extensions of DMFT, publicly available codes only start to emerge. {\tt OPENDF}~\cite{OPENDF} is the first open source implementation of the DF approach,  {\tt LadderD$\Gamma$A}~\cite{Rohringer2018}  and  {\tt AbinitioD$\Gamma$A}~\cite{Galler2017b} solve the  ladder D$\Gamma$A equations for a single orbital including a Moriya $\lambda$ correction (see Sec.~\ref{sec:dgaladder}) and multiple orbitals, respectively. {\tt AbinitioD$\Gamma$A}~\cite{Galler2017b} also includes nonlocal Coulomb interactions. {\tt parquet}~\cite{Yang2009} and {\tt Victory}~\cite{Li2017}   are program packages to solve the parquet equations for a given fully irreducible vertex $\Lambda$ as it is used, for example, in D$\Gamma$A. They differ in the way the vertex at large frequencies outside the initial frequency box is treated (periodic vs asymptotic boundary conditions for frequencies).

Let us also mention some auxiliary codes: {\tt gftools} \cite{aeantipovgftools} for manipulating Green's functions (similar tools are also included in  {\tt ALPS} and {\tt TRIQS})   and several codes for an analytical continuation with the maximum entropy method (MaxEnt): {\tt MEM}~\cite{Jarrell1996},  {\tt  maxent}~\cite{Levy2017}, {\tt OmegaMaxEnt} \cite{Bergeron2016}, and  {\tt w2dynamics}~\cite{w2dynamics,w2dynamics2018} which incorporates MaxEnt with optimization techniques developed by \onlinecite{Sandvik98b}.

\section{Conclusion and outlook}
\label{sec:conclusion}

Diagrammatic extensions of DMFT appealingly combine numerical and analytical techniques for studying strongly correlated electron systems. Local correlations are treated by the method which is arguably best at this: DMFT or a more general self-consistent (numerical) solution of an AIM. On top of this, nonlocal correlations in the self-energy and susceptibilities are constructed through Feynman diagrams. Historically this development started with methods that supplemented the local DMFT self-energy by a nonlocal one from another method such as spin-boson theory, FLEX or $GW$. These combinations are discussed in  Secs.~\ref{subsec:DMFTplus} and \ref{sec:edmftGW}, respectively. More recently, we have seen the rise of methods which more intimately connect the local and nonlocal parts. These approaches calculate a dynamical frequency-dependent local vertex by solving an impurity model numerically and derive nonlocal correlations therefrom diagrammatically. One can envisage this as raising the DMFT concept of the locality of the one-particle (irreducible) vertex (i.e., the self-energy) to the next, i.e., two-particle vertex level. All of these dynamical vertex approaches are closely related and rely on the same concept, but they differ in which local two-particle vertex is taken and how the diagrams are constructed. One may compare this to the difference between different cluster extensions of DMFT, say DCA versus CDMFT. Table~\ref{Table:alg} provides an overview  of  the various approaches.

\begin{table*}[tb]
\setlength\extrarowheight{5pt}
\begin{center}
  \begin{tabular}{|l | l | l | l | l|}
  \hline
    Method  & Local  vertex & Green's function &  Diagrams & Action/Functional\\
  \hline
    parquet D$\Gamma$A  [Sec.~\ref{sec:parquetDGA}]  & two-particle irreducible  $\Lambda^{\nu\nu'\omega}$ & $G_{{\mathbf k}\nu}$ & parquet & \\
    QUADRILEX [Sec.~\ref{sec:quadrilex}] &&&& $\mathcal{K}_{4}[G^{\rm loc}_\nu,G_{\sigma\sigma'}^{(2),\nu\nu'\omega}]$ ~\eqref{Eq:K2} \\
  \hline
    ladder D$\rm \Gamma$A  [Sec.~\ref{sec:dgaladder}] & 2PI in channel $r$: $\Gamma_{r}^{\nu\nu'\omega}$ & $G_{{\mathbf k}\nu}$ &  ladder&  --- \\
  \hline
    DF [Sec.~\ref{sec:funcint}] &   one-particle reducible $F^{\nu\nu'\omega}$  & $\widetilde{G}_{0, {{\mathbf k}\nu}}$ &2nd order, ladder,&  $\widetilde{\pazocal{S}}[\tilde{c}^+,\tilde{c}]$  ~\eqref{equ:defdualaction} \\
    & & & parquet & \\
  \hline
    1PI  [Sec.~\ref{subsec:1pi}] &  one-particle irreducible$^{\rm \ref{note1PI}}$ $F^{\nu\nu'\omega}$ & $\widetilde{G}_{0, {{\mathbf k}\nu}}$, $G_{\nu}^{\rm loc}$ & ladder & $\mathcal{S}^{\text{1PI}}$  ~\eqref{equ:cap4action1PI} \\
  \hline
    DMF$^2$RG  [Sec.~\ref{subsec:DMFT2fRG}] &  one-particle irreducible$^{\rm \ref{note1PI}}$ $F^{\nu\nu'\omega}$ & $G_{_\mathit{\Lambda}, {{\mathbf k}\nu}}$, $S_{_\mathit{\Lambda},{{\mathbf k}\nu}}$ & RG flow in $\mathit{\Lambda}$& $\pazocal{S}_\mathit{\Lambda} [c^+,c]$  ~\eqref{equ:actionconnect}\\
  \hline
  \hline
    TRILEX [Sec.~\ref{sec:TRILEX}] & three-leg vertex $\gamma_{\nu \omega}$ & ${G}_{\kv\nu}$, ${W}_{\qv\omega}$ & Hedin Eqs.~\eqref{eq:HedinPi} & $\mathcal{K}_{3}[G_\nu,W_{\omega},\chi_{\nu\omega}]$ ~\eqref{eq:k3trilex} \\
  \hline
    DB [Sec.~\ref{sec:db}] &   $F^{\nu\nu'\omega}$, $\gamma_{\nu \omega}$ & $\widetilde{G}_{0,\kv\nu}$, $\widetilde{W}_{0,\qv\omega}$ & 2nd order, ladder &  $\widetilde{\mathcal{S}}[\widetilde{c}^+,\widetilde{c},\widetilde{\rho}^*,\widetilde{\rho}]$  ~\eqref{Eq:DBaction} \\
  \hline
  \end{tabular}
\end{center}
    \caption{Summary of the various closely related diagrammatic extensions of DMFT. The {\em first column} denotes the method and the {\em second column} the local vertex function  that serves as a starting point; these are the different two-particle vertices defined in Sec.~\ref{sec:formalism}\footnote{1PI and DMF$^2$RG expand in terms of 1PI vertices, which for the {\em two-particle  vertex} happens to be identical to the full vertex $F$ used in DF. Even if truncated at the two-particle level,  the difference in the  expansion scheme leads none\-the\-less to distinct 1PI and DF methods; see Sec.~\ref{subsec:1pi}.\label{note1PI}} and  the  bosonic three-leg vertex $\gamma_{\nu \omega}$. The {\em third column} identifies the Green's function lines connecting these local building blocks via the Feynman diagrams of the {\em fourth column}.\footnote{$G_{\nu}^{\rm loc}$ is the local propagator of the reference system and  $\widetilde{G}_{0, {{\mathbf k}\nu}}$ the nonlocal lattice propagator constructed with the local self-energy of the reference system  [Eqs.~\eqref{equ:dualgaussfinal}, \eqref{equ:cap4nambubare}, and \eqref{eq:tildeG0}]; $\widetilde{W}_{0,\qv\omega}$ is a corresponding bosonic (interaction) propagator  [Eq.~\eqref{eq:tildeW0}].} The {\em last column} denotes the fundamental functional $\mathcal{K}$ or the action ${\mathcal{S}}$ of the functional integral the method is based on. Here DF and DB introduce dual fermionic $\tilde{c}^{(\dagger)}$ and bosonic variables  $\tilde{\rho}^{(*)}$. For further details we refer the reader to the corresponding sections.}
  \label{Table:alg}
\end{table*}

In Sec.~\ref{Sec:results}, we extensively compared these vertex approaches with each other and against other state-of-the-art approaches such as cluster extensions of DMFT and lattice QMC. Application to the Hubbard model has shown a qualitatively consistent picture for the self-energy, susceptibility, and its phase diagram, e.g., regarding coherent quasiparticle excitations and the suppression of long-range order in comparison with DMFT. The methods have been shown to reproduce highly nontrivial effects such as the pseudogap. In some cases where benchmarks are available, we have even seen quantitative agreement. In general, one finds that, while the second-order diagram already contains nontrivial effects of dynamical short-range correlations, ladder diagrams are important for a quantitative description. The fluctuation diagnostics  of the self-energy can be exploited to choose which ladder should be considered. As expected, the one-dimensional case turned out to be the most challenging for an extension of (dynamical) mean field theory. However it is encouraging that the corresponding results are improved substantially by using the vertex on a cluster of sites as a starting point. The application to the (extended) Hubbard, Falicov-Kimball, Kondo lattice and Anderson models demonstrates that these methods are very versatile.

A unique feature of diagrammatic extensions is that they combine the nonperturbative physics of local correlations \`a la DMFT with truly long-ranged correlations over hundreds or thousands of sites. In addition, they obviously do not suffer from a sign problem. We have seen results which would be difficult or even impossible to obtain using other methods. Examples are the numerical calculation of (quantum) critical exponents of models for strongly correlated electrons, the absence of a Mott transition in the two-dimensional Hubbard model or the rich phase diagram of the frustrated triangular-lattice Hubbard model. Another highlight is the renormalization of the dispersion and spectral weight transfer of collective modes. The application to the Kondo lattice model revealed an interplay of local Kondo physics and long-range antiferromagnetic correlations around the quantum critical point. These results establish the complementarity of diagrammatic extensions to other many-body methods, in particular, cluster extensions.

After roughly ten years of development, we have seen only the tip of the iceberg of diagrammatic extensions of DMFT. The field is very dynamic and many new results emerged even during the preparation of this review. We expect to see many more applications in the future. This is driven by a growing community of users of these methods and techniques as well as by methodological advances. These have already allowed the treatment of nonlocal interactions in AbinitioD$\Gamma$A or DB. Recent advances in impurity solvers have triggered a first realistic multiorbital  AbinitioD$\Gamma$A calculation  for SrVO$_3$, which offers the prospects of  improving even upon $GW$+DMFT. Indeed diagrammatic extensions of DMFT  are much more promising for {\em ab initio} materials calculations than cluster extensions of DMFT, which are numerically feasible only for a few cluster sites if multiple  orbitals need to be taken into account. Another forthcoming topic is the calculation of vertex corrections to the electrical conductivity. Using the real-space formulation of these methods one can imagine applications to ultracold atoms in harmonic trapping potentials or topological insulators.

Thanks to  new technical developments, the calculation of the local three-particle vertex has become possible. This allowed for the calculation of selected diagrams and  an estimate of the error when the DF vertex is truncated at the two-particle level. Further diagrams need to be derived and considered, and a similar analysis still needs to be done for 1PI and D$\Gamma$A. But the first results already show that a systematic improvement of the diagrammatic extensions of DMFT is feasible, albeit  to a limited extent. Besides, this development may lead to new insight and the discovery of physical effects originating from three-particle excitations. On the other hand, diagrammatic extensions that expand around a DCA or CDMFT cluster as a starting point have been shown to be doable and it is clear that such calculations will become more prevalent in the future. The combination of diagrammatic extensions with the functional renormalization group is very appealing from a theoretical point of view, which should be further explored. The DMF$^2$RG has shown first encouraging results.

Last but not least, we believe that the reviewed diagrammatic approaches offer a new route to the thriving field of  strongly correlated electron systems out of equilibrium.

In the past, these methods were used by a number of specialized groups. Because these methods deal with vertex functions they are technically more involved than other approaches. This barrier will become much less important in the near future, not least due to efforts to release codes into the public domain. We are convinced that these methods will become standard tools in the research of strongly correlated systems and hope that this review will encourage our readers to use one of these methods in their research.

\paragraph*{Acknowledgment}
First we would thank our close and long-term co-workers with whom we explored the hitherto unknown quantum fields that extend beyond  DMFT: S.~Andergassen,  S.~Brener, A.~Galler, P.~Gunacker, G.~Li, W.~Metzner, T.~Ribic, G.~Sangiovanni, T.~Sch\"afer, E.~Stepanov, A.~Tagliavini, C.~Taranto, P.~Thunstr\"om, J.~Tomczak, A.~Valli, E.G.C.P.~van~Loon, and N.~Wentzell.
We also profited tremendously from discussions with T.~Ayral, A.~Georges, J.~Gukelberger, E.~Gull, O. Gunnarsson, D.~Hirschmeier, V.~Jani\v{s}, M.~Jarrell, C.~Jung, A.~Kauch, J.~Kaufmann, S.~Kirchner, M.~Kitatani, A.~Koga, E.~Kozik, F.~Krien, J.~Kune\v{s}, H.~Monien, J.~Otsuki, O.~Parcollet,  K.~R.~Patton, P.~Pudleiner, P.~Ribeiro, M.~Ringel,  N.~Takemori, A.-M.S.~Tremblay, D.~Vollhardt, M.~Wallerberger, and P.~Werner.
This work has been supported in part by the European Research Council  under the European Union's Seventh Framework Program (FP/2007-2013) through ERC Grant No.\ 306447 (K. H.); the Austrian Science  Fund (FWF) through  SFB ViCoM F41 (A. T., K. H.) and projects I 2794-N35 and I-610 (A. T.); the Deutsche Forschungsgemeinschaft (DFG) through research unit FOR1346 comprising FWF I-1395 (A. T., A. I. L, K. H.); the Russian Federation through theme ``Quant'' AAAA-A18-118020190095-4 of FASO (A. A. K.); and the joint Russian Science Foundation (RSF)/DFG Grant No. 16-42-01057 / LI 1413/9-1 (G. R., A. I. L., A. N. R.).

\bibliography{main}

\begin{thebibliography}{390}%
\makeatletter
\providecommand \@ifxundefined [1]{%
 \@ifx{#1\undefined}
}%
\providecommand \@ifnum [1]{%
 \ifnum #1\expandafter \@firstoftwo
 \else \expandafter \@secondoftwo
 \fi
}%
\providecommand \@ifx [1]{%
 \ifx #1\expandafter \@firstoftwo
 \else \expandafter \@secondoftwo
 \fi
}%
\providecommand \natexlab [1]{#1}%
\providecommand \enquote  [1]{``#1''}%
\providecommand \bibnamefont  [1]{#1}%
\providecommand \bibfnamefont [1]{#1}%
\providecommand \citenamefont [1]{#1}%
\providecommand \href@noop [0]{\@secondoftwo}%
\providecommand \href [0]{\begingroup \@sanitize@url \@href}%
\providecommand \@href[1]{\@@startlink{#1}\@@href}%
\providecommand \@@href[1]{\endgroup#1\@@endlink}%
\providecommand \@sanitize@url [0]{\catcode `\\12\catcode `\$12\catcode
  `\&12\catcode `\#12\catcode `\^12\catcode `\_12\catcode `\%12\relax}%
\providecommand \@@startlink[1]{}%
\providecommand \@@endlink[0]{}%
\providecommand \url  [0]{\begingroup\@sanitize@url \@url }%
\providecommand \@url [1]{\endgroup\@href {#1}{\urlprefix }}%
\providecommand \urlprefix  [0]{URL }%
\providecommand \Eprint [0]{\href }%
\providecommand \doibase [0]{http://dx.doi.org/}%
\providecommand \selectlanguage [0]{\@gobble}%
\providecommand \bibinfo  [0]{\@secondoftwo}%
\providecommand \bibfield  [0]{\@secondoftwo}%
\providecommand \translation [1]{[#1]}%
\providecommand \BibitemOpen [0]{}%
\providecommand \bibitemStop [0]{}%
\providecommand \bibitemNoStop [0]{.\EOS\space}%
\providecommand \EOS [0]{\spacefactor3000\relax}%
\providecommand \BibitemShut  [1]{\csname bibitem#1\endcsname}%
\let\auto@bib@innerbib\@empty
\bibitem [{\citenamefont {Abrikosov}\ \emph {et~al.}(1975)\citenamefont
  {Abrikosov}, \citenamefont {Gorkov},\ and\ \citenamefont
  {Dzyaloshinski}}]{Abrikosov1975}%
  \BibitemOpen
  \bibfield  {author} {\bibinfo {author} {\bibnamefont {Abrikosov},
  \bibfnamefont {A.~A.}}, \bibinfo {author} {\bibfnamefont {L.~P.}\
  \bibnamefont {Gorkov}}, \ and\ \bibinfo {author} {\bibfnamefont {I.~E.}\
  \bibnamefont {Dzyaloshinski}}} (\bibinfo {year} {1975}),\ \href@noop {}
  {\emph {\bibinfo {title} {Methods of Quantum Field Theory in Statistical
  Physics}}}\ (\bibinfo  {publisher} {Dover},\ \bibinfo {address} {New
  York})\BibitemShut {NoStop}%
\bibitem [{\citenamefont {Aichhorn}\ \emph {et~al.}(2004)\citenamefont
  {Aichhorn}, \citenamefont {Evertz}, \citenamefont {von~der Linden},\ and\
  \citenamefont {Potthoff}}]{Aichhorn07}%
  \BibitemOpen
  \bibfield  {author} {\bibinfo {author} {\bibnamefont {Aichhorn},
  \bibfnamefont {M.}}, \bibinfo {author} {\bibfnamefont {H.~G.}\ \bibnamefont
  {Evertz}}, \bibinfo {author} {\bibfnamefont {W.}~\bibnamefont {von~der
  Linden}}, \ and\ \bibinfo {author} {\bibfnamefont {M.}~\bibnamefont
  {Potthoff}}} (\bibinfo {year} {2004}),\ \href {\doibase
  10.1103/PhysRevB.70.235107} {\bibfield  {journal} {\bibinfo  {journal} {Phys.
  Rev. B}\ }\textbf {\bibinfo {volume} {70}},\ \bibinfo {pages}
  {235107}}\BibitemShut {NoStop}%
\bibitem [{\citenamefont {Anderson}(1958)}]{Anderson1958}%
  \BibitemOpen
  \bibfield  {author} {\bibinfo {author} {\bibnamefont {Anderson},
  \bibfnamefont {P.~W.}}} (\bibinfo {year} {1958}),\ \href {\doibase
  10.1103/PhysRev.109.1492} {\bibfield  {journal} {\bibinfo  {journal} {Phys.
  Rev.}\ }\textbf {\bibinfo {volume} {109}}~(\bibinfo {number} {5}),\ \bibinfo
  {pages} {1492}}\BibitemShut {NoStop}%
\bibitem [{\citenamefont {Anderson}(1997)}]{Anderson1997}%
  \BibitemOpen
  \bibfield  {author} {\bibinfo {author} {\bibnamefont {Anderson},
  \bibfnamefont {P.~W.}}} (\bibinfo {year} {1997}),\ \href@noop {} {\emph
  {\bibinfo {title} {The Theory of Superconductivity in the High-T$_c$
  Cuprates}}}\ (\bibinfo  {publisher} {Princeton University Press, Princeton,
  NJ})\BibitemShut {NoStop}%
\bibitem [{\citenamefont {Antipov}(2013)}]{aeantipovgftools}%
  \BibitemOpen
  \bibfield  {author} {\bibinfo {author} {\bibnamefont {Antipov}, \bibfnamefont
  {A.~A.}}} (\bibinfo {year} {2013}),\ \href
  {http://aeantipov.github.io/gftools} {\enquote {\bibinfo {title} {Gftools},}\
  }\BibitemShut {NoStop}%
\bibitem [{\citenamefont {Antipov}\ \emph {et~al.}(2014)\citenamefont
  {Antipov}, \citenamefont {Gull},\ and\ \citenamefont
  {Kirchner}}]{Antipov2014}%
  \BibitemOpen
  \bibfield  {author} {\bibinfo {author} {\bibnamefont {Antipov}, \bibfnamefont
  {A.~E.}}, \bibinfo {author} {\bibfnamefont {E.}~\bibnamefont {Gull}}, \ and\
  \bibinfo {author} {\bibfnamefont {S.}~\bibnamefont {Kirchner}}} (\bibinfo
  {year} {2014}),\ \href {\doibase 10.1103/PhysRevLett.112.226401} {\bibfield
  {journal} {\bibinfo  {journal} {Phys. Rev. Lett.}\ }\textbf {\bibinfo
  {volume} {112}},\ \bibinfo {pages} {226401}}\BibitemShut {NoStop}%
\bibitem [{\citenamefont {Antipov}\ \emph {et~al.}(2016)\citenamefont
  {Antipov}, \citenamefont {Javanmard}, \citenamefont {Ribeiro},\ and\
  \citenamefont {Kirchner}}]{Antipov2016}%
  \BibitemOpen
  \bibfield  {author} {\bibinfo {author} {\bibnamefont {Antipov}, \bibfnamefont
  {A.~E.}}, \bibinfo {author} {\bibfnamefont {Y.}~\bibnamefont {Javanmard}},
  \bibinfo {author} {\bibfnamefont {P.}~\bibnamefont {Ribeiro}}, \ and\
  \bibinfo {author} {\bibfnamefont {S.}~\bibnamefont {Kirchner}}} (\bibinfo
  {year} {2016}),\ \href {\doibase 10.1103/PhysRevLett.117.146601} {\bibfield
  {journal} {\bibinfo  {journal} {Phys. Rev. Lett.}\ }\textbf {\bibinfo
  {volume} {117}},\ \bibinfo {pages} {146601}}\BibitemShut {NoStop}%
\bibitem [{\citenamefont {Antipov}\ and\ \citenamefont
  {Krivenko}(2015)}]{Antipov2015a}%
  \BibitemOpen
  \bibfield  {author} {\bibinfo {author} {\bibnamefont {Antipov}, \bibfnamefont
  {A.~E.}}, \ and\ \bibinfo {author} {\bibfnamefont {I.}~\bibnamefont
  {Krivenko}}} (\bibinfo {year} {2015}),\ \href {\doibase 10.5281/zenodo.17900}
  {\enquote {\bibinfo {title} {pomerol: 1.1},}\ }\BibitemShut {NoStop}%
\bibitem [{\citenamefont {Antipov}\ \emph {et~al.}(2015)\citenamefont
  {Antipov}, \citenamefont {LeBlanc},\ and\ \citenamefont {Gull}}]{OPENDF}%
  \BibitemOpen
  \bibfield  {author} {\bibinfo {author} {\bibnamefont {Antipov}, \bibfnamefont
  {A.~E.}}, \bibinfo {author} {\bibfnamefont {J.~P.}\ \bibnamefont {LeBlanc}},
  \ and\ \bibinfo {author} {\bibfnamefont {E.}~\bibnamefont {Gull}}} (\bibinfo
  {year} {2015}),\ \href {\doibase 10.1016/j.phpro.2015.07.107} {\bibfield
  {journal} {\bibinfo  {journal} {Physics Procedia}\ }\textbf {\bibinfo
  {volume} {68}},\ \bibinfo {pages} {43}}\BibitemShut {NoStop}%
\bibitem [{\citenamefont {Antipov}\ \emph {et~al.}(2011)\citenamefont
  {Antipov}, \citenamefont {Rubtsov}, \citenamefont {Katsnelson},\ and\
  \citenamefont {Lichtenstein}}]{Antipov2011}%
  \BibitemOpen
  \bibfield  {author} {\bibinfo {author} {\bibnamefont {Antipov}, \bibfnamefont
  {A.~E.}}, \bibinfo {author} {\bibfnamefont {A.~N.}\ \bibnamefont {Rubtsov}},
  \bibinfo {author} {\bibfnamefont {M.~I.}\ \bibnamefont {Katsnelson}}, \ and\
  \bibinfo {author} {\bibfnamefont {A.~I.}\ \bibnamefont {Lichtenstein}}}
  (\bibinfo {year} {2011}),\ \href {\doibase 10.1103/PhysRevB.83.115126}
  {\bibfield  {journal} {\bibinfo  {journal} {Phys. Rev. B}\ }\textbf {\bibinfo
  {volume} {83}},\ \bibinfo {pages} {115126}}\BibitemShut {NoStop}%
\bibitem [{\citenamefont {Aoki}\ \emph {et~al.}(2014)\citenamefont {Aoki},
  \citenamefont {Tsuji}, \citenamefont {Eckstein}, \citenamefont {Kollar},
  \citenamefont {Oka},\ and\ \citenamefont {Werner}}]{Aoki2014}%
  \BibitemOpen
  \bibfield  {author} {\bibinfo {author} {\bibnamefont {Aoki}, \bibfnamefont
  {H.}}, \bibinfo {author} {\bibfnamefont {N.}~\bibnamefont {Tsuji}}, \bibinfo
  {author} {\bibfnamefont {M.}~\bibnamefont {Eckstein}}, \bibinfo {author}
  {\bibfnamefont {M.}~\bibnamefont {Kollar}}, \bibinfo {author} {\bibfnamefont
  {T.}~\bibnamefont {Oka}}, \ and\ \bibinfo {author} {\bibfnamefont
  {P.}~\bibnamefont {Werner}}} (\bibinfo {year} {2014}),\ \href
  {http://link.aps.org/doi/10.1103/RevModPhys.86.779} {\bibfield  {journal}
  {\bibinfo  {journal} {Rev. Mod. Phys.}\ }\textbf {\bibinfo {volume} {86}},\
  \bibinfo {pages} {779}}\BibitemShut {NoStop}%
\bibitem [{\citenamefont {Arita}\ \emph {et~al.}(2004)\citenamefont {Arita},
  \citenamefont {Onari}, \citenamefont {Kuroki},\ and\ \citenamefont
  {Aoki}}]{Arita2004}%
  \BibitemOpen
  \bibfield  {author} {\bibinfo {author} {\bibnamefont {Arita}, \bibfnamefont
  {R.}}, \bibinfo {author} {\bibfnamefont {S.}~\bibnamefont {Onari}}, \bibinfo
  {author} {\bibfnamefont {K.}~\bibnamefont {Kuroki}}, \ and\ \bibinfo {author}
  {\bibfnamefont {H.}~\bibnamefont {Aoki}}} (\bibinfo {year} {2004}),\ \href
  {\doibase 10.1103/PhysRevLett.92.247006} {\bibfield  {journal} {\bibinfo
  {journal} {Phys. Rev. Lett.}\ }\textbf {\bibinfo {volume} {92}},\ \bibinfo
  {pages} {247006}}\BibitemShut {NoStop}%
\bibitem [{\citenamefont {Aryasetiawan}\ and\ \citenamefont
  {Gunnarsson}(1998)}]{Aryasetiawan1998}%
  \BibitemOpen
  \bibfield  {author} {\bibinfo {author} {\bibnamefont {Aryasetiawan},
  \bibfnamefont {F.}}, \ and\ \bibinfo {author} {\bibfnamefont
  {O.}~\bibnamefont {Gunnarsson}}} (\bibinfo {year} {1998}),\ \href
  {http://stacks.iop.org/0034-4885/61/i=3/a=002} {\bibfield  {journal}
  {\bibinfo  {journal} {Reports on Progress in Physics}\ }\textbf {\bibinfo
  {volume} {61}}~(\bibinfo {number} {3}),\ \bibinfo {pages} {237}}\BibitemShut
  {NoStop}%
\bibitem [{\citenamefont {Aryasetiawan}\ \emph {et~al.}(2004)\citenamefont
  {Aryasetiawan}, \citenamefont {Imada}, \citenamefont {Georges}, \citenamefont
  {Kotliar}, \citenamefont {Biermann},\ and\ \citenamefont
  {Lichtenstein}}]{Aryasetiawan2004}%
  \BibitemOpen
  \bibfield  {author} {\bibinfo {author} {\bibnamefont {Aryasetiawan},
  \bibfnamefont {F.}}, \bibinfo {author} {\bibfnamefont {M.}~\bibnamefont
  {Imada}}, \bibinfo {author} {\bibfnamefont {A.}~\bibnamefont {Georges}},
  \bibinfo {author} {\bibfnamefont {G.}~\bibnamefont {Kotliar}}, \bibinfo
  {author} {\bibfnamefont {S.}~\bibnamefont {Biermann}}, \ and\ \bibinfo
  {author} {\bibfnamefont {A.~I.}\ \bibnamefont {Lichtenstein}}} (\bibinfo
  {year} {2004}),\ \href {\doibase 10.1103/PhysRevB.70.195104} {\bibfield
  {journal} {\bibinfo  {journal} {Phys. Rev. B}\ }\textbf {\bibinfo {volume}
  {70}},\ \bibinfo {pages} {195104}}\BibitemShut {NoStop}%
\bibitem [{\citenamefont {Assaad}(1999)}]{Assaad1999}%
  \BibitemOpen
  \bibfield  {author} {\bibinfo {author} {\bibnamefont {Assaad}, \bibfnamefont
  {F.~F.}}} (\bibinfo {year} {1999}),\ \href {\doibase
  10.1103/PhysRevLett.83.796} {\bibfield  {journal} {\bibinfo  {journal} {Phys.
  Rev. Lett.}\ }\textbf {\bibinfo {volume} {83}},\ \bibinfo {pages}
  {796}}\BibitemShut {NoStop}%
\bibitem [{\citenamefont {Avella}\ \emph {et~al.}(2001)\citenamefont {Avella},
  \citenamefont {Mancini},\ and\ \citenamefont {M\"unzner}}]{Avella2001}%
  \BibitemOpen
  \bibfield  {author} {\bibinfo {author} {\bibnamefont {Avella}, \bibfnamefont
  {A.}}, \bibinfo {author} {\bibfnamefont {F.}~\bibnamefont {Mancini}}, \ and\
  \bibinfo {author} {\bibfnamefont {R.}~\bibnamefont {M\"unzner}}} (\bibinfo
  {year} {2001}),\ \href {\doibase 10.1103/PhysRevB.63.245117} {\bibfield
  {journal} {\bibinfo  {journal} {Phys. Rev. B}\ }\textbf {\bibinfo {volume}
  {63}},\ \bibinfo {pages} {245117}}\BibitemShut {NoStop}%
\bibitem [{\citenamefont {Ayral}\ \emph {et~al.}(2013)\citenamefont {Ayral},
  \citenamefont {Biermann},\ and\ \citenamefont {Werner}}]{Ayral2013}%
  \BibitemOpen
  \bibfield  {author} {\bibinfo {author} {\bibnamefont {Ayral}, \bibfnamefont
  {T.}}, \bibinfo {author} {\bibfnamefont {S.}~\bibnamefont {Biermann}}, \ and\
  \bibinfo {author} {\bibfnamefont {P.}~\bibnamefont {Werner}}} (\bibinfo
  {year} {2013}),\ \href {\doibase 10.1103/PhysRevB.87.125149} {\bibfield
  {journal} {\bibinfo  {journal} {Phys. Rev. B}\ }\textbf {\bibinfo {volume}
  {87}},\ \bibinfo {pages} {125149}}\BibitemShut {NoStop}%
\bibitem [{\citenamefont {Ayral}\ \emph
  {et~al.}(2017{\natexlab{a}})\citenamefont {Ayral}, \citenamefont {Biermann},
  \citenamefont {Werner},\ and\ \citenamefont {Boehnke}}]{Ayral2017a}%
  \BibitemOpen
  \bibfield  {author} {\bibinfo {author} {\bibnamefont {Ayral}, \bibfnamefont
  {T.}}, \bibinfo {author} {\bibfnamefont {S.}~\bibnamefont {Biermann}},
  \bibinfo {author} {\bibfnamefont {P.}~\bibnamefont {Werner}}, \ and\ \bibinfo
  {author} {\bibfnamefont {L.}~\bibnamefont {Boehnke}}} (\bibinfo {year}
  {2017}{\natexlab{a}}),\ \href {\doibase 10.1103/PhysRevB.95.245130}
  {\bibfield  {journal} {\bibinfo  {journal} {Phys. Rev. B}\ }\textbf {\bibinfo
  {volume} {95}},\ \bibinfo {pages} {245130}}\BibitemShut {NoStop}%
\bibitem [{\citenamefont {Ayral}\ \emph
  {et~al.}(2017{\natexlab{b}})\citenamefont {Ayral}, \citenamefont {Vu\ifmmode
  \check{c}\else \v{c}\fi{}i\ifmmode \check{c}\else
  \v{c}\fi{}evi\ifmmode~\acute{c}\else \'{c}\fi{}},\ and\ \citenamefont
  {Parcollet}}]{Ayral2017}%
  \BibitemOpen
  \bibfield  {author} {\bibinfo {author} {\bibnamefont {Ayral}, \bibfnamefont
  {T.}}, \bibinfo {author} {\bibfnamefont {J.}~\bibnamefont {Vu\ifmmode
  \check{c}\else \v{c}\fi{}i\ifmmode \check{c}\else
  \v{c}\fi{}evi\ifmmode~\acute{c}\else \'{c}\fi{}}}, \ and\ \bibinfo {author}
  {\bibfnamefont {O.}~\bibnamefont {Parcollet}}} (\bibinfo {year}
  {2017}{\natexlab{b}}),\ \href {\doibase 10.1103/PhysRevLett.119.166401}
  {\bibfield  {journal} {\bibinfo  {journal} {Phys. Rev. Lett.}\ }\textbf
  {\bibinfo {volume} {119}},\ \bibinfo {pages} {166401}}\BibitemShut {NoStop}%
\bibitem [{\citenamefont {Ayral}\ and\ \citenamefont
  {Parcollet}(2015)}]{Ayral2015}%
  \BibitemOpen
  \bibfield  {author} {\bibinfo {author} {\bibnamefont {Ayral}, \bibfnamefont
  {T.}}, \ and\ \bibinfo {author} {\bibfnamefont {O.}~\bibnamefont
  {Parcollet}}} (\bibinfo {year} {2015}),\ \href
  {http://link.aps.org/doi/10.1103/PhysRevB.92.115109} {\bibfield  {journal}
  {\bibinfo  {journal} {Phys Rev. B}\ }\textbf {\bibinfo {volume} {92}},\
  \bibinfo {pages} {115109}}\BibitemShut {NoStop}%
\bibitem [{\citenamefont {Ayral}\ and\ \citenamefont
  {Parcollet}(2016{\natexlab{a}})}]{Ayral2016}%
  \BibitemOpen
  \bibfield  {author} {\bibinfo {author} {\bibnamefont {Ayral}, \bibfnamefont
  {T.}}, \ and\ \bibinfo {author} {\bibfnamefont {O.}~\bibnamefont
  {Parcollet}}} (\bibinfo {year} {2016}{\natexlab{a}}),\ \href {\doibase
  10.1103/PhysRevB.94.075159} {\bibfield  {journal} {\bibinfo  {journal} {Phys.
  Rev. B}\ }\textbf {\bibinfo {volume} {94}},\ \bibinfo {pages}
  {075159}}\BibitemShut {NoStop}%
\bibitem [{\citenamefont {Ayral}\ and\ \citenamefont
  {Parcollet}(2016{\natexlab{b}})}]{Ayral2016a}%
  \BibitemOpen
  \bibfield  {author} {\bibinfo {author} {\bibnamefont {Ayral}, \bibfnamefont
  {T.}}, \ and\ \bibinfo {author} {\bibfnamefont {O.}~\bibnamefont
  {Parcollet}}} (\bibinfo {year} {2016}{\natexlab{b}}),\ \href {\doibase
  10.1103/PhysRevB.93.235124} {\bibfield  {journal} {\bibinfo  {journal} {Phys.
  Rev. B}\ }\textbf {\bibinfo {volume} {93}},\ \bibinfo {pages}
  {235124}}\BibitemShut {NoStop}%
\bibitem [{\citenamefont {Ayral}\ \emph {et~al.}(2012)\citenamefont {Ayral},
  \citenamefont {Werner},\ and\ \citenamefont {Biermann}}]{Ayral2012}%
  \BibitemOpen
  \bibfield  {author} {\bibinfo {author} {\bibnamefont {Ayral}, \bibfnamefont
  {T.}}, \bibinfo {author} {\bibfnamefont {P.}~\bibnamefont {Werner}}, \ and\
  \bibinfo {author} {\bibfnamefont {S.}~\bibnamefont {Biermann}}} (\bibinfo
  {year} {2012}),\ \href {\doibase 10.1103/PhysRevLett.109.226401} {\bibfield
  {journal} {\bibinfo  {journal} {Phys. Rev. Lett.}\ }\textbf {\bibinfo
  {volume} {109}},\ \bibinfo {pages} {226401}}\BibitemShut {NoStop}%
\bibitem [{\citenamefont {Baranov}(2008)}]{Baranov2008}%
  \BibitemOpen
  \bibfield  {author} {\bibinfo {author} {\bibnamefont {Baranov}, \bibfnamefont
  {M.}}} (\bibinfo {year} {2008}),\ \href {\doibase
  http://dx.doi.org/10.1016/j.physrep.2008.04.007} {\bibfield  {journal}
  {\bibinfo  {journal} {Physics Reports}\ }\textbf {\bibinfo {volume}
  {464}}~(\bibinfo {number} {3}),\ \bibinfo {pages} {71 }}\BibitemShut
  {NoStop}%
\bibitem [{\citenamefont {Baranov}\ \emph {et~al.}(2012)\citenamefont
  {Baranov}, \citenamefont {Dalmonte}, \citenamefont {Pupillo},\ and\
  \citenamefont {Zoller}}]{Zoller2012}%
  \BibitemOpen
  \bibfield  {author} {\bibinfo {author} {\bibnamefont {Baranov}, \bibfnamefont
  {M.~A.}}, \bibinfo {author} {\bibfnamefont {M.}~\bibnamefont {Dalmonte}},
  \bibinfo {author} {\bibfnamefont {G.}~\bibnamefont {Pupillo}}, \ and\
  \bibinfo {author} {\bibfnamefont {P.}~\bibnamefont {Zoller}}} (\bibinfo
  {year} {2012}),\ \href {\doibase 10.1021/cr2003568} {\bibfield  {journal}
  {\bibinfo  {journal} {Chemical Reviews}\ }\textbf {\bibinfo {volume}
  {112}}~(\bibinfo {number} {9}),\ \bibinfo {pages} {5012}}\BibitemShut
  {NoStop}%
\bibitem [{\citenamefont {Bauer}\ \emph {et~al.}(2011)\citenamefont {Bauer},
  \citenamefont {Carr}, \citenamefont {Evertz}, \citenamefont {Feiguin},
  \citenamefont {Freire}, \citenamefont {Fuchs}, \citenamefont {Gamper},
  \citenamefont {Gukelberger}, \citenamefont {Gull}, \citenamefont {Guertler},
  \citenamefont {Hehn}, \citenamefont {Igarashi}, \citenamefont {Isakov},
  \citenamefont {Koop}, \citenamefont {Ma}, \citenamefont {Mates},
  \citenamefont {Matsuo}, \citenamefont {Parcollet}, \citenamefont
  {Pawłowski}, \citenamefont {Picon}, \citenamefont {Pollet}, \citenamefont
  {Santos}, \citenamefont {Scarola}, \citenamefont {Schollwöck}, \citenamefont
  {Silva}, \citenamefont {Surer}, \citenamefont {Todo}, \citenamefont {Trebst},
  \citenamefont {Troyer}, \citenamefont {Wall}, \citenamefont {Werner},\ and\
  \citenamefont {Wessel}}]{ALPS2}%
  \BibitemOpen
  \bibfield  {author} {\bibinfo {author} {\bibnamefont {Bauer}, \bibfnamefont
  {B.}}, \bibinfo {author} {\bibfnamefont {L.~D.}\ \bibnamefont {Carr}},
  \bibinfo {author} {\bibfnamefont {H.~G.}\ \bibnamefont {Evertz}}, \bibinfo
  {author} {\bibfnamefont {A.}~\bibnamefont {Feiguin}}, \bibinfo {author}
  {\bibfnamefont {J.}~\bibnamefont {Freire}}, \bibinfo {author} {\bibfnamefont
  {S.}~\bibnamefont {Fuchs}}, \bibinfo {author} {\bibfnamefont
  {L.}~\bibnamefont {Gamper}}, \bibinfo {author} {\bibfnamefont
  {J.}~\bibnamefont {Gukelberger}}, \bibinfo {author} {\bibfnamefont
  {E.}~\bibnamefont {Gull}}, \bibinfo {author} {\bibfnamefont {S.}~\bibnamefont
  {Guertler}}, \bibinfo {author} {\bibfnamefont {A.}~\bibnamefont {Hehn}},
  \bibinfo {author} {\bibfnamefont {R.}~\bibnamefont {Igarashi}}, \bibinfo
  {author} {\bibfnamefont {S.~V.}\ \bibnamefont {Isakov}}, \bibinfo {author}
  {\bibfnamefont {D.}~\bibnamefont {Koop}}, \bibinfo {author} {\bibfnamefont
  {P.~N.}\ \bibnamefont {Ma}}, \bibinfo {author} {\bibfnamefont
  {P.}~\bibnamefont {Mates}}, \bibinfo {author} {\bibfnamefont
  {H.}~\bibnamefont {Matsuo}}, \bibinfo {author} {\bibfnamefont
  {O.}~\bibnamefont {Parcollet}}, \bibinfo {author} {\bibfnamefont
  {G.}~\bibnamefont {Pawłowski}}, \bibinfo {author} {\bibfnamefont {J.~D.}\
  \bibnamefont {Picon}}, \bibinfo {author} {\bibfnamefont {L.}~\bibnamefont
  {Pollet}}, \bibinfo {author} {\bibfnamefont {E.}~\bibnamefont {Santos}},
  \bibinfo {author} {\bibfnamefont {V.~W.}\ \bibnamefont {Scarola}}, \bibinfo
  {author} {\bibfnamefont {U.}~\bibnamefont {Schollwöck}}, \bibinfo {author}
  {\bibfnamefont {C.}~\bibnamefont {Silva}}, \bibinfo {author} {\bibfnamefont
  {B.}~\bibnamefont {Surer}}, \bibinfo {author} {\bibfnamefont
  {S.}~\bibnamefont {Todo}}, \bibinfo {author} {\bibfnamefont {S.}~\bibnamefont
  {Trebst}}, \bibinfo {author} {\bibfnamefont {M.}~\bibnamefont {Troyer}},
  \bibinfo {author} {\bibfnamefont {M.~L.}\ \bibnamefont {Wall}}, \bibinfo
  {author} {\bibfnamefont {P.}~\bibnamefont {Werner}}, \ and\ \bibinfo {author}
  {\bibfnamefont {S.}~\bibnamefont {Wessel}}} (\bibinfo {year} {2011}),\ \href
  {http://stacks.iop.org/1742-5468/2011/i=05/a=P05001} {\bibfield  {journal}
  {\bibinfo  {journal} {Journal of Statistical Mechanics: Theory and
  Experiment}\ }\textbf {\bibinfo {volume} {2011}}~(\bibinfo {number} {05}),\
  \bibinfo {pages} {P05001}}\BibitemShut {NoStop}%
\bibitem [{\citenamefont {Baym}\ and\ \citenamefont
  {Kadanoff}(1961)}]{BaymKadanoff}%
  \BibitemOpen
  \bibfield  {author} {\bibinfo {author} {\bibnamefont {Baym}, \bibfnamefont
  {G.}}, \ and\ \bibinfo {author} {\bibfnamefont {L.~P.}\ \bibnamefont
  {Kadanoff}}} (\bibinfo {year} {1961}),\ \href {\doibase
  10.1103/PhysRev.124.287} {\bibfield  {journal} {\bibinfo  {journal} {Phys.
  Rev.}\ }\textbf {\bibinfo {volume} {124}},\ \bibinfo {pages}
  {287}}\BibitemShut {NoStop}%
\bibitem [{\citenamefont {Bednorz}\ and\ \citenamefont
  {M\"uller}(1986)}]{Bednorz1986}%
  \BibitemOpen
  \bibfield  {author} {\bibinfo {author} {\bibnamefont {Bednorz}, \bibfnamefont
  {J.~G.}}, \ and\ \bibinfo {author} {\bibfnamefont {K.~A.}\ \bibnamefont
  {M\"uller}}} (\bibinfo {year} {1986}),\ \href@noop {} {\bibfield  {journal}
  {\bibinfo  {journal} {Zeitschrift f\"ur Physik B Condensed Matter}\ }\textbf
  {\bibinfo {volume} {64}},\ \bibinfo {pages} {189}}\BibitemShut {NoStop}%
\bibitem [{\citenamefont {Bergeron}\ and\ \citenamefont
  {Tremblay}(2016)}]{Bergeron2016}%
  \BibitemOpen
  \bibfield  {author} {\bibinfo {author} {\bibnamefont {Bergeron},
  \bibfnamefont {D.}}, \ and\ \bibinfo {author} {\bibfnamefont {A.-M.~S.}\
  \bibnamefont {Tremblay}}} (\bibinfo {year} {2016}),\ \href {\doibase
  10.1103/PhysRevE.94.023303} {\bibfield  {journal} {\bibinfo  {journal} {Phys.
  Rev. E}\ }\textbf {\bibinfo {volume} {94}},\ \bibinfo {pages}
  {023303}}\BibitemShut {NoStop}%
\bibitem [{\citenamefont {Bickers}(2004)}]{Bickers2004}%
  \BibitemOpen
  \bibfield  {author} {\bibinfo {author} {\bibnamefont {Bickers}, \bibfnamefont
  {N.~E.}}} (\bibinfo {year} {2004}),\ \enquote {\bibinfo {title} {Theoretical
  methods for strongly correlated electrons},}\ Chap.~\bibinfo {chapter} {6}\
  (\bibinfo  {publisher} {Springer-Verlag New York Berlin Heidelbert})\ pp.\
  \bibinfo {pages} {237--296}\BibitemShut {NoStop}%
\bibitem [{\citenamefont {Bickers}\ and\ \citenamefont
  {White}(1991)}]{Bickers1991}%
  \BibitemOpen
  \bibfield  {author} {\bibinfo {author} {\bibnamefont {Bickers}, \bibfnamefont
  {N.~E.}}, \ and\ \bibinfo {author} {\bibfnamefont {S.~R.}\ \bibnamefont
  {White}}} (\bibinfo {year} {1991}),\ \href {\doibase
  10.1103/PhysRevB.43.8044} {\bibfield  {journal} {\bibinfo  {journal} {Phys.
  Rev. B}\ }\textbf {\bibinfo {volume} {43}},\ \bibinfo {pages}
  {8044}}\BibitemShut {NoStop}%
\bibitem [{\citenamefont {Biermann}\ \emph {et~al.}(2003)\citenamefont
  {Biermann}, \citenamefont {Aryasetiawan},\ and\ \citenamefont
  {Georges}}]{Biermann2003}%
  \BibitemOpen
  \bibfield  {author} {\bibinfo {author} {\bibnamefont {Biermann},
  \bibfnamefont {S.}}, \bibinfo {author} {\bibfnamefont {F.}~\bibnamefont
  {Aryasetiawan}}, \ and\ \bibinfo {author} {\bibfnamefont {A.}~\bibnamefont
  {Georges}}} (\bibinfo {year} {2003}),\ \href {\doibase
  10.1103/PhysRevLett.90.086402} {\bibfield  {journal} {\bibinfo  {journal}
  {Phys. Rev. Lett.}\ }\textbf {\bibinfo {volume} {90}},\ \bibinfo {pages}
  {086402}}\BibitemShut {NoStop}%
\bibitem [{\citenamefont {Biermann}\ \emph {et~al.}(2005)\citenamefont
  {Biermann}, \citenamefont {Poteryaev}, \citenamefont {I.Lichtenstein},\ and\
  \citenamefont {Georges}}]{Biermann2005}%
  \BibitemOpen
  \bibfield  {author} {\bibinfo {author} {\bibnamefont {Biermann},
  \bibfnamefont {S.}}, \bibinfo {author} {\bibfnamefont {A.}~\bibnamefont
  {Poteryaev}}, \bibinfo {author} {\bibfnamefont {A.}~\bibnamefont
  {I.Lichtenstein}}, \ and\ \bibinfo {author} {\bibfnamefont {A.}~\bibnamefont
  {Georges}}} (\bibinfo {year} {2005}),\ \href@noop {} {\bibfield  {journal}
  {\bibinfo  {journal} {Phys. Rev. Lett.}\ }\textbf {\bibinfo {volume} {94}},\
  \bibinfo {pages} {026404}}\BibitemShut {NoStop}%
\bibitem [{\citenamefont {Bloch}\ \emph {et~al.}(2012)\citenamefont {Bloch},
  \citenamefont {Dalibard},\ and\ \citenamefont {Nascimb\`{e}ne}}]{Bloch2012}%
  \BibitemOpen
  \bibfield  {author} {\bibinfo {author} {\bibnamefont {Bloch}, \bibfnamefont
  {I.}}, \bibinfo {author} {\bibfnamefont {J.}~\bibnamefont {Dalibard}}, \ and\
  \bibinfo {author} {\bibfnamefont {S.}~\bibnamefont {Nascimb\`{e}ne}}}
  (\bibinfo {year} {2012}),\ \href {\doibase 10.1038/nphys2259} {\bibfield
  {journal} {\bibinfo  {journal} {Nat. Phys.}\ }\textbf {\bibinfo {volume}
  {8}},\ \bibinfo {pages} {267}}\BibitemShut {NoStop}%
\bibitem [{\citenamefont {Bloch}\ \emph {et~al.}(2008)\citenamefont {Bloch},
  \citenamefont {Dalibard},\ and\ \citenamefont {Zwerger}}]{Bloch2008}%
  \BibitemOpen
  \bibfield  {author} {\bibinfo {author} {\bibnamefont {Bloch}, \bibfnamefont
  {I.}}, \bibinfo {author} {\bibfnamefont {J.}~\bibnamefont {Dalibard}}, \ and\
  \bibinfo {author} {\bibfnamefont {W.}~\bibnamefont {Zwerger}}} (\bibinfo
  {year} {2008}),\ \href {\doibase 10.1103/RevModPhys.80.885} {\bibfield
  {journal} {\bibinfo  {journal} {Rev. Mod. Phys.}\ }\textbf {\bibinfo {volume}
  {80}},\ \bibinfo {pages} {885}}\BibitemShut {NoStop}%
\bibitem [{\citenamefont {Bl\"umer}(2002)}]{Blumer2002}%
  \BibitemOpen
  \bibfield  {author} {\bibinfo {author} {\bibnamefont {Bl\"umer},
  \bibfnamefont {N.}}} (\bibinfo {year} {2002}),\ \emph {\bibinfo {title}
  {Mott-Hubbard Metal-Insulator Transition and Optical Conductivity in High
  Dimensions}},\ \href@noop {} {Ph.D. thesis}\ (\bibinfo  {school}
  {Universit\"at Augsburg})\BibitemShut {NoStop}%
\bibitem [{\citenamefont {Boehnke}\ \emph {et~al.}(2011)\citenamefont
  {Boehnke}, \citenamefont {Hafermann}, \citenamefont {Ferrero}, \citenamefont
  {Lechermann},\ and\ \citenamefont {Parcollet}}]{Boehnke2011}%
  \BibitemOpen
  \bibfield  {author} {\bibinfo {author} {\bibnamefont {Boehnke}, \bibfnamefont
  {L.}}, \bibinfo {author} {\bibfnamefont {H.}~\bibnamefont {Hafermann}},
  \bibinfo {author} {\bibfnamefont {M.}~\bibnamefont {Ferrero}}, \bibinfo
  {author} {\bibfnamefont {F.}~\bibnamefont {Lechermann}}, \ and\ \bibinfo
  {author} {\bibfnamefont {O.}~\bibnamefont {Parcollet}}} (\bibinfo {year}
  {2011}),\ \href {\doibase 10.1103/PhysRevB.84.075145} {\bibfield  {journal}
  {\bibinfo  {journal} {Phys. Rev. B}\ }\textbf {\bibinfo {volume} {84}},\
  \bibinfo {pages} {075145}}\BibitemShut {NoStop}%
\bibitem [{\citenamefont {Boehnke}\ and\ \citenamefont
  {Lechermann}(2012)}]{Boehnke2012}%
  \BibitemOpen
  \bibfield  {author} {\bibinfo {author} {\bibnamefont {Boehnke}, \bibfnamefont
  {L.}}, \ and\ \bibinfo {author} {\bibfnamefont {F.}~\bibnamefont
  {Lechermann}}} (\bibinfo {year} {2012}),\ \href {\doibase
  10.1103/PhysRevB.85.115128} {\bibfield  {journal} {\bibinfo  {journal} {Phys.
  Rev. B}\ }\textbf {\bibinfo {volume} {85}},\ \bibinfo {pages}
  {115128}}\BibitemShut {NoStop}%
\bibitem [{\citenamefont {Boehnke}\ \emph {et~al.}(2016)\citenamefont
  {Boehnke}, \citenamefont {Nilsson}, \citenamefont {Aryasetiawan},\ and\
  \citenamefont {Werner}}]{Boehnke2016}%
  \BibitemOpen
  \bibfield  {author} {\bibinfo {author} {\bibnamefont {Boehnke}, \bibfnamefont
  {L.}}, \bibinfo {author} {\bibfnamefont {F.}~\bibnamefont {Nilsson}},
  \bibinfo {author} {\bibfnamefont {F.}~\bibnamefont {Aryasetiawan}}, \ and\
  \bibinfo {author} {\bibfnamefont {P.}~\bibnamefont {Werner}}} (\bibinfo
  {year} {2016}),\ \href {\doibase 10.1103/PhysRevB.94.201106} {\bibfield
  {journal} {\bibinfo  {journal} {Phys. Rev. B}\ }\textbf {\bibinfo {volume}
  {94}},\ \bibinfo {pages} {201106}}\BibitemShut {NoStop}%
\bibitem [{\citenamefont {Boies}\ \emph {et~al.}(1995)\citenamefont {Boies},
  \citenamefont {Bourbonnais},\ and\ \citenamefont {Tremblay}}]{Boies1995}%
  \BibitemOpen
  \bibfield  {author} {\bibinfo {author} {\bibnamefont {Boies}, \bibfnamefont
  {D.}}, \bibinfo {author} {\bibfnamefont {C.}~\bibnamefont {Bourbonnais}}, \
  and\ \bibinfo {author} {\bibfnamefont {A.~M.~S.}\ \bibnamefont {Tremblay}}}
  (\bibinfo {year} {1995}),\ \href {\doibase 10.1103/PhysRevLett.74.968}
  {\bibfield  {journal} {\bibinfo  {journal} {Phys. Rev. Lett.}\ }\textbf
  {\bibinfo {volume} {74}},\ \bibinfo {pages} {968}}\BibitemShut {NoStop}%
\bibitem [{\citenamefont {Bolech}\ \emph {et~al.}(2003)\citenamefont {Bolech},
  \citenamefont {Kancharla},\ and\ \citenamefont {Kotliar}}]{Bolech03}%
  \BibitemOpen
  \bibfield  {author} {\bibinfo {author} {\bibnamefont {Bolech}, \bibfnamefont
  {C.~J.}}, \bibinfo {author} {\bibfnamefont {S.~S.}\ \bibnamefont
  {Kancharla}}, \ and\ \bibinfo {author} {\bibfnamefont {G.}~\bibnamefont
  {Kotliar}}} (\bibinfo {year} {2003}),\ \href {\doibase
  10.1103/PhysRevB.67.075110} {\bibfield  {journal} {\bibinfo  {journal} {Phys.
  Rev. B}\ }\textbf {\bibinfo {volume} {67}},\ \bibinfo {pages}
  {075110}}\BibitemShut {NoStop}%
\bibitem [{\citenamefont {Borejsza}\ and\ \citenamefont
  {Dupuis}(2004)}]{Borejsza2004}%
  \BibitemOpen
  \bibfield  {author} {\bibinfo {author} {\bibnamefont {Borejsza},
  \bibfnamefont {K.}}, \ and\ \bibinfo {author} {\bibfnamefont
  {N.}~\bibnamefont {Dupuis}}} (\bibinfo {year} {2004}),\ \href@noop {}
  {\bibfield  {journal} {\bibinfo  {journal} {Phys. Rev. B}\ }\textbf {\bibinfo
  {volume} {69}},\ \bibinfo {pages} {085119}}\BibitemShut {NoStop}%
\bibitem [{\citenamefont {{Borejsza, K.}}\ and\ \citenamefont {{Dupuis,
  N.}}(2003)}]{Borejsza2003}%
  \BibitemOpen
  \bibfield  {author} {\bibinfo {author} {\bibnamefont {{Borejsza, K.}},}, \
  and\ \bibinfo {author} {\bibnamefont {{Dupuis, N.}}}} (\bibinfo {year}
  {2003}),\ \href {\doibase 10.1209/epl/i2003-00584-7} {\bibfield  {journal}
  {\bibinfo  {journal} {Europhys. Lett.}\ }\textbf {\bibinfo {volume}
  {63}}~(\bibinfo {number} {5}),\ \bibinfo {pages} {722}}\BibitemShut {NoStop}%
\bibitem [{\citenamefont {Brandt}\ and\ \citenamefont
  {Mielsch}(1989)}]{Brandt1989}%
  \BibitemOpen
  \bibfield  {author} {\bibinfo {author} {\bibnamefont {Brandt}, \bibfnamefont
  {U.}}, \ and\ \bibinfo {author} {\bibfnamefont {C.}~\bibnamefont {Mielsch}}}
  (\bibinfo {year} {1989}),\ \href {\doibase 10.1007/BF01321824} {\bibfield
  {journal} {\bibinfo  {journal} {Zeitschrift f{\"{u}}r Phys. B Condens.
  Matter}\ }\textbf {\bibinfo {volume} {75}}~(\bibinfo {number} {3}),\ \bibinfo
  {pages} {365}}\BibitemShut {NoStop}%
\bibitem [{\citenamefont {Brandt}\ and\ \citenamefont
  {Mielsch}(1990)}]{Brandt1990}%
  \BibitemOpen
  \bibfield  {author} {\bibinfo {author} {\bibnamefont {Brandt}, \bibfnamefont
  {U.}}, \ and\ \bibinfo {author} {\bibfnamefont {C.}~\bibnamefont {Mielsch}}}
  (\bibinfo {year} {1990}),\ \href {\doibase 10.1007/BF01406598} {\bibfield
  {journal} {\bibinfo  {journal} {Zeitschrift f{\"{u}}r Phys. B Condens.
  Matter}\ }\textbf {\bibinfo {volume} {79}}~(\bibinfo {number} {2}),\ \bibinfo
  {pages} {295}}\BibitemShut {NoStop}%
\bibitem [{\citenamefont {Brandt}\ and\ \citenamefont
  {Mielsch}(1991)}]{Brandt1991}%
  \BibitemOpen
  \bibfield  {author} {\bibinfo {author} {\bibnamefont {Brandt}, \bibfnamefont
  {U.}}, \ and\ \bibinfo {author} {\bibfnamefont {C.}~\bibnamefont {Mielsch}}}
  (\bibinfo {year} {1991}),\ \href {\doibase 10.1007/BF01313984} {\bibfield
  {journal} {\bibinfo  {journal} {Zeitschrift f{\"{u}}r Phys. B Condens.
  Matter}\ }\textbf {\bibinfo {volume} {82}}~(\bibinfo {number} {1}),\ \bibinfo
  {pages} {37}}\BibitemShut {NoStop}%
\bibitem [{\citenamefont {Brandt}\ and\ \citenamefont
  {Schmidt}(1986)}]{Brandt1986}%
  \BibitemOpen
  \bibfield  {author} {\bibinfo {author} {\bibnamefont {Brandt}, \bibfnamefont
  {U.}}, \ and\ \bibinfo {author} {\bibfnamefont {R.}~\bibnamefont {Schmidt}}}
  (\bibinfo {year} {1986}),\ \href {\doibase 10.1007/BF01312577} {\bibfield
  {journal} {\bibinfo  {journal} {Zeitschrift f{\"u}r Phys. B Condens. Matter}\
  }\textbf {\bibinfo {volume} {63}}~(\bibinfo {number} {1}),\ \bibinfo {pages}
  {45}}\BibitemShut {NoStop}%
\bibitem [{\citenamefont {Brener}\ \emph {et~al.}(2008)\citenamefont {Brener},
  \citenamefont {Hafermann}, \citenamefont {Rubtsov}, \citenamefont
  {Katsnelson},\ and\ \citenamefont {Lichtenstein}}]{Brener08}%
  \BibitemOpen
  \bibfield  {author} {\bibinfo {author} {\bibnamefont {Brener}, \bibfnamefont
  {S.}}, \bibinfo {author} {\bibfnamefont {H.}~\bibnamefont {Hafermann}},
  \bibinfo {author} {\bibfnamefont {A.~N.}\ \bibnamefont {Rubtsov}}, \bibinfo
  {author} {\bibfnamefont {M.~I.}\ \bibnamefont {Katsnelson}}, \ and\ \bibinfo
  {author} {\bibfnamefont {A.~I.}\ \bibnamefont {Lichtenstein}}} (\bibinfo
  {year} {2008}),\ \href {\doibase 10.1103/PhysRevB.77.195105} {\bibfield
  {journal} {\bibinfo  {journal} {Phys. Rev. B}\ }\textbf {\bibinfo {volume}
  {77}}~(\bibinfo {number} {19}),\ \bibinfo {eid} {195105}}\BibitemShut
  {NoStop}%
\bibitem [{\citenamefont {Bulla}\ \emph {et~al.}(2008)\citenamefont {Bulla},
  \citenamefont {Costi},\ and\ \citenamefont {Pruschke}}]{Bulla2008}%
  \BibitemOpen
  \bibfield  {author} {\bibinfo {author} {\bibnamefont {Bulla}, \bibfnamefont
  {R.}}, \bibinfo {author} {\bibfnamefont {T.}~\bibnamefont {Costi}}, \ and\
  \bibinfo {author} {\bibfnamefont {T.}~\bibnamefont {Pruschke}}} (\bibinfo
  {year} {2008}),\ \href {\doibase 10.1103/RevModPhys.80.395} {\bibfield
  {journal} {\bibinfo  {journal} {Rev. Mod. Phys.}\ }\textbf {\bibinfo {volume}
  {80}},\ \bibinfo {pages} {395}}\BibitemShut {NoStop}%
\bibitem [{\citenamefont {Bulla}\ \emph {et~al.}(1998)\citenamefont {Bulla},
  \citenamefont {Hewson},\ and\ \citenamefont {Pruschke}}]{Bulla1998}%
  \BibitemOpen
  \bibfield  {author} {\bibinfo {author} {\bibnamefont {Bulla}, \bibfnamefont
  {R.}}, \bibinfo {author} {\bibfnamefont {A.~C.}\ \bibnamefont {Hewson}}, \
  and\ \bibinfo {author} {\bibfnamefont {T.}~\bibnamefont {Pruschke}}}
  (\bibinfo {year} {1998}),\ \href
  {http://stacks.iop.org/0953-8984/10/i=37/a=021} {\bibfield  {journal}
  {\bibinfo  {journal} {Journal of Physics: Condensed Matter}\ }\textbf
  {\bibinfo {volume} {10}}~(\bibinfo {number} {37}),\ \bibinfo {pages}
  {8365}}\BibitemShut {NoStop}%
\bibitem [{\citenamefont {Byczuk}\ \emph {et~al.}(2005)\citenamefont {Byczuk},
  \citenamefont {Hofstetter},\ and\ \citenamefont {Vollhardt}}]{Byczuk2005}%
  \BibitemOpen
  \bibfield  {author} {\bibinfo {author} {\bibnamefont {Byczuk}, \bibfnamefont
  {K.}}, \bibinfo {author} {\bibfnamefont {W.}~\bibnamefont {Hofstetter}}, \
  and\ \bibinfo {author} {\bibfnamefont {D.}~\bibnamefont {Vollhardt}}}
  (\bibinfo {year} {2005}),\ \href {\doibase 10.1103/PhysRevLett.94.056404}
  {\bibfield  {journal} {\bibinfo  {journal} {Phys. Rev. Lett.}\ }\textbf
  {\bibinfo {volume} {94}},\ \bibinfo {pages} {056404}}\BibitemShut {NoStop}%
\bibitem [{\citenamefont {Byczuk}\ \emph {et~al.}(2009)\citenamefont {Byczuk},
  \citenamefont {Hofstetter}, \citenamefont {Yu},\ and\ \citenamefont
  {Vollhardt}}]{Byczuk2009}%
  \BibitemOpen
  \bibfield  {author} {\bibinfo {author} {\bibnamefont {Byczuk}, \bibfnamefont
  {K.}}, \bibinfo {author} {\bibfnamefont {W.}~\bibnamefont {Hofstetter}},
  \bibinfo {author} {\bibfnamefont {U.}~\bibnamefont {Yu}}, \ and\ \bibinfo
  {author} {\bibfnamefont {D.}~\bibnamefont {Vollhardt}}} (\bibinfo {year}
  {2009}),\ \href {\doibase 10.1140/epjst/e2010-01215-2} {\bibfield  {journal}
  {\bibinfo  {journal} {Eur. Phys. J. Spec. Top.}\ }\textbf {\bibinfo {volume}
  {180}}~(\bibinfo {number} {1}),\ \bibinfo {pages} {135}}\BibitemShut
  {NoStop}%
\bibitem [{\citenamefont {Capone}\ and\ \citenamefont
  {Kotliar}(2006)}]{Capone2006}%
  \BibitemOpen
  \bibfield  {author} {\bibinfo {author} {\bibnamefont {Capone}, \bibfnamefont
  {M.}}, \ and\ \bibinfo {author} {\bibfnamefont {G.}~\bibnamefont {Kotliar}}}
  (\bibinfo {year} {2006}),\ \href {\doibase 10.1103/PhysRevB.74.054513}
  {\bibfield  {journal} {\bibinfo  {journal} {Phys. Rev. B}\ }\textbf {\bibinfo
  {volume} {74}},\ \bibinfo {pages} {054513}}\BibitemShut {NoStop}%
\bibitem [{\citenamefont {Castellani}\ \emph {et~al.}(1979)\citenamefont
  {Castellani}, \citenamefont {Castro}, \citenamefont {Feinberg},\ and\
  \citenamefont {Ranninger}}]{Castellani1979}%
  \BibitemOpen
  \bibfield  {author} {\bibinfo {author} {\bibnamefont {Castellani},
  \bibfnamefont {C.}}, \bibinfo {author} {\bibfnamefont {C.~D.}\ \bibnamefont
  {Castro}}, \bibinfo {author} {\bibfnamefont {D.}~\bibnamefont {Feinberg}}, \
  and\ \bibinfo {author} {\bibfnamefont {J.}~\bibnamefont {Ranninger}}}
  (\bibinfo {year} {1979}),\ \href {\doibase 10.1103/PhysRevLett.43.1957}
  {\bibfield  {journal} {\bibinfo  {journal} {Phys. Rev. Lett.}\ }\textbf
  {\bibinfo {volume} {43}},\ \bibinfo {pages} {1957}}\BibitemShut {NoStop}%
\bibitem [{\citenamefont {Castellani}\ \emph {et~al.}(1992)\citenamefont
  {Castellani}, \citenamefont {Di~Castro},\ and\ \citenamefont
  {Metzner}}]{Castellani1992}%
  \BibitemOpen
  \bibfield  {author} {\bibinfo {author} {\bibnamefont {Castellani},
  \bibfnamefont {C.}}, \bibinfo {author} {\bibfnamefont {C.}~\bibnamefont
  {Di~Castro}}, \ and\ \bibinfo {author} {\bibfnamefont {W.}~\bibnamefont
  {Metzner}}} (\bibinfo {year} {1992}),\ \href {\doibase
  10.1103/PhysRevLett.69.1703} {\bibfield  {journal} {\bibinfo  {journal}
  {Phys. Rev. Lett.}\ }\textbf {\bibinfo {volume} {69}},\ \bibinfo {pages}
  {1703}}\BibitemShut {NoStop}%
\bibitem [{\citenamefont {Chakravarty}\ \emph {et~al.}(2001)\citenamefont
  {Chakravarty}, \citenamefont {Laughlin}, \citenamefont {Morr},\ and\
  \citenamefont {Nayak}}]{Chakravarty2001}%
  \BibitemOpen
  \bibfield  {author} {\bibinfo {author} {\bibnamefont {Chakravarty},
  \bibfnamefont {S.}}, \bibinfo {author} {\bibfnamefont {R.~B.}\ \bibnamefont
  {Laughlin}}, \bibinfo {author} {\bibfnamefont {D.~K.}\ \bibnamefont {Morr}},
  \ and\ \bibinfo {author} {\bibfnamefont {C.}~\bibnamefont {Nayak}}} (\bibinfo
  {year} {2001}),\ \href {\doibase 10.1103/PhysRevB.63.094503} {\bibfield
  {journal} {\bibinfo  {journal} {Phys. Rev. B}\ }\textbf {\bibinfo {volume}
  {63}},\ \bibinfo {pages} {094503}}\BibitemShut {NoStop}%
\bibitem [{\citenamefont {Chalupa}\ \emph {et~al.}(2017)\citenamefont
  {Chalupa}, \citenamefont {Gunacker}, \citenamefont {Sch{\"a}fer},
  \citenamefont {Held},\ and\ \citenamefont {Toschi}}]{Chalupa2017}%
  \BibitemOpen
  \bibfield  {author} {\bibinfo {author} {\bibnamefont {Chalupa}, \bibfnamefont
  {P.}}, \bibinfo {author} {\bibfnamefont {P.}~\bibnamefont {Gunacker}},
  \bibinfo {author} {\bibfnamefont {T.}~\bibnamefont {Sch{\"a}fer}}, \bibinfo
  {author} {\bibfnamefont {K.}~\bibnamefont {Held}}, \ and\ \bibinfo {author}
  {\bibfnamefont {A.}~\bibnamefont {Toschi}}} (\bibinfo {year} {2017}),\
  \href@noop {} {\ }\Eprint {http://arxiv.org/abs/1712.04171}
  {arXiv:1712.04171} \BibitemShut {NoStop}%
\bibitem [{\citenamefont {Chitra}\ and\ \citenamefont
  {Kotliar}(2000)}]{Chitra00}%
  \BibitemOpen
  \bibfield  {author} {\bibinfo {author} {\bibnamefont {Chitra}, \bibfnamefont
  {R.}}, \ and\ \bibinfo {author} {\bibfnamefont {G.}~\bibnamefont {Kotliar}}}
  (\bibinfo {year} {2000}),\ \href {\doibase 10.1103/PhysRevLett.84.3678}
  {\bibfield  {journal} {\bibinfo  {journal} {Phys. Rev. Lett.}\ }\textbf
  {\bibinfo {volume} {84}},\ \bibinfo {pages} {3678}}\BibitemShut {NoStop}%
\bibitem [{\citenamefont {Chitra}\ and\ \citenamefont
  {Kotliar}(2001)}]{Chitra01}%
  \BibitemOpen
  \bibfield  {author} {\bibinfo {author} {\bibnamefont {Chitra}, \bibfnamefont
  {R.}}, \ and\ \bibinfo {author} {\bibfnamefont {G.}~\bibnamefont {Kotliar}}}
  (\bibinfo {year} {2001}),\ \href {\doibase 10.1103/PhysRevB.63.115110}
  {\bibfield  {journal} {\bibinfo  {journal} {Phys. Rev. B}\ }\textbf {\bibinfo
  {volume} {63}},\ \bibinfo {pages} {115110}}\BibitemShut {NoStop}%
\bibitem [{\citenamefont {Dagotto}(1994)}]{Dagotto94}%
  \BibitemOpen
  \bibfield  {author} {\bibinfo {author} {\bibnamefont {Dagotto}, \bibfnamefont
  {E.}}} (\bibinfo {year} {1994}),\ \href {\doibase 10.1103/RevModPhys.66.763}
  {\bibfield  {journal} {\bibinfo  {journal} {Rev. Mod. Phys.}\ }\textbf
  {\bibinfo {volume} {66}},\ \bibinfo {pages} {763}}\BibitemShut {NoStop}%
\bibitem [{\citenamefont {Dai}\ \emph {et~al.}(2005)\citenamefont {Dai},
  \citenamefont {Haule},\ and\ \citenamefont {Kotliar}}]{Dai2005}%
  \BibitemOpen
  \bibfield  {author} {\bibinfo {author} {\bibnamefont {Dai}, \bibfnamefont
  {X.}}, \bibinfo {author} {\bibfnamefont {K.}~\bibnamefont {Haule}}, \ and\
  \bibinfo {author} {\bibfnamefont {G.}~\bibnamefont {Kotliar}}} (\bibinfo
  {year} {2005}),\ \href {\doibase 10.1103/PhysRevB.72.045111} {\bibfield
  {journal} {\bibinfo  {journal} {Phys. Rev. B}\ }\textbf {\bibinfo {volume}
  {72}},\ \bibinfo {pages} {045111}}\BibitemShut {NoStop}%
\bibitem [{\citenamefont {Dar\'e}\ and\ \citenamefont
  {Albinet}(2000)}]{Dare2000}%
  \BibitemOpen
  \bibfield  {author} {\bibinfo {author} {\bibnamefont {Dar\'e}, \bibfnamefont
  {A.-M.}}, \ and\ \bibinfo {author} {\bibfnamefont {G.}~\bibnamefont
  {Albinet}}} (\bibinfo {year} {2000}),\ \href {\doibase
  10.1103/PhysRevB.61.4567} {\bibfield  {journal} {\bibinfo  {journal} {Phys.
  Rev. B}\ }\textbf {\bibinfo {volume} {61}},\ \bibinfo {pages}
  {4567}}\BibitemShut {NoStop}%
\bibitem [{\citenamefont {Das}\ \emph {et~al.}(2011)\citenamefont {Das},
  \citenamefont {Sangiovanni}, \citenamefont {Valli}, \citenamefont {Held},\
  and\ \citenamefont {Saha-Dasgupta}}]{Das2011}%
  \BibitemOpen
  \bibfield  {author} {\bibinfo {author} {\bibnamefont {Das}, \bibfnamefont
  {H.}}, \bibinfo {author} {\bibfnamefont {G.}~\bibnamefont {Sangiovanni}},
  \bibinfo {author} {\bibfnamefont {A.}~\bibnamefont {Valli}}, \bibinfo
  {author} {\bibfnamefont {K.}~\bibnamefont {Held}}, \ and\ \bibinfo {author}
  {\bibfnamefont {T.}~\bibnamefont {Saha-Dasgupta}}} (\bibinfo {year} {2011}),\
  \href {\doibase 10.1103/PhysRevLett.107.197202} {\bibfield  {journal}
  {\bibinfo  {journal} {Phys. Rev. Lett.}\ }\textbf {\bibinfo {volume} {107}},\
  \bibinfo {pages} {197202}}\BibitemShut {NoStop}%
\bibitem [{\citenamefont {Davoudi}\ and\ \citenamefont
  {Tremblay}(2007)}]{Davoudi07}%
  \BibitemOpen
  \bibfield  {author} {\bibinfo {author} {\bibnamefont {Davoudi}, \bibfnamefont
  {B.}}, \ and\ \bibinfo {author} {\bibfnamefont {A.-M.~S.}\ \bibnamefont
  {Tremblay}}} (\bibinfo {year} {2007}),\ \href {\doibase
  10.1103/PhysRevB.76.085115} {\bibfield  {journal} {\bibinfo  {journal} {Phys.
  Rev. B}\ }\textbf {\bibinfo {volume} {76}},\ \bibinfo {pages}
  {085115}}\BibitemShut {NoStop}%
\bibitem [{\citenamefont {{De Dominicis}}(1962)}]{DeDominicis1962}%
  \BibitemOpen
  \bibfield  {author} {\bibinfo {author} {\bibnamefont {{De Dominicis}},
  \bibfnamefont {C.}}} (\bibinfo {year} {1962}),\ \href {\doibase
  10.1063/1.1724313} {\bibfield  {journal} {\bibinfo  {journal} {J. Math.
  Phys.}\ }\textbf {\bibinfo {volume} {3}}~(\bibinfo {number} {5}),\ \bibinfo
  {pages} {983}}\BibitemShut {NoStop}%
\bibitem [{\citenamefont {{De Dominicis}}\ and\ \citenamefont
  {Martin}(1964{\natexlab{a}})}]{DeDominicis1964}%
  \BibitemOpen
  \bibfield  {author} {\bibinfo {author} {\bibnamefont {{De Dominicis}},
  \bibfnamefont {C.}}, \ and\ \bibinfo {author} {\bibfnamefont {P.~C.}\
  \bibnamefont {Martin}}} (\bibinfo {year} {1964}{\natexlab{a}}),\ \href
  {\doibase 10.1063/1.1704062} {\bibfield  {journal} {\bibinfo  {journal} {J.
  Math. Phys.}\ }\textbf {\bibinfo {volume} {5}}~(\bibinfo {number} {1}),\
  \bibinfo {pages} {14}}\BibitemShut {NoStop}%
\bibitem [{\citenamefont {{De Dominicis}}\ and\ \citenamefont
  {Martin}(1964{\natexlab{b}})}]{DeDominicis1964b}%
  \BibitemOpen
  \bibfield  {author} {\bibinfo {author} {\bibnamefont {{De Dominicis}},
  \bibfnamefont {C.}}, \ and\ \bibinfo {author} {\bibfnamefont {P.~C.}\
  \bibnamefont {Martin}}} (\bibinfo {year} {1964}{\natexlab{b}}),\ \href
  {\doibase 10.1063/1.1704064} {\bibfield  {journal} {\bibinfo  {journal} {J.
  Math. Phys.}\ }\textbf {\bibinfo {volume} {5}}~(\bibinfo {number} {1}),\
  \bibinfo {pages} {31}}\BibitemShut {NoStop}%
\bibitem [{\citenamefont {Diatlov}\ \emph {et~al.}(1957)\citenamefont
  {Diatlov}, \citenamefont {Sudakov},\ and\ \citenamefont
  {Ter-Martirosian}}]{Diatlov1957}%
  \BibitemOpen
  \bibfield  {author} {\bibinfo {author} {\bibnamefont {Diatlov}, \bibfnamefont
  {I.~T.}}, \bibinfo {author} {\bibfnamefont {V.~V.}\ \bibnamefont {Sudakov}},
  \ and\ \bibinfo {author} {\bibfnamefont {K.~A.}\ \bibnamefont
  {Ter-Martirosian}}} (\bibinfo {year} {1957}),\ \href@noop {} {\bibfield
  {journal} {\bibinfo  {journal} {Sov. Phys. JETP}\ }\textbf {\bibinfo {volume}
  {5}},\ \bibinfo {pages} {631}}\BibitemShut {NoStop}%
\bibitem [{\citenamefont {Dobrosavljevi{\'{c}}}\ \emph
  {et~al.}(2003)\citenamefont {Dobrosavljevi{\'{c}}}, \citenamefont {Pastor},\
  and\ \citenamefont {Nikoli{\'{c}}}}]{Dobrosavljevic2003}%
  \BibitemOpen
  \bibfield  {author} {\bibinfo {author} {\bibnamefont {Dobrosavljevi{\'{c}}},
  \bibfnamefont {V.}}, \bibinfo {author} {\bibfnamefont {A.~A.}\ \bibnamefont
  {Pastor}}, \ and\ \bibinfo {author} {\bibfnamefont {B.~K.}\ \bibnamefont
  {Nikoli{\'{c}}}}} (\bibinfo {year} {2003}),\ \href {\doibase
  10.1209/epl/i2003-00364-5} {\bibfield  {journal} {\bibinfo  {journal}
  {Europhys. Lett.}\ }\textbf {\bibinfo {volume} {62}}~(\bibinfo {number}
  {1}),\ \bibinfo {pages} {76}}\BibitemShut {NoStop}%
\bibitem [{\citenamefont {Eckstein}\ and\ \citenamefont
  {Kollar}(2008)}]{Eckstein2008}%
  \BibitemOpen
  \bibfield  {author} {\bibinfo {author} {\bibnamefont {Eckstein},
  \bibfnamefont {M.}}, \ and\ \bibinfo {author} {\bibfnamefont
  {M.}~\bibnamefont {Kollar}}} (\bibinfo {year} {2008}),\ \href {\doibase
  10.1103/PhysRevLett.100.120404} {\bibfield  {journal} {\bibinfo  {journal}
  {Phys. Rev. Lett.}\ }\textbf {\bibinfo {volume} {100}}~(\bibinfo {number}
  {12}),\ \bibinfo {pages} {120404}}\BibitemShut {NoStop}%
\bibitem [{\citenamefont {Eckstein}\ \emph {et~al.}(2009)\citenamefont
  {Eckstein}, \citenamefont {Kollar},\ and\ \citenamefont
  {Werner}}]{Eckstein2009}%
  \BibitemOpen
  \bibfield  {author} {\bibinfo {author} {\bibnamefont {Eckstein},
  \bibfnamefont {M.}}, \bibinfo {author} {\bibfnamefont {M.}~\bibnamefont
  {Kollar}}, \ and\ \bibinfo {author} {\bibfnamefont {P.}~\bibnamefont
  {Werner}}} (\bibinfo {year} {2009}),\ \href {\doibase
  10.1103/PhysRevLett.103.056403} {\bibfield  {journal} {\bibinfo  {journal}
  {Phys. Rev. Lett.}\ }\textbf {\bibinfo {volume} {103}},\ \bibinfo {pages}
  {056403}}\BibitemShut {NoStop}%
\bibitem [{\citenamefont {Ekuma}\ \emph {et~al.}(2014)\citenamefont {Ekuma},
  \citenamefont {Terletska}, \citenamefont {Tam}, \citenamefont {Meng},
  \citenamefont {Moreno},\ and\ \citenamefont {Jarrell}}]{Ekuma2014a}%
  \BibitemOpen
  \bibfield  {author} {\bibinfo {author} {\bibnamefont {Ekuma}, \bibfnamefont
  {C.~E.}}, \bibinfo {author} {\bibfnamefont {H.}~\bibnamefont {Terletska}},
  \bibinfo {author} {\bibfnamefont {K.-M.}\ \bibnamefont {Tam}}, \bibinfo
  {author} {\bibfnamefont {Z.-Y.}\ \bibnamefont {Meng}}, \bibinfo {author}
  {\bibfnamefont {J.}~\bibnamefont {Moreno}}, \ and\ \bibinfo {author}
  {\bibfnamefont {M.}~\bibnamefont {Jarrell}}} (\bibinfo {year} {2014}),\ \href
  {\doibase 10.1103/PhysRevB.89.081107} {\bibfield  {journal} {\bibinfo
  {journal} {Phys. Rev. B}\ }\textbf {\bibinfo {volume} {89}}~(\bibinfo
  {number} {8}),\ \bibinfo {pages} {081107}}\BibitemShut {NoStop}%
\bibitem [{\citenamefont {Faleev}\ \emph {et~al.}(2004)\citenamefont {Faleev},
  \citenamefont {van Schilfgaarde},\ and\ \citenamefont {Kotani}}]{Faleev2004}%
  \BibitemOpen
  \bibfield  {author} {\bibinfo {author} {\bibnamefont {Faleev}, \bibfnamefont
  {S.~V.}}, \bibinfo {author} {\bibfnamefont {M.}~\bibnamefont {van
  Schilfgaarde}}, \ and\ \bibinfo {author} {\bibfnamefont {T.}~\bibnamefont
  {Kotani}}} (\bibinfo {year} {2004}),\ \href {\doibase
  10.1103/PhysRevLett.93.126406} {\bibfield  {journal} {\bibinfo  {journal}
  {Phys. Rev. Lett.}\ }\textbf {\bibinfo {volume} {93}},\ \bibinfo {pages}
  {126406}}\BibitemShut {NoStop}%
\bibitem [{\citenamefont {Falicov}\ and\ \citenamefont
  {Kimball}(1969)}]{Falicov1969}%
  \BibitemOpen
  \bibfield  {author} {\bibinfo {author} {\bibnamefont {Falicov}, \bibfnamefont
  {L.~M.}}, \ and\ \bibinfo {author} {\bibfnamefont {J.~C.}\ \bibnamefont
  {Kimball}}} (\bibinfo {year} {1969}),\ \href {\doibase
  10.1103/PhysRevLett.22.997} {\bibfield  {journal} {\bibinfo  {journal} {Phys.
  Rev. Lett.}\ }\textbf {\bibinfo {volume} {22}}~(\bibinfo {number} {19}),\
  \bibinfo {pages} {997}}\BibitemShut {NoStop}%
\bibitem [{\citenamefont {Florens}(2007)}]{Florens2007}%
  \BibitemOpen
  \bibfield  {author} {\bibinfo {author} {\bibnamefont {Florens}, \bibfnamefont
  {S.}}} (\bibinfo {year} {2007}),\ \href {\doibase
  10.1103/PhysRevLett.99.046402} {\bibfield  {journal} {\bibinfo  {journal}
  {Phys. Rev. Lett.}\ }\textbf {\bibinfo {volume} {99}},\ \bibinfo {pages}
  {046402}}\BibitemShut {NoStop}%
\bibitem [{\citenamefont {Fratino}\ \emph {et~al.}(2017)\citenamefont
  {Fratino}, \citenamefont {S\'emon}, \citenamefont {Charlebois}, \citenamefont
  {Sordi},\ and\ \citenamefont {Tremblay}}]{Fratino2017}%
  \BibitemOpen
  \bibfield  {author} {\bibinfo {author} {\bibnamefont {Fratino}, \bibfnamefont
  {L.}}, \bibinfo {author} {\bibfnamefont {P.}~\bibnamefont {S\'emon}},
  \bibinfo {author} {\bibfnamefont {M.}~\bibnamefont {Charlebois}}, \bibinfo
  {author} {\bibfnamefont {G.}~\bibnamefont {Sordi}}, \ and\ \bibinfo {author}
  {\bibfnamefont {A.-M.~S.}\ \bibnamefont {Tremblay}}} (\bibinfo {year}
  {2017}),\ \href {\doibase 10.1103/PhysRevB.95.235109} {\bibfield  {journal}
  {\bibinfo  {journal} {Phys. Rev. B}\ }\textbf {\bibinfo {volume} {95}},\
  \bibinfo {pages} {235109}}\BibitemShut {NoStop}%
\bibitem [{\citenamefont {Freericks}\ and\ \citenamefont
  {Zlati\ifmmode~\acute{c}\else \'{c}\fi{}}(2003)}]{Freericks2003}%
  \BibitemOpen
  \bibfield  {author} {\bibinfo {author} {\bibnamefont {Freericks},
  \bibfnamefont {J.~K.}}, \ and\ \bibinfo {author} {\bibfnamefont
  {V.}~\bibnamefont {Zlati\ifmmode~\acute{c}\else \'{c}\fi{}}}} (\bibinfo
  {year} {2003}),\ \href {\doibase 10.1103/RevModPhys.75.1333} {\bibfield
  {journal} {\bibinfo  {journal} {Rev. Mod. Phys.}\ }\textbf {\bibinfo {volume}
  {75}},\ \bibinfo {pages} {1333}}\BibitemShut {NoStop}%
\bibitem [{\citenamefont {Freericks}\ and\ \citenamefont
  {Lema{\'{n}}ski}(2000)}]{Freericks2000c}%
  \BibitemOpen
  \bibfield  {author} {\bibinfo {author} {\bibnamefont {Freericks},
  \bibfnamefont {J.~K.}}, \ and\ \bibinfo {author} {\bibfnamefont
  {R.}~\bibnamefont {Lema{\'{n}}ski}}} (\bibinfo {year} {2000}),\ \href
  {\doibase 10.1103/PhysRevB.61.13438} {\bibfield  {journal} {\bibinfo
  {journal} {Phys. Rev. B}\ }\textbf {\bibinfo {volume} {61}}~(\bibinfo
  {number} {20}),\ \bibinfo {pages} {13438}}\BibitemShut {NoStop}%
\bibitem [{\citenamefont {Freericks}\ and\ \citenamefont
  {Miller}(2000)}]{Freericks2000a}%
  \BibitemOpen
  \bibfield  {author} {\bibinfo {author} {\bibnamefont {Freericks},
  \bibfnamefont {J.~K.}}, \ and\ \bibinfo {author} {\bibfnamefont
  {P.}~\bibnamefont {Miller}}} (\bibinfo {year} {2000}),\ \href {\doibase
  10.1103/PhysRevB.62.10022} {\bibfield  {journal} {\bibinfo  {journal} {Phys.
  Rev. B}\ }\textbf {\bibinfo {volume} {62}}~(\bibinfo {number} {15}),\
  \bibinfo {pages} {10022}}\BibitemShut {NoStop}%
\bibitem [{\citenamefont {Fuchs}\ \emph {et~al.}(2011)\citenamefont {Fuchs},
  \citenamefont {Gull}, \citenamefont {Troyer}, \citenamefont {Jarrell},\ and\
  \citenamefont {Pruschke}}]{Fuchs2011a}%
  \BibitemOpen
  \bibfield  {author} {\bibinfo {author} {\bibnamefont {Fuchs}, \bibfnamefont
  {S.}}, \bibinfo {author} {\bibfnamefont {E.}~\bibnamefont {Gull}}, \bibinfo
  {author} {\bibfnamefont {M.}~\bibnamefont {Troyer}}, \bibinfo {author}
  {\bibfnamefont {M.}~\bibnamefont {Jarrell}}, \ and\ \bibinfo {author}
  {\bibfnamefont {T.}~\bibnamefont {Pruschke}}} (\bibinfo {year} {2011}),\
  \href {\doibase 10.1103/PhysRevB.83.235113} {\bibfield  {journal} {\bibinfo
  {journal} {Phys. Rev. B}\ }\textbf {\bibinfo {volume} {83}},\ \bibinfo
  {pages} {235113}}\BibitemShut {NoStop}%
\bibitem [{\citenamefont {Gaenko}\ \emph {et~al.}(2017)\citenamefont {Gaenko},
  \citenamefont {Antipov}, \citenamefont {Carcassi}, \citenamefont {Chen},
  \citenamefont {Chen}, \citenamefont {Dong}, \citenamefont {Gamper},
  \citenamefont {Gukelberger}, \citenamefont {Igarashi}, \citenamefont
  {Iskakov}, \citenamefont {Könz}, \citenamefont {LeBlanc}, \citenamefont
  {Levy}, \citenamefont {Ma}, \citenamefont {Paki}, \citenamefont {Shinaoka},
  \citenamefont {Todo}, \citenamefont {Troyer},\ and\ \citenamefont
  {Gull}}]{ALPSCORE}%
  \BibitemOpen
  \bibfield  {author} {\bibinfo {author} {\bibnamefont {Gaenko}, \bibfnamefont
  {A.}}, \bibinfo {author} {\bibfnamefont {A.}~\bibnamefont {Antipov}},
  \bibinfo {author} {\bibfnamefont {G.}~\bibnamefont {Carcassi}}, \bibinfo
  {author} {\bibfnamefont {T.}~\bibnamefont {Chen}}, \bibinfo {author}
  {\bibfnamefont {X.}~\bibnamefont {Chen}}, \bibinfo {author} {\bibfnamefont
  {Q.}~\bibnamefont {Dong}}, \bibinfo {author} {\bibfnamefont {L.}~\bibnamefont
  {Gamper}}, \bibinfo {author} {\bibfnamefont {J.}~\bibnamefont {Gukelberger}},
  \bibinfo {author} {\bibfnamefont {R.}~\bibnamefont {Igarashi}}, \bibinfo
  {author} {\bibfnamefont {S.}~\bibnamefont {Iskakov}}, \bibinfo {author}
  {\bibfnamefont {M.}~\bibnamefont {Könz}}, \bibinfo {author} {\bibfnamefont
  {J.}~\bibnamefont {LeBlanc}}, \bibinfo {author} {\bibfnamefont
  {R.}~\bibnamefont {Levy}}, \bibinfo {author} {\bibfnamefont {P.}~\bibnamefont
  {Ma}}, \bibinfo {author} {\bibfnamefont {J.}~\bibnamefont {Paki}}, \bibinfo
  {author} {\bibfnamefont {H.}~\bibnamefont {Shinaoka}}, \bibinfo {author}
  {\bibfnamefont {S.}~\bibnamefont {Todo}}, \bibinfo {author} {\bibfnamefont
  {M.}~\bibnamefont {Troyer}}, \ and\ \bibinfo {author} {\bibfnamefont
  {E.}~\bibnamefont {Gull}}} (\bibinfo {year} {2017}),\ \href {\doibase
  http://dx.doi.org/10.1016/j.cpc.2016.12.009} {\bibfield  {journal} {\bibinfo
  {journal} {Computer Physics Communications}\ }\textbf {\bibinfo {volume}
  {213}},\ \bibinfo {pages} {235 }}\BibitemShut {NoStop}%
\bibitem [{\citenamefont {Galler}\ \emph {et~al.}(2018)\citenamefont {Galler},
  \citenamefont {Kaufmann}, \citenamefont {Gunacker}, \citenamefont {Pickem},
  \citenamefont {Thunström}, \citenamefont {Tomczak},\ and\ \citenamefont
  {Held}}]{Galler2017a}%
  \BibitemOpen
  \bibfield  {author} {\bibinfo {author} {\bibnamefont {Galler}, \bibfnamefont
  {A.}}, \bibinfo {author} {\bibfnamefont {J.}~\bibnamefont {Kaufmann}},
  \bibinfo {author} {\bibfnamefont {P.}~\bibnamefont {Gunacker}}, \bibinfo
  {author} {\bibfnamefont {M.}~\bibnamefont {Pickem}}, \bibinfo {author}
  {\bibfnamefont {P.}~\bibnamefont {Thunström}}, \bibinfo {author}
  {\bibfnamefont {J.~M.}\ \bibnamefont {Tomczak}}, \ and\ \bibinfo {author}
  {\bibfnamefont {K.}~\bibnamefont {Held}}} (\bibinfo {year} {2018}),\ \href
  {\doibase 10.7566/JPSJ.87.041004} {\bibfield  {journal} {\bibinfo  {journal}
  {Journal of the Physical Society of Japan}\ }\textbf {\bibinfo {volume}
  {87}}~(\bibinfo {number} {4}),\ \bibinfo {pages} {041004}}\BibitemShut
  {NoStop}%
\bibitem [{\citenamefont {Galler}\ \emph
  {et~al.}(2017{\natexlab{a}})\citenamefont {Galler}, \citenamefont
  {Thunstr\"om}, \citenamefont {Gunacker}, \citenamefont {Tomczak},\ and\
  \citenamefont {Held}}]{Galler2016}%
  \BibitemOpen
  \bibfield  {author} {\bibinfo {author} {\bibnamefont {Galler}, \bibfnamefont
  {A.}}, \bibinfo {author} {\bibfnamefont {P.}~\bibnamefont {Thunstr\"om}},
  \bibinfo {author} {\bibfnamefont {P.}~\bibnamefont {Gunacker}}, \bibinfo
  {author} {\bibfnamefont {J.~M.}\ \bibnamefont {Tomczak}}, \ and\ \bibinfo
  {author} {\bibfnamefont {K.}~\bibnamefont {Held}}} (\bibinfo {year}
  {2017}{\natexlab{a}}),\ \href {\doibase 10.1103/PhysRevB.95.115107}
  {\bibfield  {journal} {\bibinfo  {journal} {Phys. Rev. B}\ }\textbf {\bibinfo
  {volume} {95}},\ \bibinfo {pages} {115107}}\BibitemShut {NoStop}%
\bibitem [{\citenamefont {Galler}\ \emph
  {et~al.}(2017{\natexlab{b}})\citenamefont {Galler}, \citenamefont
  {{Thunstr{\"o}m}}, \citenamefont {{Kaufmann}}, \citenamefont {{Pickem}},
  \citenamefont {{Tomczak}},\ and\ \citenamefont {{Held}}}]{Galler2017b}%
  \BibitemOpen
  \bibfield  {author} {\bibinfo {author} {\bibnamefont {Galler}, \bibfnamefont
  {A.}}, \bibinfo {author} {\bibfnamefont {P.}~\bibnamefont {{Thunstr{\"o}m}}},
  \bibinfo {author} {\bibfnamefont {J.}~\bibnamefont {{Kaufmann}}}, \bibinfo
  {author} {\bibfnamefont {M.}~\bibnamefont {{Pickem}}}, \bibinfo {author}
  {\bibfnamefont {J.~M.}\ \bibnamefont {{Tomczak}}}, \ and\ \bibinfo {author}
  {\bibfnamefont {K.}~\bibnamefont {{Held}}}} (\bibinfo {year}
  {2017}{\natexlab{b}}),\ \href {https://arxiv.org/abs/1710.06651} {\ }\Eprint
  {http://arxiv.org/abs/1710.06651} {arXiv:1710.06651} \BibitemShut {NoStop}%
\bibitem [{\citenamefont {Gebhard}(1997)}]{Gebhard1997}%
  \BibitemOpen
  \bibfield  {author} {\bibinfo {author} {\bibnamefont {Gebhard}, \bibfnamefont
  {F.}}} (\bibinfo {year} {1997}),\ \href@noop {} {\emph {\bibinfo {title} {The
  Mott Metal-insulator transition}}}\ (\bibinfo  {publisher} {Springer-Verlag
  (Berlin)})\BibitemShut {NoStop}%
\bibitem [{\citenamefont {Georges}\ and\ \citenamefont
  {Kotliar}(1992)}]{Georges1992a}%
  \BibitemOpen
  \bibfield  {author} {\bibinfo {author} {\bibnamefont {Georges}, \bibfnamefont
  {A.}}, \ and\ \bibinfo {author} {\bibfnamefont {G.}~\bibnamefont {Kotliar}}}
  (\bibinfo {year} {1992}),\ \href {\doibase 10.1103/PhysRevB.45.6479}
  {\bibfield  {journal} {\bibinfo  {journal} {Phys. Rev. B}\ }\textbf {\bibinfo
  {volume} {45}},\ \bibinfo {pages} {6479}}\BibitemShut {NoStop}%
\bibitem [{\citenamefont {Georges}\ \emph {et~al.}(1996)\citenamefont
  {Georges}, \citenamefont {Kotliar}, \citenamefont {Krauth},\ and\
  \citenamefont {Rozenberg}}]{Georges1996}%
  \BibitemOpen
  \bibfield  {author} {\bibinfo {author} {\bibnamefont {Georges}, \bibfnamefont
  {A.}}, \bibinfo {author} {\bibfnamefont {G.}~\bibnamefont {Kotliar}},
  \bibinfo {author} {\bibfnamefont {W.}~\bibnamefont {Krauth}}, \ and\ \bibinfo
  {author} {\bibfnamefont {M.~J.}\ \bibnamefont {Rozenberg}}} (\bibinfo {year}
  {1996}),\ \href {\doibase 10.1103/RevModPhys.68.13} {\bibfield  {journal}
  {\bibinfo  {journal} {Rev. Mod. Phys.}\ }\textbf {\bibinfo {volume}
  {68}}~(\bibinfo {number} {1}),\ \bibinfo {pages} {13}}\BibitemShut {NoStop}%
\bibitem [{\citenamefont {Greif}\ \emph {et~al.}(2015)\citenamefont {Greif},
  \citenamefont {Jotzu}, \citenamefont {Messer}, \citenamefont {Desbuquois},\
  and\ \citenamefont {Esslinger}}]{Greif2015}%
  \BibitemOpen
  \bibfield  {author} {\bibinfo {author} {\bibnamefont {Greif}, \bibfnamefont
  {D.}}, \bibinfo {author} {\bibfnamefont {G.}~\bibnamefont {Jotzu}}, \bibinfo
  {author} {\bibfnamefont {M.}~\bibnamefont {Messer}}, \bibinfo {author}
  {\bibfnamefont {R.}~\bibnamefont {Desbuquois}}, \ and\ \bibinfo {author}
  {\bibfnamefont {T.}~\bibnamefont {Esslinger}}} (\bibinfo {year} {2015}),\
  \href {\doibase 10.1103/PhysRevLett.115.260401} {\bibfield  {journal}
  {\bibinfo  {journal} {Phys. Rev. Lett.}\ }\textbf {\bibinfo {volume}
  {115}}~(\bibinfo {number} {26}),\ \bibinfo {pages} {260401}}\BibitemShut
  {NoStop}%
\bibitem [{\citenamefont {Grempel}\ and\ \citenamefont
  {Si}(2003)}]{Grempel2003}%
  \BibitemOpen
  \bibfield  {author} {\bibinfo {author} {\bibnamefont {Grempel}, \bibfnamefont
  {D.~R.}}, \ and\ \bibinfo {author} {\bibfnamefont {Q.}~\bibnamefont {Si}}}
  (\bibinfo {year} {2003}),\ \href {\doibase 10.1103/PhysRevLett.91.026401}
  {\bibfield  {journal} {\bibinfo  {journal} {Phys. Rev. Lett.}\ }\textbf
  {\bibinfo {volume} {91}},\ \bibinfo {pages} {026401}}\BibitemShut {NoStop}%
\bibitem [{\citenamefont {Gukelberger}\ \emph {et~al.}(2015)\citenamefont
  {Gukelberger}, \citenamefont {Huang},\ and\ \citenamefont
  {Werner}}]{Gukelberger2015}%
  \BibitemOpen
  \bibfield  {author} {\bibinfo {author} {\bibnamefont {Gukelberger},
  \bibfnamefont {J.}}, \bibinfo {author} {\bibfnamefont {L.}~\bibnamefont
  {Huang}}, \ and\ \bibinfo {author} {\bibfnamefont {P.}~\bibnamefont
  {Werner}}} (\bibinfo {year} {2015}),\ \href {\doibase
  10.1103/PhysRevB.91.235114} {\bibfield  {journal} {\bibinfo  {journal} {Phys.
  Rev. B}\ }\textbf {\bibinfo {volume} {91}},\ \bibinfo {pages}
  {235114}}\BibitemShut {NoStop}%
\bibitem [{\citenamefont {Gukelberger}\ \emph {et~al.}(2017)\citenamefont
  {Gukelberger}, \citenamefont {Kozik},\ and\ \citenamefont
  {Hafermann}}]{Gukelberger2016}%
  \BibitemOpen
  \bibfield  {author} {\bibinfo {author} {\bibnamefont {Gukelberger},
  \bibfnamefont {J.}}, \bibinfo {author} {\bibfnamefont {E.}~\bibnamefont
  {Kozik}}, \ and\ \bibinfo {author} {\bibfnamefont {H.}~\bibnamefont
  {Hafermann}}} (\bibinfo {year} {2017}),\ \href {\doibase
  10.1103/PhysRevB.96.035152} {\bibfield  {journal} {\bibinfo  {journal} {Phys.
  Rev. B}\ }\textbf {\bibinfo {volume} {96}},\ \bibinfo {pages}
  {035152}}\BibitemShut {NoStop}%
\bibitem [{\citenamefont {Gull}\ \emph
  {et~al.}(2011{\natexlab{a}})\citenamefont {Gull}, \citenamefont {Millis},
  \citenamefont {Lichtenstein}, \citenamefont {Rubtsov}, \citenamefont
  {Troyer},\ and\ \citenamefont {Werner}}]{Gull2011a}%
  \BibitemOpen
  \bibfield  {author} {\bibinfo {author} {\bibnamefont {Gull}, \bibfnamefont
  {E.}}, \bibinfo {author} {\bibfnamefont {A.~J.}\ \bibnamefont {Millis}},
  \bibinfo {author} {\bibfnamefont {A.~I.}\ \bibnamefont {Lichtenstein}},
  \bibinfo {author} {\bibfnamefont {A.~N.}\ \bibnamefont {Rubtsov}}, \bibinfo
  {author} {\bibfnamefont {M.}~\bibnamefont {Troyer}}, \ and\ \bibinfo {author}
  {\bibfnamefont {P.}~\bibnamefont {Werner}}} (\bibinfo {year}
  {2011}{\natexlab{a}}),\ \href {\doibase 10.1103/RevModPhys.83.349} {\bibfield
   {journal} {\bibinfo  {journal} {Rev. Mod. Phys.}\ }\textbf {\bibinfo
  {volume} {83}}~(\bibinfo {number} {2}),\ \bibinfo {pages} {349}}\BibitemShut
  {NoStop}%
\bibitem [{\citenamefont {Gull}\ \emph {et~al.}(2013)\citenamefont {Gull},
  \citenamefont {Parcollet},\ and\ \citenamefont {Millis}}]{Gull2013}%
  \BibitemOpen
  \bibfield  {author} {\bibinfo {author} {\bibnamefont {Gull}, \bibfnamefont
  {E.}}, \bibinfo {author} {\bibfnamefont {O.}~\bibnamefont {Parcollet}}, \
  and\ \bibinfo {author} {\bibfnamefont {A.~J.}\ \bibnamefont {Millis}}}
  (\bibinfo {year} {2013}),\ \href {\doibase 10.1103/PhysRevLett.110.216405}
  {\bibfield  {journal} {\bibinfo  {journal} {Phys. Rev. Lett.}\ }\textbf
  {\bibinfo {volume} {110}},\ \bibinfo {pages} {216405}}\BibitemShut {NoStop}%
\bibitem [{\citenamefont {Gull}\ \emph
  {et~al.}(2011{\natexlab{b}})\citenamefont {Gull}, \citenamefont {Staar},
  \citenamefont {Fuchs}, \citenamefont {Nukala}, \citenamefont {Summers},
  \citenamefont {Pruschke}, \citenamefont {Schulthess},\ and\ \citenamefont
  {Maier}}]{Gull2011}%
  \BibitemOpen
  \bibfield  {author} {\bibinfo {author} {\bibnamefont {Gull}, \bibfnamefont
  {E.}}, \bibinfo {author} {\bibfnamefont {P.}~\bibnamefont {Staar}}, \bibinfo
  {author} {\bibfnamefont {S.}~\bibnamefont {Fuchs}}, \bibinfo {author}
  {\bibfnamefont {P.}~\bibnamefont {Nukala}}, \bibinfo {author} {\bibfnamefont
  {M.~S.}\ \bibnamefont {Summers}}, \bibinfo {author} {\bibfnamefont
  {T.}~\bibnamefont {Pruschke}}, \bibinfo {author} {\bibfnamefont {T.~C.}\
  \bibnamefont {Schulthess}}, \ and\ \bibinfo {author} {\bibfnamefont
  {T.}~\bibnamefont {Maier}}} (\bibinfo {year} {2011}{\natexlab{b}}),\ \href
  {\doibase 10.1103/PhysRevB.83.075122} {\bibfield  {journal} {\bibinfo
  {journal} {Phys. Rev. B}\ }\textbf {\bibinfo {volume} {83}},\ \bibinfo
  {pages} {075122}}\BibitemShut {NoStop}%
\bibitem [{\citenamefont {Gull}\ \emph
  {et~al.}(2011{\natexlab{c}})\citenamefont {Gull}, \citenamefont {Werner},
  \citenamefont {Fuchs}, \citenamefont {Surer}, \citenamefont {Pruschke},\ and\
  \citenamefont {Troyer}}]{Gull2011b}%
  \BibitemOpen
  \bibfield  {author} {\bibinfo {author} {\bibnamefont {Gull}, \bibfnamefont
  {E.}}, \bibinfo {author} {\bibfnamefont {P.}~\bibnamefont {Werner}}, \bibinfo
  {author} {\bibfnamefont {S.}~\bibnamefont {Fuchs}}, \bibinfo {author}
  {\bibfnamefont {B.}~\bibnamefont {Surer}}, \bibinfo {author} {\bibfnamefont
  {T.}~\bibnamefont {Pruschke}}, \ and\ \bibinfo {author} {\bibfnamefont
  {M.}~\bibnamefont {Troyer}}} (\bibinfo {year} {2011}{\natexlab{c}}),\ \href
  {\doibase 10.1016/j.cpc.2010.12.050} {\bibfield  {journal} {\bibinfo
  {journal} {Computer Physics Communications}\ }\textbf {\bibinfo {volume}
  {182}}~(\bibinfo {number} {4}),\ \bibinfo {pages} {1078 }}\BibitemShut
  {NoStop}%
\bibitem [{\citenamefont {Gull}\ \emph {et~al.}(2007)\citenamefont {Gull},
  \citenamefont {Werner}, \citenamefont {Millis},\ and\ \citenamefont
  {Troyer}}]{Gull2007}%
  \BibitemOpen
  \bibfield  {author} {\bibinfo {author} {\bibnamefont {Gull}, \bibfnamefont
  {E.}}, \bibinfo {author} {\bibfnamefont {P.}~\bibnamefont {Werner}}, \bibinfo
  {author} {\bibfnamefont {A.}~\bibnamefont {Millis}}, \ and\ \bibinfo {author}
  {\bibfnamefont {M.}~\bibnamefont {Troyer}}} (\bibinfo {year} {2007}),\ \href
  {\doibase 10.1103/PhysRevB.76.235123} {\bibfield  {journal} {\bibinfo
  {journal} {Phys. Rev. B}\ }\textbf {\bibinfo {volume} {76}},\ \bibinfo
  {pages} {235123}}\BibitemShut {NoStop}%
\bibitem [{\citenamefont {Gull}\ \emph {et~al.}(2008)\citenamefont {Gull},
  \citenamefont {Werner}, \citenamefont {Parcollet},\ and\ \citenamefont
  {Troyer}}]{Gull2008a}%
  \BibitemOpen
  \bibfield  {author} {\bibinfo {author} {\bibnamefont {Gull}, \bibfnamefont
  {E.}}, \bibinfo {author} {\bibfnamefont {P.}~\bibnamefont {Werner}}, \bibinfo
  {author} {\bibfnamefont {O.}~\bibnamefont {Parcollet}}, \ and\ \bibinfo
  {author} {\bibfnamefont {M.}~\bibnamefont {Troyer}}} (\bibinfo {year}
  {2008}),\ \href {http://stacks.iop.org/0295-5075/82/i=5/a=57003} {\bibfield
  {journal} {\bibinfo  {journal} {EPL (Europhysics Letters)}\ }\textbf
  {\bibinfo {volume} {82}}~(\bibinfo {number} {5}),\ \bibinfo {pages}
  {57003}}\BibitemShut {NoStop}%
\bibitem [{\citenamefont {Gunacker}\ \emph {et~al.}(2015)\citenamefont
  {Gunacker}, \citenamefont {Wallerberger}, \citenamefont {Gull}, \citenamefont
  {Hausoel}, \citenamefont {Sangiovanni},\ and\ \citenamefont
  {Held}}]{Gunacker15}%
  \BibitemOpen
  \bibfield  {author} {\bibinfo {author} {\bibnamefont {Gunacker},
  \bibfnamefont {P.}}, \bibinfo {author} {\bibfnamefont {M.}~\bibnamefont
  {Wallerberger}}, \bibinfo {author} {\bibfnamefont {E.}~\bibnamefont {Gull}},
  \bibinfo {author} {\bibfnamefont {A.}~\bibnamefont {Hausoel}}, \bibinfo
  {author} {\bibfnamefont {G.}~\bibnamefont {Sangiovanni}}, \ and\ \bibinfo
  {author} {\bibfnamefont {K.}~\bibnamefont {Held}}} (\bibinfo {year} {2015}),\
  \href {\doibase 10.1103/PhysRevB.92.155102} {\bibfield  {journal} {\bibinfo
  {journal} {Phys. Rev. B}\ }\textbf {\bibinfo {volume} {92}},\ \bibinfo
  {pages} {155102}}\BibitemShut {NoStop}%
\bibitem [{\citenamefont {Gunacker}\ \emph {et~al.}(2016)\citenamefont
  {Gunacker}, \citenamefont {Wallerberger}, \citenamefont {Ribic},
  \citenamefont {Hausoel}, \citenamefont {Sangiovanni},\ and\ \citenamefont
  {Held}}]{Gunacker2016}%
  \BibitemOpen
  \bibfield  {author} {\bibinfo {author} {\bibnamefont {Gunacker},
  \bibfnamefont {P.}}, \bibinfo {author} {\bibfnamefont {M.}~\bibnamefont
  {Wallerberger}}, \bibinfo {author} {\bibfnamefont {T.}~\bibnamefont {Ribic}},
  \bibinfo {author} {\bibfnamefont {A.}~\bibnamefont {Hausoel}}, \bibinfo
  {author} {\bibfnamefont {G.}~\bibnamefont {Sangiovanni}}, \ and\ \bibinfo
  {author} {\bibfnamefont {K.}~\bibnamefont {Held}}} (\bibinfo {year} {2016}),\
  \href {\doibase 10.1103/PhysRevB.94.125153} {\bibfield  {journal} {\bibinfo
  {journal} {Phys. Rev. B}\ }\textbf {\bibinfo {volume} {94}},\ \bibinfo
  {pages} {125153}}\BibitemShut {NoStop}%
\bibitem [{\citenamefont {Gunnarsson}\ \emph {et~al.}(2017)\citenamefont
  {Gunnarsson}, \citenamefont {Rohringer}, \citenamefont {Sch\"afer},
  \citenamefont {Sangiovanni},\ and\ \citenamefont {Toschi}}]{Gunnarsson2017}%
  \BibitemOpen
  \bibfield  {author} {\bibinfo {author} {\bibnamefont {Gunnarsson},
  \bibfnamefont {O.}}, \bibinfo {author} {\bibfnamefont {G.}~\bibnamefont
  {Rohringer}}, \bibinfo {author} {\bibfnamefont {T.}~\bibnamefont
  {Sch\"afer}}, \bibinfo {author} {\bibfnamefont {G.}~\bibnamefont
  {Sangiovanni}}, \ and\ \bibinfo {author} {\bibfnamefont {A.}~\bibnamefont
  {Toschi}}} (\bibinfo {year} {2017}),\ \href {\doibase
  10.1103/PhysRevLett.119.056402} {\bibfield  {journal} {\bibinfo  {journal}
  {Phys. Rev. Lett.}\ }\textbf {\bibinfo {volume} {119}},\ \bibinfo {pages}
  {056402}}\BibitemShut {NoStop}%
\bibitem [{\citenamefont {Gunnarsson}\ \emph {et~al.}(2015)\citenamefont
  {Gunnarsson}, \citenamefont {Sch\"afer}, \citenamefont {LeBlanc},
  \citenamefont {Gull}, \citenamefont {Merino}, \citenamefont {Sangiovanni},
  \citenamefont {Rohringer},\ and\ \citenamefont {Toschi}}]{Gunnarsson2015}%
  \BibitemOpen
  \bibfield  {author} {\bibinfo {author} {\bibnamefont {Gunnarsson},
  \bibfnamefont {O.}}, \bibinfo {author} {\bibfnamefont {T.}~\bibnamefont
  {Sch\"afer}}, \bibinfo {author} {\bibfnamefont {J.~P.~F.}\ \bibnamefont
  {LeBlanc}}, \bibinfo {author} {\bibfnamefont {E.}~\bibnamefont {Gull}},
  \bibinfo {author} {\bibfnamefont {J.}~\bibnamefont {Merino}}, \bibinfo
  {author} {\bibfnamefont {G.}~\bibnamefont {Sangiovanni}}, \bibinfo {author}
  {\bibfnamefont {G.}~\bibnamefont {Rohringer}}, \ and\ \bibinfo {author}
  {\bibfnamefont {A.}~\bibnamefont {Toschi}}} (\bibinfo {year} {2015}),\ \href
  {\doibase 10.1103/PhysRevLett.114.236402} {\bibfield  {journal} {\bibinfo
  {journal} {Phys. Rev. Lett.}\ }\textbf {\bibinfo {volume} {114}},\ \bibinfo
  {pages} {236402}}\BibitemShut {NoStop}%
\bibitem [{\citenamefont {Gunnarsson}\ \emph {et~al.}(2016)\citenamefont
  {Gunnarsson}, \citenamefont {Sch\"afer}, \citenamefont {LeBlanc},
  \citenamefont {Merino}, \citenamefont {Sangiovanni}, \citenamefont
  {Rohringer},\ and\ \citenamefont {Toschi}}]{Gunnarsson2016}%
  \BibitemOpen
  \bibfield  {author} {\bibinfo {author} {\bibnamefont {Gunnarsson},
  \bibfnamefont {O.}}, \bibinfo {author} {\bibfnamefont {T.}~\bibnamefont
  {Sch\"afer}}, \bibinfo {author} {\bibfnamefont {J.~P.~F.}\ \bibnamefont
  {LeBlanc}}, \bibinfo {author} {\bibfnamefont {J.}~\bibnamefont {Merino}},
  \bibinfo {author} {\bibfnamefont {G.}~\bibnamefont {Sangiovanni}}, \bibinfo
  {author} {\bibfnamefont {G.}~\bibnamefont {Rohringer}}, \ and\ \bibinfo
  {author} {\bibfnamefont {A.}~\bibnamefont {Toschi}}} (\bibinfo {year}
  {2016}),\ \href {\doibase 10.1103/PhysRevB.93.245102} {\bibfield  {journal}
  {\bibinfo  {journal} {Phys. Rev. B}\ }\textbf {\bibinfo {volume} {93}},\
  \bibinfo {pages} {245102}}\BibitemShut {NoStop}%
\bibitem [{\citenamefont {Haase}\ \emph {et~al.}(2017)\citenamefont {Haase},
  \citenamefont {Yang}, \citenamefont {Pruschke}, \citenamefont {Moreno},\ and\
  \citenamefont {Jarrell}}]{Haase2016}%
  \BibitemOpen
  \bibfield  {author} {\bibinfo {author} {\bibnamefont {Haase}, \bibfnamefont
  {P.}}, \bibinfo {author} {\bibfnamefont {S.-X.}\ \bibnamefont {Yang}},
  \bibinfo {author} {\bibfnamefont {T.}~\bibnamefont {Pruschke}}, \bibinfo
  {author} {\bibfnamefont {J.}~\bibnamefont {Moreno}}, \ and\ \bibinfo {author}
  {\bibfnamefont {M.}~\bibnamefont {Jarrell}}} (\bibinfo {year} {2017}),\ \href
  {\doibase 10.1103/PhysRevB.95.045130} {\bibfield  {journal} {\bibinfo
  {journal} {Phys. Rev. B}\ }\textbf {\bibinfo {volume} {95}},\ \bibinfo
  {pages} {045130}}\BibitemShut {NoStop}%
\bibitem [{\citenamefont {Hafermann}(2010)}]{Hafermannphd}%
  \BibitemOpen
  \bibfield  {author} {\bibinfo {author} {\bibnamefont {Hafermann},
  \bibfnamefont {H.}}} (\bibinfo {year} {2010}),\ \emph {\bibinfo {title}
  {Numerical Approaches to Spatial Correlations in Strongly Interacting Fermion
  Systems}},\ \href@noop {} {Ph.D. thesis}\ (\bibinfo  {school} {University of
  Hamburg})\BibitemShut {NoStop}%
\bibitem [{\citenamefont {Hafermann}(2014)}]{Hafermann2014}%
  \BibitemOpen
  \bibfield  {author} {\bibinfo {author} {\bibnamefont {Hafermann},
  \bibfnamefont {H.}}} (\bibinfo {year} {2014}),\ \href {\doibase
  10.1103/PhysRevB.89.235128} {\bibfield  {journal} {\bibinfo  {journal} {Phys.
  Rev. B}\ }\textbf {\bibinfo {volume} {89}},\ \bibinfo {pages}
  {235128}}\BibitemShut {NoStop}%
\bibitem [{\citenamefont {Hafermann}\ \emph {et~al.}(2008)\citenamefont
  {Hafermann}, \citenamefont {Brener}, \citenamefont {Rubtsov}, \citenamefont
  {Katsnelson},\ and\ \citenamefont {Lichtenstein}}]{Hafermann2008}%
  \BibitemOpen
  \bibfield  {author} {\bibinfo {author} {\bibnamefont {Hafermann},
  \bibfnamefont {H.}}, \bibinfo {author} {\bibfnamefont {S.}~\bibnamefont
  {Brener}}, \bibinfo {author} {\bibfnamefont {A.}~\bibnamefont {Rubtsov}},
  \bibinfo {author} {\bibfnamefont {M.}~\bibnamefont {Katsnelson}}, \ and\
  \bibinfo {author} {\bibfnamefont {A.}~\bibnamefont {Lichtenstein}}} (\bibinfo
  {year} {2008}),\ \href {http://dx.doi.org/10.1134/S0021364007220134}
  {\bibfield  {journal} {\bibinfo  {journal} {JETP Letters}\ }\textbf {\bibinfo
  {volume} {86}},\ \bibinfo {pages} {677}}\BibitemShut {NoStop}%
\bibitem [{\citenamefont {Hafermann}\ \emph
  {et~al.}(2009{\natexlab{a}})\citenamefont {Hafermann}, \citenamefont {Jung},
  \citenamefont {Brener}, \citenamefont {Katsnelson}, \citenamefont {Rubtsov},\
  and\ \citenamefont {Lichtenstein}}]{Hafermann2009b}%
  \BibitemOpen
  \bibfield  {author} {\bibinfo {author} {\bibnamefont {Hafermann},
  \bibfnamefont {H.}}, \bibinfo {author} {\bibfnamefont {C.}~\bibnamefont
  {Jung}}, \bibinfo {author} {\bibfnamefont {S.}~\bibnamefont {Brener}},
  \bibinfo {author} {\bibfnamefont {M.~I.}\ \bibnamefont {Katsnelson}},
  \bibinfo {author} {\bibfnamefont {A.~N.}\ \bibnamefont {Rubtsov}}, \ and\
  \bibinfo {author} {\bibfnamefont {A.~I.}\ \bibnamefont {Lichtenstein}}}
  (\bibinfo {year} {2009}{\natexlab{a}}),\ \href
  {http://stacks.iop.org/0295-5075/85/i=2/a=27007} {\bibfield  {journal}
  {\bibinfo  {journal} {Europhysics Letters}\ }\textbf {\bibinfo {volume}
  {85}},\ \bibinfo {pages} {27007}}\BibitemShut {NoStop}%
\bibitem [{\citenamefont {Hafermann}\ \emph
  {et~al.}(2009{\natexlab{b}})\citenamefont {Hafermann}, \citenamefont
  {Kecker}, \citenamefont {Brener}, \citenamefont {Rubtsov}, \citenamefont
  {Katsnelson},\ and\ \citenamefont {Lichtenstein}}]{Hafermann2009a}%
  \BibitemOpen
  \bibfield  {author} {\bibinfo {author} {\bibnamefont {Hafermann},
  \bibfnamefont {H.}}, \bibinfo {author} {\bibfnamefont {M.}~\bibnamefont
  {Kecker}}, \bibinfo {author} {\bibfnamefont {S.}~\bibnamefont {Brener}},
  \bibinfo {author} {\bibfnamefont {A.~N.}\ \bibnamefont {Rubtsov}}, \bibinfo
  {author} {\bibfnamefont {M.~I.}\ \bibnamefont {Katsnelson}}, \ and\ \bibinfo
  {author} {\bibfnamefont {A.~I.}\ \bibnamefont {Lichtenstein}}} (\bibinfo
  {year} {2009}{\natexlab{b}}),\ \href
  {http://dx.doi.org/10.1007/s10948-008-0361-9} {\bibfield  {journal} {\bibinfo
   {journal} {J Supercond Nov Magn}\ }\textbf {\bibinfo {volume} {22}},\
  \bibinfo {pages} {45}}\BibitemShut {NoStop}%
\bibitem [{\citenamefont {Hafermann}\ \emph
  {et~al.}(2009{\natexlab{c}})\citenamefont {Hafermann}, \citenamefont {Li},
  \citenamefont {Rubtsov}, \citenamefont {Katsnelson}, \citenamefont
  {Lichtenstein},\ and\ \citenamefont {Monien}}]{Hafermann2009}%
  \BibitemOpen
  \bibfield  {author} {\bibinfo {author} {\bibnamefont {Hafermann},
  \bibfnamefont {H.}}, \bibinfo {author} {\bibfnamefont {G.}~\bibnamefont
  {Li}}, \bibinfo {author} {\bibfnamefont {A.~N.}\ \bibnamefont {Rubtsov}},
  \bibinfo {author} {\bibfnamefont {M.~I.}\ \bibnamefont {Katsnelson}},
  \bibinfo {author} {\bibfnamefont {A.~I.}\ \bibnamefont {Lichtenstein}}, \
  and\ \bibinfo {author} {\bibfnamefont {H.}~\bibnamefont {Monien}}} (\bibinfo
  {year} {2009}{\natexlab{c}}),\ \href {\doibase
  10.1103/PhysRevLett.102.206401} {\bibfield  {journal} {\bibinfo  {journal}
  {Phys. Rev. Lett.}\ }\textbf {\bibinfo {volume} {102}},\ \bibinfo {pages}
  {206401}}\BibitemShut {NoStop}%
\bibitem [{\citenamefont {Hafermann}\ \emph {et~al.}(2014)\citenamefont
  {Hafermann}, \citenamefont {van Loon}, \citenamefont {Katsnelson},
  \citenamefont {Lichtenstein},\ and\ \citenamefont
  {Parcollet}}]{Hafermann2014a}%
  \BibitemOpen
  \bibfield  {author} {\bibinfo {author} {\bibnamefont {Hafermann},
  \bibfnamefont {H.}}, \bibinfo {author} {\bibfnamefont {E.~G. C.~P.}\
  \bibnamefont {van Loon}}, \bibinfo {author} {\bibfnamefont {M.~I.}\
  \bibnamefont {Katsnelson}}, \bibinfo {author} {\bibfnamefont {A.~I.}\
  \bibnamefont {Lichtenstein}}, \ and\ \bibinfo {author} {\bibfnamefont
  {O.}~\bibnamefont {Parcollet}}} (\bibinfo {year} {2014}),\ \href {\doibase
  10.1103/PhysRevB.90.235105} {\bibfield  {journal} {\bibinfo  {journal} {Phys.
  Rev. B}\ }\textbf {\bibinfo {volume} {90}},\ \bibinfo {pages}
  {235105}}\BibitemShut {NoStop}%
\bibitem [{\citenamefont {Hafermann}\ \emph {et~al.}(2012)\citenamefont
  {Hafermann}, \citenamefont {Patton},\ and\ \citenamefont
  {Werner}}]{Hafermann2012}%
  \BibitemOpen
  \bibfield  {author} {\bibinfo {author} {\bibnamefont {Hafermann},
  \bibfnamefont {H.}}, \bibinfo {author} {\bibfnamefont {K.~R.}\ \bibnamefont
  {Patton}}, \ and\ \bibinfo {author} {\bibfnamefont {P.}~\bibnamefont
  {Werner}}} (\bibinfo {year} {2012}),\ \href {\doibase
  10.1103/PhysRevB.85.205106} {\bibfield  {journal} {\bibinfo  {journal} {Phys.
  Rev. B}\ }\textbf {\bibinfo {volume} {85}},\ \bibinfo {pages}
  {205106}}\BibitemShut {NoStop}%
\bibitem [{\citenamefont {Hafermann}\ \emph {et~al.}(2013)\citenamefont
  {Hafermann}, \citenamefont {Werner},\ and\ \citenamefont
  {Gull}}]{Hafermann2013}%
  \BibitemOpen
  \bibfield  {author} {\bibinfo {author} {\bibnamefont {Hafermann},
  \bibfnamefont {H.}}, \bibinfo {author} {\bibfnamefont {P.}~\bibnamefont
  {Werner}}, \ and\ \bibinfo {author} {\bibfnamefont {E.}~\bibnamefont {Gull}}}
  (\bibinfo {year} {2013}),\ \href {\doibase 10.1016/j.cpc.2012.12.013}
  {\bibfield  {journal} {\bibinfo  {journal} {Computer Physics Communications}\
  }\textbf {\bibinfo {volume} {184}}~(\bibinfo {number} {4}),\ \bibinfo {pages}
  {1280 }}\BibitemShut {NoStop}%
\bibitem [{\citenamefont {Hague}\ \emph {et~al.}(2004)\citenamefont {Hague},
  \citenamefont {Jarrell},\ and\ \citenamefont {Schulthess}}]{Hague04}%
  \BibitemOpen
  \bibfield  {author} {\bibinfo {author} {\bibnamefont {Hague}, \bibfnamefont
  {J.~P.}}, \bibinfo {author} {\bibfnamefont {M.}~\bibnamefont {Jarrell}}, \
  and\ \bibinfo {author} {\bibfnamefont {T.~C.}\ \bibnamefont {Schulthess}}}
  (\bibinfo {year} {2004}),\ \href {\doibase 10.1103/PhysRevB.69.165113}
  {\bibfield  {journal} {\bibinfo  {journal} {Phys. Rev. B}\ }\textbf {\bibinfo
  {volume} {69}},\ \bibinfo {pages} {165113}}\BibitemShut {NoStop}%
\bibitem [{\citenamefont {Hansmann}\ \emph {et~al.}(2016)\citenamefont
  {Hansmann}, \citenamefont {Ayral}, \citenamefont {Tejeda},\ and\
  \citenamefont {Biermann}}]{Hansmann2016}%
  \BibitemOpen
  \bibfield  {author} {\bibinfo {author} {\bibnamefont {Hansmann},
  \bibfnamefont {P.}}, \bibinfo {author} {\bibfnamefont {T.}~\bibnamefont
  {Ayral}}, \bibinfo {author} {\bibfnamefont {A.}~\bibnamefont {Tejeda}}, \
  and\ \bibinfo {author} {\bibfnamefont {S.}~\bibnamefont {Biermann}}}
  (\bibinfo {year} {{2016}}),\ \href {\doibase 10.1038/srep19728} {\bibfield
  {journal} {\bibinfo  {journal} {Scientific Reports}\ }\textbf {\bibinfo
  {volume} {{6}}},\ 10.1038/srep19728}\BibitemShut {NoStop}%
\bibitem [{\citenamefont {Hansmann}\ \emph {et~al.}(2013)\citenamefont
  {Hansmann}, \citenamefont {Ayral}, \citenamefont {Vaugier}, \citenamefont
  {Werner},\ and\ \citenamefont {Biermann}}]{Hansmann2013}%
  \BibitemOpen
  \bibfield  {author} {\bibinfo {author} {\bibnamefont {Hansmann},
  \bibfnamefont {P.}}, \bibinfo {author} {\bibfnamefont {T.}~\bibnamefont
  {Ayral}}, \bibinfo {author} {\bibfnamefont {L.}~\bibnamefont {Vaugier}},
  \bibinfo {author} {\bibfnamefont {P.}~\bibnamefont {Werner}}, \ and\ \bibinfo
  {author} {\bibfnamefont {S.}~\bibnamefont {Biermann}}} (\bibinfo {year}
  {2013}),\ \href {\doibase 10.1103/PhysRevLett.110.166401} {\bibfield
  {journal} {\bibinfo  {journal} {Phys. Rev. Lett.}\ }\textbf {\bibinfo
  {volume} {110}},\ \bibinfo {pages} {166401}}\BibitemShut {NoStop}%
\bibitem [{\citenamefont {Harland}\ \emph {et~al.}(2016)\citenamefont
  {Harland}, \citenamefont {Katsnelson},\ and\ \citenamefont
  {Lichtenstein}}]{Harland2016}%
  \BibitemOpen
  \bibfield  {author} {\bibinfo {author} {\bibnamefont {Harland}, \bibfnamefont
  {M.}}, \bibinfo {author} {\bibfnamefont {M.~I.}\ \bibnamefont {Katsnelson}},
  \ and\ \bibinfo {author} {\bibfnamefont {A.~I.}\ \bibnamefont
  {Lichtenstein}}} (\bibinfo {year} {2016}),\ \href {\doibase
  10.1103/PhysRevB.94.125133} {\bibfield  {journal} {\bibinfo  {journal} {Phys.
  Rev. B}\ }\textbf {\bibinfo {volume} {94}},\ \bibinfo {pages}
  {125133}}\BibitemShut {NoStop}%
\bibitem [{\citenamefont {Haule}(2007)}]{Haule2007}%
  \BibitemOpen
  \bibfield  {author} {\bibinfo {author} {\bibnamefont {Haule}, \bibfnamefont
  {K.}}} (\bibinfo {year} {2007}),\ \href {\doibase 10.1103/PhysRevB.75.155113}
  {\bibfield  {journal} {\bibinfo  {journal} {Phys. Rev. B}\ }\textbf {\bibinfo
  {volume} {75}},\ \bibinfo {pages} {155113}}\BibitemShut {NoStop}%
\bibitem [{\citenamefont {Haule}\ and\ \citenamefont
  {Birol}(2015)}]{Haule2015}%
  \BibitemOpen
  \bibfield  {author} {\bibinfo {author} {\bibnamefont {Haule}, \bibfnamefont
  {K.}}, \ and\ \bibinfo {author} {\bibfnamefont {T.}~\bibnamefont {Birol}}}
  (\bibinfo {year} {2015}),\ \href {\doibase 10.1103/PhysRevLett.115.256402}
  {\bibfield  {journal} {\bibinfo  {journal} {Phys. Rev. Lett.}\ }\textbf
  {\bibinfo {volume} {115}},\ \bibinfo {pages} {256402}}\BibitemShut {NoStop}%
\bibitem [{\citenamefont {Haule}\ and\ \citenamefont
  {Kotliar}(2007)}]{Haule2007b}%
  \BibitemOpen
  \bibfield  {author} {\bibinfo {author} {\bibnamefont {Haule}, \bibfnamefont
  {K.}}, \ and\ \bibinfo {author} {\bibfnamefont {G.}~\bibnamefont {Kotliar}}}
  (\bibinfo {year} {2007}),\ \href {\doibase 10.1103/PhysRevB.76.104509}
  {\bibfield  {journal} {\bibinfo  {journal} {Phys. Rev. B}\ }\textbf {\bibinfo
  {volume} {76}},\ \bibinfo {pages} {104509}}\BibitemShut {NoStop}%
\bibitem [{\citenamefont {Hedin}(1965)}]{Hedin1965}%
  \BibitemOpen
  \bibfield  {author} {\bibinfo {author} {\bibnamefont {Hedin}, \bibfnamefont
  {L.}}} (\bibinfo {year} {1965}),\ \href {\doibase 10.1103/PhysRev.139.A796}
  {\bibfield  {journal} {\bibinfo  {journal} {Phys. Rev.}\ }\textbf {\bibinfo
  {volume} {139}},\ \bibinfo {pages} {A796}}\BibitemShut {NoStop}%
\bibitem [{\citenamefont {Hedin}(1999)}]{Hedin1999}%
  \BibitemOpen
  \bibfield  {author} {\bibinfo {author} {\bibnamefont {Hedin}, \bibfnamefont
  {L.}}} (\bibinfo {year} {1999}),\ \href
  {http://stacks.iop.org/0953-8984/11/i=42/a=201} {\bibfield  {journal}
  {\bibinfo  {journal} {J. Phys.: Condens. Matter}\ }\textbf {\bibinfo {volume}
  {11}}~(\bibinfo {number} {42}),\ \bibinfo {pages} {R489}}\BibitemShut
  {NoStop}%
\bibitem [{\citenamefont {Held}(2007)}]{Held2007}%
  \BibitemOpen
  \bibfield  {author} {\bibinfo {author} {\bibnamefont {Held}, \bibfnamefont
  {K.}}} (\bibinfo {year} {2007}),\ \href {\doibase 10.1080/00018730701619647}
  {\bibfield  {journal} {\bibinfo  {journal} {Advances in Physics}\ }\textbf
  {\bibinfo {volume} {56}},\ \bibinfo {pages} {829}}\BibitemShut {NoStop}%
\bibitem [{\citenamefont {Held}(2014)}]{Held2014}%
  \BibitemOpen
  \bibfield  {author} {\bibinfo {author} {\bibnamefont {Held}, \bibfnamefont
  {K.}}} (\bibinfo {year} {2014}),\ \enquote {\bibinfo {title} {Autumn {S}chool
  on {C}orrelated {E}lectrons. {DMFT} at 25: {I}nfinite {D}imensions},}\ Chap.\
  \bibinfo {chapter} {Dynamical vertex approximation}\ (\bibinfo  {publisher}
  {Forschungszentrum J{\"u}lich})\ \bibinfo {note}
  {[arXiv:1411.5191]}\BibitemShut {NoStop}%
\bibitem [{\citenamefont {Held}\ \emph {et~al.}(2008)\citenamefont {Held},
  \citenamefont {Katanin},\ and\ \citenamefont {Toschi}}]{Held2008}%
  \BibitemOpen
  \bibfield  {author} {\bibinfo {author} {\bibnamefont {Held}, \bibfnamefont
  {K.}}, \bibinfo {author} {\bibfnamefont {A.}~\bibnamefont {Katanin}}, \ and\
  \bibinfo {author} {\bibfnamefont {A.}~\bibnamefont {Toschi}}} (\bibinfo
  {year} {2008}),\ \href
  {http://ptps.oxfordjournals.org/content/176/117.abstract} {\bibfield
  {journal} {\bibinfo  {journal} {Progress of Theoretical Physics
  (Supplement)}\ }\textbf {\bibinfo {volume} {176}},\ \bibinfo {pages}
  {117}}\BibitemShut {NoStop}%
\bibitem [{\citenamefont {Held}\ \emph {et~al.}(2006)\citenamefont {Held},
  \citenamefont {Nekrasov}, \citenamefont {Keller}, \citenamefont {Eyert},
  \citenamefont {Blümer}, \citenamefont {McMahan}, \citenamefont {Scalettar},
  \citenamefont {Pruschke}, \citenamefont {Anisimov},\ and\ \citenamefont
  {Vollhardt}}]{Held2006}%
  \BibitemOpen
  \bibfield  {author} {\bibinfo {author} {\bibnamefont {Held}, \bibfnamefont
  {K.}}, \bibinfo {author} {\bibfnamefont {I.~A.}\ \bibnamefont {Nekrasov}},
  \bibinfo {author} {\bibfnamefont {G.}~\bibnamefont {Keller}}, \bibinfo
  {author} {\bibfnamefont {V.}~\bibnamefont {Eyert}}, \bibinfo {author}
  {\bibfnamefont {N.}~\bibnamefont {Blümer}}, \bibinfo {author} {\bibfnamefont
  {A.~K.}\ \bibnamefont {McMahan}}, \bibinfo {author} {\bibfnamefont {R.~T.}\
  \bibnamefont {Scalettar}}, \bibinfo {author} {\bibfnamefont {T.}~\bibnamefont
  {Pruschke}}, \bibinfo {author} {\bibfnamefont {V.~I.}\ \bibnamefont
  {Anisimov}}, \ and\ \bibinfo {author} {\bibfnamefont {D.}~\bibnamefont
  {Vollhardt}}} (\bibinfo {year} {2006}),\ \href {\doibase
  10.1002/pssb.200642053} {\bibfield  {journal} {\bibinfo  {journal} {physica
  status solidi (b)}\ }\textbf {\bibinfo {volume} {243}}~(\bibinfo {number}
  {11}),\ \bibinfo {pages} {2599}},\ \bibinfo {note} {previously appeared as
  Psi-k Newsletter No. 56 (April 2003)}\BibitemShut {NoStop}%
\bibitem [{\citenamefont {Hertz}(1976)}]{Hertz1976}%
  \BibitemOpen
  \bibfield  {author} {\bibinfo {author} {\bibnamefont {Hertz}, \bibfnamefont
  {J.~A.}}} (\bibinfo {year} {1976}),\ \href {\doibase
  10.1103/PhysRevB.14.1165} {\bibfield  {journal} {\bibinfo  {journal} {Phys.
  Rev. B}\ }\textbf {\bibinfo {volume} {14}},\ \bibinfo {pages}
  {1165}}\BibitemShut {NoStop}%
\bibitem [{\citenamefont {Hertz}\ and\ \citenamefont
  {Edwards}(1973)}]{Hertz1973}%
  \BibitemOpen
  \bibfield  {author} {\bibinfo {author} {\bibnamefont {Hertz}, \bibfnamefont
  {J.~A.}}, \ and\ \bibinfo {author} {\bibfnamefont {D.~M.}\ \bibnamefont
  {Edwards}}} (\bibinfo {year} {1973}),\ \href
  {http://stacks.iop.org/0305-4608/3/i=12/a=018} {\bibfield  {journal}
  {\bibinfo  {journal} {J. Phys. F: Met. Phys.}\ }\textbf {\bibinfo {volume}
  {3}},\ \bibinfo {pages} {2174}}\BibitemShut {NoStop}%
\bibitem [{\citenamefont {Hettler}\ \emph {et~al.}(2000)\citenamefont
  {Hettler}, \citenamefont {Mukherjee}, \citenamefont {Jarrell},\ and\
  \citenamefont {Krishnamurthy}}]{Hettler2000}%
  \BibitemOpen
  \bibfield  {author} {\bibinfo {author} {\bibnamefont {Hettler}, \bibfnamefont
  {M.~H.}}, \bibinfo {author} {\bibfnamefont {M.}~\bibnamefont {Mukherjee}},
  \bibinfo {author} {\bibfnamefont {M.}~\bibnamefont {Jarrell}}, \ and\
  \bibinfo {author} {\bibfnamefont {H.~R.}\ \bibnamefont {Krishnamurthy}}}
  (\bibinfo {year} {2000}),\ \href {\doibase 10.1103/PhysRevB.61.12739}
  {\bibfield  {journal} {\bibinfo  {journal} {Phys. Rev. B}\ }\textbf {\bibinfo
  {volume} {61}},\ \bibinfo {pages} {12739}}\BibitemShut {NoStop}%
\bibitem [{\citenamefont {Hettler}\ \emph {et~al.}(1998)\citenamefont
  {Hettler}, \citenamefont {Tahvildar-Zadeh}, \citenamefont {Jarrell},
  \citenamefont {Pruschke},\ and\ \citenamefont {Krishnamurthy}}]{Hettler1998}%
  \BibitemOpen
  \bibfield  {author} {\bibinfo {author} {\bibnamefont {Hettler}, \bibfnamefont
  {M.~H.}}, \bibinfo {author} {\bibfnamefont {A.~N.}\ \bibnamefont
  {Tahvildar-Zadeh}}, \bibinfo {author} {\bibfnamefont {M.}~\bibnamefont
  {Jarrell}}, \bibinfo {author} {\bibfnamefont {T.}~\bibnamefont {Pruschke}}, \
  and\ \bibinfo {author} {\bibfnamefont {H.~R.}\ \bibnamefont {Krishnamurthy}}}
  (\bibinfo {year} {1998}),\ \href {\doibase 10.1103/PhysRevB.58.R7475}
  {\bibfield  {journal} {\bibinfo  {journal} {Phys. Rev. B}\ }\textbf {\bibinfo
  {volume} {58}},\ \bibinfo {pages} {R7475}}\BibitemShut {NoStop}%
\bibitem [{\citenamefont {Hirsch}\ and\ \citenamefont
  {Fye}(1986)}]{Hirsch1986}%
  \BibitemOpen
  \bibfield  {author} {\bibinfo {author} {\bibnamefont {Hirsch}, \bibfnamefont
  {J.~E.}}, \ and\ \bibinfo {author} {\bibfnamefont {R.~M.}\ \bibnamefont
  {Fye}}} (\bibinfo {year} {1986}),\ \href {\doibase
  10.1103/PhysRevLett.56.2521} {\bibfield  {journal} {\bibinfo  {journal}
  {Phys. Rev. Lett.}\ }\textbf {\bibinfo {volume} {56}},\ \bibinfo {pages}
  {2521}}\BibitemShut {NoStop}%
\bibitem [{\citenamefont {Hirschmeier}\ \emph {et~al.}(2015)\citenamefont
  {Hirschmeier}, \citenamefont {Hafermann}, \citenamefont {Gull}, \citenamefont
  {Lichtenstein},\ and\ \citenamefont {Antipov}}]{Hirschmeier2015}%
  \BibitemOpen
  \bibfield  {author} {\bibinfo {author} {\bibnamefont {Hirschmeier},
  \bibfnamefont {D.}}, \bibinfo {author} {\bibfnamefont {H.}~\bibnamefont
  {Hafermann}}, \bibinfo {author} {\bibfnamefont {E.}~\bibnamefont {Gull}},
  \bibinfo {author} {\bibfnamefont {A.~I.}\ \bibnamefont {Lichtenstein}}, \
  and\ \bibinfo {author} {\bibfnamefont {A.~E.}\ \bibnamefont {Antipov}}}
  (\bibinfo {year} {2015}),\ \href {\doibase 10.1103/PhysRevB.92.144409}
  {\bibfield  {journal} {\bibinfo  {journal} {Phys. Rev. B}\ }\textbf {\bibinfo
  {volume} {92}},\ \bibinfo {pages} {144409}}\BibitemShut {NoStop}%
\bibitem [{\citenamefont {Hirschmeier}\ \emph {et~al.}(2018)\citenamefont
  {Hirschmeier}, \citenamefont {Hafermann},\ and\ \citenamefont
  {Lichtenstein}}]{Hirschmeier2018}%
  \BibitemOpen
  \bibfield  {author} {\bibinfo {author} {\bibnamefont {Hirschmeier},
  \bibfnamefont {D.}}, \bibinfo {author} {\bibfnamefont {H.}~\bibnamefont
  {Hafermann}}, \ and\ \bibinfo {author} {\bibfnamefont {A.~I.}\ \bibnamefont
  {Lichtenstein}}} (\bibinfo {year} {2018}),\ \href {\doibase
  10.1103/PhysRevB.97.115150} {\bibfield  {journal} {\bibinfo  {journal} {Phys.
  Rev. B}\ }\textbf {\bibinfo {volume} {97}},\ \bibinfo {pages}
  {115150}}\BibitemShut {NoStop}%
\bibitem [{\citenamefont {Holm}\ and\ \citenamefont {Janke}(1993)}]{Holm1993}%
  \BibitemOpen
  \bibfield  {author} {\bibinfo {author} {\bibnamefont {Holm}, \bibfnamefont
  {C.}}, \ and\ \bibinfo {author} {\bibfnamefont {W.}~\bibnamefont {Janke}}}
  (\bibinfo {year} {1993}),\ \href {\doibase 10.1103/PhysRevB.48.936}
  {\bibfield  {journal} {\bibinfo  {journal} {Phys. Rev. B}\ }\textbf {\bibinfo
  {volume} {48}},\ \bibinfo {pages} {936}}\BibitemShut {NoStop}%
\bibitem [{\citenamefont {Huang}\ \emph {et~al.}(2014)\citenamefont {Huang},
  \citenamefont {Ayral}, \citenamefont {Biermann},\ and\ \citenamefont
  {Werner}}]{Huang14}%
  \BibitemOpen
  \bibfield  {author} {\bibinfo {author} {\bibnamefont {Huang}, \bibfnamefont
  {L.}}, \bibinfo {author} {\bibfnamefont {T.}~\bibnamefont {Ayral}}, \bibinfo
  {author} {\bibfnamefont {S.}~\bibnamefont {Biermann}}, \ and\ \bibinfo
  {author} {\bibfnamefont {P.}~\bibnamefont {Werner}}} (\bibinfo {year}
  {2014}),\ \href {\doibase 10.1103/PhysRevB.90.195114} {\bibfield  {journal}
  {\bibinfo  {journal} {Phys. Rev. B}\ }\textbf {\bibinfo {volume} {90}},\
  \bibinfo {pages} {195114}}\BibitemShut {NoStop}%
\bibitem [{\citenamefont {Huang}\ \emph {et~al.}(2015)\citenamefont {Huang},
  \citenamefont {Wang}, \citenamefont {Meng}, \citenamefont {Du}, \citenamefont
  {Werner},\ and\ \citenamefont {Dai}}]{Huang2015}%
  \BibitemOpen
  \bibfield  {author} {\bibinfo {author} {\bibnamefont {Huang}, \bibfnamefont
  {L.}}, \bibinfo {author} {\bibfnamefont {Y.}~\bibnamefont {Wang}}, \bibinfo
  {author} {\bibfnamefont {Z.~Y.}\ \bibnamefont {Meng}}, \bibinfo {author}
  {\bibfnamefont {L.}~\bibnamefont {Du}}, \bibinfo {author} {\bibfnamefont
  {P.}~\bibnamefont {Werner}}, \ and\ \bibinfo {author} {\bibfnamefont
  {X.}~\bibnamefont {Dai}}} (\bibinfo {year} {2015}),\ \href {\doibase
  10.1016/j.cpc.2015.04.020} {\bibfield  {journal} {\bibinfo  {journal}
  {Computer Physics Communications}\ }\textbf {\bibinfo {volume} {195}},\
  \bibinfo {pages} {140 }}\BibitemShut {NoStop}%
\bibitem [{\citenamefont {Hugenholtz}(1957)}]{Hugenholtz1957}%
  \BibitemOpen
  \bibfield  {author} {\bibinfo {author} {\bibnamefont {Hugenholtz},
  \bibfnamefont {N.~M.}}} (\bibinfo {year} {1957}),\ \href {\doibase
  doi:10.1016/S0031-8914(57)92950-6} {\bibfield  {journal} {\bibinfo  {journal}
  {Physica}\ }\textbf {\bibinfo {volume} {23}}~(\bibinfo {number} {1-5}),\
  \bibinfo {pages} {481}}\BibitemShut {NoStop}%
\bibitem [{\citenamefont {Imada}\ \emph {et~al.}(1998)\citenamefont {Imada},
  \citenamefont {Fujimori},\ and\ \citenamefont {Tokura}}]{Imada1998}%
  \BibitemOpen
  \bibfield  {author} {\bibinfo {author} {\bibnamefont {Imada}, \bibfnamefont
  {M.}}, \bibinfo {author} {\bibfnamefont {A.}~\bibnamefont {Fujimori}}, \ and\
  \bibinfo {author} {\bibfnamefont {Y.}~\bibnamefont {Tokura}}} (\bibinfo
  {year} {1998}),\ \href {\doibase 10.1103/RevModPhys.70.1039} {\bibfield
  {journal} {\bibinfo  {journal} {Rev. Mod. Phys.}\ }\textbf {\bibinfo {volume}
  {70}},\ \bibinfo {pages} {1039}}\BibitemShut {NoStop}%
\bibitem [{\citenamefont {Iskakov}\ \emph {et~al.}(2016)\citenamefont
  {Iskakov}, \citenamefont {Antipov},\ and\ \citenamefont
  {Gull}}]{Iskakov2016}%
  \BibitemOpen
  \bibfield  {author} {\bibinfo {author} {\bibnamefont {Iskakov}, \bibfnamefont
  {S.}}, \bibinfo {author} {\bibfnamefont {A.~E.}\ \bibnamefont {Antipov}}, \
  and\ \bibinfo {author} {\bibfnamefont {E.}~\bibnamefont {Gull}}} (\bibinfo
  {year} {2016}),\ \href {\doibase 10.1103/PhysRevB.94.035102} {\bibfield
  {journal} {\bibinfo  {journal} {Phys. Rev. B}\ }\textbf {\bibinfo {volume}
  {94}},\ \bibinfo {pages} {035102}}\BibitemShut {NoStop}%
\bibitem [{\citenamefont {Iskakov}\ \emph {et~al.}(2018)\citenamefont
  {Iskakov}, \citenamefont {Terletska},\ and\ \citenamefont
  {Gull}}]{Iskakov2018}%
  \BibitemOpen
  \bibfield  {author} {\bibinfo {author} {\bibnamefont {Iskakov}, \bibfnamefont
  {S.}}, \bibinfo {author} {\bibfnamefont {H.}~\bibnamefont {Terletska}}, \
  and\ \bibinfo {author} {\bibfnamefont {E.}~\bibnamefont {Gull}}} (\bibinfo
  {year} {2018}),\ \href {\doibase 10.1103/PhysRevB.97.125114} {\bibfield
  {journal} {\bibinfo  {journal} {Phys. Rev. B}\ }\textbf {\bibinfo {volume}
  {97}},\ \bibinfo {pages} {125114}}\BibitemShut {NoStop}%
\bibitem [{\citenamefont {Jacob}\ \emph {et~al.}(2010)\citenamefont {Jacob},
  \citenamefont {Haule},\ and\ \citenamefont {Kotliar}}]{Jacob2010}%
  \BibitemOpen
  \bibfield  {author} {\bibinfo {author} {\bibnamefont {Jacob}, \bibfnamefont
  {D.}}, \bibinfo {author} {\bibfnamefont {K.}~\bibnamefont {Haule}}, \ and\
  \bibinfo {author} {\bibfnamefont {G.}~\bibnamefont {Kotliar}}} (\bibinfo
  {year} {2010}),\ \href {\doibase 10.1103/PhysRevB.82.195115} {\bibfield
  {journal} {\bibinfo  {journal} {Phys. Rev. B}\ }\textbf {\bibinfo {volume}
  {82}},\ \bibinfo {pages} {195115}}\BibitemShut {NoStop}%
\bibitem [{\citenamefont {Jani{\v{s}}}(1991)}]{Janis1991}%
  \BibitemOpen
  \bibfield  {author} {\bibinfo {author} {\bibnamefont {Jani{\v{s}}},
  \bibfnamefont {V.}}} (\bibinfo {year} {1991}),\ \href {\doibase
  10.1007/BF01309423} {\bibfield  {journal} {\bibinfo  {journal} {Zeitschrift
  f{\"u}r Physik B Condensed Matter}\ }\textbf {\bibinfo {volume}
  {83}}~(\bibinfo {number} {2}),\ \bibinfo {pages} {227}}\BibitemShut {NoStop}%
\bibitem [{\citenamefont {Jani{\v{s}}}(1998)}]{Janis1998}%
  \BibitemOpen
  \bibfield  {author} {\bibinfo {author} {\bibnamefont {Jani{\v{s}}},
  \bibfnamefont {V.}}} (\bibinfo {year} {1998}),\ \href
  {https://arxiv.org/abs/cond-mat/9806118} {\ }\Eprint
  {http://arxiv.org/abs/cond-mat/9806118} {arXiv:cond-mat/9806118} \BibitemShut
  {NoStop}%
\bibitem [{\citenamefont {Jani{\v{s}}}(1999{\natexlab{a}})}]{Janis1999a}%
  \BibitemOpen
  \bibfield  {author} {\bibinfo {author} {\bibnamefont {Jani{\v{s}}},
  \bibfnamefont {V.}}} (\bibinfo {year} {1999}{\natexlab{a}}),\ \href {\doibase
  10.1103/PhysRevLett.83.2781} {\bibfield  {journal} {\bibinfo  {journal}
  {Phys. Rev. Lett.}\ }\textbf {\bibinfo {volume} {83}},\ \bibinfo {pages}
  {2781}}\BibitemShut {NoStop}%
\bibitem [{\citenamefont {Jani{\v{s}}}(1999{\natexlab{b}})}]{Janis1999}%
  \BibitemOpen
  \bibfield  {author} {\bibinfo {author} {\bibnamefont {Jani{\v{s}}},
  \bibfnamefont {V.}}} (\bibinfo {year} {1999}{\natexlab{b}}),\ \href {\doibase
  10.1103/PhysRevB.60.11345} {\bibfield  {journal} {\bibinfo  {journal} {Phys.
  Rev. B}\ }\textbf {\bibinfo {volume} {60}},\ \bibinfo {pages}
  {11345}}\BibitemShut {NoStop}%
\bibitem [{\citenamefont {Jani{\v{s}}}\ and\ \citenamefont
  {Koloren\v{c}}(2004)}]{Janis2004}%
  \BibitemOpen
  \bibfield  {author} {\bibinfo {author} {\bibnamefont {Jani{\v{s}}},
  \bibfnamefont {V.}}, \ and\ \bibinfo {author} {\bibfnamefont
  {J.}~\bibnamefont {Koloren\v{c}}}} (\bibinfo {year} {2004}),\ \href {\doibase
  10.1142/S0217984904007591} {\bibfield  {journal} {\bibinfo  {journal} {Mod.
  Phys. Lett. B}\ }\textbf {\bibinfo {volume} {18}},\ \bibinfo {pages}
  {1051}}\BibitemShut {NoStop}%
\bibitem [{\citenamefont {Jani{\v{s}}}\ and\ \citenamefont
  {Koloren\v{c}}(2016)}]{Janis2016}%
  \BibitemOpen
  \bibfield  {author} {\bibinfo {author} {\bibnamefont {Jani{\v{s}}},
  \bibfnamefont {V.}}, \ and\ \bibinfo {author} {\bibfnamefont
  {J.}~\bibnamefont {Koloren\v{c}}}} (\bibinfo {year} {2016}),\ \href {\doibase
  10.1140/epjb/e2016-70188-1} {\bibfield  {journal} {\bibinfo  {journal} {Eur.
  Phys. J. B}\ }\textbf {\bibinfo {volume} {89}},\ \bibinfo {pages}
  {170}}\BibitemShut {NoStop}%
\bibitem [{\citenamefont {Jani{\v{s}}}\ and\ \citenamefont
  {Pokorn\'y}(2010)}]{Janis2010}%
  \BibitemOpen
  \bibfield  {author} {\bibinfo {author} {\bibnamefont {Jani{\v{s}}},
  \bibfnamefont {V.}}, \ and\ \bibinfo {author} {\bibfnamefont
  {V.}~\bibnamefont {Pokorn\'y}}} (\bibinfo {year} {2010}),\ \href {\doibase
  10.1103/PhysRevB.81.165103} {\bibfield  {journal} {\bibinfo  {journal} {Phys.
  Rev. B}\ }\textbf {\bibinfo {volume} {81}},\ \bibinfo {pages}
  {165103}}\BibitemShut {NoStop}%
\bibitem [{\citenamefont {Jani{\v{s}}}\ and\ \citenamefont
  {Vollhardt}(1992)}]{Janis1992}%
  \BibitemOpen
  \bibfield  {author} {\bibinfo {author} {\bibnamefont {Jani{\v{s}}},
  \bibfnamefont {V.}}, \ and\ \bibinfo {author} {\bibfnamefont
  {D.}~\bibnamefont {Vollhardt}}} (\bibinfo {year} {1992}),\ \href {\doibase
  10.1103/PhysRevB.46.15712} {\bibfield  {journal} {\bibinfo  {journal} {Phys.
  Rev. B}\ }\textbf {\bibinfo {volume} {46}},\ \bibinfo {pages}
  {15712}}\BibitemShut {NoStop}%
\bibitem [{\citenamefont {Jani\v{s}}(2001)}]{Janis2001}%
  \BibitemOpen
  \bibfield  {author} {\bibinfo {author} {\bibnamefont {Jani\v{s}},
  \bibfnamefont {V.}}} (\bibinfo {year} {2001}),\ \href {\doibase
  10.1103/PhysRevB.64.115115} {\bibfield  {journal} {\bibinfo  {journal} {Phys.
  Rev. B}\ }\textbf {\bibinfo {volume} {64}},\ \bibinfo {pages}
  {115115}}\BibitemShut {NoStop}%
\bibitem [{\citenamefont {Jani\v{s}}\ \emph {et~al.}(2017)\citenamefont
  {Jani\v{s}}, \citenamefont {Kauch},\ and\ \citenamefont
  {Pokorn\'y}}]{Janis2017}%
  \BibitemOpen
  \bibfield  {author} {\bibinfo {author} {\bibnamefont {Jani\v{s}},
  \bibfnamefont {V.}}, \bibinfo {author} {\bibfnamefont {A.}~\bibnamefont
  {Kauch}}, \ and\ \bibinfo {author} {\bibfnamefont {V.}~\bibnamefont
  {Pokorn\'y}}} (\bibinfo {year} {2017}),\ \href {\doibase
  10.1103/PhysRevB.95.045108} {\bibfield  {journal} {\bibinfo  {journal} {Phys.
  Rev. B}\ }\textbf {\bibinfo {volume} {95}},\ \bibinfo {pages}
  {045108}}\BibitemShut {NoStop}%
\bibitem [{\citenamefont {Jani\v{s}}\ and\ \citenamefont
  {Koloren\v{c}}(2005)}]{Janis2005}%
  \BibitemOpen
  \bibfield  {author} {\bibinfo {author} {\bibnamefont {Jani\v{s}},
  \bibfnamefont {V.}}, \ and\ \bibinfo {author} {\bibfnamefont
  {J.}~\bibnamefont {Koloren\v{c}}}} (\bibinfo {year} {2005}),\ \href {\doibase
  10.1103/PhysRevB.71.245106} {\bibfield  {journal} {\bibinfo  {journal} {Phys.
  Rev. B}\ }\textbf {\bibinfo {volume} {71}},\ \bibinfo {pages}
  {245106}}\BibitemShut {NoStop}%
\bibitem [{\citenamefont {Jani\v{s}}\ and\ \citenamefont
  {Pokorn\'y}(2014)}]{Janis2014}%
  \BibitemOpen
  \bibfield  {author} {\bibinfo {author} {\bibnamefont {Jani\v{s}},
  \bibfnamefont {V.}}, \ and\ \bibinfo {author} {\bibfnamefont
  {V.}~\bibnamefont {Pokorn\'y}}} (\bibinfo {year} {2014}),\ \href {\doibase
  10.1103/PhysRevB.90.045143} {\bibfield  {journal} {\bibinfo  {journal} {Phys.
  Rev. B}\ }\textbf {\bibinfo {volume} {90}},\ \bibinfo {pages}
  {045143}}\BibitemShut {NoStop}%
\bibitem [{\citenamefont {Jani\v{s}}\ and\ \citenamefont
  {Vollhardt}(1992)}]{Janis92}%
  \BibitemOpen
  \bibfield  {author} {\bibinfo {author} {\bibnamefont {Jani\v{s}},
  \bibfnamefont {V.}}, \ and\ \bibinfo {author} {\bibfnamefont
  {D.}~\bibnamefont {Vollhardt}}} (\bibinfo {year} {1992}),\ \href {\doibase
  10.1142/S0217979292000438} {\bibfield  {journal} {\bibinfo  {journal} {Int.
  J. Mod. Phys. B}\ }\textbf {\bibinfo {volume} {06}},\ \bibinfo {pages}
  {731}}\BibitemShut {NoStop}%
\bibitem [{\citenamefont {Jarrell}(1992)}]{Jarrell1992}%
  \BibitemOpen
  \bibfield  {author} {\bibinfo {author} {\bibnamefont {Jarrell}, \bibfnamefont
  {M.}}} (\bibinfo {year} {1992}),\ \href {\doibase 10.1103/PhysRevLett.69.168}
  {\bibfield  {journal} {\bibinfo  {journal} {Phys. Rev. Lett.}\ }\textbf
  {\bibinfo {volume} {69}}~(\bibinfo {number} {1}),\ \bibinfo {pages}
  {168}}\BibitemShut {NoStop}%
\bibitem [{\citenamefont {Jarrell}\ and\ \citenamefont
  {Gubernatis}(1996)}]{Jarrell1996}%
  \BibitemOpen
  \bibfield  {author} {\bibinfo {author} {\bibnamefont {Jarrell}, \bibfnamefont
  {M.}}, \ and\ \bibinfo {author} {\bibfnamefont {J.~E.}\ \bibnamefont
  {Gubernatis}}} (\bibinfo {year} {1996}),\ \href {\doibase DOI:
  10.1016/0370-1573(95)00074-7} {\bibfield  {journal} {\bibinfo  {journal}
  {Physics Reports}\ }\textbf {\bibinfo {volume} {269}}~(\bibinfo {number}
  {3}),\ \bibinfo {pages} {133 }}\BibitemShut {NoStop}%
\bibitem [{\citenamefont {Jarrell}\ and\ \citenamefont
  {Krishnamurthy}(2001)}]{Jarrell2001a}%
  \BibitemOpen
  \bibfield  {author} {\bibinfo {author} {\bibnamefont {Jarrell}, \bibfnamefont
  {M.}}, \ and\ \bibinfo {author} {\bibfnamefont {H.~R.}\ \bibnamefont
  {Krishnamurthy}}} (\bibinfo {year} {2001}),\ \href {\doibase
  10.1103/PhysRevB.63.125102} {\bibfield  {journal} {\bibinfo  {journal} {Phys.
  Rev. B}\ }\textbf {\bibinfo {volume} {63}},\ \bibinfo {pages}
  {125102}}\BibitemShut {NoStop}%
\bibitem [{\citenamefont {{Jiang}}\ \emph {et~al.}(2017)\citenamefont
  {{Jiang}}, \citenamefont {{H ahner}}, \citenamefont {{Schulthess}},\ and\
  \citenamefont {{Maier}}}]{Jiang2017}%
  \BibitemOpen
  \bibfield  {author} {\bibinfo {author} {\bibnamefont {{Jiang}}, \bibfnamefont
  {M.}}, \bibinfo {author} {\bibfnamefont {U.~R.}\ \bibnamefont {{H ahner}}},
  \bibinfo {author} {\bibfnamefont {T.~C.}\ \bibnamefont {{Schulthess}}}, \
  and\ \bibinfo {author} {\bibfnamefont {T.~A.}\ \bibnamefont {{Maier}}}}
  (\bibinfo {year} {2017}),\ \href {https://arxiv.org/abs/1707.06093} {\
  }\Eprint {http://arxiv.org/abs/1707.06093} {arXiv:1707.06093} \BibitemShut
  {NoStop}%
\bibitem [{\citenamefont {Jotzu}\ \emph {et~al.}(2015)\citenamefont {Jotzu},
  \citenamefont {Messer}, \citenamefont {G\"org}, \citenamefont {Greif},
  \citenamefont {Desbuquois},\ and\ \citenamefont {Esslinger}}]{Jotzu2015}%
  \BibitemOpen
  \bibfield  {author} {\bibinfo {author} {\bibnamefont {Jotzu}, \bibfnamefont
  {G.}}, \bibinfo {author} {\bibfnamefont {M.}~\bibnamefont {Messer}}, \bibinfo
  {author} {\bibfnamefont {F.}~\bibnamefont {G\"org}}, \bibinfo {author}
  {\bibfnamefont {D.}~\bibnamefont {Greif}}, \bibinfo {author} {\bibfnamefont
  {R.}~\bibnamefont {Desbuquois}}, \ and\ \bibinfo {author} {\bibfnamefont
  {T.}~\bibnamefont {Esslinger}}} (\bibinfo {year} {2015}),\ \href {\doibase
  10.1103/PhysRevLett.115.073002} {\bibfield  {journal} {\bibinfo  {journal}
  {Phys. Rev. Lett.}\ }\textbf {\bibinfo {volume} {115}},\ \bibinfo {pages}
  {073002}}\BibitemShut {NoStop}%
\bibitem [{\citenamefont {Jung}(2010)}]{Jung2010}%
  \BibitemOpen
  \bibfield  {author} {\bibinfo {author} {\bibnamefont {Jung}, \bibfnamefont
  {C.}}} (\bibinfo {year} {2010}),\ \emph {\bibinfo {title} {Superperturbation
  theory for correlated fermions}},\ \href@noop {} {Ph.D. thesis}\ (\bibinfo
  {school} {University of Hamburg})\BibitemShut {NoStop}%
\bibitem [{\citenamefont {Jung}\ \emph {et~al.}(2012)\citenamefont {Jung},
  \citenamefont {Lieder}, \citenamefont {Brener}, \citenamefont {Hafermann},
  \citenamefont {Baxevanis}, \citenamefont {Chudnovskiy}, \citenamefont
  {Rubtsov}, \citenamefont {Katsnelson},\ and\ \citenamefont
  {Lichtenstein}}]{Jung2012}%
  \BibitemOpen
  \bibfield  {author} {\bibinfo {author} {\bibnamefont {Jung}, \bibfnamefont
  {C.}}, \bibinfo {author} {\bibfnamefont {A.}~\bibnamefont {Lieder}}, \bibinfo
  {author} {\bibfnamefont {S.}~\bibnamefont {Brener}}, \bibinfo {author}
  {\bibfnamefont {H.}~\bibnamefont {Hafermann}}, \bibinfo {author}
  {\bibfnamefont {B.}~\bibnamefont {Baxevanis}}, \bibinfo {author}
  {\bibfnamefont {A.}~\bibnamefont {Chudnovskiy}}, \bibinfo {author}
  {\bibfnamefont {A.}~\bibnamefont {Rubtsov}}, \bibinfo {author} {\bibfnamefont
  {M.}~\bibnamefont {Katsnelson}}, \ and\ \bibinfo {author} {\bibfnamefont
  {A.}~\bibnamefont {Lichtenstein}}} (\bibinfo {year} {2012}),\ \href {\doibase
  10.1002/andp.201100045} {\bibfield  {journal} {\bibinfo  {journal} {Annalen
  der Physik}\ }\textbf {\bibinfo {volume} {524}}~(\bibinfo {number} {1}),\
  \bibinfo {pages} {49}}\BibitemShut {NoStop}%
\bibitem [{\citenamefont {Jung}\ \emph {et~al.}(2011)\citenamefont {Jung},
  \citenamefont {Wilhelm}, \citenamefont {Hafermann}, \citenamefont {Brener},\
  and\ \citenamefont {Lichtenstein}}]{Jung2011}%
  \BibitemOpen
  \bibfield  {author} {\bibinfo {author} {\bibnamefont {Jung}, \bibfnamefont
  {C.}}, \bibinfo {author} {\bibfnamefont {A.}~\bibnamefont {Wilhelm}},
  \bibinfo {author} {\bibfnamefont {H.}~\bibnamefont {Hafermann}}, \bibinfo
  {author} {\bibfnamefont {S.}~\bibnamefont {Brener}}, \ and\ \bibinfo {author}
  {\bibfnamefont {A.}~\bibnamefont {Lichtenstein}}} (\bibinfo {year} {2011}),\
  \href {\doibase 10.1002/andp.201100043} {\bibfield  {journal} {\bibinfo
  {journal} {Annalen der Physik}\ }\textbf {\bibinfo {volume} {523}}~(\bibinfo
  {number} {8-9}),\ \bibinfo {pages} {706}}\BibitemShut {NoStop}%
\bibitem [{\citenamefont {Kajueter}(1996)}]{Kajueter96}%
  \BibitemOpen
  \bibfield  {author} {\bibinfo {author} {\bibnamefont {Kajueter},
  \bibfnamefont {H.}}} (\bibinfo {year} {1996}),\ \href@noop {} {Ph.D. thesis}\
  (\bibinfo  {school} {Rutgers University})\BibitemShut {NoStop}%
\bibitem [{\citenamefont {Kananenka}\ \emph {et~al.}(2015)\citenamefont
  {Kananenka}, \citenamefont {Gull},\ and\ \citenamefont {Zgid}}]{Kamenka2015}%
  \BibitemOpen
  \bibfield  {author} {\bibinfo {author} {\bibnamefont {Kananenka},
  \bibfnamefont {A.~A.}}, \bibinfo {author} {\bibfnamefont {E.}~\bibnamefont
  {Gull}}, \ and\ \bibinfo {author} {\bibfnamefont {D.}~\bibnamefont {Zgid}}}
  (\bibinfo {year} {2015}),\ \href {\doibase 10.1103/PhysRevB.91.121111}
  {\bibfield  {journal} {\bibinfo  {journal} {Phys. Rev. B}\ }\textbf {\bibinfo
  {volume} {91}},\ \bibinfo {pages} {121111}}\BibitemShut {NoStop}%
\bibitem [{\citenamefont {Katanin}(2005)}]{Katanin2005}%
  \BibitemOpen
  \bibfield  {author} {\bibinfo {author} {\bibnamefont {Katanin}, \bibfnamefont
  {A.~A.}}} (\bibinfo {year} {2005}),\ \href
  {http://dx.doi.org/10.1103/PhysRevB.72.035111} {\bibfield  {journal}
  {\bibinfo  {journal} {Phys. Rev. B}\ }\textbf {\bibinfo {volume} {72}},\
  \bibinfo {pages} {035111}}\BibitemShut {NoStop}%
\bibitem [{\citenamefont {Katanin}(2013)}]{Katanin2013}%
  \BibitemOpen
  \bibfield  {author} {\bibinfo {author} {\bibnamefont {Katanin}, \bibfnamefont
  {A.~A.}}} (\bibinfo {year} {2013}),\ \href
  {http://stacks.iop.org/1751-8121/46/i=4/a=045002} {\bibfield  {journal}
  {\bibinfo  {journal} {J. Phys. A: Math. Theor.}\ }\textbf {\bibinfo {volume}
  {46}}~(\bibinfo {number} {4}),\ \bibinfo {pages} {045002}}\BibitemShut
  {NoStop}%
\bibitem [{\citenamefont {Katanin}(2015)}]{Katanin2015}%
  \BibitemOpen
  \bibfield  {author} {\bibinfo {author} {\bibnamefont {Katanin}, \bibfnamefont
  {A.~A.}}} (\bibinfo {year} {2015}),\ \href {\doibase
  10.1134/S1063776115050039} {\bibfield  {journal} {\bibinfo  {journal} {JETP}\
  }\textbf {\bibinfo {volume} {120}},\ \bibinfo {pages} {1085}}\BibitemShut
  {NoStop}%
\bibitem [{\citenamefont {Katanin}(2016)}]{Katanin2016}%
  \BibitemOpen
  \bibfield  {author} {\bibinfo {author} {\bibnamefont {Katanin}, \bibfnamefont
  {A.~A.}}} (\bibinfo {year} {2016}),\ \href {http://arxiv.org/abs/1604.01702}
  {\ }\Eprint {http://arxiv.org/abs/1604.01702} {arXiv:1604.01702} \BibitemShut
  {NoStop}%
\bibitem [{\citenamefont {Katanin}\ \emph {et~al.}(2009)\citenamefont
  {Katanin}, \citenamefont {Toschi},\ and\ \citenamefont {Held}}]{Katanin2009}%
  \BibitemOpen
  \bibfield  {author} {\bibinfo {author} {\bibnamefont {Katanin}, \bibfnamefont
  {A.~A.}}, \bibinfo {author} {\bibfnamefont {A.}~\bibnamefont {Toschi}}, \
  and\ \bibinfo {author} {\bibfnamefont {K.}~\bibnamefont {Held}}} (\bibinfo
  {year} {2009}),\ \href {\doibase 10.1103/PhysRevB.80.075104} {\bibfield
  {journal} {\bibinfo  {journal} {Phys. Rev. B}\ }\textbf {\bibinfo {volume}
  {80}},\ \bibinfo {pages} {075104}}\BibitemShut {NoStop}%
\bibitem [{\citenamefont {Katsnelson}\ \emph {et~al.}(2008)\citenamefont
  {Katsnelson}, \citenamefont {Irkhin}, \citenamefont {Chioncel}, \citenamefont
  {Lichtenstein},\ and\ \citenamefont {de~Groot}}]{Katsnelson2008}%
  \BibitemOpen
  \bibfield  {author} {\bibinfo {author} {\bibnamefont {Katsnelson},
  \bibfnamefont {M.~I.}}, \bibinfo {author} {\bibfnamefont {V.~Y.}\
  \bibnamefont {Irkhin}}, \bibinfo {author} {\bibfnamefont {L.}~\bibnamefont
  {Chioncel}}, \bibinfo {author} {\bibfnamefont {A.~I.}\ \bibnamefont
  {Lichtenstein}}, \ and\ \bibinfo {author} {\bibfnamefont {R.~A.}\
  \bibnamefont {de~Groot}}} (\bibinfo {year} {2008}),\ \href {\doibase
  10.1103/RevModPhys.80.315} {\bibfield  {journal} {\bibinfo  {journal} {Rev.
  Mod. Phys.}\ }\textbf {\bibinfo {volume} {80}},\ \bibinfo {pages}
  {315}}\BibitemShut {NoStop}%
\bibitem [{\citenamefont {Kaufmann}\ \emph {et~al.}(2017)\citenamefont
  {Kaufmann}, \citenamefont {Gunacker},\ and\ \citenamefont
  {Held}}]{Kaufmann2017}%
  \BibitemOpen
  \bibfield  {author} {\bibinfo {author} {\bibnamefont {Kaufmann},
  \bibfnamefont {J.}}, \bibinfo {author} {\bibfnamefont {P.}~\bibnamefont
  {Gunacker}}, \ and\ \bibinfo {author} {\bibfnamefont {K.}~\bibnamefont
  {Held}}} (\bibinfo {year} {2017}),\ \href {\doibase
  10.1103/PhysRevB.96.035114} {\bibfield  {journal} {\bibinfo  {journal} {Phys.
  Rev. B}\ }\textbf {\bibinfo {volume} {96}},\ \bibinfo {pages}
  {035114}}\BibitemShut {NoStop}%
\bibitem [{\citenamefont {Keiter}\ and\ \citenamefont
  {Leuders}(2000)}]{Keiter2000}%
  \BibitemOpen
  \bibfield  {author} {\bibinfo {author} {\bibnamefont {Keiter}, \bibfnamefont
  {H.}}, \ and\ \bibinfo {author} {\bibfnamefont {T.}~\bibnamefont {Leuders}}}
  (\bibinfo {year} {2000}),\ \href
  {http://stacks.iop.org/0295-5075/49/i=6/a=801} {\bibfield  {journal}
  {\bibinfo  {journal} {EPL (Europhysics Letters)}\ }\textbf {\bibinfo {volume}
  {49}}~(\bibinfo {number} {6}),\ \bibinfo {pages} {801}}\BibitemShut {NoStop}%
\bibitem [{\citenamefont {Kennedy}\ and\ \citenamefont
  {Lieb}(1986)}]{Kennedy1986}%
  \BibitemOpen
  \bibfield  {author} {\bibinfo {author} {\bibnamefont {Kennedy}, \bibfnamefont
  {T.}}, \ and\ \bibinfo {author} {\bibfnamefont {E.~H.}\ \bibnamefont {Lieb}}}
  (\bibinfo {year} {1986}),\ \href {\doibase 10.1016/0378-4371(86)90188-3}
  {\bibfield  {journal} {\bibinfo  {journal} {Physica A}\ }\textbf {\bibinfo
  {volume} {138}}~(\bibinfo {number} {1-2}),\ \bibinfo {pages}
  {320}}\BibitemShut {NoStop}%
\bibitem [{\citenamefont {Kent}\ \emph {et~al.}(2005)\citenamefont {Kent},
  \citenamefont {Jarrell}, \citenamefont {Maier},\ and\ \citenamefont
  {Pruschke}}]{Kent2005}%
  \BibitemOpen
  \bibfield  {author} {\bibinfo {author} {\bibnamefont {Kent}, \bibfnamefont
  {P.~R.~C.}}, \bibinfo {author} {\bibfnamefont {M.}~\bibnamefont {Jarrell}},
  \bibinfo {author} {\bibfnamefont {T.~A.}\ \bibnamefont {Maier}}, \ and\
  \bibinfo {author} {\bibfnamefont {T.}~\bibnamefont {Pruschke}}} (\bibinfo
  {year} {2005}),\ \href {\doibase 10.1103/PhysRevB.72.060411} {\bibfield
  {journal} {\bibinfo  {journal} {Phys. Rev. B}\ }\textbf {\bibinfo {volume}
  {72}},\ \bibinfo {pages} {060411}}\BibitemShut {NoStop}%
\bibitem [{\citenamefont {Kinza}\ and\ \citenamefont
  {Honerkamp}(2013)}]{Kinza2013}%
  \BibitemOpen
  \bibfield  {author} {\bibinfo {author} {\bibnamefont {Kinza}, \bibfnamefont
  {M.}}, \ and\ \bibinfo {author} {\bibfnamefont {C.}~\bibnamefont
  {Honerkamp}}} (\bibinfo {year} {2013}),\ \href {\doibase
  10.1103/PhysRevB.88.195136} {\bibfield  {journal} {\bibinfo  {journal} {Phys.
  Rev. B}\ }\textbf {\bibinfo {volume} {88}},\ \bibinfo {pages}
  {195136}}\BibitemShut {NoStop}%
\bibitem [{\citenamefont {Kitatani}\ \emph {et~al.}(2018)\citenamefont
  {Kitatani}, \citenamefont {Sch\"afer}, \citenamefont {Aoki},\ and\
  \citenamefont {Held}}]{Kitatani2017}%
  \BibitemOpen
  \bibfield  {author} {\bibinfo {author} {\bibnamefont {Kitatani},
  \bibfnamefont {M.}}, \bibinfo {author} {\bibfnamefont {T.}~\bibnamefont
  {Sch\"afer}}, \bibinfo {author} {\bibfnamefont {H.}~\bibnamefont {Aoki}}, \
  and\ \bibinfo {author} {\bibfnamefont {K.}~\bibnamefont {Held}}} (\bibinfo
  {year} {2018}),\ \href@noop {} {\ }\Eprint
  {http://arxiv.org/abs/arXiv:1801.05991} {arXiv:1801.05991} \BibitemShut
  {NoStop}%
\bibitem [{\citenamefont {Kitatani}\ \emph {et~al.}(2015)\citenamefont
  {Kitatani}, \citenamefont {Tsuji},\ and\ \citenamefont
  {Aoki}}]{Kitatani2015}%
  \BibitemOpen
  \bibfield  {author} {\bibinfo {author} {\bibnamefont {Kitatani},
  \bibfnamefont {M.}}, \bibinfo {author} {\bibfnamefont {N.}~\bibnamefont
  {Tsuji}}, \ and\ \bibinfo {author} {\bibfnamefont {H.}~\bibnamefont {Aoki}}}
  (\bibinfo {year} {2015}),\ \href
  {http://link.aps.org/doi/10.1103/PhysRevB.92.085104} {\bibfield  {journal}
  {\bibinfo  {journal} {Phys. Rev. B}\ }\textbf {\bibinfo {volume} {92}},\
  \bibinfo {pages} {085104}}\BibitemShut {NoStop}%
\bibitem [{\citenamefont {Kohiki}\ \emph {et~al.}(2000)\citenamefont {Kohiki},
  \citenamefont {Arai}, \citenamefont {Yoshikawa}, \citenamefont {Fukushima},
  \citenamefont {Oku},\ and\ \citenamefont {Waseda}}]{Kohiki2000}%
  \BibitemOpen
  \bibfield  {author} {\bibinfo {author} {\bibnamefont {Kohiki}, \bibfnamefont
  {S.}}, \bibinfo {author} {\bibfnamefont {M.}~\bibnamefont {Arai}}, \bibinfo
  {author} {\bibfnamefont {H.}~\bibnamefont {Yoshikawa}}, \bibinfo {author}
  {\bibfnamefont {S.}~\bibnamefont {Fukushima}}, \bibinfo {author}
  {\bibfnamefont {M.}~\bibnamefont {Oku}}, \ and\ \bibinfo {author}
  {\bibfnamefont {Y.}~\bibnamefont {Waseda}}} (\bibinfo {year} {2000}),\ \href
  {\doibase 10.1103/PhysRevB.62.7964} {\bibfield  {journal} {\bibinfo
  {journal} {Phys. Rev. B}\ }\textbf {\bibinfo {volume} {62}},\ \bibinfo
  {pages} {7964}}\BibitemShut {NoStop}%
\bibitem [{\citenamefont {Kotliar}\ \emph {et~al.}(2006)\citenamefont
  {Kotliar}, \citenamefont {Savrasov}, \citenamefont {Haule}, \citenamefont
  {Oudovenko}, \citenamefont {Parcollet},\ and\ \citenamefont
  {Marianetti}}]{Kotliar2006}%
  \BibitemOpen
  \bibfield  {author} {\bibinfo {author} {\bibnamefont {Kotliar}, \bibfnamefont
  {G.}}, \bibinfo {author} {\bibfnamefont {S.~Y.}\ \bibnamefont {Savrasov}},
  \bibinfo {author} {\bibfnamefont {K.}~\bibnamefont {Haule}}, \bibinfo
  {author} {\bibfnamefont {V.~S.}\ \bibnamefont {Oudovenko}}, \bibinfo {author}
  {\bibfnamefont {O.}~\bibnamefont {Parcollet}}, \ and\ \bibinfo {author}
  {\bibfnamefont {C.~A.}\ \bibnamefont {Marianetti}}} (\bibinfo {year}
  {2006}),\ \href {\doibase 10.1103/RevModPhys.78.865} {\bibfield  {journal}
  {\bibinfo  {journal} {Rev. Mod. Phys.}\ }\textbf {\bibinfo {volume} {78}},\
  \bibinfo {pages} {865}}\BibitemShut {NoStop}%
\bibitem [{\citenamefont {Kotliar}\ \emph {et~al.}(2001)\citenamefont
  {Kotliar}, \citenamefont {Savrasov}, \citenamefont {P\'alsson},\ and\
  \citenamefont {Biroli}}]{Kotliar2001}%
  \BibitemOpen
  \bibfield  {author} {\bibinfo {author} {\bibnamefont {Kotliar}, \bibfnamefont
  {G.}}, \bibinfo {author} {\bibfnamefont {S.~Y.}\ \bibnamefont {Savrasov}},
  \bibinfo {author} {\bibfnamefont {G.}~\bibnamefont {P\'alsson}}, \ and\
  \bibinfo {author} {\bibfnamefont {G.}~\bibnamefont {Biroli}}} (\bibinfo
  {year} {2001}),\ \href {\doibase 10.1103/PhysRevLett.87.186401} {\bibfield
  {journal} {\bibinfo  {journal} {Phys. Rev. Lett.}\ }\textbf {\bibinfo
  {volume} {87}},\ \bibinfo {pages} {186401}}\BibitemShut {NoStop}%
\bibitem [{\citenamefont {Kotliar}\ and\ \citenamefont
  {Vollhardt}(2004)}]{Kotliar2004}%
  \BibitemOpen
  \bibfield  {author} {\bibinfo {author} {\bibnamefont {Kotliar}, \bibfnamefont
  {G.}}, \ and\ \bibinfo {author} {\bibfnamefont {D.}~\bibnamefont
  {Vollhardt}}} (\bibinfo {year} {2004}),\ \href@noop {} {\bibfield  {journal}
  {\bibinfo  {journal} {Physics Today}\ }\textbf {\bibinfo {volume} {57}},\
  \bibinfo {pages} {53}}\BibitemShut {NoStop}%
\bibitem [{\citenamefont {Kotov}\ \emph {et~al.}(2012)\citenamefont {Kotov},
  \citenamefont {Uchoa}, \citenamefont {Pereira}, \citenamefont {Guinea},\ and\
  \citenamefont {Castro~Neto}}]{Kutov2012}%
  \BibitemOpen
  \bibfield  {author} {\bibinfo {author} {\bibnamefont {Kotov}, \bibfnamefont
  {V.~N.}}, \bibinfo {author} {\bibfnamefont {B.}~\bibnamefont {Uchoa}},
  \bibinfo {author} {\bibfnamefont {V.~M.}\ \bibnamefont {Pereira}}, \bibinfo
  {author} {\bibfnamefont {F.}~\bibnamefont {Guinea}}, \ and\ \bibinfo {author}
  {\bibfnamefont {A.~H.}\ \bibnamefont {Castro~Neto}}} (\bibinfo {year}
  {2012}),\ \href {\doibase 10.1103/RevModPhys.84.1067} {\bibfield  {journal}
  {\bibinfo  {journal} {Rev. Mod. Phys.}\ }\textbf {\bibinfo {volume} {84}},\
  \bibinfo {pages} {1067}}\BibitemShut {NoStop}%
\bibitem [{\citenamefont {Kozik}\ \emph {et~al.}(2013)\citenamefont {Kozik},
  \citenamefont {Burovski}, \citenamefont {Scarola},\ and\ \citenamefont
  {Troyer}}]{Kozik2013}%
  \BibitemOpen
  \bibfield  {author} {\bibinfo {author} {\bibnamefont {Kozik}, \bibfnamefont
  {E.}}, \bibinfo {author} {\bibfnamefont {E.}~\bibnamefont {Burovski}},
  \bibinfo {author} {\bibfnamefont {V.~W.}\ \bibnamefont {Scarola}}, \ and\
  \bibinfo {author} {\bibfnamefont {M.}~\bibnamefont {Troyer}}} (\bibinfo
  {year} {2013}),\ \href {\doibase 10.1103/PhysRevB.87.205102} {\bibfield
  {journal} {\bibinfo  {journal} {Phys. Rev. B}\ }\textbf {\bibinfo {volume}
  {87}},\ \bibinfo {pages} {205102}}\BibitemShut {NoStop}%
\bibitem [{\citenamefont {Kozik}\ \emph {et~al.}(2015)\citenamefont {Kozik},
  \citenamefont {Ferrero},\ and\ \citenamefont {Georges}}]{Kozik2015}%
  \BibitemOpen
  \bibfield  {author} {\bibinfo {author} {\bibnamefont {Kozik}, \bibfnamefont
  {E.}}, \bibinfo {author} {\bibfnamefont {M.}~\bibnamefont {Ferrero}}, \ and\
  \bibinfo {author} {\bibfnamefont {A.}~\bibnamefont {Georges}}} (\bibinfo
  {year} {2015}),\ \href {\doibase 10.1103/PhysRevLett.114.156402} {\bibfield
  {journal} {\bibinfo  {journal} {Phys. Rev. Lett.}\ }\textbf {\bibinfo
  {volume} {114}},\ \bibinfo {pages} {156402}}\BibitemShut {NoStop}%
\bibitem [{\citenamefont {Krien}\ \emph {et~al.}(2017)\citenamefont {Krien},
  \citenamefont {van Loon}, \citenamefont {Hafermann}, \citenamefont {Otsuki},
  \citenamefont {Katsnelson},\ and\ \citenamefont {Lichtenstein}}]{Krien2017}%
  \BibitemOpen
  \bibfield  {author} {\bibinfo {author} {\bibnamefont {Krien}, \bibfnamefont
  {F.}}, \bibinfo {author} {\bibfnamefont {E.~G. C.~P.}\ \bibnamefont {van
  Loon}}, \bibinfo {author} {\bibfnamefont {H.}~\bibnamefont {Hafermann}},
  \bibinfo {author} {\bibfnamefont {J.}~\bibnamefont {Otsuki}}, \bibinfo
  {author} {\bibfnamefont {M.~I.}\ \bibnamefont {Katsnelson}}, \ and\ \bibinfo
  {author} {\bibfnamefont {A.~I.}\ \bibnamefont {Lichtenstein}}} (\bibinfo
  {year} {2017}),\ \href {\doibase 10.1103/PhysRevB.96.075155} {\bibfield
  {journal} {\bibinfo  {journal} {Phys. Rev. B}\ }\textbf {\bibinfo {volume}
  {96}},\ \bibinfo {pages} {075155}}\BibitemShut {NoStop}%
\bibitem [{\citenamefont {Krivenko}\ \emph {et~al.}(2010)\citenamefont
  {Krivenko}, \citenamefont {Rubtsov}, \citenamefont {Katsnelson},\ and\
  \citenamefont {Lichtenstein}}]{Krivenko2010}%
  \BibitemOpen
  \bibfield  {author} {\bibinfo {author} {\bibnamefont {Krivenko},
  \bibfnamefont {I.}}, \bibinfo {author} {\bibfnamefont {A.}~\bibnamefont
  {Rubtsov}}, \bibinfo {author} {\bibfnamefont {M.}~\bibnamefont {Katsnelson}},
  \ and\ \bibinfo {author} {\bibfnamefont {A.}~\bibnamefont {Lichtenstein}}}
  (\bibinfo {year} {2010}),\ \href {\doibase 10.1134/S0021364010060123}
  {\bibfield  {journal} {\bibinfo  {journal} {JETP Letters}\ }\textbf {\bibinfo
  {volume} {91}}~(\bibinfo {number} {6}),\ \bibinfo {pages} {319}}\BibitemShut
  {NoStop}%
\bibitem [{\citenamefont {Kroha}(1990)}]{Kroha1990}%
  \BibitemOpen
  \bibfield  {author} {\bibinfo {author} {\bibnamefont {Kroha}, \bibfnamefont
  {J.}}} (\bibinfo {year} {1990}),\ \href {\doibase
  10.1016/0378-4371(90)90055-W} {\bibfield  {journal} {\bibinfo  {journal}
  {Physica A}\ }\textbf {\bibinfo {volume} {167}}~(\bibinfo {number} {1}),\
  \bibinfo {pages} {231}}\BibitemShut {NoStop}%
\bibitem [{\citenamefont {Kuchinskii}\ \emph {et~al.}(2010)\citenamefont
  {Kuchinskii}, \citenamefont {Kuleeva}, \citenamefont {Nekrasov},\ and\
  \citenamefont {Sadovskii}}]{Kuchinskii2010}%
  \BibitemOpen
  \bibfield  {author} {\bibinfo {author} {\bibnamefont {Kuchinskii},
  \bibfnamefont {E.~Z.}}, \bibinfo {author} {\bibfnamefont {N.~A.}\
  \bibnamefont {Kuleeva}}, \bibinfo {author} {\bibfnamefont {I.~A.}\
  \bibnamefont {Nekrasov}}, \ and\ \bibinfo {author} {\bibfnamefont {M.~V.}\
  \bibnamefont {Sadovskii}}} (\bibinfo {year} {2010}),\ \href {\doibase
  10.1134/S1063776110020160} {\bibfield  {journal} {\bibinfo  {journal}
  {Journal of Experimental and Theoretical Physics}\ }\textbf {\bibinfo
  {volume} {110}}~(\bibinfo {number} {2}),\ \bibinfo {pages} {325}}\BibitemShut
  {NoStop}%
\bibitem [{\citenamefont {Kuchinskii}\ \emph {et~al.}(2005)\citenamefont
  {Kuchinskii}, \citenamefont {Nekrasov},\ and\ \citenamefont
  {Sadovskii}}]{Kuchinskii2005}%
  \BibitemOpen
  \bibfield  {author} {\bibinfo {author} {\bibnamefont {Kuchinskii},
  \bibfnamefont {E.~Z.}}, \bibinfo {author} {\bibfnamefont {I.~A.}\
  \bibnamefont {Nekrasov}}, \ and\ \bibinfo {author} {\bibfnamefont {M.~V.}\
  \bibnamefont {Sadovskii}}} (\bibinfo {year} {2005}),\ \href {\doibase
  10.1134/1.2121814} {\bibfield  {journal} {\bibinfo  {journal} {JETP Letters}\
  }\textbf {\bibinfo {volume} {82}},\ \bibinfo {pages} {198}}\BibitemShut
  {NoStop}%
\bibitem [{\citenamefont {Kuchinskii}\ \emph {et~al.}(2006)\citenamefont
  {Kuchinskii}, \citenamefont {Nekrasov},\ and\ \citenamefont
  {Sadovskii}}]{Kuchinskii2006}%
  \BibitemOpen
  \bibfield  {author} {\bibinfo {author} {\bibnamefont {Kuchinskii},
  \bibfnamefont {E.~Z.}}, \bibinfo {author} {\bibfnamefont {I.~A.}\
  \bibnamefont {Nekrasov}}, \ and\ \bibinfo {author} {\bibfnamefont {M.~V.}\
  \bibnamefont {Sadovskii}}} (\bibinfo {year} {2006}),\ \href
  {http://dx.doi.org/10.1063/1.2199442} {\bibfield  {journal} {\bibinfo
  {journal} {Low Temp. Phys.}\ }\textbf {\bibinfo {volume} {32}},\ \bibinfo
  {pages} {398}}\BibitemShut {NoStop}%
\bibitem [{\citenamefont {Kuchinskii}\ \emph {et~al.}(2007)\citenamefont
  {Kuchinskii}, \citenamefont {Nekrasov},\ and\ \citenamefont
  {Sadovskii}}]{Kuchinskii2007}%
  \BibitemOpen
  \bibfield  {author} {\bibinfo {author} {\bibnamefont {Kuchinskii},
  \bibfnamefont {E.~Z.}}, \bibinfo {author} {\bibfnamefont {I.~A.}\
  \bibnamefont {Nekrasov}}, \ and\ \bibinfo {author} {\bibfnamefont {M.~V.}\
  \bibnamefont {Sadovskii}}} (\bibinfo {year} {2007}),\ \href {\doibase
  10.1103/PhysRevB.75.115102} {\bibfield  {journal} {\bibinfo  {journal} {Phys.
  Rev. B}\ }\textbf {\bibinfo {volume} {75}},\ \bibinfo {pages}
  {115102}}\BibitemShut {NoStop}%
\bibitem [{\citenamefont {Kuchinskii}\ \emph {et~al.}(2009)\citenamefont
  {Kuchinskii}, \citenamefont {Nekrasov},\ and\ \citenamefont
  {Sadovskii}}]{Kuchinskii2009}%
  \BibitemOpen
  \bibfield  {author} {\bibinfo {author} {\bibnamefont {Kuchinskii},
  \bibfnamefont {E.~Z.}}, \bibinfo {author} {\bibfnamefont {I.~A.}\
  \bibnamefont {Nekrasov}}, \ and\ \bibinfo {author} {\bibfnamefont {M.~V.}\
  \bibnamefont {Sadovskii}}} (\bibinfo {year} {2009}),\ \href {\doibase
  10.1103/PhysRevB.80.115124} {\bibfield  {journal} {\bibinfo  {journal} {Phys.
  Rev. B}\ }\textbf {\bibinfo {volume} {80}},\ \bibinfo {pages}
  {115124}}\BibitemShut {NoStop}%
\bibitem [{\citenamefont {Kuchinskii}\ \emph {et~al.}(2012)\citenamefont
  {Kuchinskii}, \citenamefont {Nekrasov},\ and\ \citenamefont
  {Sadovskii}}]{Kuchinskii2012}%
  \BibitemOpen
  \bibfield  {author} {\bibinfo {author} {\bibnamefont {Kuchinskii},
  \bibfnamefont {E.~Z.}}, \bibinfo {author} {\bibfnamefont {I.~A.}\
  \bibnamefont {Nekrasov}}, \ and\ \bibinfo {author} {\bibfnamefont {M.~V.}\
  \bibnamefont {Sadovskii}}} (\bibinfo {year} {2012}),\ \href
  {http://stacks.iop.org/1063-7869/55/i=4/a=R01} {\bibfield  {journal}
  {\bibinfo  {journal} {Physics-Uspekhi}\ }\textbf {\bibinfo {volume}
  {55}}~(\bibinfo {number} {4}),\ \bibinfo {pages} {325}}\BibitemShut {NoStop}%
\bibitem [{\citenamefont {Kuchinskii}\ and\ \citenamefont
  {Sadovskii}(1999)}]{Sadovskii1999}%
  \BibitemOpen
  \bibfield  {author} {\bibinfo {author} {\bibnamefont {Kuchinskii},
  \bibfnamefont {E.~Z.}}, \ and\ \bibinfo {author} {\bibfnamefont {M.~V.}\
  \bibnamefont {Sadovskii}}} (\bibinfo {year} {1999}),\ \href
  {http://dx.doi.org/10.1134/1.558879} {\bibfield  {journal} {\bibinfo
  {journal} {JETP}\ }\textbf {\bibinfo {volume} {88}},\ \bibinfo {pages}
  {968}}\BibitemShut {NoStop}%
\bibitem [{\citenamefont {Kugler}\ and\ \citenamefont {von
  Delft}(2018{\natexlab{a}})}]{Kugler2018a}%
  \BibitemOpen
  \bibfield  {author} {\bibinfo {author} {\bibnamefont {Kugler}, \bibfnamefont
  {F.~B.}}, \ and\ \bibinfo {author} {\bibfnamefont {J.}~\bibnamefont {von
  Delft}}} (\bibinfo {year} {2018}{\natexlab{a}}),\ \href {\doibase
  10.1103/PhysRevB.97.035162} {\bibfield  {journal} {\bibinfo  {journal} {Phys.
  Rev. B}\ }\textbf {\bibinfo {volume} {97}},\ \bibinfo {pages}
  {035162}}\BibitemShut {NoStop}%
\bibitem [{\citenamefont {Kugler}\ and\ \citenamefont {von
  Delft}(2018{\natexlab{b}})}]{Kugler2018}%
  \BibitemOpen
  \bibfield  {author} {\bibinfo {author} {\bibnamefont {Kugler}, \bibfnamefont
  {F.~B.}}, \ and\ \bibinfo {author} {\bibfnamefont {J.}~\bibnamefont {von
  Delft}}} (\bibinfo {year} {2018}{\natexlab{b}}),\ \href {\doibase
  10.1103/PhysRevLett.120.057403} {\bibfield  {journal} {\bibinfo  {journal}
  {Phys. Rev. Lett.}\ }\textbf {\bibinfo {volume} {120}},\ \bibinfo {pages}
  {057403}}\BibitemShut {NoStop}%
\bibitem [{\citenamefont {Kune\v{s}}(2011)}]{Kunes2011}%
  \BibitemOpen
  \bibfield  {author} {\bibinfo {author} {\bibnamefont {Kune\v{s}},
  \bibfnamefont {J.}}} (\bibinfo {year} {2011}),\ \href {\doibase
  10.1103/PhysRevB.83.085102} {\bibfield  {journal} {\bibinfo  {journal} {Phys.
  Rev. B}\ }\textbf {\bibinfo {volume} {83}},\ \bibinfo {pages}
  {085102}}\BibitemShut {NoStop}%
\bibitem [{\citenamefont {Kusunose}(2006)}]{Kusunose2006}%
  \BibitemOpen
  \bibfield  {author} {\bibinfo {author} {\bibnamefont {Kusunose},
  \bibfnamefont {H.}}} (\bibinfo {year} {2006}),\ \href {\doibase
  10.1143/JPSJ.75.054713} {\bibfield  {journal} {\bibinfo  {journal} {J. Phys.
  Soc. Jpn.}\ }\textbf {\bibinfo {volume} {75}}~(\bibinfo {number} {5}),\
  \bibinfo {pages} {054713}}\BibitemShut {NoStop}%
\bibitem [{\citenamefont {Kyung}\ \emph {et~al.}(2003)\citenamefont {Kyung},
  \citenamefont {Landry},\ and\ \citenamefont {Tremblay}}]{Kyung2003}%
  \BibitemOpen
  \bibfield  {author} {\bibinfo {author} {\bibnamefont {Kyung}, \bibfnamefont
  {B.}}, \bibinfo {author} {\bibfnamefont {J.-S.}\ \bibnamefont {Landry}}, \
  and\ \bibinfo {author} {\bibfnamefont {A.-M.~S.}\ \bibnamefont {Tremblay}}}
  (\bibinfo {year} {2003}),\ \href {\doibase 10.1103/PhysRevB.68.174502}
  {\bibfield  {journal} {\bibinfo  {journal} {Phys. Rev. B}\ }\textbf {\bibinfo
  {volume} {68}},\ \bibinfo {pages} {174502}}\BibitemShut {NoStop}%
\bibitem [{\citenamefont {Lahaye}\ \emph {et~al.}(2009)\citenamefont {Lahaye},
  \citenamefont {Menotti}, \citenamefont {Santos}, \citenamefont {Lewenstein},\
  and\ \citenamefont {Pfau}}]{Lahaye2009}%
  \BibitemOpen
  \bibfield  {author} {\bibinfo {author} {\bibnamefont {Lahaye}, \bibfnamefont
  {T.}}, \bibinfo {author} {\bibfnamefont {C.}~\bibnamefont {Menotti}},
  \bibinfo {author} {\bibfnamefont {L.}~\bibnamefont {Santos}}, \bibinfo
  {author} {\bibfnamefont {M.}~\bibnamefont {Lewenstein}}, \ and\ \bibinfo
  {author} {\bibfnamefont {T.}~\bibnamefont {Pfau}}} (\bibinfo {year} {2009}),\
  \href {http://stacks.iop.org/0034-4885/72/i=12/a=126401} {\bibfield
  {journal} {\bibinfo  {journal} {Reports on Progress in Physics}\ }\textbf
  {\bibinfo {volume} {72}}~(\bibinfo {number} {12}),\ \bibinfo {pages}
  {126401}}\BibitemShut {NoStop}%
\bibitem [{\citenamefont {Laubach}\ \emph {et~al.}(2015)\citenamefont
  {Laubach}, \citenamefont {Thomale}, \citenamefont {Platt}, \citenamefont
  {Hanke},\ and\ \citenamefont {Li}}]{Laubach2015}%
  \BibitemOpen
  \bibfield  {author} {\bibinfo {author} {\bibnamefont {Laubach}, \bibfnamefont
  {M.}}, \bibinfo {author} {\bibfnamefont {R.}~\bibnamefont {Thomale}},
  \bibinfo {author} {\bibfnamefont {C.}~\bibnamefont {Platt}}, \bibinfo
  {author} {\bibfnamefont {W.}~\bibnamefont {Hanke}}, \ and\ \bibinfo {author}
  {\bibfnamefont {G.}~\bibnamefont {Li}}} (\bibinfo {year} {2015}),\ \href
  {\doibase 10.1103/PhysRevB.91.245125} {\bibfield  {journal} {\bibinfo
  {journal} {Phys. Rev. B}\ }\textbf {\bibinfo {volume} {91}},\ \bibinfo
  {pages} {245125}}\BibitemShut {NoStop}%
\bibitem [{\citenamefont {LeBlanc}\ \emph {et~al.}(2017)\citenamefont
  {LeBlanc}, \citenamefont {Chen}, \citenamefont {Levy}, \citenamefont
  {Antipov}, \citenamefont {Millis},\ and\ \citenamefont
  {Gull}}]{LeBlanc2016_2}%
  \BibitemOpen
  \bibfield  {author} {\bibinfo {author} {\bibnamefont {LeBlanc}, \bibfnamefont
  {J.}}, \bibinfo {author} {\bibfnamefont {X.}~\bibnamefont {Chen}}, \bibinfo
  {author} {\bibfnamefont {R.}~\bibnamefont {Levy}}, \bibinfo {author}
  {\bibfnamefont {A.~E.}\ \bibnamefont {Antipov}}, \bibinfo {author}
  {\bibfnamefont {A.}~\bibnamefont {Millis}}, \ and\ \bibinfo {author}
  {\bibfnamefont {E.}~\bibnamefont {Gull}}} (\bibinfo {year} {2017}),\
  \href@noop {} {\bibinfo  {journal} {(unpublished)}\ }\BibitemShut {NoStop}%
\bibitem [{\citenamefont {LeBlanc}\ \emph {et~al.}(2015)\citenamefont
  {LeBlanc}, \citenamefont {Antipov}, \citenamefont {Becca}, \citenamefont
  {Bulik}, \citenamefont {Chan}, \citenamefont {Chung}, \citenamefont {Deng},
  \citenamefont {Ferrero}, \citenamefont {Henderson}, \citenamefont
  {Jim\'enez-Hoyos}, \citenamefont {Kozik}, \citenamefont {Liu}, \citenamefont
  {Millis}, \citenamefont {Prokof'ev}, \citenamefont {Qin}, \citenamefont
  {Scuseria}, \citenamefont {Shi}, \citenamefont {Svistunov}, \citenamefont
  {Tocchio}, \citenamefont {Tupitsyn}, \citenamefont {White}, \citenamefont
  {Zhang}, \citenamefont {Zheng}, \citenamefont {Zhu},\ and\ \citenamefont
  {Gull}}]{LeBlanc2015}%
  \BibitemOpen
\bibfield  {journal} {  }\bibfield  {author} {\bibinfo {author} {\bibnamefont
  {LeBlanc}, \bibfnamefont {J.~P.~F.}}, \bibinfo {author} {\bibfnamefont
  {A.~E.}\ \bibnamefont {Antipov}}, \bibinfo {author} {\bibfnamefont
  {F.}~\bibnamefont {Becca}}, \bibinfo {author} {\bibfnamefont {I.~W.}\
  \bibnamefont {Bulik}}, \bibinfo {author} {\bibfnamefont {G.~K.-L.}\
  \bibnamefont {Chan}}, \bibinfo {author} {\bibfnamefont {C.-M.}\ \bibnamefont
  {Chung}}, \bibinfo {author} {\bibfnamefont {Y.}~\bibnamefont {Deng}},
  \bibinfo {author} {\bibfnamefont {M.}~\bibnamefont {Ferrero}}, \bibinfo
  {author} {\bibfnamefont {T.~M.}\ \bibnamefont {Henderson}}, \bibinfo {author}
  {\bibfnamefont {C.~A.}\ \bibnamefont {Jim\'enez-Hoyos}}, \bibinfo {author}
  {\bibfnamefont {E.}~\bibnamefont {Kozik}}, \bibinfo {author} {\bibfnamefont
  {X.-W.}\ \bibnamefont {Liu}}, \bibinfo {author} {\bibfnamefont {A.~J.}\
  \bibnamefont {Millis}}, \bibinfo {author} {\bibfnamefont {N.~V.}\
  \bibnamefont {Prokof'ev}}, \bibinfo {author} {\bibfnamefont {M.}~\bibnamefont
  {Qin}}, \bibinfo {author} {\bibfnamefont {G.~E.}\ \bibnamefont {Scuseria}},
  \bibinfo {author} {\bibfnamefont {H.}~\bibnamefont {Shi}}, \bibinfo {author}
  {\bibfnamefont {B.~V.}\ \bibnamefont {Svistunov}}, \bibinfo {author}
  {\bibfnamefont {L.~F.}\ \bibnamefont {Tocchio}}, \bibinfo {author}
  {\bibfnamefont {I.~S.}\ \bibnamefont {Tupitsyn}}, \bibinfo {author}
  {\bibfnamefont {S.~R.}\ \bibnamefont {White}}, \bibinfo {author}
  {\bibfnamefont {S.}~\bibnamefont {Zhang}}, \bibinfo {author} {\bibfnamefont
  {B.-X.}\ \bibnamefont {Zheng}}, \bibinfo {author} {\bibfnamefont
  {Z.}~\bibnamefont {Zhu}}, \ and\ \bibinfo {author} {\bibfnamefont
  {E.}~\bibnamefont {Gull}} (\bibinfo {collaboration} {Simons Collaboration on
  the Many-Electron Problem})} (\bibinfo {year} {2015}),\ \href {\doibase
  10.1103/PhysRevX.5.041041} {\bibfield  {journal} {\bibinfo  {journal} {Phys.
  Rev. X}\ }\textbf {\bibinfo {volume} {5}},\ \bibinfo {pages}
  {041041}}\BibitemShut {NoStop}%
\bibitem [{\citenamefont {Lee}\ \emph {et~al.}(2012)\citenamefont {Lee},
  \citenamefont {Foyevtsova}, \citenamefont {Ferber}, \citenamefont {Aichhorn},
  \citenamefont {Jeschke},\ and\ \citenamefont {Valent\'{\i}}}]{Lee2012}%
  \BibitemOpen
  \bibfield  {author} {\bibinfo {author} {\bibnamefont {Lee}, \bibfnamefont
  {H.}}, \bibinfo {author} {\bibfnamefont {K.}~\bibnamefont {Foyevtsova}},
  \bibinfo {author} {\bibfnamefont {J.}~\bibnamefont {Ferber}}, \bibinfo
  {author} {\bibfnamefont {M.}~\bibnamefont {Aichhorn}}, \bibinfo {author}
  {\bibfnamefont {H.~O.}\ \bibnamefont {Jeschke}}, \ and\ \bibinfo {author}
  {\bibfnamefont {R.}~\bibnamefont {Valent\'{\i}}}} (\bibinfo {year} {2012}),\
  \href {\doibase 10.1103/PhysRevB.85.165103} {\bibfield  {journal} {\bibinfo
  {journal} {Phys. Rev. B}\ }\textbf {\bibinfo {volume} {85}},\ \bibinfo
  {pages} {165103}}\BibitemShut {NoStop}%
\bibitem [{\citenamefont {Lee}\ \emph {et~al.}(2008)\citenamefont {Lee},
  \citenamefont {Li},\ and\ \citenamefont {Monien}}]{Lee2008}%
  \BibitemOpen
  \bibfield  {author} {\bibinfo {author} {\bibnamefont {Lee}, \bibfnamefont
  {H.}}, \bibinfo {author} {\bibfnamefont {G.}~\bibnamefont {Li}}, \ and\
  \bibinfo {author} {\bibfnamefont {H.}~\bibnamefont {Monien}}} (\bibinfo
  {year} {2008}),\ \href {\doibase 10.1103/PhysRevB.78.205117} {\bibfield
  {journal} {\bibinfo  {journal} {Phys. Rev. B}\ }\textbf {\bibinfo {volume}
  {78}},\ \bibinfo {pages} {205117}}\BibitemShut {NoStop}%
\bibitem [{\citenamefont {Levy}\ \emph {et~al.}(2017)\citenamefont {Levy},
  \citenamefont {LeBlanc},\ and\ \citenamefont {Gull}}]{Levy2017}%
  \BibitemOpen
  \bibfield  {author} {\bibinfo {author} {\bibnamefont {Levy}, \bibfnamefont
  {R.}}, \bibinfo {author} {\bibfnamefont {J.}~\bibnamefont {LeBlanc}}, \ and\
  \bibinfo {author} {\bibfnamefont {E.}~\bibnamefont {Gull}}} (\bibinfo {year}
  {2017}),\ \href {\doibase 10.1016/j.cpc.2017.01.018} {\bibfield  {journal}
  {\bibinfo  {journal} {Computer Physics Communications}\ ,\ }}\bibinfo {note}
  {(in print)}\BibitemShut {NoStop}%
\bibitem [{\citenamefont {Lewenstein}\ \emph {et~al.}(2012)\citenamefont
  {Lewenstein}, \citenamefont {Sanpera},\ and\ \citenamefont
  {Ahufinger}}]{Lewenstein2012}%
  \BibitemOpen
  \bibfield  {author} {\bibinfo {author} {\bibnamefont {Lewenstein},
  \bibfnamefont {M.}}, \bibinfo {author} {\bibfnamefont {A.}~\bibnamefont
  {Sanpera}}, \ and\ \bibinfo {author} {\bibfnamefont {V.}~\bibnamefont
  {Ahufinger}}} (\bibinfo {year} {2012}),\ \href@noop {} {\emph {\bibinfo
  {title} {Ultracold Atoms in Optical Lattices: Simulating Quantum Many-Body
  Systems}}}\ (\bibinfo  {publisher} {Oxford University Press,
  Oxford})\BibitemShut {NoStop}%
\bibitem [{\citenamefont {Li}(2015)}]{Li2015}%
  \BibitemOpen
  \bibfield  {author} {\bibinfo {author} {\bibnamefont {Li}, \bibfnamefont
  {G.}}} (\bibinfo {year} {2015}),\ \href {\doibase 10.1103/PhysRevB.91.165134}
  {\bibfield  {journal} {\bibinfo  {journal} {Phys. Rev. B}\ }\textbf {\bibinfo
  {volume} {91}},\ \bibinfo {pages} {165134}}\BibitemShut {NoStop}%
\bibitem [{\citenamefont {Li}\ \emph {et~al.}(2014)\citenamefont {Li},
  \citenamefont {Antipov}, \citenamefont {Rubtsov}, \citenamefont {Kirchner},\
  and\ \citenamefont {Hanke}}]{Li2014}%
  \BibitemOpen
  \bibfield  {author} {\bibinfo {author} {\bibnamefont {Li}, \bibfnamefont
  {G.}}, \bibinfo {author} {\bibfnamefont {A.~E.}\ \bibnamefont {Antipov}},
  \bibinfo {author} {\bibfnamefont {A.~N.}\ \bibnamefont {Rubtsov}}, \bibinfo
  {author} {\bibfnamefont {S.}~\bibnamefont {Kirchner}}, \ and\ \bibinfo
  {author} {\bibfnamefont {W.}~\bibnamefont {Hanke}}} (\bibinfo {year}
  {2014}),\ \href {\doibase 10.1103/PhysRevB.89.161118} {\bibfield  {journal}
  {\bibinfo  {journal} {Phys. Rev. B}\ }\textbf {\bibinfo {volume} {89}},\
  \bibinfo {pages} {161118}}\BibitemShut {NoStop}%
\bibitem [{\citenamefont {{Li}}\ \emph {et~al.}(2017)\citenamefont {{Li}},
  \citenamefont {{Kauch}}, \citenamefont {{Pudleiner}},\ and\ \citenamefont
  {{Held}}}]{Li2017}%
  \BibitemOpen
  \bibfield  {author} {\bibinfo {author} {\bibnamefont {{Li}}, \bibfnamefont
  {G.}}, \bibinfo {author} {\bibfnamefont {A.}~\bibnamefont {{Kauch}}},
  \bibinfo {author} {\bibfnamefont {P.}~\bibnamefont {{Pudleiner}}}, \ and\
  \bibinfo {author} {\bibfnamefont {K.}~\bibnamefont {{Held}}}} (\bibinfo
  {year} {2017}),\ \href {https://arxiv.org/abs/1708.07457} {\ }\Eprint
  {http://arxiv.org/abs/1708.07457} {arXiv:1708.07457} \BibitemShut {NoStop}%
\bibitem [{\citenamefont {Li}\ \emph {et~al.}(2011)\citenamefont {Li},
  \citenamefont {Laubach}, \citenamefont {Fleszar},\ and\ \citenamefont
  {Hanke}}]{Li2011}%
  \BibitemOpen
  \bibfield  {author} {\bibinfo {author} {\bibnamefont {Li}, \bibfnamefont
  {G.}}, \bibinfo {author} {\bibfnamefont {M.}~\bibnamefont {Laubach}},
  \bibinfo {author} {\bibfnamefont {A.}~\bibnamefont {Fleszar}}, \ and\
  \bibinfo {author} {\bibfnamefont {W.}~\bibnamefont {Hanke}}} (\bibinfo {year}
  {2011}),\ \href {\doibase 10.1103/PhysRevB.83.041104} {\bibfield  {journal}
  {\bibinfo  {journal} {Phys. Rev. B}\ }\textbf {\bibinfo {volume} {83}},\
  \bibinfo {pages} {041104}}\BibitemShut {NoStop}%
\bibitem [{\citenamefont {Li}\ \emph {et~al.}(2008)\citenamefont {Li},
  \citenamefont {Lee},\ and\ \citenamefont {Monien}}]{Li2008}%
  \BibitemOpen
  \bibfield  {author} {\bibinfo {author} {\bibnamefont {Li}, \bibfnamefont
  {G.}}, \bibinfo {author} {\bibfnamefont {H.}~\bibnamefont {Lee}}, \ and\
  \bibinfo {author} {\bibfnamefont {H.}~\bibnamefont {Monien}}} (\bibinfo
  {year} {2008}),\ \href {\doibase 10.1103/PhysRevB.78.195105} {\bibfield
  {journal} {\bibinfo  {journal} {Phys. Rev. B}\ }\textbf {\bibinfo {volume}
  {78}},\ \bibinfo {pages} {195105}}\BibitemShut {NoStop}%
\bibitem [{\citenamefont {Li}\ \emph {et~al.}(2016)\citenamefont {Li},
  \citenamefont {Wentzell}, \citenamefont {Pudleiner}, \citenamefont
  {Thunstr\"om},\ and\ \citenamefont {Held}}]{Li2016}%
  \BibitemOpen
  \bibfield  {author} {\bibinfo {author} {\bibnamefont {Li}, \bibfnamefont
  {G.}}, \bibinfo {author} {\bibfnamefont {N.}~\bibnamefont {Wentzell}},
  \bibinfo {author} {\bibfnamefont {P.}~\bibnamefont {Pudleiner}}, \bibinfo
  {author} {\bibfnamefont {P.}~\bibnamefont {Thunstr\"om}}, \ and\ \bibinfo
  {author} {\bibfnamefont {K.}~\bibnamefont {Held}}} (\bibinfo {year} {2016}),\
  \href {\doibase 10.1103/PhysRevB.93.165103} {\bibfield  {journal} {\bibinfo
  {journal} {Phys. Rev. B}\ }\textbf {\bibinfo {volume} {93}},\ \bibinfo
  {pages} {165103}}\BibitemShut {NoStop}%
\bibitem [{\citenamefont {Lichtenstein}\ and\ \citenamefont
  {Katsnelson}(2000)}]{Lichtenstein2000}%
  \BibitemOpen
  \bibfield  {author} {\bibinfo {author} {\bibnamefont {Lichtenstein},
  \bibfnamefont {A.~I.}}, \ and\ \bibinfo {author} {\bibfnamefont {M.~I.}\
  \bibnamefont {Katsnelson}}} (\bibinfo {year} {2000}),\ \href {\doibase
  10.1103/PhysRevB.62.R9283} {\bibfield  {journal} {\bibinfo  {journal} {Phys.
  Rev. B}\ }\textbf {\bibinfo {volume} {62}},\ \bibinfo {pages}
  {R9283}}\BibitemShut {NoStop}%
\bibitem [{\citenamefont {Lieb}\ and\ \citenamefont {Wu}(1968)}]{Lieb1968}%
  \BibitemOpen
  \bibfield  {author} {\bibinfo {author} {\bibnamefont {Lieb}, \bibfnamefont
  {E.~H.}}, \ and\ \bibinfo {author} {\bibfnamefont {F.~Y.}\ \bibnamefont
  {Wu}}} (\bibinfo {year} {1968}),\ \href {\doibase
  10.1103/PhysRevLett.20.1445} {\bibfield  {journal} {\bibinfo  {journal}
  {Phys. Rev. Lett.}\ }\textbf {\bibinfo {volume} {20}},\ \bibinfo {pages}
  {1445}}\BibitemShut {NoStop}%
\bibitem [{\citenamefont {Lifshitz}(1960)}]{Lifshitz1960}%
  \BibitemOpen
  \bibfield  {author} {\bibinfo {author} {\bibnamefont {Lifshitz},
  \bibfnamefont {I.~M.}}} (\bibinfo {year} {1960}),\ \href@noop {} {\bibfield
  {journal} {\bibinfo  {journal} {Sov. Phys. JEPT}\ }\textbf {\bibinfo {volume}
  {11}},\ \bibinfo {pages} {1130}}\BibitemShut {NoStop}%
\bibitem [{\citenamefont {Liu}\ and\ \citenamefont {Wang}(2015)}]{Liu2015}%
  \BibitemOpen
  \bibfield  {author} {\bibinfo {author} {\bibnamefont {Liu}, \bibfnamefont
  {Y.-H.}}, \ and\ \bibinfo {author} {\bibfnamefont {L.}~\bibnamefont {Wang}}}
  (\bibinfo {year} {2015}),\ \href
  {http://journals.aps.org/prb/abstract/10.1103/PhysRevB.92.235129} {\bibfield
  {journal} {\bibinfo  {journal} {Phys. Rev. B}\ }\textbf {\bibinfo {volume}
  {92}}~(\bibinfo {number} {23}),\ \bibinfo {pages} {235129}}\BibitemShut
  {NoStop}%
\bibitem [{\citenamefont {v.~L\"ohneysen}\ \emph {et~al.}(2007)\citenamefont
  {v.~L\"ohneysen}, \citenamefont {Rosch}, \citenamefont {Vojta},\ and\
  \citenamefont {W\"olfle}}]{Loehneysen2007}%
  \BibitemOpen
  \bibfield  {author} {\bibinfo {author} {\bibnamefont {v.~L\"ohneysen},
  \bibfnamefont {H.}}, \bibinfo {author} {\bibfnamefont {A.}~\bibnamefont
  {Rosch}}, \bibinfo {author} {\bibfnamefont {M.}~\bibnamefont {Vojta}}, \ and\
  \bibinfo {author} {\bibfnamefont {P.}~\bibnamefont {W\"olfle}}} (\bibinfo
  {year} {2007}),\ \href {\doibase 10.1103/RevModPhys.79.1015} {\bibfield
  {journal} {\bibinfo  {journal} {Rev. Mod. Phys.}\ }\textbf {\bibinfo {volume}
  {79}},\ \bibinfo {pages} {1015}}\BibitemShut {NoStop}%
\bibitem [{\citenamefont {van Loon}\ \emph {et~al.}(2018)\citenamefont {van
  Loon}, \citenamefont {Hafermann},\ and\ \citenamefont
  {Katsnelson}}]{vanLoon2017}%
  \BibitemOpen
  \bibfield  {author} {\bibinfo {author} {\bibnamefont {van Loon},
  \bibfnamefont {E.~G. C.~P.}}, \bibinfo {author} {\bibfnamefont
  {H.}~\bibnamefont {Hafermann}}, \ and\ \bibinfo {author} {\bibfnamefont
  {M.}~\bibnamefont {Katsnelson}}} (\bibinfo {year} {2018}),\ \href {\doibase
  https://doi.org/10.1103/PhysRevB.97.085125} {\bibfield  {journal} {\bibinfo
  {journal} {Phys Rev. B}\ }\textbf {\bibinfo {volume} {97}},\ \bibinfo {pages}
  {085125}}\BibitemShut {NoStop}%
\bibitem [{\citenamefont {van Loon}\ \emph
  {et~al.}(2015{\natexlab{a}})\citenamefont {van Loon}, \citenamefont
  {Hafermann}, \citenamefont {Lichtenstein},\ and\ \citenamefont
  {Katsnelson}}]{vanLoon2015}%
  \BibitemOpen
  \bibfield  {author} {\bibinfo {author} {\bibnamefont {van Loon},
  \bibfnamefont {E.~G. C.~P.}}, \bibinfo {author} {\bibfnamefont
  {H.}~\bibnamefont {Hafermann}}, \bibinfo {author} {\bibfnamefont {A.~I.}\
  \bibnamefont {Lichtenstein}}, \ and\ \bibinfo {author} {\bibfnamefont
  {M.~I.}\ \bibnamefont {Katsnelson}}} (\bibinfo {year} {2015}{\natexlab{a}}),\
  \href {\doibase 10.1103/PhysRevB.92.085106} {\bibfield  {journal} {\bibinfo
  {journal} {Phys. Rev. B}\ }\textbf {\bibinfo {volume} {92}},\ \bibinfo
  {pages} {085106}}\BibitemShut {NoStop}%
\bibitem [{\citenamefont {van Loon}\ \emph
  {et~al.}(2014{\natexlab{a}})\citenamefont {van Loon}, \citenamefont
  {Hafermann}, \citenamefont {Lichtenstein}, \citenamefont {Rubtsov},\ and\
  \citenamefont {Katsnelson}}]{vanLoon2014}%
  \BibitemOpen
  \bibfield  {author} {\bibinfo {author} {\bibnamefont {van Loon},
  \bibfnamefont {E.~G. C.~P.}}, \bibinfo {author} {\bibfnamefont
  {H.}~\bibnamefont {Hafermann}}, \bibinfo {author} {\bibfnamefont {A.~I.}\
  \bibnamefont {Lichtenstein}}, \bibinfo {author} {\bibfnamefont {A.~N.}\
  \bibnamefont {Rubtsov}}, \ and\ \bibinfo {author} {\bibfnamefont {M.~I.}\
  \bibnamefont {Katsnelson}}} (\bibinfo {year} {2014}{\natexlab{a}}),\ \href
  {\doibase 10.1103/PhysRevLett.113.246407} {\bibfield  {journal} {\bibinfo
  {journal} {Phys. Rev. Lett.}\ }\textbf {\bibinfo {volume} {113}},\ \bibinfo
  {pages} {246407}}\BibitemShut {NoStop}%
\bibitem [{\citenamefont {van Loon}\ \emph
  {et~al.}(2016{\natexlab{a}})\citenamefont {van Loon}, \citenamefont
  {Katsnelson}, \citenamefont {Chomaz},\ and\ \citenamefont
  {Lemeshko}}]{vanLoon2016-2}%
  \BibitemOpen
  \bibfield  {author} {\bibinfo {author} {\bibnamefont {van Loon},
  \bibfnamefont {E.~G. C.~P.}}, \bibinfo {author} {\bibfnamefont {M.~I.}\
  \bibnamefont {Katsnelson}}, \bibinfo {author} {\bibfnamefont
  {L.}~\bibnamefont {Chomaz}}, \ and\ \bibinfo {author} {\bibfnamefont
  {M.}~\bibnamefont {Lemeshko}}} (\bibinfo {year} {2016}{\natexlab{a}}),\ \href
  {\doibase 10.1103/PhysRevB.93.195145} {\bibfield  {journal} {\bibinfo
  {journal} {Phys. Rev. B}\ }\textbf {\bibinfo {volume} {93}},\ \bibinfo
  {pages} {195145}}\BibitemShut {NoStop}%
\bibitem [{\citenamefont {van Loon}\ \emph
  {et~al.}(2015{\natexlab{b}})\citenamefont {van Loon}, \citenamefont
  {Katsnelson},\ and\ \citenamefont {Lemeshko}}]{vanLoon2015-2}%
  \BibitemOpen
  \bibfield  {author} {\bibinfo {author} {\bibnamefont {van Loon},
  \bibfnamefont {E.~G. C.~P.}}, \bibinfo {author} {\bibfnamefont {M.~I.}\
  \bibnamefont {Katsnelson}}, \ and\ \bibinfo {author} {\bibfnamefont
  {M.}~\bibnamefont {Lemeshko}}} (\bibinfo {year} {2015}{\natexlab{b}}),\ \href
  {\doibase 10.1103/PhysRevB.92.081106} {\bibfield  {journal} {\bibinfo
  {journal} {Phys. Rev. B}\ }\textbf {\bibinfo {volume} {92}},\ \bibinfo
  {pages} {081106}}\BibitemShut {NoStop}%
\bibitem [{\citenamefont {van Loon}\ \emph
  {et~al.}(2016{\natexlab{b}})\citenamefont {van Loon}, \citenamefont {Krien},
  \citenamefont {Hafermann}, \citenamefont {Stepanov}, \citenamefont
  {Lichtenstein},\ and\ \citenamefont {Katsnelson}}]{vanLoon2016}%
  \BibitemOpen
  \bibfield  {author} {\bibinfo {author} {\bibnamefont {van Loon},
  \bibfnamefont {E.~G. C.~P.}}, \bibinfo {author} {\bibfnamefont
  {F.}~\bibnamefont {Krien}}, \bibinfo {author} {\bibfnamefont
  {H.}~\bibnamefont {Hafermann}}, \bibinfo {author} {\bibfnamefont {E.~A.}\
  \bibnamefont {Stepanov}}, \bibinfo {author} {\bibfnamefont {A.~I.}\
  \bibnamefont {Lichtenstein}}, \ and\ \bibinfo {author} {\bibfnamefont
  {M.~I.}\ \bibnamefont {Katsnelson}}} (\bibinfo {year} {2016}{\natexlab{b}}),\
  \href {\doibase 10.1103/PhysRevB.93.155162} {\bibfield  {journal} {\bibinfo
  {journal} {Phys. Rev. B}\ }\textbf {\bibinfo {volume} {93}},\ \bibinfo
  {pages} {155162}}\BibitemShut {NoStop}%
\bibitem [{\citenamefont {van Loon}\ \emph
  {et~al.}(2014{\natexlab{b}})\citenamefont {van Loon}, \citenamefont
  {Lichtenstein}, \citenamefont {Katsnelson}, \citenamefont {Parcollet},\ and\
  \citenamefont {Hafermann}}]{vanLoon2014a}%
  \BibitemOpen
  \bibfield  {author} {\bibinfo {author} {\bibnamefont {van Loon},
  \bibfnamefont {E.~G. C.~P.}}, \bibinfo {author} {\bibfnamefont {A.~I.}\
  \bibnamefont {Lichtenstein}}, \bibinfo {author} {\bibfnamefont {M.~I.}\
  \bibnamefont {Katsnelson}}, \bibinfo {author} {\bibfnamefont
  {O.}~\bibnamefont {Parcollet}}, \ and\ \bibinfo {author} {\bibfnamefont
  {H.}~\bibnamefont {Hafermann}}} (\bibinfo {year} {2014}{\natexlab{b}}),\
  \href {\doibase 10.1103/PhysRevB.90.235135} {\bibfield  {journal} {\bibinfo
  {journal} {Phys. Rev. B}\ }\textbf {\bibinfo {volume} {90}},\ \bibinfo
  {pages} {235135}}\BibitemShut {NoStop}%
\bibitem [{\citenamefont {Mahan}(2000)}]{Mahan2000}%
  \BibitemOpen
  \bibfield  {author} {\bibinfo {author} {\bibnamefont {Mahan}, \bibfnamefont
  {G.~D.}}} (\bibinfo {year} {2000}),\ \href@noop {} {\emph {\bibinfo {title}
  {Many-Particle Physics}}}\ (\bibinfo  {publisher} {Kluwer Academic/Plenum
  Publishers, New York})\BibitemShut {NoStop}%
\bibitem [{\citenamefont {Maier}\ \emph
  {et~al.}(2005{\natexlab{a}})\citenamefont {Maier}, \citenamefont {Jarrell},
  \citenamefont {Pruschke},\ and\ \citenamefont {Hettler}}]{Maier2005}%
  \BibitemOpen
  \bibfield  {author} {\bibinfo {author} {\bibnamefont {Maier}, \bibfnamefont
  {T.}}, \bibinfo {author} {\bibfnamefont {M.}~\bibnamefont {Jarrell}},
  \bibinfo {author} {\bibfnamefont {T.}~\bibnamefont {Pruschke}}, \ and\
  \bibinfo {author} {\bibfnamefont {M.~H.}\ \bibnamefont {Hettler}}} (\bibinfo
  {year} {2005}{\natexlab{a}}),\ \href {\doibase 10.1103/RevModPhys.77.1027}
  {\bibfield  {journal} {\bibinfo  {journal} {Rev. Mod. Phys.}\ }\textbf
  {\bibinfo {volume} {77}},\ \bibinfo {pages} {1027}}\BibitemShut {NoStop}%
\bibitem [{\citenamefont {Maier}\ \emph
  {et~al.}(2005{\natexlab{b}})\citenamefont {Maier}, \citenamefont {Jarrell},
  \citenamefont {Schulthess}, \citenamefont {Kent},\ and\ \citenamefont
  {White}}]{Maier2005a}%
  \BibitemOpen
  \bibfield  {author} {\bibinfo {author} {\bibnamefont {Maier}, \bibfnamefont
  {T.~A.}}, \bibinfo {author} {\bibfnamefont {M.}~\bibnamefont {Jarrell}},
  \bibinfo {author} {\bibfnamefont {T.~C.}\ \bibnamefont {Schulthess}},
  \bibinfo {author} {\bibfnamefont {P.~R.~C.}\ \bibnamefont {Kent}}, \ and\
  \bibinfo {author} {\bibfnamefont {J.~B.}\ \bibnamefont {White}}} (\bibinfo
  {year} {2005}{\natexlab{b}}),\ \href {\doibase 10.1103/PhysRevLett.95.237001}
  {\bibfield  {journal} {\bibinfo  {journal} {Phys. Rev. Lett.}\ }\textbf
  {\bibinfo {volume} {95}},\ \bibinfo {pages} {237001}}\BibitemShut {NoStop}%
\bibitem [{\citenamefont {Maier}\ \emph {et~al.}(2006)\citenamefont {Maier},
  \citenamefont {Jarrell},\ and\ \citenamefont {Scalapino}}]{Maier2006}%
  \BibitemOpen
  \bibfield  {author} {\bibinfo {author} {\bibnamefont {Maier}, \bibfnamefont
  {T.~A.}}, \bibinfo {author} {\bibfnamefont {M.~S.}\ \bibnamefont {Jarrell}},
  \ and\ \bibinfo {author} {\bibfnamefont {D.~J.}\ \bibnamefont {Scalapino}}}
  (\bibinfo {year} {2006}),\ \href {\doibase 10.1103/PhysRevLett.96.047005}
  {\bibfield  {journal} {\bibinfo  {journal} {Phys. Rev. Lett.}\ }\textbf
  {\bibinfo {volume} {96}},\ \bibinfo {pages} {047005}}\BibitemShut {NoStop}%
\bibitem [{\citenamefont {Mancini}(2000)}]{Mancini2000}%
  \BibitemOpen
  \bibfield  {author} {\bibinfo {author} {\bibnamefont {Mancini}, \bibfnamefont
  {F.}}} (\bibinfo {year} {2000}),\ \href
  {http://stacks.iop.org/0295-5075/50/i=2/a=229} {\bibfield  {journal}
  {\bibinfo  {journal} {EPL (Europhysics Letters)}\ }\textbf {\bibinfo {volume}
  {50}}~(\bibinfo {number} {2}),\ \bibinfo {pages} {229}}\BibitemShut {NoStop}%
\bibitem [{\citenamefont {Ma{\'{s}}ka}\ and\ \citenamefont
  {Czajka}(2005)}]{Maska2005}%
  \BibitemOpen
  \bibfield  {author} {\bibinfo {author} {\bibnamefont {Ma{\'{s}}ka},
  \bibfnamefont {M.~M.}}, \ and\ \bibinfo {author} {\bibfnamefont
  {K.}~\bibnamefont {Czajka}}} (\bibinfo {year} {2005}),\ \href {\doibase
  10.1002/pssb.200460067} {\bibfield  {journal} {\bibinfo  {journal} {Phys.
  Status Solidi B}\ }\textbf {\bibinfo {volume} {242}}~(\bibinfo {number}
  {2}),\ \bibinfo {pages} {479}}\BibitemShut {NoStop}%
\bibitem [{\citenamefont {Ma{\'{s}}ka}\ and\ \citenamefont
  {Czajka}(2006)}]{Maska2006}%
  \BibitemOpen
  \bibfield  {author} {\bibinfo {author} {\bibnamefont {Ma{\'{s}}ka},
  \bibfnamefont {M.~M.}}, \ and\ \bibinfo {author} {\bibfnamefont
  {K.}~\bibnamefont {Czajka}}} (\bibinfo {year} {2006}),\ \href {\doibase
  10.1103/PhysRevB.74.035109} {\bibfield  {journal} {\bibinfo  {journal} {Phys.
  Rev. B}\ }\textbf {\bibinfo {volume} {74}}~(\bibinfo {number} {3}),\ \bibinfo
  {pages} {035109}}\BibitemShut {NoStop}%
\bibitem [{\citenamefont {Merino}\ and\ \citenamefont
  {Gunnarsson}(2014)}]{Merino2014}%
  \BibitemOpen
  \bibfield  {author} {\bibinfo {author} {\bibnamefont {Merino}, \bibfnamefont
  {J.}}, \ and\ \bibinfo {author} {\bibfnamefont {O.}~\bibnamefont
  {Gunnarsson}}} (\bibinfo {year} {2014}),\ \href {\doibase
  10.1103/PhysRevB.89.245130} {\bibfield  {journal} {\bibinfo  {journal} {Phys.
  Rev. B}\ }\textbf {\bibinfo {volume} {89}},\ \bibinfo {pages}
  {245130}}\BibitemShut {NoStop}%
\bibitem [{\citenamefont {Merker}\ \emph {et~al.}(2013)\citenamefont {Merker},
  \citenamefont {Kirchner}, \citenamefont {Mu\~noz},\ and\ \citenamefont
  {Costi}}]{Merker2013}%
  \BibitemOpen
  \bibfield  {author} {\bibinfo {author} {\bibnamefont {Merker}, \bibfnamefont
  {L.}}, \bibinfo {author} {\bibfnamefont {S.}~\bibnamefont {Kirchner}},
  \bibinfo {author} {\bibfnamefont {E.}~\bibnamefont {Mu\~noz}}, \ and\
  \bibinfo {author} {\bibfnamefont {T.~A.}\ \bibnamefont {Costi}}} (\bibinfo
  {year} {2013}),\ \href {\doibase 10.1103/PhysRevB.87.165132} {\bibfield
  {journal} {\bibinfo  {journal} {Phys. Rev. B}\ }\textbf {\bibinfo {volume}
  {87}},\ \bibinfo {pages} {165132}}\BibitemShut {NoStop}%
\bibitem [{\citenamefont {Mermin}\ and\ \citenamefont
  {Wagner}(1966)}]{Mermin1966}%
  \BibitemOpen
  \bibfield  {author} {\bibinfo {author} {\bibnamefont {Mermin}, \bibfnamefont
  {N.~D.}}, \ and\ \bibinfo {author} {\bibfnamefont {H.}~\bibnamefont
  {Wagner}}} (\bibinfo {year} {1966}),\ \href {\doibase
  10.1103/PhysRevLett.17.1307} {\bibfield  {journal} {\bibinfo  {journal}
  {Phys. Rev. Lett.}\ }\textbf {\bibinfo {volume} {17}},\ \bibinfo {pages}
  {1307}}\BibitemShut {NoStop}%
\bibitem [{\citenamefont {Metzner}(1991)}]{Metzner1991}%
  \BibitemOpen
  \bibfield  {author} {\bibinfo {author} {\bibnamefont {Metzner}, \bibfnamefont
  {W.}}} (\bibinfo {year} {1991}),\ \href {\doibase 10.1103/PhysRevB.43.8549}
  {\bibfield  {journal} {\bibinfo  {journal} {Phys. Rev. B}\ }\textbf {\bibinfo
  {volume} {43}},\ \bibinfo {pages} {8549}}\BibitemShut {NoStop}%
\bibitem [{\citenamefont {Metzner}\ \emph {et~al.}(2012)\citenamefont
  {Metzner}, \citenamefont {Salmhofer}, \citenamefont {Honerkamp},
  \citenamefont {Meden},\ and\ \citenamefont {Sch\"onhammer}}]{Metzner2012}%
  \BibitemOpen
  \bibfield  {author} {\bibinfo {author} {\bibnamefont {Metzner}, \bibfnamefont
  {W.}}, \bibinfo {author} {\bibfnamefont {M.}~\bibnamefont {Salmhofer}},
  \bibinfo {author} {\bibfnamefont {C.}~\bibnamefont {Honerkamp}}, \bibinfo
  {author} {\bibfnamefont {V.}~\bibnamefont {Meden}}, \ and\ \bibinfo {author}
  {\bibfnamefont {K.}~\bibnamefont {Sch\"onhammer}}} (\bibinfo {year} {2012}),\
  \href {\doibase 10.1103/RevModPhys.84.299} {\bibfield  {journal} {\bibinfo
  {journal} {Rev. Mod. Phys.}\ }\textbf {\bibinfo {volume} {84}},\ \bibinfo
  {pages} {299}}\BibitemShut {NoStop}%
\bibitem [{\citenamefont {Metzner}\ and\ \citenamefont
  {Vollhardt}(1989)}]{Metzner1989}%
  \BibitemOpen
  \bibfield  {author} {\bibinfo {author} {\bibnamefont {Metzner}, \bibfnamefont
  {W.}}, \ and\ \bibinfo {author} {\bibfnamefont {D.}~\bibnamefont
  {Vollhardt}}} (\bibinfo {year} {1989}),\ \href {\doibase
  10.1103/PhysRevLett.62.324} {\bibfield  {journal} {\bibinfo  {journal} {Phys.
  Rev. Lett.}\ }\textbf {\bibinfo {volume} {62}},\ \bibinfo {pages}
  {324}}\BibitemShut {NoStop}%
\bibitem [{\citenamefont {Mielke}\ and\ \citenamefont
  {Tasaki}(1993)}]{Mielke1993}%
  \BibitemOpen
  \bibfield  {author} {\bibinfo {author} {\bibnamefont {Mielke}, \bibfnamefont
  {A.}}, \ and\ \bibinfo {author} {\bibfnamefont {H.}~\bibnamefont {Tasaki}}}
  (\bibinfo {year} {1993}),\ \href@noop {} {\bibfield  {journal} {\bibinfo
  {journal} {Commun. Math. Phys.}\ }\textbf {\bibinfo {volume} {158}},\
  \bibinfo {pages} {341}}\BibitemShut {NoStop}%
\bibitem [{\citenamefont {Millis}(1993)}]{Millis1993}%
  \BibitemOpen
  \bibfield  {author} {\bibinfo {author} {\bibnamefont {Millis}, \bibfnamefont
  {A.~J.}}} (\bibinfo {year} {1993}),\ \href {\doibase
  10.1103/PhysRevB.48.7183} {\bibfield  {journal} {\bibinfo  {journal} {Phys.
  Rev. B}\ }\textbf {\bibinfo {volume} {48}},\ \bibinfo {pages}
  {7183}}\BibitemShut {NoStop}%
\bibitem [{\citenamefont {Miyake}\ \emph {et~al.}(2013)\citenamefont {Miyake},
  \citenamefont {Martins}, \citenamefont {Sakuma},\ and\ \citenamefont
  {Aryasetiawan}}]{Miyake13}%
  \BibitemOpen
  \bibfield  {author} {\bibinfo {author} {\bibnamefont {Miyake}, \bibfnamefont
  {T.}}, \bibinfo {author} {\bibfnamefont {C.}~\bibnamefont {Martins}},
  \bibinfo {author} {\bibfnamefont {R.}~\bibnamefont {Sakuma}}, \ and\ \bibinfo
  {author} {\bibfnamefont {F.}~\bibnamefont {Aryasetiawan}}} (\bibinfo {year}
  {2013}),\ \href {\doibase 10.1103/PhysRevB.87.115110} {\bibfield  {journal}
  {\bibinfo  {journal} {Phys. Rev. B}\ }\textbf {\bibinfo {volume} {87}},\
  \bibinfo {pages} {115110}}\BibitemShut {NoStop}%
\bibitem [{\citenamefont {Montorsi}(1992)}]{Hubbcollbook}%
  \BibitemOpen
  \bibinfo {editor} {\bibnamefont {Montorsi}, \bibfnamefont {A.}},\ Ed.
  (\bibinfo {year} {1992}),\ \href@noop {} {\emph {\bibinfo {title} {The
  Hubbard Model - A Reprint Volume}}}\ (\bibinfo  {publisher} {World
  Scientific, Singapore})\BibitemShut {NoStop}%
\bibitem [{\citenamefont {Moreo}(1993)}]{Moreo1993}%
  \BibitemOpen
  \bibfield  {author} {\bibinfo {author} {\bibnamefont {Moreo}, \bibfnamefont
  {A.}}} (\bibinfo {year} {1993}),\ \href {\doibase 10.1103/PhysRevB.48.3380}
  {\bibfield  {journal} {\bibinfo  {journal} {Phys. Rev. B}\ }\textbf {\bibinfo
  {volume} {48}}~(\bibinfo {number} {5}),\ \bibinfo {pages} {3380}}\BibitemShut
  {NoStop}%
\bibitem [{\citenamefont {Morita}\ \emph {et~al.}(2002)\citenamefont {Morita},
  \citenamefont {Watanabe},\ and\ \citenamefont {Imada}}]{Morita2002}%
  \BibitemOpen
  \bibfield  {author} {\bibinfo {author} {\bibnamefont {Morita}, \bibfnamefont
  {H.}}, \bibinfo {author} {\bibfnamefont {S.}~\bibnamefont {Watanabe}}, \ and\
  \bibinfo {author} {\bibfnamefont {M.}~\bibnamefont {Imada}}} (\bibinfo {year}
  {2002}),\ \href {\doibase 10.1143/JPSJ.71.2109} {\bibfield  {journal}
  {\bibinfo  {journal} {Journal of the Physical Society of Japan}\ }\textbf
  {\bibinfo {volume} {71}}~(\bibinfo {number} {9}),\ \bibinfo {pages}
  {2109}}\BibitemShut {NoStop}%
\bibitem [{\citenamefont {Moriya}(1985)}]{Moriya1985}%
  \BibitemOpen
  \bibfield  {author} {\bibinfo {author} {\bibnamefont {Moriya}, \bibfnamefont
  {T.}}} (\bibinfo {year} {1985}),\ \href@noop {} {\emph {\bibinfo {title}
  {Spin Fluctuations in Itinerant Electron Magnetism}}},\ edited by\ \bibinfo
  {editor} {\bibfnamefont {M.}~\bibnamefont {Cardona}}, \bibinfo {editor}
  {\bibfnamefont {P.}~\bibnamefont {Fulde}}, \ and\ \bibinfo {editor}
  {\bibfnamefont {H.-J.}\ \bibnamefont {Queisser}}\ (\bibinfo  {publisher}
  {Springer Verlag, Berlin, Heidelberg})\BibitemShut {NoStop}%
\bibitem [{\citenamefont {Moukouri}\ \emph {et~al.}(2000)\citenamefont
  {Moukouri}, \citenamefont {Allen}, \citenamefont {Lemay}, \citenamefont
  {Kyung}, \citenamefont {Poulin}, \citenamefont {Vilk},\ and\ \citenamefont
  {Tremblay}}]{Moukouri2000}%
  \BibitemOpen
  \bibfield  {author} {\bibinfo {author} {\bibnamefont {Moukouri},
  \bibfnamefont {S.}}, \bibinfo {author} {\bibfnamefont {S.}~\bibnamefont
  {Allen}}, \bibinfo {author} {\bibfnamefont {F.}~\bibnamefont {Lemay}},
  \bibinfo {author} {\bibfnamefont {B.}~\bibnamefont {Kyung}}, \bibinfo
  {author} {\bibfnamefont {D.}~\bibnamefont {Poulin}}, \bibinfo {author}
  {\bibfnamefont {Y.~M.}\ \bibnamefont {Vilk}}, \ and\ \bibinfo {author}
  {\bibfnamefont {A.-M.~S.}\ \bibnamefont {Tremblay}}} (\bibinfo {year}
  {2000}),\ \href {\doibase 10.1103/PhysRevB.61.7887} {\bibfield  {journal}
  {\bibinfo  {journal} {Phys. Rev. B}\ }\textbf {\bibinfo {volume} {61}},\
  \bibinfo {pages} {7887}}\BibitemShut {NoStop}%
\bibitem [{\citenamefont {Moukouri}\ and\ \citenamefont
  {Jarrell}(2001)}]{Moukouri2001}%
  \BibitemOpen
  \bibfield  {author} {\bibinfo {author} {\bibnamefont {Moukouri},
  \bibfnamefont {S.}}, \ and\ \bibinfo {author} {\bibfnamefont
  {M.}~\bibnamefont {Jarrell}}} (\bibinfo {year} {2001}),\ \href {\doibase
  10.1103/PhysRevLett.87.167010} {\bibfield  {journal} {\bibinfo  {journal}
  {Phys. Rev. Lett.}\ }\textbf {\bibinfo {volume} {87}},\ \bibinfo {pages}
  {167010}}\BibitemShut {NoStop}%
\bibitem [{\citenamefont {Mu\~noz}\ \emph {et~al.}(2013)\citenamefont
  {Mu\~noz}, \citenamefont {Bolech},\ and\ \citenamefont
  {Kirchner}}]{Munoz2013}%
  \BibitemOpen
  \bibfield  {author} {\bibinfo {author} {\bibnamefont {Mu\~noz}, \bibfnamefont
  {E.}}, \bibinfo {author} {\bibfnamefont {C.~J.}\ \bibnamefont {Bolech}}, \
  and\ \bibinfo {author} {\bibfnamefont {S.}~\bibnamefont {Kirchner}}}
  (\bibinfo {year} {2013}),\ \href {\doibase 10.1103/PhysRevLett.110.016601}
  {\bibfield  {journal} {\bibinfo  {journal} {Phys. Rev. Lett.}\ }\textbf
  {\bibinfo {volume} {110}},\ \bibinfo {pages} {016601}}\BibitemShut {NoStop}%
\bibitem [{\citenamefont {M{\"u}ller-Hartmann}(1989)}]{Muller-Hartmann1988}%
  \BibitemOpen
  \bibfield  {author} {\bibinfo {author} {\bibnamefont {M{\"u}ller-Hartmann},
  \bibfnamefont {E.}}} (\bibinfo {year} {1989}),\ \href {\doibase
  10.1007/BF01311397} {\bibfield  {journal} {\bibinfo  {journal} {Zeitschrift
  f{\"u}r Physik B Condensed Matter}\ }\textbf {\bibinfo {volume}
  {74}}~(\bibinfo {number} {4}),\ \bibinfo {pages} {507}}\BibitemShut {NoStop}%
\bibitem [{\citenamefont {Negele}\ and\ \citenamefont
  {Orland}(1998)}]{Negele1998}%
  \BibitemOpen
  \bibfield  {author} {\bibinfo {author} {\bibnamefont {Negele}, \bibfnamefont
  {J.~W.}}, \ and\ \bibinfo {author} {\bibfnamefont {H.}~\bibnamefont
  {Orland}}} (\bibinfo {year} {1998}),\ \href@noop {} {\emph {\bibinfo {title}
  {Quantum Many-Particle Systems}}},\ edited by\ \bibinfo {editor}
  {\bibfnamefont {D.}~\bibnamefont {Pines}}\ (\bibinfo  {publisher} {Taylor \&
  Francis Inc},\ \bibinfo {address} {Reading, MA, United States})\BibitemShut
  {NoStop}%
\bibitem [{\citenamefont {Nekrasov}\ \emph {et~al.}(2008)\citenamefont
  {Nekrasov}, \citenamefont {Kokorina}, \citenamefont {Kuchinskii},
  \citenamefont {Pchelkina},\ and\ \citenamefont {Sadovskii}}]{Nekrasov2008}%
  \BibitemOpen
  \bibfield  {author} {\bibinfo {author} {\bibnamefont {Nekrasov},
  \bibfnamefont {I.}}, \bibinfo {author} {\bibfnamefont {E.}~\bibnamefont
  {Kokorina}}, \bibinfo {author} {\bibfnamefont {E.}~\bibnamefont
  {Kuchinskii}}, \bibinfo {author} {\bibfnamefont {Z.}~\bibnamefont
  {Pchelkina}}, \ and\ \bibinfo {author} {\bibfnamefont {M.}~\bibnamefont
  {Sadovskii}}} (\bibinfo {year} {2008}),\ \href {\doibase
  10.1016/j.jpcs.2008.06.120} {\bibfield  {journal} {\bibinfo  {journal} {J.
  Phys. Chem. Solids}\ }\textbf {\bibinfo {volume} {69}}~(\bibinfo {number}
  {12}),\ \bibinfo {pages} {3269}}\BibitemShut {NoStop}%
\bibitem [{\citenamefont {Nekrasov}\ \emph {et~al.}(2011)\citenamefont
  {Nekrasov}, \citenamefont {Kuchinskii},\ and\ \citenamefont
  {Sadovskii}}]{Nekrasov2011}%
  \BibitemOpen
  \bibfield  {author} {\bibinfo {author} {\bibnamefont {Nekrasov},
  \bibfnamefont {I.}}, \bibinfo {author} {\bibfnamefont {E.}~\bibnamefont
  {Kuchinskii}}, \ and\ \bibinfo {author} {\bibfnamefont {M.}~\bibnamefont
  {Sadovskii}}} (\bibinfo {year} {2011}),\ \href {\doibase
  10.1016/j.jpcs.2010.10.081} {\bibfield  {journal} {\bibinfo  {journal} {J.
  Phys. Chem. Solids}\ }\textbf {\bibinfo {volume} {72}}~(\bibinfo {number}
  {5}),\ \bibinfo {pages} {371}}\BibitemShut {NoStop}%
\bibitem [{\citenamefont {Ohashi}\ \emph {et~al.}(2008)\citenamefont {Ohashi},
  \citenamefont {Momoi}, \citenamefont {Tsunetsugu},\ and\ \citenamefont
  {Kawakami}}]{Ohashi2008}%
  \BibitemOpen
  \bibfield  {author} {\bibinfo {author} {\bibnamefont {Ohashi}, \bibfnamefont
  {T.}}, \bibinfo {author} {\bibfnamefont {T.}~\bibnamefont {Momoi}}, \bibinfo
  {author} {\bibfnamefont {H.}~\bibnamefont {Tsunetsugu}}, \ and\ \bibinfo
  {author} {\bibfnamefont {N.}~\bibnamefont {Kawakami}}} (\bibinfo {year}
  {2008}),\ \href {\doibase 10.1103/PhysRevLett.100.076402} {\bibfield
  {journal} {\bibinfo  {journal} {Phys. Rev. Lett.}\ }\textbf {\bibinfo
  {volume} {100}},\ \bibinfo {pages} {076402}}\BibitemShut {NoStop}%
\bibitem [{\citenamefont {Osipov}\ and\ \citenamefont
  {Rubtsov}(2013)}]{Osipov2013}%
  \BibitemOpen
  \bibfield  {author} {\bibinfo {author} {\bibnamefont {Osipov}, \bibfnamefont
  {A.~A.}}, \ and\ \bibinfo {author} {\bibfnamefont {A.~N.}\ \bibnamefont
  {Rubtsov}}} (\bibinfo {year} {2013}),\ \href {\doibase
  10.1088/1367-2630/15/7/075016} {\bibfield  {journal} {\bibinfo  {journal}
  {New J. Phys.}\ }\textbf {\bibinfo {volume} {15}}~(\bibinfo {number} {7}),\
  \bibinfo {pages} {075016}}\BibitemShut {NoStop}%
\bibitem [{\citenamefont {Otsuki}(2013)}]{Otsuki2013}%
  \BibitemOpen
  \bibfield  {author} {\bibinfo {author} {\bibnamefont {Otsuki}, \bibfnamefont
  {J.}}} (\bibinfo {year} {2013}),\ \href {\doibase 10.1103/PhysRevB.87.125102}
  {\bibfield  {journal} {\bibinfo  {journal} {Phys. Rev. B}\ }\textbf {\bibinfo
  {volume} {87}},\ \bibinfo {pages} {125102}}\BibitemShut {NoStop}%
\bibitem [{\citenamefont {Otsuki}(2015)}]{Otsuki2015}%
  \BibitemOpen
  \bibfield  {author} {\bibinfo {author} {\bibnamefont {Otsuki}, \bibfnamefont
  {J.}}} (\bibinfo {year} {2015}),\ \href {\doibase
  10.1103/PhysRevLett.115.036404} {\bibfield  {journal} {\bibinfo  {journal}
  {Phys. Rev. Lett.}\ }\textbf {\bibinfo {volume} {115}},\ \bibinfo {pages}
  {036404}}\BibitemShut {NoStop}%
\bibitem [{\citenamefont {Otsuki}\ \emph {et~al.}(2014)\citenamefont {Otsuki},
  \citenamefont {Hafermann},\ and\ \citenamefont {Lichtenstein}}]{Otsuki2014}%
  \BibitemOpen
  \bibfield  {author} {\bibinfo {author} {\bibnamefont {Otsuki}, \bibfnamefont
  {J.}}, \bibinfo {author} {\bibfnamefont {H.}~\bibnamefont {Hafermann}}, \
  and\ \bibinfo {author} {\bibfnamefont {A.~I.}\ \bibnamefont {Lichtenstein}}}
  (\bibinfo {year} {2014}),\ \href {\doibase 10.1103/PhysRevB.90.235132}
  {\bibfield  {journal} {\bibinfo  {journal} {Phys. Rev. B}\ }\textbf {\bibinfo
  {volume} {90}},\ \bibinfo {pages} {235132}}\BibitemShut {NoStop}%
\bibitem [{\citenamefont {Otsuki}\ \emph {et~al.}(2009)\citenamefont {Otsuki},
  \citenamefont {Kusunose},\ and\ \citenamefont {Kuramoto}}]{Otsuki2009}%
  \BibitemOpen
  \bibfield  {author} {\bibinfo {author} {\bibnamefont {Otsuki}, \bibfnamefont
  {J.}}, \bibinfo {author} {\bibfnamefont {H.}~\bibnamefont {Kusunose}}, \ and\
  \bibinfo {author} {\bibfnamefont {Y.}~\bibnamefont {Kuramoto}}} (\bibinfo
  {year} {2009}),\ \href {\doibase 10.1103/PhysRevLett.102.017202} {\bibfield
  {journal} {\bibinfo  {journal} {Phys. Rev. Lett.}\ }\textbf {\bibinfo
  {volume} {102}},\ \bibinfo {pages} {017202}}\BibitemShut {NoStop}%
\bibitem [{\citenamefont {Otsuki}\ \emph {et~al.}(2017)\citenamefont {Otsuki},
  \citenamefont {Ohzeki}, \citenamefont {Shinaoka},\ and\ \citenamefont
  {Yoshimi}}]{Otsuki2017}%
  \BibitemOpen
  \bibfield  {author} {\bibinfo {author} {\bibnamefont {Otsuki}, \bibfnamefont
  {J.}}, \bibinfo {author} {\bibfnamefont {M.}~\bibnamefont {Ohzeki}}, \bibinfo
  {author} {\bibfnamefont {H.}~\bibnamefont {Shinaoka}}, \ and\ \bibinfo
  {author} {\bibfnamefont {K.}~\bibnamefont {Yoshimi}}} (\bibinfo {year}
  {2017}),\ \href {\doibase 10.1103/PhysRevE.95.061302} {\bibfield  {journal}
  {\bibinfo  {journal} {Phys. Rev. E}\ }\textbf {\bibinfo {volume} {95}},\
  \bibinfo {pages} {061302}}\BibitemShut {NoStop}%
\bibitem [{\citenamefont {Pairault}\ \emph {et~al.}(1998)\citenamefont
  {Pairault}, \citenamefont {S\'en\'echal},\ and\ \citenamefont
  {Tremblay}}]{Pairault1998}%
  \BibitemOpen
  \bibfield  {author} {\bibinfo {author} {\bibnamefont {Pairault},
  \bibfnamefont {S.}}, \bibinfo {author} {\bibfnamefont {D.}~\bibnamefont
  {S\'en\'echal}}, \ and\ \bibinfo {author} {\bibfnamefont {A.-M.~S.}\
  \bibnamefont {Tremblay}}} (\bibinfo {year} {1998}),\ \href {\doibase
  10.1103/PhysRevLett.80.5389} {\bibfield  {journal} {\bibinfo  {journal}
  {Phys. Rev. Lett.}\ }\textbf {\bibinfo {volume} {80}},\ \bibinfo {pages}
  {5389}}\BibitemShut {NoStop}%
\bibitem [{\citenamefont {Pairault}\ \emph {et~al.}(2000)\citenamefont
  {Pairault}, \citenamefont {S\'en\'echal},\ and\ \citenamefont
  {Tremblay}}]{Pairault2000}%
  \BibitemOpen
  \bibfield  {author} {\bibinfo {author} {\bibnamefont {Pairault},
  \bibfnamefont {S.}}, \bibinfo {author} {\bibfnamefont {D.}~\bibnamefont
  {S\'en\'echal}}, \ and\ \bibinfo {author} {\bibfnamefont {A.-M.~S.}\
  \bibnamefont {Tremblay}}} (\bibinfo {year} {2000}),\ \href
  {http://dx.doi.org/10.1007/s100510070253} {\bibfield  {journal} {\bibinfo
  {journal} {Eur. Phys. J. B}\ }\textbf {\bibinfo {volume} {16}},\ \bibinfo
  {pages} {85}}\BibitemShut {NoStop}%
\bibitem [{\citenamefont {Parcollet}\ \emph {et~al.}(2015)\citenamefont
  {Parcollet}, \citenamefont {Ferrero}, \citenamefont {Ayral}, \citenamefont
  {Hafermann}, \citenamefont {Krivenko}, \citenamefont {Messio},\ and\
  \citenamefont {Seth}}]{TRIQS}%
  \BibitemOpen
  \bibfield  {author} {\bibinfo {author} {\bibnamefont {Parcollet},
  \bibfnamefont {O.}}, \bibinfo {author} {\bibfnamefont {M.}~\bibnamefont
  {Ferrero}}, \bibinfo {author} {\bibfnamefont {T.}~\bibnamefont {Ayral}},
  \bibinfo {author} {\bibfnamefont {H.}~\bibnamefont {Hafermann}}, \bibinfo
  {author} {\bibfnamefont {I.}~\bibnamefont {Krivenko}}, \bibinfo {author}
  {\bibfnamefont {L.}~\bibnamefont {Messio}}, \ and\ \bibinfo {author}
  {\bibfnamefont {P.}~\bibnamefont {Seth}}} (\bibinfo {year} {2015}),\ \href
  {\doibase http://dx.doi.org/10.1016/j.cpc.2015.04.023} {\bibfield  {journal}
  {\bibinfo  {journal} {Computer Physics Communications}\ }\textbf {\bibinfo
  {volume} {196}},\ \bibinfo {pages} {398 }}\BibitemShut {NoStop}%
\bibitem [{\citenamefont {Park}\ \emph {et~al.}(2008)\citenamefont {Park},
  \citenamefont {Haule},\ and\ \citenamefont {Kotliar}}]{Park2008}%
  \BibitemOpen
  \bibfield  {author} {\bibinfo {author} {\bibnamefont {Park}, \bibfnamefont
  {H.}}, \bibinfo {author} {\bibfnamefont {K.}~\bibnamefont {Haule}}, \ and\
  \bibinfo {author} {\bibfnamefont {G.}~\bibnamefont {Kotliar}}} (\bibinfo
  {year} {2008}),\ \href {\doibase 10.1103/PhysRevLett.101.186403} {\bibfield
  {journal} {\bibinfo  {journal} {Phys. Rev. Lett.}\ }\textbf {\bibinfo
  {volume} {101}},\ \bibinfo {pages} {186403}}\BibitemShut {NoStop}%
\bibitem [{\citenamefont {Parragh}\ \emph {et~al.}(2012)\citenamefont
  {Parragh}, \citenamefont {Toschi}, \citenamefont {Held},\ and\ \citenamefont
  {Sangiovanni}}]{w2dynamics}%
  \BibitemOpen
  \bibfield  {author} {\bibinfo {author} {\bibnamefont {Parragh}, \bibfnamefont
  {N.}}, \bibinfo {author} {\bibfnamefont {A.}~\bibnamefont {Toschi}}, \bibinfo
  {author} {\bibfnamefont {K.}~\bibnamefont {Held}}, \ and\ \bibinfo {author}
  {\bibfnamefont {G.}~\bibnamefont {Sangiovanni}}} (\bibinfo {year} {2012}),\
  \href {\doibase 10.1103/PhysRevB.86.155158} {\bibfield  {journal} {\bibinfo
  {journal} {Phys. Rev. B}\ }\textbf {\bibinfo {volume} {86}},\ \bibinfo
  {pages} {155158}}\BibitemShut {NoStop}%
\bibitem [{\citenamefont {Pavarini}\ \emph {et~al.}(2014)\citenamefont
  {Pavarini}, \citenamefont {Koch}, \citenamefont {Vollhardt},\ and\
  \citenamefont {Lichtenstein}}]{DMFT25}%
  \BibitemOpen
  \bibfield  {author} {\bibinfo {author} {\bibnamefont {Pavarini},
  \bibfnamefont {E.}}, \bibinfo {author} {\bibfnamefont {E.}~\bibnamefont
  {Koch}}, \bibinfo {author} {\bibfnamefont {D.}~\bibnamefont {Vollhardt}}, \
  and\ \bibinfo {author} {\bibfnamefont {A.}~\bibnamefont {Lichtenstein}}}
  (\bibinfo {year} {2014}),\ \href {https://juser.fz-juelich.de/record/155829}
  {\emph {\bibinfo {title} {{DMFT} at 25: {I}nfinite {D}imensions}}},\ \bibinfo
  {series} {Reihe Modeling and Simulation 4}, Vol.~\bibinfo {volume} {4}\
  (\bibinfo  {publisher} {Forschungszentrum J{\"u}lich Zentralbibliothek,
  Verlag (J{\"u}lich)},\ \bibinfo {address} {Jülich})\BibitemShut {NoStop}%
\bibitem [{\citenamefont {Philipp}\ \emph {et~al.}(2017)\citenamefont
  {Philipp}, \citenamefont {Wallerberger}, \citenamefont {Gunacker},\ and\
  \citenamefont {Held}}]{Philipp2016}%
  \BibitemOpen
  \bibfield  {author} {\bibinfo {author} {\bibnamefont {Philipp}, \bibfnamefont
  {M.-T.}}, \bibinfo {author} {\bibfnamefont {M.}~\bibnamefont {Wallerberger}},
  \bibinfo {author} {\bibfnamefont {P.}~\bibnamefont {Gunacker}}, \ and\
  \bibinfo {author} {\bibfnamefont {K.}~\bibnamefont {Held}}} (\bibinfo {year}
  {2017}),\ \href {\doibase 10.1140/epjb/e2017-80115-7} {\bibfield  {journal}
  {\bibinfo  {journal} {The European Physical Journal B}\ }\textbf {\bibinfo
  {volume} {90}}~(\bibinfo {number} {6}),\ \bibinfo {pages} {114}}\BibitemShut
  {NoStop}%
\bibitem [{\citenamefont {Pixley}\ \emph {et~al.}(2015)\citenamefont {Pixley},
  \citenamefont {Cai},\ and\ \citenamefont {Si}}]{Pixley2015}%
  \BibitemOpen
  \bibfield  {author} {\bibinfo {author} {\bibnamefont {Pixley}, \bibfnamefont
  {J.~H.}}, \bibinfo {author} {\bibfnamefont {A.}~\bibnamefont {Cai}}, \ and\
  \bibinfo {author} {\bibfnamefont {Q.}~\bibnamefont {Si}}} (\bibinfo {year}
  {2015}),\ \href {\doibase 10.1103/PhysRevB.91.125127} {\bibfield  {journal}
  {\bibinfo  {journal} {Phys. Rev. B}\ }\textbf {\bibinfo {volume} {91}},\
  \bibinfo {pages} {125127}}\BibitemShut {NoStop}%
\bibitem [{\citenamefont {Pokorny}\ and\ \citenamefont
  {Janis}(2013)}]{Pokorny2013}%
  \BibitemOpen
  \bibfield  {author} {\bibinfo {author} {\bibnamefont {Pokorny}, \bibfnamefont
  {V.}}, \ and\ \bibinfo {author} {\bibfnamefont {V.}~\bibnamefont {Janis}}}
  (\bibinfo {year} {2013}),\ \href {\doibase 10.1088/0953-8984/25/17/175502}
  {\bibfield  {journal} {\bibinfo  {journal} {J. Phys. Condens. Matter}\
  }\textbf {\bibinfo {volume} {25}}~(\bibinfo {number} {2}),\ \bibinfo {pages}
  {175502}}\BibitemShut {NoStop}%
\bibitem [{\citenamefont {Pollet}\ \emph {et~al.}(2011)\citenamefont {Pollet},
  \citenamefont {Prokof'ev},\ and\ \citenamefont {Svistunov}}]{Pollet2011}%
  \BibitemOpen
  \bibfield  {author} {\bibinfo {author} {\bibnamefont {Pollet}, \bibfnamefont
  {L.}}, \bibinfo {author} {\bibfnamefont {N.~V.}\ \bibnamefont {Prokof'ev}}, \
  and\ \bibinfo {author} {\bibfnamefont {B.~V.}\ \bibnamefont {Svistunov}}}
  (\bibinfo {year} {2011}),\ \href {\doibase 10.1103/PhysRevB.83.161103}
  {\bibfield  {journal} {\bibinfo  {journal} {Phys. Rev. B}\ }\textbf {\bibinfo
  {volume} {83}},\ \bibinfo {pages} {161103}}\BibitemShut {NoStop}%
\bibitem [{\citenamefont {Potthoff}(2006)}]{Potthoff2006}%
  \BibitemOpen
  \bibfield  {author} {\bibinfo {author} {\bibnamefont {Potthoff},
  \bibfnamefont {M.}}} (\bibinfo {year} {2006}),\ \href {\doibase
  10.5488/CMP.9.3.557} {\bibfield  {journal} {\bibinfo  {journal} {Cond. Mat.
  Phys.}\ }\textbf {\bibinfo {volume} {9}},\ \bibinfo {pages}
  {557}}\BibitemShut {NoStop}%
\bibitem [{\citenamefont {Pruschke}\ \emph {et~al.}(2001)\citenamefont
  {Pruschke}, \citenamefont {Metzner},\ and\ \citenamefont
  {Vollhardt}}]{Pruschke2001}%
  \BibitemOpen
  \bibfield  {author} {\bibinfo {author} {\bibnamefont {Pruschke},
  \bibfnamefont {T.}}, \bibinfo {author} {\bibfnamefont {W.}~\bibnamefont
  {Metzner}}, \ and\ \bibinfo {author} {\bibfnamefont {D.}~\bibnamefont
  {Vollhardt}}} (\bibinfo {year} {2001}),\ \href
  {http://stacks.iop.org/0953-8984/13/i=42/a=306} {\bibfield  {journal}
  {\bibinfo  {journal} {Journal of Physics: Condensed Matter}\ }\textbf
  {\bibinfo {volume} {13}}~(\bibinfo {number} {42}),\ \bibinfo {pages}
  {9455}}\BibitemShut {NoStop}%
\bibitem [{\citenamefont {Pudleiner}\ \emph {et~al.}(2016)\citenamefont
  {Pudleiner}, \citenamefont {Sch\"afer}, \citenamefont {Rost}, \citenamefont
  {Li}, \citenamefont {Held},\ and\ \citenamefont {Bl\"umer}}]{Pudleiner2016}%
  \BibitemOpen
  \bibfield  {author} {\bibinfo {author} {\bibnamefont {Pudleiner},
  \bibfnamefont {P.}}, \bibinfo {author} {\bibfnamefont {T.}~\bibnamefont
  {Sch\"afer}}, \bibinfo {author} {\bibfnamefont {D.}~\bibnamefont {Rost}},
  \bibinfo {author} {\bibfnamefont {G.}~\bibnamefont {Li}}, \bibinfo {author}
  {\bibfnamefont {K.}~\bibnamefont {Held}}, \ and\ \bibinfo {author}
  {\bibfnamefont {N.}~\bibnamefont {Bl\"umer}}} (\bibinfo {year} {2016}),\
  \href {\doibase 10.1103/PhysRevB.93.195134} {\bibfield  {journal} {\bibinfo
  {journal} {Phys. Rev. B}\ }\textbf {\bibinfo {volume} {93}},\ \bibinfo
  {pages} {195134}}\BibitemShut {NoStop}%
\bibitem [{\citenamefont {Ramirez}\ \emph {et~al.}(1970)\citenamefont
  {Ramirez}, \citenamefont {Falicov},\ and\ \citenamefont
  {Kimball}}]{Ramirez1970}%
  \BibitemOpen
  \bibfield  {author} {\bibinfo {author} {\bibnamefont {Ramirez}, \bibfnamefont
  {R.}}, \bibinfo {author} {\bibfnamefont {L.}~\bibnamefont {Falicov}}, \ and\
  \bibinfo {author} {\bibfnamefont {J.}~\bibnamefont {Kimball}}} (\bibinfo
  {year} {1970}),\ \href {\doibase 10.1103/PhysRevB.2.3383} {\bibfield
  {journal} {\bibinfo  {journal} {Phys. Rev. B}\ }\textbf {\bibinfo {volume}
  {2}}~(\bibinfo {number} {8}),\ \bibinfo {pages} {3383}}\BibitemShut {NoStop}%
\bibitem [{\citenamefont {Reymbaut}\ \emph {et~al.}(2016)\citenamefont
  {Reymbaut}, \citenamefont {Charlebois}, \citenamefont {Asiani}, \citenamefont
  {Fratino}, \citenamefont {S\'emon}, \citenamefont {Sordi},\ and\
  \citenamefont {Tremblay}}]{Reymbaut2016}%
  \BibitemOpen
  \bibfield  {author} {\bibinfo {author} {\bibnamefont {Reymbaut},
  \bibfnamefont {A.}}, \bibinfo {author} {\bibfnamefont {M.}~\bibnamefont
  {Charlebois}}, \bibinfo {author} {\bibfnamefont {M.~F.}\ \bibnamefont
  {Asiani}}, \bibinfo {author} {\bibfnamefont {L.}~\bibnamefont {Fratino}},
  \bibinfo {author} {\bibfnamefont {P.}~\bibnamefont {S\'emon}}, \bibinfo
  {author} {\bibfnamefont {G.}~\bibnamefont {Sordi}}, \ and\ \bibinfo {author}
  {\bibfnamefont {A.-M.~S.}\ \bibnamefont {Tremblay}}} (\bibinfo {year}
  {2016}),\ \href {\doibase 10.1103/PhysRevB.94.155146} {\bibfield  {journal}
  {\bibinfo  {journal} {Phys. Rev. B}\ }\textbf {\bibinfo {volume} {94}},\
  \bibinfo {pages} {155146}}\BibitemShut {NoStop}%
\bibitem [{\citenamefont {Ribic}\ \emph
  {et~al.}(2017{\natexlab{a}})\citenamefont {Ribic}, \citenamefont {Gunacker},
  \citenamefont {Iskakov}, \citenamefont {Wallerberger}, \citenamefont
  {Rohringer}, \citenamefont {Rubtsov}, \citenamefont {Gull},\ and\
  \citenamefont {Held}}]{Ribic2017b}%
  \BibitemOpen
  \bibfield  {author} {\bibinfo {author} {\bibnamefont {Ribic}, \bibfnamefont
  {T.}}, \bibinfo {author} {\bibfnamefont {P.}~\bibnamefont {Gunacker}},
  \bibinfo {author} {\bibfnamefont {S.}~\bibnamefont {Iskakov}}, \bibinfo
  {author} {\bibfnamefont {M.}~\bibnamefont {Wallerberger}}, \bibinfo {author}
  {\bibfnamefont {G.}~\bibnamefont {Rohringer}}, \bibinfo {author}
  {\bibfnamefont {A.~N.}\ \bibnamefont {Rubtsov}}, \bibinfo {author}
  {\bibfnamefont {E.}~\bibnamefont {Gull}}, \ and\ \bibinfo {author}
  {\bibfnamefont {K.}~\bibnamefont {Held}}} (\bibinfo {year}
  {2017}{\natexlab{a}}),\ \href {\doibase 10.1103/PhysRevB.96.235127}
  {\bibfield  {journal} {\bibinfo  {journal} {Phys. Rev. B}\ }\textbf {\bibinfo
  {volume} {96}},\ \bibinfo {pages} {235127}}\BibitemShut {NoStop}%
\bibitem [{\citenamefont {Ribic}\ \emph {et~al.}(2016)\citenamefont {Ribic},
  \citenamefont {Rohringer},\ and\ \citenamefont {Held}}]{Ribic2016}%
  \BibitemOpen
  \bibfield  {author} {\bibinfo {author} {\bibnamefont {Ribic}, \bibfnamefont
  {T.}}, \bibinfo {author} {\bibfnamefont {G.}~\bibnamefont {Rohringer}}, \
  and\ \bibinfo {author} {\bibfnamefont {K.}~\bibnamefont {Held}}} (\bibinfo
  {year} {2016}),\ \href {\doibase 10.1103/PhysRevB.93.195105} {\bibfield
  {journal} {\bibinfo  {journal} {Phys. Rev. B}\ }\textbf {\bibinfo {volume}
  {93}}~(\bibinfo {number} {19}),\ \bibinfo {pages} {195105}}\BibitemShut
  {NoStop}%
\bibitem [{\citenamefont {Ribic}\ \emph
  {et~al.}(2017{\natexlab{b}})\citenamefont {Ribic}, \citenamefont
  {Rohringer},\ and\ \citenamefont {Held}}]{Ribic2016b}%
  \BibitemOpen
  \bibfield  {author} {\bibinfo {author} {\bibnamefont {Ribic}, \bibfnamefont
  {T.}}, \bibinfo {author} {\bibfnamefont {G.}~\bibnamefont {Rohringer}}, \
  and\ \bibinfo {author} {\bibfnamefont {K.}~\bibnamefont {Held}}} (\bibinfo
  {year} {2017}{\natexlab{b}}),\ \href {\doibase 10.1103/PhysRevB.95.155130}
  {\bibfield  {journal} {\bibinfo  {journal} {Phys. Rev. B}\ }\textbf {\bibinfo
  {volume} {95}},\ \bibinfo {pages} {155130}}\BibitemShut {NoStop}%
\bibitem [{\citenamefont {van Roekeghem}\ \emph {et~al.}(2014)\citenamefont
  {van Roekeghem}, \citenamefont {Ayral}, \citenamefont {Tomczak},
  \citenamefont {Casula}, \citenamefont {Xu}, \citenamefont {Ding},
  \citenamefont {Ferrero}, \citenamefont {Parcollet}, \citenamefont {Jiang},\
  and\ \citenamefont {Biermann}}]{paris_sex}%
  \BibitemOpen
  \bibfield  {author} {\bibinfo {author} {\bibnamefont {van Roekeghem},
  \bibfnamefont {A.}}, \bibinfo {author} {\bibfnamefont {T.}~\bibnamefont
  {Ayral}}, \bibinfo {author} {\bibfnamefont {J.~M.}\ \bibnamefont {Tomczak}},
  \bibinfo {author} {\bibfnamefont {M.}~\bibnamefont {Casula}}, \bibinfo
  {author} {\bibfnamefont {N.}~\bibnamefont {Xu}}, \bibinfo {author}
  {\bibfnamefont {H.}~\bibnamefont {Ding}}, \bibinfo {author} {\bibfnamefont
  {M.}~\bibnamefont {Ferrero}}, \bibinfo {author} {\bibfnamefont
  {O.}~\bibnamefont {Parcollet}}, \bibinfo {author} {\bibfnamefont
  {H.}~\bibnamefont {Jiang}}, \ and\ \bibinfo {author} {\bibfnamefont
  {S.}~\bibnamefont {Biermann}}} (\bibinfo {year} {2014}),\ \href {\doibase
  10.1103/PhysRevLett.113.266403} {\bibfield  {journal} {\bibinfo  {journal}
  {Phys. Rev. Lett.}\ }\textbf {\bibinfo {volume} {113}},\ \bibinfo {pages}
  {266403}}\BibitemShut {NoStop}%
\bibitem [{\citenamefont {Rohringer}(2013)}]{Rohringer2013a}%
  \BibitemOpen
  \bibfield  {author} {\bibinfo {author} {\bibnamefont {Rohringer},
  \bibfnamefont {G.}}} (\bibinfo {year} {2013}),\ \emph {\bibinfo {title} {New
  routes towards a theoretical treatment of nonlocal electronic
  correlations}},\ \href@noop {} {Ph.D. thesis}\ (\bibinfo  {school} {Vienna
  University of Technology})\BibitemShut {NoStop}%
\bibitem [{\citenamefont {Rohringer}\ \emph {et~al.}(2018)\citenamefont
  {Rohringer}, \citenamefont {Katanin}, \citenamefont {Sch{\"a}fer},
  \citenamefont {Hausoel}, \citenamefont {Held},\ and\ \citenamefont
  {Toschi}}]{Rohringer2018}%
  \BibitemOpen
  \bibfield  {author} {\bibinfo {author} {\bibnamefont {Rohringer},
  \bibfnamefont {G.}}, \bibinfo {author} {\bibfnamefont {A.}~\bibnamefont
  {Katanin}}, \bibinfo {author} {\bibfnamefont {T.}~\bibnamefont
  {Sch{\"a}fer}}, \bibinfo {author} {\bibfnamefont {A.}~\bibnamefont
  {Hausoel}}, \bibinfo {author} {\bibfnamefont {K.}~\bibnamefont {Held}}, \
  and\ \bibinfo {author} {\bibfnamefont {A.}~\bibnamefont {Toschi}}} (\bibinfo
  {year} {2018}),\ \href {https://github.com/ladderDGA/ladderDGA} {\ }\Eprint
  {http://arxiv.org/abs/github.com/ladderDGA} {github.com/ladderDGA}
  \BibitemShut {NoStop}%
\bibitem [{\citenamefont {Rohringer}\ and\ \citenamefont
  {Toschi}(2016)}]{Rohringer2016}%
  \BibitemOpen
  \bibfield  {author} {\bibinfo {author} {\bibnamefont {Rohringer},
  \bibfnamefont {G.}}, \ and\ \bibinfo {author} {\bibfnamefont
  {A.}~\bibnamefont {Toschi}}} (\bibinfo {year} {2016}),\ \href {\doibase
  10.1103/PhysRevB.94.125144} {\bibfield  {journal} {\bibinfo  {journal} {Phys.
  Rev. B}\ }\textbf {\bibinfo {volume} {94}},\ \bibinfo {pages}
  {125144}}\BibitemShut {NoStop}%
\bibitem [{\citenamefont {Rohringer}\ \emph {et~al.}(2013)\citenamefont
  {Rohringer}, \citenamefont {Toschi}, \citenamefont {Hafermann}, \citenamefont
  {Held}, \citenamefont {Anisimov},\ and\ \citenamefont
  {Katanin}}]{Rohringer2013}%
  \BibitemOpen
  \bibfield  {author} {\bibinfo {author} {\bibnamefont {Rohringer},
  \bibfnamefont {G.}}, \bibinfo {author} {\bibfnamefont {A.}~\bibnamefont
  {Toschi}}, \bibinfo {author} {\bibfnamefont {H.}~\bibnamefont {Hafermann}},
  \bibinfo {author} {\bibfnamefont {K.}~\bibnamefont {Held}}, \bibinfo {author}
  {\bibfnamefont {V.~I.}\ \bibnamefont {Anisimov}}, \ and\ \bibinfo {author}
  {\bibfnamefont {A.~A.}\ \bibnamefont {Katanin}}} (\bibinfo {year} {2013}),\
  \href {http://link.aps.org/doi/10.1103/PhysRevB.88.115112} {\bibfield
  {journal} {\bibinfo  {journal} {Phys. Rev. B}\ }\textbf {\bibinfo {volume}
  {88}},\ \bibinfo {pages} {115112}}\BibitemShut {NoStop}%
\bibitem [{\citenamefont {Rohringer}\ \emph {et~al.}(2011)\citenamefont
  {Rohringer}, \citenamefont {Toschi}, \citenamefont {Katanin},\ and\
  \citenamefont {Held}}]{Rohringer2011}%
  \BibitemOpen
  \bibfield  {author} {\bibinfo {author} {\bibnamefont {Rohringer},
  \bibfnamefont {G.}}, \bibinfo {author} {\bibfnamefont {A.}~\bibnamefont
  {Toschi}}, \bibinfo {author} {\bibfnamefont {A.}~\bibnamefont {Katanin}}, \
  and\ \bibinfo {author} {\bibfnamefont {K.}~\bibnamefont {Held}}} (\bibinfo
  {year} {2011}),\ \href {\doibase 10.1103/PhysRevLett.107.256402} {\bibfield
  {journal} {\bibinfo  {journal} {Phys. Rev. Lett.}\ }\textbf {\bibinfo
  {volume} {107}},\ \bibinfo {pages} {256402}}\BibitemShut {NoStop}%
\bibitem [{\citenamefont {Rohringer}\ \emph {et~al.}(2012)\citenamefont
  {Rohringer}, \citenamefont {Valli},\ and\ \citenamefont
  {Toschi}}]{Rohringer2012}%
  \BibitemOpen
  \bibfield  {author} {\bibinfo {author} {\bibnamefont {Rohringer},
  \bibfnamefont {G.}}, \bibinfo {author} {\bibfnamefont {A.}~\bibnamefont
  {Valli}}, \ and\ \bibinfo {author} {\bibfnamefont {A.}~\bibnamefont
  {Toschi}}} (\bibinfo {year} {2012}),\ \href {\doibase
  10.1103/PhysRevB.86.125114} {\bibfield  {journal} {\bibinfo  {journal} {Phys.
  Rev. B}\ }\textbf {\bibinfo {volume} {86}},\ \bibinfo {pages}
  {125114}}\BibitemShut {NoStop}%
\bibitem [{\citenamefont {Rossi}\ \emph {et~al.}(2016)\citenamefont {Rossi},
  \citenamefont {Werner}, \citenamefont {Prokof'ev},\ and\ \citenamefont
  {Svistunov}}]{Rossi2016}%
  \BibitemOpen
  \bibfield  {author} {\bibinfo {author} {\bibnamefont {Rossi}, \bibfnamefont
  {R.}}, \bibinfo {author} {\bibfnamefont {F.}~\bibnamefont {Werner}}, \bibinfo
  {author} {\bibfnamefont {N.}~\bibnamefont {Prokof'ev}}, \ and\ \bibinfo
  {author} {\bibfnamefont {B.}~\bibnamefont {Svistunov}}} (\bibinfo {year}
  {2016}),\ \href {\doibase 10.1103/PhysRevB.93.161102} {\bibfield  {journal}
  {\bibinfo  {journal} {Phys. Rev. B}\ }\textbf {\bibinfo {volume} {93}},\
  \bibinfo {pages} {161102}}\BibitemShut {NoStop}%
\bibitem [{\citenamefont {Rost}\ \emph {et~al.}(2012)\citenamefont {Rost},
  \citenamefont {Gorelik}, \citenamefont {Assaad},\ and\ \citenamefont
  {Bl\"umer}}]{Rost2012}%
  \BibitemOpen
  \bibfield  {author} {\bibinfo {author} {\bibnamefont {Rost}, \bibfnamefont
  {D.}}, \bibinfo {author} {\bibfnamefont {E.~V.}\ \bibnamefont {Gorelik}},
  \bibinfo {author} {\bibfnamefont {F.}~\bibnamefont {Assaad}}, \ and\ \bibinfo
  {author} {\bibfnamefont {N.}~\bibnamefont {Bl\"umer}}} (\bibinfo {year}
  {2012}),\ \href {\doibase 10.1103/PhysRevB.86.155109} {\bibfield  {journal}
  {\bibinfo  {journal} {Phys. Rev. B}\ }\textbf {\bibinfo {volume} {86}},\
  \bibinfo {pages} {155109}}\BibitemShut {NoStop}%
\bibitem [{\citenamefont {Rubtsov}(2006)}]{Rubtsov2006}%
  \BibitemOpen
  \bibfield  {author} {\bibinfo {author} {\bibnamefont {Rubtsov}, \bibfnamefont
  {A.~N.}}} (\bibinfo {year} {2006}),\ \href
  {http://arxiv.org/abs/cond-mat/0601333} {\ }\Eprint
  {http://arxiv.org/abs/cond-mat/0601333} {arXiv:cond-mat/0601333} \BibitemShut
  {NoStop}%
\bibitem [{\citenamefont {Rubtsov}\ \emph {et~al.}(2008)\citenamefont
  {Rubtsov}, \citenamefont {Katsnelson},\ and\ \citenamefont
  {Lichtenstein}}]{Rubtsov2008}%
  \BibitemOpen
  \bibfield  {author} {\bibinfo {author} {\bibnamefont {Rubtsov}, \bibfnamefont
  {A.~N.}}, \bibinfo {author} {\bibfnamefont {M.~I.}\ \bibnamefont
  {Katsnelson}}, \ and\ \bibinfo {author} {\bibfnamefont {A.~I.}\ \bibnamefont
  {Lichtenstein}}} (\bibinfo {year} {2008}),\ \href {\doibase
  10.1103/PhysRevB.77.033101} {\bibfield  {journal} {\bibinfo  {journal} {Phys.
  Rev. B}\ }\textbf {\bibinfo {volume} {77}},\ \bibinfo {pages}
  {033101}}\BibitemShut {NoStop}%
\bibitem [{\citenamefont {Rubtsov}\ \emph {et~al.}(2012)\citenamefont
  {Rubtsov}, \citenamefont {Katsnelson},\ and\ \citenamefont
  {Lichtenstein}}]{Rubtsov12}%
  \BibitemOpen
  \bibfield  {author} {\bibinfo {author} {\bibnamefont {Rubtsov}, \bibfnamefont
  {A.~N.}}, \bibinfo {author} {\bibfnamefont {M.~I.}\ \bibnamefont
  {Katsnelson}}, \ and\ \bibinfo {author} {\bibfnamefont {A.~I.}\ \bibnamefont
  {Lichtenstein}}} (\bibinfo {year} {2012}),\ \href {\doibase
  10.1016/j.aop.2012.01.002} {\bibfield  {journal} {\bibinfo  {journal} {Ann.
  Phys.}\ }\textbf {\bibinfo {volume} {327}}~(\bibinfo {number} {5}),\ \bibinfo
  {pages} {1320}}\BibitemShut {NoStop}%
\bibitem [{\citenamefont {Rubtsov}\ \emph {et~al.}(2009)\citenamefont
  {Rubtsov}, \citenamefont {Katsnelson}, \citenamefont {Lichtenstein},\ and\
  \citenamefont {Georges}}]{Rubtsov2009}%
  \BibitemOpen
  \bibfield  {author} {\bibinfo {author} {\bibnamefont {Rubtsov}, \bibfnamefont
  {A.~N.}}, \bibinfo {author} {\bibfnamefont {M.~I.}\ \bibnamefont
  {Katsnelson}}, \bibinfo {author} {\bibfnamefont {A.~I.}\ \bibnamefont
  {Lichtenstein}}, \ and\ \bibinfo {author} {\bibfnamefont {A.}~\bibnamefont
  {Georges}}} (\bibinfo {year} {2009}),\ \href {\doibase
  10.1103/PhysRevB.79.045133} {\bibfield  {journal} {\bibinfo  {journal} {Phys.
  Rev. B}\ }\textbf {\bibinfo {volume} {79}}~(\bibinfo {number} {4}),\ \bibinfo
  {pages} {045133}}\BibitemShut {NoStop}%
\bibitem [{\citenamefont {Rubtsov}\ \emph {et~al.}(2005)\citenamefont
  {Rubtsov}, \citenamefont {Savkin},\ and\ \citenamefont
  {Lichtenstein}}]{Rubtsov2005}%
  \BibitemOpen
  \bibfield  {author} {\bibinfo {author} {\bibnamefont {Rubtsov}, \bibfnamefont
  {A.~N.}}, \bibinfo {author} {\bibfnamefont {V.~V.}\ \bibnamefont {Savkin}}, \
  and\ \bibinfo {author} {\bibfnamefont {A.~I.}\ \bibnamefont {Lichtenstein}}}
  (\bibinfo {year} {2005}),\ \href {\doibase 10.1103/PhysRevB.72.035122}
  {\bibfield  {journal} {\bibinfo  {journal} {Phys. Rev. B}\ }\textbf {\bibinfo
  {volume} {72}},\ \bibinfo {pages} {035122}}\BibitemShut {NoStop}%
\bibitem [{\citenamefont {Sadovskii}\ \emph {et~al.}(2005)\citenamefont
  {Sadovskii}, \citenamefont {Nekrasov}, \citenamefont {Kuchinskii},
  \citenamefont {Pruschke},\ and\ \citenamefont {Anisimov}}]{Sadovskii2005}%
  \BibitemOpen
  \bibfield  {author} {\bibinfo {author} {\bibnamefont {Sadovskii},
  \bibfnamefont {M.~V.}}, \bibinfo {author} {\bibfnamefont {I.~A.}\
  \bibnamefont {Nekrasov}}, \bibinfo {author} {\bibfnamefont {E.~Z.}\
  \bibnamefont {Kuchinskii}}, \bibinfo {author} {\bibfnamefont
  {T.}~\bibnamefont {Pruschke}}, \ and\ \bibinfo {author} {\bibfnamefont
  {V.~I.}\ \bibnamefont {Anisimov}}} (\bibinfo {year} {2005}),\ \href
  {http://link.aps.org/doi/10.1103/PhysRevB.72.155105} {\bibfield  {journal}
  {\bibinfo  {journal} {Phys. Rev. B}\ }\textbf {\bibinfo {volume} {72}},\
  \bibinfo {pages} {155105}}\BibitemShut {NoStop}%
\bibitem [{\citenamefont {Sahebsara}\ and\ \citenamefont
  {S\'en\'echal}(2008)}]{Sahebsara2008}%
  \BibitemOpen
  \bibfield  {author} {\bibinfo {author} {\bibnamefont {Sahebsara},
  \bibfnamefont {P.}}, \ and\ \bibinfo {author} {\bibfnamefont
  {D.}~\bibnamefont {S\'en\'echal}}} (\bibinfo {year} {2008}),\ \href {\doibase
  10.1103/PhysRevLett.100.136402} {\bibfield  {journal} {\bibinfo  {journal}
  {Phys. Rev. Lett.}\ }\textbf {\bibinfo {volume} {100}},\ \bibinfo {pages}
  {136402}}\BibitemShut {NoStop}%
\bibitem [{\citenamefont {Sakai}\ \emph {et~al.}(2009)\citenamefont {Sakai},
  \citenamefont {Motome},\ and\ \citenamefont {Imada}}]{Sakai2009}%
  \BibitemOpen
  \bibfield  {author} {\bibinfo {author} {\bibnamefont {Sakai}, \bibfnamefont
  {S.}}, \bibinfo {author} {\bibfnamefont {Y.}~\bibnamefont {Motome}}, \ and\
  \bibinfo {author} {\bibfnamefont {M.}~\bibnamefont {Imada}}} (\bibinfo {year}
  {2009}),\ \href {\doibase 10.1103/PhysRevLett.102.056404} {\bibfield
  {journal} {\bibinfo  {journal} {Phys. Rev. Lett.}\ }\textbf {\bibinfo
  {volume} {102}},\ \bibinfo {pages} {056404}}\BibitemShut {NoStop}%
\bibitem [{\citenamefont {Sakai}\ \emph {et~al.}(2012)\citenamefont {Sakai},
  \citenamefont {Sangiovanni}, \citenamefont {Civelli}, \citenamefont {Motome},
  \citenamefont {Held},\ and\ \citenamefont {Imada}}]{Sakai12}%
  \BibitemOpen
  \bibfield  {author} {\bibinfo {author} {\bibnamefont {Sakai}, \bibfnamefont
  {S.}}, \bibinfo {author} {\bibfnamefont {G.}~\bibnamefont {Sangiovanni}},
  \bibinfo {author} {\bibfnamefont {M.}~\bibnamefont {Civelli}}, \bibinfo
  {author} {\bibfnamefont {Y.}~\bibnamefont {Motome}}, \bibinfo {author}
  {\bibfnamefont {K.}~\bibnamefont {Held}}, \ and\ \bibinfo {author}
  {\bibfnamefont {M.}~\bibnamefont {Imada}}} (\bibinfo {year} {2012}),\ \href
  {\doibase 10.1103/PhysRevB.85.035102} {\bibfield  {journal} {\bibinfo
  {journal} {Phys. Rev. B}\ }\textbf {\bibinfo {volume} {85}},\ \bibinfo
  {pages} {035102}}\BibitemShut {NoStop}%
\bibitem [{\citenamefont {Sandvik}(1998{\natexlab{a}})}]{Sandvik98}%
  \BibitemOpen
  \bibfield  {author} {\bibinfo {author} {\bibnamefont {Sandvik}, \bibfnamefont
  {A.~W.}}} (\bibinfo {year} {1998}{\natexlab{a}}),\ \href {\doibase
  10.1103/PhysRevLett.80.5196} {\bibfield  {journal} {\bibinfo  {journal}
  {Phys. Rev. Lett.}\ }\textbf {\bibinfo {volume} {80}},\ \bibinfo {pages}
  {5196}}\BibitemShut {NoStop}%
\bibitem [{\citenamefont {Sandvik}(1998{\natexlab{b}})}]{Sandvik98b}%
  \BibitemOpen
  \bibfield  {author} {\bibinfo {author} {\bibnamefont {Sandvik}, \bibfnamefont
  {A.~W.}}} (\bibinfo {year} {1998}{\natexlab{b}}),\ \href {\doibase
  10.1103/PhysRevB.57.10287} {\bibfield  {journal} {\bibinfo  {journal} {Phys.
  Rev. B}\ }\textbf {\bibinfo {volume} {57}},\ \bibinfo {pages}
  {10287}}\BibitemShut {NoStop}%
\bibitem [{\citenamefont {Sarker}(1988)}]{Sarker1988}%
  \BibitemOpen
  \bibfield  {author} {\bibinfo {author} {\bibnamefont {Sarker}, \bibfnamefont
  {S.~K.}}} (\bibinfo {year} {1988}),\ \href
  {http://stacks.iop.org/0022-3719/21/i=18/a=002} {\bibfield  {journal}
  {\bibinfo  {journal} {J. Phys. C: Solid State Physics}\ }\textbf {\bibinfo
  {volume} {21}}~(\bibinfo {number} {18}),\ \bibinfo {pages}
  {L667}}\BibitemShut {NoStop}%
\bibitem [{\citenamefont {Scalapino}(2012)}]{Scalapino12}%
  \BibitemOpen
  \bibfield  {author} {\bibinfo {author} {\bibnamefont {Scalapino},
  \bibfnamefont {D.~J.}}} (\bibinfo {year} {2012}),\ \href {\doibase
  10.1103/RevModPhys.84.1383} {\bibfield  {journal} {\bibinfo  {journal} {Rev.
  Mod. Phys.}\ }\textbf {\bibinfo {volume} {84}},\ \bibinfo {pages}
  {1383}}\BibitemShut {NoStop}%
\bibitem [{\citenamefont {Sch\"afer}\ \emph
  {et~al.}(2016{\natexlab{a}})\citenamefont {Sch\"afer}, \citenamefont
  {Ciuchi}, \citenamefont {Wallerberger}, \citenamefont {Thunstr{\"o}m},
  \citenamefont {Gunnarsson}, \citenamefont {Sangiovanni}, \citenamefont
  {Rohringer},\ and\ \citenamefont {Toschi}}]{Schaefer2016c}%
  \BibitemOpen
  \bibfield  {author} {\bibinfo {author} {\bibnamefont {Sch\"afer},
  \bibfnamefont {T.}}, \bibinfo {author} {\bibfnamefont {S.}~\bibnamefont
  {Ciuchi}}, \bibinfo {author} {\bibfnamefont {M.}~\bibnamefont
  {Wallerberger}}, \bibinfo {author} {\bibfnamefont {P.}~\bibnamefont
  {Thunstr{\"o}m}}, \bibinfo {author} {\bibfnamefont {O.}~\bibnamefont
  {Gunnarsson}}, \bibinfo {author} {\bibfnamefont {G.}~\bibnamefont
  {Sangiovanni}}, \bibinfo {author} {\bibfnamefont {G.}~\bibnamefont
  {Rohringer}}, \ and\ \bibinfo {author} {\bibfnamefont {A.}~\bibnamefont
  {Toschi}}} (\bibinfo {year} {2016}{\natexlab{a}}),\ \href {\doibase
  10.1103/PhysRevB.94.235108} {\bibfield  {journal} {\bibinfo  {journal} {Phys.
  Rev. B}\ }\textbf {\bibinfo {volume} {94}},\ \bibinfo {pages}
  {235108}}\BibitemShut {NoStop}%
\bibitem [{\citenamefont {Sch\"afer}\ \emph
  {et~al.}(2015{\natexlab{a}})\citenamefont {Sch\"afer}, \citenamefont {Geles},
  \citenamefont {Rost}, \citenamefont {Rohringer}, \citenamefont {Arrigoni},
  \citenamefont {Held}, \citenamefont {Bl\"umer}, \citenamefont {Aichhorn},\
  and\ \citenamefont {Toschi}}]{Schaefer2015-2}%
  \BibitemOpen
  \bibfield  {author} {\bibinfo {author} {\bibnamefont {Sch\"afer},
  \bibfnamefont {T.}}, \bibinfo {author} {\bibfnamefont {F.}~\bibnamefont
  {Geles}}, \bibinfo {author} {\bibfnamefont {D.}~\bibnamefont {Rost}},
  \bibinfo {author} {\bibfnamefont {G.}~\bibnamefont {Rohringer}}, \bibinfo
  {author} {\bibfnamefont {E.}~\bibnamefont {Arrigoni}}, \bibinfo {author}
  {\bibfnamefont {K.}~\bibnamefont {Held}}, \bibinfo {author} {\bibfnamefont
  {N.}~\bibnamefont {Bl\"umer}}, \bibinfo {author} {\bibfnamefont
  {M.}~\bibnamefont {Aichhorn}}, \ and\ \bibinfo {author} {\bibfnamefont
  {A.}~\bibnamefont {Toschi}}} (\bibinfo {year} {2015}{\natexlab{a}}),\ \href
  {\doibase 10.1103/PhysRevB.91.125109} {\bibfield  {journal} {\bibinfo
  {journal} {Phys. Rev. B}\ }\textbf {\bibinfo {volume} {91}},\ \bibinfo
  {pages} {125109}}\BibitemShut {NoStop}%
\bibitem [{\citenamefont {Sch\"afer}\ \emph {et~al.}(2017)\citenamefont
  {Sch\"afer}, \citenamefont {Katanin}, \citenamefont {Held},\ and\
  \citenamefont {Toschi}}]{Schaefer2016}%
  \BibitemOpen
  \bibfield  {author} {\bibinfo {author} {\bibnamefont {Sch\"afer},
  \bibfnamefont {T.}}, \bibinfo {author} {\bibfnamefont {A.~A.}\ \bibnamefont
  {Katanin}}, \bibinfo {author} {\bibfnamefont {K.}~\bibnamefont {Held}}, \
  and\ \bibinfo {author} {\bibfnamefont {A.}~\bibnamefont {Toschi}}} (\bibinfo
  {year} {2017}),\ \href {\doibase 10.1103/PhysRevLett.119.046402} {\bibfield
  {journal} {\bibinfo  {journal} {Phys. Rev. Lett.}\ }\textbf {\bibinfo
  {volume} {119}},\ \bibinfo {pages} {046402}}\BibitemShut {NoStop}%
\bibitem [{\citenamefont {Sch\"afer}\ \emph {et~al.}(2013)\citenamefont
  {Sch\"afer}, \citenamefont {Rohringer}, \citenamefont {Gunnarsson},
  \citenamefont {Ciuchi}, \citenamefont {Sangiovanni},\ and\ \citenamefont
  {Toschi}}]{Schafer2013}%
  \BibitemOpen
  \bibfield  {author} {\bibinfo {author} {\bibnamefont {Sch\"afer},
  \bibfnamefont {T.}}, \bibinfo {author} {\bibfnamefont {G.}~\bibnamefont
  {Rohringer}}, \bibinfo {author} {\bibfnamefont {O.}~\bibnamefont
  {Gunnarsson}}, \bibinfo {author} {\bibfnamefont {S.}~\bibnamefont {Ciuchi}},
  \bibinfo {author} {\bibfnamefont {G.}~\bibnamefont {Sangiovanni}}, \ and\
  \bibinfo {author} {\bibfnamefont {A.}~\bibnamefont {Toschi}}} (\bibinfo
  {year} {2013}),\ \href {\doibase 10.1103/PhysRevLett.110.246405} {\bibfield
  {journal} {\bibinfo  {journal} {Phys. Rev. Lett.}\ }\textbf {\bibinfo
  {volume} {110}},\ \bibinfo {pages} {246405}}\BibitemShut {NoStop}%
\bibitem [{\citenamefont {Sch\"afer}\ \emph
  {et~al.}(2016{\natexlab{b}})\citenamefont {Sch\"afer}, \citenamefont
  {Toschi},\ and\ \citenamefont {Held}}]{Schaefer2015-3}%
  \BibitemOpen
  \bibfield  {author} {\bibinfo {author} {\bibnamefont {Sch\"afer},
  \bibfnamefont {T.}}, \bibinfo {author} {\bibfnamefont {A.}~\bibnamefont
  {Toschi}}, \ and\ \bibinfo {author} {\bibfnamefont {K.}~\bibnamefont {Held}}}
  (\bibinfo {year} {2016}{\natexlab{b}}),\ \href {\doibase
  10.1016/j.jmmm.2015.07.103} {\bibfield  {journal} {\bibinfo  {journal}
  {Journal of Magnetism and Magnetic Materials}\ }\textbf {\bibinfo {volume}
  {400}},\ \bibinfo {pages} {107}}\BibitemShut {NoStop}%
\bibitem [{\citenamefont {Sch\"afer}\ \emph
  {et~al.}(2015{\natexlab{b}})\citenamefont {Sch\"afer}, \citenamefont
  {Toschi},\ and\ \citenamefont {Tomczak}}]{Schaefer2015}%
  \BibitemOpen
  \bibfield  {author} {\bibinfo {author} {\bibnamefont {Sch\"afer},
  \bibfnamefont {T.}}, \bibinfo {author} {\bibfnamefont {A.}~\bibnamefont
  {Toschi}}, \ and\ \bibinfo {author} {\bibfnamefont {J.~M.}\ \bibnamefont
  {Tomczak}}} (\bibinfo {year} {2015}{\natexlab{b}}),\ \href {\doibase
  10.1103/PhysRevB.91.121107} {\bibfield  {journal} {\bibinfo  {journal} {Phys.
  Rev. B}\ }\textbf {\bibinfo {volume} {91}},\ \bibinfo {pages}
  {121107}}\BibitemShut {NoStop}%
\bibitem [{\citenamefont {Schauerte}\ and\ \citenamefont {van
  Dongen}(2002)}]{Schauerte2002}%
  \BibitemOpen
  \bibfield  {author} {\bibinfo {author} {\bibnamefont {Schauerte},
  \bibfnamefont {T.}}, \ and\ \bibinfo {author} {\bibfnamefont {P.~G.~J.}\
  \bibnamefont {van Dongen}}} (\bibinfo {year} {2002}),\ \href {\doibase
  10.1103/PhysRevB.65.081105} {\bibfield  {journal} {\bibinfo  {journal} {Phys.
  Rev. B}\ }\textbf {\bibinfo {volume} {65}},\ \bibinfo {pages}
  {081105}}\BibitemShut {NoStop}%
\bibitem [{\citenamefont {Schiller}\ and\ \citenamefont
  {Ingersent}(1995)}]{Schiller1995}%
  \BibitemOpen
  \bibfield  {author} {\bibinfo {author} {\bibnamefont {Schiller},
  \bibfnamefont {A.}}, \ and\ \bibinfo {author} {\bibfnamefont
  {K.}~\bibnamefont {Ingersent}}} (\bibinfo {year} {1995}),\ \href {\doibase
  10.1103/PhysRevLett.75.113} {\bibfield  {journal} {\bibinfo  {journal} {Phys.
  Rev. Lett.}\ }\textbf {\bibinfo {volume} {75}},\ \bibinfo {pages}
  {113}}\BibitemShut {NoStop}%
\bibitem [{\citenamefont {Schmalian}\ \emph {et~al.}(1999)\citenamefont
  {Schmalian}, \citenamefont {Pines},\ and\ \citenamefont
  {Stojkovi\'{c}}}]{Schmalian1999}%
  \BibitemOpen
  \bibfield  {author} {\bibinfo {author} {\bibnamefont {Schmalian},
  \bibfnamefont {J.}}, \bibinfo {author} {\bibfnamefont {D.}~\bibnamefont
  {Pines}}, \ and\ \bibinfo {author} {\bibfnamefont {B.}~\bibnamefont
  {Stojkovi\'{c}}}} (\bibinfo {year} {1999}),\ \href
  {http://dx.doi.org/10.1103/PhysRevB.60.667} {\bibfield  {journal} {\bibinfo
  {journal} {Phys. Rev. B}\ }\textbf {\bibinfo {volume} {60}},\ \bibinfo
  {pages} {667}}\BibitemShut {NoStop}%
\bibitem [{\citenamefont {Schwartz}\ and\ \citenamefont
  {Siggia}(1972)}]{Schwartz1972}%
  \BibitemOpen
  \bibfield  {author} {\bibinfo {author} {\bibnamefont {Schwartz},
  \bibfnamefont {L.}}, \ and\ \bibinfo {author} {\bibfnamefont
  {E.}~\bibnamefont {Siggia}}} (\bibinfo {year} {1972}),\ \href {\doibase
  10.1103/PhysRevB.5.383} {\bibfield  {journal} {\bibinfo  {journal} {Phys.
  Rev. B}\ }\textbf {\bibinfo {volume} {5}},\ \bibinfo {pages}
  {383}}\BibitemShut {NoStop}%
\bibitem [{\citenamefont {S\'emon}\ and\ \citenamefont
  {Tremblay}(2012)}]{Semon2012}%
  \BibitemOpen
  \bibfield  {author} {\bibinfo {author} {\bibnamefont {S\'emon}, \bibfnamefont
  {P.}}, \ and\ \bibinfo {author} {\bibfnamefont {A.-M.~S.}\ \bibnamefont
  {Tremblay}}} (\bibinfo {year} {2012}),\ \href {\doibase
  10.1103/PhysRevB.85.201101} {\bibfield  {journal} {\bibinfo  {journal} {Phys.
  Rev. B}\ }\textbf {\bibinfo {volume} {85}},\ \bibinfo {pages}
  {201101}}\BibitemShut {NoStop}%
\bibitem [{\citenamefont {Sengupta}\ and\ \citenamefont
  {Georges}(1995)}]{Sengupta1995}%
  \BibitemOpen
  \bibfield  {author} {\bibinfo {author} {\bibnamefont {Sengupta},
  \bibfnamefont {A.~M.}}, \ and\ \bibinfo {author} {\bibfnamefont
  {A.}~\bibnamefont {Georges}}} (\bibinfo {year} {1995}),\ \href {\doibase
  10.1103/PhysRevB.52.10295} {\bibfield  {journal} {\bibinfo  {journal} {Phys.
  Rev. B}\ }\textbf {\bibinfo {volume} {52}},\ \bibinfo {pages}
  {10295}}\BibitemShut {NoStop}%
\bibitem [{\citenamefont {Seth}\ \emph {et~al.}(2016)\citenamefont {Seth},
  \citenamefont {Krivenko}, \citenamefont {Ferrero},\ and\ \citenamefont
  {Parcollet}}]{Seth2016}%
  \BibitemOpen
  \bibfield  {author} {\bibinfo {author} {\bibnamefont {Seth}, \bibfnamefont
  {P.}}, \bibinfo {author} {\bibfnamefont {I.}~\bibnamefont {Krivenko}},
  \bibinfo {author} {\bibfnamefont {M.}~\bibnamefont {Ferrero}}, \ and\
  \bibinfo {author} {\bibfnamefont {O.}~\bibnamefont {Parcollet}}} (\bibinfo
  {year} {2016}),\ \href {\doibase 10.1016/j.cpc.2015.10.023} {\bibfield
  {journal} {\bibinfo  {journal} {Computer Physics Communications}\ }\textbf
  {\bibinfo {volume} {200}},\ \bibinfo {pages} {274 }}\BibitemShut {NoStop}%
\bibitem [{\citenamefont {Shinaoka}\ \emph
  {et~al.}(2017{\natexlab{a}})\citenamefont {Shinaoka}, \citenamefont {Gull},\
  and\ \citenamefont {Werner}}]{Shinaoka2016}%
  \BibitemOpen
  \bibfield  {author} {\bibinfo {author} {\bibnamefont {Shinaoka},
  \bibfnamefont {H.}}, \bibinfo {author} {\bibfnamefont {E.}~\bibnamefont
  {Gull}}, \ and\ \bibinfo {author} {\bibfnamefont {P.}~\bibnamefont {Werner}}}
  (\bibinfo {year} {2017}{\natexlab{a}}),\ \href {\doibase
  http://dx.doi.org/10.1016/j.cpc.2017.01.003} {\bibfield  {journal} {\bibinfo
  {journal} {Computer Physics Communications}\ ,\ }}\bibinfo {note} {(in
  print)}\BibitemShut {NoStop}%
\bibitem [{\citenamefont {Shinaoka}\ \emph
  {et~al.}(2017{\natexlab{b}})\citenamefont {Shinaoka}, \citenamefont {Otsuki},
  \citenamefont {Ohzeki},\ and\ \citenamefont {Yoshimi}}]{Shinaoka2017}%
  \BibitemOpen
  \bibfield  {author} {\bibinfo {author} {\bibnamefont {Shinaoka},
  \bibfnamefont {H.}}, \bibinfo {author} {\bibfnamefont {J.}~\bibnamefont
  {Otsuki}}, \bibinfo {author} {\bibfnamefont {M.}~\bibnamefont {Ohzeki}}, \
  and\ \bibinfo {author} {\bibfnamefont {K.}~\bibnamefont {Yoshimi}}} (\bibinfo
  {year} {2017}{\natexlab{b}}),\ \href {\doibase 10.1103/PhysRevB.96.035147}
  {\bibfield  {journal} {\bibinfo  {journal} {Phys. Rev. B}\ }\textbf {\bibinfo
  {volume} {96}},\ \bibinfo {pages} {035147}}\BibitemShut {NoStop}%
\bibitem [{\citenamefont {Shvaika}(2000)}]{Shvaika2000}%
  \BibitemOpen
  \bibfield  {author} {\bibinfo {author} {\bibnamefont {Shvaika}, \bibfnamefont
  {A.~M.}}} (\bibinfo {year} {2000}),\ \href {\doibase
  10.1016/S0921-4534(00)00435-4} {\bibfield  {journal} {\bibinfo  {journal}
  {Physica C}\ }\textbf {\bibinfo {volume} {341-348}},\ \bibinfo {pages}
  {177}}\BibitemShut {NoStop}%
\bibitem [{\citenamefont {Si}\ and\ \citenamefont {Smith}(1996)}]{Si1996}%
  \BibitemOpen
  \bibfield  {author} {\bibinfo {author} {\bibnamefont {Si}, \bibfnamefont
  {Q.}}, \ and\ \bibinfo {author} {\bibfnamefont {J.~L.}\ \bibnamefont
  {Smith}}} (\bibinfo {year} {1996}),\ \href {\doibase
  10.1103/PhysRevLett.77.3391} {\bibfield  {journal} {\bibinfo  {journal}
  {Phys. Rev. Lett.}\ }\textbf {\bibinfo {volume} {77}},\ \bibinfo {pages}
  {3391}}\BibitemShut {NoStop}%
\bibitem [{\citenamefont {Slezak}\ \emph {et~al.}(2009)\citenamefont {Slezak},
  \citenamefont {M.}, \citenamefont {Maier},\ and\ \citenamefont
  {Deisz}}]{Slezak2009}%
  \BibitemOpen
  \bibfield  {author} {\bibinfo {author} {\bibnamefont {Slezak}, \bibfnamefont
  {C.}}, \bibinfo {author} {\bibfnamefont {J.}~\bibnamefont {M.}}, \bibinfo
  {author} {\bibfnamefont {T.}~\bibnamefont {Maier}}, \ and\ \bibinfo {author}
  {\bibfnamefont {J.}~\bibnamefont {Deisz}}} (\bibinfo {year} {2009}),\ \href
  {http://stacks.iop.org/0953-8984/21/i=43/a=435604} {\bibfield  {journal}
  {\bibinfo  {journal} {J. Phys.: Condens. Matter}\ }\textbf {\bibinfo {volume}
  {21}},\ \bibinfo {pages} {435604}}\BibitemShut {NoStop}%
\bibitem [{\citenamefont {Smith}\ and\ \citenamefont {Si}(2000)}]{Smith00}%
  \BibitemOpen
  \bibfield  {author} {\bibinfo {author} {\bibnamefont {Smith}, \bibfnamefont
  {J.~L.}}, \ and\ \bibinfo {author} {\bibfnamefont {Q.}~\bibnamefont {Si}}}
  (\bibinfo {year} {2000}),\ \href {\doibase 10.1103/PhysRevB.61.5184}
  {\bibfield  {journal} {\bibinfo  {journal} {Phys. Rev. B}\ }\textbf {\bibinfo
  {volume} {61}},\ \bibinfo {pages} {5184}}\BibitemShut {NoStop}%
\bibitem [{\citenamefont {Smith}(1992)}]{Smith1992}%
  \BibitemOpen
  \bibfield  {author} {\bibinfo {author} {\bibnamefont {Smith}, \bibfnamefont
  {R.~A.}}} (\bibinfo {year} {1992}),\ \href {\doibase
  10.1103/PhysRevA.46.4586} {\bibfield  {journal} {\bibinfo  {journal} {Phys.
  Rev. A}\ }\textbf {\bibinfo {volume} {46}},\ \bibinfo {pages}
  {4586}}\BibitemShut {NoStop}%
\bibitem [{\citenamefont {Sordi}\ \emph {et~al.}(2011)\citenamefont {Sordi},
  \citenamefont {Haule},\ and\ \citenamefont {Tremblay}}]{Sordi2011}%
  \BibitemOpen
  \bibfield  {author} {\bibinfo {author} {\bibnamefont {Sordi}, \bibfnamefont
  {G.}}, \bibinfo {author} {\bibfnamefont {K.}~\bibnamefont {Haule}}, \ and\
  \bibinfo {author} {\bibfnamefont {A.-M.~S.}\ \bibnamefont {Tremblay}}}
  (\bibinfo {year} {2011}),\ \href {\doibase 10.1103/PhysRevB.84.075161}
  {\bibfield  {journal} {\bibinfo  {journal} {Phys. Rev. B}\ }\textbf {\bibinfo
  {volume} {84}},\ \bibinfo {pages} {075161}}\BibitemShut {NoStop}%
\bibitem [{\citenamefont {Sordi}\ \emph {et~al.}(2012)\citenamefont {Sordi},
  \citenamefont {S\'emon}, \citenamefont {Haule},\ and\ \citenamefont
  {Tremblay}}]{Sordi2012}%
  \BibitemOpen
  \bibfield  {author} {\bibinfo {author} {\bibnamefont {Sordi}, \bibfnamefont
  {G.}}, \bibinfo {author} {\bibfnamefont {P.}~\bibnamefont {S\'emon}},
  \bibinfo {author} {\bibfnamefont {K.}~\bibnamefont {Haule}}, \ and\ \bibinfo
  {author} {\bibfnamefont {A.-M.~S.}\ \bibnamefont {Tremblay}}} (\bibinfo
  {year} {2012}),\ \href {\doibase 10.1103/PhysRevLett.108.216401} {\bibfield
  {journal} {\bibinfo  {journal} {Phys. Rev. Lett.}\ }\textbf {\bibinfo
  {volume} {108}},\ \bibinfo {pages} {216401}}\BibitemShut {NoStop}%
\bibitem [{\citenamefont {Soven}(1969)}]{Soven1969}%
  \BibitemOpen
  \bibfield  {author} {\bibinfo {author} {\bibnamefont {Soven}, \bibfnamefont
  {P.}}} (\bibinfo {year} {1969}),\ \href {\doibase 10.1103/PhysRev.178.1136}
  {\bibfield  {journal} {\bibinfo  {journal} {Phys. Rev.}\ }\textbf {\bibinfo
  {volume} {178}}~(\bibinfo {number} {3}),\ \bibinfo {pages}
  {1136}}\BibitemShut {NoStop}%
\bibitem [{\citenamefont {Staar}\ \emph {et~al.}(2012)\citenamefont {Staar},
  \citenamefont {Maier},\ and\ \citenamefont {Schulthess}}]{Staar2012}%
  \BibitemOpen
  \bibfield  {author} {\bibinfo {author} {\bibnamefont {Staar}, \bibfnamefont
  {P.}}, \bibinfo {author} {\bibfnamefont {T.~A.}\ \bibnamefont {Maier}}, \
  and\ \bibinfo {author} {\bibfnamefont {T.~C.}\ \bibnamefont {Schulthess}}}
  (\bibinfo {year} {2012}),\ \href
  {http://stacks.iop.org/1742-6596/402/i=1/a=012015} {\bibfield  {journal}
  {\bibinfo  {journal} {Journal of Physics: Conference Series}\ }\textbf
  {\bibinfo {volume} {402}}~(\bibinfo {number} {1}),\ \bibinfo {pages}
  {012015}}\BibitemShut {NoStop}%
\bibitem [{\citenamefont {Staar}\ \emph {et~al.}(2013)\citenamefont {Staar},
  \citenamefont {Maier}, \citenamefont {Summers}, \citenamefont {Fourestey},
  \citenamefont {Solca},\ and\ \citenamefont {Schulthess}}]{Staar2013a}%
  \BibitemOpen
  \bibfield  {author} {\bibinfo {author} {\bibnamefont {Staar}, \bibfnamefont
  {P.}}, \bibinfo {author} {\bibfnamefont {T.~A.}\ \bibnamefont {Maier}},
  \bibinfo {author} {\bibfnamefont {M.~S.}\ \bibnamefont {Summers}}, \bibinfo
  {author} {\bibfnamefont {G.}~\bibnamefont {Fourestey}}, \bibinfo {author}
  {\bibfnamefont {R.}~\bibnamefont {Solca}}, \ and\ \bibinfo {author}
  {\bibfnamefont {T.~C.}\ \bibnamefont {Schulthess}}} (\bibinfo {year}
  {2013}),\ in\ \href {\doibase 10.1145/2503210.2503282} {\emph {\bibinfo
  {booktitle} {Proceedings of the International Conference on High Performance
  Computing, Networking, Storage and Analysis}}},\ \bibinfo {series and number}
  {SC '13}\ (\bibinfo  {publisher} {ACM},\ \bibinfo {address} {New York, NY,
  USA})\ pp.\ \bibinfo {pages} {1:1--1:11}\BibitemShut {NoStop}%
\bibitem [{\citenamefont {Stan}\ \emph {et~al.}(2015)\citenamefont {Stan},
  \citenamefont {Romaniello}, \citenamefont {Rigamonti}, \citenamefont
  {Reining},\ and\ \citenamefont {Berger}}]{Stan2015}%
  \BibitemOpen
  \bibfield  {author} {\bibinfo {author} {\bibnamefont {Stan}, \bibfnamefont
  {A.}}, \bibinfo {author} {\bibfnamefont {P.}~\bibnamefont {Romaniello}},
  \bibinfo {author} {\bibfnamefont {S.}~\bibnamefont {Rigamonti}}, \bibinfo
  {author} {\bibfnamefont {L.}~\bibnamefont {Reining}}, \ and\ \bibinfo
  {author} {\bibfnamefont {J.~A.}\ \bibnamefont {Berger}}} (\bibinfo {year}
  {2015}),\ \href {http://stacks.iop.org/1367-2630/17/i=9/a=093045} {\bibfield
  {journal} {\bibinfo  {journal} {New J. Phys.}\ }\textbf {\bibinfo {volume}
  {17}}~(\bibinfo {number} {9}),\ \bibinfo {pages} {093045}}\BibitemShut
  {NoStop}%
\bibitem [{\citenamefont {Stanescu}\ and\ \citenamefont
  {Kotliar}(2004)}]{Stanescu2004}%
  \BibitemOpen
  \bibfield  {author} {\bibinfo {author} {\bibnamefont {Stanescu},
  \bibfnamefont {T.~D.}}, \ and\ \bibinfo {author} {\bibfnamefont
  {G.}~\bibnamefont {Kotliar}}} (\bibinfo {year} {2004}),\ \href {\doibase
  10.1103/PhysRevB.70.205112} {\bibfield  {journal} {\bibinfo  {journal} {Phys.
  Rev. B}\ }\textbf {\bibinfo {volume} {70}},\ \bibinfo {pages}
  {205112}}\BibitemShut {NoStop}%
\bibitem [{\citenamefont {Stepanov}\ \emph
  {et~al.}(2016{\natexlab{a}})\citenamefont {Stepanov}, \citenamefont {Huber},
  \citenamefont {van Loon}, \citenamefont {Lichtenstein},\ and\ \citenamefont
  {Katsnelson}}]{Stepanov2016a}%
  \BibitemOpen
  \bibfield  {author} {\bibinfo {author} {\bibnamefont {Stepanov},
  \bibfnamefont {E.~A.}}, \bibinfo {author} {\bibfnamefont {A.}~\bibnamefont
  {Huber}}, \bibinfo {author} {\bibfnamefont {E.~G. C.~P.}\ \bibnamefont {van
  Loon}}, \bibinfo {author} {\bibfnamefont {A.~I.}\ \bibnamefont
  {Lichtenstein}}, \ and\ \bibinfo {author} {\bibfnamefont {M.~I.}\
  \bibnamefont {Katsnelson}}} (\bibinfo {year} {2016}{\natexlab{a}}),\ \href
  {\doibase 10.1103/PhysRevB.94.205110} {\bibfield  {journal} {\bibinfo
  {journal} {Phys. Rev. B}\ }\textbf {\bibinfo {volume} {94}},\ \bibinfo
  {pages} {205110}}\BibitemShut {NoStop}%
\bibitem [{\citenamefont {Stepanov}\ \emph
  {et~al.}(2016{\natexlab{b}})\citenamefont {Stepanov}, \citenamefont {van
  Loon}, \citenamefont {Katanin}, \citenamefont {Lichtenstein}, \citenamefont
  {Katsnelson},\ and\ \citenamefont {Rubtsov}}]{Stepanov2016}%
  \BibitemOpen
  \bibfield  {author} {\bibinfo {author} {\bibnamefont {Stepanov},
  \bibfnamefont {E.~A.}}, \bibinfo {author} {\bibfnamefont {E.~G. C.~P.}\
  \bibnamefont {van Loon}}, \bibinfo {author} {\bibfnamefont {A.~A.}\
  \bibnamefont {Katanin}}, \bibinfo {author} {\bibfnamefont {A.~I.}\
  \bibnamefont {Lichtenstein}}, \bibinfo {author} {\bibfnamefont {M.~I.}\
  \bibnamefont {Katsnelson}}, \ and\ \bibinfo {author} {\bibfnamefont {A.~N.}\
  \bibnamefont {Rubtsov}}} (\bibinfo {year} {2016}{\natexlab{b}}),\ \href
  {\doibase 10.1103/PhysRevB.93.045107} {\bibfield  {journal} {\bibinfo
  {journal} {Phys. Rev. B}\ }\textbf {\bibinfo {volume} {93}},\ \bibinfo
  {pages} {045107}}\BibitemShut {NoStop}%
\bibitem [{\citenamefont {Sun}\ and\ \citenamefont {Kotliar}(2002)}]{Sun02}%
  \BibitemOpen
  \bibfield  {author} {\bibinfo {author} {\bibnamefont {Sun}, \bibfnamefont
  {P.}}, \ and\ \bibinfo {author} {\bibfnamefont {G.}~\bibnamefont {Kotliar}}}
  (\bibinfo {year} {2002}),\ \href {\doibase 10.1103/PhysRevB.66.085120}
  {\bibfield  {journal} {\bibinfo  {journal} {Phys. Rev. B}\ }\textbf {\bibinfo
  {volume} {66}},\ \bibinfo {pages} {085120}}\BibitemShut {NoStop}%
\bibitem [{\citenamefont {Sun}\ and\ \citenamefont {Kotliar}(2004)}]{Sun2004}%
  \BibitemOpen
  \bibfield  {author} {\bibinfo {author} {\bibnamefont {Sun}, \bibfnamefont
  {P.}}, \ and\ \bibinfo {author} {\bibfnamefont {G.}~\bibnamefont {Kotliar}}}
  (\bibinfo {year} {2004}),\ \href
  {http://dx.doi.org/10.1103/PhysRevLett.92.196402} {\bibfield  {journal}
  {\bibinfo  {journal} {Phys. Rev. Lett.}\ }\textbf {\bibinfo {volume} {92}},\
  \bibinfo {pages} {196402}}\BibitemShut {NoStop}%
\bibitem [{\citenamefont {Sweep}\ \emph {et~al.}(2013)\citenamefont {Sweep},
  \citenamefont {Rubtsov},\ and\ \citenamefont {Katsnelson}}]{Sweep2013}%
  \BibitemOpen
  \bibfield  {author} {\bibinfo {author} {\bibnamefont {Sweep}, \bibfnamefont
  {J.}}, \bibinfo {author} {\bibfnamefont {A.}~\bibnamefont {Rubtsov}}, \ and\
  \bibinfo {author} {\bibfnamefont {M.}~\bibnamefont {Katsnelson}}} (\bibinfo
  {year} {2013}),\ \href {\doibase 10.1134/S0021364013200149} {\bibfield
  {journal} {\bibinfo  {journal} {JETP Letters}\ }\textbf {\bibinfo {volume}
  {98}}~(\bibinfo {number} {7}),\ \bibinfo {pages} {427}}\BibitemShut {NoStop}%
\bibitem [{\citenamefont {Tagliavini}\ \emph {et~al.}(2018)\citenamefont
  {Tagliavini}, \citenamefont {Hummel}, \citenamefont {Wentzell}, \citenamefont
  {Andergassen}, \citenamefont {Toschi},\ and\ \citenamefont
  {Rohringer}}]{Tagliavini2018}%
  \BibitemOpen
  \bibfield  {author} {\bibinfo {author} {\bibnamefont {Tagliavini},
  \bibfnamefont {A.}}, \bibinfo {author} {\bibfnamefont {S.}~\bibnamefont
  {Hummel}}, \bibinfo {author} {\bibfnamefont {N.}~\bibnamefont {Wentzell}},
  \bibinfo {author} {\bibfnamefont {S.}~\bibnamefont {Andergassen}}, \bibinfo
  {author} {\bibfnamefont {A.}~\bibnamefont {Toschi}}, \ and\ \bibinfo {author}
  {\bibfnamefont {G.}~\bibnamefont {Rohringer}}} (\bibinfo {year} {2018}),\
  \href@noop {} {\bibinfo  {journal} {arXiv:1803.03036}\ }\BibitemShut
  {NoStop}%
\bibitem [{\citenamefont {Tahvildar-Zadeh}\ \emph {et~al.}(1997)\citenamefont
  {Tahvildar-Zadeh}, \citenamefont {Freericks},\ and\ \citenamefont
  {Jarrell}}]{Jarrell1997}%
  \BibitemOpen
\bibfield  {journal} {  }\bibfield  {author} {\bibinfo {author} {\bibnamefont
  {Tahvildar-Zadeh}, \bibfnamefont {A.~N.}}, \bibinfo {author} {\bibfnamefont
  {J.~K.}\ \bibnamefont {Freericks}}, \ and\ \bibinfo {author} {\bibfnamefont
  {M.}~\bibnamefont {Jarrell}}} (\bibinfo {year} {1997}),\ \href {\doibase
  10.1103/PhysRevB.55.942} {\bibfield  {journal} {\bibinfo  {journal} {Phys.
  Rev. B}\ }\textbf {\bibinfo {volume} {55}},\ \bibinfo {pages}
  {942}}\BibitemShut {NoStop}%
\bibitem [{\citenamefont {Takahashi}(1977)}]{Takahashi77}%
  \BibitemOpen
  \bibfield  {author} {\bibinfo {author} {\bibnamefont {Takahashi},
  \bibfnamefont {M.}}} (\bibinfo {year} {1977}),\ \href
  {http://stacks.iop.org/0022-3719/10/i=8/a=031} {\bibfield  {journal}
  {\bibinfo  {journal} {J. Phys. C: Solid State Physics}\ }\textbf {\bibinfo
  {volume} {10}}~(\bibinfo {number} {8}),\ \bibinfo {pages} {1289}}\BibitemShut
  {NoStop}%
\bibitem [{\citenamefont {Takemori}(2016)}]{Takemori2016a}%
  \BibitemOpen
  \bibfield  {author} {\bibinfo {author} {\bibnamefont {Takemori},
  \bibfnamefont {N.}}} (\bibinfo {year} {2016}),\ \emph {\bibinfo {title}
  {Strong electron correlation effects in a quasiperiodic lattice}},\
  \href@noop {} {Ph.D. thesis}\ (\bibinfo  {school} {Tokyo Institute of
  Technology})\BibitemShut {NoStop}%
\bibitem [{\citenamefont {Takemori}\ \emph {et~al.}(2016)\citenamefont
  {Takemori}, \citenamefont {Koga},\ and\ \citenamefont
  {Hafermann}}]{Takemori2016}%
  \BibitemOpen
  \bibfield  {author} {\bibinfo {author} {\bibnamefont {Takemori},
  \bibfnamefont {N.}}, \bibinfo {author} {\bibfnamefont {A.}~\bibnamefont
  {Koga}}, \ and\ \bibinfo {author} {\bibfnamefont {H.}~\bibnamefont
  {Hafermann}}} (\bibinfo {year} {2016}),\ \href
  {http://stacks.iop.org/1742-6596/683/i=1/a=012040} {\bibfield  {journal}
  {\bibinfo  {journal} {J. Phys.: Conference Series}\ }\textbf {\bibinfo
  {volume} {683}}~(\bibinfo {number} {1}),\ \bibinfo {pages}
  {012040}}\BibitemShut {NoStop}%
\bibitem [{\citenamefont {Tam}\ \emph {et~al.}(2013)\citenamefont {Tam},
  \citenamefont {Fotso}, \citenamefont {Yang}, \citenamefont {Lee},
  \citenamefont {Moreno}, \citenamefont {Ramanujam},\ and\ \citenamefont
  {Jarrell}}]{Tam2013}%
  \BibitemOpen
  \bibfield  {author} {\bibinfo {author} {\bibnamefont {Tam}, \bibfnamefont
  {K.-M.}}, \bibinfo {author} {\bibfnamefont {H.}~\bibnamefont {Fotso}},
  \bibinfo {author} {\bibfnamefont {S.-X.}\ \bibnamefont {Yang}}, \bibinfo
  {author} {\bibfnamefont {T.-W.}\ \bibnamefont {Lee}}, \bibinfo {author}
  {\bibfnamefont {J.}~\bibnamefont {Moreno}}, \bibinfo {author} {\bibfnamefont
  {J.}~\bibnamefont {Ramanujam}}, \ and\ \bibinfo {author} {\bibfnamefont
  {M.}~\bibnamefont {Jarrell}}} (\bibinfo {year} {2013}),\ \href {\doibase
  10.1103/PhysRevE.87.013311} {\bibfield  {journal} {\bibinfo  {journal} {Phys.
  Rev. E}\ }\textbf {\bibinfo {volume} {87}},\ \bibinfo {pages}
  {013311}}\BibitemShut {NoStop}%
\bibitem [{\citenamefont {Tanaka}(2016)}]{TanakaPC2016}%
  \BibitemOpen
  \bibfield  {author} {\bibinfo {author} {\bibnamefont {Tanaka},}} (\bibinfo
  {year} {2016}),\ \href@noop {} {}\bibinfo {howpublished} {private
  communication}\BibitemShut {NoStop}%
\bibitem [{\citenamefont {Tarantino}\ \emph {et~al.}(2017)\citenamefont
  {Tarantino}, \citenamefont {Romaniello}, \citenamefont {Berger},\ and\
  \citenamefont {Reining}}]{Tarantino2017}%
  \BibitemOpen
  \bibfield  {author} {\bibinfo {author} {\bibnamefont {Tarantino},
  \bibfnamefont {W.}}, \bibinfo {author} {\bibfnamefont {P.}~\bibnamefont
  {Romaniello}}, \bibinfo {author} {\bibfnamefont {J.~A.}\ \bibnamefont
  {Berger}}, \ and\ \bibinfo {author} {\bibfnamefont {L.}~\bibnamefont
  {Reining}}} (\bibinfo {year} {2017}),\ \href {\doibase
  10.1103/PhysRevB.96.045124} {\bibfield  {journal} {\bibinfo  {journal} {Phys.
  Rev. B}\ }\textbf {\bibinfo {volume} {96}},\ \bibinfo {pages}
  {045124}}\BibitemShut {NoStop}%
\bibitem [{\citenamefont {Taranto}\ \emph {et~al.}(2014)\citenamefont
  {Taranto}, \citenamefont {Andergassen}, \citenamefont {Bauer}, \citenamefont
  {Held}, \citenamefont {Katanin}, \citenamefont {Metzner}, \citenamefont
  {Rohringer},\ and\ \citenamefont {Toschi}}]{Taranto2014}%
  \BibitemOpen
  \bibfield  {author} {\bibinfo {author} {\bibnamefont {Taranto}, \bibfnamefont
  {C.}}, \bibinfo {author} {\bibfnamefont {S.}~\bibnamefont {Andergassen}},
  \bibinfo {author} {\bibfnamefont {J.}~\bibnamefont {Bauer}}, \bibinfo
  {author} {\bibfnamefont {K.}~\bibnamefont {Held}}, \bibinfo {author}
  {\bibfnamefont {A.}~\bibnamefont {Katanin}}, \bibinfo {author} {\bibfnamefont
  {W.}~\bibnamefont {Metzner}}, \bibinfo {author} {\bibfnamefont
  {G.}~\bibnamefont {Rohringer}}, \ and\ \bibinfo {author} {\bibfnamefont
  {A.}~\bibnamefont {Toschi}}} (\bibinfo {year} {2014}),\ \href {\doibase
  10.1103/PhysRevLett.112.196402} {\bibfield  {journal} {\bibinfo  {journal}
  {Phys. Rev. Lett.}\ }\textbf {\bibinfo {volume} {112}},\ \bibinfo {pages}
  {196402}}\BibitemShut {NoStop}%
\bibitem [{\citenamefont {Taylor}(1967)}]{Taylor1967}%
  \BibitemOpen
  \bibfield  {author} {\bibinfo {author} {\bibnamefont {Taylor}, \bibfnamefont
  {D.~W.}}} (\bibinfo {year} {1967}),\ \href {\doibase
  10.1103/PhysRev.156.1017} {\bibfield  {journal} {\bibinfo  {journal} {Phys.
  Rev.}\ }\textbf {\bibinfo {volume} {156}},\ \bibinfo {pages}
  {1017}}\BibitemShut {NoStop}%
\bibitem [{\citenamefont {Terasaki}\ \emph {et~al.}(1997)\citenamefont
  {Terasaki}, \citenamefont {Sasago},\ and\ \citenamefont
  {Uchinokura}}]{Terasaki1997}%
  \BibitemOpen
  \bibfield  {author} {\bibinfo {author} {\bibnamefont {Terasaki},
  \bibfnamefont {I.}}, \bibinfo {author} {\bibfnamefont {Y.}~\bibnamefont
  {Sasago}}, \ and\ \bibinfo {author} {\bibfnamefont {K.}~\bibnamefont
  {Uchinokura}}} (\bibinfo {year} {1997}),\ \href {\doibase
  10.1103/PhysRevB.56.R12685} {\bibfield  {journal} {\bibinfo  {journal} {Phys.
  Rev. B}\ }\textbf {\bibinfo {volume} {56}},\ \bibinfo {pages}
  {R12685}}\BibitemShut {NoStop}%
\bibitem [{\citenamefont {Terletska}\ \emph {et~al.}(2017)\citenamefont
  {Terletska}, \citenamefont {Chen},\ and\ \citenamefont
  {Gull}}]{Terletska2017}%
  \BibitemOpen
  \bibfield  {author} {\bibinfo {author} {\bibnamefont {Terletska},
  \bibfnamefont {H.}}, \bibinfo {author} {\bibfnamefont {T.}~\bibnamefont
  {Chen}}, \ and\ \bibinfo {author} {\bibfnamefont {E.}~\bibnamefont {Gull}}}
  (\bibinfo {year} {2017}),\ \href {\doibase 10.1103/PhysRevB.95.115149}
  {\bibfield  {journal} {\bibinfo  {journal} {Phys. Rev. B}\ }\textbf {\bibinfo
  {volume} {95}},\ \bibinfo {pages} {115149}}\BibitemShut {NoStop}%
\bibitem [{\citenamefont {Terletska}\ \emph {et~al.}(2014)\citenamefont
  {Terletska}, \citenamefont {Ekuma}, \citenamefont {Moore}, \citenamefont
  {Tam}, \citenamefont {Moreno},\ and\ \citenamefont
  {Jarrell}}]{Terletska2014}%
  \BibitemOpen
  \bibfield  {author} {\bibinfo {author} {\bibnamefont {Terletska},
  \bibfnamefont {H.}}, \bibinfo {author} {\bibfnamefont {C.~E.}\ \bibnamefont
  {Ekuma}}, \bibinfo {author} {\bibfnamefont {C.}~\bibnamefont {Moore}},
  \bibinfo {author} {\bibfnamefont {K.-M.}\ \bibnamefont {Tam}}, \bibinfo
  {author} {\bibfnamefont {J.}~\bibnamefont {Moreno}}, \ and\ \bibinfo {author}
  {\bibfnamefont {M.}~\bibnamefont {Jarrell}}} (\bibinfo {year} {2014}),\ \href
  {\doibase 10.1103/PhysRevB.90.094208} {\bibfield  {journal} {\bibinfo
  {journal} {Phys. Rev. B}\ }\textbf {\bibinfo {volume} {90}}~(\bibinfo
  {number} {9}),\ \bibinfo {pages} {094208}}\BibitemShut {NoStop}%
\bibitem [{\citenamefont {Terletska}\ \emph {et~al.}(2013)\citenamefont
  {Terletska}, \citenamefont {Yang}, \citenamefont {Meng}, \citenamefont
  {Moreno},\ and\ \citenamefont {Jarrell}}]{Terletska2013}%
  \BibitemOpen
  \bibfield  {author} {\bibinfo {author} {\bibnamefont {Terletska},
  \bibfnamefont {H.}}, \bibinfo {author} {\bibfnamefont {S.-X.}\ \bibnamefont
  {Yang}}, \bibinfo {author} {\bibfnamefont {Z.~Y.}\ \bibnamefont {Meng}},
  \bibinfo {author} {\bibfnamefont {J.}~\bibnamefont {Moreno}}, \ and\ \bibinfo
  {author} {\bibfnamefont {M.}~\bibnamefont {Jarrell}}} (\bibinfo {year}
  {2013}),\ \href {\doibase 10.1103/PhysRevB.87.134208} {\bibfield  {journal}
  {\bibinfo  {journal} {Phys. Rev. B}\ }\textbf {\bibinfo {volume}
  {87}}~(\bibinfo {number} {13}),\ \bibinfo {pages} {134208}}\BibitemShut
  {NoStop}%
\bibitem [{\citenamefont {Thouless}(1974)}]{Thouless1974}%
  \BibitemOpen
  \bibfield  {author} {\bibinfo {author} {\bibnamefont {Thouless},
  \bibfnamefont {D.}}} (\bibinfo {year} {1974}),\ \href {\doibase
  10.1016/0370-1573(74)90029-5} {\bibfield  {journal} {\bibinfo  {journal}
  {Phys. Rep.}\ }\textbf {\bibinfo {volume} {13}}~(\bibinfo {number} {3}),\
  \bibinfo {pages} {93}}\BibitemShut {NoStop}%
\bibitem [{\citenamefont {Tomczak}(2015)}]{jmt_sces14}%
  \BibitemOpen
  \bibfield  {author} {\bibinfo {author} {\bibnamefont {Tomczak}, \bibfnamefont
  {J.~M.}}} (\bibinfo {year} {2015}),\ \href
  {http://iopscience.iop.org/article/10.1088/1742-6596/592/1/012055} {\bibfield
   {journal} {\bibinfo  {journal} {J. Phys.: Conference Series}\ }\textbf
  {\bibinfo {volume} {592}}~(\bibinfo {number} {1}),\ \bibinfo {pages}
  {012055}}\BibitemShut {NoStop}%
\bibitem [{\citenamefont {Tomczak}\ \emph
  {et~al.}(2012{\natexlab{a}})\citenamefont {Tomczak}, \citenamefont {Casula},
  \citenamefont {Miyake}, \citenamefont {Aryasetiawan},\ and\ \citenamefont
  {Biermann}}]{Tomczak2012}%
  \BibitemOpen
  \bibfield  {author} {\bibinfo {author} {\bibnamefont {Tomczak}, \bibfnamefont
  {J.~M.}}, \bibinfo {author} {\bibfnamefont {M.}~\bibnamefont {Casula}},
  \bibinfo {author} {\bibfnamefont {T.}~\bibnamefont {Miyake}}, \bibinfo
  {author} {\bibfnamefont {F.}~\bibnamefont {Aryasetiawan}}, \ and\ \bibinfo
  {author} {\bibfnamefont {S.}~\bibnamefont {Biermann}}} (\bibinfo {year}
  {2012}{\natexlab{a}}),\ \href
  {http://stacks.iop.org/0295-5075/100/i=6/a=67001} {\bibfield  {journal}
  {\bibinfo  {journal} {EPL (Europhysics Letters)}\ }\textbf {\bibinfo {volume}
  {100}}~(\bibinfo {number} {6}),\ \bibinfo {pages} {67001}}\BibitemShut
  {NoStop}%
\bibitem [{\citenamefont {Tomczak}\ \emph {et~al.}(2014)\citenamefont
  {Tomczak}, \citenamefont {Casula}, \citenamefont {Miyake},\ and\
  \citenamefont {Biermann}}]{Tomczak2014}%
  \BibitemOpen
  \bibfield  {author} {\bibinfo {author} {\bibnamefont {Tomczak}, \bibfnamefont
  {J.~M.}}, \bibinfo {author} {\bibfnamefont {M.}~\bibnamefont {Casula}},
  \bibinfo {author} {\bibfnamefont {T.}~\bibnamefont {Miyake}}, \ and\ \bibinfo
  {author} {\bibfnamefont {S.}~\bibnamefont {Biermann}}} (\bibinfo {year}
  {2014}),\ \href {\doibase 10.1103/PhysRevB.90.165138} {\bibfield  {journal}
  {\bibinfo  {journal} {Phys. Rev. B}\ }\textbf {\bibinfo {volume} {90}},\
  \bibinfo {pages} {165138}}\BibitemShut {NoStop}%
\bibitem [{\citenamefont {Tomczak}\ \emph {et~al.}(2017)\citenamefont
  {Tomczak}, \citenamefont {Liu}, \citenamefont {Toschi}, \citenamefont
  {Kresse},\ and\ \citenamefont {Held}}]{Tomczak2017}%
  \BibitemOpen
  \bibfield  {author} {\bibinfo {author} {\bibnamefont {Tomczak}, \bibfnamefont
  {J.~M.}}, \bibinfo {author} {\bibfnamefont {P.}~\bibnamefont {Liu}}, \bibinfo
  {author} {\bibfnamefont {A.}~\bibnamefont {Toschi}}, \bibinfo {author}
  {\bibfnamefont {G.}~\bibnamefont {Kresse}}, \ and\ \bibinfo {author}
  {\bibfnamefont {K.}~\bibnamefont {Held}}} (\bibinfo {year} {2017}),\ \href
  {\doibase 10.1140/epjst/e2017-70053-1} {\bibfield  {journal} {\bibinfo
  {journal} {The European Physical Journal Special Topics}\ }\textbf {\bibinfo
  {volume} {226}}~(\bibinfo {number} {11}),\ \bibinfo {pages}
  {2565}}\BibitemShut {NoStop}%
\bibitem [{\citenamefont {Tomczak}\ \emph
  {et~al.}(2012{\natexlab{b}})\citenamefont {Tomczak}, \citenamefont {van
  Schilfgaarde},\ and\ \citenamefont {Kotliar}}]{jmt_pnict}%
  \BibitemOpen
  \bibfield  {author} {\bibinfo {author} {\bibnamefont {Tomczak}, \bibfnamefont
  {J.~M.}}, \bibinfo {author} {\bibfnamefont {M.}~\bibnamefont {van
  Schilfgaarde}}, \ and\ \bibinfo {author} {\bibfnamefont {G.}~\bibnamefont
  {Kotliar}}} (\bibinfo {year} {2012}{\natexlab{b}}),\ \href {\doibase
  10.1103/PhysRevLett.109.237010} {\bibfield  {journal} {\bibinfo  {journal}
  {Phys. Rev. Lett.}\ }\textbf {\bibinfo {volume} {109}},\ \bibinfo {pages}
  {237010}}\BibitemShut {NoStop}%
\bibitem [{\citenamefont {Toschi}\ \emph {et~al.}(2007)\citenamefont {Toschi},
  \citenamefont {Katanin},\ and\ \citenamefont {Held}}]{Toschi2007}%
  \BibitemOpen
  \bibfield  {author} {\bibinfo {author} {\bibnamefont {Toschi}, \bibfnamefont
  {A.}}, \bibinfo {author} {\bibfnamefont {A.~A.}\ \bibnamefont {Katanin}}, \
  and\ \bibinfo {author} {\bibfnamefont {K.}~\bibnamefont {Held}}} (\bibinfo
  {year} {2007}),\ \href {\doibase 10.1103/PhysRevB.75.045118} {\bibfield
  {journal} {\bibinfo  {journal} {Phys Rev. B}\ }\textbf {\bibinfo {volume}
  {75}},\ \bibinfo {pages} {045118}}\BibitemShut {NoStop}%
\bibitem [{\citenamefont {Toschi}\ \emph {et~al.}(2011)\citenamefont {Toschi},
  \citenamefont {Rohringer}, \citenamefont {Katanin},\ and\ \citenamefont
  {Held}}]{Toschi2011}%
  \BibitemOpen
  \bibfield  {author} {\bibinfo {author} {\bibnamefont {Toschi}, \bibfnamefont
  {A.}}, \bibinfo {author} {\bibfnamefont {G.}~\bibnamefont {Rohringer}},
  \bibinfo {author} {\bibfnamefont {A.}~\bibnamefont {Katanin}}, \ and\
  \bibinfo {author} {\bibfnamefont {K.}~\bibnamefont {Held}}} (\bibinfo {year}
  {2011}),\ \href {\doibase 10.1002/andp.201100036} {\bibfield  {journal}
  {\bibinfo  {journal} {Annalen der Physik}\ }\textbf {\bibinfo {volume}
  {523}}~(\bibinfo {number} {8-9}),\ \bibinfo {pages} {698}}\BibitemShut
  {NoStop}%
\bibitem [{\citenamefont {Valli}\ \emph
  {et~al.}(2015{\natexlab{a}})\citenamefont {Valli}, \citenamefont {Das},
  \citenamefont {Sangiovanni}, \citenamefont {Saha-Dasgupta},\ and\
  \citenamefont {Held}}]{Valli2015a}%
  \BibitemOpen
  \bibfield  {author} {\bibinfo {author} {\bibnamefont {Valli}, \bibfnamefont
  {A.}}, \bibinfo {author} {\bibfnamefont {H.}~\bibnamefont {Das}}, \bibinfo
  {author} {\bibfnamefont {G.}~\bibnamefont {Sangiovanni}}, \bibinfo {author}
  {\bibfnamefont {T.}~\bibnamefont {Saha-Dasgupta}}, \ and\ \bibinfo {author}
  {\bibfnamefont {K.}~\bibnamefont {Held}}} (\bibinfo {year}
  {2015}{\natexlab{a}}),\ \href {\doibase 10.1103/PhysRevB.92.115143}
  {\bibfield  {journal} {\bibinfo  {journal} {Phys. Rev. B}\ }\textbf {\bibinfo
  {volume} {92}},\ \bibinfo {pages} {115143}}\BibitemShut {NoStop}%
\bibitem [{\citenamefont {Valli}\ \emph {et~al.}(2010)\citenamefont {Valli},
  \citenamefont {Sangiovanni}, \citenamefont {Gunnarsson}, \citenamefont
  {Toschi},\ and\ \citenamefont {Held}}]{Valli2010}%
  \BibitemOpen
  \bibfield  {author} {\bibinfo {author} {\bibnamefont {Valli}, \bibfnamefont
  {A.}}, \bibinfo {author} {\bibfnamefont {G.}~\bibnamefont {Sangiovanni}},
  \bibinfo {author} {\bibfnamefont {O.}~\bibnamefont {Gunnarsson}}, \bibinfo
  {author} {\bibfnamefont {A.}~\bibnamefont {Toschi}}, \ and\ \bibinfo {author}
  {\bibfnamefont {K.}~\bibnamefont {Held}}} (\bibinfo {year} {2010}),\ \href
  {\doibase 10.1103/PhysRevLett.104.246402} {\bibfield  {journal} {\bibinfo
  {journal} {Phys. Rev. Lett.}\ }\textbf {\bibinfo {volume} {104}},\ \bibinfo
  {pages} {246402}}\BibitemShut {NoStop}%
\bibitem [{\citenamefont {Valli}\ \emph {et~al.}(2012)\citenamefont {Valli},
  \citenamefont {Sangiovanni}, \citenamefont {Toschi},\ and\ \citenamefont
  {Held}}]{Valli2012}%
  \BibitemOpen
  \bibfield  {author} {\bibinfo {author} {\bibnamefont {Valli}, \bibfnamefont
  {A.}}, \bibinfo {author} {\bibfnamefont {G.}~\bibnamefont {Sangiovanni}},
  \bibinfo {author} {\bibfnamefont {A.}~\bibnamefont {Toschi}}, \ and\ \bibinfo
  {author} {\bibfnamefont {K.}~\bibnamefont {Held}}} (\bibinfo {year} {2012}),\
  \href {\doibase 10.1103/PhysRevB.86.115418} {\bibfield  {journal} {\bibinfo
  {journal} {Phys. Rev. B}\ }\textbf {\bibinfo {volume} {86}},\ \bibinfo
  {pages} {115418}}\BibitemShut {NoStop}%
\bibitem [{\citenamefont {Valli}\ \emph
  {et~al.}(2015{\natexlab{b}})\citenamefont {Valli}, \citenamefont {Sch\"afer},
  \citenamefont {Thunstr\"om}, \citenamefont {Rohringer}, \citenamefont
  {Andergassen}, \citenamefont {Sangiovanni}, \citenamefont {Held},\ and\
  \citenamefont {Toschi}}]{Valli2015}%
  \BibitemOpen
  \bibfield  {author} {\bibinfo {author} {\bibnamefont {Valli}, \bibfnamefont
  {A.}}, \bibinfo {author} {\bibfnamefont {T.}~\bibnamefont {Sch\"afer}},
  \bibinfo {author} {\bibfnamefont {P.}~\bibnamefont {Thunstr\"om}}, \bibinfo
  {author} {\bibfnamefont {G.}~\bibnamefont {Rohringer}}, \bibinfo {author}
  {\bibfnamefont {S.}~\bibnamefont {Andergassen}}, \bibinfo {author}
  {\bibfnamefont {G.}~\bibnamefont {Sangiovanni}}, \bibinfo {author}
  {\bibfnamefont {K.}~\bibnamefont {Held}}, \ and\ \bibinfo {author}
  {\bibfnamefont {A.}~\bibnamefont {Toschi}}} (\bibinfo {year}
  {2015}{\natexlab{b}}),\ \href {\doibase 10.1103/PhysRevB.91.115115}
  {\bibfield  {journal} {\bibinfo  {journal} {Phys. Rev. B}\ }\textbf {\bibinfo
  {volume} {91}},\ \bibinfo {pages} {115115}}\BibitemShut {NoStop}%
\bibitem [{\citenamefont {Veki\'c}\ and\ \citenamefont
  {White}(1993)}]{Vekic1993}%
  \BibitemOpen
  \bibfield  {author} {\bibinfo {author} {\bibnamefont {Veki\'c}, \bibfnamefont
  {M.}}, \ and\ \bibinfo {author} {\bibfnamefont {S.~R.}\ \bibnamefont
  {White}}} (\bibinfo {year} {1993}),\ \href {\doibase
  10.1103/PhysRevB.47.1160} {\bibfield  {journal} {\bibinfo  {journal} {Phys.
  Rev. B}\ }\textbf {\bibinfo {volume} {47}},\ \bibinfo {pages}
  {1160}}\BibitemShut {NoStop}%
\bibitem [{\citenamefont {Vilk}\ \emph {et~al.}(1994)\citenamefont {Vilk},
  \citenamefont {Chen},\ and\ \citenamefont {Tremblay}}]{Vilk1994}%
  \BibitemOpen
  \bibfield  {author} {\bibinfo {author} {\bibnamefont {Vilk}, \bibfnamefont
  {Y.~M.}}, \bibinfo {author} {\bibfnamefont {L.}~\bibnamefont {Chen}}, \ and\
  \bibinfo {author} {\bibfnamefont {A.-M.~S.}\ \bibnamefont {Tremblay}}}
  (\bibinfo {year} {1994}),\ \href {\doibase 10.1103/PhysRevB.49.13267}
  {\bibfield  {journal} {\bibinfo  {journal} {Phys. Rev. B}\ }\textbf {\bibinfo
  {volume} {49}},\ \bibinfo {pages} {13267}}\BibitemShut {NoStop}%
\bibitem [{\citenamefont {Vilk}\ and\ \citenamefont
  {Tremblay}(1996)}]{Vilk1996}%
  \BibitemOpen
  \bibfield  {author} {\bibinfo {author} {\bibnamefont {Vilk}, \bibfnamefont
  {Y.~M.}}, \ and\ \bibinfo {author} {\bibfnamefont {A.-M.~S.}\ \bibnamefont
  {Tremblay}}} (\bibinfo {year} {1996}),\ \href
  {http://stacks.iop.org/0295-5075/33/i=2/a=159} {\bibfield  {journal}
  {\bibinfo  {journal} {EPL (Europhysics Letters)}\ }\textbf {\bibinfo {volume}
  {33}}~(\bibinfo {number} {2}),\ \bibinfo {pages} {159}}\BibitemShut {NoStop}%
\bibitem [{\citenamefont {Vilk}\ and\ \citenamefont
  {Tremblay}(1997)}]{Vilk1997}%
  \BibitemOpen
  \bibfield  {author} {\bibinfo {author} {\bibnamefont {Vilk}, \bibfnamefont
  {Y.~M.}}, \ and\ \bibinfo {author} {\bibfnamefont {A.-M.~S.}\ \bibnamefont
  {Tremblay}}} (\bibinfo {year} {1997}),\ \href {\doibase 10.1051/jp1:1997135}
  {\bibfield  {journal} {\bibinfo  {journal} {J. Phys. I France}\ }\textbf
  {\bibinfo {volume} {7}}~(\bibinfo {number} {11}),\ \bibinfo {pages}
  {1309}}\BibitemShut {NoStop}%
\bibitem [{\citenamefont {Vlaming}\ and\ \citenamefont
  {Vollhardt}(1992)}]{Vlaming1992}%
  \BibitemOpen
  \bibfield  {author} {\bibinfo {author} {\bibnamefont {Vlaming}, \bibfnamefont
  {R.}}, \ and\ \bibinfo {author} {\bibfnamefont {D.}~\bibnamefont
  {Vollhardt}}} (\bibinfo {year} {1992}),\ \href {\doibase
  10.1103/PhysRevB.45.4637} {\bibfield  {journal} {\bibinfo  {journal} {Phys.
  Rev. B}\ }\textbf {\bibinfo {volume} {45}},\ \bibinfo {pages}
  {4637}}\BibitemShut {NoStop}%
\bibitem [{\citenamefont {Vollhardt}\ \emph {et~al.}(1998)\citenamefont
  {Vollhardt}, \citenamefont {Bl{\"{u}}mer}, \citenamefont {Held},
  \citenamefont {Kollar}, \citenamefont {Schlipf}, \citenamefont {Ulmke},\ and\
  \citenamefont {Wahle}}]{Vollhardt98a}%
  \BibitemOpen
  \bibfield  {author} {\bibinfo {author} {\bibnamefont {Vollhardt},
  \bibfnamefont {D.}}, \bibinfo {author} {\bibfnamefont {N.}~\bibnamefont
  {Bl{\"{u}}mer}}, \bibinfo {author} {\bibfnamefont {K.}~\bibnamefont {Held}},
  \bibinfo {author} {\bibfnamefont {M.}~\bibnamefont {Kollar}}, \bibinfo
  {author} {\bibfnamefont {J.}~\bibnamefont {Schlipf}}, \bibinfo {author}
  {\bibfnamefont {M.}~\bibnamefont {Ulmke}}, \ and\ \bibinfo {author}
  {\bibfnamefont {J.}~\bibnamefont {Wahle}}} (\bibinfo {year} {1998}),\ \href
  {\doibase 10.1007/BFb0107631} {\bibfield  {journal} {\bibinfo  {journal}
  {Advances In Solid State Physics}\ }\textbf {\bibinfo {volume} {38}},\
  \bibinfo {pages} {383}}\BibitemShut {NoStop}%
\bibitem [{\citenamefont {Vu\v{c}i\v{c}evi\'c}\ \emph
  {et~al.}(2017)\citenamefont {Vu\v{c}i\v{c}evi\'c}, \citenamefont {Ayral},\
  and\ \citenamefont {Parcollet}}]{Vucicevic2017}%
  \BibitemOpen
  \bibfield  {author} {\bibinfo {author} {\bibnamefont {Vu\v{c}i\v{c}evi\'c},
  \bibfnamefont {J.}}, \bibinfo {author} {\bibfnamefont {T.}~\bibnamefont
  {Ayral}}, \ and\ \bibinfo {author} {\bibfnamefont {O.}~\bibnamefont
  {Parcollet}}} (\bibinfo {year} {2017}),\ \href {\doibase
  10.1103/PhysRevB.96.104504} {\bibfield  {journal} {\bibinfo  {journal} {Phys.
  Rev. B}\ }\textbf {\bibinfo {volume} {96}},\ \bibinfo {pages}
  {104504}}\BibitemShut {NoStop}%
\bibitem [{\citenamefont {Vu\v{c}i\v{c}evi\'c}\ \emph
  {et~al.}(2018)\citenamefont {Vu\v{c}i\v{c}evi\'c}, \citenamefont {Wentzell},
  \citenamefont {Ferrero},\ and\ \citenamefont {Parcollet}}]{Vucicevic2018}%
  \BibitemOpen
  \bibfield  {author} {\bibinfo {author} {\bibnamefont {Vu\v{c}i\v{c}evi\'c},
  \bibfnamefont {J.}}, \bibinfo {author} {\bibfnamefont {N.}~\bibnamefont
  {Wentzell}}, \bibinfo {author} {\bibfnamefont {M.}~\bibnamefont {Ferrero}}, \
  and\ \bibinfo {author} {\bibfnamefont {O.}~\bibnamefont {Parcollet}}}
  (\bibinfo {year} {2018}),\ \href {\doibase 10.1103/PhysRevB.97.125141}
  {\bibfield  {journal} {\bibinfo  {journal} {Phys. Rev. B}\ }\textbf {\bibinfo
  {volume} {97}},\ \bibinfo {pages} {125141}}\BibitemShut {NoStop}%
\bibitem [{\citenamefont {Wahle}\ \emph {et~al.}(1998)\citenamefont {Wahle},
  \citenamefont {Bl\"umer}, \citenamefont {Schlipf}, \citenamefont {Held},\
  and\ \citenamefont {Vollhardt}}]{Wahle1998}%
  \BibitemOpen
  \bibfield  {author} {\bibinfo {author} {\bibnamefont {Wahle}, \bibfnamefont
  {J.}}, \bibinfo {author} {\bibfnamefont {N.}~\bibnamefont {Bl\"umer}},
  \bibinfo {author} {\bibfnamefont {J.}~\bibnamefont {Schlipf}}, \bibinfo
  {author} {\bibfnamefont {K.}~\bibnamefont {Held}}, \ and\ \bibinfo {author}
  {\bibfnamefont {D.}~\bibnamefont {Vollhardt}}} (\bibinfo {year} {1998}),\
  \href {\doibase 10.1103/PhysRevB.58.12749} {\bibfield  {journal} {\bibinfo
  {journal} {Phys. Rev. B}\ }\textbf {\bibinfo {volume} {58}},\ \bibinfo
  {pages} {12749}}\BibitemShut {NoStop}%
\bibitem [{\citenamefont {Wallerberger}\ \emph {et~al.}(2018)\citenamefont
  {Wallerberger}, \citenamefont {Hausoel}, \citenamefont {Gunacker},
  \citenamefont {Goth}, \citenamefont {Parragh}, \citenamefont {Held},\ and\
  \citenamefont {Sangiovanni}}]{w2dynamics2018}%
  \BibitemOpen
  \bibfield  {author} {\bibinfo {author} {\bibnamefont {Wallerberger},
  \bibfnamefont {M.}}, \bibinfo {author} {\bibfnamefont {A.}~\bibnamefont
  {Hausoel}}, \bibinfo {author} {\bibfnamefont {P.}~\bibnamefont {Gunacker}},
  \bibinfo {author} {\bibfnamefont {F.}~\bibnamefont {Goth}}, \bibinfo {author}
  {\bibfnamefont {N.}~\bibnamefont {Parragh}}, \bibinfo {author} {\bibfnamefont
  {K.}~\bibnamefont {Held}}, \ and\ \bibinfo {author} {\bibfnamefont
  {P.}~\bibnamefont {Sangiovanni}}} (\bibinfo {year} {2018}),\ \href@noop {} {\
  }\Eprint {http://arxiv.org/abs/arXiv:1801.10209} {arXiv:1801.10209}
  \BibitemShut {NoStop}%
\bibitem [{\citenamefont {Wang}\ \emph {et~al.}(2004)\citenamefont {Wang},
  \citenamefont {Zheng}, \citenamefont {Wu}, \citenamefont {Ma}, \citenamefont
  {Xiang}, \citenamefont {Jin},\ and\ \citenamefont {Mandrus}}]{Wang2004}%
  \BibitemOpen
  \bibfield  {author} {\bibinfo {author} {\bibnamefont {Wang}, \bibfnamefont
  {N.~L.}}, \bibinfo {author} {\bibfnamefont {P.}~\bibnamefont {Zheng}},
  \bibinfo {author} {\bibfnamefont {D.}~\bibnamefont {Wu}}, \bibinfo {author}
  {\bibfnamefont {Y.~C.}\ \bibnamefont {Ma}}, \bibinfo {author} {\bibfnamefont
  {T.}~\bibnamefont {Xiang}}, \bibinfo {author} {\bibfnamefont {R.~Y.}\
  \bibnamefont {Jin}}, \ and\ \bibinfo {author} {\bibfnamefont
  {D.}~\bibnamefont {Mandrus}}} (\bibinfo {year} {2004}),\ \href {\doibase
  10.1103/PhysRevLett.93.237007} {\bibfield  {journal} {\bibinfo  {journal}
  {Phys. Rev. Lett.}\ }\textbf {\bibinfo {volume} {93}},\ \bibinfo {pages}
  {237007}}\BibitemShut {NoStop}%
\bibitem [{\citenamefont {Wehling}\ \emph {et~al.}(2011)\citenamefont
  {Wehling}, \citenamefont {\ifmmode \mbox{\c{S}}\else \c{S}\fi{}a\ifmmode
  \mbox{\c{s}}\else \c{s}\fi{}\ifmmode \imath \else \i
  \fi{}o\ifmmode~\breve{g}\else \u{g}\fi{}lu}, \citenamefont {Friedrich},
  \citenamefont {Lichtenstein}, \citenamefont {Katsnelson},\ and\ \citenamefont
  {Bl\"ugel}}]{Wehling11}%
  \BibitemOpen
  \bibfield  {author} {\bibinfo {author} {\bibnamefont {Wehling}, \bibfnamefont
  {T.~O.}}, \bibinfo {author} {\bibfnamefont {E.}~\bibnamefont {\ifmmode
  \mbox{\c{S}}\else \c{S}\fi{}a\ifmmode \mbox{\c{s}}\else \c{s}\fi{}\ifmmode
  \imath \else \i \fi{}o\ifmmode~\breve{g}\else \u{g}\fi{}lu}}, \bibinfo
  {author} {\bibfnamefont {C.}~\bibnamefont {Friedrich}}, \bibinfo {author}
  {\bibfnamefont {A.~I.}\ \bibnamefont {Lichtenstein}}, \bibinfo {author}
  {\bibfnamefont {M.~I.}\ \bibnamefont {Katsnelson}}, \ and\ \bibinfo {author}
  {\bibfnamefont {S.}~\bibnamefont {Bl\"ugel}}} (\bibinfo {year} {2011}),\
  \href {\doibase 10.1103/PhysRevLett.106.236805} {\bibfield  {journal}
  {\bibinfo  {journal} {Phys. Rev. Lett.}\ }\textbf {\bibinfo {volume} {106}},\
  \bibinfo {pages} {236805}}\BibitemShut {NoStop}%
\bibitem [{\citenamefont {Wentzell}\ \emph {et~al.}(2016)\citenamefont
  {Wentzell}, \citenamefont {Li}, \citenamefont {Tagliavini}, \citenamefont
  {Taranto}, \citenamefont {Rohringer}, \citenamefont {Held}, \citenamefont
  {Toschi},\ and\ \citenamefont {Andergassen}}]{Wentzell2016}%
  \BibitemOpen
  \bibfield  {author} {\bibinfo {author} {\bibnamefont {Wentzell},
  \bibfnamefont {N.}}, \bibinfo {author} {\bibfnamefont {G.}~\bibnamefont
  {Li}}, \bibinfo {author} {\bibfnamefont {A.}~\bibnamefont {Tagliavini}},
  \bibinfo {author} {\bibfnamefont {C.}~\bibnamefont {Taranto}}, \bibinfo
  {author} {\bibfnamefont {G.}~\bibnamefont {Rohringer}}, \bibinfo {author}
  {\bibfnamefont {K.}~\bibnamefont {Held}}, \bibinfo {author} {\bibfnamefont
  {A.}~\bibnamefont {Toschi}}, \ and\ \bibinfo {author} {\bibfnamefont
  {S.}~\bibnamefont {Andergassen}}} (\bibinfo {year} {2016}),\ \href@noop {} {\
  }\Eprint {http://arxiv.org/abs/1610.06520} {arXiv:1610.06520} \BibitemShut
  {NoStop}%
\bibitem [{\citenamefont {Wentzell}\ \emph {et~al.}(2015)\citenamefont
  {Wentzell}, \citenamefont {Taranto}, \citenamefont {Katanin}, \citenamefont
  {Toschi},\ and\ \citenamefont {Andergassen}}]{Wentzell2015}%
  \BibitemOpen
  \bibfield  {author} {\bibinfo {author} {\bibnamefont {Wentzell},
  \bibfnamefont {N.}}, \bibinfo {author} {\bibfnamefont {C.}~\bibnamefont
  {Taranto}}, \bibinfo {author} {\bibfnamefont {A.}~\bibnamefont {Katanin}},
  \bibinfo {author} {\bibfnamefont {A.}~\bibnamefont {Toschi}}, \ and\ \bibinfo
  {author} {\bibfnamefont {S.}~\bibnamefont {Andergassen}}} (\bibinfo {year}
  {2015}),\ \href {\doibase 10.1103/PhysRevB.91.045120} {\bibfield  {journal}
  {\bibinfo  {journal} {Phys. Rev. B}\ }\textbf {\bibinfo {volume} {91}},\
  \bibinfo {pages} {045120}}\BibitemShut {NoStop}%
\bibitem [{\citenamefont {Werner}(2013)}]{WernerPC13}%
  \BibitemOpen
  \bibfield  {author} {\bibinfo {author} {\bibnamefont {Werner}, \bibfnamefont
  {P.}}} (\bibinfo {year} {2013}),\ \href@noop {} {}\bibinfo {howpublished}
  {private communication}\BibitemShut {NoStop}%
\bibitem [{\citenamefont {Werner}\ \emph {et~al.}(2012)\citenamefont {Werner},
  \citenamefont {Casula}, \citenamefont {Miyake}, \citenamefont {Aryasetiawan},
  \citenamefont {Millis},\ and\ \citenamefont {Biermann}}]{Werner2012}%
  \BibitemOpen
  \bibfield  {author} {\bibinfo {author} {\bibnamefont {Werner}, \bibfnamefont
  {P.}}, \bibinfo {author} {\bibfnamefont {M.}~\bibnamefont {Casula}}, \bibinfo
  {author} {\bibfnamefont {T.}~\bibnamefont {Miyake}}, \bibinfo {author}
  {\bibfnamefont {F.}~\bibnamefont {Aryasetiawan}}, \bibinfo {author}
  {\bibfnamefont {A.~J.}\ \bibnamefont {Millis}}, \ and\ \bibinfo {author}
  {\bibfnamefont {S.}~\bibnamefont {Biermann}}} (\bibinfo {year} {2012}),\
  \href {\doibase 10.1038/nphys2250} {\bibfield  {journal} {\bibinfo  {journal}
  {Nat. Phys.}\ }\textbf {\bibinfo {volume} {8}},\ \bibinfo {pages}
  {331}}\BibitemShut {NoStop}%
\bibitem [{\citenamefont {Werner}\ \emph {et~al.}(2006)\citenamefont {Werner},
  \citenamefont {Comanac}, \citenamefont {de' Medici}, \citenamefont {Troyer},\
  and\ \citenamefont {Millis}}]{Werner2006}%
  \BibitemOpen
  \bibfield  {author} {\bibinfo {author} {\bibnamefont {Werner}, \bibfnamefont
  {P.}}, \bibinfo {author} {\bibfnamefont {A.}~\bibnamefont {Comanac}},
  \bibinfo {author} {\bibfnamefont {L.}~\bibnamefont {de' Medici}}, \bibinfo
  {author} {\bibfnamefont {M.}~\bibnamefont {Troyer}}, \ and\ \bibinfo {author}
  {\bibfnamefont {A.~J.}\ \bibnamefont {Millis}}} (\bibinfo {year} {2006}),\
  \href {\doibase 10.1103/PhysRevLett.97.076405} {\bibfield  {journal}
  {\bibinfo  {journal} {Phys. Rev. Lett.}\ }\textbf {\bibinfo {volume} {97}},\
  \bibinfo {pages} {076405}}\BibitemShut {NoStop}%
\bibitem [{\citenamefont {Werner}\ and\ \citenamefont
  {Millis}(2006)}]{Werner2006a}%
  \BibitemOpen
  \bibfield  {author} {\bibinfo {author} {\bibnamefont {Werner}, \bibfnamefont
  {P.}}, \ and\ \bibinfo {author} {\bibfnamefont {A.~J.}\ \bibnamefont
  {Millis}}} (\bibinfo {year} {2006}),\ \href {\doibase
  10.1103/PhysRevB.74.155107} {\bibfield  {journal} {\bibinfo  {journal} {Phys.
  Rev. B}\ }\textbf {\bibinfo {volume} {74}},\ \bibinfo {pages}
  {155107}}\BibitemShut {NoStop}%
\bibitem [{\citenamefont {Werner}\ and\ \citenamefont
  {Millis}(2007)}]{Werner07}%
  \BibitemOpen
  \bibfield  {author} {\bibinfo {author} {\bibnamefont {Werner}, \bibfnamefont
  {P.}}, \ and\ \bibinfo {author} {\bibfnamefont {A.~J.}\ \bibnamefont
  {Millis}}} (\bibinfo {year} {2007}),\ \href {\doibase
  10.1103/PhysRevLett.99.146404} {\bibfield  {journal} {\bibinfo  {journal}
  {Phys. Rev. Lett.}\ }\textbf {\bibinfo {volume} {99}},\ \bibinfo {pages}
  {146404}}\BibitemShut {NoStop}%
\bibitem [{\citenamefont {Werner}\ and\ \citenamefont
  {Millis}(2010)}]{Werner10}%
  \BibitemOpen
  \bibfield  {author} {\bibinfo {author} {\bibnamefont {Werner}, \bibfnamefont
  {P.}}, \ and\ \bibinfo {author} {\bibfnamefont {A.~J.}\ \bibnamefont
  {Millis}}} (\bibinfo {year} {2010}),\ \href {\doibase
  10.1103/PhysRevLett.104.146401} {\bibfield  {journal} {\bibinfo  {journal}
  {Phys. Rev. Lett.}\ }\textbf {\bibinfo {volume} {104}},\ \bibinfo {pages}
  {146401}}\BibitemShut {NoStop}%
\bibitem [{\citenamefont {Wilhelm}\ \emph {et~al.}(2015)\citenamefont
  {Wilhelm}, \citenamefont {Lechermann}, \citenamefont {Hafermann},
  \citenamefont {Katsnelson},\ and\ \citenamefont
  {Lichtenstein}}]{Wilhelm2015}%
  \BibitemOpen
  \bibfield  {author} {\bibinfo {author} {\bibnamefont {Wilhelm}, \bibfnamefont
  {A.}}, \bibinfo {author} {\bibfnamefont {F.}~\bibnamefont {Lechermann}},
  \bibinfo {author} {\bibfnamefont {H.}~\bibnamefont {Hafermann}}, \bibinfo
  {author} {\bibfnamefont {M.~I.}\ \bibnamefont {Katsnelson}}, \ and\ \bibinfo
  {author} {\bibfnamefont {A.~I.}\ \bibnamefont {Lichtenstein}}} (\bibinfo
  {year} {2015}),\ \href {\doibase 10.1103/PhysRevB.91.155114} {\bibfield
  {journal} {\bibinfo  {journal} {Phys. Rev. B}\ }\textbf {\bibinfo {volume}
  {91}},\ \bibinfo {pages} {155114}}\BibitemShut {NoStop}%
\bibitem [{\citenamefont {Wissgott}\ \emph {et~al.}(2011)\citenamefont
  {Wissgott}, \citenamefont {Toschi}, \citenamefont {Sangiovanni},\ and\
  \citenamefont {Held}}]{Wissgott2011}%
  \BibitemOpen
  \bibfield  {author} {\bibinfo {author} {\bibnamefont {Wissgott},
  \bibfnamefont {P.}}, \bibinfo {author} {\bibfnamefont {A.}~\bibnamefont
  {Toschi}}, \bibinfo {author} {\bibfnamefont {G.}~\bibnamefont {Sangiovanni}},
  \ and\ \bibinfo {author} {\bibfnamefont {K.}~\bibnamefont {Held}}} (\bibinfo
  {year} {2011}),\ \href {\doibase 10.1103/PhysRevB.84.085129} {\bibfield
  {journal} {\bibinfo  {journal} {Phys. Rev. B}\ }\textbf {\bibinfo {volume}
  {84}},\ \bibinfo {pages} {085129}}\BibitemShut {NoStop}%
\bibitem [{\citenamefont {Wissgott}\ \emph {et~al.}(2010)\citenamefont
  {Wissgott}, \citenamefont {Toschi}, \citenamefont {Usui}, \citenamefont
  {Kuroki},\ and\ \citenamefont {Held}}]{Wissgott2010}%
  \BibitemOpen
  \bibfield  {author} {\bibinfo {author} {\bibnamefont {Wissgott},
  \bibfnamefont {P.}}, \bibinfo {author} {\bibfnamefont {A.}~\bibnamefont
  {Toschi}}, \bibinfo {author} {\bibfnamefont {H.}~\bibnamefont {Usui}},
  \bibinfo {author} {\bibfnamefont {K.}~\bibnamefont {Kuroki}}, \ and\ \bibinfo
  {author} {\bibfnamefont {K.}~\bibnamefont {Held}}} (\bibinfo {year} {2010}),\
  \href {\doibase 10.1103/PhysRevB.82.201106} {\bibfield  {journal} {\bibinfo
  {journal} {Phys. Rev. B}\ }\textbf {\bibinfo {volume} {82}},\ \bibinfo
  {pages} {201106(R)}}\BibitemShut {NoStop}%
\bibitem [{\citenamefont {Yang}\ \emph {et~al.}(2010)\citenamefont {Yang},
  \citenamefont {L\"auchli}, \citenamefont {Mila},\ and\ \citenamefont
  {Schmidt}}]{Yang2010}%
  \BibitemOpen
  \bibfield  {author} {\bibinfo {author} {\bibnamefont {Yang}, \bibfnamefont
  {H.-Y.}}, \bibinfo {author} {\bibfnamefont {A.~M.}\ \bibnamefont
  {L\"auchli}}, \bibinfo {author} {\bibfnamefont {F.}~\bibnamefont {Mila}}, \
  and\ \bibinfo {author} {\bibfnamefont {K.~P.}\ \bibnamefont {Schmidt}}}
  (\bibinfo {year} {2010}),\ \href {\doibase 10.1103/PhysRevLett.105.267204}
  {\bibfield  {journal} {\bibinfo  {journal} {Phys. Rev. Lett.}\ }\textbf
  {\bibinfo {volume} {105}},\ \bibinfo {pages} {267204}}\BibitemShut {NoStop}%
\bibitem [{\citenamefont {Yang}\ \emph {et~al.}(2011)\citenamefont {Yang},
  \citenamefont {Fotso}, \citenamefont {Hafermann}, \citenamefont {Tam},
  \citenamefont {Moreno}, \citenamefont {Pruschke},\ and\ \citenamefont
  {Jarrell}}]{Yang2011a}%
  \BibitemOpen
  \bibfield  {author} {\bibinfo {author} {\bibnamefont {Yang}, \bibfnamefont
  {S.-X.}}, \bibinfo {author} {\bibfnamefont {H.}~\bibnamefont {Fotso}},
  \bibinfo {author} {\bibfnamefont {H.}~\bibnamefont {Hafermann}}, \bibinfo
  {author} {\bibfnamefont {K.-M.}\ \bibnamefont {Tam}}, \bibinfo {author}
  {\bibfnamefont {J.}~\bibnamefont {Moreno}}, \bibinfo {author} {\bibfnamefont
  {T.}~\bibnamefont {Pruschke}}, \ and\ \bibinfo {author} {\bibfnamefont
  {M.}~\bibnamefont {Jarrell}}} (\bibinfo {year} {2011}),\ \href {\doibase
  10.1103/PhysRevB.84.155106} {\bibfield  {journal} {\bibinfo  {journal} {Phys.
  Rev. B}\ }\textbf {\bibinfo {volume} {84}},\ \bibinfo {pages}
  {155106}}\BibitemShut {NoStop}%
\bibitem [{\citenamefont {Yang}\ \emph {et~al.}(2009)\citenamefont {Yang},
  \citenamefont {Fotso}, \citenamefont {Liu}, \citenamefont {Maier},
  \citenamefont {Tomko}, \citenamefont {D'Azevedo}, \citenamefont {Scalettar},
  \citenamefont {Pruschke},\ and\ \citenamefont {Jarrell}}]{Yang2009}%
  \BibitemOpen
  \bibfield  {author} {\bibinfo {author} {\bibnamefont {Yang}, \bibfnamefont
  {S.~X.}}, \bibinfo {author} {\bibfnamefont {H.}~\bibnamefont {Fotso}},
  \bibinfo {author} {\bibfnamefont {J.}~\bibnamefont {Liu}}, \bibinfo {author}
  {\bibfnamefont {T.~A.}\ \bibnamefont {Maier}}, \bibinfo {author}
  {\bibfnamefont {K.}~\bibnamefont {Tomko}}, \bibinfo {author} {\bibfnamefont
  {E.~F.}\ \bibnamefont {D'Azevedo}}, \bibinfo {author} {\bibfnamefont {R.~T.}\
  \bibnamefont {Scalettar}}, \bibinfo {author} {\bibfnamefont {T.}~\bibnamefont
  {Pruschke}}, \ and\ \bibinfo {author} {\bibfnamefont {M.}~\bibnamefont
  {Jarrell}}} (\bibinfo {year} {2009}),\ \href {\doibase
  10.1103/PhysRevE.80.046706} {\bibfield  {journal} {\bibinfo  {journal} {Phys.
  Rev. E}\ }\textbf {\bibinfo {volume} {80}},\ \bibinfo {pages}
  {046706}}\BibitemShut {NoStop}%
\bibitem [{\citenamefont {Yang}\ \emph {et~al.}(2014)\citenamefont {Yang},
  \citenamefont {Haase}, \citenamefont {Terletska}, \citenamefont {Meng},
  \citenamefont {Pruschke}, \citenamefont {Moreno},\ and\ \citenamefont
  {Jarrell}}]{Yang2014}%
  \BibitemOpen
  \bibfield  {author} {\bibinfo {author} {\bibnamefont {Yang}, \bibfnamefont
  {S.-X.}}, \bibinfo {author} {\bibfnamefont {P.}~\bibnamefont {Haase}},
  \bibinfo {author} {\bibfnamefont {H.}~\bibnamefont {Terletska}}, \bibinfo
  {author} {\bibfnamefont {Z.~Y.}\ \bibnamefont {Meng}}, \bibinfo {author}
  {\bibfnamefont {T.}~\bibnamefont {Pruschke}}, \bibinfo {author}
  {\bibfnamefont {J.}~\bibnamefont {Moreno}}, \ and\ \bibinfo {author}
  {\bibfnamefont {M.}~\bibnamefont {Jarrell}}} (\bibinfo {year} {2014}),\ \href
  {\doibase 10.1103/PhysRevB.89.195116} {\bibfield  {journal} {\bibinfo
  {journal} {Phys. Rev. B}\ }\textbf {\bibinfo {volume} {89}},\ \bibinfo
  {pages} {195116}}\BibitemShut {NoStop}%
\bibitem [{\citenamefont {Yang}\ \emph {et~al.}(2013)\citenamefont {Yang},
  \citenamefont {Terletska}, \citenamefont {Meng}, \citenamefont {Moreno},\
  and\ \citenamefont {Jarrell}}]{Yang2013}%
  \BibitemOpen
  \bibfield  {author} {\bibinfo {author} {\bibnamefont {Yang}, \bibfnamefont
  {S.-X.}}, \bibinfo {author} {\bibfnamefont {H.}~\bibnamefont {Terletska}},
  \bibinfo {author} {\bibfnamefont {Z.~Y.}\ \bibnamefont {Meng}}, \bibinfo
  {author} {\bibfnamefont {J.}~\bibnamefont {Moreno}}, \ and\ \bibinfo {author}
  {\bibfnamefont {M.}~\bibnamefont {Jarrell}}} (\bibinfo {year} {2013}),\ \href
  {\doibase 10.1103/PhysRevE.88.063306} {\bibfield  {journal} {\bibinfo
  {journal} {Phys. Rev. E}\ }\textbf {\bibinfo {volume} {88}},\ \bibinfo
  {pages} {063306}}\BibitemShut {NoStop}%
\bibitem [{\citenamefont {Yoshioka}\ \emph {et~al.}(2009)\citenamefont
  {Yoshioka}, \citenamefont {Koga},\ and\ \citenamefont
  {Kawakami}}]{Yoshioka2009}%
  \BibitemOpen
  \bibfield  {author} {\bibinfo {author} {\bibnamefont {Yoshioka},
  \bibfnamefont {T.}}, \bibinfo {author} {\bibfnamefont {A.}~\bibnamefont
  {Koga}}, \ and\ \bibinfo {author} {\bibfnamefont {N.}~\bibnamefont
  {Kawakami}}} (\bibinfo {year} {2009}),\ \href {\doibase
  10.1103/PhysRevLett.103.036401} {\bibfield  {journal} {\bibinfo  {journal}
  {Phys. Rev. Lett.}\ }\textbf {\bibinfo {volume} {103}},\ \bibinfo {pages}
  {036401}}\BibitemShut {NoStop}%
\bibitem [{\citenamefont {Yudin}\ \emph {et~al.}(2014)\citenamefont {Yudin},
  \citenamefont {Hirschmeier}, \citenamefont {Hafermann}, \citenamefont
  {Eriksson}, \citenamefont {Lichtenstein},\ and\ \citenamefont
  {Katsnelson}}]{Yudin2014}%
  \BibitemOpen
  \bibfield  {author} {\bibinfo {author} {\bibnamefont {Yudin}, \bibfnamefont
  {D.}}, \bibinfo {author} {\bibfnamefont {D.}~\bibnamefont {Hirschmeier}},
  \bibinfo {author} {\bibfnamefont {H.}~\bibnamefont {Hafermann}}, \bibinfo
  {author} {\bibfnamefont {O.}~\bibnamefont {Eriksson}}, \bibinfo {author}
  {\bibfnamefont {A.~I.}\ \bibnamefont {Lichtenstein}}, \ and\ \bibinfo
  {author} {\bibfnamefont {M.~I.}\ \bibnamefont {Katsnelson}}} (\bibinfo {year}
  {2014}),\ \href {\doibase 10.1103/PhysRevLett.112.070403} {\bibfield
  {journal} {\bibinfo  {journal} {Phys. Rev. Lett.}\ }\textbf {\bibinfo
  {volume} {112}},\ \bibinfo {pages} {070403}}\BibitemShut {NoStop}%
\bibitem [{\citenamefont {Zgid}\ and\ \citenamefont {Gull}(2017)}]{Zgid2017}%
  \BibitemOpen
  \bibfield  {author} {\bibinfo {author} {\bibnamefont {Zgid}, \bibfnamefont
  {D.}}, \ and\ \bibinfo {author} {\bibfnamefont {E.}~\bibnamefont {Gull}}}
  (\bibinfo {year} {2017}),\ \href
  {http://stacks.iop.org/1367-2630/19/i=2/a=023047} {\bibfield  {journal}
  {\bibinfo  {journal} {New Journal of Physics}\ }\textbf {\bibinfo {volume}
  {19}}~(\bibinfo {number} {2}),\ \bibinfo {pages} {023047}}\BibitemShut
  {NoStop}%
\bibitem [{\citenamefont {Zhang}\ and\ \citenamefont
  {Callaway}(1989)}]{Zhang89}%
  \BibitemOpen
  \bibfield  {author} {\bibinfo {author} {\bibnamefont {Zhang}, \bibfnamefont
  {Y.}}, \ and\ \bibinfo {author} {\bibfnamefont {J.}~\bibnamefont {Callaway}}}
  (\bibinfo {year} {1989}),\ \href {\doibase 10.1103/PhysRevB.39.9397}
  {\bibfield  {journal} {\bibinfo  {journal} {Phys. Rev. B}\ }\textbf {\bibinfo
  {volume} {39}},\ \bibinfo {pages} {9397}}\BibitemShut {NoStop}%
\bibitem [{\citenamefont {Zhu}\ \emph {et~al.}(2003)\citenamefont {Zhu},
  \citenamefont {Grempel},\ and\ \citenamefont {Si}}]{Zhu2003}%
  \BibitemOpen
  \bibfield  {author} {\bibinfo {author} {\bibnamefont {Zhu}, \bibfnamefont
  {J.-X.}}, \bibinfo {author} {\bibfnamefont {D.~R.}\ \bibnamefont {Grempel}},
  \ and\ \bibinfo {author} {\bibfnamefont {Q.}~\bibnamefont {Si}}} (\bibinfo
  {year} {2003}),\ \href {\doibase 10.1103/PhysRevLett.91.156404} {\bibfield
  {journal} {\bibinfo  {journal} {Phys. Rev. Lett.}\ }\textbf {\bibinfo
  {volume} {91}},\ \bibinfo {pages} {156404}}\BibitemShut {NoStop}%
\bibitem [{\citenamefont {{\v{Z}}onda}\ \emph {et~al.}(2009)\citenamefont
  {{\v{Z}}onda}, \citenamefont {Farka{\v{s}}ovsk{\'{y}}},\ and\ \citenamefont
  {{\v{C}}en{\v{c}}arikov{\'{a}}}}]{Zonda2009}%
  \BibitemOpen
  \bibfield  {author} {\bibinfo {author} {\bibnamefont {{\v{Z}}onda},
  \bibfnamefont {M.}}, \bibinfo {author} {\bibfnamefont {P.}~\bibnamefont
  {Farka{\v{s}}ovsk{\'{y}}}}, \ and\ \bibinfo {author} {\bibfnamefont
  {H.}~\bibnamefont {{\v{C}}en{\v{c}}arikov{\'{a}}}}} (\bibinfo {year}
  {2009}),\ \href
  {http://linkinghub.elsevier.com/retrieve/pii/S0038109809005328} {\bibfield
  {journal} {\bibinfo  {journal} {Solid State Commun.}\ }\textbf {\bibinfo
  {volume} {149}}~(\bibinfo {number} {45-46}),\ \bibinfo {pages}
  {1997}}\BibitemShut {NoStop}%
\end{thebibliography}%

\end{document}